Universidad Complutense de Madrid
Facultad de Ciencias Físicas

# Tesis Doctoral

## Discos Keplerianos y flujos en estrellas binarias post-AGB

## Keplerian disks and outflows in binary post-AGB stars

Memoria para optar al grado de Doctor
presentada por

**Iván Gallardo Cava**

Directores

**Valentín Bujarrabal Fernández**
**Javier Alcolea Jiménez**

UNIVERSIDAD
COMPLUTENSE
MADRID

**DECLARACIÓN DE AUTORÍA Y ORIGINALIDAD DE LA TESIS PRESENTADA PARA OBTENER EL TÍTULO DE DOCTOR**

D./Dña. Iván Gallardo Cava,
estudiante en el Programa de Doctorado Astrofísica,
de la Facultad de Ciencias Físicas de la Universidad Complutense de Madrid, como autor/a de la tesis presentada para la obtención del título de Doctor y titulada:

Keplerian disks and outflows in binary post-AGB stars

y dirigida por: Valentín Bujarrabal Fernández y Javier Alcolea Jiménez

**DECLARO QUE:**

La tesis es una obra original que no infringe los derechos de propiedad intelectual ni los derechos de propiedad industrial u otros, de acuerdo con el ordenamiento jurídico vigente, en particular, la Ley de Propiedad Intelectual (R.D. legislativo 1/1996, de 12 de abril, por el que se aprueba el texto refundido de la Ley de Propiedad Intelectual, modificado por la Ley 2/2019, de 1 de marzo, regularizando, aclarando y armonizando las disposiciones legales vigentes sobre la materia), en particular, las disposiciones referidas al derecho de cita.

Del mismo modo, asumo frente a la Universidad cualquier responsabilidad que pudiera derivarse de la autoría o falta de originalidad del contenido de la tesis presentada de conformidad con el ordenamiento jurídico vigente.

En Madrid, a 29 de octubre de 2022

Fdo.: GALLARDO CAVA, IVAN (FIRMA)

Firmado digitalmente por GALLARDO CAVA, IVAN (FIRMA)
Fecha: 2022.10.29 15:25:57 +02'00'

Esta DECLARACIÓN DE AUTORÍA Y ORIGINALIDAD debe ser insertada en la primera página de la tesis presentada para la obtención del título de Doctor.

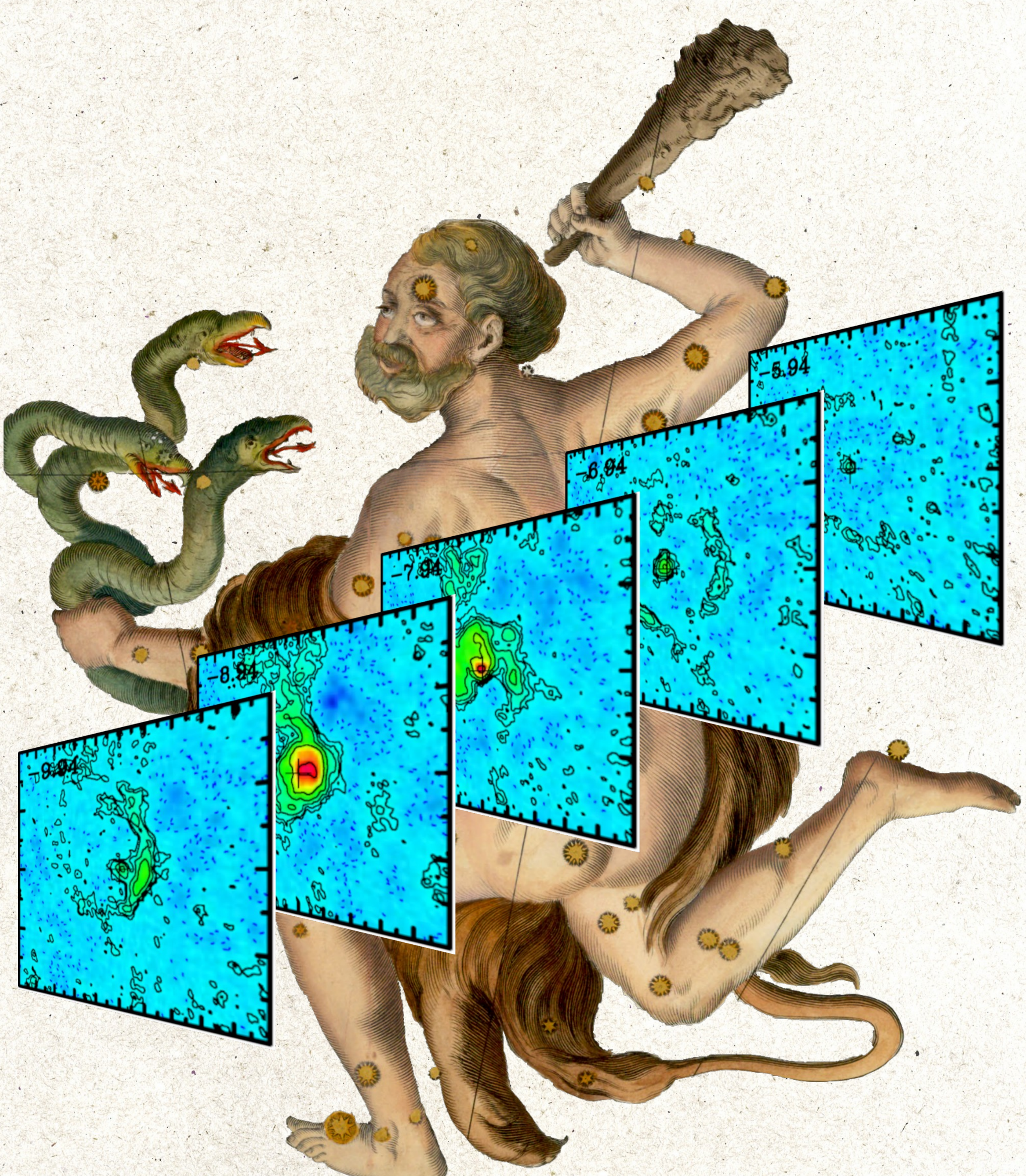

# KEPLERIAN DISKS AND OUTFLOWS IN BINARY POST-AGB STARS

IVÁN GALLARDO CAVA

# Keplerian disks and outflows in binary post-AGB stars

The spectacular evolution from AGB circumstellar envelopes to post-AGB nebulae takes place in a very short time ($\sim$1000 years). The accepted scenario to explain this evolution implies that material is accreted by a companion from a rotating disk, followed by the launching of fast jets. There is a class of binary post-AGB stars that systematically show evidence of the presence of disks. All of them present a remarkable near-infrared excess and the narrow CO line profiles characteristic of rotating disks. Their spectral energy distributions reveal the presence of hot dust close to the stellar system, and its disk-like shape has been confirmed by interferometric infrared data. These disks must be stable structures, because their infrared spectra reveal the presence of highly processed grains.

Before this work, the presence of both rotating disk and expanding gas had been confirmed in the Red Rectangle, IW Carinae, and IRAS 08544–4431. All of them account for $\sim$90% of the total nebular mass located in the rotating component, even if they have different disk mass values. The rest of the mass corresponds to bipolar low-velocity outflows. Most likely, these outflows are disk winds composed by gas and dust extracted from the rotating component. This thesis presents a comprehensive study at millimetre wavelengths. On the one hand, we perform a detailed kinetic study of these sources through NOEMA interferometric observations and complex models. On the other hand, we study the chemistry of these objects, thanks to our sensitive single-dish observations.

We present interferometric maps of CO for AC Herculis, 89 Herculis, IRAS 19125+0343, and R Scuti. Thanks to our new models and detailed analyses of our new maps of AC Her, we know that there is a weak and low-mass expanding component surrounding the Keplerian disk, whose mass represents $\leq$5% of the total nebular mass. Consequently, the disk of this source, as in the case of the previously studied sources, clearly dominates the whole nebula. Our observational data reveal that some sources present nebulae whose outflows are larger and more prominent compared to those previously studied. This is the case of R Sct, IRAS 19125+0343, and 89 Her. Our maps and models reveal that these sources show nebulae that are dominated by the expanding component, instead of the disk, which contains 65–75% of the total molecular mass. Based on their peculiar CO line profiles, there are three other sources that seem to be dominated by their outflow: IRAS 20056+1834, IRAS 18123+0511, and AI Cmi. Therefore, thanks to our results, we can differentiate two kinds of disk-containing nebulae around binary post-AGB stars: the disk- and the outow-dominated sources.

The chemistry of these objects has always been a conundrum. We present a deep survey of millimetre-wave molecular lines in nebulae around binary post-AGB stars. We performed our single-dish observations with the 30m IRAM and 40m Yebes telescopes, which has been the first systematic study of its kind. Thanks to our survey and analysis, we infer that the molecular content and molecular abundances are low in our sources, specially in those ones where the disk is the dominant component of the nebula. Additionally, we determine that the nebulae around AC Her, the Red Rectangle, AI CMi, IRAS 20056+1834, and HD 52961 are O-rich environments, while the nebula around 89 Her is C-rich.

The results of this thesis have been obtained through didifferent observation methods, but their union yields a comprehensive study of the molecular gas present in these sources. Hopefully, this Ph.D. thesis will become a reference for future studies of molecular gas in nebulae around binary post-AGB stars.

“ Con constancia y tenacidad se consigue lo que se desea;

la palabra imposible no tiene significado ”

*Para Belén, mi señora esposa.*

# Contents

*Se dará tiempo al tiempo, que suele dar dulce salida a muchas amargas dificultades.*

Miguel de Cervantes Saavedra – La Gitanilla

# Agradecimientos

Tenía 8 años cuando me regalaron el que sería mi primer libro científico: "1001 Secretos del espacio". Se trata de un libro infantil que sembró la semilla de mi interés por la astrofísica. Las incontables lecturas que realicé de ese libro me acercaban tanto a las estrellas como perfilaban mi vocación. Mi interés en el Universo hizo que siempre tuviera un claro interés por este campo y que durante toda mi formación académica persiguiera su estudio, motivo por el que estudié física y astrofísica en la Universidad.

Por circunstancias adversas, costó mucho encontrar un trabajo de doctorado y financiación para satisfacer mis deseos. Pero cuando encontré este doctorado mi alegría fue máxima. Como consecuencia de mi trabajo de investigación, presento esta tesis doctoral, que es el fruto de años de trabajo y dedicación y cuyo desarrollo ha sido muy duro. Sin embargo, puedo jurar que he disfrutado cada día que he dedicado a la astrofísica y que me siento muy orgulloso de mi contribución a la ciencia.

Agradezco profundamente haber podido cumplir el sueño de doctorarme a mis directores: Valentín Bujarrabal y Javier Alcolea. He tenido la suerte de doctorarme con dos titanes de la astronomía como vosotros. Gracias por haber confiado en mí para realizar este trabajo, gracias por haberme guiado desde el inicio de mi carrera científica y gracias por vuestra inifinita paciencia, cariño y dedicación para lograr este objetivo.

Quiero agradecer a Arancha el cariño con el que trata los mapas interferométricos, haciendo que la pérdida de flujo sea algo desconocido.

Quiero agradecer también a toda la familia que conforma el Observatorio Astronómico Nacional por lo que me ha aportado estos años, desde el guardia en su garita, hasta su director, pasando por todos sus miembros.

Gracias a la Universidad Complutense de Madrid, mi alma mater, donde estudié mi carrera en Física y mi máster en Astrofísica. Quiero agradecer especialmente a David Montes, con quien me inicié en el campo de la investigación, su atención y cariño.

Un doctorado no se puede hacer sin financiación, gracias al proyecto BES2017-080616 del Ministerio de Ciencia e Innovación.

Quiero dedicar esta tesis a Belén, mi mujer, por absolutamente todo. A mis padres, Sotero y Gema, mi papoides y mi mamoides, gracias por vuestro eterno e incondicional apoyo y sacrificio. A mi hermano Roberto, mi Enanoides, por entenderme y por tu fundamental apoyo. A Víctor Manuel, Ros y Víctor Enrique, por vuestro apoyo y vuestro amor, gracias por estar en mi vida. Gracias familia, gracias tutis.

A mi tío Julio, por actuar más como un hermano mayor, por tu cariño, por brindarme tu experiencia y por un Raymond Chang que valió una carrera en Física.

A Mini y Fobos, mis pequeños coautores. Vuestro amor y compañía han sido fundamentales para la consumación de esta tesis doctoral.

A mi amigo Miguel, gracias por tu ayuda y por ponerme en contacto de nuevo con la ciencia, gracias por todo. A mi cervantino y shakespiriano amigo Héctor, por aparecer en mi vida y por tu ayuda en el sprint final. A mi amigo Antonio, futuro doctor en química, gracias por tu sincero interés en mi trabajo. A mi amigo Dani, por ser mi amigo y haber escuchado durante tantas horas mis charlas sobre ciencia. A mis amigos Víctor y Guillermo, por no haber perdido nunca la confianza en mí.

Vuelvo de nuevo contigo, Belén, sin ti no hubiera llegado hasta aquí. Desde el primer examen de la carrera hasta la defensa de esta tesis doctoral, recorrer este largo y duro camino habría sido imposible sin ti a mi lado. Gracias por las alas que me has dado para llegar a las estrellas, esta tesis es para ti.

# Resumen

La evolución de las estrellas de masa intermedia ($0.8 < M_* < 8\,M_\odot$) en las últimas fases de su vida es muy rápida. En estas fases la estrella evoluciona desde la Rama Asintótica de las Gigantes (AGB, por sus siglas en inglés) sobre la fase de estrella post-AGB hacia la etapa de estrella central de una nebulosa planetaria (CSPN por sus siglas en inglés). Al final de la etapa AGB, la estrella muestra grandes tasas de pérdida de masa, hasta $10^{-4}\,M_\odot\,a^{-1}$. La estrella empieza la fase post-AGB después de haber expulsado la mayor parte de su manto. La nebulosa que hay alrededor de una estrella post-AGB es conocida como nebulosa pre-Planetaria (pPNe) y la mayoría de ellas muestra usualmente flujos bipolares de alta velocidad.

Nos centramos en una cierta clase de estrellas post-AGB que forman parte de un sistema binario. Estos sistemas binarios presentan periodos orbitales entre 100 y 3 000 días y una separación menor que 5 AU, esquivando (o atravesando rápidamente) la fase de envoltura común. Estas fuentes muestran indicios de la presencia de discos alrededor del sistema binario, como son su notable exceso en el infrarrojo cercano (NIR) presente en su distribución espectral de energía (SED). Este exceso a longitudes de onda infrarrojas es indicativo de la presencia de polvo caliente en las regiones más internas del disco. Los discos se forman debido a interacciones del sistema binario. Hay varios indicios de que se trata de estructuras estables, como es la detección de dinámica Kepleriana o la presencia de granos de polvo muy procesados.

Hay otra componente gaseosa de la nebulosa que rodea el disco y que presenta expansión y velocidad lenta. Antes de este trabajo, la presencia de discos rotantes y gas en expansión había sido confirmada en el Rectángulo Rojo, IW Carinae e IRAS 08544−4431. Todos ellos presentan $\sim 90\%$ de su masa nebular total en la componente rotante, a pesar de que los discos presentan diferentes valores para su masa. El resto de la masa corresponde con flujos bipolares de velocidad lenta. Estos flujos son probablemente vientos de disco, consistentes en gas y polvo, extraídos de la componente rotante.

Esta tesis presenta un estudio muy completo de este tipo de objetos en ondas milimétricas. Por una parte, hacemos un estudio muy detallado de la cinemática de estos objetos a través de observaciones interferométricas de NOEMA y complejos modelos. Por otra parte, estudiamos la química de estas fuentes gracias a nuestras observaciones de antena única.

Presentamos mapas de las líneas de emisión de $^{12}CO$ y $^{13}CO$ $J = 2 - 1$ para cuatro fuentes: AC Herculis, 89 Herculis, IRAS 19125+0343 y R Scuti. Gracias a nuestros nuevos modelos y análisis muy detallados de nuestros mapas de AC Her, sabemos que hay una componente débil en expansión y de masa muy pequeña rodeando el disco Kepleriano, que representa $\lesssim 5\%$ de la masa nebular total. Por lo tanto, el disco de esta fuente, como en el caso de las fuentes anteriormente estudiadas, claramente domina toda la nebulosa. Nuestros datos observacionales revelan que algunas de las fuentes presentan nebulosas que son más grandes y más prominentes comparadas con las neb-ulosas de los objetos ya estudiados. Éste es el caso de R Sct e IRAS 19125+0343. Nuestros mapas y nuestros modelos revelan que estas fuentes presentan nebulosas que están dominadas por la componente en expansión, en lugar de por el disco rotante, y que contienen $\sim$75% de la masa total molecular. Gracias a nuestros resultados, podemos diferenciar dos tipos de nebulosas que contienen discos alrededor de estrellas post-AGB binarias: aquellas dominadas por el disco rotante y aquellas dominadas por los flujos bipolares.

Nuestros mapas de 89 Her revelan la presencia de un disco sin resolver y de una componente en expansión con forma de reloj de arena. En un principio, esta nebulosa fue clasificada como fuente intermedia, ya que la masa del disco y la masa del viento eran aproximadamente iguales. Sin embargo, la masa del viento estaba subestimada porque estos mapas presentaban una pérdida de flujo significativa. Para resolver este problema, realizamos observaciones on-the-fly con el radiotelescopio milimétrico de 30 m de IRAM para detectar todo el flujo de esta fuente. Mezclamos estos mapa con los mapas de NOEMA que ya teníamos, para obtener mapas combinados de alta resolución que contuviesen todo el flujo de la fuente. Gracias a nuestros mapas combinados y el consiguiente nuevo modelo, ahora sabemos que la estructura con forma de reloj de arena es más grande y más masiva de lo que pensábamos, haciendo que 89 Her sea el último miembro en añadirse a la subclase de fuentes dominadas por el flujo. Hay otras tres fuentes que también parecen pertenecer a esta subclase de acuerdo con sus peculiares perfiles de CO: IRAS 20056+1834, IRAS 18123+0511 y AI CMi.

La química de este tipo de objetos era prácticamente desconocida. Presentamos la búsqueda de líneas espectrales más profunda llevada a cabo en ondas milimétricas en este tipo de estrellas post-AGB binarias. Estas observaciones de antena única se realizaron con los telescopios de 30 m del IRAM y 40 m del IGN, y se trata del primer estudio sistemático de este tipo. Gracias a nuestra profunda búsqueda y nuestro análisis, sabemos que el contenido molecular de estas fuentes, así como sus abundancias moleculares, es pequeño (excepto para CO, que muestra una abundancia normal), sobre todo en aquellas fuentes en las que el disco es la componente dominante de la nebulosa. Adicionalmente, podemos clasificar la química de alguno de nuestros objetos como rica en oxígeno o rica en carbono. De esta manera, las nebulosas alrededor de AC Her, el Rectángulo Rojo, AI CMi, IRAS 20056+1834, HD 52961 y R Sct son medios oxigenados, mientras que la nebulosa alrededor de 89 Her presenta una química carbonada.

Los resultados de esta tesis han sido obtenidos a través de distintos métodos de observación, pero la unión de ellos proporciona un estudio completo del gas molecular presente en estas fuentes. Esta tesis doctoral servirá como referencia para futuros estudios del gas molecular en nebulosas alrededor de estrellas binarias post-AGB.

# Abstract

The evolution of intermediate mass stars ($0.8 < M_* < 8\,M_\odot$) in their last stages is fast. In these phases the star evolves from the asymptotic giant branch (AGB) over the post-AGB star phase towards the central star of planetary nebula (CSPN) phase. At the end of the AGB phase, the star exhibits high mass-loss rates, up to $10^{-4}\,M_\odot\,a^{-1}$. The star begins the post-AGB stage after having expelled most of its stellar envelope. The nebula around a post-AGB star is known as a pre-planetary nebula (pPN), and most of them typically exhibit fast bipolar outflows.

We focus on a certain class of post-AGB stars that are part of a binary system. These binary systems present orbital periods between 100 and 3 000 days and separations smaller than 5 AU, avoiding (or rapidly going through) the common envelope phase. Their remarkable near-infrared excess in the spectral energy distribution (SED) is an observational probe for the presence of disks around the binary system. This excess at IR wavelengths is indicative of the presence of hot dust produced by the inner regions of these disks. They are formed during the binary interaction. They are stable structures and there are several evidences that prove it, such as the detection of Keplerian dynamics or the presence of highly processed dust grains.

There is also another component that surrounds the disk and presents expansion and slow velocity. Before this work, the presence of both rotating disk and expanding gas had been confirmed in the Red Rectangle, IW Carinae, and IRAS 08544−4431. All of them account for $\sim$90% of the total nebular mass located in the rotating component, even if they have different disk mass values. The rest of the mass corresponds to bipolar low-velocity outflows. Most likely, these outflows are disk winds composed by gas and dust extracted from the rotating component.

This thesis presents a comprehensive study at millimetre wavelengths. On the one hand, we perform a detailed kinetic study of these objects through NOEMA interferometric observations and complex models. On the other hand, we study the chemistry of these sources, thanks to our sensitive single-dish observations.

We present maps of $^{12}$CO and $^{13}$CO $J = 2-1$ emission lines for four objects: AC Herculis, 89 Herculis, IRAS 19125+0343, and R Scuti. Thanks to our new models and very detailed analyses of our new maps of AC Her, we know that there is a weak and very low-mass expanding component surrounding the Keplerian disk, whose mass represents $\lesssim$5% of the total nebular mass. Consequently, the disk of this source, as in the case of the previously studied sources, clearly dominates the whole nebula. Our observational data reveal that some sources might present nebulae whose outflows are larger and more prominent compared to those previously studied. This is the case of R Sct and IRAS 19125+0343. Our maps and models reveal that these sources show nebulae that are dominated by the expanding component, instead of the rotating disk, which contains $\sim$75% of the total molecular mass. Thanks to our results, we can differentiate two kinds of disk-containing nebulae around binary post-AGB stars: the

disk- and the outflow-dominated sources.

Our maps of 89 Her reveal the presence of an unresolved disk and an hourglass-like outflowing component. At first, this nebula was classified as an intermediate source, because the mass of the disk and the mass of the outflow were similar. Nevertheless, the mass of the outflow was underestimated, because these maps presented a significant amount of flux loss. To solve this problem, we perform on-the-fly observations with the 30 m IRAM telescope to detect all the flux of this source. We merge these 30 m IRAM maps with our previous NOEMA maps to obtain high resolution maps containing all the flux of the source. Thanks to our combined maps and and our new top-motch model, we infer that the hourglass-like structure is larger and more massive than we previously thought, making 89 Her the last member of the outflow-dominated subclass. There are three other sources that seem to be outflow-dominated based on their peculiar CO line profiles: IRAS 20056+1834, IRAS 18123+0511, and AI CMi.

The chemistry of these objects has always been a conundrum. We present a deep survey of millimetre-wave molecular lines in nebulae around binary post-AGB stars. We performed our single-dish observations with the 30 m IRAM and 40 m IGN telescopes, which has been the first systematic study of its kind Thanks to our survey and analysis, we infer that the molecular content and molecular abundances are low in our sources (except for CO, which presents a relatively normal abundance), specially in those ones where the disk is the dominant component of the nebula. Additionally, we classify the chemistry of some of our sources as O- or C-rich. In this way, we determine that the nebulae around AC Her, the Red Rectangle, AI CMi, IRAS 20056+1834, HD 52961, and R Sct are O-rich environments, while the nebula around 89 Her presents a C-rich chemistry.

The results of this thesis have been obtained through different observational methods, but their union yields a comprehensive study of the molecular gas present in these sources. Hopefully, this Ph.D. thesis will become a reference for future studies of molecular gas in nebulae around binary post-AGB stars.

# List of acronyms

| | |
|---|---|
| **AGB** | Asymptotic Giant Branch |
| **AGB-CSE** | Circumstellar Envelope around Asymptotic Giant Branch |
| **ALMA** | Atacama Large Millimeter/submillimeter Array |
| **BWFN** | Full Beam between First Nulls |
| **CLASS** | Continuum and Line Analysis Single-dish Software |
| **CLIC** | Continuum and Line Interferometer Calibration |
| **CSE** | Circumstellar Envelope |
| **CSPN** | Central Star of Planetary Nebula |
| **e-PMS** | early-PMS |
| **FIR** | Far-Infrared |
| **FUV** | Far-Ultraviolet |
| **GILDAS** | Grenoble Image and Line Data Software |
| **GREG** | Grenoble Graphic |
| **HB** | Horizontal Branch |
| **HR** | Hertzsprung-Russell |
| **HPBW** | Half Power Beam Width |
| **IR** | Infrared |
| **IRAM** | Institut de Radioastronomie Millimétrique |
| **ISM** | Interstellar Medium |
| **LTE** | Local Thermodynamic Equilibrium |
| **MAPPING** | A GILDAS Software |
| **MB** | Main Beam |
| **MS** | Main Sequence |
| **NIR** | Near-Infrared |
| **NOEMA** | NOrthern Extended Millimeter Array |
| **OTF** | On-The-Fly |
| **PDR** | Photodissociation Region |
| **PMS** | Pre-Main Sequence |
| **Post-AGB** | Post Asymptotic Giant Branch |
| **PPD** | Protoplanetary Disk |
| **pPN** | pre-Planetary Nebula |
| **PN** | Planetary Nebula |
| **PSF** | Point-Spread Function |
| **RGB** | Red Giant Branch |
| **SED** | Spectral Energy Distribution |
| **TP-AGB** | Thermally Pulsing Asymptotic Giant Branch |
| **UV** | Ultraviolet |

# List of Figures

# List of Tables

# Part I

# Introduction

*I find your lack of faith disturbing.*
Darth Vader – Star Wars: Episode IV

# 1

# The evolution of stars and their circumstellar envelopes

*In this chapter we summarize the evolution of low- and intermediate-mass stars and their envelopes, from the onset of the main sequence until they become post-AGB stars surrounded by a pre-planetary nebula. We focus on the physical and chemical properties of the circumstellar envelopes around AGB and post-AGB stars. The final part of this chapter is devoted to the particular class of sources that have been studied in this thesis: binary post-AGB stars and their envelopes.*

## 1.1 Intermediate-mass stars

### 1.1.1 The long way to the AGB phase

It is considered that an astronomical source is a star when it first appears in the stellar birthplace, a position in the Hertzsprung-Russell (HR) diagram where the young star becomes optically visible because it has a sufficiently high surface temperature. This is the earliest phase of stellar evolution, the so-called pre-main-sequence (PMS) phase (see Fig. 1.1). PMS stars first move vertically in the HR diagram along the Hayashi tracks during $10-15$ Ma. The stellar luminosity decreases due to contractions of the PMS star by gravitational collapse (no nuclear reactions). This stage is also known as the early-PMS (e-PMS) phase. After that, the PMS stars move horizontally to the left along the Henyey tracks, when the surface temperature rapidly increases and the luminosity is slightly enhanced, until they finally reach the main sequence (MS). The position and shape of these tracks strongly depend on the initial stellar mass. Stars with masses between $3-10\,M_\odot$ are directly born onto the MS without PMS evolution.

Once the core temperature and density are suitable, the core generates thermal energy through nuclear reactions that transform hydrogen into helium. This produces a radiation pressure that opposes the gravitational pull, preventing the star from collapsing. The star is now in the MS stage of the HR diagram (see Fig. 1.2). Main-sequence lifetimes depend on the stellar mass. Intermediate mass stars ($0.8 < M_* < 8\,M_\odot$) spend most of their lifetimes ($\sim 90\%$) on the MS, as opposed to high-mass stars, because they need a smaller amount of fuel to maintain the balance between pressure and gravity.

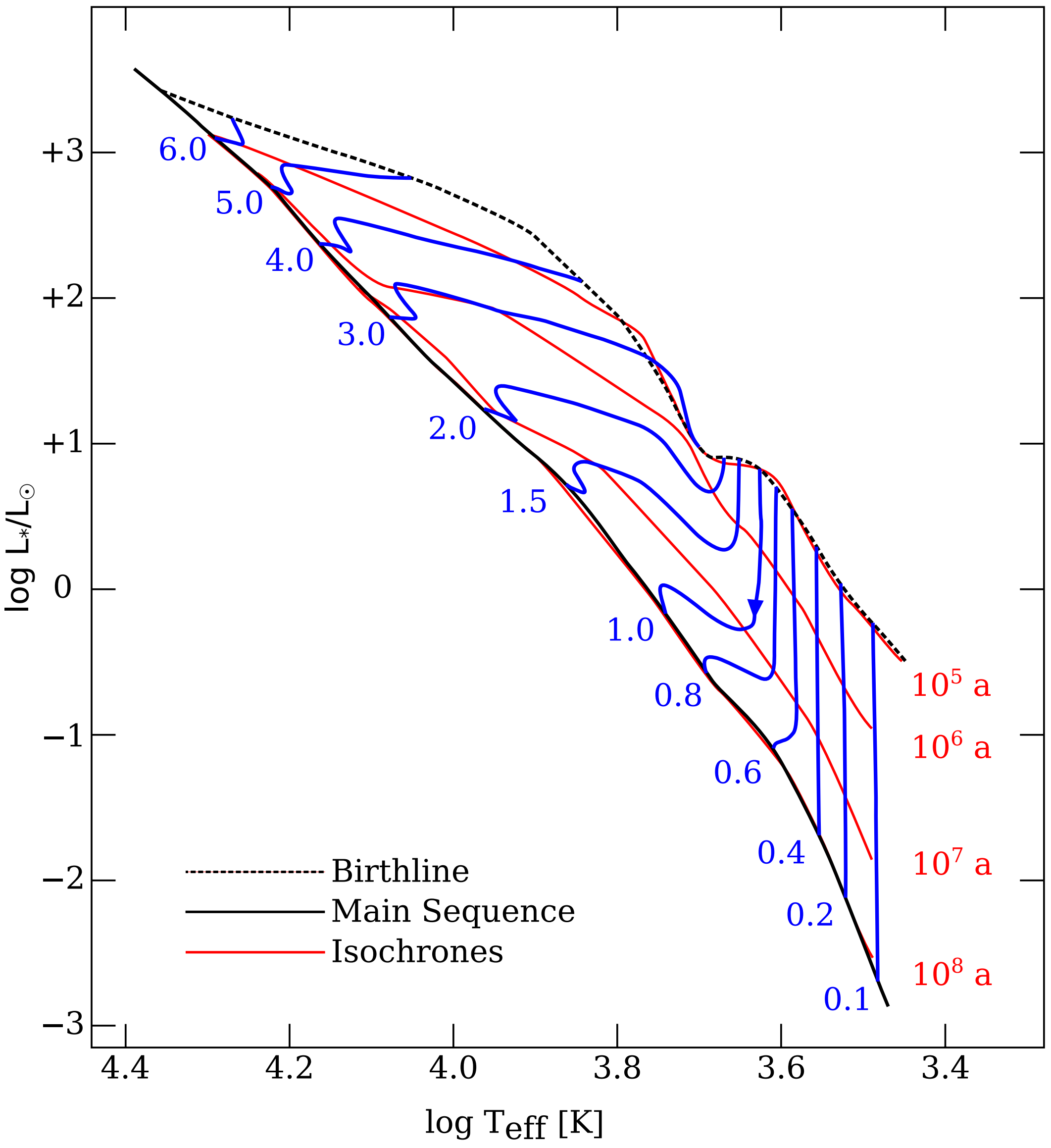

Figure 1.1: Pre-main-sequence evolutionary tracks. Each track is labeled by its stellar mass, expressed in solar units ($M_\odot$). The red curves are isochrones, expressed in years (a). Figure adapted from Stahler and Palla (2004).

During the MS phase, helium begins to concentrate in the stellar core where it starts to compress, becoming a degenerate gas. As a consequence of the compression of the core, the outer layers expand. As a consequence of the core contraction, the outer layers expand (this is the so-called mirror principle), which implies an increase in the stellar luminosity and a decrease in the effective temperature.

During the MS, H burning proceeds in increasingly outer layers, while the He degenerate core gains mass while gradually contracts and increases its temperature.

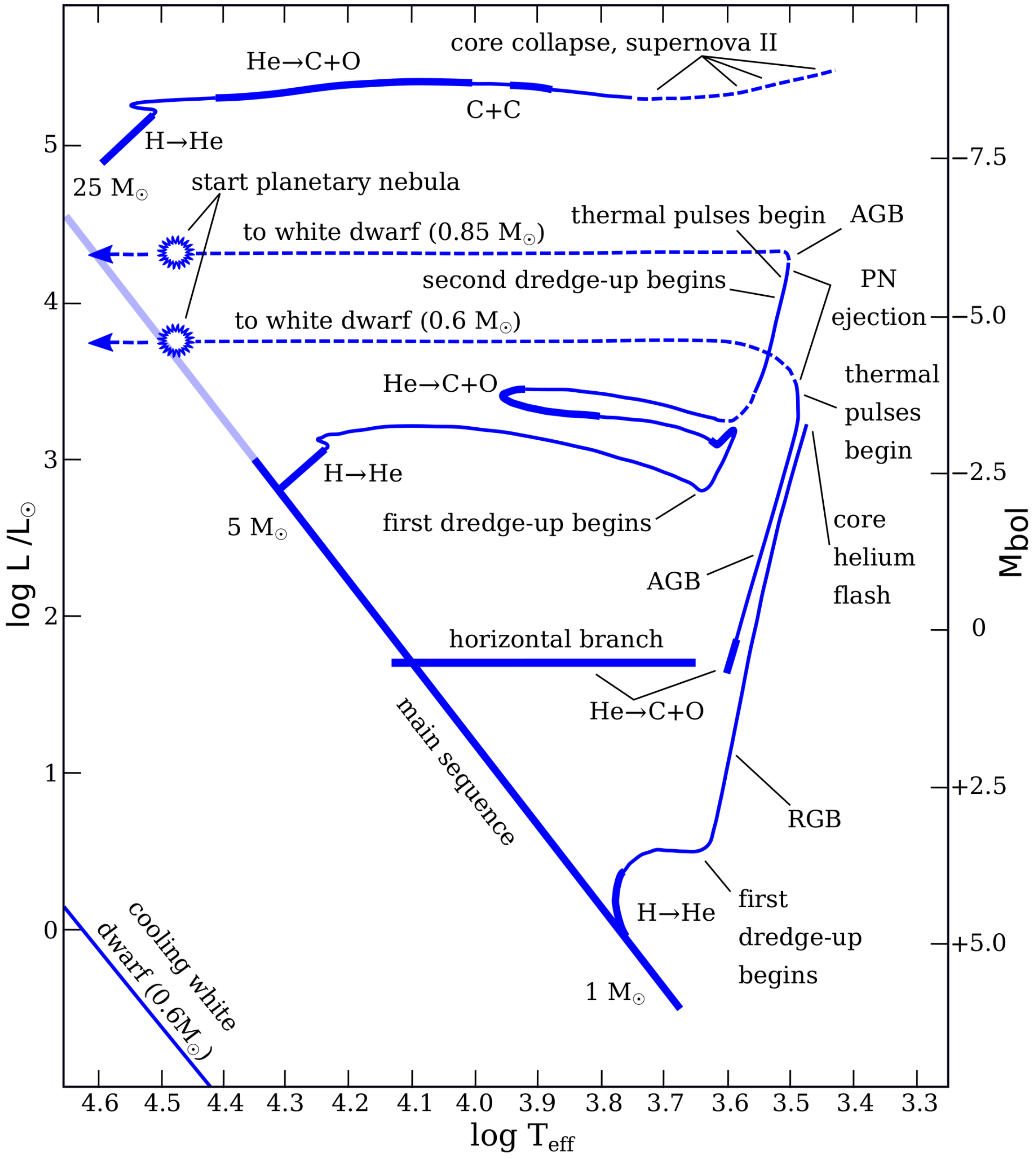

Figure 1.2: HR diagram showing evolutionary tracks for stellar models of various initial masses: 1, 5, and 25 $M_\odot$. The stellar luminosity is expressed in units of solar luminosity ($L_\odot$) and the effective temperature ($T_e$) in K. Figure adapted from Iben (1985).

Because of the increasing opacity of the outermost non-burning layers, the energy transport is mainly convective, which is effective during this stage. The convective layer extends towards the stellar core and it succeeds at transporting material throughout the different stellar layers. These phenomena are also known as dredge-ups.

Dredge-ups are essential to understand the chemical composition of stellar surfaces and the circumstellar envelopes that develop in later evolutionary stages. The star leaves the main sequence when when H has been consumed from the stellar core. At this point, the expansion of the outermost layers is accelerated. This significant increase in the stellar size implies a decrease in the surface temperature and an increase in the luminosity, depending on the stellar mass. The star is now ascending along the red giant branch (RGB).

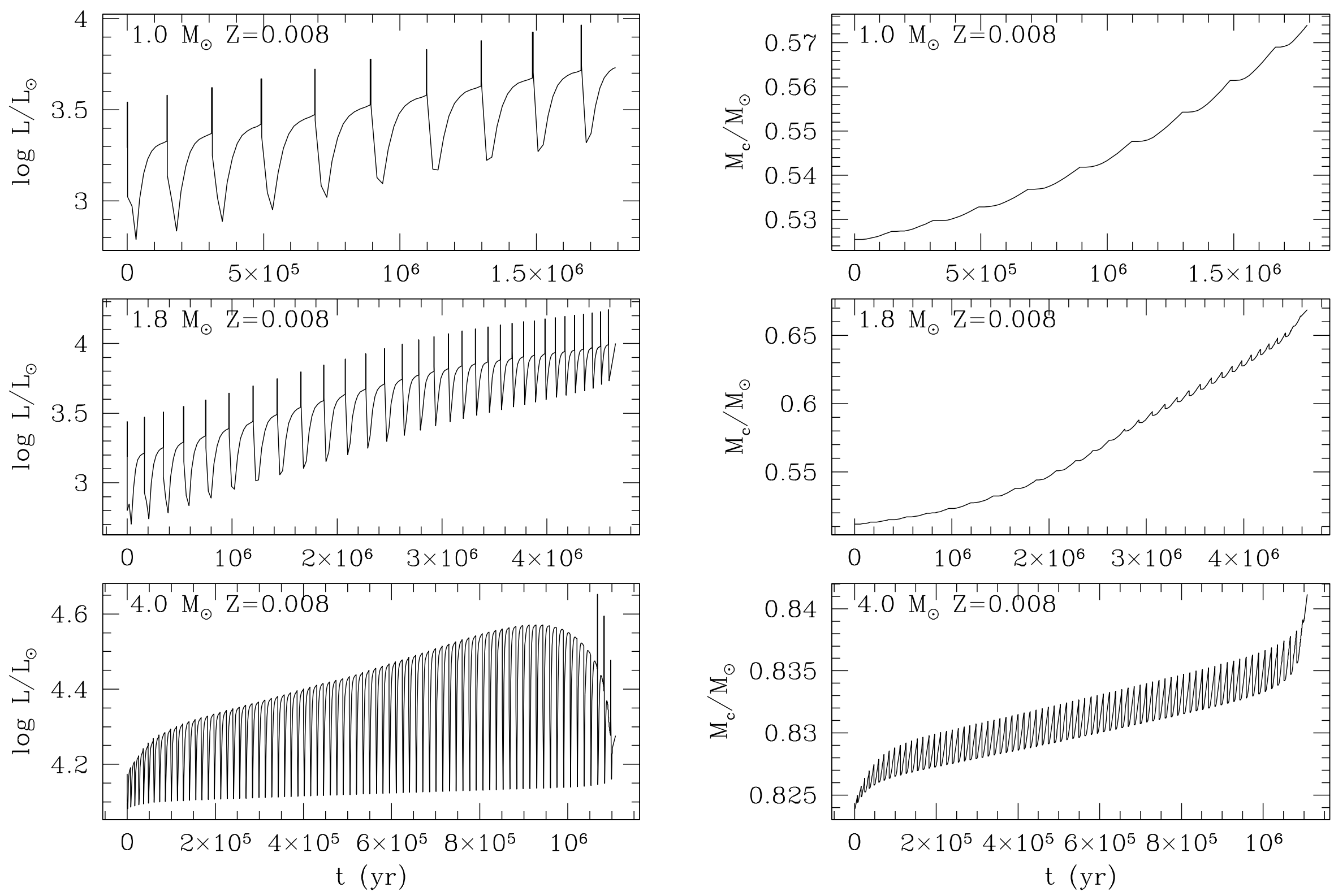

Figure 1.3: Evolution of the stellar luminosity (*left* panels) and core mass (*right* panels) along the TP-AGB stage of models with different initial masses. Figures taken from Marigo and Girardi (2007).

When the temperature is high enough to start burning helium ($\sim 10^8$ K) in stars with masses $\lesssim 2.25\,M_\odot$, the core is no longer degenerate (the thermal pressure overcomes the degeneracy pressure) and the nucleus expands in a quasi-explosive manner (helium flash). Nuclear reactions in this phase convert helium nuclei into carbon and oxygen. The star contracts and it enters in a phase of stability, named horizontal branch (HB), where the energy of the star comes from He burning in the core and from H burning in the outer layers. In the case of stars with masses $\gtrsim 2.25\,M_\odot$, their core never reaches a degenerate state, so the helium flash does not take place; helium burning, though, occurs in a less abrupt way. During the evolution along the RGB phase, the products of stellar nucleosynthesis derived from the CNO cycle, such as He and N, appear on the stellar surface because of convective mixing that transfers material to the external layers (first dredge-up).

Once helium has been exhausted in the stellar core, carbon and oxygen are the main nuclear species in these innermost regions. Given that the star has no more fuel to burn, the lack of radiation pressure causes the core to contract again, resulting in a new expansion of the outer layers (following the mirror principle). The inert core is surrounded by a He-rich layer and a yet outer H-rich layer that ignite alternatively. The red giant is now in the asymptotic giant branch (AGB). In stars with masses $\sim 4\,M_\odot$, H burning stops and the convective layer extends again towards the center of the star, where He is being consumed (second dredge-up).

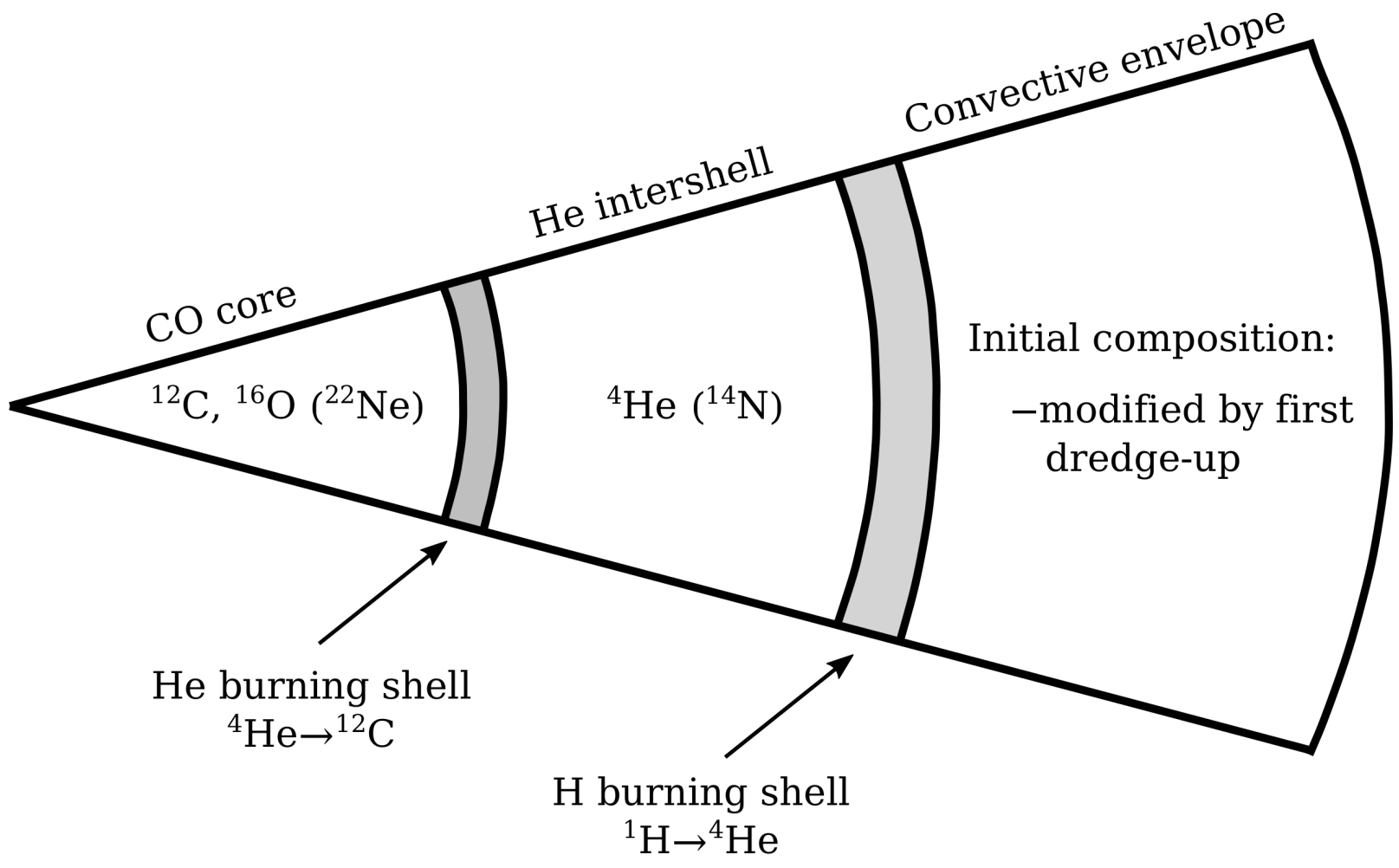

Figure 1.4: Structure and main composition of a star in the AGB phase. Figure adapted from Lattanzio (2003).

He burning in the new expanding layers halts H burning in the outermost layers. This He burning regime is unstable because it expands and cools down the outermost layers; consequently, He burning stops, and H starts to burn again in these layers. H burning yields He, which marks the onset of a new He burning regime. This phase of stellar evolution is known as the thermally pulsing asymptotic giant branch (TP-AGB) and these burning processes, with periods of more than 1 000 years, produce thermal pulses (see Fig. 1.3). Due to convective mixing, the elements resulting from He burning are transported to the outer layers (third dredge-up).

Additionally, there is another kind of pulsation that takes place when the opacity increases with temperature for AGB stars. The period of this type of pulsation is $300 - 600$ days. This is why these stars are known as Long Period Variables (LPV). This pulsation occurs as a consequence of H (and He) ionization, being sensitive to the temperature. Small increases in the temperature, caused by the gas compression, will induce significant increases in the density of free electrons. This density effect wins over the temperature effect and causes recombination, which increases the opacity, causing these layers to expand and cool down. This process is known as the kappa-mechanism, which causes instability in the stellar atmosphere and, in turn, stellar pulsations with rising/infalling velocities of tens of $\mathrm{km\,s^{-1}}$, comparable to the escape velocity. This also results into changes in the stellar radius of about 50%, and thus, into significant changes in the effective temperature and in the optical magnitude. Note that, for effective temperatures typical of AGB stars, in optical wavelengths, the stellar magnitude falls in the Wien regime of the spectrum.

The effective temperature of an AGB star lies between $\sim$2 800 and 3 600 K. An AGB star has a layered structure composed of a carbon/oxygen (C/O) core, a slowly expanding outer layer that is stratified into a layer of burning He, a layer of burning H that yields He, and an extended H layer in the outermost regions (see Fig. 1.4).

The pulsation of AGB stars is an oscillation around a hydrodynamic equilibrium state. The pulsation period is not the same for all the layers because these pulsations affect them differently at increasing distances from the stellar center. The outermost layers suffer less gravitational attraction than the inner layers, so they present longer pulsation periods. This generates shock waves that propagate through the stellar atmo-sphere, which lift the outer layers to create a relatively cool layer of "levitating" matter around the star. In these layers, there is an effective formation of simple molecules and dust grains that are dynamically coupled with the gas. We find three kinds of environments depending on the C/O ratio: C-type stars with $C/O > 1$, M-type stars with $C/O < 1$, and S-type stars with $C/O \approx 1$. The composition of these dust grains is determined by the abundances of the stellar atmosphere. Amorphous carbon grains are found in C-rich envelopes, while silicate particles are found in O-rich environments.

Due to their opacity, the grains are accelerated and taken away from the star by radiation pressure, dragging the gas along. This marks the onset of a copious mass loss (with rates between $10^{-7} - 10^{-5}\,M_\odot\,a^{-1}$ and $10^{-4}\,M_\odot\,a^{-1}$), which tends to increase with time, and leads to the formation of a circumstellar envelope (CSE) around the star. In the case of a high mass loss rate, the CSE becomes so dense and opaque that the star is not visible anymore (at optical wavelengths). When the mass loss finishes, the AGB phase ends, leaving a C/O stellar degenerate core and a CSE around it.

In the case of massive stars ($M_* > 8\,M_\odot$), the central temperatures and densities are sufficiently high for the C/O core to burn, producing new elements in the star. These stars are initially powered by main sequence H burning, followed by He, C, O, and Si burning (resulting into iron-peak elements) while the star moves horizontally to the right in the HR diagram (see Fig. 1.2). The lifespan of each burning process is shorter than that of the previous one; while hydrogen burning lasts for millions of years, silicon burning takes place in only a month. When each fuel is depleted, the star contracts due to energy losses, as the efficiency of energy production is reduced in each successive burning phase. Every burning process leaves unburned ashes. Thus, the core gradually becomes layered in a similar way to an onion. This process takes place until the nuclear reactions create a core composed of $^{56}$Fe, where the binding energy per nucleon reaches a maximum and no more nuclear energy can be produced via fusion reactions. Eventually, the star collapses into its $^{56}$Fe core, which can no longer support itself, and the stellar demise takes place as a core-collapse supernova explosion.

### 1.1.2 Post-AGB stars and last phases

During the last stages of the AGB phase, the stellar surface temperature reaches 3 000 K, and the mass loss increases from $10^{-7} - 10^{-5}$ to $10^{-4}\,M_\odot\,a^{-1}$. All the material escaping from the star generates an expanding circumstellar envelope with velocities of $10 - 20\,\mathrm{km\,s^{-1}}$ (Castro-Carrizo et al., 2010). The copious ejection of stellar matter ends a few thousand years after the mass loss rate reaches its maximum. This marks the end of the AGB phase. The star evolves into the post-AGB stage, maintaining a constant luminosity $\sim 10^4\,L_\odot$, while its radius decreases and its surface temperature constantly increases (from 3 000 to 30 000 K, see Fig. 1.2). Due to the expansion of the circumstellar envelope and the low mass loss, the central shells of the circumstellar envelope contract, decreasing the opacity and making the post-AGB star visible in optical wavelengths.

The circumstellar envelope becomes a pre-planetary nebula (pPN), which is a short-lived stage during the rapid evolution of the star between the late AGB and the subsequent blue dwarf surrounded by a planetary nebula (PN). These objects show conspicuous departures from spherical symmetry, often-times with a symmetry axis that is developed on extremely short time scales, seemingly at the end of the AGB evolution, when the mass loss rate decreases abruptly (Balick and Frank, 2002). The rate at which a post-AGB star increases its temperature also affects the formation of the subsequent PN, and mainly depends on the initial stellar mass (Miller Bertolami, 2016). The aim of this thesis is to study a specific kind of post-AGB stars, which will be described in more detail in Chapter 1.4. The post-AGB stellar surface temperature continues to increase until the star becomes a blue dwarf (see e.g. Adams et al., 2005), with an effective temperature higher than 30 000 K and a radius $\sim 10^{11}$ cm. Due to the high temperature of the central star inside the planetary nebula (CSPN) and the ultraviolet radiation from the interstellar medium, the envelope around the central star is strongly ionized. Finally, the star cannot generate thermonuclear reactions and its temperature and luminosity decrease, while its radius remains constant. This is the last evolutionary stage of a (old) star: a white dwarf.

## 1.2 Circumstellar envelopes around AGB stars

In order to understand the physical and chemical properties of pPNe, it is fundamental to know and to comprehend the properties of their progenitors: the circumstellar envelopes around AGB stars (AGB-CSEs). We divide this topic into two parts: the physical conditions of the AGB-CSE, mainly the expansion and the thermodynamics of the envelope, and its chemistry.

### 1.2.1 Circumstellar envelope expansion

Stellar pulsations have important effects on the dynamics of the outermost layers of AGB stars (see Sect. 1.1). The escape velocity is relatively small ($\sim 20\,\mathrm{km\,s^{-1}}$) so, when the pulsation period is large, the velocity of the shallower material is close to the escape velocity. Pulsations in AGB stars create a layer of “levitating” matter around the star at a distance of several stellar radii, oscillating without escaping from the star (Hinkle et al., 1982). This is the first step of mass loss in AGB stars, which will appear as a result of radiation pressure acting onto dust grains, helped by pulsation shocks (see Höfner and Olofsson, 2018).

Dust grains condense under non-equilibrium conditions, with the condensation temperature acting as a threshold. The efficiency of the dust grain growth is determined by the predominant chemical abundances and densities. The high densities found in regions of the stellar photosphere with sufficiently low temperatures allow dust to condense. The rate of grain growth is comparable to the stellar pulsation (see e.g. Höfner and Olofsson, 2018). The temperature of the dust grains found in CSEs is set by two processes: heating, by absorption of stellar photons, and cooling, by thermal emission (instead of collisions with gas particles). There is a strong dependence of the dust grains properties with the surrounding temperature, where low temperatures ($\sim 1\,600 - 2\,000$ K) are crucial for the grains to condense (see e.g. Bladh et al., 2015). There are several theories about the first step of grain formation, known as nucleation. In the case of C-rich environments, the classical nucleation theory (see Feder et al., 1966; Katz, 1970; Salpeter, 1974) suggests the presence of amorphous carbon grains. In the case of O-rich environments, the nature of the first condensates is still debated, while wind models usually avoid the problem of nucleation by assuming the existence of tiny seed nuclei (see e.g. Gobrecht et al., 2016). They are supposed to nucleate in $Al_2O_3$ grains that are transparent but suffer radiation pressure by scattering (not by absorption), remaining below the condensation temperature (see e.g. Höfner et al., 2016).

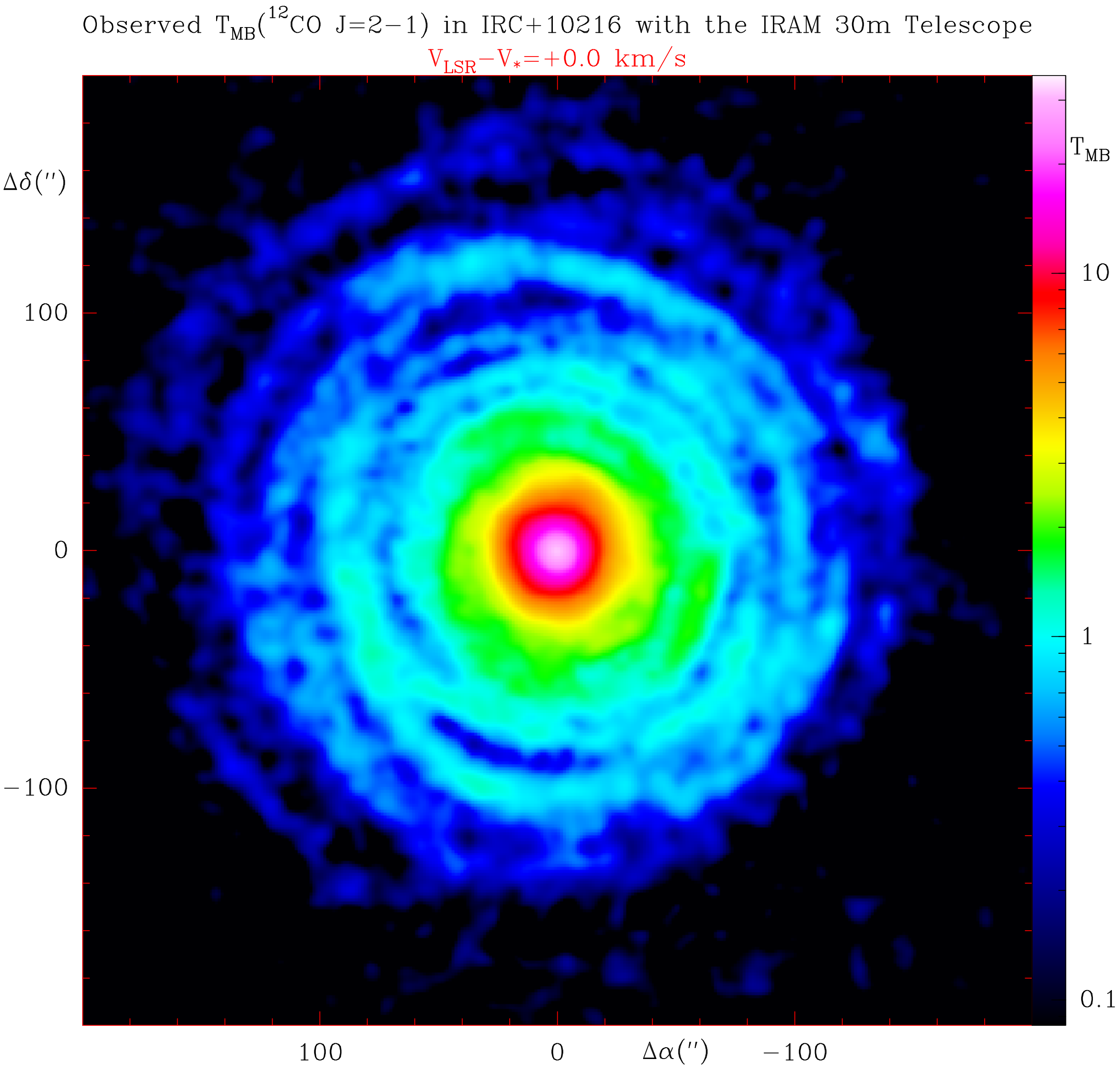

Figure 1.5: $^{12}$CO $J = 2-1$ line brightness temperature (in main beam scale) observed at the systemic velocity in IRC+10216. CSEs around AGB stars usually present a large-scale spherical shape. A spiral structure is also present because this source is a binary star with its orbital plane near to a face-on view. Figure taken from Cernicharo et al. (2015).

Dust grains are formed at $5-10$ stellar radius and transfer linear momentum to the gas in an effective manner, which implies the expansion of the envelope (Pijpers and Hearn, 1989; Pijpers and Habing, 1989). The equations for gas and dust expansion are (see Kwok, 1975; Goldreich and Scoville, 1976; Alcolea, 1993, for a complete description):

$$m_{\rm g} v_{\rm g} \frac{{\rm d}v_{\rm g}}{{\rm d}r} = -m_{\rm g} \frac{1}{\rho_{\rm g}} \frac{{\rm d}P_{\rm g}}{{\rm d}r} - m_{\rm g} \frac{GM}{r^2} + F_{\rm drag} \quad \text{for gas,} \tag{1.1}$$

$$m_{\rm d} v_{\rm d} \frac{{\rm d}v_{\rm d}}{{\rm d}r} = \frac{\bar{Q}\sigma_{\rm d} L}{4\pi r^2 c} - m_{\rm d} \frac{GM}{r^2} - F'_{\rm drag} \quad \text{for dust,} \tag{1.2}$$

where $G$, $c$, $M$, and $L$ are the universal gravitational constant, the speed of light in

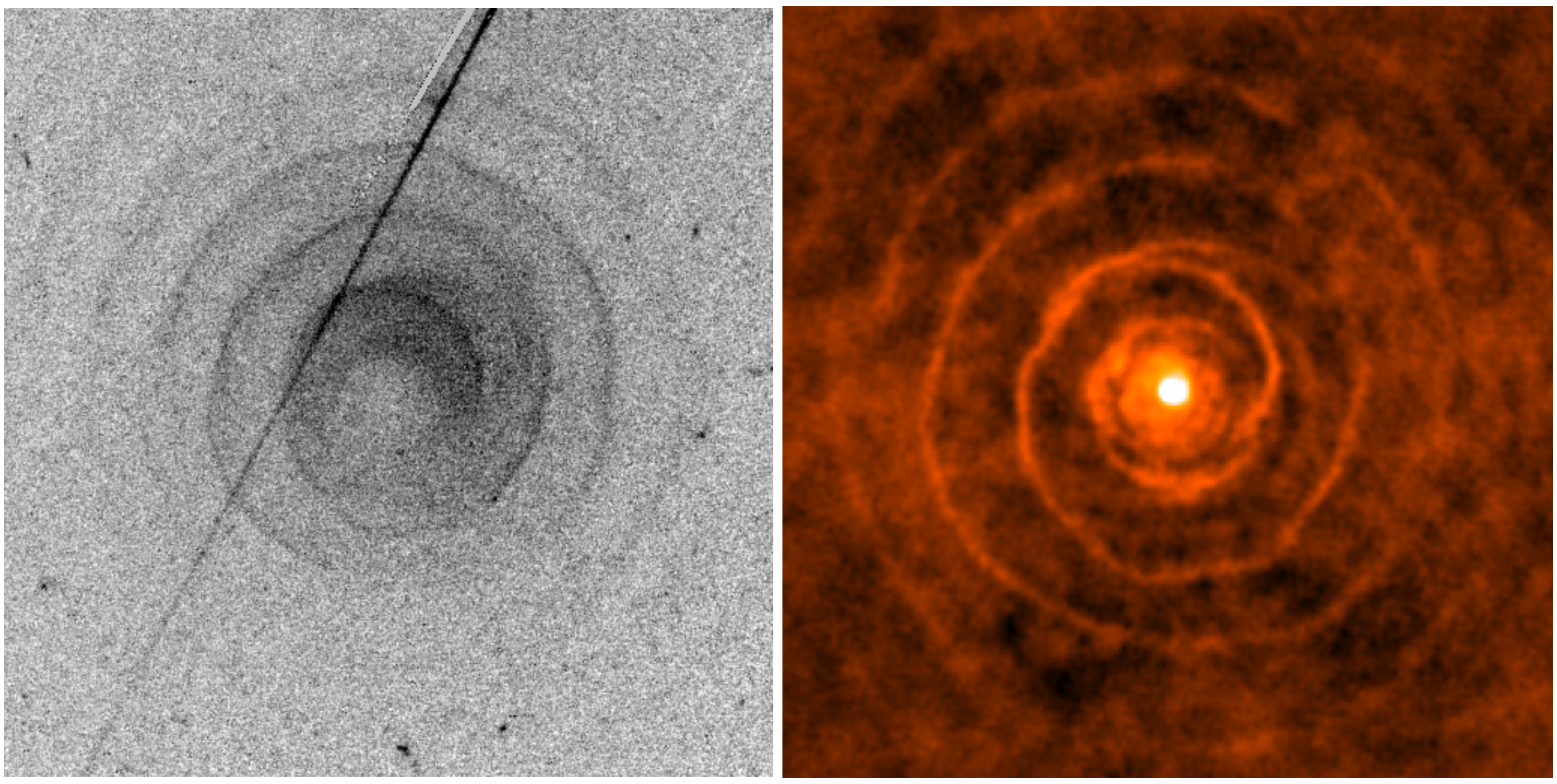

Figure 1.6: Images of LL Pegasi (AFGL 3068) observed with the *Hubble Space Telescope*, taken with the ACS instrument (*left*), and with ALMA in the $^{12}$CO $J = 2-1$ emission line (*right*). The pattern of the CSE serves as a probe of the central binary system. The spiral-shell closely resembles a true Archimedes spiral. *Left* figure taken from Mauron and Huggins (2006) and *right* figure adapted from Kim et al. (2017).

vacuum, the mass, and the luminosity of the star, respectively. $m_{\rm g}$, $v_{\rm g}$, and $\rho_{\rm g}$ are the mass, velocity, and density of the gas. Moreover, $m_{\rm d}$, $v_{\rm d}$, and $\sigma_{\rm d}$ are the mass, velocity, and the cross section of the dust grains, respectively. $\bar{Q}$ is the average efficiency that represents the linear momentum transferred from photons to dust grains. Additionally, $F_{\rm drag}$ and $F'_{\rm drag}$ are the drag forces produced by collision with gas molecules, which can be expressed as:

$$F_{\rm drag} \sim m_{\rm g}\sigma_{\rm d}n_{\rm d}v\sqrt{v^2 + c_{\rm s}^2}, \tag{1.3}$$

$$F'_{\rm drag} \sim m_{\rm g}\sigma_{\rm d}n_{\rm g}v\sqrt{v^2 + c_{\rm s}^2}, \tag{1.4}$$

where $v = v_{\rm d} - v_{\rm g}$, $c_{\rm s}$ are the drift velocity and the speed of sound, respectively. If we assume that the gas density, with respect to the dust density, does not depend on the distance:

$$\frac{\rm d}{{\rm d}r}\frac{n_{\rm d}}{n_{\rm g}} = 0 \rightarrow \frac{n_{\rm d}}{n_{\rm g}} = \text{constant}, \tag{1.5}$$

and taking into account the continuity equation:

$$\frac{\rm d}{{\rm d}r}\rho_{\rm g}v_{\rm g}r^2 = 0 \rightarrow \dot{M} = 4\pi\rho_{\rm g}v_{\rm g}r^2 = 4\pi n_{\rm g}m_{\rm g}v_{\rm g}r^2, \tag{1.6}$$

we can solve Eqs. 1.1 and 1.2. Thus, we can write the gas and the dust expansion velocities as:

$$v_{\rm g}(r) = \sqrt{v_\infty^2 - \frac{2}{r}\left(\frac{\bar{Q}L\sigma_{\rm d}n_{\rm g}}{4\pi m_{\rm g}n_{\rm g}c} - GM\right)}, \tag{1.7}$$

$$v_{\rm d}(r) = v_{\rm g}(r) + \sqrt{\frac{\bar{Q}Lv_{\rm g}}{\dot{M}c}}, \tag{1.8}$$

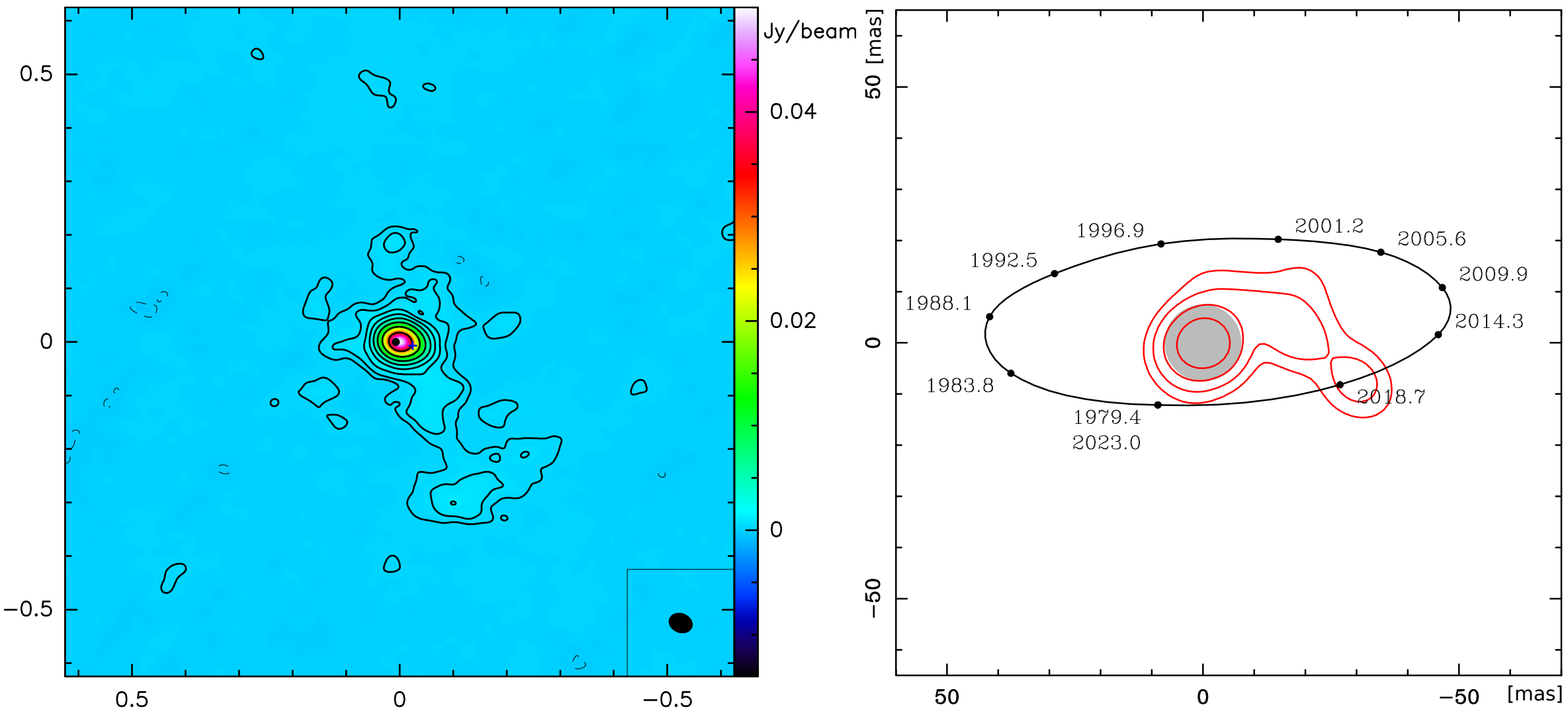

Figure 1.7: *Left*: ALMA map of the continuum at 0.9 mm in R Aqr. The mass loss material is located at the plane of the orbit. *Right*: Relative movement of both stars. Figures adapted from Bujarrabal et al. (2018a).

where $v_\infty$ represents the terminal expansion rate of the circumstellar envelope. The symmetry of the envelopes around AGB stars is, in principle, spherical (see Fig. 1.5), and the velocity field is radial and isotropic, with expansion velocities between 5 and 30 km s$^{-1}$ (see e.g. Neri et al., 1998).

This reasoning applies to the simplest case: a single star with a spherical envelope, and with a moderate rotation velocity. There is a high acceleration in the areas closest to the star and a velocity that asymptotically approaches the final velocity. Most recently, spiral features have been observed in CSEs (see Fig. 1.6). The observation of these structures can be explained by the presence of a binary (or multiple) system, whose stars are not far apart, and whose movement around the center of mass of the system produces these spirals in the envelope (Mauron and Huggins, 2006; Mauron et al., 2013; Kim et al., 2017). However, if the interaction in the binary system is stronger, the spherical symmetry is broken and the mass loss is located at the plane of the orbit (see e.g. Bujarrabal et al., 2018a, 2021). The objects in our analysis represent an even more extreme case, since they present stable equatorial structures with Keplerian rotation around the binary system: circumbinary disks. It is not strange that our objects (binary post-AGB stars) represent more extreme cases, because their orbital periods are smaller, 100 to 3 000 days, while in the case of R Aqr it is 42 years, see Fig. 1.7 (similar orbital periods are found in other AGBs).

There is a scenario where the presence of a companion could imply the development of a circumbinary disk around the stellar system. This is supported by the estimated excess in the angular momentum, because it is necessary to form a rotating disk from previously ejected material that is expelled with an insignificant amount of angular moment. This is the case of $L_2$ Pup, the second nearest AGB star (behind R Doradus), a semi-regular pulsating variable star and with a compact binary companion (see Kervella et al., 2016; Homan et al., 2017). The rotating structure is confirmed by its interferometric PV diagrams, which reveal the Keplerian and sub-Keplerian domains in the disk, see Fig. 1.8. This rotating structure around a binary AGB star could be a precursor of our objects of study.

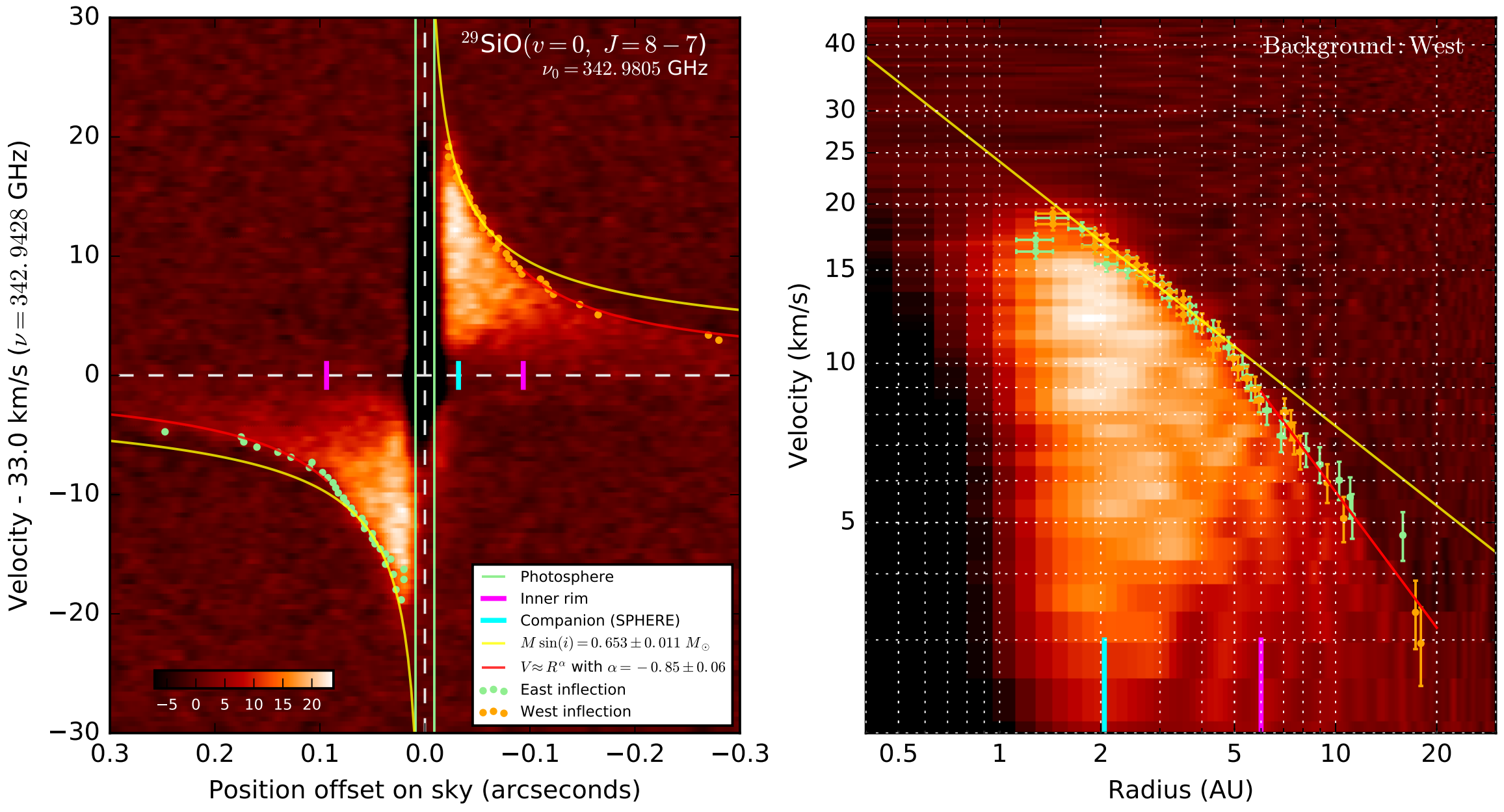

Figure 1.8: $L_2$ Pup PV diagram from ALMA maps of $^{29}$SiO $J = 8 - 7$ with Cartesian coordinate axes (*left*) and with logarithmic coordinates (right) that reveals the Keplerian domain of the disk rotation. The yellow curve corresponds to an inner cavity, while the red one represents a power law fit to the non-Keplerian regime. The color scale is in mJy beam$^{-1}$. Figure taken from Kervella et al. (2016).

### 1.2.2 Circumstellar envelope thermodynamics

In the case of a star with no companion and a spherical CSE, the gas temperature can be derived from the first law of thermodynamics:

$$dU = dQ + dW \rightarrow \dot{u} = \frac{P_g}{\rho_g^2}\dot{\rho}_g - \dot{q}, \tag{1.9}$$

where d$U$ denotes the change in the internal energy of a closed system, $Q$ is the amount of thermal energy supplied to the system, and $W$ is the amount of thermodynamic work done by the system on its surroundings. $P_g$ and $\rho_g$ are the gas pressure and density, respectively. The temperature of the circumstellar gas, $T_g$, can be derived from the energy balance at a macroscopic level between the heating (or absorbed heat, $H$) and the cooling (or produced heat, $C$) processes of the molecules that compose the gas (see e.g. Rowan-Robinson and Harris, 1982; Groenewegen, 1994; Justtanont et al., 1994). We can express the heat and work terms as:

$$\dot{q} = \frac{C - H}{m_g n_g}, \tag{1.10}$$

$$\frac{P_g}{\rho_g^2}\dot{\rho}_g = -\frac{kT_g}{m_g}\left(\frac{dv_g}{dr} + \frac{2v_g}{r}\right). \tag{1.11}$$

Assuming a velocity gradient for the medium, and combining the previous equations, wthe temperature gradient can be derived as follows:

$$\frac{dT_g}{dr} = -\frac{4T_g}{3r}\left(\frac{1}{2}\frac{d\ln v_g}{d\ln r} + 1\right) + \frac{8\pi r^3 m_g (H - C)}{3kT_g\dot{M}}. \tag{1.12}$$

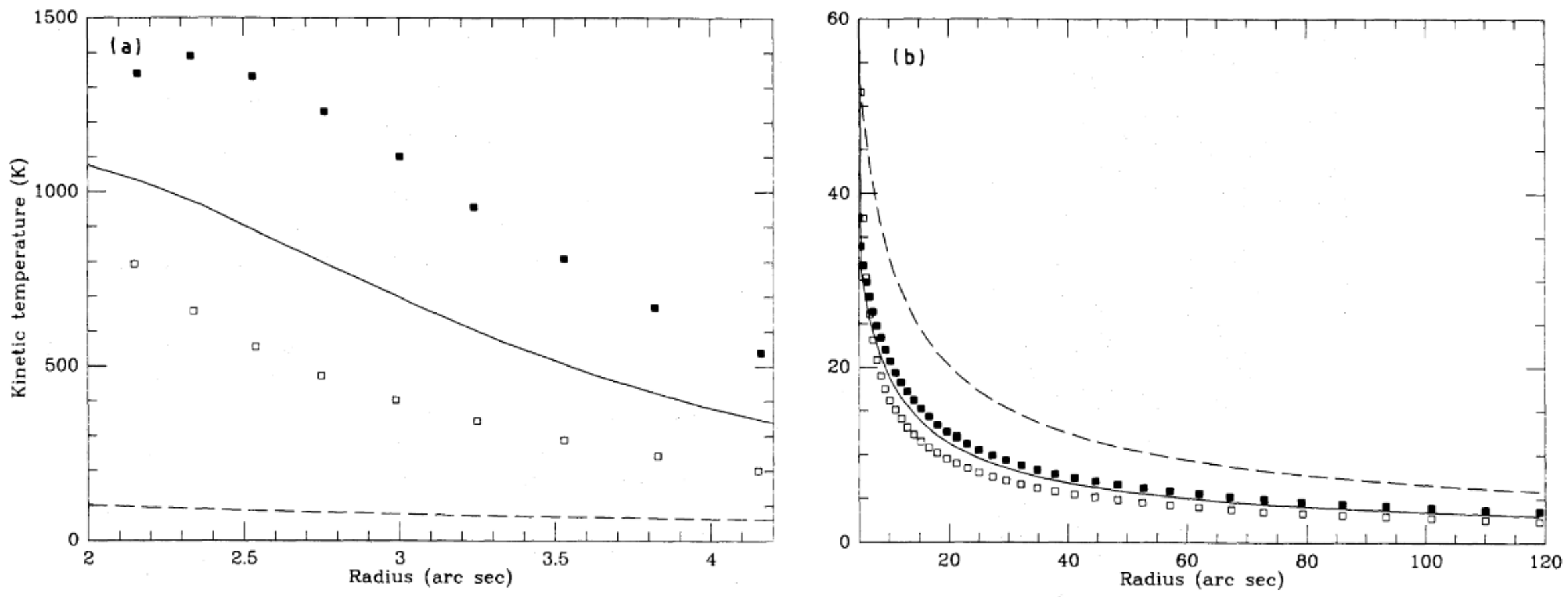

Figure 1.9: Radial gas temperature distribution for IRC+10216 in the inner region (a) and in the outer zones (b). The solid line represents the temperature for a distance of 200 pc, open squares for a distance of 100 pc, and filled squares for a distance to the sources of 300 pc. Figure taken from Truong-Bach et al. (1991).

The first term accounts for the adiabatic expansion. $H$ and $C$ are the gained and lost energies by the gas, respectively. The heating processes are dominated by collisions between gas molecules with other molecules and dust grains. On the contrary, the most common mechanism for gas cooling is radiative, consisting of photon emission in molecular and atomic transitions.

We focus on the adiabatic expansion term, which is responsible for a faster cooling. The temperature gradient can be derived assuming $\mathrm{d}\ln v_{\mathrm{g}}/\mathrm{d}\ln r = 0$ ($v_{\mathrm{g}} = \mathrm{cte}$):

$$\frac{\mathrm{d}T_{\mathrm{g}}}{\mathrm{d}r} = -\frac{4T_{\mathrm{g}}}{3r} \rightarrow T_{\mathrm{g}} = T_0 \left(\frac{r}{r_0}\right)^{-4/3}. \tag{1.13}$$

This equation indicates that there is a large variation in $T_{\mathrm{g}}$ between the regions close to the star and the outer areas of the envelope. The temperature gradient can be derived with a velocity term proportional to the distance to the center ($\mathrm{d}\ln v_{\mathrm{g}}/\mathrm{d}\ln r = 1$):

$$\frac{\mathrm{d}T_{\mathrm{g}}}{\mathrm{d}r} = -2\frac{T_{\mathrm{g}}}{r} \rightarrow T_{\mathrm{g}} = T_0 \left(\frac{r}{r_0}\right)^{-2}, \tag{1.14}$$

which produces even larger variations in the temperature between the inner and the outer regions when compared to the case of a constant velocity expansion. This kind of temperature gradient is usually found in nebulae around post-AGB stars.

In a more general case, this is the procedure for calculating the gas temperature. By contrast, the temperature of the dust grains, $T_{\mathrm{d}}$, is mainly determined by the type and amount of radiation and the opacity of the envelope. The main energy exchange processes between dust grains and gas are the absorption and emission of radiation and collisions (due to the effective coupling between gas and dust). Under these conditions, the dust temperature can be expressed as:

$$T_{\mathrm{d}} \propto \left(\frac{1}{r^2}\right)^{-\frac{1}{4+\alpha}}, \tag{1.15}$$

where $\alpha \in [1, 2]$. See e.g. Andriesse (1974); Draine and Lee (1984) for details.

Under these circumstances, sources are very large, so the gas temperature varies significantly between the inner regions and the outermost layers. The envelope material close to the stellar photosphere presents temperatures $\sim 1\,500$ K. As the circumstellar matter flows away from the central AGB star and beyond the dust grains formation zone, the temperature decreases such that, at the outermost layer of the shell, the temperature is $\sim 25$ K (see Fig. 1.9 and Truong-Bach et al., 1991; Ziurys, 2006). In the case of the dust grains, they are created at $\sim 3-15\,R_*$. The dust temperature varies from $\sim 1\,300$ K at $3\,R_*$, to $\sim 500$ K at $15\,R_*$, and $\sim 300$ K at $40\,R_*$ (see e.g. Keady et al., 1988; Men'shchikov et al., 2001).

### 1.2.3 Circumstellar chemistry

Active chemical processes take place in the circumstellar medium (and previously, in the stellar atmosphere). As a consequence, a large number of molecular species are created. To understand the chemical composition of CSEs, it is necessary to introduce the concept of reaction rate ($k$), which is the velocity at which a chemical reaction takes place. For an exothermic reaction:

$$\mathrm{R} + \mathrm{S} \rightarrow \mathrm{P} + \mathrm{Q}, \tag{1.16}$$

$$v_\mathrm{r} = -\frac{\mathrm{d[R]}}{\mathrm{d}t} = -\frac{\mathrm{d[S]}}{\mathrm{d}t} = \frac{\mathrm{d[P]}}{\mathrm{d}t} = \frac{\mathrm{d[Q]}}{\mathrm{d}t} = k\mathrm{[R][S]}, \tag{1.17}$$

where $v_\mathrm{r}$ is the reaction velocity (or reaction rate) and [R] represents the abundance of the chemical species R. When the reaction rate is constant, the Arrhenius equation gives its dependence with the absolute temperature:

$$k(T) = A\exp\left(-\frac{E_\mathrm{a}}{kT}\right), \tag{1.18}$$

where $E_\mathrm{a}$ is the activation energy for the reaction, and $A$ is the probability of the reaction to occur when two atoms or molecules collide if the potential barrier is overcome. In the case of reactions involving stable molecules, $E_\mathrm{a}/k \gtrsim 10\,000$ K; in the case of reactions with a stable molecule and an unstable molecule or atom, $0 \lesssim E_\mathrm{a}/k \lesssim 5\,000$ K, and, in the case of reactions between two radicals, $E_\mathrm{a}/k \gtrsim 0$ K . Typically, $A \,\epsilon\, [10^{-11} - 10^{-10}]\,\mathrm{cm^3\,s^{-1}}$.

#### Equilibrium chemistry

The temperatures found in the atmospheres of AGB stars are lower than the dissociation energies of stable molecules, which implies that most of the gas is in a molecular state. Equilibrium chemistry is characterized by a high abundance of stable molecules, while radicals are very rare. CO is an important molecule because of its high stability, therefore its abundance at chemical equilibrium is conspicuous. This molecule is so abundant that CO is expected to be produced without leaving any traces of C or O, and therefore, almost all the available C or O nuclei will be converted to CO. There are three different kinds of circumstellar envelopes depending on the C/O ratio: M-type stars with $C/O < 1$, S-type stars with $C/O \approx 1$, and C-type stars with $C/O > 1$. In the case of O-rich envelopes, all the available C forms CO, while any surplus O forms O-bearing molecules, such as SiO, $SO_2$, SO, $H_2O$,..., while C-bearing molecules, other than CO, are rare. In the case of C-rich envelopes, all the available O forms CO, and any remaining C forms C-bearing molecules, such as HCN, $HC_{2n+1}N$, $C_2S$, CS, often alongside SiS. See e.g. Engels (1979); Olofsson et al. (1993); Velilla Prieto et al. (2017).

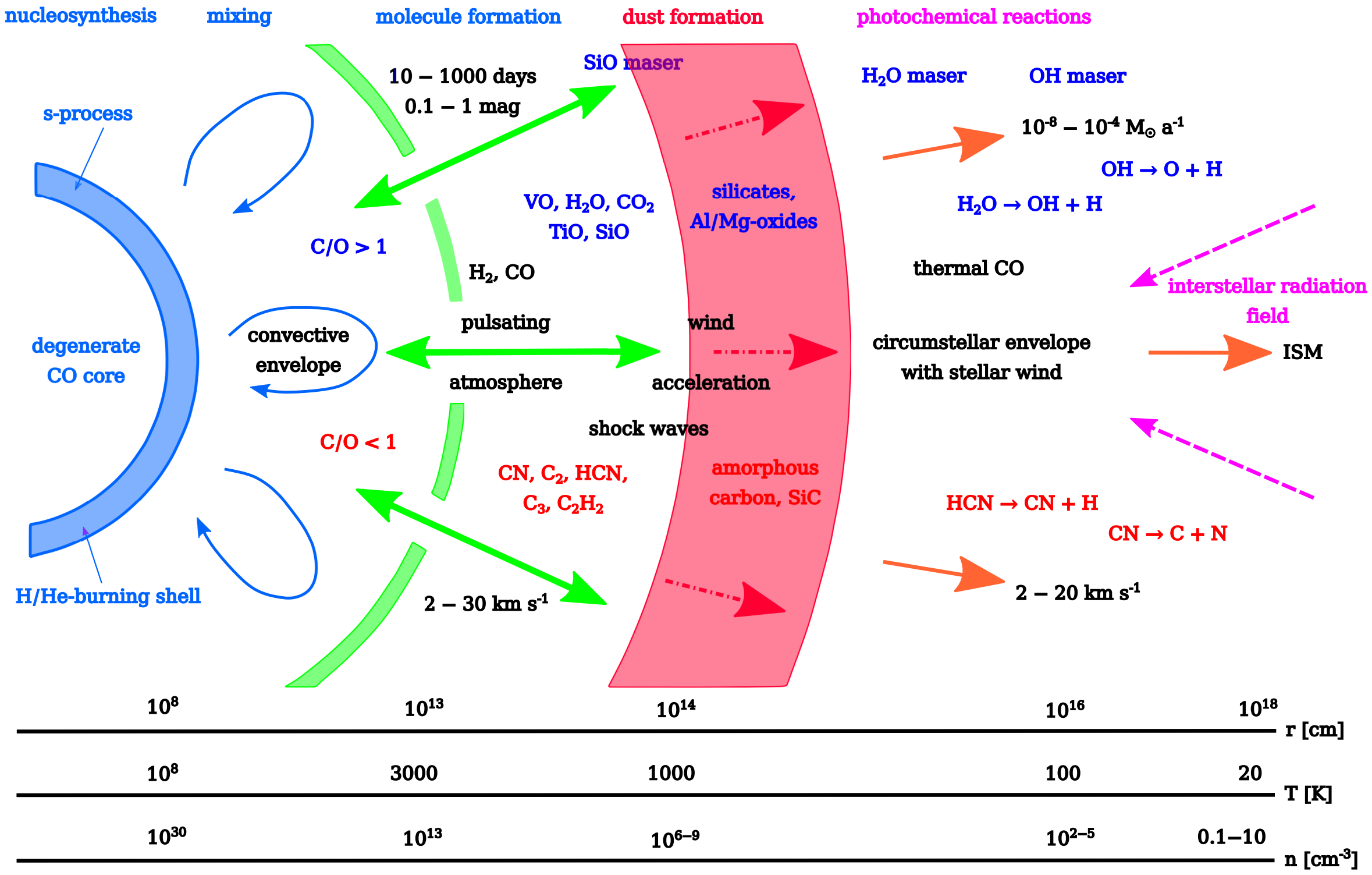

Figure 1.10: Sketch of the structure of an AGB star from its centre to its outermost layers showing various physical and chemical processes that occur in the different regions. Figure adapted from Uttenthaler (2007).

The circumstellar material is continuously expanding, and the physical conditions vary depending on the distance from the star. The characteristic timescale of this process is determined by the macroscopic evolution due to the envelope expansion:

$$t_\mathrm{c} \sim \frac{r}{v}, \tag{1.19}$$

where $v$ is the terminal expansion rate of the circumstellar envelope (see Eq. 1.7). If the chemical reactions reach equilibrium at timescales shorter than $t_\mathrm{c}$, a succession of states in chemical equilibrium take place at great distances from the star. These are known as fast reactions. However, if a chemical reaction is slower than the rate at which physical conditions change, chemical equilibrium will not be reached, as the envelope expands and the chemical abundances are "frozen" at the values corresponding to the last chemical equilibrium, despite the temperature gradient in the radial coordinate. This occurs at distances $>10^{14}$ cm. In real cases, chemical equilibrium is only reached in the atmosphere of the evolved star and in the inner layers of the envelope (except for radicals and ions whose reaction rates are much larger).

Table 1.1 Molecular abundances in C- and O-rich CSEs around standard AGB stars, where shocks are not important.

| O-rich CSEs | | | C-rich CSEs | | |
|---|---|---|---|---|---|
| Molecule | $X$ | $R_{\rm out}$ [cm] | Molecule | $X$ | $R_{\rm out}$ [cm] |
| $^{12}$CO | $3\times10^{-4}$ | $>10^{16}$ | $^{12}$CO | $8\times10^{-4}$ | $>10^{16}$ |
| $^{13}$CO | $10^{-5}$ | $\gtrsim10^{16}$ | $^{13}$CO | $2\times10^{-5}$ | $>10^{16}$ |
| $H_2O$ | $2\times10^{-4}$ | $1{-}3\times10^{16}$ | HCN | $3\times10^{-5}$ | $1{-}5\times10^{16}$ |
| OH | $10^{-4}$ | $1{-}5\times10^{16}$ | H$^{13}$CN | $10^{-6}$ | $1{-}3\times10^{16}$ |
| HCN | $4\times10^{-7}$ | $1{-}3\times10^{16}$ | CN | $5\times10^{-6}$ | $1{-}8\times10^{16}$ |
| CN | $10^{-7}{-}10^{8}$ | $1{-}3\times10^{16}$ | $C_2H_2$ | $3\times10^{-4}$ | $1{-}3\times10^{16}$ |
| $HC_3N$ | $<2\times10^{-7}$ | $1{-}3\times10^{16}$ | $HC_3N$ | $10^{-6}$ | $1{-}5\times10^{16}$ |
| CS | $10^{-7}$ | $1{-}3\times10^{16}$ | CS | $10^{-6}$ | $1{-}3\times10^{16}$ |
| SiS | $7\times10^{-7}$ | $1{-}3\times10^{16}$ | $C_4H$ | $2\times10^{-6}$ | $1{-}5\times10^{16}$ |
| SiO | $10^{-5}$ | $1{-}3\times10^{15}$ | SiS | $2\times10^{-6}$ | $1{-}3\times10^{16}$ |
| SO | $10^{-6}$ | $10^{16}$ | SiO | $5\times10^{-7}$ | $10^{16}$ |

**Notes.** Abundances and outer radius taken from Bujarrabal et al. (1994b,a); Nyman et al. (1993); Guelin et al. (1993); Dayal and Bieging (1995).

### Non-equilibrium chemistry

There are also chemical processes that occur under non-equilibrium conditions, such as the accretion of matter on dust grains and photodissociation effects.

When a molecule in a gas with $<1\,000$ K encounters a dust grain, it is highly likely to stick to the grain. For temperature values above $1\,000$ K the dust sublimates. Accretion in grains extracts most of the solidifiable material from the gas (see e.g. Schirrmacher et al., 2003). This solidification process is efficient for some species that account for a high fraction of the grains, such as silicon derivatives. Consequently, molecules like SiO are easily incorporated into dust grains, and the abundance of gaseous SiO must decrease from the star as SiO molecules in the outflowing stellar wind are incorporated into the dust grains (see e.g. González Delgado et al., 2003; Massalkhi et al., 2020). To summarize, silicon disappears from the gas practically at the same time when the dust grains are formed. Other interesting species, such as sulfur, manganese, or iron solidify before expansion in the constant-velocity phase. Amorphous carbon dust formed out of primary carbon is dominant in C-rich envelopes, while silicon carbide and magnesium sulphide are minor widespread dust components (see e.g. Zhukovska and Gail, 2008). Nevertheless, CO remains in the gas phase, because this molecule presents a dissociation energy higher than that of the corresponding crystalline states.

Interstellar radiation is characterized by a temperature $\sim10\,000$ K, but it is not well described by a black body radiation, so dissociation by UV photons is a non-equilibrium process. It destroys stable molecules created during equilibrium chemistry in the outermost layers of the circumstellar envelope. Thus, interstellar UV photons produce radicals and ions, which will also be destroyed afterwards. This process produces either OH from $H_2O$ in O-rich environments or CN from HCN in C-rich environments (see Fig. 1.10). The $H_2O$ abundance decreases at distances $\gtrsim10^{16}$ cm, while the OH abundance increases until its value is equal to that of $H_2O$. Nevertheless, OH soon begins to photodissociate, so this molecule is located in a relatively thin layer in the outermost region of the envelope. The presence of radicals in these regions leads to fast reactions, producing additional chemical species. In Table 1.1, we present the abundances and the outer radius that delimit the spatial distribution of each molecular species in a circumstellar envelope. CO is a stable molecule and the most resistant to photodissociation. Alongside with its large abundance, this is the reason why it is used to estimate the outer radius of CSEs. Finally, all molecules are photodissociated at $\gtrsim 10^{18}$ cm.

The molecules produced in photodissociation regions (PDR) are usually found in non-equilibrium chemistry environments. The chemistry of these regions is dominated by far-ultraviolet (FUV) radiation. In objects with even higher temperatures, like PNe, the ionized region it is often referred to as XDR (X-ray dominated region), where the chemical processes are dominated by X-ray radiation.

The chemical reactions that take place in PDR regions are similar to those in the outermost regions of an envelope, where ISM radiation produces photoionization (see Fig. 1.10), but PDR are located in the inner layers of the envelope and are due to the high temperature of the star that is becoming a central star of a PN (CSPN). In general, the formation of radicals in PDR, like CN, and ions, like $HCO^+$, $N_2H^+$, $SO^+$, $CO^+$ is favored.

These peculiar chemical reactions are important to explain, for instance, the presence of HCN in O-rich environments. This is the case of the Red Rectangle, in which $H^{13}CN$, C I, and C II are detected (Bujarrabal et al., 2016).

# 1.3 pre-Planetary Nebulae

## 1.3.1 Physical conditions

A pre-planetary Nebula (pPN) is the nebula around a post-AGB star. The previous stage of pPNe are AGB-CSEs, and the following and final phase are PNe. Post-AGB stars are objects transitioning between the AGB phase and blue/white dwarfs (the nuclei of PNe, see Sect. 1.1.2). The effective temperature of these post-AGB stars varies between 3 000 and 30 000 K (see Fig. 1.2).

During the post-AGB phase, the kinematics of the envelope differ from those of the previous stage, because the system symmetry changes from spherical to axial, polar, or even irregular. Apparently, this is caused by a strong interaction with a companion star. Circumbinary structures are produced when the interaction with the companion is stable. However, in the case of AGB stars, the binarity may imply spiral features in the CSE (see Fig. 1.6). Most of the spatially resolved pPNe show a clear axial symmetry and a bipolar structure. This kind of shape is the rule rather than the exception, with many of them showing multipolar structure (Balick and Frank, 2002; Sahai et al., 2009). In general, pPNe develop their nebular shape by fast and collimated outflows that are injected into slow winds ejected during the AGB stage (Sahai and Trauger, 1998; Höfner and Olofsson, 2018). Regarding the launching of these post-AGB collimated bipolar outflows, there are several mechanisms that have been postulated, such as the presence of magnetic fields, the presence of high-density equatorial structures, or the effect of binarity (see De Marco, 2009; Sahai et al., 2011).

The mass loss becomes non spherical during the late-AGB and/or the early post-AGB phase (see e.g. Trammell and Goodrich, 2002). These objects might present complex structures such as jets, loops, bubbles, and knots. The presence of jets during the early post-AGB star phase could be crucial in the development of complex structures in pPNe and PNe (Sahai and Trauger, 1998). Those envelopes that seem to be spherical may have a nebular axis perpendicular to the sky plane, hiding the axial symmetry and the bipolar outflow (see e.g. Sahai et al., 1999).

These nebulae present a mass over $0.1\,M_\odot$, with a typical value of around $1\,M_\odot$, and a typical mass loss of $\sim 10^{-4}\,M_\odot\,a^{-1}$ Bujarrabal et al. (2001). Different kinematic components can be identified in these pPNe. On the one hand, there is a massive component with a spheroidal- or ring-shape around the central star that presents velocities similar to those found in envelopes around AGB stars, implying a radial expansion ($\sim 20\,\mathrm{km\,s^{-1}}$, see Castro-Carrizo et al., 2010). On the other hand, the second identified structure, which is usually less massive, is composed of fast bipolar outflows ($\sim 100\,\mathrm{km\,s^{-1}}$ or more) and its matter is distributed along a symmetry axis. This component is probably powered by post-AGB ejections and systematically shows an extremely high amount of kinetic energy and linear momentum (with values $10^{45} - 10^{47}$ erg and $10^{38} - 5\times10^{39}\,\mathrm{g\,cm\,s^{-1}}$, respectively, see Bujarrabal et al., 2001). The kinematics and sizes ($\sim 10^{17}$ cm) found in pPNe suggest that these fast components have ages around $10^3$ years. In some cases, the spherical AGB component is not seen, as in M 1−92 or M 2−56; by contrast, CRL 618 and CRL 2688 are clear examples showing both components.

Masses and velocities may differ across different nebulae. When compared to the CSEs around AGB stars, pPNe present complex physical (geometry) and chemical (thermodynamics that include shock heating factors, which can be difficult to model) conditions.

Some pPNe present spectacular structures, such as M 1−92 (Fig. 1.11). The dominant component of this nebula can be seen in molecular emission, with a mass of $1\,M_\odot$ and temperatures of around 20 K. These temperatures are probably related with adiabatic cooling, but the low probability of shock heating complicates the thermodynamics. These low temperatures are usually found in pPNe. This component has a structure composed by two hollow shells, joined in their central zone, that forms a high-density ring. The dominant component of the gas velocity is parallel to the perpendicular axis of the ring and reaches $70\,\mathrm{km\,s^{-1}}$. See Bujarrabal et al. (1998a) for further details. Another interesting source is OH 231+4.2 (Fig. 1.12). The mass-dominant component of this nebula is large, elongated, and cold. The expanding material has been observed via CO line emission and via scattered light in optical wavelengths; it presents axial velocities of around $300\,\mathrm{km\,s^{-1}}$. Hot bubble-like structures can be seen in this nebula through observations of atomic lines (H$\alpha$ and N II). The southern bubble presents typical density values of around $50\,\mathrm{cm^{-3}}$ and it is known to be expanding with a velocity of $\sim 400$ kms. Additionally, NaCl, KCl, and $H_2O$ observations reveal a rotating circumbinary disk present in the innermost region of the nebula. This is the first time that a rotating structure has been observed in such a massive nebula. Additional details can be found in Sánchez Contreras et al. (2000); Bujarrabal et al. (2002); Sánchez Contreras et al. (2022).

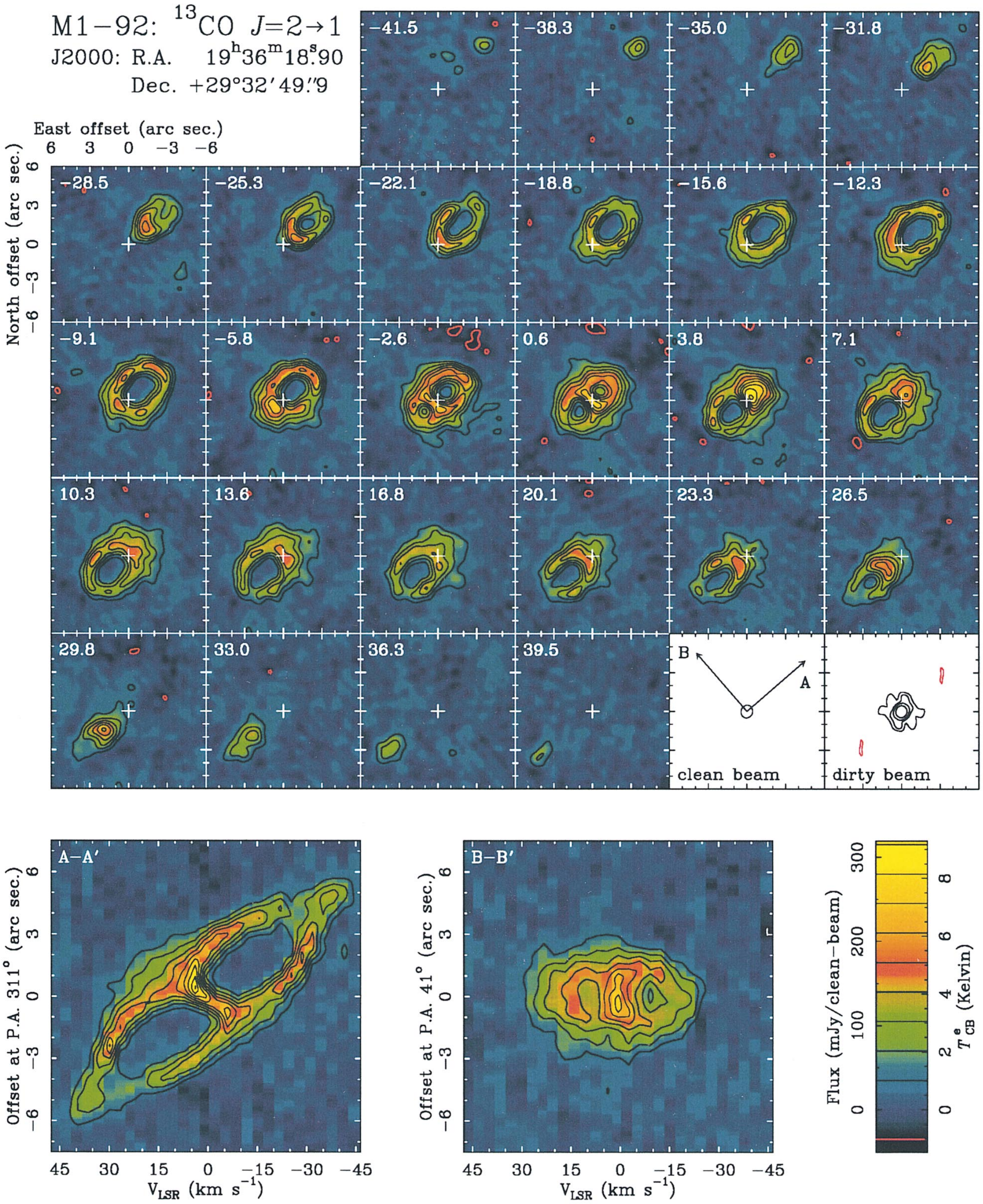

Figure 1.11: Plateau de Bure observations of M 1−92 in $^{13}$CO $J = 2-1$ emission (*Top*) and relevant PV diagrams along the axis and equatorial directions (*Bottom*). Figures taken from Bujarrabal et al. (1998a).

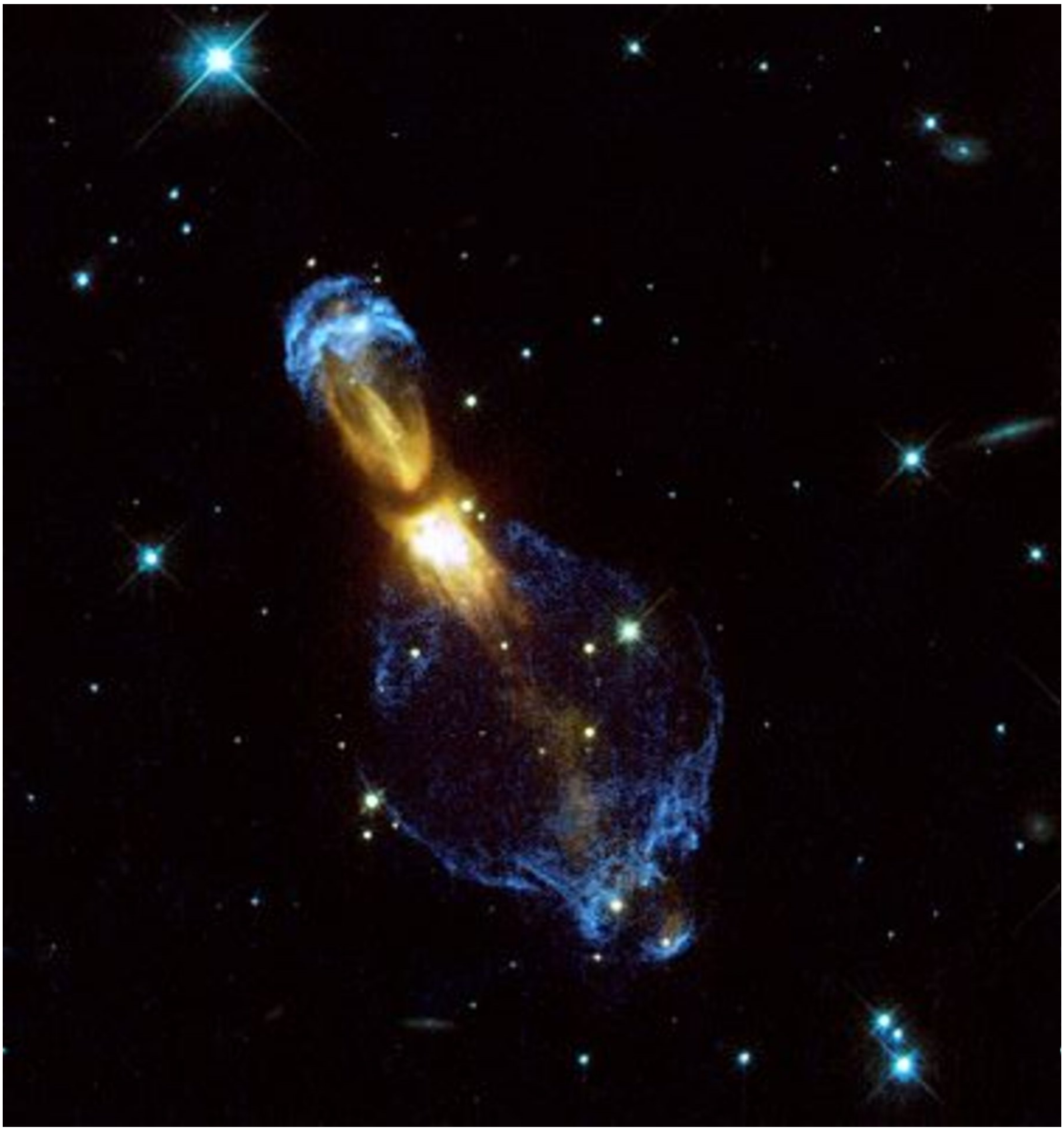

Figure 1.12: Observations of the pPN OH 231.8+4.2 (also known as the Calabash Nebula). This figure was taken using the *Hubble Space Telescope* with the WFPC2 instrument in four different filters: light from 791 nm is displayed in red, 675 nm in green, while combined light from H$\alpha$ at 656 nm and N II at 658 nm are shown as blue. Figure adapted from Bujarrabal et al. (2002).

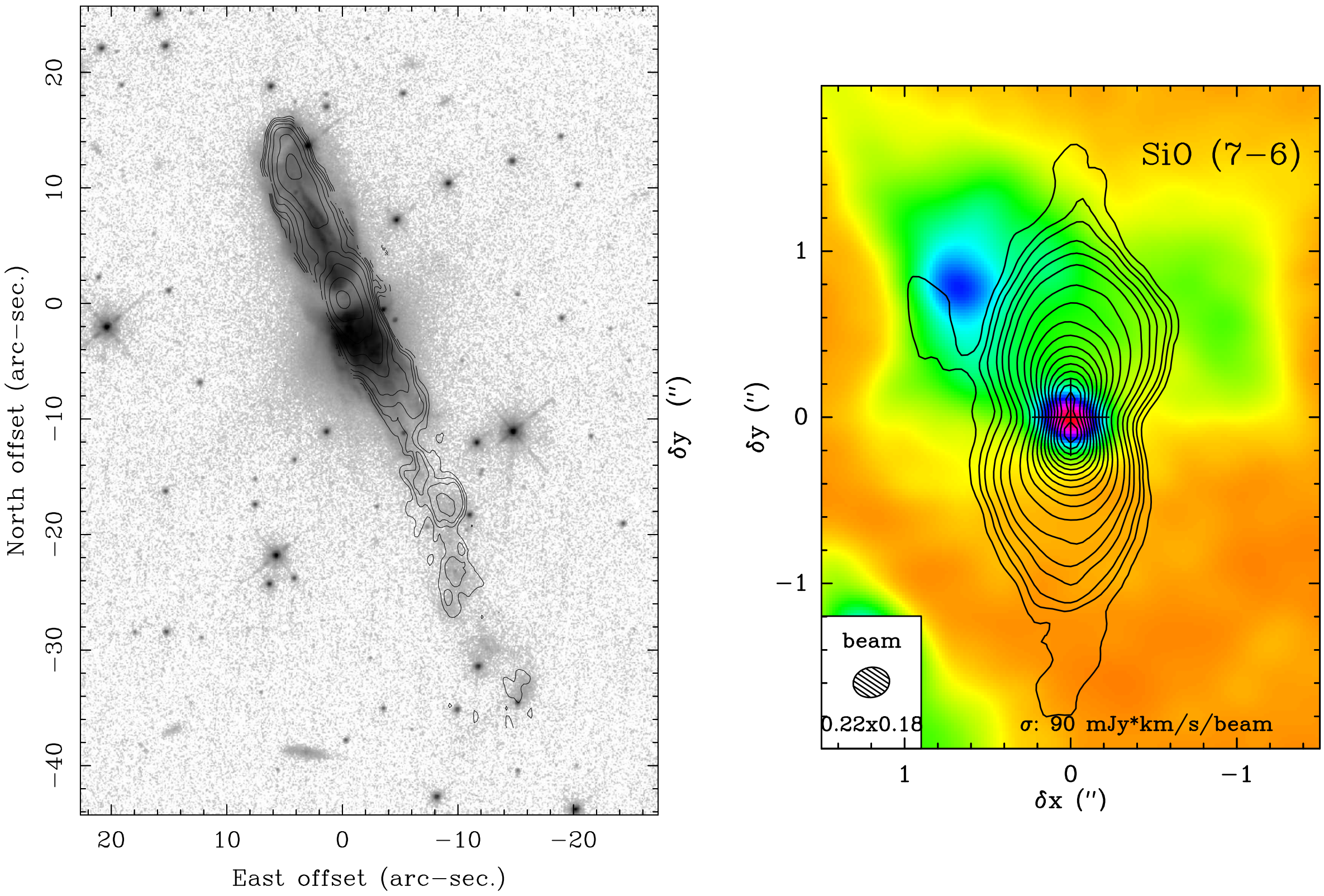

Figure 1.13: *Left*: NOEMA observations of OH 231.8 + 4.2 in $^{12}$CO $J = 2 - 1$ emission (velocity integrated) showing the molecular outflow (Alcolea et al., 2001), superposed on an *HST* /WFPC2 image (Bujarrabal et al., 2002). Figure adapted from Velilla Prieto et al. (2015). *Right*: ALMA observations of the innermost region of OH 231.8 + 4.2 in SiO $J = 7 - 6$ emission (velocity integrated). The background image is the 330 GHz-continuum map. Figure taken from Sánchez Contreras et al. (2018).

### 1.3.2 Molecular lines

Molecular emission is often intense in pPNe, so observations of molecular lines yield useful information about these sources. This molecular emission is detectable trough rotational transitions in the "radio window". See Chapter 2 for a description of molecular radio-emission. The molecular composition of a star changes from the AGB to the post-AGB phase, insomuch as some low-abundance molecules are abundant in pPNe, such as SO, $SO_2$, $HCO^+$, or HNC (see e.g. Sánchez Contreras et al., 2000). The nebulae that surround post-AGB stars present a high molecular diversity when compared to CSEs. The most abundant molecules in pPNe are $^{12}$CO, $^{13}$CO, SiO, and HCN (see Bujarrabal et al., 1994b,a), and each molecule provides information about different nebular regions. The most abundant and most important of these molecules are $^{12}$CO and $^{13}$CO. The radiation emitted by the post-AGB star is not photodissociative, and CO is be the best trace of the pPN. $^{12}$CO produces intense lines that are often optically thick, which makes it difficult to interpret them. However, $^{13}$CO produces optically thin, but weaker lines. Theoretical models of the physical and chemical conditions in AGB-CSEs are relatively straightforward, due to the simple geometry, thermodynamics, and dynamics of the envelope. In contrast, in pPNe, the geometry and kinematics are complex, and the thermodynamics includes complicated shock-boosting terms. These kinds of nebulae are typically characterized by the presence of shocks, although the presence of stable molecules around these regions, such as CO, is frequent. There are archetypal molecules of shock-accelerated regions, like SiO or $HCO^+$ (see e.g. Sánchez Contreras et al., 2000; Bublitz et al., 2019). Shocks enhance the molecular diversity in these regions; it is possible to detect more complex molecules, such as $CH_3OH$, or even $H_2O$, in C-rich environments. In addition, the presence of maser emission is not as usual as in the envelopes around AGB stars. This happens because the emission zone of these lines, often-times relatively close to the star, tends to disappear in these stages of the stellar evolutionary phase. Maser emission seems to be more frequent in those pPNe that are younger, such as the presence of SiO maser emission in OH 231.8+4.2 (Sánchez Contreras et al., 2000), revealing the presence of gas close to the star (see Fig. 1.13). $H_2O$ and OH maser emission (at 22.2 and 1.62 GHz, respectively) have been also detected in nebulae around post-AGB stars (Lewis, 1989; Yoon et al., 2014). These three different kinds of masers allow us to study the structure of the nebula, because the SiO maser emission comes from the inner layers, the 22 GHz $H_2O$ maser emission is usually found from 5 to 50 stellar radius, and the 1 612 MHz OH maser comes from the outer layers of the envelope, at distances around 1 000 to 2 000 AU. Additionally, due to the evolution of the central star, the SiO maser is the first one to disappear, followed by the $H_2O$ maser, and finally, the OH maser fades away. Therefore, the study of these masers is essential to analyze the structure of the nebulae.

### 1.3.3 Atomic lines

The central stars in pPNe, and especially, in PNe, present sufficiently high temperatures to generate excitation regions close to the central star. These energetic photons can dissociate molecules or even atoms. These ionized atoms can also be observed in other hot regions dominated by shocks due to the presence of highly collimated jets in the nebula. Recombination lines are useful to trace excitation regions that are ionized by the stellar radiation. In optical wavelengths, the most widely used recombination lines are H$\alpha$ and H$\beta$. Forbidden lines of excited and ionized heavy atoms are useful to trace excitation regions dominated by shocks (Kwok, 1993). These forbidden lines are [N II] at 6 548 and 6 584 Å, [S II] at 6 717 and 6 731 Å, [O I] at 6 300 and 6 306 Å, [O II] at 3 727Å, and [O III] at 4 959 and 5 007 Å. The H$\alpha$ recombination line and some relevant forbidden lines can be detected in M 1−92 distributed along the nebular axis (see Fig. 1.14).

Atomic lines can be detected at far infrared (FIR) wavelengths that correspond to fine structure lines[1]. These FIR transitions are observed mainly with space telescopes, such as *ISO*, *IRAS*, and *Herschell*. Fine-structure transition lines are good tracers of low- and intermediate- excitation regions, with temperatures between 100 and 2 000 K, where the gas might remain in a low excitation state. These intermediate regions between molecular gas and photoionized gas are called photodissociation regions (PDRs), and they appear during the post-AGB evolution as an intermediate stage between the cold AGB-CSEs with high molecular richness and the hot atomic ionized gas present in old PNe. Some good PDR tracers are C II and O I. C I is also associated with PDRs.

[1]Complete electron shells do not interact with electrons from incomplete shells. These incomplete shells are the ones that generate the fine structure transitions. The Hamiltonian of this system can be expressed as the sum of a central field $H_{\rm o}$, an electrostatic repulsion of the outer electrons $H_{\rm es}$, and a spin-orbit interaction associated with the coupling between the angular momentum ($\vec{L}$) and the spin of the system ($\vec{S}$), $H_{\rm so}$: $H = H_{\rm o} + H_{\rm es} + H_{\rm so}$. Any change in the coupling $\vec{L} \cdot \vec{S}$ produces a change in the source energy, and thus,a transition known as a fine-structure transition.

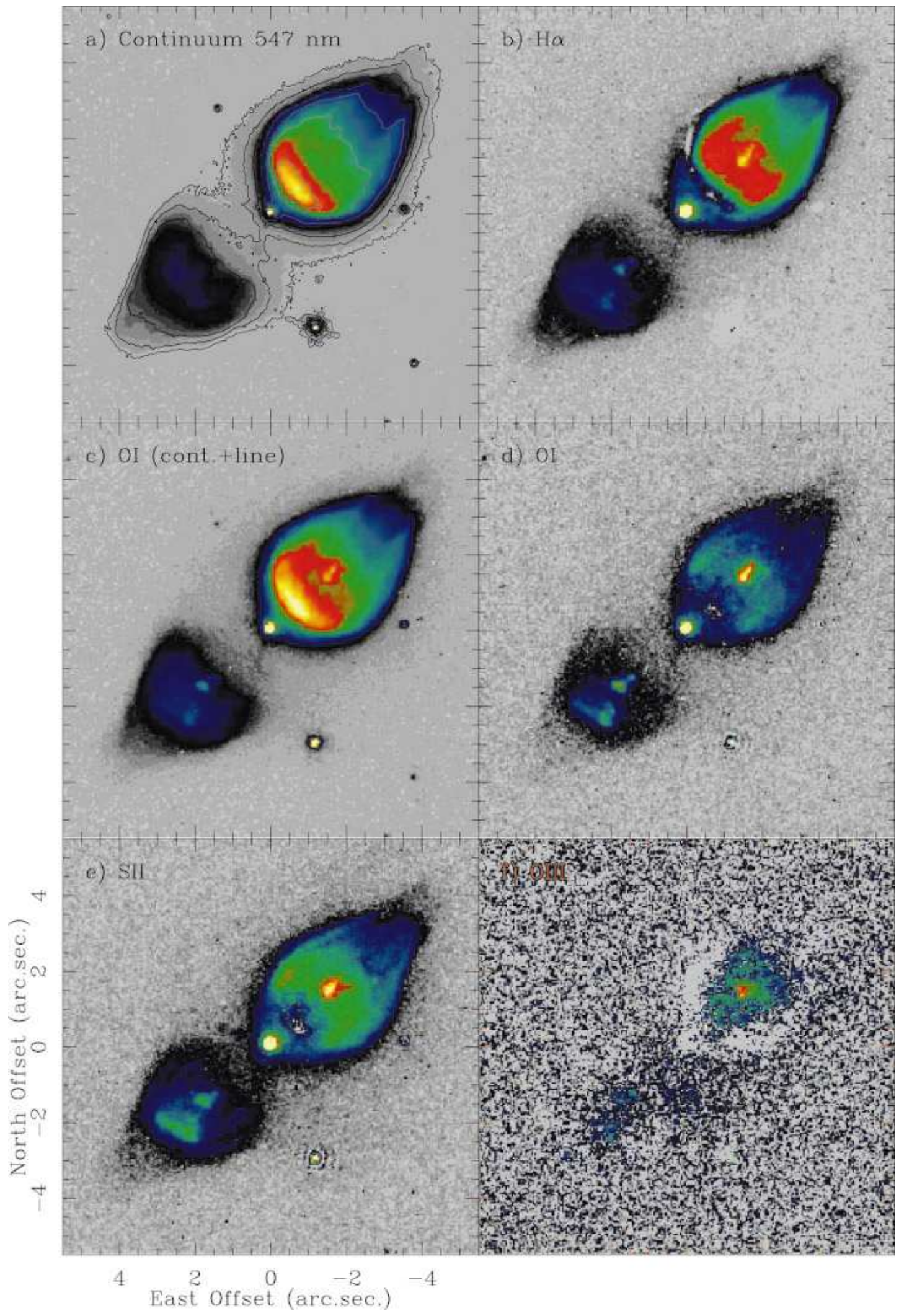

Figure 1.14: *Hubble Space Telescope* observations of the pre-planetary nebula M 1−92 displaying atomic emission: (a) 5 500 Å continuum, (b) Hα emission after subtraction of scattered light, (c) [O I] 6 300 Å emission before subtraction of scattered light, (d) [O I] 6 300 Å emission after subtraction of scattered light, (e) [S II] 6 700 Å emission after continuum subtraction, and (f) [O III] 5 000 Å emission after continuum subtraction. Figure taken from Bujarrabal et al. (1998b).

Deep studies reveal that atomic lines have been detected in pPNe whose central stars have $T_{\rm eff}$ > 10 000 K. Nevertheless, there are no detections of atomic lines in pPNe whose central stars have $T_{\rm eff}$ < 10 000 K, inasmuch as PDR emission requires higher temperatures (see Fong et al., 2001; Castro-Carrizo et al., 2001, for further details). For example, the detection of $H^{13}CN$, C I, and C II in the pre-planetary nebula of the Red Rectangle is consistent with the presence of a photodissociation region (Bujarrabal et al., 2016). Additionally, PDRs have been found in some young PNe, such as NGC 7027.

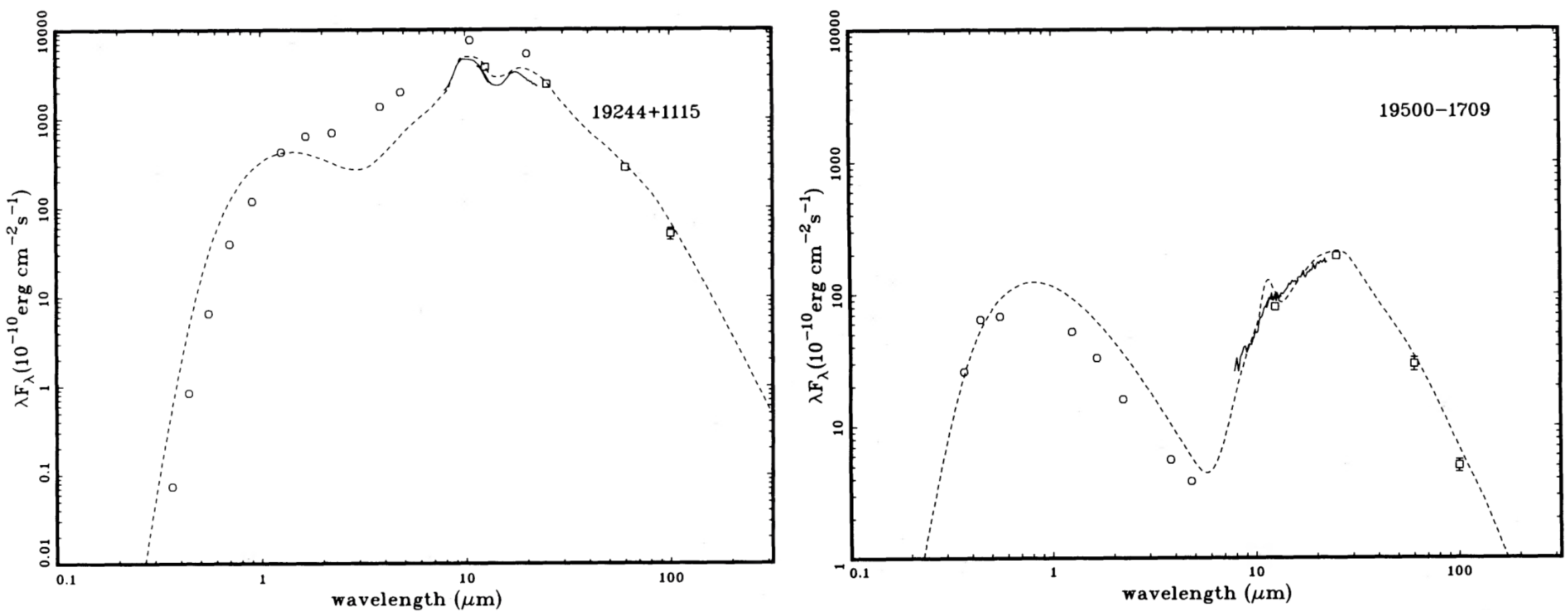

Figure 1.15: SEDs of IRAS 19244+1115 (*left*) and IRAS 19500−1709 (*right*). Both pPNe present IR excess in their SED. The spectra can be separated into two components, one due to the star and another one due to the dust emission from previously absorbed stellar radiation. The age of a pPNe can be inferred from the separation of these two peaks. IRAS 19244+1115 is a relatively young source, since the separation is small, meaning that there is still warm dust close to the star. On the contrary, in IRAS 19500−1709, the separation is larger, meaning that there is no warm dust due to the expansion of the inner envelope layers over time; consequently, it is a relatively evolved pPN. Figures taken from Hrivnak et al. (1989).

### 1.3.4 Continuum emission

The continuum emission from a pPNe can originate from two different mechanisms, either thermal emission or (photon) scattering by dust grains.

The innermost regions of a pPNe drift away from the central star and cool down over time, therefore the thermal emission of dust grains occurs at relatively long wavelengths. A larger separation between the two maxima in the SED, one due to the stellar emission and another one due to dust grain emission, can be identified in old pPNe. On the contrary, if the maxima are close, the nebula is probably young (see Fig. 1.15). Images of light emitted by grains in the IR or mm-waves waves typically show that the emitting regions are small. An estimate of the nebular dust mass and the inner radius of the envelope can be derived from SED models.

In the case of light scattered by dust, the images are much more spectacular because of the high resolution of space telescopes (see for example Fig. 1.12). The scattered light is crucial because it is independent of the excitation state of the medium and traces the entire structure of the nebula. However, this type of observations are limited by the extinction of the inner regions of the pPNe.

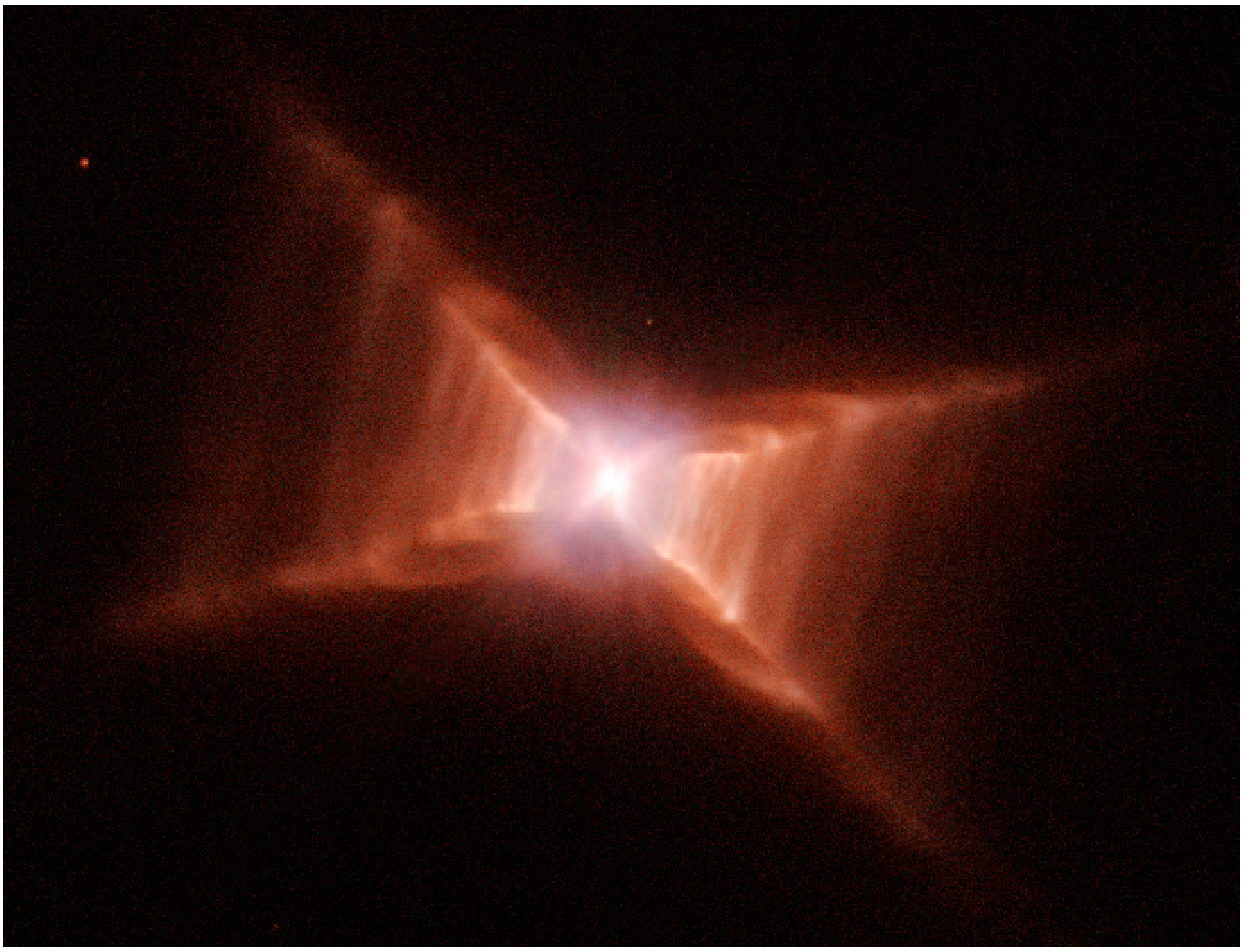

Figure 1.16: Observations of HD 44179 (the Red Rectangle) taken with the WFPC2 instrument on board *HST* in different filters: [O III] at 5 041.12 Å, He I and Na I at 5 893.51 Å, and [O I] at 6 187.50 and 6 306.44 Å. Figure taken from Van Winckel et al. (2002).

## 1.4 Binary post-AGB stars

The aim of this thesis is to study a specific kind of post-AGB stars, those that are part of a binary system and are surrounded by a circumbinary disk with Keplerian dynamics (see Fig. 1.16). The impact of binarity is crucial because it definitely alters the intrinsic properties of the post-AGB star, such as its mass loss, the kind of envelope around it and its morphology, and the formation of dust, both in terms of the formation rate and the chemical composition of the grains (see Waelkens et al., 1996; Van Winckel, 2003; De Marco and Izzard, 2017). These sources are a particular class of objects, and they are the first ones where the existence of rotating disks has been studied in detail. These rotating disks have been postulated as the key part in explaining the formation of collimated bipolar jets archetypal of the post-AGB phase. Although the nebulae of my thesis do not present such jets, understanding the role that these disks play is fundamental to comprehend the formation and shape of pPNe.

The orbital period of the binary system is crucial for understanding the future of the envelope around it. If the future post-AGB star companion is located at a distance lower than 2 AU, it is tidally attracted and the binary system may undergo a common envelope (CE) phase where the main star evolves along the RGB or even the AGB phase, expanding before eventually overflowing its Roche lobe and engulfing the second star (Paczynski, 1976). The small separation between the components of the binary system might lead to a stellar merger but, if this phenomenon is avoided, the orbital period of the system will be $\lesssim 100$ days. On the contrary, if the orbital separation is wider, the interaction between the stars occurs via wind accretion, and the main star evolves off the giant branch. Then, the secondary star reaches the giant branch, and the interaction could take place again, but with a compact companion. According to population synthesis models, this scenario yields orbital periods between $\sim 3\,000$ and $10^6$ days. Thus, there is a clear bimodal distribution in theoretical predictions (see Fig. 1.18 and Nie et al., 2012; Han et al., 2002, 2003; Pols, 2004; Van Winckel, 2018). Nevertheless, the binary post-AGB stars surrounded by Keplerian disks in our sample present orbital periods ranging from 100 to 3 000 days (derived from their radial-velocity curves; see Fig. 1.18 and Oomen et al., 2018; Van Winckel et al., 2014), so they lie in the middle of that bimodal distribution (see Figs. 1.18 and 1.19 *left*). The orbits of these binary post-AGB stars reveal that the common envelope stage was rapidly traversed or even avoided (see e.g. Kluska et al., 2022). Consequently, the angular momentum is transferred to circumstellar layers due to the interaction of the stellar system, and therefore, a disk is created (see Fig. 1.17).

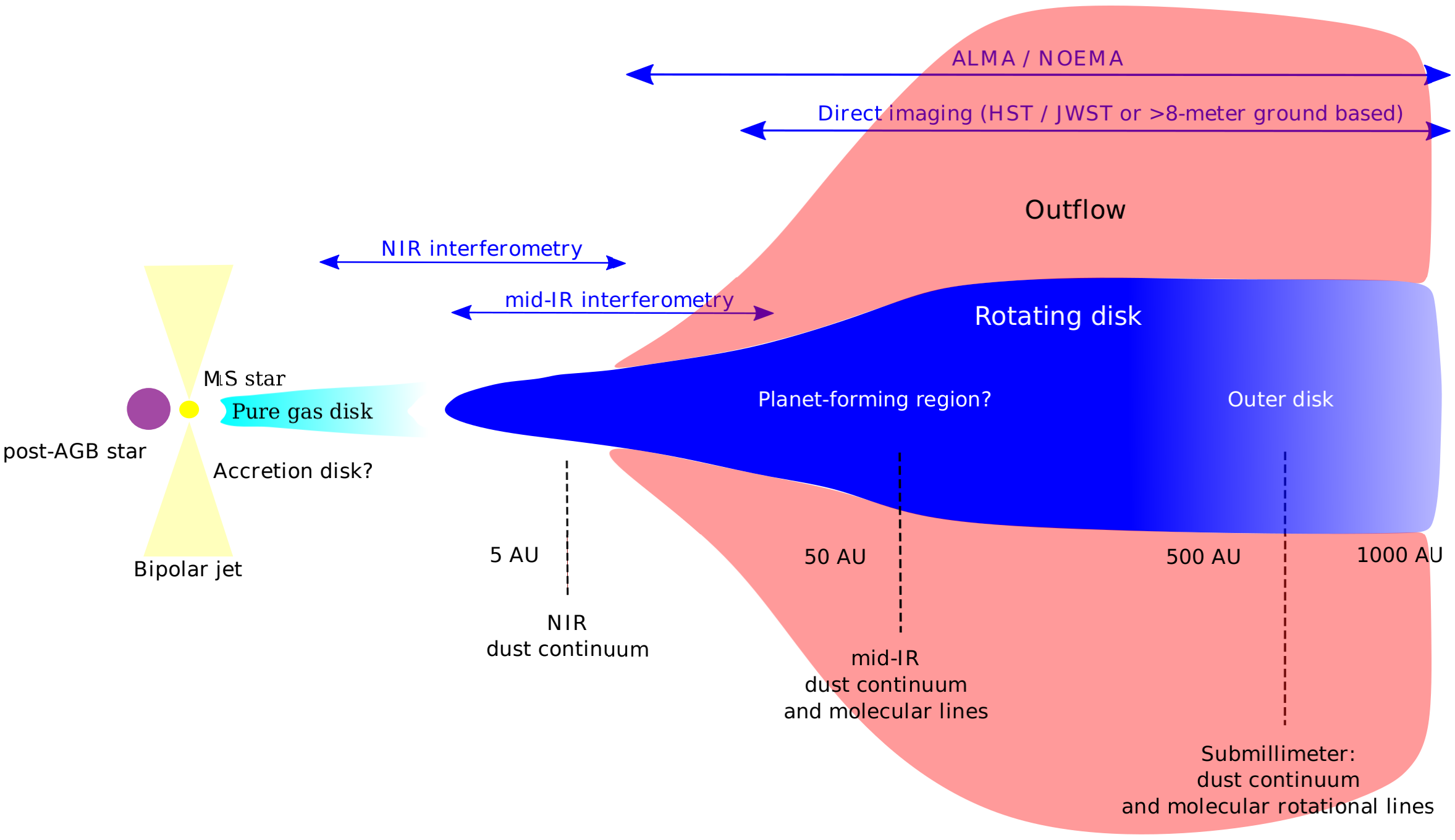

Figure 1.17: Sketch of a binary post-AGB star surrounded by a disk. We see a post-AGB star that belongs to a binary system surrounded by a disk with Keplerian (or quasi-Keplerian) dynamics and extended outflows of gas escaping from the disk. The presence of an accretion disk is possible due to the accreted material from the circumbinary disk by the companion star. Inner regions of the disk ($< 30$ AU) can be observed at NIR and mid-IR wavelengths through interferometric observations with the Very Large Telescope Interferometer (VLTI). These regions show the innermost and warmest zones of the disk (see e.g. Hillen et al., 2017; Kluska et al., 2019). The IR spectra of these sources reveal the presence of highly processed grains (see e.g. Sahai et al., 2011). The intermediate and outermost regions of the disk ($\sim 30 - 1\,000$ AU) can be observed through interferometric millimeter(mm)-wave maps of CO lines and dust continuum, with ALMA or NOEMA (see e.g. Bujarrabal et al., 2018b). They can also be observed through single-dish to find emission of molecular rotational lines.

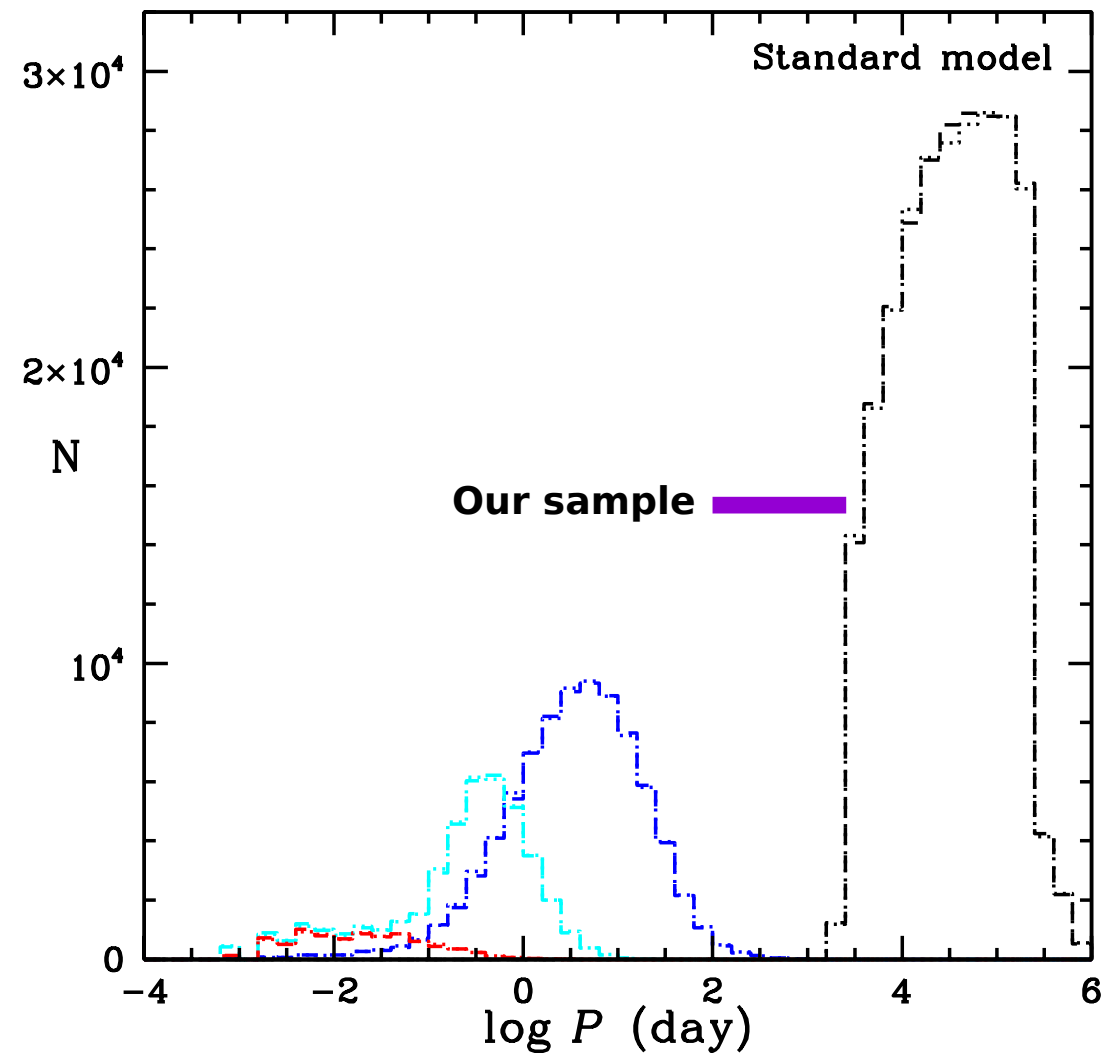

Figure 1.18: Orbital periods of binary planetary nebula nuclei, PNNe (black), close binary PNNe (blue), post-RGB and post-EAGB binaries (cyan), and double degenerate secondaries (red). Predictions of the evolutionary fate of binary systems is made with the standard model used by Nie et al. (2012). They clearly show a bimodal distribution. The orbital periods of our sample of binary post-AGB stars are represented with a horizontal purple line. Figure adapted from Nie et al. (2012).

The binary post-AGB stars in our analysis present a separation between 0.4 – 6 AU, and most of these binaries have a separation smaller than ∼ 4 AU (see Fig. 1.20 *left* and Van Winckel, 2018; Oomen et al., 2018). Another important orbital parameter, directly related to the angular momentum of the system, is the eccentricity of the orbit ($e$). The representation of $e$ as a function of the period is a relevant tool to understand the properties of a population of binary post-AGB stars (see Fig. 1.20 *middle* and *right*). This diagram shows that a relevant fraction of the sample (30%) have perfectly circular orbits, while the remaining 70% of the binary systems present eccentricities ranging up to 0.7. These binary systems share many traits when compared to other types of binaries that have undergone an interaction with a red giant star (those binary systems that present small orbital periods in the bimodal distribution of Fig. 1.18). However, and according to their relatively low orbital periods, our binary post-AGB stars have probably avoided the common envelope phase, perhaps because of the loss of angular momentum or eccentricity-pumping[2] by the Keplerian disks. Therefore, additional knowledge about these sources is necessary to understand these eccentricities and orbital periods (Van Winckel, 2018; Oomen et al., 2018; Hrivnak et al., 2017).

The post-AGB stars that are part of our binary systems have a high luminosity (1 000 – 14 000 $L_{\odot}$, see Bujarrabal et al., 2013a), intermediate spectral types (F - G), and significant FIR excess, indicative of the presence of stellar ejecta. In addition, they show a remarkable near infrared (NIR) excess in their spectral energy distributions (SEDs) due to thermal emission of relatively hot dust (see Fig. 1.19 *right*). Their IR spectra reveal the presence of highly processed grains (Sahai et al., 2011). Today, it is known that these characteristics of the SED from this kind of sources are indicative of the presence of stable circumbinary disks (de Ruyter et al., 2006; Hillen et al., 2013; Gezer et al., 2015; Oomen et al., 2018).

[2]Phase-dependent mass loss on the AGB might pump the eccentricity of the orbit by tidally enhancing the mass loss rate at the periastron of the orbit (Bonačić Marinović et al., 2008).

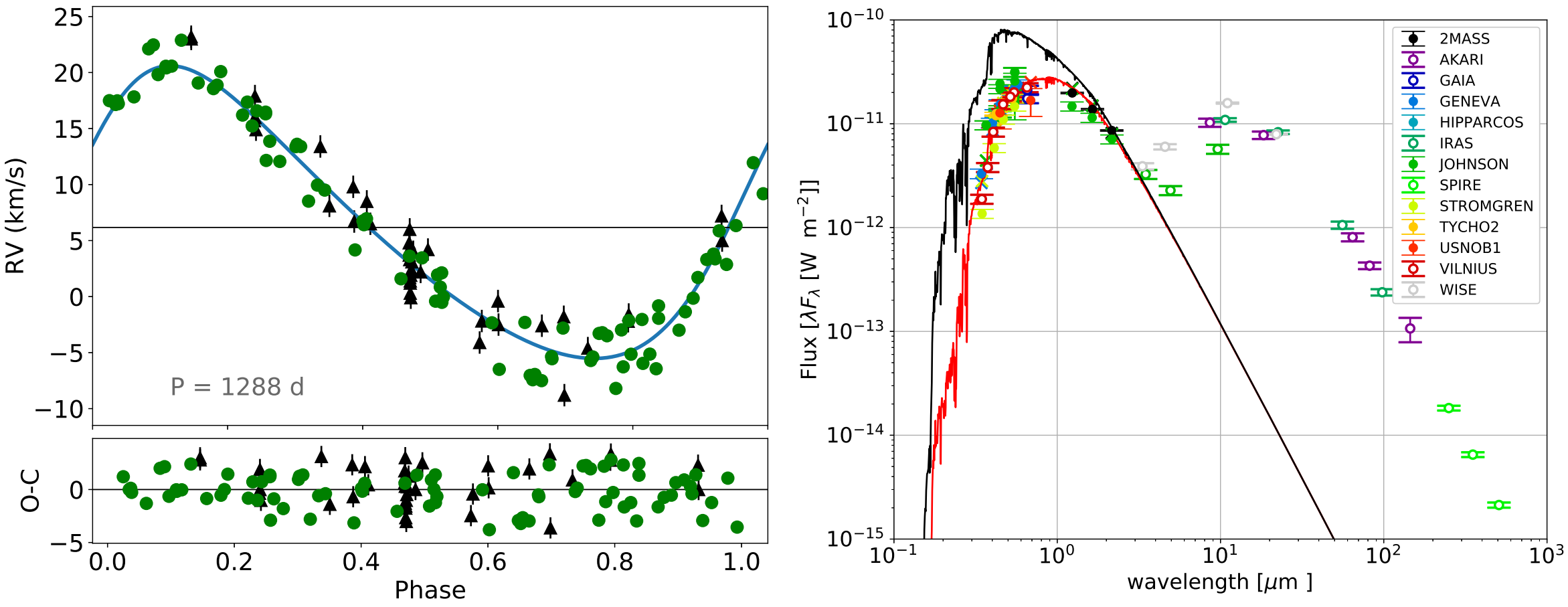

Figure 1.19: *Left*: Radial-velocity curve of HD 52961, one of the binary post-AGB stars of the sample. The green circles show data points taken with the HERMES spectrograph, while the black triangles represent older data. The horizontal black line represents the systemic velocity of the source. The residuals after subtraction of the orbital best model fit (blue line) are shown below the radial-velocity curve (O-C graph). *Right*: SED of AC Her, a well-known binary post-AGB star. The black curve represents the atmospheric model, the red one is the reddened model, and the symbols correspond to the observed fluxes in different photometry surveys. Figures taken from Oomen et al. (2018).

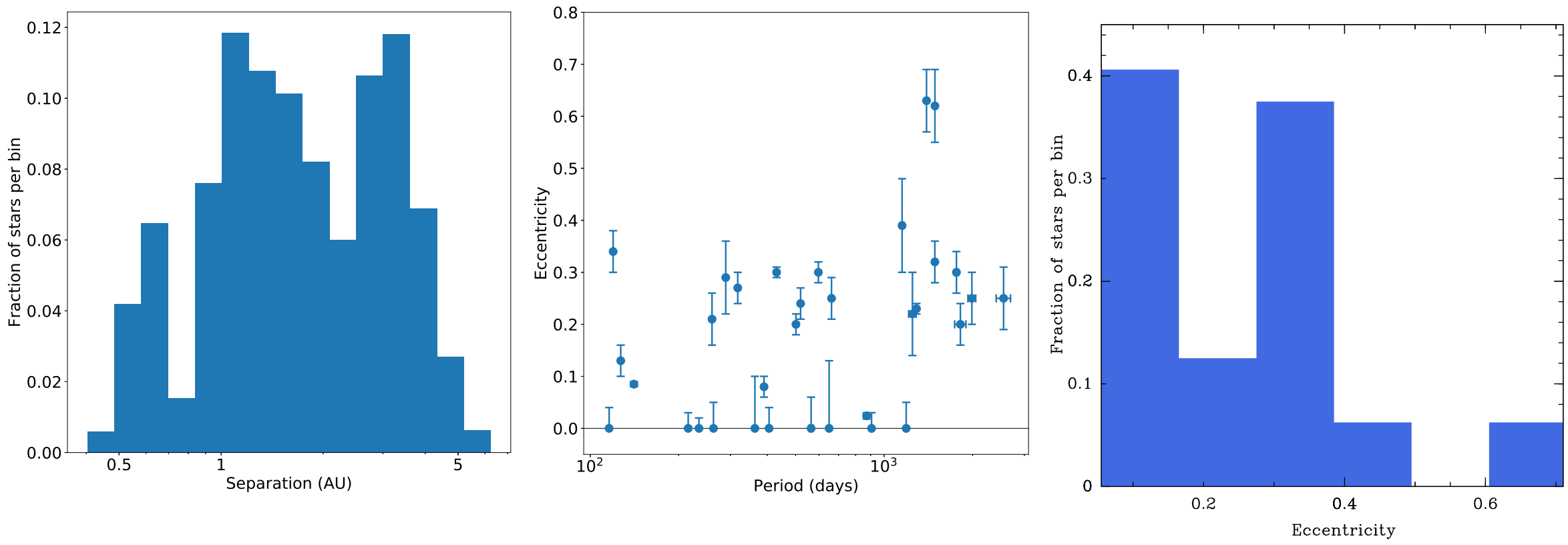

Figure 1.20: *Left*: Histogram of the distribution of separations in astronomical units for binary post-AGB stars. The Y-axis represents the fraction of binaries in that bin with a total number of 15 bins. The binary systems of the sample present a separation between $0.4-6$ AU. *Middle*: The $e-\log P$ diagram of a sample of binary post-AGB stars. Note that 30% of the binary systems have orbits indistinguishable from circular orbits ($e \simeq 0$), while the remaining 70% of these systems present eccentricities ranging up to 0.7. Figures taken from Oomen et al. (2018). *Right*: Histogram of the distribution of eccentricities for this sample. The Y-axis represents the fraction of sources in that bin with a total number of 6 bins.

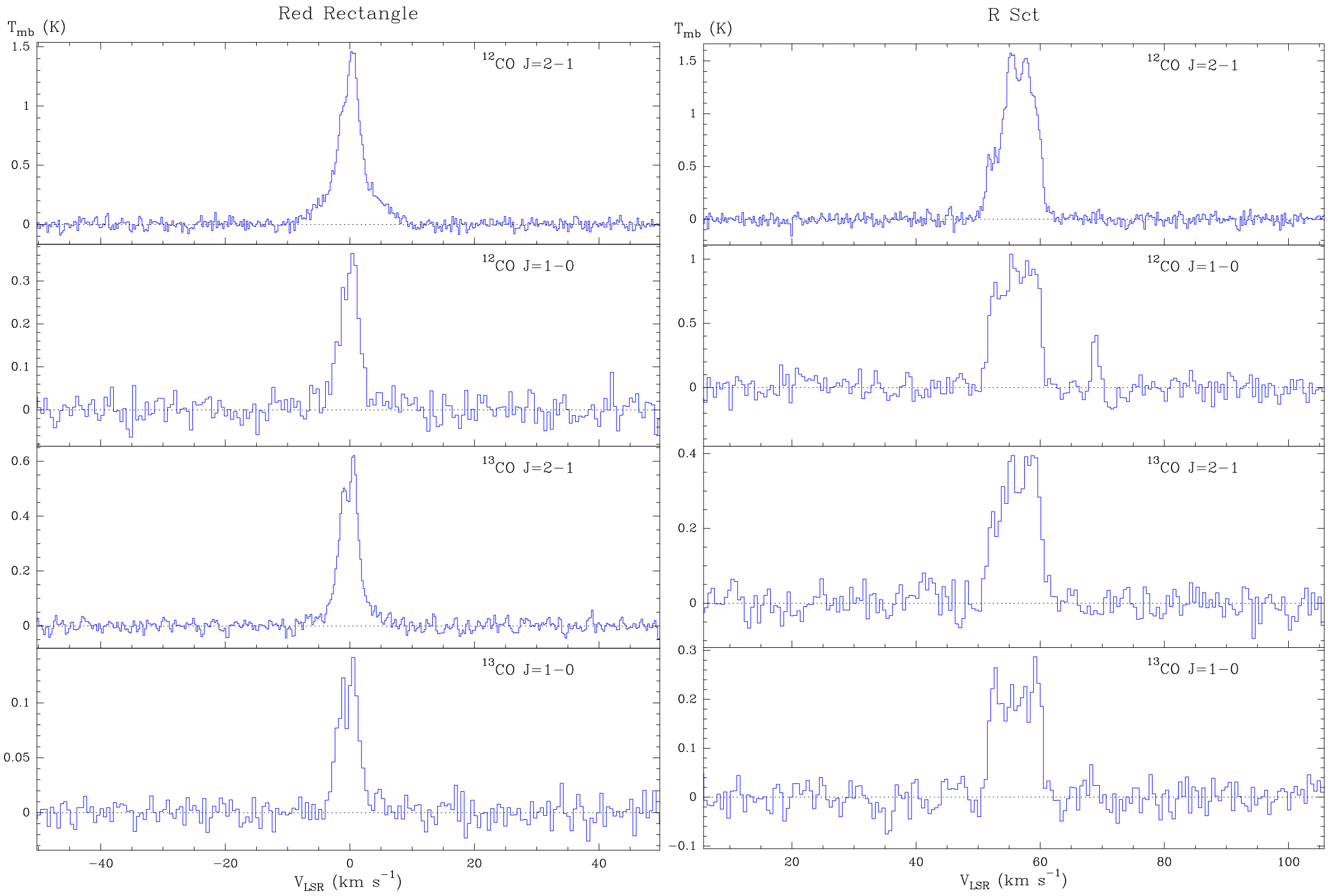

Figure 1.21: $^{12}$CO and $^{13}$CO profiles observed with the 30 m IRAM telescope in the Red Rectangle (*left*) and in R Sct (*right*). We find two kinds of spectra in these sources: narrow CO line profiles with relatively wide wings characteristic of (almost) pure rotating disks, such as in the Red Rectangle, and composite CO line profiles that include a narrow component, which represents emission from the Keplerian disk, and intense wings, which represent a significant contribution from an expanding component, such as in R Sct. Figures adapted from Bujarrabal et al. (2013a).

### 1.4.1 The nebulae around binary post-AGB stars

In Fig. 1.16 we present an image of the best-studied source among this wide class of post-AGB nebulae: the Red Rectangle. It consists of an equatorial rotating disk orbiting an evolved close binary star, with a period of about 320 days and a spectacular X-shaped axisymmetric nebula in expansion that can be studied in several wavelengths (see Men'shchikov et al., 2002; Cohen et al., 2004; Bujarrabal et al., 2016).

The rotating disks that surround binary post-AGB stars have been studied mainly in infrared and radio wavelengths. These disks can be spatially resolved, from the inner edge to the outer radius, thanks to very precise and accurate interferometric studies.

#### Analysis of Radio data

The CO emission from these sources has been observed via single-dish (30 m IRAM and APEX), yielding low-angular resolution data, and via mm-interferometers (NOEMA and ALMA telescopes), allowing for high-resolution mapping. Binary post-AGB stars have been thoroughly studied by means of $^{12}$CO and $^{13}$CO, $J = 3 - 2$, $J = 2 - 1$, and $J = 1 - 0$ lines, via single-dish observations (Bujarrabal et al., 2013a). They systematically show narrow CO line profiles and relatively wide wings, archetypal of rotating disks (see Fig. 1.21 *left*). These line profiles are similar to those found in young stars surrounded by rotating disks made of interstellar medium remnants (see e.g. Bujarrabal et al., 2005; Guilloteau et al., 2013). On the contrary, composite CO line profiles that include a narrow component, which likely represents emission from the rotating disk, and intense wings that come from extended and massive outflows surrounding the disks (see Fig. 1.21 *right*), have also been found. Thus, these binary post-AGB stars are surrounded by Keplerian disks and by an extended expanding component that can be even more massive than the disks. Even in relatively narrow profiles, like those of the Red Rectangle, a significant contribution of the emission from the outflow has been identified (Bujarrabal and Alcolea, 2013; Bujarrabal et al., 2013b), showing that the presence of more or less massive outflows is widespread.

One of the most relevant probes that confirms the stability of these disks comes from the detection of Keplerian dynamics (see Fig. 1.22). The CO emission of several sources has been thoroughly studied using the NOEMA and ALMA interferometers. Thanks to precise mm-interferometric maps and detailed models, the structure and the mass of the nebula (disk and outflow) can be estimated. The velocity field can be inferred as well, because the outflow presents an expanding velocity and the rotating disk presents Keplerian (and quasi-Keplerian) dynamics that allow us to calculate the mass of the stellar system (see Figs. 1.22 and 1.23).

The nebula around a binary post-AGB star is composed of a rotating and an expanding component. The expanding component is likely a disk wind consisting of gas escaping from the rotating component. The origin of this disk wind may be a low-velocity magnetohydrodynamical gas ejection from the outer regions of the disk. The estimated mass-loss rates lead to disk lifetimes between 5 000 and 20 000 years (Bujarrabal et al., 2016). Before the work of this thesis, only four disk-nebulae around binary post-AGB stars had been thoroughly studied in radio lines: the Red Rectangle (Bujarrabal et al., 2013b, 2016), AC Her (Bujarrabal et al., 2015), IW Car (Bujarrabal et al., 2017), and IRAS 08544−4431 (Bujarrabal et al., 2018b). We find typical disk-radii of $10^{15} - 10^{16}$ cm and disk-masses of $10^{-3} - 10^{-2}$ $M_{\odot}$ that account for around 90% of the total nebular mass. The derived velocity, density and temperature of the rotating disk of the Red Rectangle can be seen in Fig. 1.22 *right*.

It is also important to observe the dust continuum emission. The segregation of dust grains of different sizes depends on the distance from the equator. Large dust grains appear in stable disks and are expected to be located closer to the equator (both in disks around post-AGB stars and in disks around young stars, see e.g. Williams and Cieza, 2011; Lin, 2019). Dust grains around $\sim 1$ mm are abundant in disks around post-AGB stars and precede the creation of larger bodies (see e.g. Völschow et al., 2014; Nixon et al., 2018; Kluska et al., 2022; Scicluna et al., 2020). However, observational data are not detailed enough to discriminate between grain segregation and the whole dusty disk structures (see Fig. 1.23 and see e.g. Bujarrabal et al., 2016).

Additionally, the presence of molecular emission in these sources is poor. Before this PhD work, apart from CO, only $H^{13}CN$ emission had been discovered, alongside with the atomic lines C I and C II (in the Red Rectangle, see Bujarrabal et al., 2016). The detection of these lines is consistent with the presence of a PDR, because they are good tracers of these regions (see e.g. Agúndez et al., 2008).

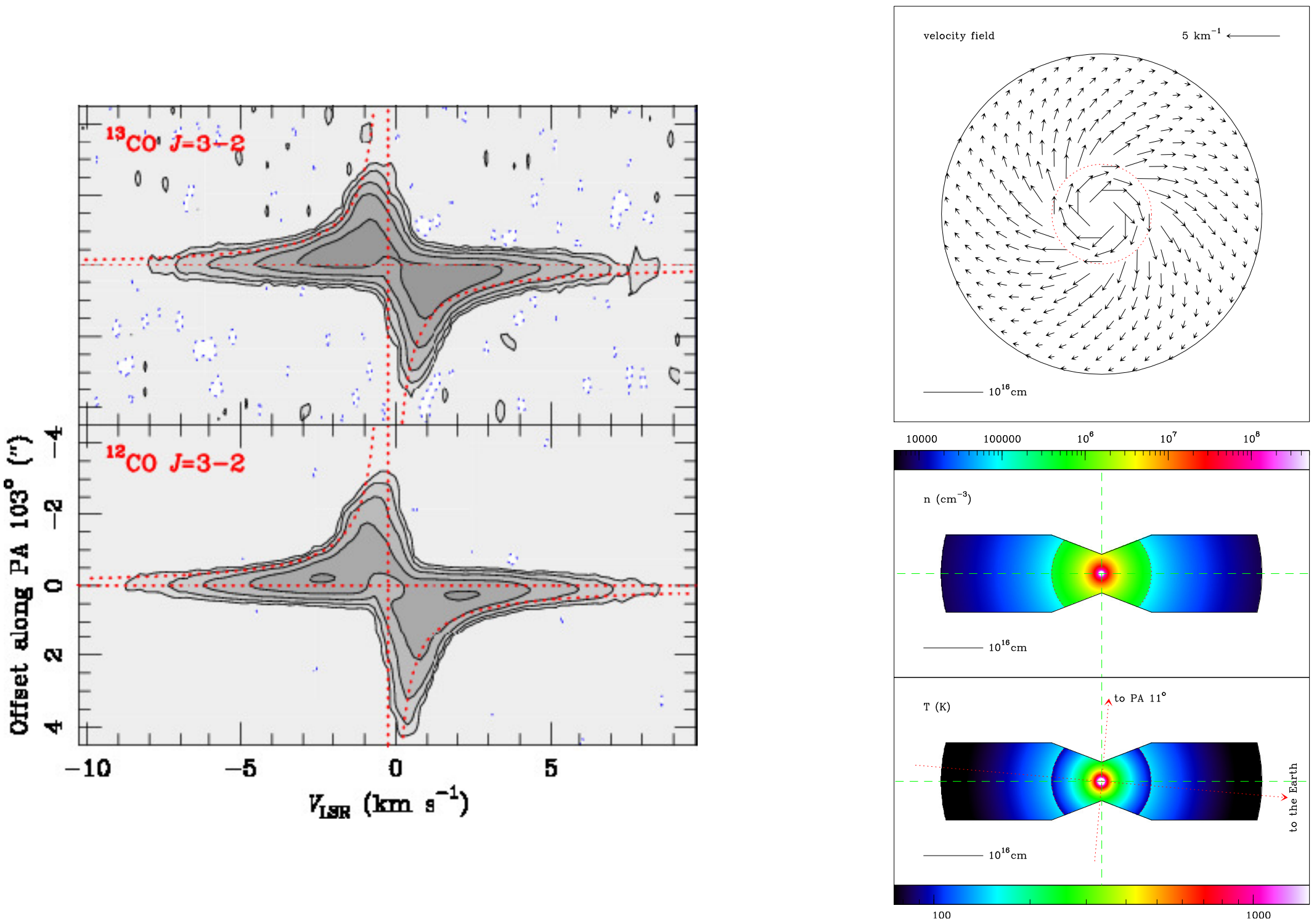

Figure 1.22: *Left*: PV diagrams along the equatorial direction of the ALMA observations of the Red Rectangle in $^{12}$CO and $^{13}$CO $J = 3-2$ emission. Horizontal and vertical lines indicate the central position and systemic velocity of the nebula; hyperbolae in red dots show the distributions of the emission of gas in Keplerian rotation. Figure taken from Bujarrabal et al. (2016). *Right*: Keplerian and sub-Keplerian velocity, density, and temperature distribution of the Red Rectangle disk. The red circle indicates the value of the Keplerian radius, from which the dynamical regime varies. Figure taken from Bujarrabal and Alcolea (2013).

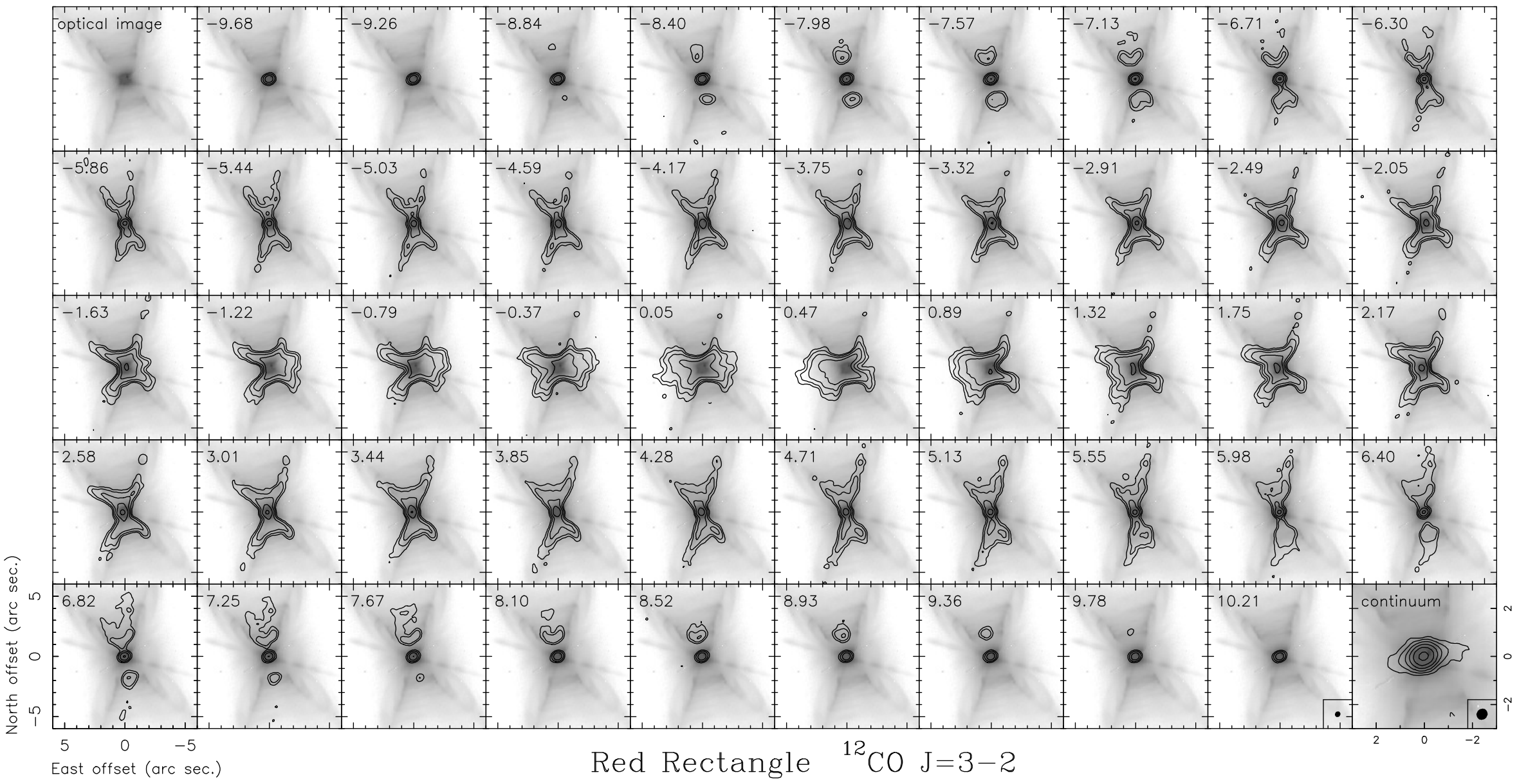

Figure 1.23: ALMA observations of the Red Rectangle in $^{12}$CO $J = 3-2$ emission, superposed on an *HST*/WFPC2 optical image. The panel at the bottom right corner shows the 0.85 mm continuum image zoomed by a factor 2 and the inserts in the last two panels show the beam width. Figure taken from Bujarrabal et al. (2013b).

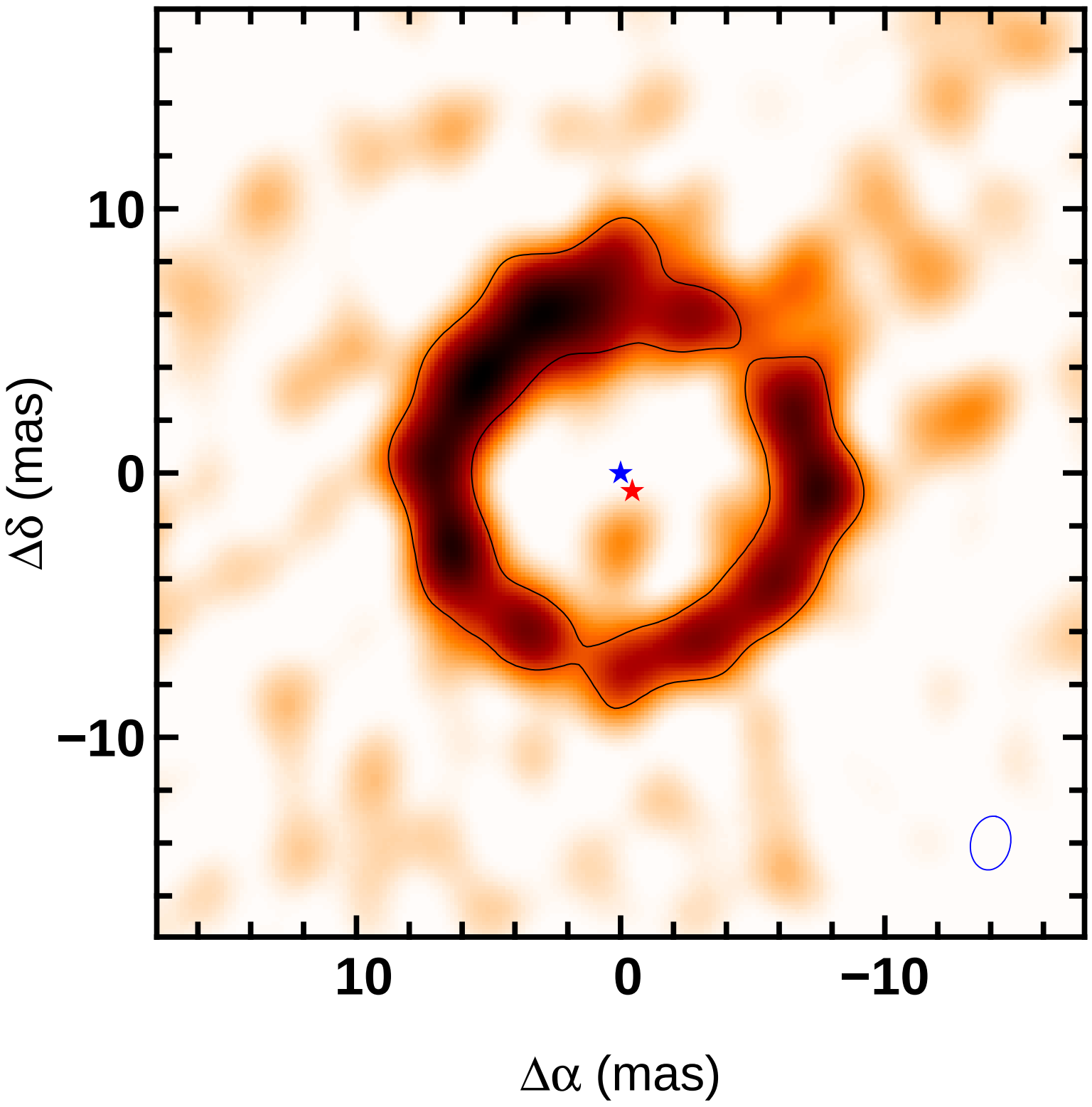

Figure 1.24: Circumbinary disk of IRAS 08544−4431 in the $H$-band (1.65 $\mu$m) with PIONIER at the VLTI. The blue and red stars indicate the positions of the primary and secondary stars, respectively. The blue ellipse represents the beam size. Figure taken from Kluska et al. (2018).

### Analysis of IR data

Observations at infrared wavelengths are useful tools to reveal the most compact regions of these disks. We have already discussed the results from IR photometry. The most compact and innermost regions of the circumbinary disks around post-AGB stars are also observable by means of near- and mid-IR long-baseline interferometric observations. These interferometric data are compared with detailed physical modeling (see Deroo et al., 2006, 2007; Hillen et al., 2013, 2014, 2015, 2016, 2017; Kluska et al., 2018, 2019). Pieces of information derived from the combination of observational data and theoretical models reveal that the structure of the circumbinary disk is that of a dusty settled disk. The growth of dust grains is significant and, alongside with the observational evidence of dust grain processing in the form of a high degree of crystallization, it is a strong indicator of long-term stability (see e.g. de Ruyter et al., 2005; Gielen et al., 2008). Dust grains as large as 1 mm are abundant in post-AGB disks and could be the precursors of larger solid bodies. Additionally, it is thought that planetoids and even a second generation of planets can be formed during the lifetimes of post-AGB disks in the presence of instabilities or dust traps, and could systematically appear around post-AGB stars and white dwarfs (see e.g. Nixon et al., 2018; Scicluna et al., 2020; Kluska et al., 2022).

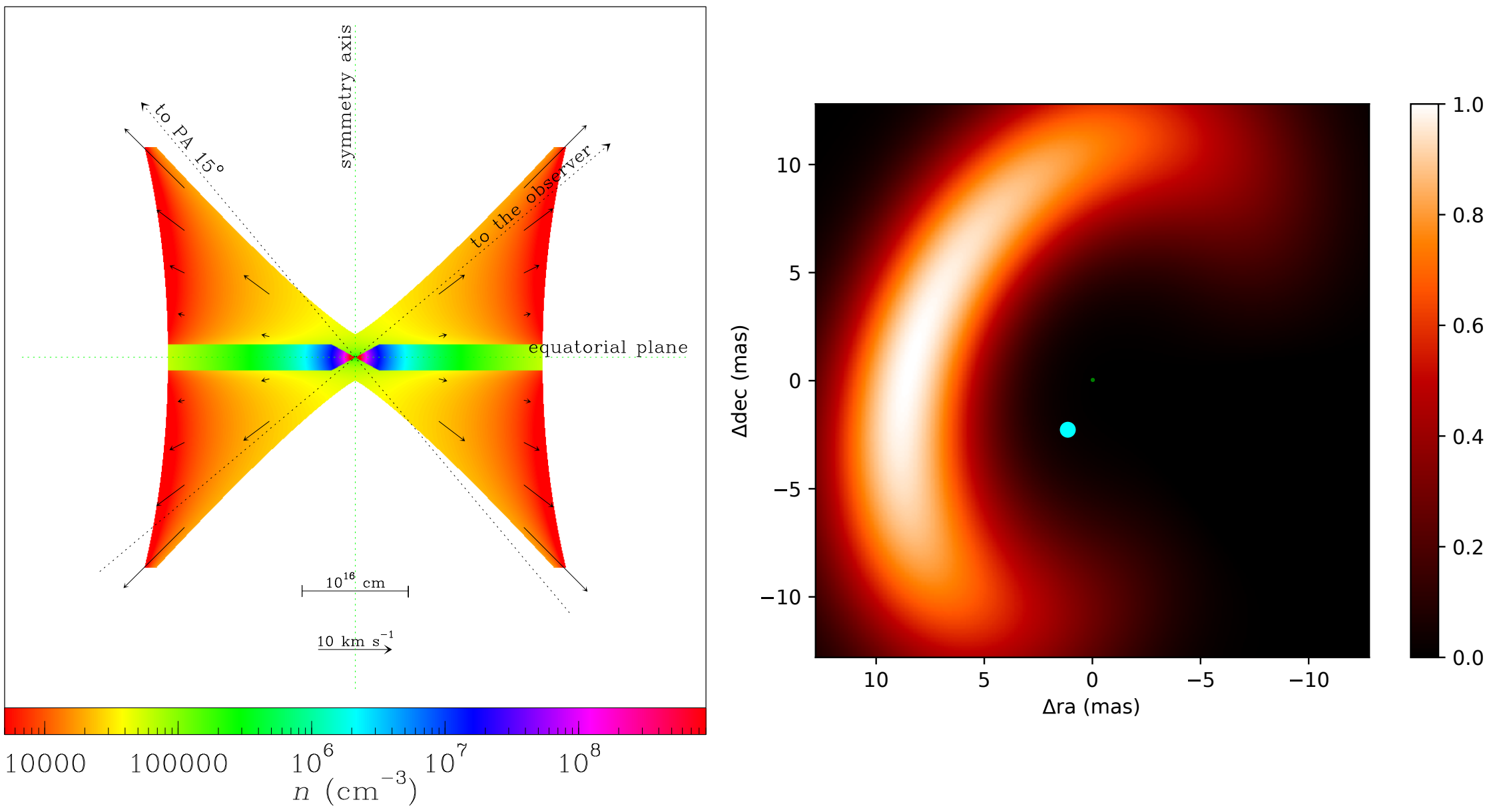

Figure 1.25: Comparative of the disk of IW Car studied both in infrared and millimeter wavelengths. *Left*: ALMA best-fit model for the Keplerian disk and the outflow, from CO $J = 3-2$ ALMA observations, which shows the structure and distribution of the density and the expansion velocity of the outflow. The inclination of the circumbinary disk with respect to the line of sight is represented by the red dashed line. Figure taken from Bujarrabal et al. (2017). *Right*: VLTI/PIONIER best-fit model image of the inner rim in the circumbinary disk, expressed in normalized flux units. Assuming a distance of 1 kpc, the size of the innermost region of the disk is around $3 \times 10^{13}$ cm, which is significantly smaller than the whole size of the disk (*right*). The green star represents the primary star, while the cyan star represents the secondary one. Figure adapted from Kluska et al. (2019).

### Global view

The study of these sources in mm and in near- and mid-IR wavelengths is crucial to understand the formation, evolution, and demise of these Keplerian disks. The observation of these objects in IR-wavelengths allows to analyze the most compact and warmest regions of the disk, which are the closest parts to the binary system. On the contrary, the observation of binary post-AGB nebulae at radio wavelengths permits to comprehend more distant and colder regions of the disk. The Keplerian dynamics of the disk allow to estimate the total mass of the stellar system. Additionally, the outflowing and expanding component that surrounds the Keplerian disk can be studied at mm-wavelengths.

In Fig. 1.25 we display the nebula around IW Car, studied both in infrared and radio wavelengths. Fig. 1.25 *left* shows the entire disk that surrounds the binary system. The outflowing and expanding component that surrounds the circumbinary disk is also present. Fig. 1.25 *right* shows the innermost region of the disk, which is the warmest part. The size of the inner rim of the disk is notably smaller compared to the whole size of the rotating component. Spatially resolved IR observations of the inner rim of the disk permit the estimation of the inner radius, height, even the eccentricity of the orbit of the binary system.

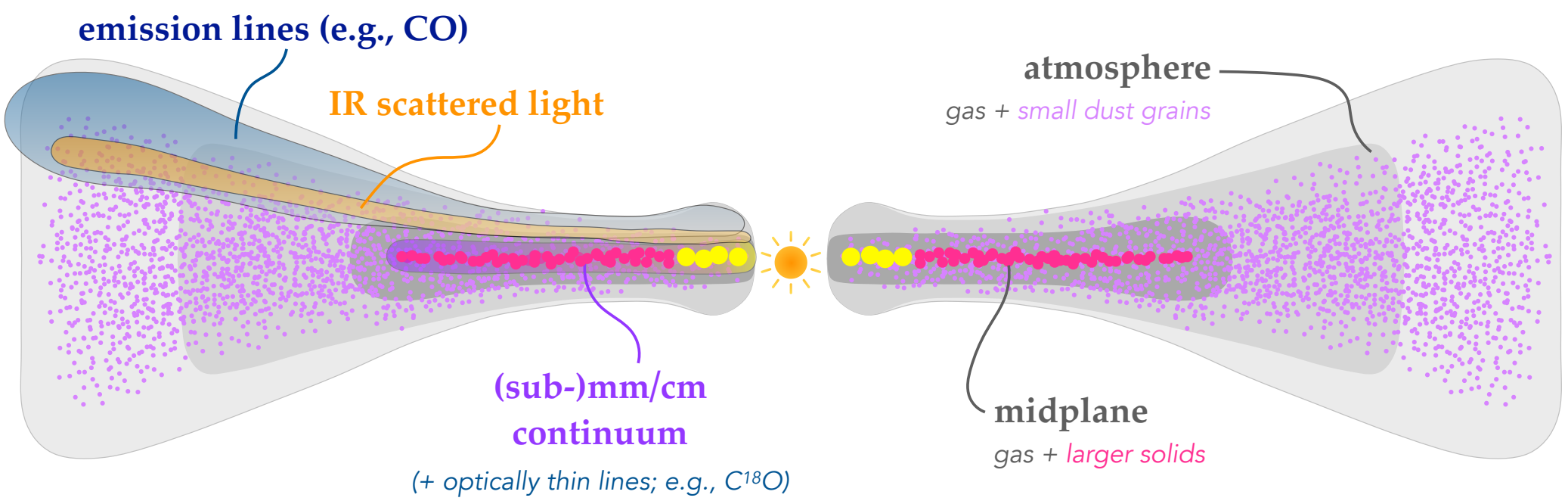

Figure 1.26: Diagram showing the protoplanetary disk structure. The gas is represented in gray scale, while solids are represented with exaggerated sizes and colours. The whole disk is composed of a midplane (gas and larger solids) and a atmosphere (gas + small dust grains). The left side of the sketch shows the approximate locations of emission tracers. Figure taken from Williams and Cieza (2011).

### Comparison of disks of different origin

The physical and chemical conditions of rotating disks around the binary post-AGB stars in our sample are barely known. Nevertheless, there are similar disks around young binary and single stars. Their formation, evolution, physical, and chemical properties differ from those of our disk-containing nebulae; they are known as protoplanetary disks (PPDs). See Williams and Cieza (2011); Andrews (2020); Dutrey et al. (2014); Henning and Semenov (2013) as general references.

PPDs are simple consequences of angular momentum conservation during the formation of a low-mass star through gravitational collapse. At the beginning, these disks funnel material onto the low-mass star, but the accretion rate decreases and a small amount of material persists, as the surrounding molecular core is exhausted. The disk mass limits the protostellar accretion rates, and the disk density distribution defines the angular momentum transfer that determines the mass of the star. These flattened and rotating disks are composed of cool dust and gas, extending from 10 to a few hundreds AU, and they orbit around young low-mass stars after their birth. PPDs can be described as rotating dusty gaseous structures transporting a net amount of mass towards the central star and the angular momentum outwards (see Fig. 1.26). By contrast, in our disk-containing nebulae around evolved stars, the presence of a stellar companion is absolutely necessary, since the ejected shells would not have enough angular momentum to form the disks otherwise. Our Keplerian disks can be larger than PPDs, because their size could reach 1 000 AU.

The shape and kinetics of PPDs and post-AGB disks may look similar, but they show significant differences. In both kinds of objects there is a remarkable increase of the height over the equator with respect to the distance to the axis. In the case of the "flaring disks" often-times found around young stars (see e.g. Guilloteau and Dutrey, 1998), that increase is due to the expected variation of the vertical dispersion in passive Keplerian disks under hydrostatic equilibrium, with all the gas in those flaring disks undergoing Keplerian rotation. However, what we see far from the equator in our post-AGB disks is outflowing gas, which likely escapes from the rotating disk as a "disk wind" powered by magnetocentrifugal effects (a phenomenon that is also observed in well-known PPDs); see e.g. Ferreira et al. (2006), their Fig. 1 is enlightening. There is a significant jump in the density between both components in post-AGBs, as the least dense gas is present in the outflow. In addition, a slow expansion is detected in the equatorial directions; the rotation regime is under-Keplerian in most of the disk.

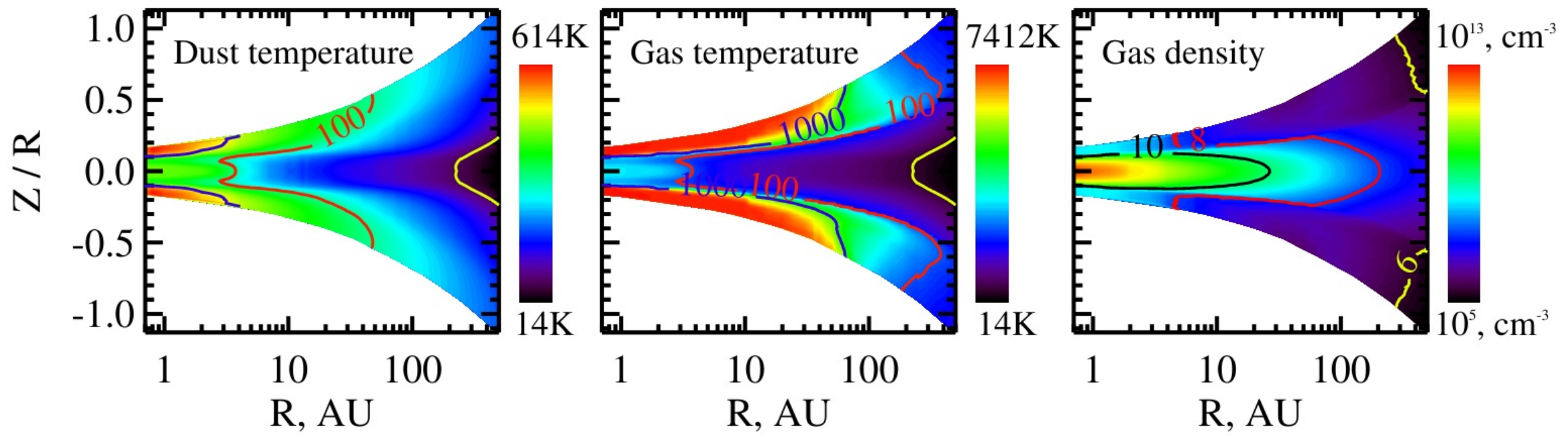

Figure 1.27: Density and temperature distribution for the gas and dust of a protoplanetary disk. Figure taken from Henning and Semenov (2013).

PPDs are characterized by a relatively fast evolution, since their lifetime is $\sim 10^6$ a. During this time span, a fraction of the material is accreted onto the star, whereas more material is lost through the outflows and photoevaporation, and the rest may condense into small bodies or planetesimals (since PPDs are the essential reservoirs of material for planet formation). These small bodies will evolve during the life of the disk from micron-sized dust particles, to pebbles, and finally to planets. In our study, the lifetime of the post-AGB disks ranges between 5 000 and 20 000 a, which is a hundred times shorter than that of PPDs. However, the presence of highly processed dust grains has been confirmed in our disks, and the formation of planetoids (or even a second generation of planets) could be possible; it is under debate.

High-resolution radio observations of gas and dust in PPDs provide fundamental information on their mass distribution and dynamical states. These rotating structures are characterized by a strong radial, and even vertical, temperature and density gradient (see Fig. 1.27). The gas temperature can vary from a few tens to $\lesssim 10\,000$ K, and up to 1 000 K in the case of the dust. The gas density of PPDs varies between $10^5$ and $10^{13}\,\mathrm{cm}^{-3}$. These densities and sizes yield disk mass values between 0.3 and 2.5 $M_\odot$. By contrast, our post-AGB disks present typical gas densities of $10^5 - 10^8\,\mathrm{cm}^{-3}$ and temperature values of $20 - 200$ K (see Fig. 1.22 *right*). Under these conditions, our circumbinary disks tend to present a molecular mass of $10^{-3} - 10^{-2}\,M_\odot$.

Since $H_2$ cannot be used as a tracer of the disk gas, it is necessary to study the most abundant molecules after $H_2$. This way, mm and sub-mm observations have improved the knowledge of the chemical composition of PPDs. Apart from CO, numerous molecules have been detected in these kinds of disks, such as: $^{13}CO$, $HCO^+$, $H^{13}CO^+$, $DCO^+$, CS, HCN, $HC_3N$, DCN, CN, $H_2CO$, $N_2H^+$, $C_2H$, $C_3H_2$, in addition to many other molecules of recent discovery. On the contrary, previously to this work, the molecular content of our disk-containing nebulae (and standard pPNe) was practically unknown, apart from CO (and $^{13}CO$). Now, and thanks to our deep survey, it has been inferred that some of our sources present weak emission in a few molecular lines: $C^{17}O$, $C^{18}O$, SiO (thermal and maser), $^{29}SiO$, $^{30}SiO$, SO, $SO_2$, CS, HCN, $H^{13}CN$, SiS, $H_2O$ (maser), OH (maser). This is the total molecular inventory of our disk-containing nebulae. The molecular richness of our sources is low when compared to that of PPDs, which have a higher molecular abundance, especially in ions and radicals (and of course deuterated species); see e.g. Guilloteau et al. (2016).

Table 1.2 Comparison of the main magnitudes between our rotating disks and PPDs.

| Disks | $M_{\rm disk}$ [$M_\odot$] | $R$ [AU] | $n$ [cm$^{-3}$] | T [K] | $t_{\rm lt}$ [a] | Chemistry |
|---|---|---|---|---|---|---|
| Circumbinary disks | $10^{-3} - 10^{-2}$ | $\lesssim$1000 | $10^5 - 10^8$ | $20 - 200$ | $10^4$ | Poor |
| PPDs | $0.3 - 2.5$ | $10 - 10^2$ | $10^5 - 10^{13}$ | $10 - 10^4$ | $10^6$ | |

**Notes.** $M_{\rm disk}$ is the mass of the disk, $R$ is the radius, $n$ is the density, $T$ is the temperature, and $t_{\rm lt}$ represents the lifetime of the disk.

In conclusion, our circumbinary disks around binary post-AGB stars and PPDs are both rotating structures; they may look similar at a first sight, but present significant differences in practically all the relevant physical and chemical quantities, such as origin, lifetime, size, density, temperature, mass, or chemical abundances (see Table 1.2). Additionally, we note that the rotating disks in our study had never been thoroughly analyzed, specially from a theoretical point of view.

### This thesis work

The rotating disks around binary post-AGB stars are objects that have not been extensively studied (especially from a theoretical point of view). Until the work derived from this doctoral thesis, only four disk-containing nebulae around binary post-AGB stars had had a thorough dynamical analysis. All of them are rotating disks surrounded by weak outflows. Nevertheless, there are other disk-containing nebulae where the outflow may represent most of the total molecular mass. Thus, some of the main goals of this doctoral thesis are:

1. Completing the study of the gas properties of these disks-containing nebulae via detailed observations and modeling of more sources.

2. Analyzing the chemical properties of these objects through their molecular emission.

This way, we have considerably expanded our sample of studied sources through interferometric maps. This work reveals the existence of two kinds of nebulae around binary post-AGB stars that clearly conform a bimodal distribution: those dominated by the rotating disk, and those where the outflow is the dominant component. Between $85 - 95\%$ of the total mass in disk-dominated sources is located in the rotating structure. On the contrary, between $25 - 35\%$ of the total mass in outflow-dominated sources is located in the disk. The sizes of their expanding components tend to be significantly larger than those of the disk-dominated sources. These expanding components, mostly composed of cold gas, could present different shapes, such as an extended hourglass-like outflow, or an elongated structure with large cavities in the symmetry axis. This conclusion is supported by interferometric maps in mm wavelengths and by complex models. See Chapters 4, 6, and 7.

Additionally, the molecular content of these kind of sources was practically unknown (apart from CO) until our study. In this work, we present a deep survey of mm wavelength lines in binary post-AGB stars to search for molecules, which is a pioneer study in this field. Moreover, this survey significantly contributes to the search of molecules in standard pre-planetary nebulae, since the chemical studies on pPNe are relatively scarce (even the chemistry in PNe is poorly studied). Our sample can be classified as C-/O-rich via integrated intensities of pairs of molecules (other than CO). Moreover, thanks to our detailed analysis, the molecular richness of these sources and their abundances have been studied. See Chapters 5 and 6.

The work presented in this Ph.D. thesis significantly contributes to increase the knowledge of these disk-containing nebulae around binary post-AGB stars. In addition, the work derived from this doctoral thesis has been published in four refereed papers (see Gallardo Cava et al., 2021, 2022c,b,a).

*All it takes is one bad day to reduce the sanest man alive to lunacy. That's how far the world is from where I am. Just one bad day.*

Joker – The killing joke

# 2

# Radiation mechanisms

*This chapter aims to introduce the theory that support the work of this thesis. We describe the radiative transfer process and the physical conditions of the excitation of molecular lines. We also describe the physical conditions under which maser amplification occurs.*

## 2.1 Molecular quantum mechanics

Molecular spectroscopy studies the energetic changes that a molecule experiences when interacting with matter and/or radiation. As a consequence of this interaction, molecules can experience changes in their electronic, vibrational and rotational configuration. We consider a simple case of a diatomic and rigid molecule with masses $m_1$ and $m_2$ and internuclear distance $r$. Hence, the moment of inertia is:

$$I = \frac{m_1 m_2}{m_1 + m_2} r^2 = \mu r^2, \tag{2.1}$$

where $\mu$ is known as the reduced mass. The rotational energy ($E_{\mathrm{rot}}$) can be expressed as a function of the angular velocity ($\omega$):

$$E_{\mathrm{rot}} = \frac{1}{2} I \omega^2. \tag{2.2}$$

The angular momentum ($L$) of a rotating molecule is related to its moment of inertia $I$:

$$L = I\omega = \mu r^2 \omega. \tag{2.3}$$

Therefore, the rotational energy can be expressed as:

$$E_{\mathrm{rot}} = \frac{L^2}{2\mu r^2} = \frac{L^2}{2I}. \tag{2.4}$$

The energy levels of a rigid rotor molecule can be determined by solving the Schrödinger equation:

$$\hat{H}\psi = E\psi = i\hbar\dot{\psi}, \tag{2.5}$$

where $\hat{H}$ is the Hamiltonian operator:

$$\hat{H} = -\frac{\hbar^2}{2\mu}\nabla^2, \tag{2.6}$$

and $\psi$ is the molecular (rovibronic, rotation-vibration-electronic) wave function that describes the system and, according to the Born-Oppenheimer approximation, can be written as a single product of the electronic part ($\psi_{\rm e}$), the vibrational part ($\psi_{\rm v}$), the rotational part ($\psi_{\rm r}$), and the nuclear spin part ($\psi_{\rm ns}$):

$$\psi = \psi_{\rm e}\psi_{\rm v}\psi_{\rm r}\psi_{\rm ns}. \tag{2.7}$$

It can be applied in the most common case: the diatomic molecules. Thus, the decomposition of the total energy in additive terms allows us to solve for each term independently:

$$E_{\rm total} = E_{\rm electronic} + E_{\rm vibrational} + E_{\rm rotational} + E_{\rm nuclear\,spin}, \tag{2.8}$$

with $E_{\rm electronic} \gg E_{\rm vibrational} \gg E_{\rm rotational} \gg E_{\rm nuclear\,spin}$. The nuclear spin energy is so small that it is often omitted.

In this Ph.D. thesis, we focus on the rotational transitions. Therefore, applying the Hamiltonian to the rotational eigenstates, $|\psi_r\rangle = |J, M_J\rangle$, gives the following expression:

$$\hat{H}|J, M_J\rangle = \frac{L^2}{2\mu r^2}|J, M_J\rangle, \tag{2.9}$$

where $J$ and $M_J$ are the rotational angular momentum quantum number and the quantum number for rotation about $z$ in the space-fixed frame, respectively. Additionally, knowing that:

$$L^2|J, M_J\rangle = J(J+1)\hbar^2|J, M_J\rangle, \tag{2.10}$$

the Eq. 2.9 becomes:

$$\hat{H}|J, M_J\rangle = \frac{J(J+1)\hbar^2}{2\mu r^2}|J, M_J\rangle. \tag{2.11}$$

Finally, the eigenvalues of the rotational term corresponding to the rigid rotor energies are:

$$E_J = \frac{\hbar^2}{2I}J(J+1) = BJ(J+1), \tag{2.12}$$

where $B$ is the rotational constant and $J = 0, 1, \ldots$, with $\Delta J = \pm 1$, according to the quantum selection rule. Furthermore, for each $J$, there are $2J+1$ possible values of $M_J$: $M_J = 0, \pm 1, \pm 2, \ldots, \pm J$.

When a molecule decays from the rotational level $J$ to $J-1$, the rotational energy loss results in the emission of a photon, whose energy is directly proportional to the level $J$ and equal to the following:

$$\Delta E_{\rm rot} = \frac{\hbar^2}{I}J. \tag{2.13}$$

Taking into account the energy-frequency relation of Planck, we can calculate the frequency at which a photon is emitted from a rotating molecule level $J$ to $J-1$ level:

$$\nu_{J\to J-1} = \frac{\Delta E_{\mathrm{rot}}}{h} = \frac{\hbar}{2\pi I}J. \tag{2.14}$$

Thus, these energy changes can be observed at different wavelengths: electronic transitions are observable in ultraviolet and optical wavelengths; vibrational transitions are detectable in infrared frequencies; rotational transitions at radio wavelengths.

In a more general way, molecules are not rigid rotors and when they are rotating, the moment of inertia of this molecule increases and decreasing the value of the rotational constant $B$. Therefore, a centrifugal distortion term is added to the rotational energy levels of the diatomic molecule, and Eq. 2.12 becomes:

$$E_J = BJ(J+1) - DJ^2(J+1)^2, \tag{2.15}$$

where $D$ is the centrifugal distortion constant and it is positive for diatomic molecules.

## 2.2 Molecular lines in circumstellar gas

In the case of the sources studied in this thesis, nebulae around binary post-AGB stars, the study of molecular lines provides more intrinsic information about these objects (see Sects. 1.2 and 1.3 for further details). These kind of observations are crucial to study the different envelope layers. Observations of atomic lines are mainly performed with infrared satellites, such as *ISO*, *IRAS*, and *Herschell*. Atomic emission only appears in evolved sources, in which the central star may form an inner region of atomic gas, because of photodissociation. Most of the gas in envelopes around AGB stars or post-AGB stars is in molecular form. Circumstellar envelopes can be observed at different wavelengths:

- FIR:
  - Emission of fine-structure atomic lines.
  - Re-emitted light by the dust grains, heated by the stellar light.
- Radio and FIR: Emission of molecular lines, mainly rotational.
- NIR and visible:
  - Diffusion by circumstellar dust grains of light emitted by the central star.
  - Absorption by atoms or circumstellar molecules in the inner layers of the envelope.
  - Absorption by dust grains of the stellar continuum that it is seen in the form of extinction.
- Visible: Light diffused by atoms. Includes resonant diffusion and fluorescence.

### 2.2.1 Radiative transfer

The variation of the intensity of a light ray passing through a material medium can be expressed with the radiative transfer equation, that it is expressed as:

$$\frac{\mathrm{d}I_\nu}{\mathrm{d}s} = -I_\nu \kappa_\nu + j_\nu, \tag{2.16}$$

where $I_\nu$ is the radiation intensity, expressed as the energy per unit of time, frequency, and solid angle, that passes through the unit surface perpendicular to its direction of propagation; $ds$ is the differential element of length in the direction of propagation; $\kappa_\nu$ and $j_\nu$ are the absorption and emissivity coefficients of the medium, respectively. In the case of thermodynamic equilibrium, at a certain radiation equilibrium temperature $T$, $I$ is the intensity of the blackbody emission and it depends only on the frequency and the temperature:

$$\frac{\mathrm{d}I_\nu}{\mathrm{d}s} = 0 \rightarrow I_\nu = B_\nu(T) = \frac{2h\nu^3}{c^2}\frac{1}{\exp\left(\frac{h\nu}{kT}\right) - 1}. \tag{2.17}$$

We deduce the general relationship between the absorption and emission coefficients for an equilibrium temperature $T$, which means that whatever their values take, the ratio of emissivity and absorption must be equal to the function of Planck in any medium in thermal equilibrium. This rule is also known as the law of Kirchoff:

$$B_\nu(T) = \frac{j_\nu}{\kappa_\nu}. \tag{2.18}$$

We can define $B_\nu$ as the source function $S$, even if there is no thermodynamic equilibrium:

$$S_\nu = \frac{j_\nu}{\kappa_\nu}. \tag{2.19}$$

We can define the optical depth (or opacity, $\tau$) as the transparency measure of the medium:

$$\mathrm{d}\tau_\nu = \mathrm{d}\tau(\nu) = -\kappa_\nu\, \mathrm{d}s. \tag{2.20}$$

This coefficient is directly related to the medium density and the ability to absorb radiation of that emitting medium. We can distinguish optically thin ($\tau \ll 1$, transparent) or optically thick mediums ($\tau \gg 1$, opaque). Consequently, we find a solution to the radiative equation that expresses the intensity between two arbitrary points $s_0$ and $s$:

$$I_\nu = I_\nu(s_0)\, \mathrm{e}^{-[\tau(s)-\tau(s_0)]} + \int_{s_0}^{s} S_\nu\, \mathrm{e}^{-[\tau(s)-\tau(x)]}\, \mathrm{d}\tau(x). \tag{2.21}$$

Assuming $s_0$ as the origin, we can express Eq. 2.21 as:

$$I_\nu = I_\nu(0)\, \mathrm{e}^{-\tau(s)} + \int_{0}^{s} S_\nu\, \mathrm{e}^{-[\tau(s)-\tau(x)]}\, \mathrm{d}\tau(x). \tag{2.22}$$

Additionally, assuming a homogeneous medium:

$$I_\nu = I_\nu(0)\, \mathrm{e}^{-\tau_\nu(s)} + S_\nu\left(1 - \mathrm{e}^{-\tau_\nu(s)}\right) = I_\nu(0)\, \mathrm{e}^{-\kappa_\nu s} + S_\nu\left(1 - \mathrm{e}^{-\kappa_\nu s}\right). \tag{2.23}$$

Thus, the intensity of incident radiation is exponentially attenuated as it passes through the material ($\kappa_\nu > 0$) and it is reinforced by the medium emissivity. Thus, if a material medium is not transparent, it emits. However, in the case of $\kappa_\nu < 0$ this intensity is exponentially amplified as it passes through material (maser emission, see Sect. 2.2.3).

### 2.2.2 Excitation of molecular lines

Consider that the medium into which the radiation propagates can be simplified to a two-energy level system: upper (u) and lower (l). The absorption and emission coefficients can be expressed, as a function of the frequency of transition, as follows:

$$\kappa_\nu = (n_\mathrm{l} B_\mathrm{lu} - n_\mathrm{u} B_\mathrm{ul}) \frac{h\nu}{4\pi} \phi(\nu), \tag{2.24}$$

$$j_\nu = n_\mathrm{u} A_\mathrm{ul} \frac{hv}{4\pi} \phi(\nu), \tag{2.25}$$

where $n_\mathrm{l}$ and $n_\mathrm{u}$ represent the population of the upper and lower levels, respectively, and $\phi(\nu)$ is the normalized transition profile to the unit $\left(\int \phi(\nu)\, \mathrm{d}\nu = 1\right)$. The quantum processes involved in the radiative excitation of the molecule can be statistically described by the spontaneous emission, induced emission and absorption Einstein coefficients. The spontaneous emission coefficient ($A_\mathrm{ul}$) represents the probability that a molecule emits a photon spontaneously when decaying from the upper to the lower level. The induced emission coefficient ($B_\mathrm{ul}$) represents the probability that a molecule emits a photon decaying in a stimulated way. Finally, the absorption coefficient ($B_\mathrm{lu}$) is the probability that a molecule absorbs a photon and transitioning from the lower to the upper level. The three Einstein coefficients are not independent of each other and are expressed as:

$$A_\mathrm{ul} = \frac{8\pi h\nu^3}{c^2} B_\mathrm{ul}, \tag{2.26}$$

$$B_\mathrm{lu} = B_\mathrm{lu} \frac{g_\mathrm{u}}{g_\mathrm{l}} = A_\mathrm{ul} \frac{c^2}{8\pi h\nu^3} \frac{g_\mathrm{u}}{g_\mathrm{l}}. \tag{2.27}$$

We can define the transition excitation temperature ($T_\mathrm{ex}$) as the temperature that satisfies the following equation:

$$\frac{x_\mathrm{u}}{x_\mathrm{l}} = \exp\left(-\frac{h\nu_\mathrm{ul}}{kT_\mathrm{ex}}\right), \tag{2.28}$$

where $x_\mathrm{u}$ and $x_\mathrm{l}$ are the real population of the upper and lower levels and satisfy that $n_\mathrm{i} = x_\mathrm{i} g_\mathrm{i}$, where $g_\mathrm{i}$ is the statistical weight of the upper and lower level.

Note that we are using the same profile in frequencies for $j_\nu$ and $\kappa_\nu$ (known as complete redistribution hypothesis, see e.g. Sukhorukov and Leenaarts, 2017). The local width of the profiles exclusively depends on the microscopic velocity dispersion, because of the microturbulence. The profiles of $j_\nu$ and $\kappa_\nu$ comes from the velocity distribution at the considered energy levels. The complete distribution hypothesis is valid in those cases in which the population of the levels is collisional, because each level can be independently maxwellianized. This approximation is valid in most radiative transfer models (except for maser emission, for instance).

Thus, we only have to estimate the population levels using the statistical equilibrium equations, which are derived from population rates of a given level i, compared with another level j, with energies $E_\mathrm{i}$ and $E_\mathrm{j}$, due to radiative or collisional transitions:

$$\frac{\mathrm{d}n_\mathrm{i}}{\mathrm{d}t} = -\sum_{\mathrm{j}\neq\mathrm{i}} R_\mathrm{ij} n_\mathrm{i} + \sum_{\mathrm{j}\neq\mathrm{i}} R_\mathrm{ji} n_\mathrm{j} - \sum_{\mathrm{j}\neq\mathrm{i}} C_\mathrm{ij} n_\mathrm{i} + \sum_{\mathrm{j}\neq\mathrm{i}} C_\mathrm{ji} n_\mathrm{j}. \tag{2.29}$$

$R_{\rm ij}$ represents the probability of a radiative transition from level i to level j, while $C'_{\rm ij}$ is the probability of a transition from level i to level j after a collision, and can be expressed as follows:

$$C_{\rm lu} = C_{\rm ul}\frac{g_{\rm u}}{g_{\rm l}}\exp\left(-\frac{\Delta E}{kT_{\rm K}}\right). \tag{2.30}$$

It is derived from the principle of microreversibility and is expressed as a function of the kinetic temperature ($T_{\rm K}$). Both $R_{\rm ij}$ and $C'_{\rm ij}$ are expressed in $\rm s^{-1}$.

In evolved sources, the transition times between levels are shorter than evolution times of the physical and chemical conditions, so we can assume that the populations of the levels of a specie A are in equilibrium:

$$\frac{{\rm d}n_{\rm i}}{{\rm d}t} = 0 \rightarrow \sum_{\rm i} n_{\rm i} = n({\rm A}). \tag{2.31}$$

According to Eq. 2.29, the molecular excitation depends on the radiation and the collisions. The incoming molecular radiation depends on the photons arriving around the molecule, and its main origins are the central star, the background radiation, and dust grains. Those molecules with higher dipole moment are more affected by radiation than by collisions and tend to present higher $T_{\rm ex}$. In the case of $T_{\rm ex} = T_{\rm K}$ and a collision-dominated high-density regime, the excitation of the molecule thermalizes. Thermalization is facilitated if the transitions are opaque (see Eqs. 2.28 and 2.30). The critical column density for a gas to thermalize depends on the dipole moment of the molecule. Thus, a higher dipole moment implies that the excitation of the molecule thermalizes at high density values.

If a molecule is thermalized, it is easier to study it, because with one temperature value all its transitions are studied. CO is quickly thermalized and is abundant, so it is useful to trace low-density regions. At density values $\gtrsim 10^4\,{\rm cm^{-3}}$, the relevant CO lines are essentially thermalized. However, other molecules with higher dipole moment trace denser gas.

### The Large Velocity Gradient approximation

To exactly resolve these previous equations is too complex and in most of the cases the Large Velocity Gradient approximation is used. The Large Velocity Gradient, or LVG approach, is the most widely used approximation to radiative transfer (Sobolev, 1960). Each region is assumed to be independent of others. Therefore, regions are studied locally and the equations are solved for each region. It is applied to clouds with a large macroscopic velocity gradient, so the velocity variation between relatively close points is greater than the local velocity dispersion. Thus, a point only interacts radiatively with its immediate surroundings. Under this assumption, the opacity can be expressed as (see Castor, 1970):

$$\tau = \frac{\tau_0(\nu)}{1+\mu^2\frac{{\rm d}\ln v}{{\rm d}\ln r}-1} = \frac{\tau_0(\nu)}{1+\mu^2(\epsilon-1)}, \tag{2.32}$$

where $\epsilon$ is the logarithmic gradient of velocity, $\mu$ represents the cosine of the angle formed by the line of sight and the radial direction, and $\tau_0$ is a function that depends on the frequency:

$$\tau_0(\nu) = g_{\rm u}(x_{\rm l}-x_{\rm u})\frac{c^2 A_{\rm ul}}{8\pi\nu^2}\frac{rc}{\nu v(r)}. \tag{2.33}$$

Under the LVG assumptions, the averaged intensity at all angles and frequencies can be written as:

$$\bar{J} = S(r)\,(1 - \beta(r)) + \beta_c(r) I_c(r), \tag{2.34}$$

where $I_c$ is the averaged bright of the source and $\beta_c$ can be expressed in different ways depending on the continuum source. The parameter $\beta$ is the escape probability and can be defined as the chance of a photon escaping from the gas shell that we are considering. We find $\beta = 0$ if the source is opaque, and $\beta < 1$ otherwise:

$$\beta(r) = \int_0^1 \frac{1 - \mathrm{e}^{-\tau(\mu,r)}}{\tau(\mu, r)}\,\mathrm{d}\mu. \tag{2.35}$$

Normally, a constant gradient, $\epsilon = 1$, is applied when the actual gradient is unknown. In this case, we can rewrite $\beta$ as:

$$\beta = \frac{1 - \mathrm{e}^{-\tau_0}}{\tau_0} = \begin{cases} 1 - \tau_0/2 \rightarrow 1, & \text{if } \tau \ll 1 \\ 1/\tau_0 \rightarrow 0, & \text{if } \tau \gg 1 \end{cases}. \tag{2.36}$$

The LVG approach can present a logarithmic velocity gradient $\epsilon \lesssim 0.1$ in the outer layers of the envelope. Under this condition, the treatment of some effects, such as stellar radiation absorption, may not be very accurate. Nevertheless, the LVG method, which takes into account the main factors of radiative transfer, produces solutions that are easier to obtain and to interpret. Therefore, this approach is widely used in the analysis of molecular emission in CSEs.

### Column density

The column density measures the number of molecules in the source per unit area along the line of sight. We can express this magnitude as the number of molecules at the energy level u integrated over the pathlength d$s$ (see e.g. Mangum and Shirley, 2015):

$$N_\mathrm{u} = \int n_\mathrm{u} \mathrm{d}s. \tag{2.37}$$

The opacity, which is strongly related to the column density, can be expressed in the upper transition state $N_\mathrm{u}$ as:

$$\tau = \frac{c^2}{8\pi\nu^2}\left[\exp\left(\frac{h\nu}{kT_\mathrm{ex}}\right) - 1\right] A_\mathrm{ul}\phi_\nu N_\mathrm{u}. \tag{2.38}$$

The column density, for a particular energy level and for the total population of all energy levels, is:

$$\frac{N_\mathrm{tot}}{N_\mathrm{u}} = \frac{Q_\mathrm{rot}(T_\mathrm{ex})}{g_\mathrm{u}} \exp\left(\frac{E_\mathrm{u}}{kT_\mathrm{ex}}\right), \tag{2.39}$$

where $N_\mathrm{tot}$ is the total number of molecules per cm$^{-2}$, $g_\mathrm{u}$ and $E_\mathrm{u}$ are the statistical weight and energy of the level u, respectively, and $Q_\mathrm{rot}(T_\mathrm{ex})$ is a quantity that represents a statistical sum over all rotational energy levels in the molecule and is expressed as $Q_\mathrm{rot}(T_\mathrm{ex}) = \sum_\mathrm{u} g_\mathrm{u} \exp\left(-\frac{E_\mathrm{u}}{kT_\mathrm{ex}}\right)$.

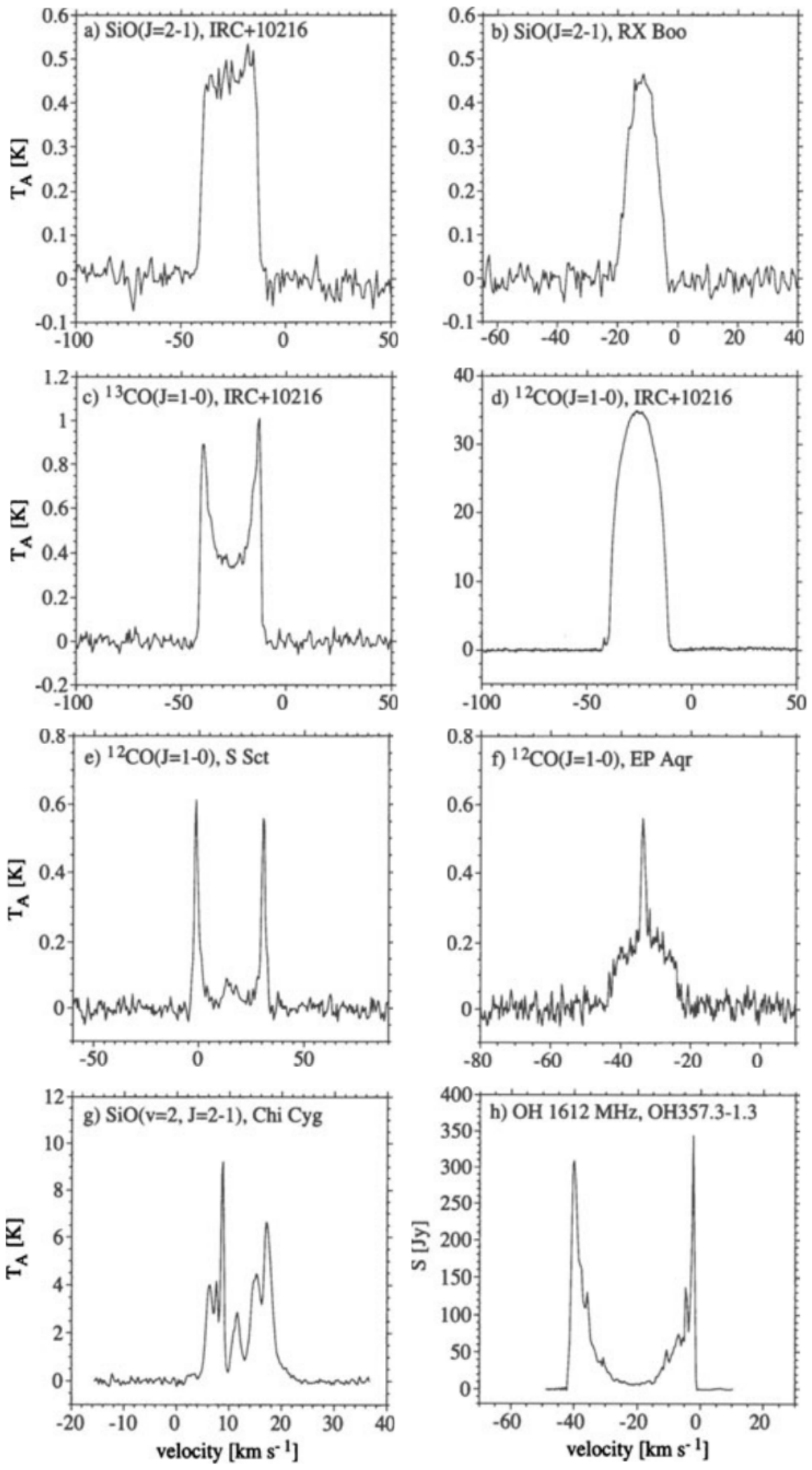

Figure 2.1: Different shapes of the line profiles in AGB-CSEs: a) optically thin, spatially unresolved emission, b) optically thick, spatially unresolved emission, c) optically thin, spatially resolved emission, d) optically thick, spatially resolved emission, e) emission from geometrically thin, spatially resolved shell, f) double-component line profile, g) maser emission line-profile, and h) maser emission-line profile. Figure taken from Habing and Olofsson (2004).

### Molecular line profiles

The observed line profiles can present different shapes (see Fig. 2.1), and their widths are mostly caused by the Doppler effect. The width of the line profile might indicate the temperature of the gas and can be indicative of the global velocity field of the molecular gas (non-symmetrical line profiles might indicate self-absorption).

In addition, the shape of the line profiles includes information about the opacity of the emitting material or even the resolution of the telescope. Assuming a point at distance $r$ with radial velocity $v(r)$, which emits in a direction defined by the cosine of the angle between the line of sight and the radial direction, $\mu = \cos\theta$, the observed velocity ($v_{\rm obs}$) depends on this radial velocity and also on the velocity of the center of gravity of the star ($v_{\rm sys}$):

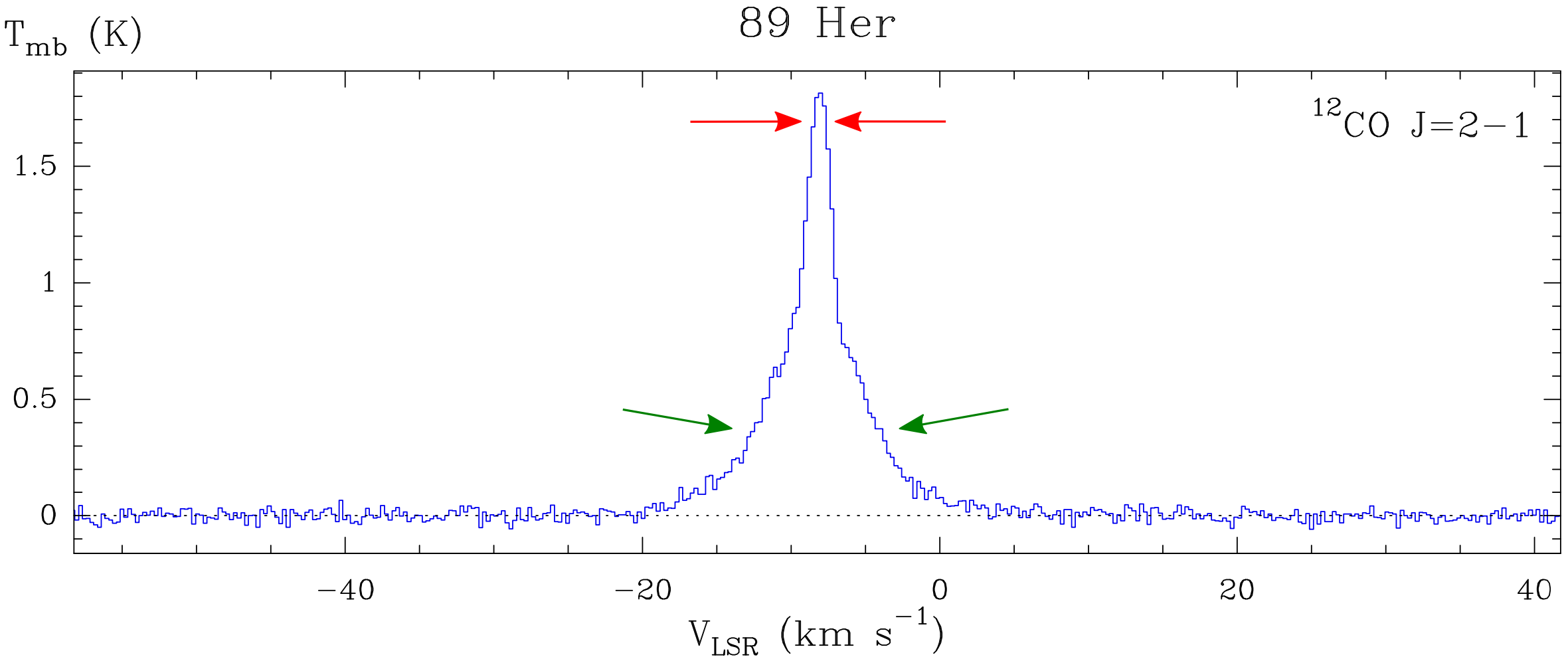

Figure 2.2: Example of the shape of the $^{12}$CO $J = 2-1$ line profile of a rotating disk and outflow around the binary post-AGB star 89 Her. The red arrows indicate the typical moderate velocities found in rotating disks. The green arrows indicate the presence of an extended and expanding component surrounding the disk. Figure adapted from Bujarrabal et al. (2013a).

$$v_{\rm obs}(r, \theta) = v_{\rm sys} - v(r)\cos\theta = v_{\rm sys} - v(r)\mu. \tag{2.40}$$

Thus, the line profile over a range of velocities is the sum of the intensities coming from each of the points emitting in that range. It is useful to assume a constant velocity in the envelope to get an idea of the different shapes of the molecular line profiles (see Fig. 2.1, and Olofsson et al., 1982; Habing and Olofsson, 2004, for further description). A parabolic profile is only produced by an unresolved and optically thick gas shell. In this case, only emission from the surface of the envelope reaches us, and the rest of the emitting material is opaque. On the contrary, the double-peaked profile is observed only towards a resolved and optically thin envelope. In this case, the envelope material is transparent and the emitting material of all layers is visible. Velocities closer to $v_{\rm sys}$ are underprivileged because the contribution of the outermost layers is not detected, resulting in profiles with two characteristic peaks at extreme velocities, $v_{\rm sys} \pm v$, and a depression at the center. Those line profiles between these extremes are produced from different combinations of optical depth and size of the source.

In the case of maser emission, what dominates is the strong amplification in the lines of sight along which the velocity variation is small. This produces only two very prominent peaks at the extreme velocities. The line profiles can be very different in the case of the OH maser emission at 1 612 MHz. The amplitude between the prominent peaks is maximum when $\mu = 0$, producing line profiles similar to the one shown in Fig. 2.1 *h*. Those profile are characteristic of envelopes with high loss mass.

Nevertheless, our disk-containing nebulae around binary post-AGB stars present spectra with narrow peaks that cannot be justified with a low-velocity expanding envelope. These narrow peaks and wide wings present in the CO line profiles are characteristic of rotating disks (see Fig. 2.2 and Sect. 1.4.1 for details). The relatively wide wings present in the spectra correspond to emission of outflows surrounding the disks.

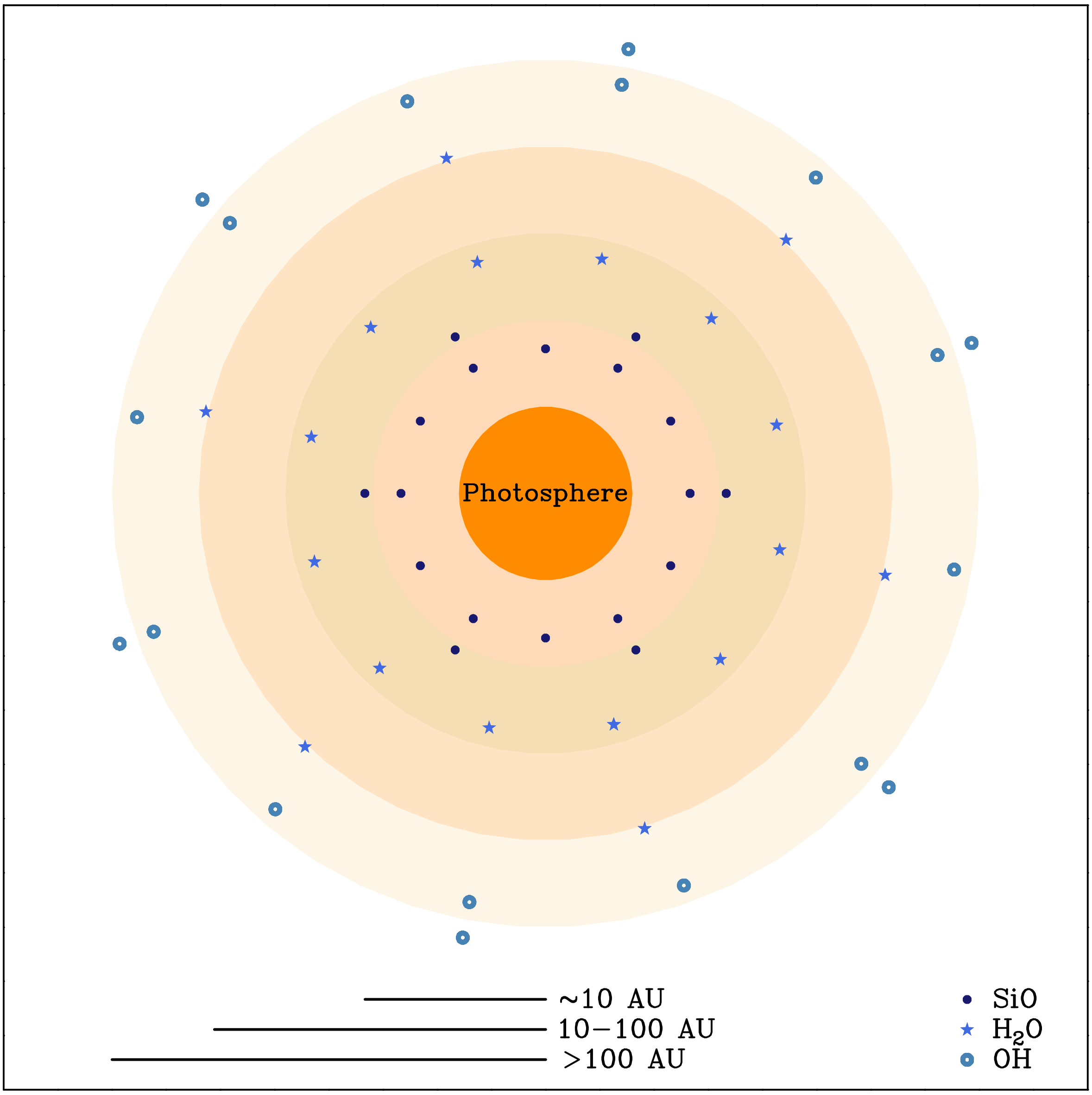

Figure 2.3: Schematic representation of an envelope showing SiO, 22 GHz $H_2O$, and 1 612 MHz OH maser emission. They are stratified in the different layers of the envelope: SiO maser is located in the innermost regions of the envelope, $H_2O$ maser takes place in regions between $10 - 100$ AU, and OH maser at distances $> 10^3$ AU.

### 2.2.3 SiO, $H_2O$, and OH maser emission

Maser emission occurs when there is population inversion ($n_2 > n_1$) and under conditions of non-equilibrium thermodynamics. In the case of population inversion, the absorption coefficient takes negative values ($\kappa_\nu < 0$). Therefore, the emission is amplified exponentially and in a coherent way (see Eq. 2.23).

SiO, $H_2O$, and OH maser emission is mainly observed in O-rich environments (see e.g. Kim et al., 2019). The maser emission of these molecules is stratified and let us study different parts of the envelope. SiO masers are located the innermost regions of the envelope, close to the star. $H_2O$ masers at 22 GHz occupy regions between 10 and 100 AU from the star. Finally, OH masers at 1 612 MHz take place in even further away regions, at distances $> 10^3$ AU (see Fig. 2.3, see e.g. Suárez et al., 2007).

SiO masers are found in envelopes around M- and S-type stars and in OH/IR stars and they are produced by far infrared pumping at 8 $\mu$m. Circumstellar SiO masers are useful for studying the innermost and denser layers of envelopes, where shocks, caused by stellar pulsations, drive the molecular material to distances at which it can condense into dust grains. This is the case of (some of) our sources. The radiation pressure on the dust grains then drives the stellar wind, implying a copious mass loss (Humphreys et al., 2002). The SiO maser phenomena occur in a few stellar radius, where silicates form, between the hot inner layers of the envelope and the cooler layers at around 4 stellar radius. In addition, SiO maser emission in envelopes tends to occur in clumpy, ring-like structures centered on the central evolved star (Diamond et al., 1994). M-type stars present strong flux variations at several wavelengths, including the SiO maser lines.

22 GHz $H_2O$ maser emission around evolved stars is usually found in approximately extended spherical from $5-50$ stellar radius. These masers are useful because they trace acceleration regions of the wind in the dust formation region but with low density (compared to SiO density regions). This maser is not detectable in a collision-dominated regime, because there is no population inversion (see Eq. 2.29). Moreover, its emission is weak if the collisions regime is low). The $H_2O$ density might be one or two orders of magnitude higher than the mean density derived from the mass loss rates derived from CO data and other observations (Humphreys et al., 2017). Some of our sources present this kind of emission.

In the case of 1 612 MHz OH masers, their emission comes from the outermost layers of the envelope around the evolved star, at distances $\sim 10^4$ AU. These masers are produced by IR pumping at 30 $\mu$m. The presence of this maser involves the molecular rupture of $H_2O$ because of the ISM radiation. The spectra usually present the characteristic double peak of expanding (quasi-)spherical envelopes, with velocities around $10-30\,\mathrm{km\,s^{-1}}$ (Bowers et al., 1989). The OH emission tend to disappear $\sim$1 000 years after the AGB star stops the mass loss (Gómez et al., 1990). This kind of maser emission has been detected in some of our sources.

The study of the presence of masers is crucial in our objects, because it allows to study the structure of the nebula (see Fig. 2.3 and Chapter 5).

## 2.3 Continuum thermal emission

Continuum emission takes place in a wide bandwidth, instead of in very specific frequencies, as in the case of molecular emission. We distinguish between thermal and non-thermal[1] emission. In the case of thermal emission, the particles that conform that source present $T > 0$ K due to the collisions between them. These particles follow a Maxwellian velocity distribution that strongly depends on the temperature.

### 2.3.1 Blackbody radiation spectrum

A blackbody is the simplest mode of thermal emission and it absorbs all incoming radiation and does not reflect any. It is a perfect absorber and emitter, because it emits efficiently at all wavelengths for a given temperature. This perfect emitter only depends on its temperature. Such blackbody radiation is described by the law of Planck, which determines the spectral energy density of the emission at each frequency ($B_\nu$) at a particular temperature:

$$B_\nu(T) = \frac{2h\nu^3}{c^2}\frac{1}{\exp\left(\frac{h\nu}{kT}\right) - 1}. \tag{2.41}$$

[1] Non-thermal processes are quite different from the thermal ones, because they do not depend on the temperature. They are associated with very energetic events or sources where particles move at near-light velocities. These processes will not be discussed, since this type of emission is not characteristic of the kind of objects that are studied in this thesis.

This expression is valid for any frequency, but there are two approximations for low- and high-frequencies. In the case of low-frequency, $h\nu \ll kT$, the exponential term can be approximated as $\exp\left(\frac{h\nu}{kT}\right) \approx 1 + \frac{h\nu}{kT}$. This regime is described by the Rayleigh-Jeans approximation, and it approximates the blackbody radiation at longer wavelengths (such as radio-wavelengths), but strongly disagrees at short wavelengths (this is known as UV catastrophe[2]):

$$B_\nu(T) = \frac{2kT}{c^2}\nu^2. \tag{2.42}$$

Nevertheless, the Wien approximation accurately describes the regime of high frequencies, but it fails for long-wavelength emission:

$$B_\nu(T) = \frac{2h\nu^3}{c^2}\exp\left(\frac{h\nu}{kT}\right)^{-1} \tag{2.43}$$

There are useful laws derived from the study of the blackbody. For a blackbody spectrum, the law of displacement of Wien states that the peak emission frequency ($\nu_{\max}$) increases linearly with absolute temperature, or in terms of wavelength, as the temperature of the body increases, the emission peak wavelength ($\lambda_{\max}$) decreases:

$$T = \frac{2.897 \times 10^{-3}\,[\mathrm{m} \cdot \mathrm{K}]}{\lambda_{\max}} = \frac{\nu_{\max}}{5.879 \times 10^{-10}\,[\mathrm{Hz} \cdot \mathrm{K}^{-1}]}. \tag{2.44}$$

Additionally, the Stefan-Boltzmann law states that the total radiant power emitted from a surface is proportional to the fourth power of its temperature, where $\sigma$ is known as the Stefan-Boltzmann constant:

$$E = \sigma T^4. \tag{2.45}$$

### 2.3.2 Gray body radiation spectrum

To satisfy the law of Planck, the blackbody opacity must be large at all wavelengths. This means that, under thermodynamic equilibrium:

$$\tau_\nu(s) \to \infty \Longleftrightarrow I_\nu(s) = S_\nu(s) = B_\nu(s) \tag{2.46}$$

Generally, sources are not perfect blackbodies (they are not completely opaque), so we usually use the term gray body radiation.

We find two examples of radiation of a thermalized gray body in the continuum: dust grains in an envelope around a star and H II regions composed of a gas of non-relativistic free electrons.

[2] The inconsistency between predictions of classical physics and observations is known as ultraviolet catastrophe.

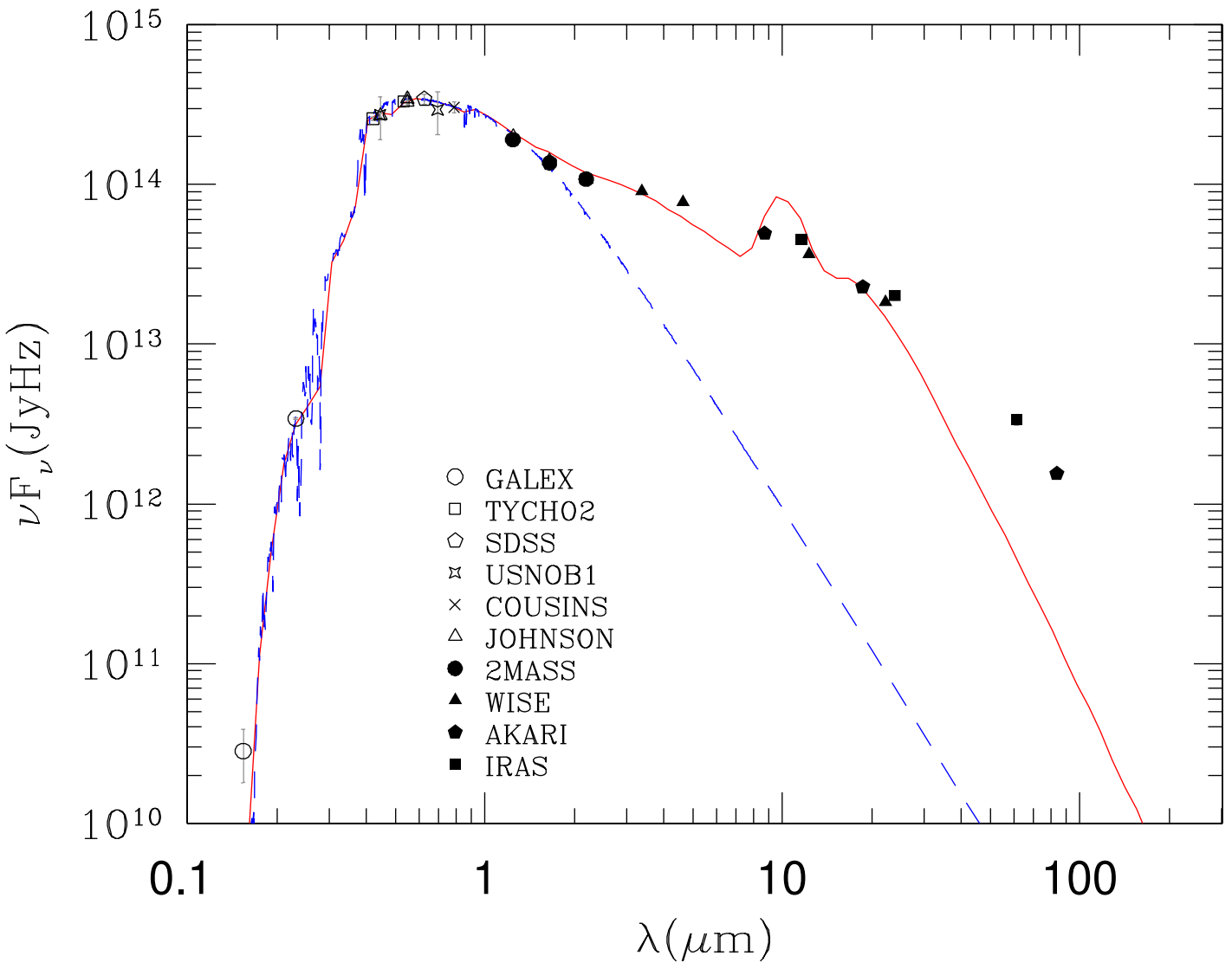

Figure 2.4: SED of BD+46°442. The symbols represent observational measurements at different wavelengths. The solid red line represents the (model) emission of the star and dusty disk system. The blue dashed line represents the theoretical emission of the star without taking into account dust emission. Figure taken from Gorlova et al. (2012).

### Dust grains

The absorption and emission dust grains coefficients are:

$$\kappa_{\nu_\mathrm{d}} = n_\mathrm{d}\sigma_\mathrm{d}Q_{\mathrm{abs},\nu}, \tag{2.47}$$

$$j_\mathrm{d} = n_\mathrm{d}\sigma_\mathrm{d}Q_{\mathrm{abs},\nu}B_\nu(T_\mathrm{d}), \tag{2.48}$$

where $T_\mathrm{d}$ is the temperature of the dust grain and $\sigma_\mathrm{d}$ is the geometric cross section, that can be approached as $\pi r_\mathrm{d}^2$ assuming that $r_\mathrm{d}$ is the dust grain radius. The theory of Mie produces extremely complex expressions in the general case, but allows for simple asymptotic expressions:

$$Q_{\mathrm{abs},\lambda} \simeq \begin{cases} 1, & \text{if } \lambda \ll r_\mathrm{d} \\ 2\pi r_\mathrm{d}\lambda^{-1}, & \text{if } \lambda \gg r_\mathrm{d} \end{cases}. \tag{2.49}$$

Thus, the absorption coefficient is constant to those wavelengths smaller than the dust grain size. On the contrary, the absorption coefficient is inversely proportional to the wavelength to smaller grains. Nevertheless, this expression is different assuming dust grains with crystalline structure:

$$Q_{\mathrm{abs},\lambda} \simeq \begin{cases} 1, & \text{if } \lambda \ll r_\mathrm{d} \\ (2\pi r_\mathrm{d})^\alpha\lambda^{-\alpha}, & \text{if } \lambda \gg r_\mathrm{d} \end{cases}, \tag{2.50}$$

where $\alpha \in [1, 2]$ depending on the composition of the grain.

In the same way, the Rayleigh theory allows to study the scattering coefficient:

$$Q_{\mathrm{sca},\lambda} \simeq \begin{cases} 1, & \text{if } \lambda \ll r_\mathrm{d} \\ 2\pi r_\mathrm{d}\lambda^{-4}, & \text{if } \lambda \gg r_\mathrm{d} \end{cases}. \tag{2.51}$$

With this, the radiation intensity $I_\nu$ of a star, with temperature $T_*$ surrounded by a dusty (and gas) shell with temperature $T_\mathrm{d}$ at a distance $d$, can be expressed through the transfer equation:

$$I_\nu(s) = B_\nu(T_*)\mathrm{e}^{-\tau_\nu(s)} + d^2\, B_\nu(T_\mathrm{d}) - \left(1 - \mathrm{e}^{-\tau_\nu(s)}\right). \tag{2.52}$$

Together with the pure blackbody emission of the star, there is another emitting component that corresponds to dust grains, resulting in a gray body spectrum (see Fig. 2.4). This is the archetypal SED of our sources, which present stables rotating disks composed of gas and dust.

Therefore, the thermodynamics of the dust grains are dominated by the absorption and emission of the radiation, because they absorb the star radiation (and also radiation of other grains) and this radiation is re-emitted with a frequency distribution that depends on the dust grain temperature.

The radiative heating and cooling rates, $H$ and $C$, respectively, in erg $\mathrm{s}^{-1}$ units, can be expressed as:

$$H = \pi r_\mathrm{d}^2 \int Q_{\mathrm{abs},\nu}\,\mathrm{d}\nu \int_{4\pi} I_\nu\,\mathrm{d}\omega, \tag{2.53}$$

$$C = 4\pi r_\mathrm{d}^2 \int Q_{\mathrm{abs},\nu} B_\nu(T_\mathrm{d})\,\mathrm{d}\nu. \tag{2.54}$$

According to this, assuming that the opacity cannot be taken into account, the dust grain temperature can be calculated as:

$$H = C(T_\mathrm{d}) \rightarrow \int_{4\pi} I_\nu\,\mathrm{d}\omega \sim \frac{\pi R_*^2}{d^2} B_\nu(T_\mathrm{d}) \rightarrow T_\mathrm{d}^4 \sim \frac{R_*^2}{4d^2} T_*^4. \tag{2.55}$$

Assuming a more realistic case for the absorption coefficient (see Eq. 2.50), the dust grain temperature:

$$\int Q_{\mathrm{abs},\nu} B_\nu(T_\mathrm{d})\,\mathrm{d}\nu \propto r_\mathrm{d}^\alpha T_\mathrm{d}^{4+\alpha} \rightarrow T_\mathrm{d} \sim \left(\frac{R_*^2 T_*^4}{d^2 r_\mathrm{d}^\alpha}\right)^{\frac{1}{4+\alpha}}. \tag{2.56}$$

Note that the dust temperature does not depend on the structure of the envelope. This magnitude depends on the distance between the grain and the star. The dust grain temperature also depends on the its size. In this way, larger grains are cooler, and therefore they emit mostly at low frequencies.

### Radiation of an ionized gas cloud

Assuming a thermalized ionized gas with temperature value $T_\mathrm{e}$, the opacity can be expressed as:

$$\tau_\nu = 8.235 \times 10^{-2}\, T_\mathrm{e}\,[\mathrm{K}]^{-1.35}\,\nu\,[\mathrm{GHz}]^{-2.1}\,a(\nu, T_\mathrm{e}) \int N_\mathrm{e}^2\,\mathrm{d}s\,[\mathrm{pc\,cm^{-6}}], \tag{2.57}$$

where $a(\nu, T_\mathrm{e}) \approx 1$ and the integral term is the emission measure ($EM$), which is the integral along the direction of radiation propagation of the square of free electron density ($N_\mathrm{e}$). The opacity $\tau_\nu$ varies as $\nu^{-2}$ in a free electron gas. Therefore, this term is important only at low frequencies corresponding to radio-waves where we can apply the Rayleigh-Jeans approximation (see Eq. 2.42).

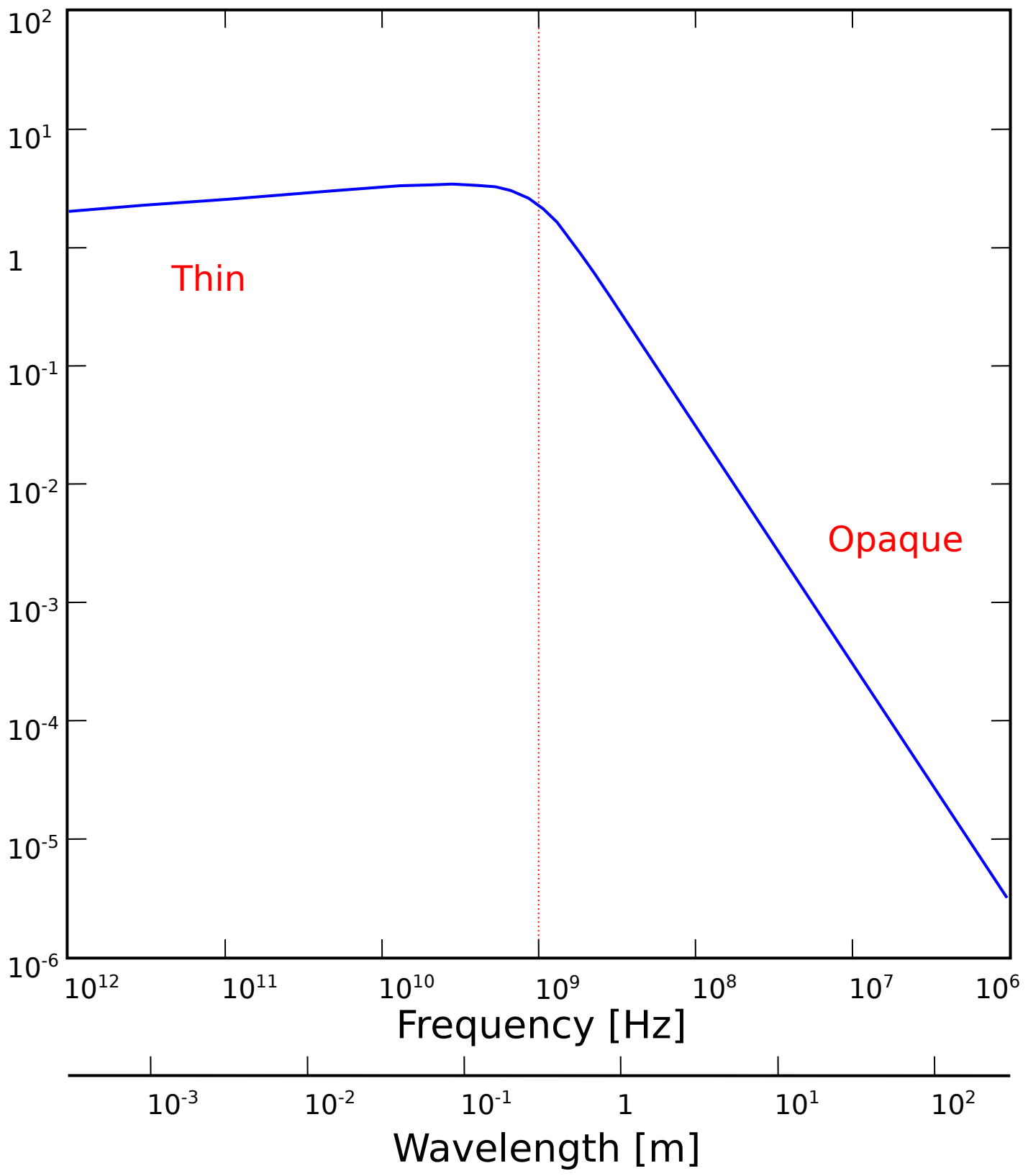

Figure 2.5: Free electron gas emission as a function of the frequency and wavelength. The thin and opaque regime is separated by a red dashed line at $\sim 10^9$ Hz.

According to Eq. 2.57, the frequency for which $\tau_\nu = 1$ is:

$$\nu_0\,[\mathrm{GHz}] = \nu\,[\mathrm{GHz}]\Big|_{\tau_\nu=1} = 0.3045\,T_\mathrm{e}\,[\mathrm{K}]^{-0.64}\left(a(\nu, T_\mathrm{e})\int N_\mathrm{e}^2\,\mathrm{d}s\,[\mathrm{pc\,cm^{-6}}]\right)^{0.48}. \quad (2.58)$$

This frequency value is proportional to the electron density and inversely proportional to the temperature, and it takes values in the order of $10^9$ Hz.

Then, the intensity of the radiation $I_\nu$ can be expressed as:

$$I_\nu = \begin{cases} 2kT_\mathrm{e}c^{-2}\nu^2, & \text{if } \tau_\nu \gg 1\ (\nu \ll \nu_0) \\ 2kT_\mathrm{e}c^{-2}\nu^2\tau_\nu, & \text{if } \tau_\nu \ll 1\ (\nu \gg \nu_0) \end{cases}. \quad (2.59)$$

On the one hand, we always find that $I_\nu \propto \nu^2$ in the opaque case. On the other hand, we find $I_\nu \propto \nu^2\tau_\nu(\nu, T_\mathrm{e}, EM)$ in the thin case. Assuming typical values of a free electron gas ($T_\mathrm{e} = 15\,000$ K and $EM = 5\times10^6\,\mathrm{pc\,cm^{-6}}$), the intensity varies as $I_\nu \propto \nu^{-0.1}$ in the thin case. See Fig. 2.5.

*Sometimes I'll start a sentence and I don't even know where it's going. I just hope I find it along the way.*

Michael Scott – The Office

# 3

# Fundamentals of radioastronomy

*This chapter introduces the theory and tools that support the observational data obtained for this Ph.D. thesis. We also introduce the main methods used to extract data through single-dish, connected interferometric observations, and on-the-fly data. We describe in detail the principles and fundamentals of these observational methods and the consequences for the interpretation of the data.*

## 3.1 Atmospheric effects

Electromagnetic waves at radio-wavelengths are detected with radiotelescopes. The atmosphere is relatively transparent in the radio window, so radio photons reach the surface of the Earth without too much attenuation. Thus, most radio telescopes are ground-based. However, in the centimeter (cm), millimeter (mm) and sub-millimeter (sub-mm) wavelength radio domains (between 20 and 600 GHz), the transmittance of the atmosphere varies significantly as a function of frequency (see Fig. 3.1). At these frequencies, the opacity of the atmosphere is mainly due to the water vapor bands (1.63, 0.92 mm, ...), $O_2$ bands (5, 2.5 mm, ...) and other molecules such as $N_2$ or $CO_2$ (with frequency values above 300 GHz). Since the water vapor content in the atmosphere is detrimental to observations at millimeter and sub-millimeter wavelengths, radiotelescopes are strategically located at dry locations and at high altitude, or even onboard aircrafts or artificial satellites, to improve observations by decreasing the effect of water vapor and the thickness of the atmosphere.

## 3.2 Reflector systems

Typically radiotelescopes work with a main reflector and a receiver system mounted at the Prime, Cassegrain, Gregorian, or Nasmyth foci (see Fig. 3.2).

Single reflector systems have the receiver mounted at the prime focus, which is the point where the rays reflected from the only mirror (M1) converge. The prime focus system has the advantage that it can be used over the full frequency range of the reflector. In contrast, the main disadvantages are that the space for and access to the feeds and receivers is strongly restricted (it is not possible to mount several receivers without damaging the entire structure by weight) and the spillover[1] noise from the pick up of ground emission decreases the sensitivity. An example of a prime focus system is the Giant Meterwave Radio Telescope (GMRT) or the very low frequency bands at MPIfR Effelsberg 100 m.

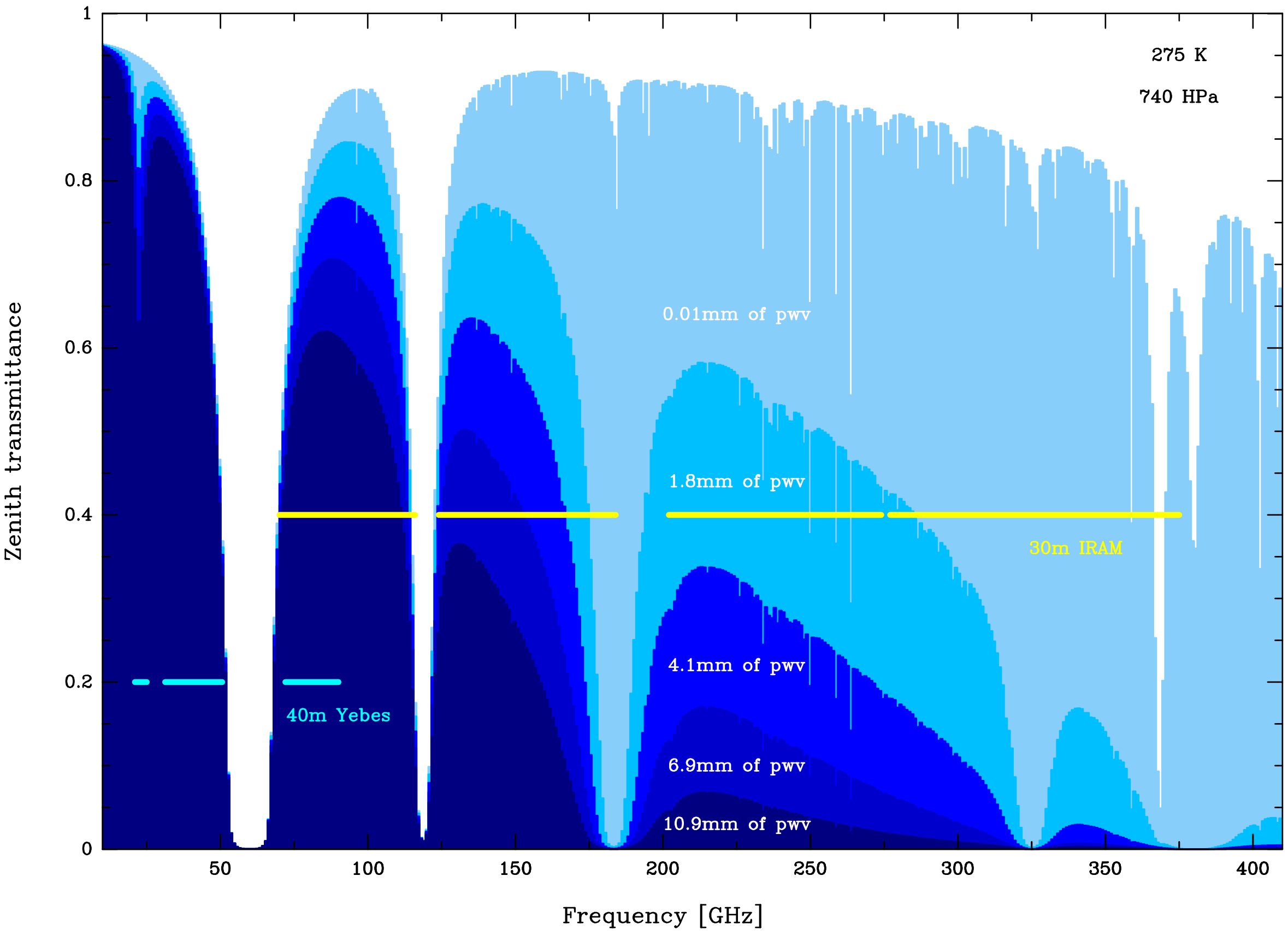

Figure 3.1: Model of zenith atmospheric transmittance at Pico Veleta (30 m IRAM), located at 2850 m of altitude, for 275 K, 740 HPa, and for 0.01, 1.8, 4.1, 6.9, and 10.9 mm of total precipitable of water vapor (*pwv*), represented as increasing colour darkness. Yellow lines represent the bandwidths of the 30 m IRAM telescope (3, 2, 1.3, and 0.8 mm receivers), cyan lines represent the bandwidths of the 40 m Yebes telescope (13, 7, and 3 mm receivers).

There are other configurations in which the receiver systems are located in the secondary focus by introducing a second reflector (M2). In the case of a Cassegrain system, a convex hyperbolic reflector is located in the converging beam immediately in front of the prime focus. This reflector sends the rays to a secondary focus that is usually situated close to the apex of the main dish. The 34 m NASA DSN telescope is an example of a Cassegrain system. A Gregorian system has an elliptical concave mirror behind the primary focus, which concentrates the signals back, close to the apex of the primary mirror. MeerKAT consists of 64 radiotelescopes of 13.5 m with an offset Gregorian configuration.

[1]Spillover effect occurs in parabolic radiotelescopes when the radiation emitted by the primary feeder located at the focus of the dish overflows the surface of the antenna dish causing a gain loss in the signal.

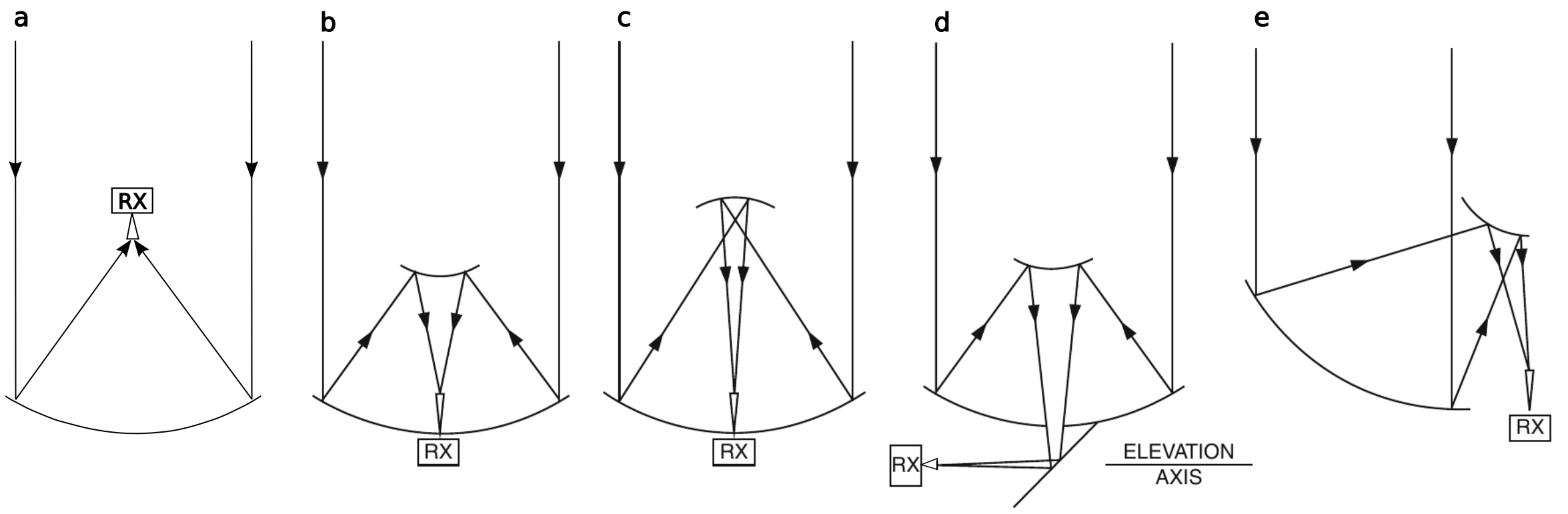

Figure 3.2: Focus arrangements of reflector radiotelescopes: a) Prime focus, b) Cassegrain focus, c) Gregorian focus, d) Nasmyth focus or Nasmyth-Cassegrain, and e) offset Cassegrain focus. The RX box represents the receiver while the small cone is the feed. Figure adapted from Wilson et al. (2013).

The Nasmyth system (or Nasmyth-Cassegrain) provides a receiver cabin outside the optical axis of the telescope thanks to a tertiary flat mirror (M3), which moves with the elevation axis of the telescope. In this configuration, the receiver remains always fixed while the telescope is pointing toward different elevations. The 30 m IRAM telescope is an example of a Nasmyth-Cassegrain system (see Fig. 3.3). Cassegrain offset systems are used to not have shadows from the secondary or support legs on the primary. These systems also avoid having a hole in the primary for the beam to pass through. The Green Bank Telescope (GBT) is an example of this kind of systems.

The multireflector systems (i.e. not primary focus systems) have the advantage of reduced spillover noise, and an easier access to and more space for feeds and receivers. In addition, the $f/D$ ratio in multiple mirror systems is 5 to 10 times larger than that of single reflector. Moreover, optical distortions are lower in this type of systems compared to primary focus systems and several receivers can be placed to receive signals of different frequencies by using, dichroic mirrors, additional mirrors or revolving structures.

High aperture efficiency requires good balance of spillover efficiency and illumination efficiency. However, in a primary focus system, the side lobe pattern of the feed extends beyond the edge of the dish, so the feed also receives thermal radiation from the ground ($\sim 300$ K). In secondary reflector systems, the power received by the feed from beyond the edge of the secondary reflector is radiation from the sky, which implies a 10 to 100 K in the case of mm-wavelengths. In low noise systems, this translates into an overall system noise temperature that is significantly lower than that of primary focus systems, where power is received from the ground. It can be quantified through the ratio of the gain of the antenna to the system noise, $G/T$ (see Sect. 3.3). See Wilson et al. (2013); Thompson et al. (2017); Baars (2007), for a complete description.

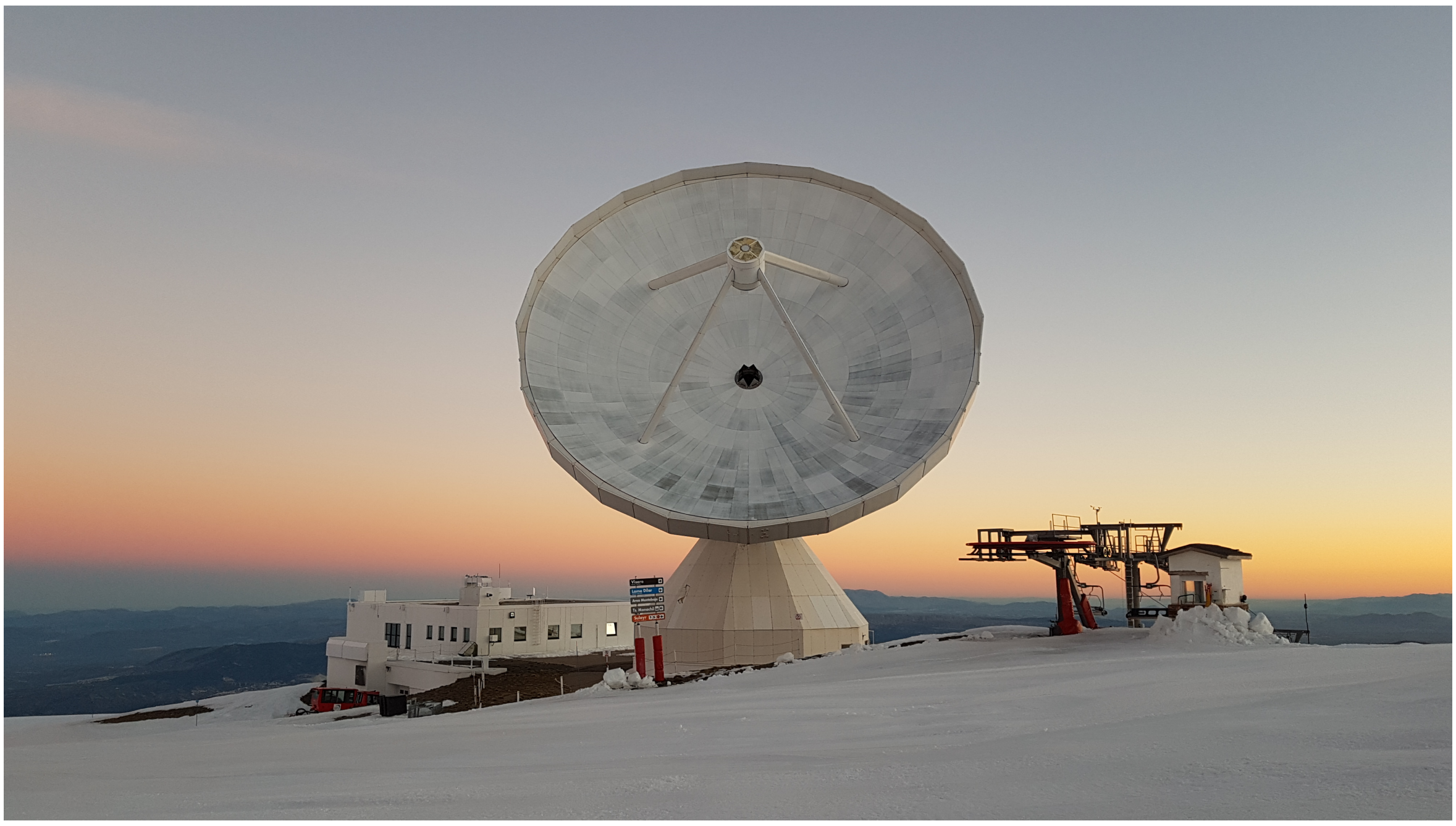

Figure 3.3: The 30 m IRAM telescope is an example of a Nasmyth-Cassegrain system (see Fig. 3.2). This telescope is one of the largest and most sensitive millimeter telescopes operating at 0.9, 1.2, 2, and 3 mm. The telescope is located on Pico Veleta in Sierra Nevada, Granada (Spain) and operates at 2 850 m above sea level. Picture taken by the author.

## 3.3 Radiotelescope parameters

The performance of a radiotelescope is characterized by relevant parameters that we describe in this section. These parameters are essential for the description of the interaction between the radiotelescope and the observed source. We mostly used the notation described in Wilson et al. (2013) and Thompson et al. (2017).

### 3.3.1 The power pattern

The normalized power pattern ($P_{\rm n}$) qualitatively shows the radiated power distribution of an antenna and is defined as the ratio of the received power in a direction $P(\theta,\varphi)$ and the maximum received power ($P_{\rm max}$). Therefore, the normalized power pattern ($P_{\rm n}$), in the far-field region[2], of an aperture antenna is given by the following equation:

$$P_{\rm n}(\theta,\varphi) = \frac{P(\theta,\varphi)}{P_{\rm max}}. \tag{3.1}$$

A typical normalized power pattern of an aperture antenna is shown in Fig. 3.4. Two magnitudes are frequently used to measure the width of the main lobe: the half power bean width (HPBW), which is the the angle between points of the main beam where the power pattern is half the maximum; and the full beam width between first nulls (BWFN), which stands for twice the angle of a direction where the power pattern falls to zero between the main lobe and the first sidelobe.

[2]The far-field of an antenna is usually defined as the region where the outgoing wavefront is planar.

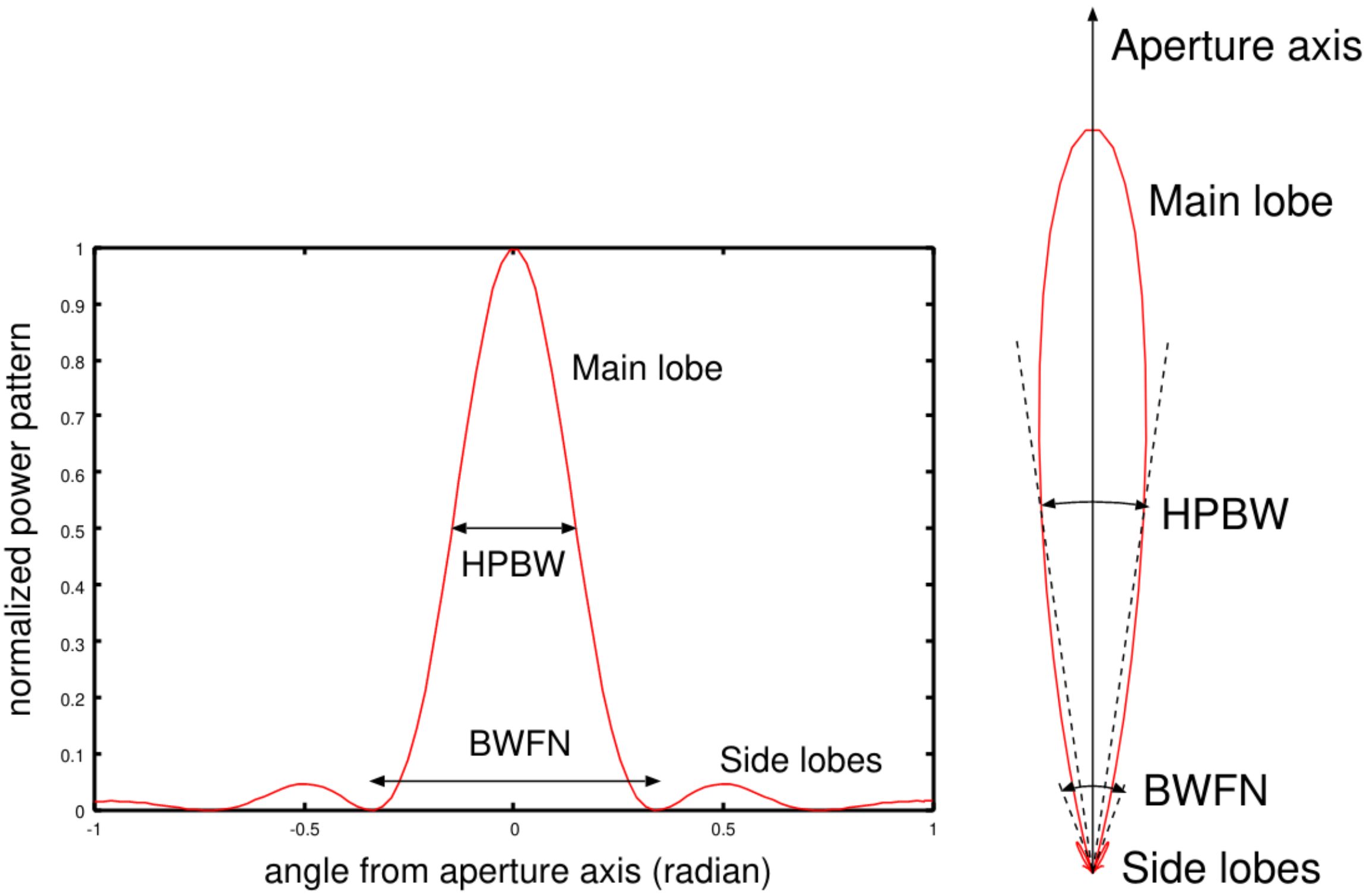

Figure 3.4: The normalized power pattern of an aperture radiotelescope in cartesian coordinate system (*left*) and in the polar coordinate system (*right*). Figure taken from Sasao and Fletcher (2009).

### 3.3.2 Beam solid angle and main beam solid angle

The beam solid angle, $\Omega_{\mathrm{A}}$, is another magnitude to measure the angular extent of the antenna beam, and it is defined as the solid angle of an ideal antenna having $P_{\mathrm{n}} = 1$ for all $\Omega_{\mathrm{A}}$ and $P_{\mathrm{n}} = 0$ elsewhere. $\Omega_{\mathrm{A}}$ is expressed in steradians and is expressed as:

$$\Omega_{\mathrm{A}} = \iint_{4\pi} P_{\mathrm{n}}(\theta, \varphi)\, \mathrm{d}\Omega = \int_0^{2\pi} \int_0^{\pi} P_{\mathrm{n}}(\theta, \varphi) \sin\theta \, \mathrm{d}\theta \, \mathrm{d}\varphi. \tag{3.2}$$

In the same way, the main beam (MB) solid angle can be quantified by the integration of the normalized power just to the extent of the main lobe:

$$\Omega_{\mathrm{MB}} = \iint_{\mathrm{MB}} P_{\mathrm{n}}(\theta, \varphi)\, \mathrm{d}\Omega. \tag{3.3}$$

The quality of an antenna depends on how well the power pattern is concentrated in the main lobe, since the main beam determines the angular resolution of the telescope. We can quantify this through the main beam efficiency (also known as beam efficiency):

$$\eta_{\mathrm{MB}} = \frac{\Omega_{\mathrm{MB}}}{\Omega_{\mathrm{A}}}, \tag{3.4}$$

where $\eta_{\mathrm{MB}}$ is an indication of the fraction of the power that is concentrated in the main beam. If it is close to 1, the antenna has a sharp main lobe and sufficiently low sidelobe levels. Antennas used in radioastronomy usually have low sidelobe levels ($\eta_{\mathrm{MB}} \approx 1 \rightarrow \Omega_{\mathrm{A}} \approx \Omega_{\mathrm{MB}}$). In Fig. 3.5, we can see the real $\eta_{\mathrm{MB}}$ values corresponding to the 30 m IRAM radiotelescope as a function of the observing frequency.

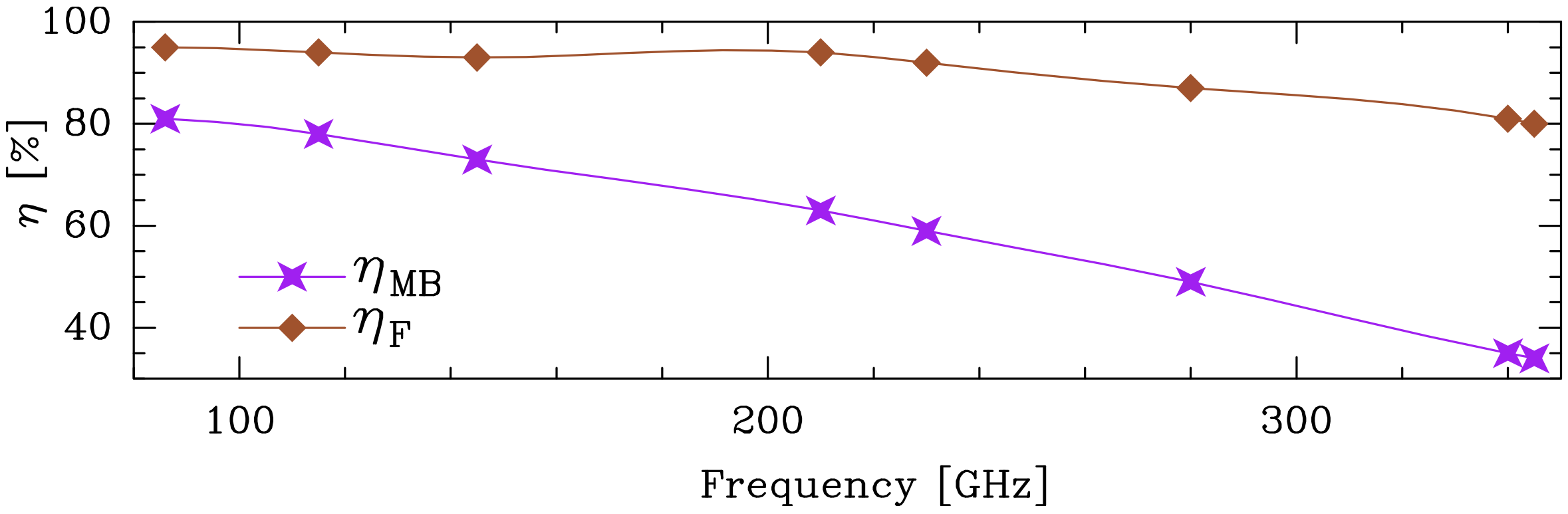

Figure 3.5: Main beam efficiency (in purple) and forward efficiency (in maroon) as a function of the observing frequency. These $\eta_{\rm MB}$ and $\eta_{\rm F}$ values correspond to the 30 m IRAM telescope.

### 3.3.3 Directivity and gain

The directivity, $D$, measures how directional the radiation pattern of an antenna is. The directivity equation is:

$$D(\theta, \varphi) = \frac{P_{\rm max}}{P} = \frac{4\pi}{\iint_{4\pi} P_{\rm n}(\theta, \varphi)\, {\rm d}\Omega} = \frac{4\pi}{\Omega_{\rm A}}. \tag{3.5}$$

Therefore, an antenna with narrow main lobe and low secondary lobe levels presents a high directivity value. Thus, an ideal antenna, with only a main lobe of infinitesimal width and no secondary lobes, would have $D = 1$. It would detect the maximum emission when pointed at the source and zero when the antenna moves any distance from the source.

Another relevant parameter is the gain, $G$, which can be defined as:

$$G(\theta, \varphi) = 4\pi \frac{P_{\rm n}(\theta, \varphi)}{\iint_{4\pi} P_{\rm n}(\theta, \varphi)\, {\rm d}\Omega} = \eta_{\rm R} D(\theta, \varphi), \tag{3.6}$$

where $\eta_{\rm R}$ is the radiation efficiency of the antenna. This parameter represents the Ohmic losses that take place in the radiotelescope, the refector, and the feed.

Finally, and according to Eqs. 3.5 and 3.6, the directivity is the maximum gain:

$$D = G_{\rm max} = \frac{4\pi}{\Omega_{\rm A}}. \tag{3.7}$$

### 3.3.4 Effective aperture

The geometric aperture, $A_{\rm g}$, is the physical area of the primary reflector or main dish. A useful parameter to estimate the power received by an antenna is the effective aperture, $A_{\rm e}$. Therefore, the effective aperture is always smaller than the geometric aperture. We can define the aperture efficiency as to the ratio of the effective aperture to the geometrical aperture:

$$\eta_{\rm A} = \frac{A_{\rm e}}{A_{\rm g}}. \tag{3.8}$$

This ratio is always smaller than 1 because of the blocking by the subreflector and the stays, the illumination taper, the spillover from the subreflector, and the Ohmic losses on the dish surface.

Additionally, a general relation for the effective aperture, in terms of the maximum antenna gain (or directivity), is:

$$A_\mathrm{e} = \frac{\lambda^2}{4\pi} G_\mathrm{max} = \frac{\lambda^2}{4\pi} D. \tag{3.9}$$

Therefore, the effective aperture and the beam solid angle are related to each other through the wavelength:

$$\lambda^2 = A_\mathrm{e} \Omega_\mathrm{A}. \tag{3.10}$$

In general, the effective aperture of an antenna decreases as the frequency (wavelength) increases (decreases). This is the reason why we cannot build large radio-telescopes efficiently operating at high frequencies.

## 3.4 Single-dish observations

### 3.4.1 Temperature scales

Let us consider an antenna with $P_\mathrm{n}(\theta, \varphi)$ that is pointing a brightness distribution source $B_\nu(\theta, \varphi)$. The total power per unit bandwidth received by the antenna is:

$$P_\nu = \frac{1}{2} A_\mathrm{e} \iint_{4\pi} B_\nu(\theta, \varphi) P_\mathrm{n}(\theta, \varphi)\, \mathrm{d}\Omega. \tag{3.11}$$

The received power can be expressed as the antenna temperature[3], $T_\mathrm{A}$, through the Johnson-Nyquist power theorem[4], which relates the output of the radiotelescope to the power of a matched resistor:

$$T_\mathrm{A} = \frac{1}{2k} A_\mathrm{e} \iint_{4\pi} B_\nu(\theta, \varphi) P_\mathrm{n}(\theta, \varphi)\, \mathrm{d}\Omega = \frac{1}{\Omega_\mathrm{A}} \int T_\mathrm{B}(\theta, \varphi) P_\mathrm{n}(\theta, \varphi)\, \mathrm{d}\Omega. \tag{3.12}$$

Thus, the measured antenna temperature is the convolution of the brightness temperature over the celestial sphere weighted by the antenna pattern and normalized by the beam solid angle of the antenna ($\Omega_\mathrm{A}$). Moreover, the brightness temperature is the thermodynamic temperature of the emitting material only for thermal emission and under the Rayleigh-Jeans limit (see Sect. 2.3).

We find different expressions for $T_\mathrm{A}$ for an extended or a compact source if we compare the beam solid angle[5] ($\Omega_\mathrm{A}$) and the source solid angle[6] ($\Omega_\mathrm{S} = \int T_\mathrm{B}(\theta, \varphi)\, \mathrm{d}\Omega$):

$$T_\mathrm{A} = \begin{cases} \frac{\Omega_\mathrm{S}}{\Omega_\mathrm{A}} T_\mathrm{B}, & \text{if} \quad \Omega_\mathrm{S} < \Omega_\mathrm{A} \\ T_\mathrm{B}, & \text{if} \quad \Omega_\mathrm{S} > \Omega_\mathrm{A} \end{cases}. \tag{3.13}$$

In other words, in the case of point sources (unresolved), the antenna temperature is equal to the source brightness temperature multiplied by the filling factor ($\Omega_\mathrm{S}/\Omega_\mathrm{A}$).

[3] The antenna temperature has nothing to do with the temperature of the antenna body, measurable by a thermometer. In fact, the antenna temperature relates the output of the antenna to the power from a matched resistor.

[4] In the Rayleigh-Jeans limit, we can exchange the brightness distribution by an equivalent brightness temperature through the Johnson-Nyquist power theorem: $P_\nu = kT_\mathrm{A}$.

[5] $\Omega_\mathrm{MB} = \frac{\pi}{4 \ln 2} \theta^2_\mathrm{HPBW} \cong 1.133\, \theta^2_\mathrm{HPBW}$.

[6] $\Omega_\mathrm{S} = 2\pi(1 - \cos\alpha) \cong \pi\alpha^2 \cong \pi \frac{r^2}{D^2} \cong \frac{S}{D^2}$, where $S$ is the size of the source and $D$ is the distance.

On the contrary, in the case of extended sources, the antenna temperature produced by a source larger than the antenna beam is equal to the source brightness temperature.

The value of the antenna temperature also includes atmospheric effects (opacity and emissivity). At a given $\nu$, $T_{\rm A}$ can be expressed as:

$$T_{\rm A} = T_{\rm B}{\rm e}^{-\tau_\nu} + T_{\rm atm}(1 - {\rm e}^{-\tau_\nu}), \tag{3.14}$$

where $\tau_\nu$ and $T_{\rm atm}$ are respectively the opacity and the Raileigh-Jeans equivalent temperature of the atmosphere. The antenna temperature corrected by atmospheric absorption is given by the expression:

$$T'_{\rm A} = T_{\rm A}{\rm e}^{\tau_\nu}. \tag{3.15}$$

We define $T^*_{\rm A}$ as the brightness temperatures of a source which fills the entire $2\pi$ sr of the forward beam pattern of the radiotelescope. Thus, taking into account this parameter, we can correct for the effects of the rear-sidelobes.

$$T^*_{\rm A} = \frac{1}{\Omega_{2\pi}} \int T_{\rm B}(\theta,\varphi)P_{\rm n}(\theta,\varphi)\,{\rm d}\Omega = \frac{\Omega_{\rm A}}{\Omega_{2\pi}} T'_{\rm A} = \frac{T'_{\rm A}}{\eta_{\rm F}}, \tag{3.16}$$

where the ratio $\Omega_{2\pi}/\Omega_{\rm A}$ is known as the forward efficiency $\eta_{\rm F}$ (Kramer, 1997). In Fig. 3.5, we see the real $\eta_{\rm F}$ values corresponding to the 30 m IRAM telescope as a function of the observing frequency.

The main beam temperature, $T_{\rm MB}$, takes into account the brightness temperature of an equivalent source just filling the main beam. So, it is the same as $T^*_{\rm A}$ but whitin the main beam instead of $2\pi$ forward sr.

$$T_{\rm MB} = \frac{1}{\Omega_{\rm MB}} \int T_{\rm B}(\theta,\varphi)P_{\rm n}(\theta,\varphi)\,{\rm d}\Omega = \frac{\Omega_{\rm A}}{\Omega_{\rm MB}} T'_{\rm A}, \tag{3.17}$$

where the ratio $\Omega_{\rm MB}/\Omega_{\rm A}$ is the main beam efficiency $\eta_{\rm MB}$ (see Sect. 3.3.2). Thus, we can relate the different temperature scales and efficiencies through the next expressions:

$$T_{\rm MB} = \frac{T'_{\rm A}}{\eta_{\rm MB}} = \frac{\eta_{\rm F}}{\eta_{\rm MB}} T^*_{\rm A}. \tag{3.18}$$

Finally, the conversion factor from $T^*_{\rm A}$ (in Kelvin) to flux ($S_\nu$, in Jy) units, which depends on the telescope, is:

$$\frac{S_\nu}{T^*_{\rm A}} = \frac{2k}{A_{\rm e}} \eta_{\rm F}. \tag{3.19}$$

### 3.4.2 Single-dish observing modes

The radio telescope not only collects radiation from the astronomical source, the received signal includes unwanted contributions characterized by the system temperature, $T_{\rm sys}$, such as system noise, spillover from the antenna radiation, background noise, or the atmosphere. Thus, the desired astronomical radiation is a fraction of the total power received. To detect the signal of interest, it must be compared with another signal that contains the exact same total power and that differs only from the first one in that it does not contain radiation from the astronomical source of interest. To obtain this value from the total received radiation, there are several observational techniques to remove as best as possible the contributions from the unwanted sources of radiation.

### Frequency switching (FSW)

This method is only valid for spectral observations. It involves switching between two frequencies while keeping the pointing on the astronomical source. This type of observation does not require any movement of the telescope, so the final spectrum will be a spectrum whose total integration time over the source will be the one that has been used ($t = t_{\mathrm{ON}}$).

$$\Delta T = \frac{T_{\mathrm{sys}}}{\sqrt{B\,t}}\sqrt{2}, \tag{3.20}$$

where $B$ is the bandwidth. The disadvantage of this observational method is that the resulting baselines are not good, since they are not flat and present a residual stationary wave (ripples) in the spectrum. This method cannot be applied to sources with a crowded spectrum or with very wide lines (and will not be used in this doctoral thesis).

### Position switching (PSW)

This method is valid for spectral and continuous observations. The received signal "on source" (ON) is compared with another signal obtained at a nearby position in the sky (OFF). In all cases the spectrum/continuum data is recorded as: (ON − OFF). Optimally, the OFF position must be at the same elevation as the ON position to ensure that the ON and OFF positions are measured through the same airmass (when possible). This observing mode is efficient if baseline ripples are a problem, because they are cancelled, and produces good line profiles. ON and OFF observations increase the total observing time, due to the the movement of the telescope and also because not all the observing time is on the astronomical source. This is reflected in the sensitivity reached in a bandwidth, $B$, after an observation time $t$ (assuming that $t = t_{\mathrm{ON}} + t_{\mathrm{OFF}}$, $t_{\mathrm{ON}} = t_{\mathrm{OFF}}$, and that the delay times are considerably lower than $t_{\mathrm{ON}}$ and $t_{\mathrm{OFF}}$):

$$\Delta T = \frac{T_{\mathrm{sys}}}{\sqrt{B\,t_{\mathrm{ON}}}} = \frac{T_{\mathrm{sys}}}{\sqrt{B\,t}}2. \tag{3.21}$$

As a consequence, the sensitivity of PSW method is a factor $\sqrt{2}$ worse compared to FSW method.

### Wobbler switching (WSW)

This method is a variant of the position switching in which the sub-reflector is switched. It is not necessary to move the entire antenna from the ON position to the OFF position. So, a higher switching frequency (0.5 to 1.0 Hz) benefits the observations, providing flat and stable baselines, and reducing the dead times between ON and OFF integrations. The displacement between ON and OFF is limited in angular distance. This observing mode is useful for compact sources (smaller than the switching angle), especially in the mm/sub-mm range. The sensitivity reached can be estimated through Eq. 3.21. In this case, by definition of wobbler switching, the delay times are significantly lower than in the classical ON − OFF mode. This is the method used in this doctoral thesis (see Chapter 5).

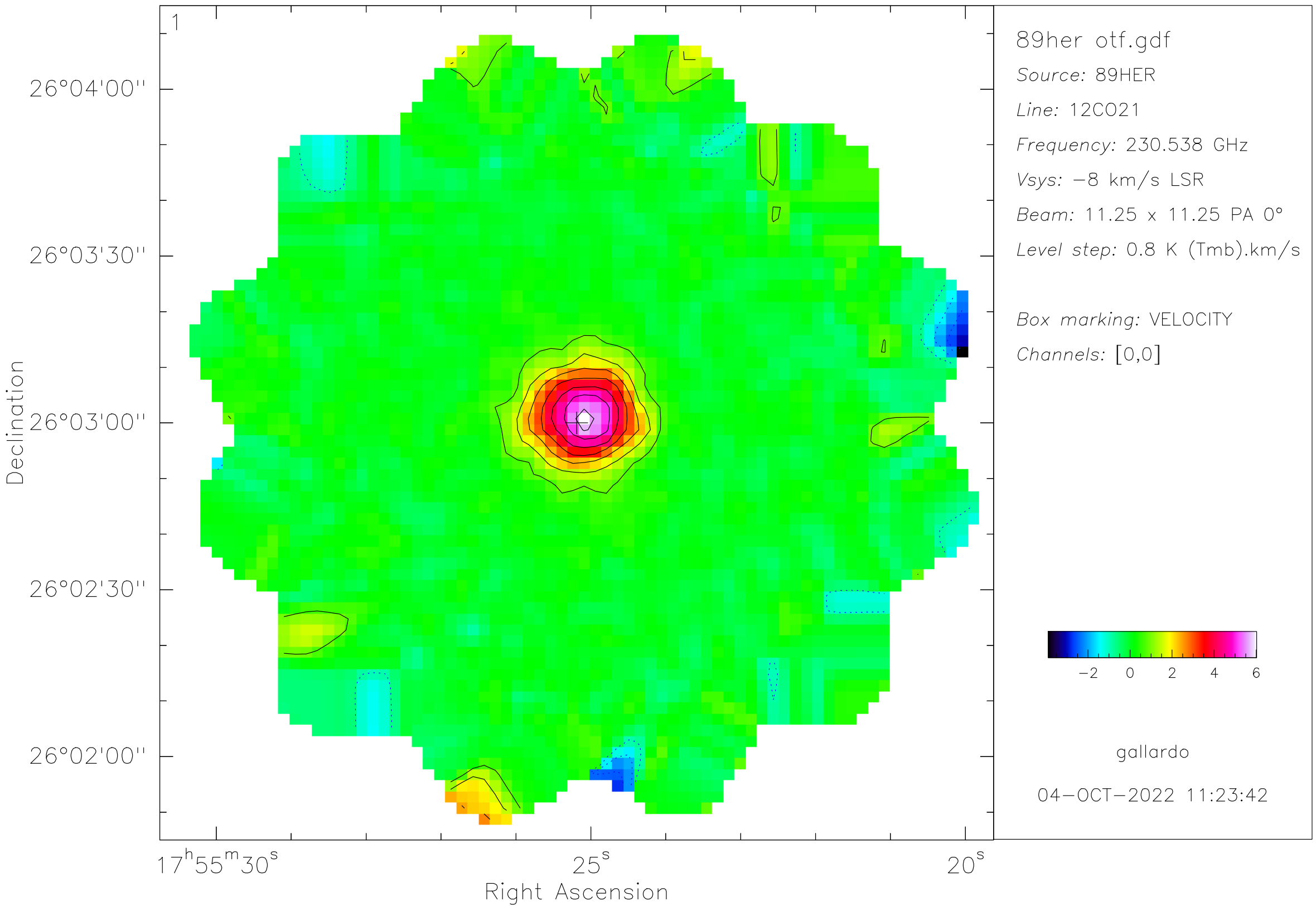

Figure 3.6: On-The-Fly map showing the scanning direction in six ways at at angles 0°, 30°, 60°, 120°, and 150°.

### On-the-fly mapping (OTF)

What we have described above is for one point in the sky. For mapping a region larger than the radiotelescope beam, this process must be performed for each of the points to be sampled, where the separation of these points will be equal to or less than the HPBW (Nyquist-Shannon sampling theorem). Nowadays, there are receivers with more than one detector (multibeam) that are capable of sampling several points simultaneously, however the use of single pixel receivers is still very widespread for which the wide sky mapping requires high integration times due to the need to sample the points one by one. This is especially true in the raster technique.

Spectral line maps were traditionally observed in a raster mode, switching the radiotelescope between the source (ON) and reference positions (OFF) for each individual raster point. This process was inefficient because of the significant overheads introduced by the movements of the antenna and initializations processes.

More recently, it has been designed a more efficient process, where the radiotelescope continuously scans (instead of tracking the individual raster points) along a straight line in the sky at a certain orientation while taking data at high cadence (a few seconds). This method is named on-the-fly (OTF) mapping. A map is then performed by scanning the sky along a series of equally spaced parallel lines oriented at certain parallactic angle, resulting in a map of rectangular region. The scanning can be performed in different parallactic angles (0°, 30°, 60°, 120°, and 150°) to minimize interleaving and weaving patterns in the final averaged total-power maps (see Fig. 3.6). This method is an extension of the position switching. Spectra are written every few of arcseconds, which corresponds to integration times of just a few seconds or even less. The derived total-power maps from OTF observations allow to study relatively large astronomical fields and to collect all the flux from the source. These total-power maps are sometimes merged with high-resolution interferometric maps that present filtered-out flux. In this way, the combined maps present the total flux and also the high-resolution provided by the interferometric observations. This single-dish observing mode has been used to observe one of our sources, 89 Her, see Sect. 3.5.8 and Chapter 7.

## 3.5 Interferometric observations

The maximum angular resolution of a telescope is defined by the diffraction limit, $\theta_\mathrm{d}$, which is determined by its diameter $D$, and by the observing wavelength $\lambda$:

$$\theta_\mathrm{d} = 1.22\,\frac{\lambda}{D}. \tag{3.22}$$

Therefore, to achieve a better resolution, telescopes must be built with larger diameters or to observe at smaller wavelengths, or a combination of both solutions. However, the construction of large antennas operating at short wavelengths presents some technical limitations, which in turn limits the spatial resolution of a radiotelescope. As an example, if an observation at 1.3 mm with a resolution of $11''$ is required, the necessary diameter is 30 m. However, to observe at the same wavelength with a resolution of $0\farcs2$, a telescope with a diameter greater than 1 600 m would be required. Faced with the impossibility of building telescopes of colossal sizes and the urgent need to achieve ever better resolutions, interferometry emerged. This technique combines the output signal from several radiotelescopes as if it came from one, where its diameter is equal to the largest baseline[7], $b$:

$$\theta_\mathrm{d} = 1.22\,\frac{\lambda}{b}. \tag{3.23}$$

Radio interferometers collect the radiation and convert it into a digital signal and generate a pattern known as interference fringes. Areas where there is light imply that the two beams of light add together in constructive interference, while dark areas imply a cancellation of the beams in destructive interference. The constructive interference occurs when the distance two waves travel differ by the even number of half wavelength, $2n\lambda/2 = n\lambda$, while the destructive interference occurs when the distance two waves travel differ by the odd number of half wavelength $(2n+1)\lambda/2$. There are several types of interferometers, and the correlation (or multiplier) interferometer is the best for radio astronomy.

The terminology used between single-dish and interferometry is quite different: one speaks of "fringes" instead of "main beam" and "sidelobes". Moreover, in the correlation of the outputs of two radiotelescopes, the fringes are centered about zero, because it improves the dynamic range of the measurements since the larger total power output of each radiotelescope is suppressed. Finer astronomical source structures can be measured by increasing the separation of two radiotelescopes when their outputs are coherently combined (see Taylor et al., 2000; Thompson et al., 2001; Condon and Ransom, 2016, for complete description).

[7]Baselines represent distances between radiotelescopes.

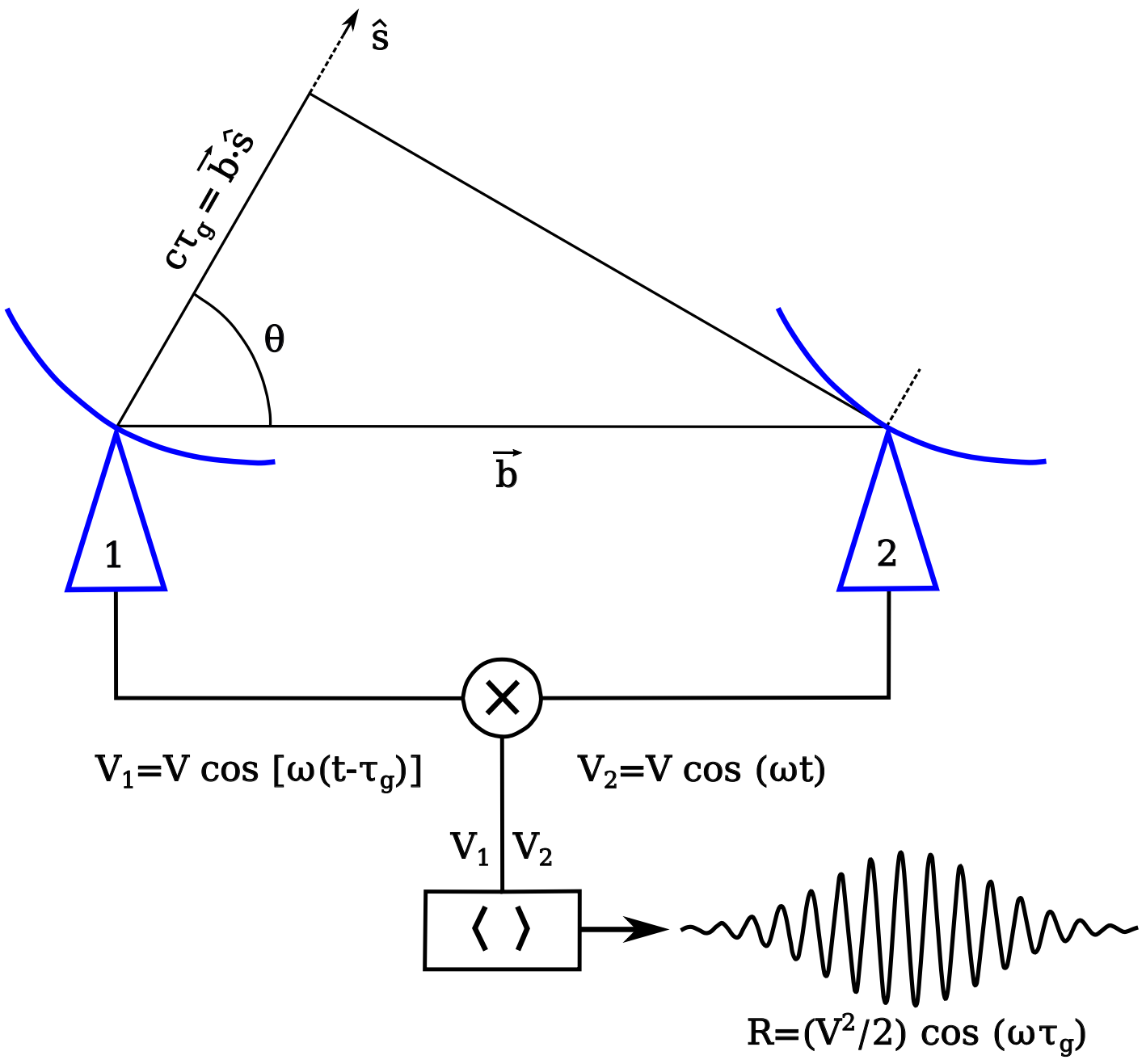

Figure 3.7: Schematic view of a radio interferometer composed of two antennas. The signals collected by the radiotelescopes ($V_1$ and $V_2$) are then sent to the correlator, which computes their averaged product, producing an single signal ($R$). Figure adapted from Condon and Ransom (2016).

### 3.5.1 Basic fundamentals of interferometry

The fundamentals of an interferometer is explained for the simplest radio interferometer following the notation of Condon and Ransom (2016). In a simple quasi-monochromatic two-element interferometer, the signal from the two radiotelescopes, separated by a distance $\vec{b}$, is correlated (multiplied and averaged). A plane electromagnetic wave of amplitude $V$ induces the voltages $V_1$ and $V_2$ at the output of the radiotelescope 1 and 2, respectively (see Fig. 3.7):

$$V_1 = V\ \cos[\omega(t - \tau_{\rm g})], \tag{3.24}$$

$$V_2 = V\ \cos(\omega t), \tag{3.25}$$

the output of both radiotelescopes is the same, but the first one is lagged in time due to the geometric delay ($\tau_{\rm g}$) because plane waves have to travel an extra distance to reach the first radiotelescope. Thus, this delay depends on the projection of the baseline onto the direction of radiation (where $\theta$ is the angle between $\vec{b}$ and $\hat{s}$, so $\vec{b} \cdot \hat{s} = b \cos\theta$):

$$\tau_{\rm g} = \frac{\vec{b} \cdot \hat{s}}{c} = \frac{b \cos\theta}{c}. \tag{3.26}$$

The signals collected by the radiotelescopes are then sent to a correlator, which computes their averaged product, producing a signal of the form:

$$R = \langle V_1, V_2 \rangle = \frac{V^2}{2} \cos(\omega \tau_{\rm g}). \tag{3.27}$$

The output of the correlator varies sinusoidally as rotation of the Earth changes the source direction relative to the baseline vector. These sinusoids are known as fringes.

The phase of the fringe is defined as:

$$\phi = \omega\tau_{\mathrm{g}} = \omega\frac{b\cos\theta}{c}. \tag{3.28}$$

The fringe phase depends on the angle that conforms the baseline vector and the unit vector in the direction of an astronomical source ($\theta$, see Fig. 3.7) as follows:

$$\frac{\mathrm{d}\phi}{\mathrm{d}\theta} = \omega\frac{b\sin\theta}{c} = 2\pi\left(\frac{b\sin\theta}{\lambda}\right). \tag{3.29}$$

### Slightly extended sources

The observed radio distribution of an slightly extended source, and smaller than the primary beamwidth, is $I_\nu(\hat{s})$. The slightly extended source can be treated as the sum of independent point sources. Therefore, the response of the interferometer ($R_{\mathrm{C}}$) is:

$$R_{\mathrm{C}} = \int I_\nu(\hat{s})\cos(\omega\tau_{\mathrm{g}})\,\mathrm{d}\Omega = \int I_\nu(\hat{s})\cos\left(2\pi\nu\frac{\vec{b}\cdot\hat{s}}{c}\right)\mathrm{d}\Omega. \tag{3.30}$$

The brightness distribution of the source can be expressed as the contribution of the even, $I_{\mathrm{E}}(\hat{s})$, and odd parts, $I_{\mathrm{O}}(\hat{s})$, which represents the symmetric and antisymmetric part of the electromagnetic wave, respectively. The Eq. 3.30 represents the even part and to detect the odd part is necessary a second correlator whose output is odd:

$$R_{\mathrm{S}} = \int I_\nu(\hat{s})\sin(\omega\tau_{\mathrm{g}})\,\mathrm{d}\Omega = \int I_\nu(\hat{s})\sin\left(2\pi\nu\frac{\vec{b}\cdot\hat{s}}{c}\right)\mathrm{d}\Omega. \tag{3.31}$$

According to the formula of Euler[8], we can combine Eqs. 3.30 and 3.31. The combination of both even and odd parts is known as complex correlator. Thus, we define the complex visibility as:

$$\Upsilon = R_{\mathrm{C}} - \mathrm{i}R_{\mathrm{S}} = A\mathrm{e}^{-\mathrm{i}\alpha}, \tag{3.32}$$

where $A$ and $\alpha$ are the visibility amplitude and the visibility phase, respectively. Both magnitudes are expressed as:

$$A = \sqrt{R_{\mathrm{C}}^2 + R_{\mathrm{S}}^2}, \tag{3.33}$$

$$\alpha = \arctan\frac{R_{\mathrm{S}}}{R_{\mathrm{C}}}. \tag{3.34}$$

Finally, according to Eqs. 3.30, 3.31, and 3.32, the complex visibility of an slightly extended source is given by the next equation:

$$\Upsilon = \int I_\nu(\hat{s})\exp\left(-2\pi\nu\mathrm{i}\frac{\vec{b}\cdot\hat{s}}{c}\right)\mathrm{d}\Omega. \tag{3.35}$$

[8] $\mathrm{e}^{\mathrm{i}\alpha} = \cos\alpha + \mathrm{i}\,\sin\alpha$.

### Finite bandwidth

The response of an interferometer must be generalized to finite bandwidths ($\Delta\nu$) centred on $\nu_0$. We can differentiate between continuum and line observations (depending on $\Delta\lambda/\lambda$). Thus, Eq. 3.35 becomes:

$$\begin{aligned}\Upsilon &= \iint_{\nu_o-\frac{\Delta\nu}{2}}^{\nu_o+\frac{\Delta\nu}{2}} I_\nu(\hat{s}) \exp\left(-2\pi\nu \mathrm{i}\frac{\vec{b}\cdot\hat{s}}{c}\right) \mathrm{d}\nu\, \mathrm{d}\Omega \\ &= \iint_{\nu_o-\frac{\Delta\nu}{2}}^{\nu_o+\frac{\Delta\nu}{2}} I_\nu(\hat{s}) \exp\left(-2\pi\nu \mathrm{i}\tau_\mathrm{g}\right) \mathrm{d}\nu\, \mathrm{d}\Omega \\ &\approx \int I_\nu(\hat{s}) \exp\left(-2\pi\nu_0 \mathrm{i}\tau_\mathrm{g}\right) \mathrm{sinc}\left(\Delta\nu\tau_\mathrm{g}\right) \mathrm{d}\Omega.\end{aligned} \tag{3.36}$$

This equation is applied to interferometers, where response is practically constant together with a constant source brightness.

Therefore, the fringe pattern has an envelope in the form of a sinc function[9]. This introduced attenuation is eliminated introducing a delay in the signal path of the reference antenna (see Condon and Ransom, 2016; Thompson et al., 2017, for further details). In any case, and in order to simplify the study, we will consider Eq. 3.35 for the study of visibilities.

## 3.5.2 Sensitivity of an interferometer

The point-source sensitivity of a two-elements interferometer is derived from the practical total-power radiometer equation[10].

$$\sigma_\mathrm{S} = \frac{\sqrt{2}kT_\mathrm{sys}}{A_\mathrm{e}\sqrt{\Delta\nu\, t_\mathrm{int}}}. \tag{3.37}$$

Under this assumption, the effective aperture of the interferometer is equal to the effective aperture of each radiotelescope, while $\Delta\nu$ and $t_\mathrm{int}$ represent the frequency resolution and the integration time, respectively.

This expression can be generalized to $N$ radiotelescopes, because an interferometer of N antennas contains $N(N-1)/2$ independent two-element interferometers. Therefore, the point-source sensitivity of an N-antennas interferometer is given by:

$$\sigma_\mathrm{S} = \frac{2kT_\mathrm{sys}}{A_\mathrm{e}\sqrt{\Delta\nu\, t_\mathrm{int} N(N-1)}}. \tag{3.38}$$

The point source sensitivity can be also expressed in a more accurate way taking into account more factors of the radiotelescopes that conform the interferometer:

$$\sigma_\mathrm{S} = \frac{2kT_\mathrm{sys}}{A_\mathrm{e}\eta_\mathrm{C}\eta_\mathrm{J}\eta_\mathrm{P}\sqrt{\Delta\nu\Delta t_\mathrm{int} n_\mathrm{P} N(N-1)}}, \tag{3.39}$$

where $n_\mathrm{P}$ is the number of linear polarizations (1 or 2), $\eta_\mathrm{C}$ is the correlator efficiency, $\eta_\mathrm{J}$ is the instrumental Jitter, and $\eta_\mathrm{P}$ is the atmospheric decorrelation.

[9] $\mathrm{sinc}\, x = \frac{\sin(\pi x)}{\pi x}$

[10] $\sigma \approx T_\mathrm{sys}\left[\frac{1}{\Delta\nu\, t_\mathrm{int}} + \left(\frac{\Delta G}{G}\right)^2\right]^{1/2}$, where $\Delta G$ represents the gain fluctuation.

### 3.5.3 Interferometer geometry

We must introduce a convenient coordinates system to solve Eq. 3.35. Therefore, we define the geometric coordinate frame for the interferometer, in which there are two fundamental vectors: $\vec{b}$ and $\hat{s}$. The first one is the baseline vector $\vec{b} = (\lambda u,\, \lambda v,\, \lambda w)$, which is expressed in units of the wavelength observed. The second one is the unit vector $\hat{s}$, which points towards origin of coordinates of the source of interest and it is expressed as a function of the direction cosines with respect to the $u$ and $v$ axes. This unit vector can be expressed as $\hat{s} = \vec{s_0} + \vec{\sigma}$, where $\vec{s_0}$ points towards the center of origin of coordinates and $\vec{\sigma}$ is a small offset vector, so $|\hat{s}| = 1$ and $|\vec{s_0}| \approx 1$ (see Fig. 3.8 *left*).

As a corollary, we can state that the visibilities only depend on the configuration/geometry of the array and the geometry of the object. With this, and knowing that $\lambda = 2\pi c/\omega$, Eq. 3.35 can be rewritten as:

$$\Upsilon\,(\vec{b}) = \iint_S P_{\mathrm{n}}(\vec{\sigma}) I(\vec{\sigma}) \exp\left(\mathrm{i}\frac{\omega}{c}\vec{b}\cdot\vec{\sigma}\right) \mathrm{d}\vec{\sigma}. \tag{3.40}$$

We define $\hat{s} = (l,\, m,\, n) = \left(l,\, m,\, \sqrt{1-l^2-m^2}\right)$. The components of the unit vector $\hat{s}$ allow to redefine $\mathrm{d}\Omega$ as:

$$\mathrm{d}\Omega = \frac{1}{\sqrt{1-l^2-m^2}} \mathrm{d}l\, \mathrm{d}m, \tag{3.41}$$

so, we express Eq. 3.40 in the $u$, $v$, $w$ coordinates as:

$$\Upsilon(u,\, v,\, w) = \int_{-\infty}^{+\infty}\int_{-\infty}^{+\infty} P_{\mathrm{n}}(l,m)\, I(l,m) \exp\left[2\pi\mathrm{i}\left(ul + vm + w\sqrt{1-l^2-m^2}\right)\right] \frac{\mathrm{d}l\, \mathrm{d}m}{\sqrt{1-l^2-m^2}}. \tag{3.42}$$

This expression is still not a proper Fourier transform, but we can get it if we consider that the third component in the phase factor is sufficiently small[11], because the third term can be obviated if it is much less than 1:

$$\Upsilon\,(u,\, v,\, w) \exp(-2\pi\mathrm{i}w) = \int_{-\infty}^{+\infty}\int_{-\infty}^{+\infty} P_{\mathrm{n}}(l,m)\, I(l,m) \exp[2\pi\mathrm{i}(ul+vm)]\, \mathrm{d}u\, \mathrm{d}v. \tag{3.43}$$

$P_{\mathrm{n}}(l,m) \approx 1$ when the source is smaller than the primary beam. Moreover, we differentiate the measurements of $\Upsilon$ to be taken on a plane (2D) or in 3D, because the term $\exp(-2\pi\mathrm{i}w) \approx 1$. Under this condition, the relation between the intensity $I$ and the visibility $\Upsilon$ becomes a 2D Fourier transform and this configuration defines an interferometer whose radiotelescopes lie on a single plane: $\Upsilon(u,\, v,\, w)\exp(-2\pi\mathrm{i}w) \cong \Upsilon(u,\, v,\, 0)$. Applying the inverse Fourier transform Eq. 3.43 becomes:

$$I'(l,\, m) = P_{\mathrm{n}}(l,\, m)\, I(l,\, m) = \int_{-\infty}^{+\infty} \Upsilon(u,\, v,\, 0) \exp[-2\pi\mathrm{i}(ul+vm)]\, \mathrm{d}u\, \mathrm{d}v, \tag{3.44}$$

where $I'(l,\, m)$ represents the intensity $I(l,\, m)$ modified by the primary beam shape $P_{\mathrm{n}}(l,\, m)$.

[11] $w\left(1-\sqrt{1-l^2-m^2}\right) = w(1-\cos\theta) \sim w\frac{\theta^2}{2} \ll 1.$

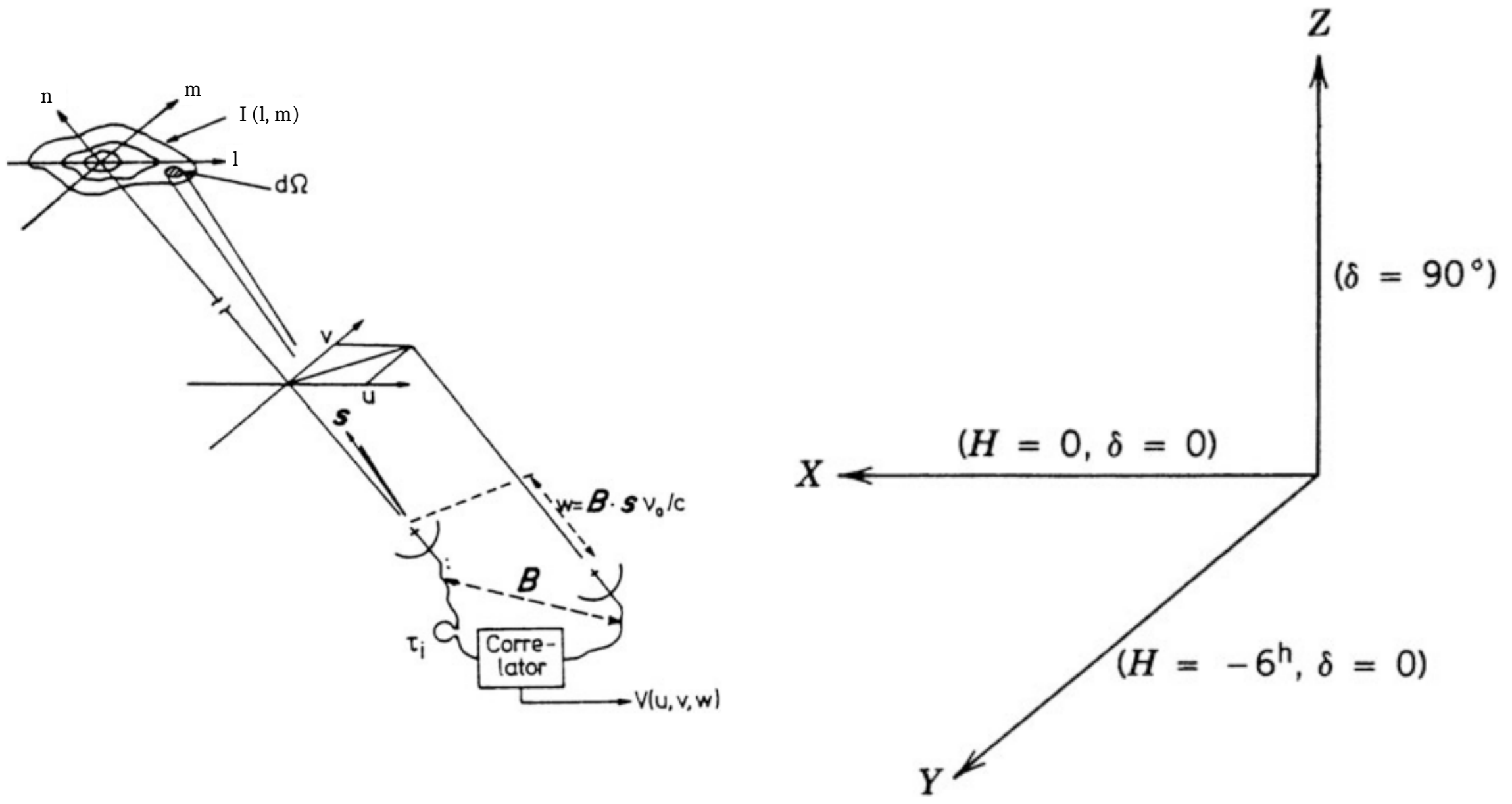

Figure 3.8: *Left*: Coordinates and geometry used in interferometry. Figure adapted from Wilson et al. (2013). *Right*: The coordinate system of relative position of radiotelescopes. Direction of the $X$-, $Y$-, and $Z$-axes are in terms of hour angle $H$ and declination $\delta$: [($0^{\mathrm{h}}$, 0°), ($-6^{\mathrm{h}}$, 0°), 0°]. Figure taken from Thompson et al. (2001).

### 3.5.4 *uv* data and aperture synthesis

The aperture synthesis is a procedure used for obtaining additional values of $\Upsilon$ without rearranging the antennas of an interferometer. It uses the rotation of Earth to produce more baselines without building new radiotelescopes, because the rotation of our planet varies the projected baseline in the $uv$-coverage of an interferometer, in which the antennas are fixed on the Earth surface (as seen from the observed sources). In particular, for a given pair of antennas configuration the values of $u$, $v$, $w$, as seen from the source, are given by the next transformation matrix:

$$\begin{bmatrix} u \\ v \\ w \end{bmatrix} = \begin{bmatrix} -\sin H & \cos H & 0 \\ \sin\delta\cos H & \sin\delta\sin H & \cos\delta \\ \cos\delta\cos H & -\cos\delta\sin H & \sin\delta \end{bmatrix} \begin{bmatrix} X \\ Y \\ Z \end{bmatrix}, \tag{3.45}$$

where $H$ and $\delta$ represent the hour angle and declination of the phase reference position, respectively. $X$, $Y$, and $Z$ represent the coordinate system where $Z$ points the North Pole, $X$ follows an axis oriented in the plane of the sky and points to $H = 0^{\mathrm{h}}$, and $Y$ is an axis contained in the celestial plane and oriented to the East, with $H = -6^{\mathrm{h}}$ (see Fig. 3.8 *right*). Thus, all confined baselines in an East-West line will remain in a single plane perpendicular to the North-South rotation axis of Earth as our planet turns daily.

Each pair of radiotelescopes of an interferometer measures just one Fourier component of the brightness distribution of the astronomical radio source. Therefore, each new baseline adds a new Fourier component to the data (and also improve the sensitivity). To be more specific, the $uv$-coverage of a $N$-antennas interferometer will be $2 \times N(N-1)/2$ concentric ellipses and each baseline adds a new Fourier component (unique fringe fitting) to the image to improve the interferometric data (see Fig. 3.9).

Thus, the combination of multiple radiotelescopes, movable radiotelescopes, and the rotation of our planet provides sufficient Fourier components in the $uv$-plane to synthesize the effect of a large aperture and to get high-resolution maps of the source.

### 3.5.5 The point spread function of an interferometer

Actually, visibilities derived from an interferometric observation are $\Upsilon(u, v)\,\zeta(u, v)$, instead of just $\Upsilon(u, v)$, where is $\zeta(u, v)$ a sampling function that checks the presence of data in the $uv$-coverage.

$$\zeta(u, v) = \begin{cases} 1, & \text{if data} \\ 0, & \text{if no data} \end{cases}. \tag{3.46}$$

So, assuming this, Eq. 3.44 becomes:

$$I''(l, m) = \int_{-\infty}^{+\infty} \Upsilon(u, v)\,\zeta(u, v)\exp[-2\pi \mathrm{i}(ul + vm)]\,\mathrm{d}u\,\mathrm{d}v. \tag{3.47}$$

When mapping through visibilities we do not get $I(u, v)$, instead we get the intensity of the object conjugated with the Fourier transform of the function $\zeta(u, v)$. In the case of infinitely small point sources located at the centre of phase, $I(u, v) = \delta(0, 0)$, the image obtained will be the inverse Fourier transform of the visibilities: $\zeta'(u, v)$.

This process is known as “mapping”, because these mathematical sequences imply passing $uvt$-data to a map, which results in what is called a dirty image of the object. This map is termed dirty because it is convolved with the point-spread function (PSF) of the interferometer, which is also known as dirty beam. See Figs. 3.9 and 3.11 (*top*). So, the final image will be:

$$I''(l, m) = I(l, m) * \mathrm{PSF}(l, m). \tag{3.48}$$

A better coverage of the $uv$-plane implies a better PSF (see Fig. 3.9). This means that the dirty image will look more like the final real image after cleaning (process to be discussed in Sect. 3.5.7).

### 3.5.6 Tapering and density weighting

There are two relevant kinds of weighting frequently used in radioastronomy: tapering and density weighting.

Tapering consists in apodizing the $uv$-coverage by the function $\exp[-\left(u^2 + v^2\right)/t_\mathrm{d}^2]$, where the factor $t_\mathrm{d}$ represents the tapering distance. This function is often a Gaussian function. The applying of a tapering distance weighting may imply to smoothing the data in the image to suppress small-scale variations in the dirty map caused by incomplete sampling on the longest baselines. As a consequence, the use of a taper may be beneficial for studying medium and relatively large structures.

The other important kind of weighting is the density weighting $W(u, v)$, because this factor is proportional to the number of $uv$-measurements of a given gridded cell. Therefore, it is possible to choose the mapping method to change the angular resolution and sensitivity of the image by selecting the appropriate weighting function $W(u, v)$:

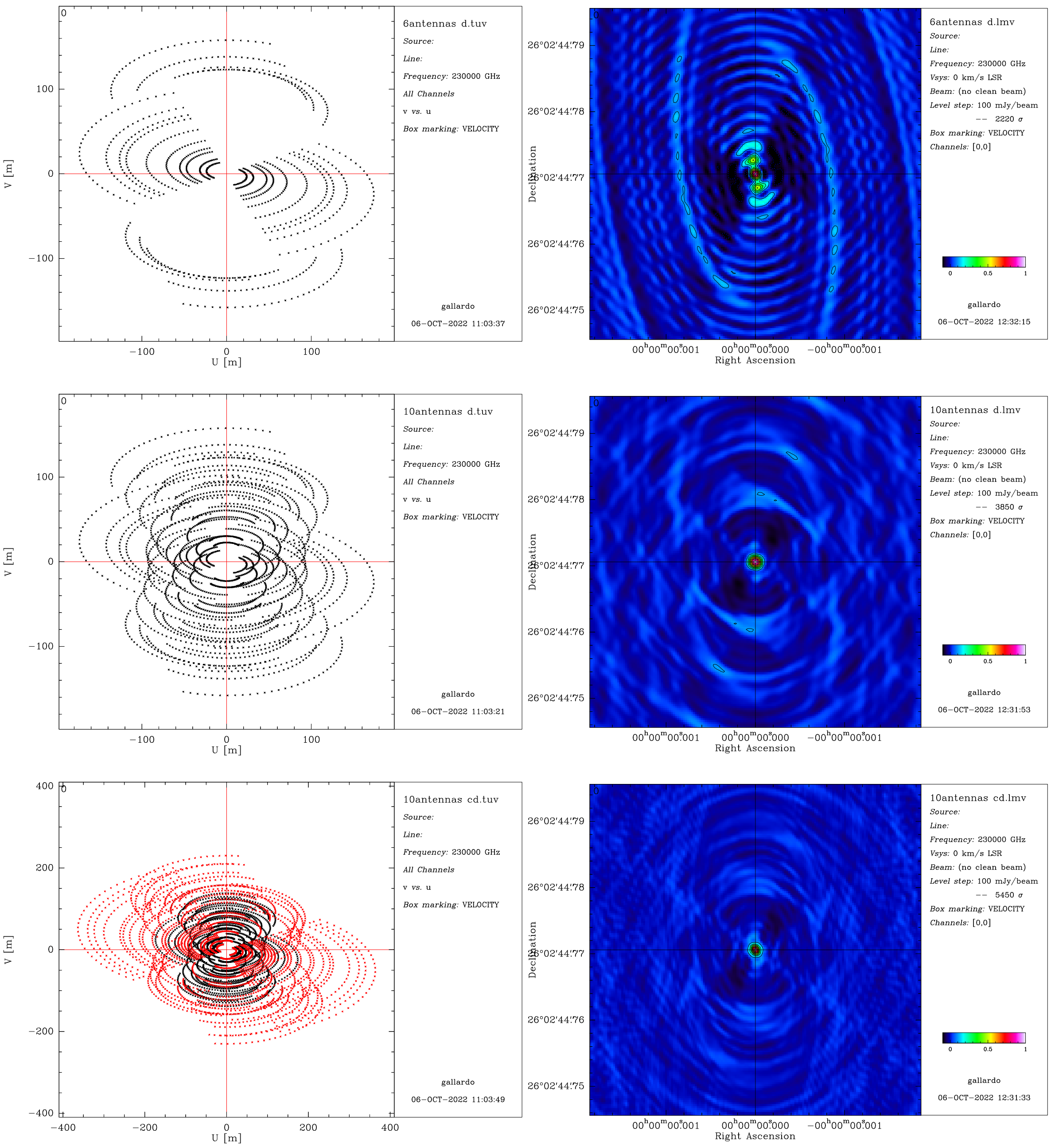

Figure 3.9: Synthetic NOEMA $uv$-coverage and dirty beam for different number of antennas and array configuration: 6 antennas in D configuration (*top*), 10 antennas in D configuration (*middle*), and 10 antennas in CD configuration (*bottom*). The scales in contours are the same in all the three cases. As we can appreciate, a better coverage of the $uv$-plane implies a smaller contribution of the secondary lobes of the dirty beam. Therefore, many antennas, and in several array configurations is better to cover the $uv$-plane.

- Natural: this method weights as $W(u, v) = 1/\sigma^2(u, v)$, where $\sigma^2(u, v)$ is variance of the noise of the $uv$-sample. This method provides the lowest noise in the final image but gives less resolution and highlights the extended structures. This is because short baselines produce more densely sample $uv$-tracks. Nevertheless, natural weighting provides more weights to the short baselines, where there are more visibilities measurements, degrading the resolution.

- Uniform: this method gives a weight inversely proportional to the sampling density function of the $uv$-plane. The weight is the same for each cell of the $uv$-plane (each cell has a typical value of $D/2$). The uniform weighting maximizes the spatial resolution. However, the noise is higher in final maps (it can be two times worse than natural weighting).

- Robust: this weighting method is between the uniform and the natural weighting and provides a compromise between them. It depends on a given threshold value $w_\mathrm{r}$. A large $w_\mathrm{r}$ provides natural weighting, while a small $w_\mathrm{r}$ gives uniform weighting.

### 3.5.7 Cleaning process

As we have seen before, the Fourier transform of our visibilities is not the image of our source, it is the image of our object convolved with the PSF (see Eq. 3.48 and Sect. 3.5.5 for details). This image is called dirty image, which shows structures due to the shape of the PSF, i.e. the interferometer response.

In order to be able to compare the image of our source with reality or with models, cleaning algorithms are developed to remove the PSF contribution through a deconvolution process (knowing $I''$ and PSF, we can obtain $I$, see Eq. 3.48). In this way, the resulting image will be called a clean image and can be obtained via the clean algorithm. The clean algorithm is based on the next fundamental steps:

1. Computing the image and the response to a point source by Fourier transformation of the visibility and the weighted transfer function.

2. To find the most intense point in the image and subtract the response to a point source (the dirty beam), including the full sidelobe pattern, centred on that position. The peak amplitude of the subtracted point-source response is equal to $\gamma$ times the corresponding image amplitude. $\gamma$ is a factor known as loop gain that presents a value of a few tenths $(0.1 - 0.2)$. Save the amplitude and the position of the component removed by inserting a $\delta$-function component into a model that will become the cleaned image.

3. The point 2 must be repeated in a iteratively way until all significant structures have been removed from the image. The user can check this condition by comparing the most intense peak with the noise level, or note when negative components begin to be deleted, or find the first time that the noise level does not decrease when subtraction is performed.

4. To replace each $\delta$-function with a clean beam function with its corresponding amplitude, or in a more accurate way, convolve the $\delta$-functions in the cleaned model with a clean-beam response. The clean beam is usually chosen as a Gaussian with a half-amplitude width equal to that of the original synthesized dirty beam (or a similar function that is free of negative values).

5. To add the residual intensities from point 3 into the clean-beam image, which is then the output. When the residual intensities are added into the clean beam image, the Fourier transform of this image is equal to the measured visibilities.

An example of applying this algorithm can be seen in Fig. 3.11. We show the result for 5, 40, 320, and 2560 iterations. The map quality is better in the case of more iterations (it is also clear on the residuals map).

There are many deconvolution processes based on the above algorithm, such as Högbom, Clark, or Multi-Scale. Högbom is the classic clean algorithm, in which points are found iteratively by searching for the peak and each point is subtracted from the full residual image using the shifted and scaled point spread function (Högbom, 1974). Clark is the most popular variant to the Högbom algorithm, because is faster. The original Högbom clean algorithm invests a lot of time looking for residuals that could be insignificant. However, Clark clean algorithm looks for a list of only the largest residuals, working with only a subregion of the beam for much time (Clark, 1980). Multi-Scale is a clean algorithm that finds emission on the largest scales of the source first, moving to finer details with each iteration (Cornwell, 2008).

### 3.5.8 Solving the filtered out flux problem

During the interferometric observations, the central regions of the $uv$-plane is never covered, $u = v = 0$, in any possible configuration. The minimum spacing of two radiotelescopes is literally limited by their size. There is a region of no sensitivity that corresponds to the space between centres of two antennas. In this way, the maximum recoverable scale is limited by 2 times the radios of the antenna, or its diameter $D$ (in $\lambda$-units). In the case of NOEMA, this factor is $2{\times}7.5\,$m. Therefore, astronomical structures with sizes larger than the maximum recoverable scale cannot be detected by the interferometer. This results in a filtered out flux problem or short-spacing problem. Theoretically, these large structures can be recovered if the signal-to-noise is infinitely large, but in practise, as we have finite S/N, these large structures cannot be recovered in a typical situation.

This undetected flux problem can be solved by the mapping of the sources with a single-dish instrument, which does not present this problem as the detection of the signal is not produced by interferometric means. In order to do this, a total power map sufficiently large must be obtained. This can be do using the OTF procedure or by other means (see Fig. 3.10 and Sect. 3.4.2). The resulting total-power maps will provide short-spacing pseudo-visibilities from $1/s$ to $D\,[\lambda]$, where $s$ is the size of the map. The reason of the size of the 30 m IRAM telescope is to provide pseudo-visibilities to this uncovered region of the NOEMA $uv$-coverage. 30 m IRAM provides resolution up to $\lambda/30\,$m, while NOEMA provides measurements down to $\lambda/15\,$m. Therefore, between $\lambda/30$ and $\lambda/15$ there is an overlapping region that can be used to test the consistency of the calibration of both instruments (single-dish and inteferometer). Thus, combined maps (interferometer + single-dish) will contain all detectable flux because they recover the filtered out flux by the interferometer of the non detected extended components. In addition, combined maps will include large areas not observable by single-dish single-pointed observations due to limitations of the beam of the radiotelescope. It is important to mention that the primary beam attenuation is automatically corrected for the interferometric data in the merging with pseudo-visibilities from the total-power data.

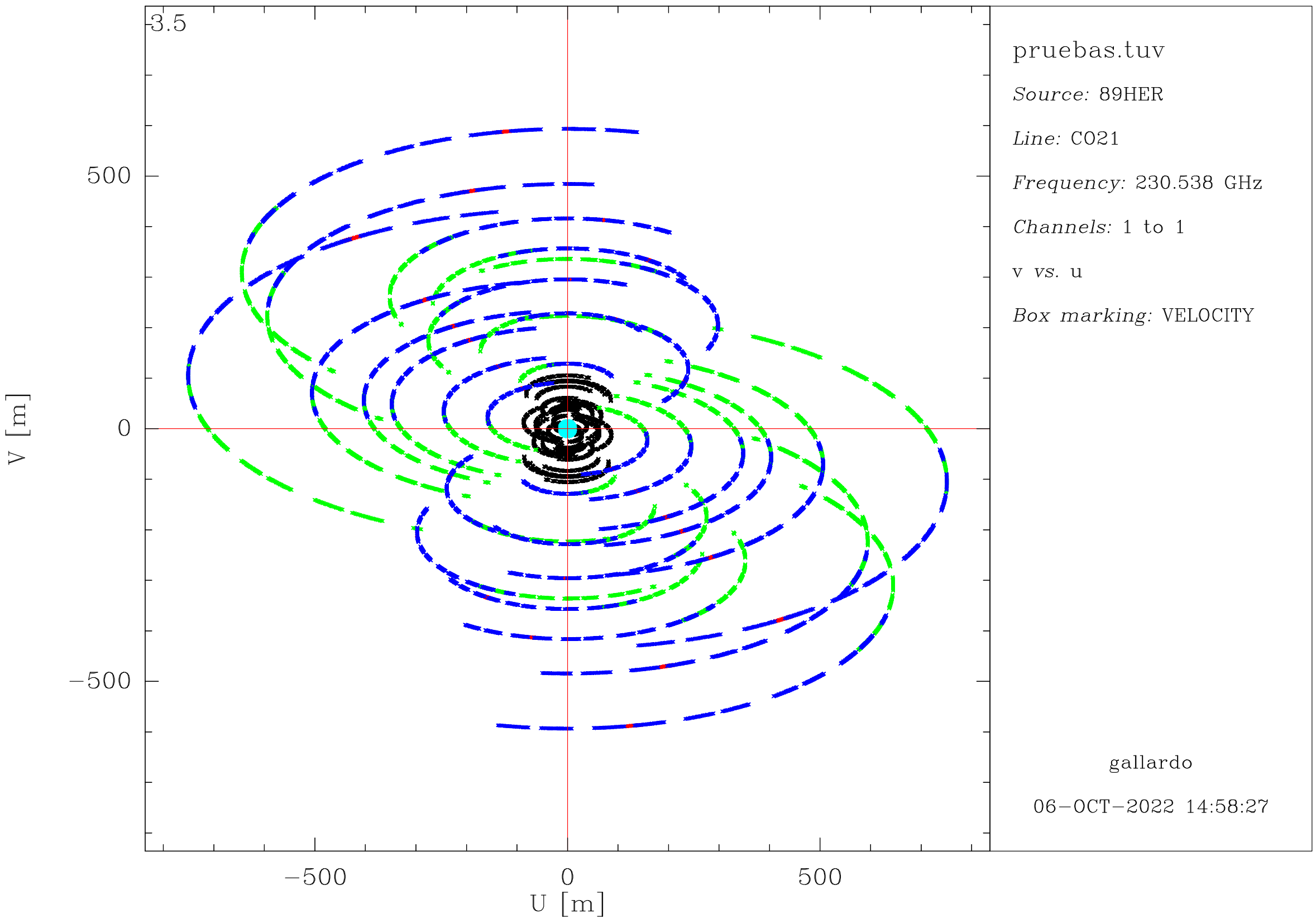

Figure 3.10: *uv*-coverage of interferometric maps in different array configurations (in blue, green, and black) that include short-spacing pseudo-visibilities at central positions of the *uv*-map (in cyan).

This method has been used in one of our sources: 89 Herculis. Our NOEMA maps suffered a significant amount of flux loss. The combination of these previous NOEMA maps and the total-power maps (derived from OTF observations) contain all detectable flux. The new combined maps recover the flux loss of the extended component filtered out by the interferometer and they also include large areas that single-dish single-pointed observations cannot detect due to the limitations of the beam of the radiotelescope. See Chapter 7, for a complete description.

### 3.5.9 Conversion from flux density to brightness temperature

The conversion from flux density to brightness temperature, for sources with sizes small compared to the telescope beam, is given by the Rayleigh-Jeans approximation (see 2.42). So, for beam shapes and a Gaussian source, this relation is:

$$T = 1.222 \times 10^3 \frac{S}{\nu^2 \theta_{\rm M} \theta_{\rm m}}, \tag{3.49}$$

where the frequency $\nu$ is expressed in GHz, $\theta_{\rm M}$ and $\theta_{\rm m}$ are the HPBW along the major and minor axes, respectively, and are expressed in arcseconds. The flux density $S$ is in mJy beam$^{-1}$, and the brightness temperature $T$ is in K. As we see, this conversion factor only depends on the observational frequency and the beam size.

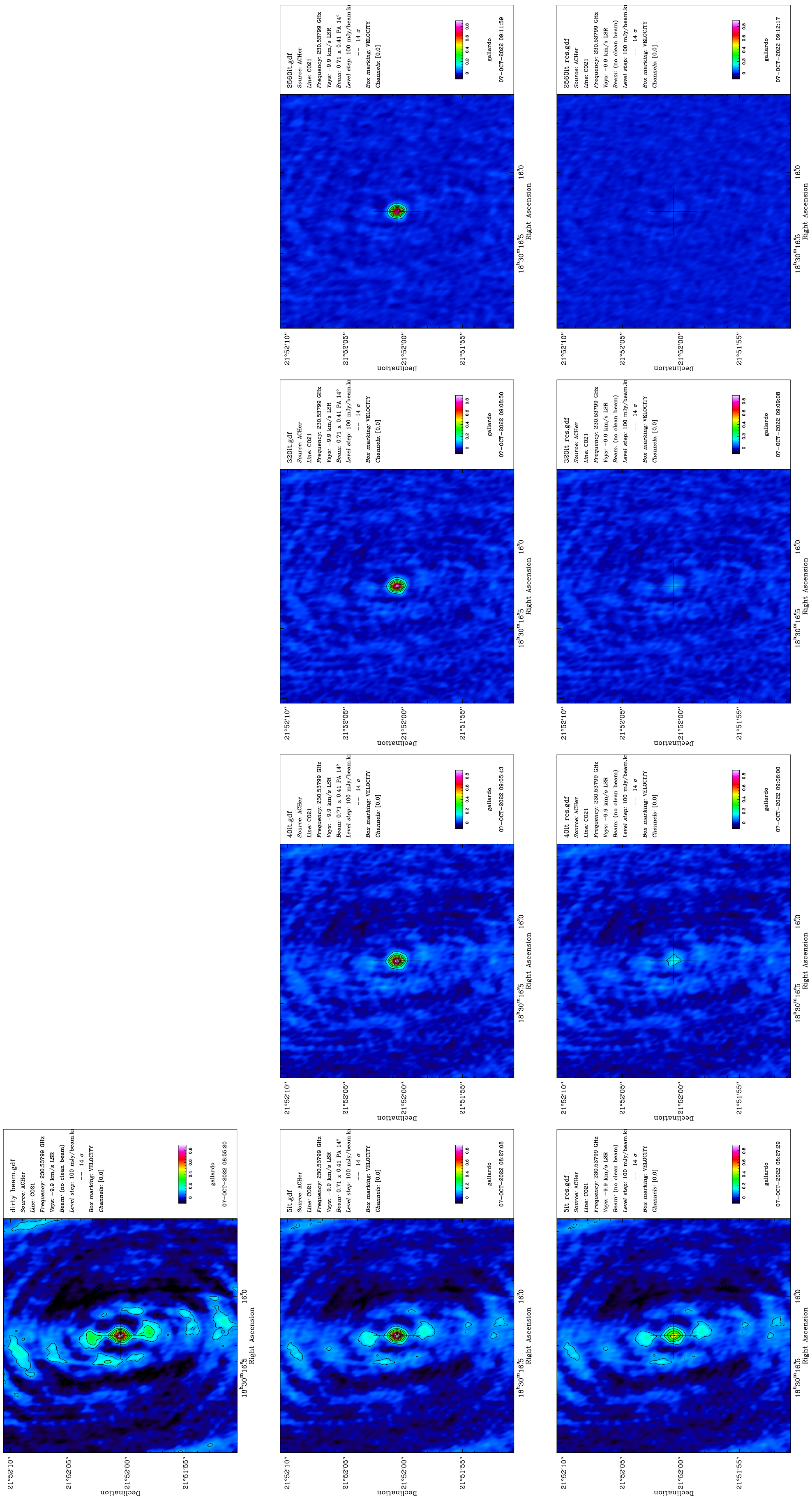

Figure 3.11: Dirty beam, final images after different iterations, and their residuals of $^{12}$CO $J = 2 - 1$ emission from AC Her. The dirty beam is shown in the *Top* panel. The final images are shown in the *Middle* panels for 5, 40, 320, and 2560 iterations of cleaning. Finally, the residuals of the dirty beam and the final images are shown in the *Bottom* panels. The scales in contours are the same in all the three kinds of maps.

# Part II

# Scientific research

*There are occasions when it pays better to fight and be beaten than not to fight at all.*

George Orwell – Homage to Catalonia

# 4

# Keplerian disks and outflows in post-AGB stars: AC Herculis, 89 Herculis, IRAS 19125+0343, and R Scuti

*There is a class of binary post-AGB stars that show indications of the presence of Keplerian disks around them. However, their detection is difficult, because it requires high spectral and spatial resolution. Only four cases have been well studied through mm-wave interferometric maps. These nebulae present a disk with Keplerian dynamics. In addition, another low-velocity expanding component that represents ∼ 10% of the total nebular mass. In contrast, there is a group of nebulae around binary post-AGB stars that also show evidences of the presence of a rotating disk, but off a large and massive outflow too, from their CO line profiles. Therefore, the study of these sources through mm-wave interferometric observations is crucial to understand the similarities and differences between these types of sources: those with and without massive outflows.*

*In this chapter, we present a detailed study of the nebulae around binary post-AGB stars. In the case of 89 Herculis, IRAS 19125+0343, and R Scuti, these nebulae are composed of a Keplerian disk and an extended and expanding component surrounding the disk. The interferometric maps (and the CO line profiles) are quite different in the disk-type and in the outflow-type nebulae and could reveal that these outflows could be even more massive than the rotating component. In the case of AC Herculis, the rotating component had been confirmed, but the presence of an outflow was completely unknown. The content of this chapter is adapted from Gallardo Cava et al. (2021).*

## Abstract

Context: There is a class of binary post-AGB stars with a remarkable near-infrared excess that are surrounded by Keplerian or quasi-Keplerian disks and extended outflows composed of gas escaping from the disk. The Keplerian dynamics had been well identified in four cases, namely the Red Rectangle, AC Her, IW Car, and IRAS 08544−4431. In these objects, the mass of the outflow represents $\sim$10% of the nebular mass, the disk being the dominant component of the nebula.

Aims: We aim to study the presence of rotating disks in sources of the same class in which the outflow seems to be the dominant component.

Methods: We present interferometric NOEMA maps of $^{12}$CO and $^{13}$CO $J = 2-1$ in 89 Her and $^{12}$CO $J = 2-1$ in AC Her, IRAS 19125+0343, and R Sct. Several properties of the nebula are obtained from the data and model fitting, including the structure, density, and temperature distributions, as well as the dynamics. We also discuss the uncertainties on the derived values.

Results: The presence of an expanding component in AC Her is doubtful, but thanks to new maps and models, we estimate an upper limit to the mass of this outflow of $\lesssim 3 \times 10^{-5}$ M$_\odot$, that is, the mass of the outflow is $\lesssim$5% of the total nebular mass. For 89 Her, we find a total nebular mass of $1.4 \times 10^{-2}$ M$_\odot$, of which $\sim$50% comes from an hourglass-shaped extended outflow. In the case of IRAS 19125+0343, the nebular mass is $1.1 \times 10^{-2}$ M$_\odot$, where the outflow contributes $\sim$70% of the total mass. The nebular mass of R Sct is $3.2 \times 10^{-2}$ M$_\odot$, of which $\sim$75% corresponds to a very extended outflow that surrounds the disk.

Conclusions: Our results for IRAS 19125+0343 and R Sct lead us to introduce a new subclass of binary post-AGB stars, for which the outflow is the dominant component of the nebula. Moreover, the outflow mass fraction found in AC Her is smaller than those found in other disk-dominated binary post-AGB stars. 89 Her would represent an intermediate case between both subclasses.

## 4.1 Introduction

Most of the protoplanetary (or pre-planetary) and planetary nebulae (pPNe and PNe) show fast bipolar outflows ($30-100$ km s$^{-1}$) with clear axial symmetry. These outflows are responsible for a good fraction of the total mass ($\sim$0.1 M$_\odot$) and carry very large amounts of linear momentum (Bujarrabal et al., 2001). The immediate precursor of the post-AGB stars, the asymptotic giant branch (AGB) stars, present spherical circumstellar envelopes, which are in isotropic expansion at moderate velocities, around $10-20$ km s$^{-1}$ (Castro-Carrizo et al., 2010). The spectacular evolution from AGB circumstellar envelopes to post-AGB nebulae takes place in a very short time ($\sim$1000 a). The accepted scenario to explain this evolution implies that material is accreted by a companion from a rotating disk, followed by the launching of very fast jets, in a process similar to that in protostars (Soker, 2002; Frank and Blackman, 2004; Blackman and Lucchini, 2014). However, the effect of binarity in post-AGB stars is still poorly understood (De Marco and Izzard, 2017).

There is a class of binary post-AGB stars that systematically show evidence of the presence of disks (Van Winckel, 2003; de Ruyter et al., 2006; Bujarrabal et al., 2013a; Hillen et al., 2017) and low initial mass (Alcolea and Bujarrabal, 1991). The observational properties of these objects were recently reviewed by Van Winckel (2018). All of them present a remarkable near-infrared (NIR) excess and the narrow CO line profiles characteristic of rotating disks. Their spectral energy distributions (SEDs) reveal the presence of hot dust close to the stellar system, and its disk-like shape has been confirmed by interferometric IR data (Hillen et al., 2017; Kluska et al., 2019). These

Table 4.1 Binary post-AGB stars observed in this work.

| Source | | Observed coordinates | | $V_{LSR}$ | $d$ | $P_{orb}$ | Sp. Type | Comments |
|---|---|---|---|---|---|---|---|---|
| GCVS name | IRAS name | J2000 | | [km s$^{-1}$] | [pc] | [d] | | |
| AC Herculis | 18281+2149 | 18:30:16.24 | +21:52:00.6 | −9.7 | 1100 | 1188.9 | F2 − K4 I | RV Tauri variable |
| 89 Herculis | 17534+2603 | 17:55:25.19 | +26:03:00.0 | −8.0 | 1000 | 289.1 | F2 I | Semiregular variable |
| – | 19125+0343 | 19:15:01.18 | +03:48:42.7 | 82.0 | 1500 | 519.7 | F2 | RV Tauri variable |
| R Scuti | 18448−0545 | 18:47:28.95 | −05:42:18.5 | 56.1 | 1000 | – | G0 − K0 I | Peculiar RV Tauri variable |

**Notes.** Distances and spectral types are adopted from Bujarrabal et al. (2013a). Velocities are derived from our observations. Orbital periods of the binary systems are taken from Oomen et al. (2018).

disks must be stable structures, because their IR spectra reveal the presence of highly processed grains (Gielen et al., 2011a; Jura, 2003; Sahai et al., 2011). Observations of $^{12}$CO and $^{13}$CO in the $J = 2-1$ and $J = 1-0$ lines have been well analyzed in sources with such a NIR excess (Bujarrabal et al., 2013a) and they show line profiles formed by a narrow single peak and relatively wide wings. These line profiles are similar to those in young stars surrounded by a rotating disk made of remnants of interstellar medium and those expected from disk-emission modelling (Bujarrabal et al., 2005; Guilloteau et al., 2013). These results indicate that the CO emission lines of our sources come from Keplerian or quasi-Keplerian disks. The systematic detection of binary systems in these objects (Oomen et al., 2018) strongly suggests that the angular momentum of the disks comes from the stellar system.

The study of Keplerian disks around post-AGB stars requires high angular- and spectral-resolution observations because of the relative small size of the rotating disks. To date, there are only four resolved cases of Keplerian rotating disks: the Red Rectangle (Bujarrabal et al., 2013b, 2016), AC Her (Bujarrabal et al., 2015), IW Car (Bujarrabal et al., 2017), and IRAS 08544−4431 (Bujarrabal et al., 2018b). All four have been very well studied through single-dish and interferometric millimeter(mm)-wave maps of CO lines. These four studied sources show CO spectra with narrow line profiles characteristic of rotating disks and weak wings. This implies that most of the material of the nebula is contained in the rotating disk. However, according to the mm-wave interferometric maps there is a second structure surrounding the disk that is less massive, contains ∼10% of the total mass, and is in expansion. This outflow is probably a disk wind consisting in material escaping from the rotating disk (see extensive discussion of the best studied source by Bujarrabal et al., 2016).

In the present work, we present NOEMA maps of $^{12}$CO and $^{13}$CO $J = 2-1$ emission lines in AC Her, 89 Her, and IRAS 19125+0343 —which are confirmed binaries—, and R Sct. R Sct is a different object, it is a very bright star in the visible and is a RV Tau variable, but its SED is different from those of the other sources in our sample, with a less prominent NIR excess. Also, its binary nature is not yet confirmed either. We discuss this source in Sect. 4.2.4. On the other hand, AC Her, 89 Her, and IRAS 19125+0343 are confirmed binaries that also belong to the same class of binary post-AGB stars with remarkable NIR excess. Our results of 89 Her, IRAS 19125+0343, and R Sct show that the extended outflowing component is even more massive than the rotating disk. This suggests that they are part of a new subclass: the outflow-dominated nebulae around post-AGB stars.

The paper is organized as follows. We present our post-AGB star sample in Sect. 4.2. Technical information of our observations is given in Sect. 4.3. We discuss the mm-wave interferometric maps in Sect. 4.4. Results from our best-fit models are shown in Sect. 4.5. Finally, we present our main conclusions in Sect. 4.6.

## 4.2 Source descriptions and previous results

We produced interferometric maps of four sources (Table 4.1) using the IRAM NOrther Extended Millimeter Array (NOEMA). All of them are identified as binary post-AGB stars (binary system including a post-AGB star) with low gravity, high luminosity, FIR-excess indicative of material ejected by the star, and the mentioned remarkable NIR excess, and have been previously studied in CO by means of single-dish observations.

We adopted the same distances used in Bujarrabal et al. (2013a). We chose these distances to facilitate the comparison with the results derived in that work and because, in the case of binary stars, the estimation of distances via parallax measurements is very delicate (Dominik et al., 2003). We discarded other options because they present large uncertainties. The best values are likely those derived from the not-yet-available *Gaia* Data Release 3 (*Gaia* DR3), which considers astrometry for binary systems. As we discuss in detail in Sect. 4.5.5, we will easily be able to scale the derived values of our best-fit models for some parameters, including size and mass, if and when a new and better value for the distance to the sources is determined.

### 4.2.1 AC Her

AC Her is a binary post-AGB star (Oomen et al., 2019); although some authors have suggested that it could rather be a post-RGB star (Hillen et al., 2015). The CO line profiles and interferometric maps of AC Her are very similar to the ones found in the Red Rectangle. According to this, AC Her must be a post-AGB star whose nebula is dominated by a Keplerian disk.

Previous mm-wave interferometric observations confirm that AC Her presents a clear disk with Keplerian dynamics, which dominates the nebula (Bujarrabal et al., 2015; Hillen et al., 2015). It presents a mass $\sim 1.5 \times 10^{-3}\,M_\odot$, densities of $10^6 - 10^4\,cm^{-3}$, and temperatures of $80 - 20\,K$. These results were obtained adopting a distance of 1600 pc. Approximately 40% of the total flux was filtered out by interferometric observations. However, no outflowing material was detected in these papers. Due to the large inner radius ($30 - 35$ AU) and the low gas-to-dust ratio, the Keplerian disk is very likely to be in an evolved state and its mass was probably larger before (Hillen et al., 2015).

### 4.2.2 89 Her

89 Her is a binary post-AGB star with a remarkable NIR-excess that implies the presence of hot dust (de Ruyter et al., 2006). It also shows significant far-infrared (FIR) emission. The presence of large grains was proposed by Shenton et al. (1995).

Single-dish observations show narrow CO lines that are very similar to the ones detected in the Red Rectangle, but with more prominent wings, which suggests a significant contribution of the outflow (Bujarrabal et al., 2013a). The source was studied in detail by Bujarrabal et al. (2007); see also Alcolea and Bujarrabal (1995) and Fong et al. (2006). Bujarrabal et al. (2007) discovered two features from NOEMA maps in

$^{12}$CO $J = 2 - 1$ in the nebula around 89 Her: an extended hourglass-shaped structure and a central clump, which probably corresponds to an unresolved disk. Near-infrared observations strongly suggest that in 89 Her the binary system is surrounded by a compact and stable circumbinary disk in Keplerian rotation, where large dust grains form and settle to the midplane (Hillen et al., 2013, 2014).

### 4.2.3 IRAS 19125+0343

IRAS 19125+0343 is a binary post-AGB star (Gielen et al., 2008). It also belongs to this class of binary post-AGB stars with remarkable NIR-excess (Oomen et al., 2018), implying the presence of hot dust. For this source, the assumed value for the distance (Table 4.1) is highly questionable because some authors adopt very different values, such as $1800 \pm 400$ pc (Gielen et al., 2008) or $4131^{+905}_{-645}$ pc (Bailer-Jones et al., 2018).

Its CO lines are narrow (Bujarrabal et al., 2013a), and so the hot dust must be located in a compact rotating disk, but they also have prominent wings. The circumbinary disk could be surrounded by an outflow containing most of the mass. The single-dish analysis yielded a nebular mass of $1.3 \times 10^{-2}$ M$_{\odot}$.

### 4.2.4 R Sct

While the first three sources mentioned above are spectroscopically confirmed binaries, this is not the case for R Sct. This bright RV Tauri star shows very irregular pulsations with variable amplitude (see Kalaee and Hasanzadeh, 2019), and only a small IR excess (Kluska et al., 2019), meaning that the SED is not clearly linked to the presence of a circumbinary disk. Hence, the star was labeled "uncertain" in the classification of post-AGB stars surrounded by disks in Gezer et al. (2015). The high amplitude of the pulsations is also reflected in the radial-velocity amplitude (Pollard et al., 1997). It is worth mentioning that a magnetic field has been detected; see e.g. Tessore et al. (2015). Moreover, Matsuura et al. (2002) suggest that it could be an AGB star in the helium-burning phase of the thermal pulse cycle.

However, it is known that R Sct is an RV Tauri type variable source and hence probably a post-AGB star. Its CO line profiles are different from those of other binary post-AGB stars studied by Bujarrabal et al. (2013a). Following this, it might be compatible with the presence of a Keplerian disk, and as Keplerian disks are only detected around binaries, it could well be binary as well, but this needs to be confirmed. This hypothesis is reinforced by interferometric data in the $H$-band showing a very compact ring (Kluska et al., 2019). Therefore, we tentatively consider R Sct as a binary post-AGB star, like the other three sources. The nebular mass derived from the single-dish studies is $5 \times 10^{-2}$ M$_{\odot}$, where $\sim$14% would correspond to the disk.

## 4.3 Observations and data reduction

Observations of the $^{12}$CO $J = 2 - 1$ rotational transition at 230.53799 GHz were carried out towards AC Her, 89 Her, IRAS 19125+0343, and R Sct with the IRAM NOEMA interferometer at Plateau de Bure (France) with six antennas. Data of the $^{13}$CO $J = 2 - 1$ transition were also obtained for 89 Her. The data calibration was performed with the CLIC software (GILDAS package). In all the observing sessions, measurements with high signal-to-noise ratio were obtained on a bright calibration

Table 4.2 Observational parameters.

| Source | Project | Observation dates | Baselines [m] | Obs. time [h] | Beam size | Sp. Resol. [km s$^{-1}$] | Noise [mJy beam$^{-1}$] |
|---|---|---|---|---|---|---|---|
| AC Herculis | W14BU | Dec. 14, Apr. 15, and Mar. 16 | 17 – 760 | 21 | 0.″35 × 0.″35 | 0.2 | $4.8 \times 10^{-3}$ |
| 89 Herculis | X073<br>W14BT | Jan. 14 to Mar. 14<br>Dec. 14 and Feb. 15 to Apr. 15 | 15 – 760 | 22 | 0.″74 × 0.″56 | 0.2 | $4.9 \times 10^{-3}$ |
| IRAS 19125+0343 | WA7D | Feb. 13 to Mar. 13 | 55 – 760 | 12 | 0.″70 × 0.″70 | 1.0 | $5.4 \times 10^{-3}$ |
| R Scuti | S15BA | Summer 15 and Autumn 15 | 18 – 240 | 12 | 3.″12 × 2.″19 | 0.5 | $2.2 \times 10^{-2}$ |

**Notes.** This table collects the main parameters of NOEMA maps of $^{12}$CO $J = 2-1$ of AC Her, IRAS 19125+0343, and R Sct. It also includes the main details of NOEMA maps of $^{13}$CO $J = 2-1$ of 89 Her.

source in order to calibrate the instrumental RF spectral response. Calibration sources close to each source were cyclically observed, and their data used to perform a first standard gain calibration. Phase self-calibration was performed later on the source continuum emission. In all the sessions, the calibration source MWC 349 was observed as the primary flux reference, with an adopted flux of 1.91 Jy at 230.5 GHz. The data four all the observed line emissions were obtained with a spectral resolution of 0.05 km s$^{-1}$. Additional bandwidth with a channel spacing of 2 – 2.5 MHz was delivered to map continuum emission, the available bandwidth being different for the different observations. More details can be found below and a summary of observational parameters can be found in Table 5.3. MAPPING, also part of GILDAS, was used for data analysis and image synthesis. Final maps were obtained after continuum-emission subtraction and analyzed following image synthesis using natural and robust weightings of the visibilities. For the relatively low resolution, all sources present continuum images that are compact and unresolved, and are only used in the reduction of the data (see Sect. 4.4).

AC Her was observed with observatory project names XB74 and W14BU. The data of project XB74 were already presented by Bujarrabal et al. (2015). In this work, we add to the previous data those of new project W14BU, which were aimed to increase the sensitivity. The new data were obtained in December 2014, April 2015, and March 2016, with baselines ranging from 17 to 760 m. Considering the addition of the two projects, a total of 21 h were obtained on source, with an available bandwidth for continuum from 228.2 to 231.8 GHz. In Fig. 4.1, we present channel maps of $^{12}$CO $J = 2-1$ emission with a resampled spectral resolution of 0.2 km s$^{-1}$ and a synthetic beam of 0.″35 × 0.″35 in size, obtained using natural weighting. Relevant position-velocity diagrams are shown in Fig. 4.2.

89 Her was observed for 11 h in November 2005 and January 2006 under project name P05E. In addition to the line emission, data were delivered for continuum in a bandwidth of 600 MHz. These data were published by Bujarrabal et al. (2007) where more details can be found. A new uv-data processing was performed using tapering in order to reduce small-scale variations in the crude map at the expense of spatial resolution. New channel maps are presented in the top panel of Fig. 4.3, which were obtained with a synthetic beam of 1.″02 × 0.″83 in size with a spectral resolution of 0.2 km s$^{-1}$, the major axis being oriented at $PA = 115°$. In addition, observations of the $^{13}$CO $J = 2-1$ line emission in 89 Her were performed under the project names X073 and W14BT. X073 data were obtained with the most extended configuration between January and March 2014, those for W14BT in December 2014, and from February to April 2015 for more compact B and D configurations, in order to attain baselines ranging from 15 to 760 m. Acquisitions were obtained on source for a total of 22 h, with a continuum bandwidth from 218.1 to 221.7 GHz with 2 MHz of channel spacing. MWC 349 was observed in all the tracks as in the primary flux calibrator, with an adopted flux of 1.86 Jy. In the bottom panel of Fig. 4.3, we present channel maps of the $^{13}$CO $J = 2-1$ line emission with a spectral resolution of 0.2 km s$^{-1}$ and a synthetic beam of $0\farcs74 \times 0\farcs56$ in size, the major axis being oriented at $PA = 28°$, obtained with natural weighting. In Fig. 4.4 we present relevant position–velocity diagrams.

Observations of IRAS 19125+0343 were made between February and March 2013 under the project name WA7D. A total of 12 h were obtained on source with extended array configuration, baselines ranging from 55 to 760 m. The interferometric visibilities were merged with zero-spacing data obtained with the 30 m IRAM telescope, which guarantees that no flux is missed in final maps. We present channel maps of the $^{12}$CO $J = 2-1$ line emission with a channel spacing of 1 km s$^{-1}$ and a synthetic beam of $0\farcs7 \times 0\farcs7$ in size (Fig. 4.6). We chose a circular beam in order to best compare with models, while the original data yield a synthetic beam of $0\farcs7 \times 0\farcs4$ with natural weighting. The spectral configuration was the same as described for AC Her. In Fig. 4.7 we present relevant position–velocity diagrams.

Observations of R Sct were performed in the summer and autumn of 2015 with the array in the most compact configurations (project name S15BA). A total of 12 h were obtained on source. The spectral configuration was the same as that described for AC Her. As R Sct line emission is considerably extended, we added to the NOEMA visibilities short-spacing pseudo-visibilities obtained from on-the-fly maps with the 30 m IRAM telescope to compensate the extended component filtered out by the interferometry. In Fig. 4.8, we present channel maps with a spectral resolution of 0.5 km s$^{-1}$ and a synthetic beam of $3\farcs12 \times 2\farcs19$ in size, with the major axis oriented at $PA = -185°$. We show relevant position–velocity diagrams in Fig. 4.9.

## 4.4 Interferometric observational results

In this section, we present the results directly obtained from the observations. We show maps per velocity channel and position–velocity (PV) diagrams along the assumed two major perpendicular directions of the nebula: along the equatorial rotating disk and along the revolution symmetry axis of the nebula.

In the cases of the nebula around 89 Her, IRAS 19125+0343, and R Sct the central rotating disk is not well resolved. However, we underline that: (a) their CO single-dish profiles are similar to those of other binary post-AGB stars (Bujarrabal et al., 2013a), in which we know that there are central disks and that their contribution to the profiles is significant, (b) position-velocity diagrams show hints of the typical PV diagram structure of disks with Keplerian dynamics, (c) the high-velocity dispersion observed in the central regions is the same as that expected from an unresolved Keplerian disk, and (d) when longer baselines are favored and inner regions are selected (even at the cost of losing flux), we find a line profile with two peaks characteristic of rotating disks (see CO line profiles of 89 Her and R Sct in Sect. 4.7.2). For these reasons, we think that these three nebulae (89 Her, IRAS 19125+0343, and R Sct) probably harbor a rotating disk with Keplerian dynamics at their center and we include such a structure in our modeling (see Sect. 4.5).

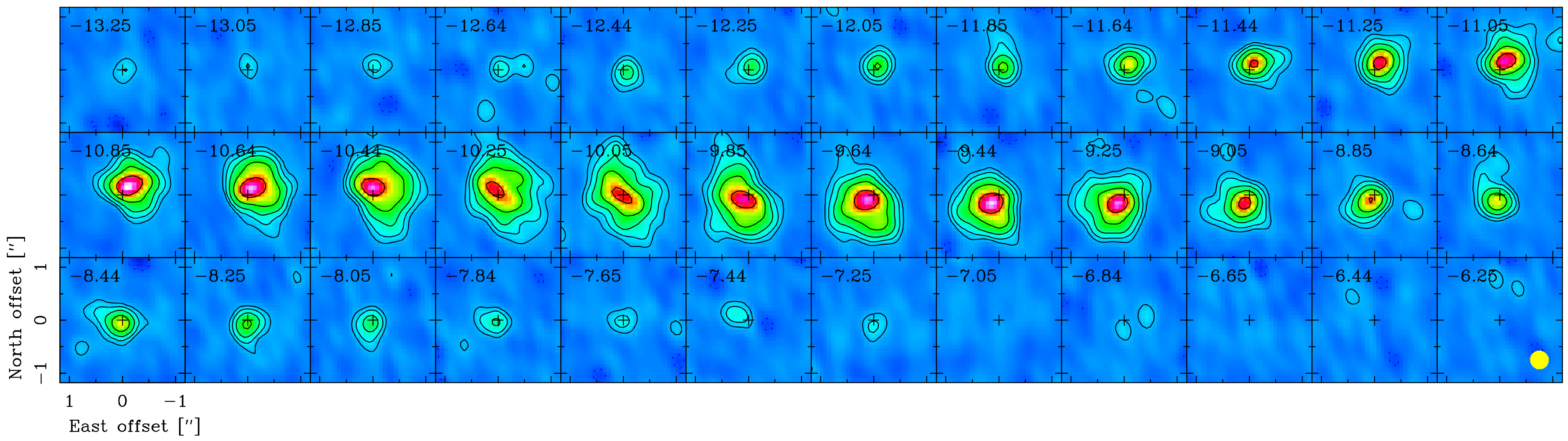

Figure 4.1: NOEMA maps per velocity channel of $^{12}$CO $J = 2 - 1$ emission from AC Her. The beam size (HPBW) is $0''.35 \times 0''.35$. The contour spacing is logarithmic: $\pm 9$, 18, 36, 76, and 144 mJy beam$^{-1}$ with a maximum emission peak of 230 mJy beam$^{-1}$. The LSR velocity is indicated in the upper left corner of each velocity-channel panel and the beam size is shown in the last panel in yellow.

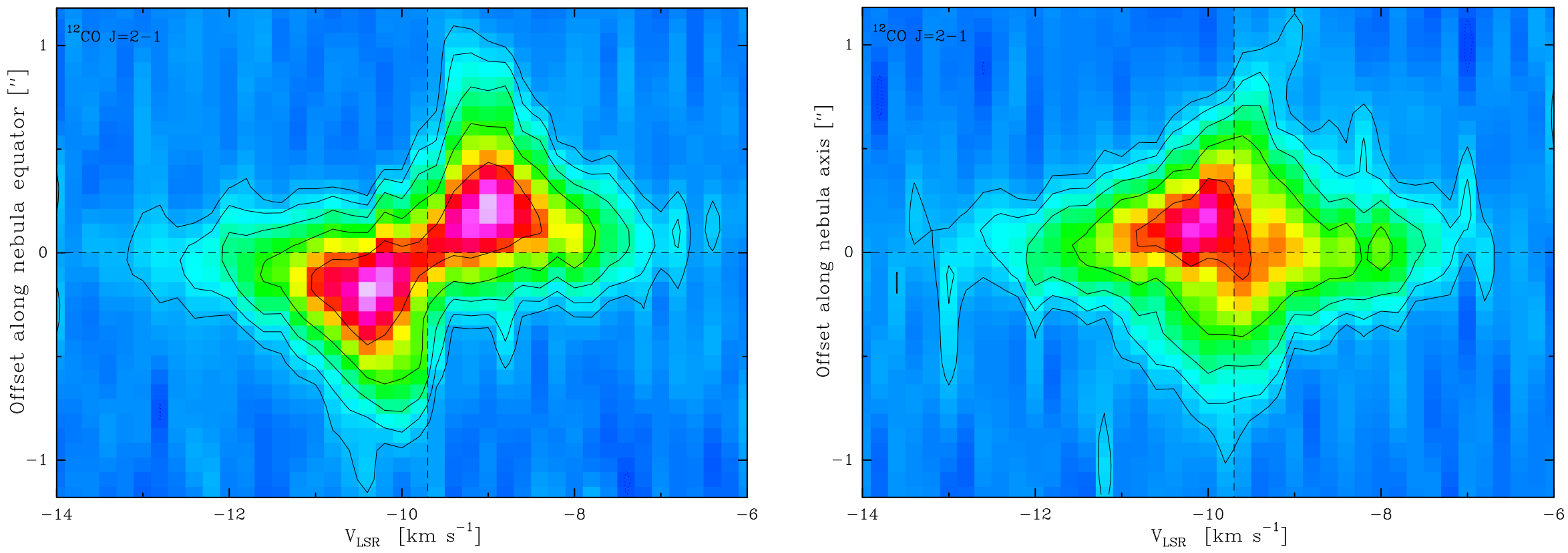

Figure 4.2: *Left:* PV diagram from our NOEMA maps of $^{12}$CO $J = 2-1$ in AC Her along the direction $PA = 136.1°$, corresponding to the nebula equator. The contour spacing is logarithmic: ± 9, 18, 36, 76, and 144 mJy beam$^{-1}$ with a maximum emission peak of 230 mJy beam$^{-1}$. The dashed lines show the approximate central position and systemic velocity. *Right*: Same as in *left* but along the perpendicular direction $PA = 46.1°$.

### 4.4.1 AC Her

The $^{12}$CO $J = 2-1$ mm-wave interferometric results are presented in Fig. 4.1. These observations are of higher resolution than those previously published (Bujarrabal et al., 2015). In addition, we also include shorter baselines to reduce the flux loss (no significant amount of flux was missed in the interferometric data; see Fig. 4.19) and to improve the sensitivity for the detection of the expanding component. A Keplerian disk was already detected in the published data; here we focus on the detection of the outflow.

In Sect. 4.7.3 we present an analysis of the emission of the outflow along different $PA$ (position angle, measured from north to east). Thanks to our detailed study, we determined that the direction better showing the Keplerian dynamics is $PA = 136.1° \pm 1.4°$ (see Sect. 4.7.3 for details); this direction is termed the "equatorial direction" hereafter.

As we can see in the left panel of Fig. 4.2, by comparison with results in Sect. 4.7.3, the PV diagram along this equatorial direction very nicely shows the characteristic signature of Keplerian rotation. The investigation of the PV diagram along the nebula axis (i.e., perpendicular to the equatorial direction) should help us to detect the presence of an axially expanding outflow. In the PV along the nebula axis in the right panel of Fig. 4.2, we can see a strong emission from the Keplerian disk in the central regions. The theoretical PV diagram along the nebula axis in pPNe in the presence of just a rotating disk shows emission with a form similar to a diamond or rhombus with similar emission in all four PV diagram quadrants. On the contrary, we see how the emission at central velocities is slightly inclined, which could be explained by the presence of a low-mass outflow surrounding the Keplerian disk. With these new observations we tentatively detect these weak vestiges of the outflow of AC Her, which may have a structure similar to the one found in similar sources like the Red Rectangle. From the detailed analysis in Sect. 4.7.3, we conclude that the outflow is tentatively detected with an emission $\lesssim 10$ mJy beam$^{-1}$ km s$^{-1}$.

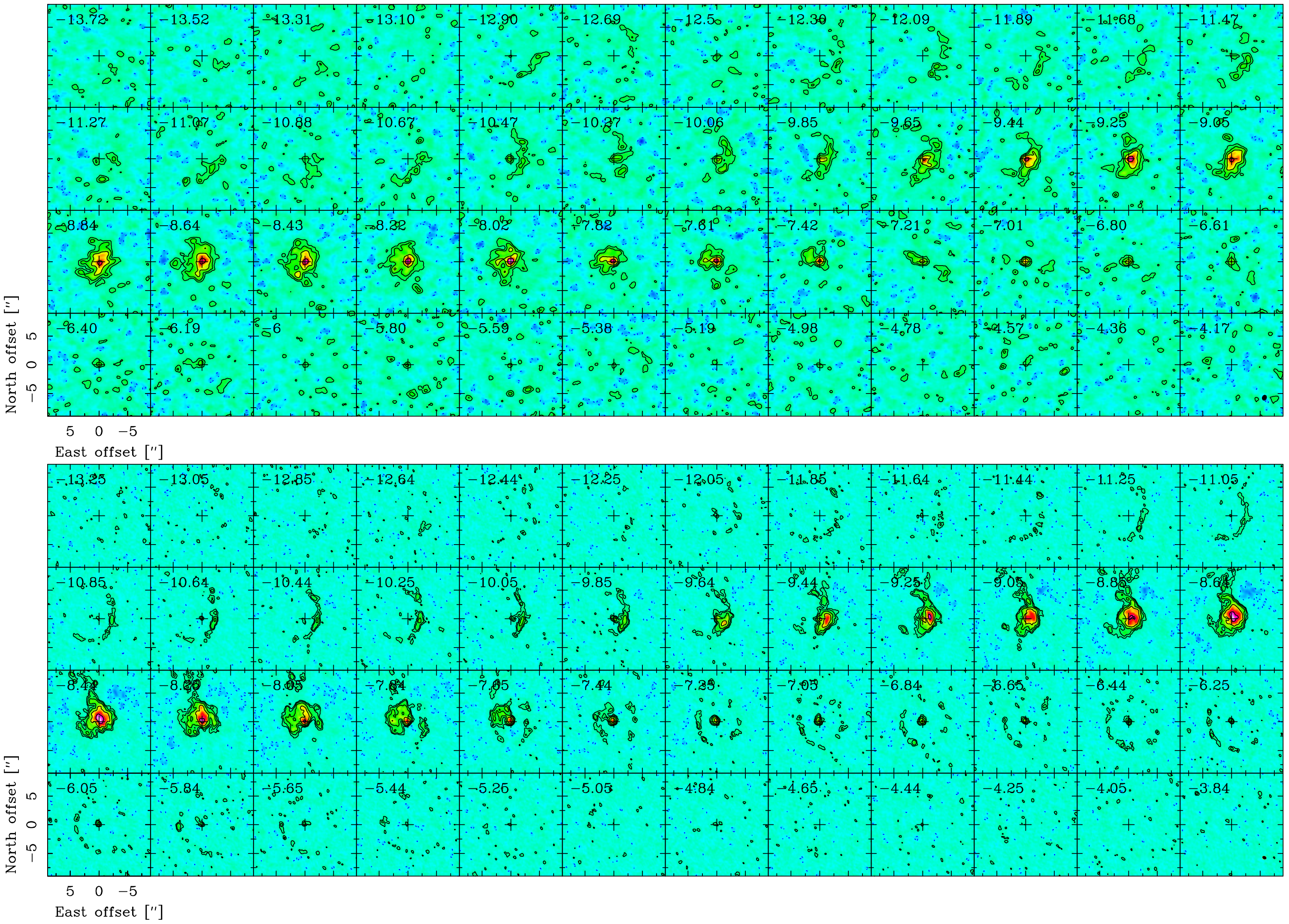

Figure 4.3: *Top*: NOEMA maps per velocity channel maps of $^{12}$CO $J = 2 - 1$ from 89 Her. The beam size (HPBW) is $1\farcs02 \times 0\farcs83$, the major axis being oriented at $PA = 115°$. The contour spacing is logarithmic: $\pm\,70$, 140, 280, and 560 mJy beam$^{-1}$ with a maximum emission peak of 870 mJy beam$^{-1}$. *Bottom*: Same as in *Top* but for $^{13}$CO $J = 2 - 1$. The beam size (HPBW) is $0\farcs74 \times 0\farcs56$, the major axis being oriented at $PA = 28°$. The contour spacing is logarithmic: $\pm\,11$, 22, 44, 88, and 144 mJy beam$^{-1}$ with a maximum emission peak of 225 mJy beam$^{-1}$. The LSR velocity is indicated in the upper left corner of each velocity-channel panel and the beam size is shown in the last panel in black.

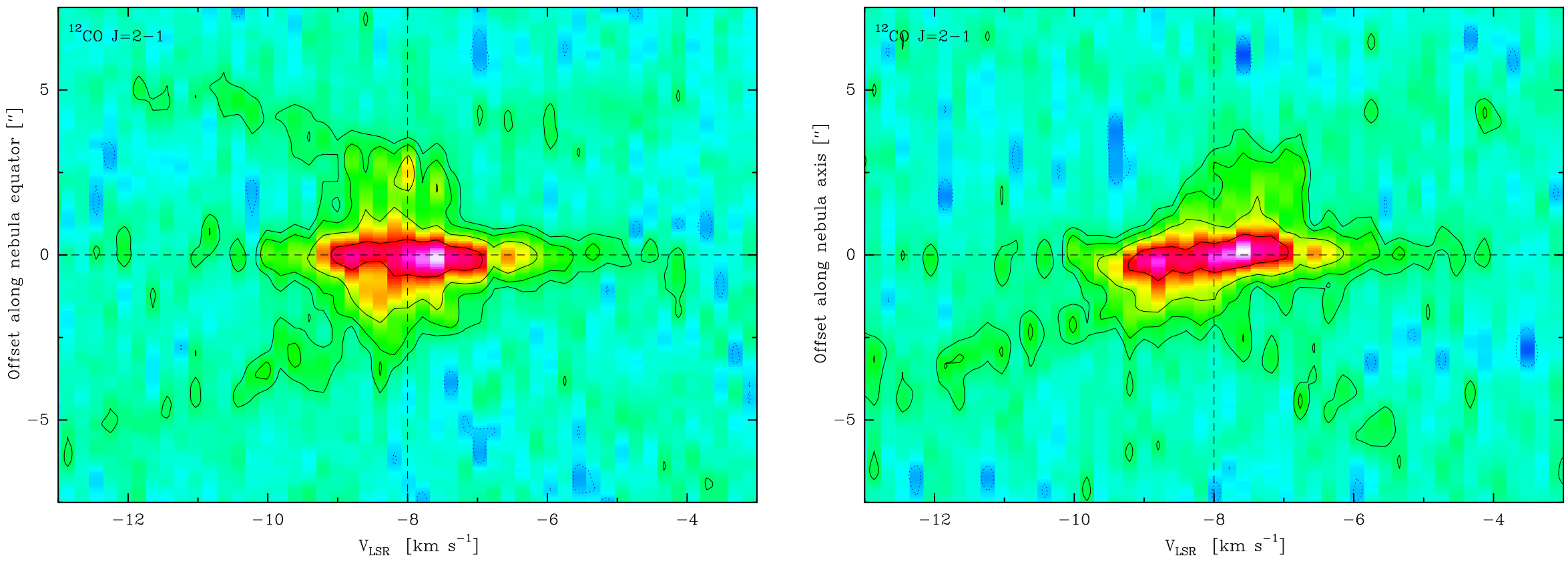

Figure 4.4: *Left:* PV diagram from our NOEMA maps of $^{12}$CO $J = 2 - 1$ in 89 Her along the direction $PA = 150°$, corresponding to the nebula equator. The contour spacing is logarithmic: $\pm$70, 140, 280, and 560 mJy beam$^{-1}$ with a maximum emission peak of 870 mJy beam$^{-1}$. The dashed lines show the approximate central position and systemic velocity. The beam is represented in the last panel. *Right*: Same as in *left* but along $PA = 60°$.

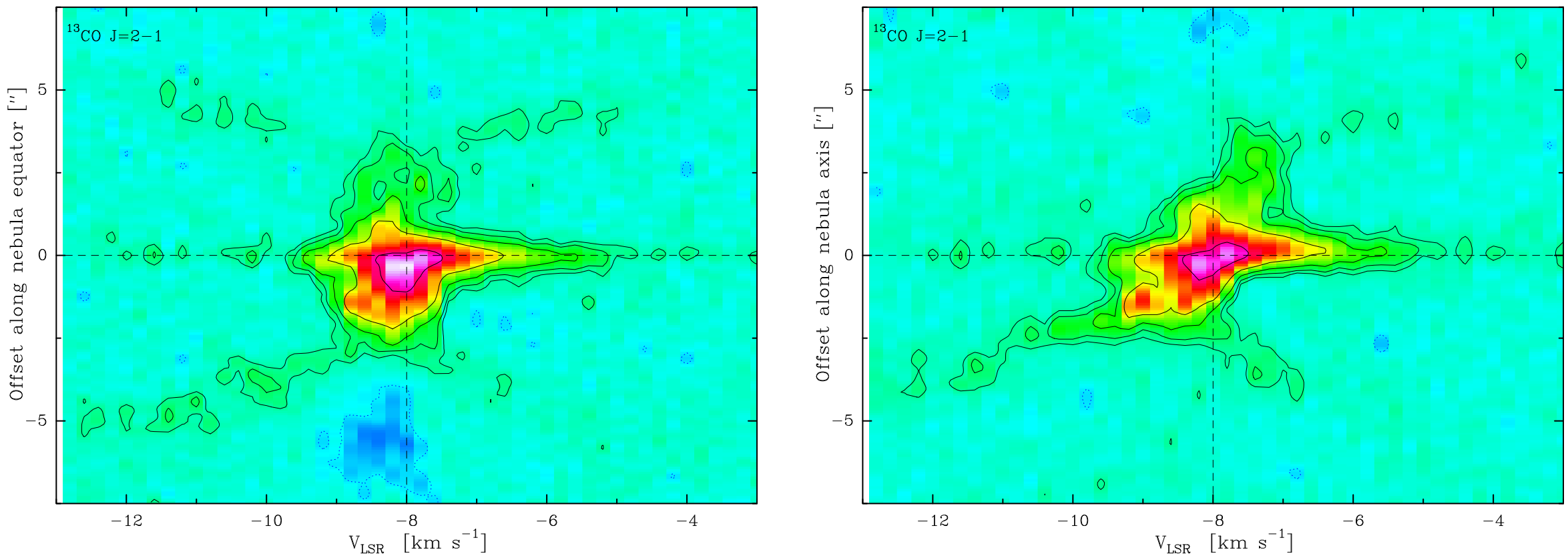

Figure 4.5: *Left:* Same as Fig. 4.4 (*left*) but for $^{13}$CO $J = 2 - 1$ emission. The contours are $\pm$11, 22, 44, 88, and 144 mJy beam$^{-1}$ with a maximum emission peak of 225 mJy beam$^{-1}$. The dashed lines show the approximate centroid in velocity and position. *Right*: Same as in *left* but along the perpendicular direction $PA = 60°$.

### 4.4.2 89 Her

We present NOEMA maps of 89 Her of $^{12}$CO and $^{13}$CO in $J = 2 - 1$ emissions. The flux loss in the interferometric observations is discussed in Sect. 4.7.1. We estimate $\sim$30% of lost flux in $^{12}$CO $J = 2 - 1$ and $\sim$50% in the wings in $^{13}$CO $J = 2 - 1$ (Figs. 4.20 and 4.21). These values exceed the relative calibration error, $\sim$30%. The presence of a certain amount of flux that is filtered out in this case is confirmed by the different profile shapes.

In these maps and PV diagrams (see Figs. 4.3, 4.4, and 4.5), we see an extended component and a central clump. The shape of the CO emission suggests the presence of an extended hourglass-like structure. This shape is also clear in PV diagrams taken along the nebula axis (see the right panels of Figs. 4.4 and 4.5). According to the angular size of the hourglass-like structure, $\sim$10$''$, and for a distance of 1000 pc, we find that the size of the nebula is at least 10 000 AU (projected in the sky), $1.5 \times 10^{17}$cm.

Inspection of the PV diagrams allows us to study the compact component of the CO emission, which seems to be a disk whose projection is elongated in the equatorial direction $PA = 150°$. The extent of the disk is not detected and must be $\leq 1''$. Position–velocity diagrams with $PA = 150°$ (see the left panels of Figs. 4 and 5) suggest that a Keplerian disk with a moderate dispersion of velocity could be responsible for that compact clump. Asymmetrical velocities are present in both maps because more emission appears at positive velocities than at negative velocities. This kind of phenomenon is often observed in our sources, probably as a result of self-absorption by cold gas in expansion located in front of the disk (e.g., Bujarrabal et al., 2018b). As argued in Sect. 4.4 and Sect. 4.7.2, we think that there must be a Keplerian disk in the inner region of the nebula of 89 Her. We find that the line profile from the central component shows the characteristic double peak of rotating disks (see Fig. 4.23). However, we stress that these two peaks are barely detected and we cannot accurately discern the emission of the central disk from the emission and absorption from the very inner and dense gas of the outflow.

### 4.4.3 IRAS 19125+0343

We present combined NOEMA and 30 m maps of IRAS 19125+0343 of $^{12}$CO $J = 2-1$ emission in Fig. 4.6 (see Sect. 4.3 for details). Maps of the 30 m reveals that the source is compact and it presents a smaller size than the beam. The main beam of NOEMA is even bigger than that of the 30 m, and so we are sure that there is no flux loss in the combined presented maps and all components of the source are detected. Due to the small size of the source (see Fig. 4.6), the brightness distribution is barely resolved in channel maps. The extent of the source is better appreciated in PV diagrams (see Fig. 4.7 *left* and *right*), but we must be aware of the synthetic beam. The linear dependence of the position with the velocity suggests a nebula elongated along $PA = 50°$, which would be the direction of the symmetry axis projection in the plane of the sky, with expansion velocity increasing with distance (a typical field in expanding nebulae around post-AGB stars and pPNe). A velocity gradient is confirmed by our model (see Sect. 4.5.3). The observed velocity gradient cannot be caused by the rotating disk, because the velocity seems to increase with the distance to the center. If the observed velocity were attributed to rotation this would imply an extremely large central mass. From the PV diagram in the right panel of Fig. 4.7 and taking into account the distance (1500 pc; see Table 4.1), the supposed central mass would be $> 50\,M_{\odot}$. This result is not acceptable. Therefore, the observed velocity gradient must have its origin in the axial expansion of the outflow. We do not have clear evidence of rotation from the inner region, which is not resolved. However, the moderate dispersion of velocity of the inner region and the characteristic single-dish CO profiles lead us to suggest that there must be a rotating component in the inner part of the nebula.

What is more interesting is the strong emission from the extended component that surrounds the inner region of the nebula (right panel of Fig. 4.7). The outflow contribution to the total emission is dominant, and its mass could be at least similar to that of the rotating disk. This is not a surprise because the wings of the single-dish profiles of IRAS 19125+0343 are very intense in $^{12}$CO $J = 2-1$.

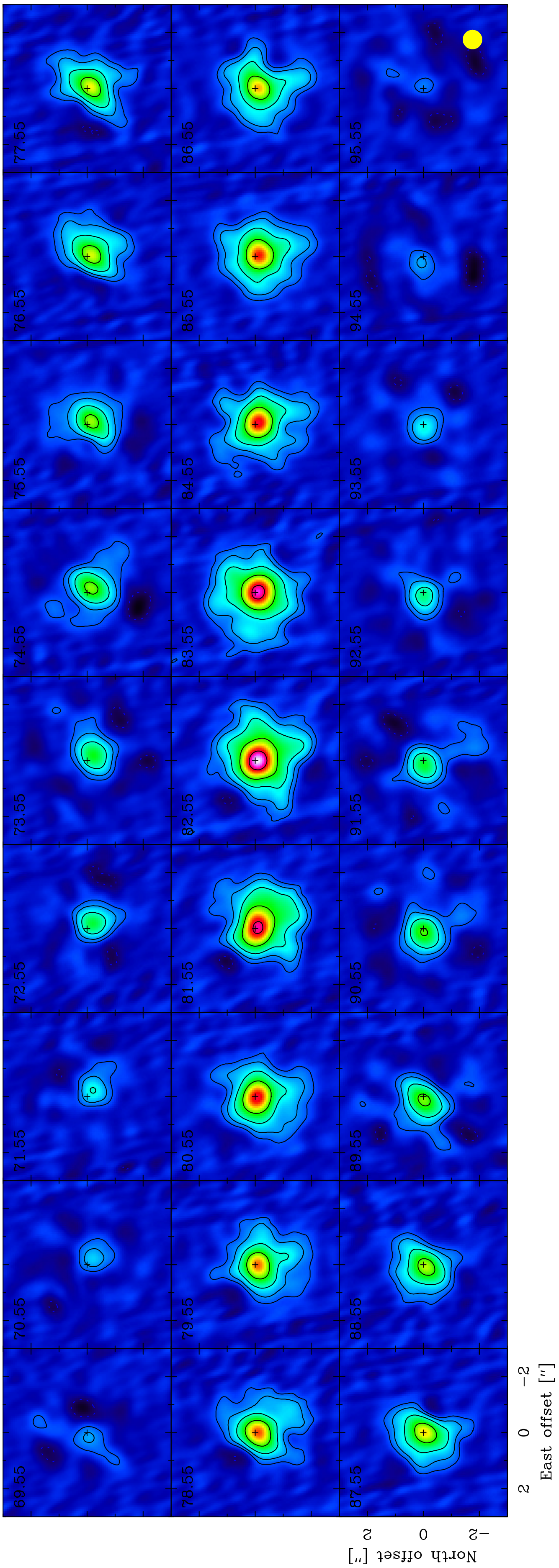

Figure 4.6: Maps per velocity channel of $^{12}$CO $J = 2 - 1$ emission from IRAS 19125+0343. The beam size (HPBW) is $0.''70 \times 0.''70$. The contour spacing is logarithmic: ± 20, 40, 80, 160, and 320 mJy beam$^{-1}$ with a maximum emission peak is 413 mJy beam$^{-1}$. The LSR velocity is indicated in the upper left corner of each velocity-channel panel and the beam size is shown in the last panel in yellow.

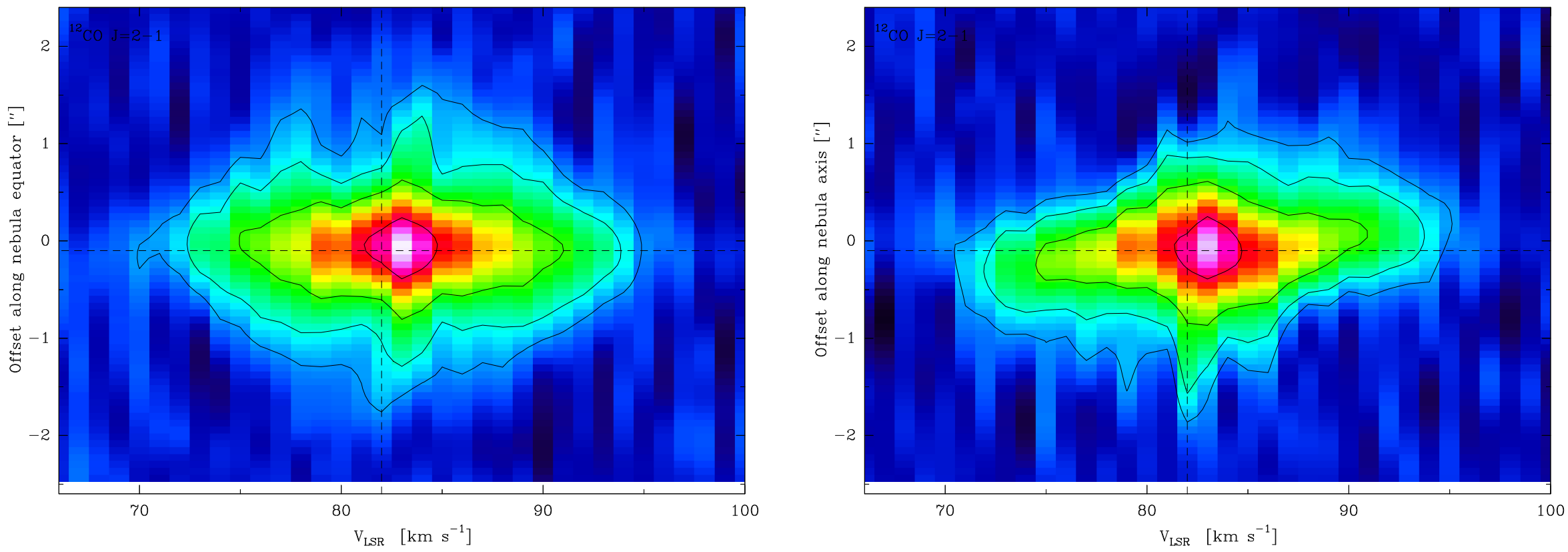

Figure 4.7: *Left:* PV diagram from our maps of $^{12}$CO $J = 2-1$ in IRAS 19125+0343 along the direction $PA = -40°$, corresponding to the nebula equator. The contour spacing is logarithmic: $\pm 40$, 80, 160, and 320 mJy beam$^{-1}$ with a maximum emission peak of 413 mJy beam$^{-1}$. The dashed lines show the approximate central position and systemic velocity. *Right*: Same as in *left* but along the perpendicular direction $PA = 50°$.

In the case of IRAS 19125+0343, and according to Sect. 4.7.2, there is no sign of the characteristic double peak in the line profile from the central compact component (see Fig. 4.23). We are convinced that the inclination of the disk with respect to the line of sight prevents us from seeing that effect. This is expected because AC Her for example does not present the double peak in its single-dish CO line profiles (Bujarrabal et al., 2013a), and this source basically consists in a Keplerian disk and an inclination similar to that of IRAS 19125+0343 (see Sects. 4.5.1, 4.5.3, and Bujarrabal et al., 2015).

### 4.4.4 R Sct

We present combined NOEMA and 30 m maps of R Sct in $^{12}$CO $J = 2-1$ emission in Fig. 4.8 (see Sect. 4.3 for details). Our NOEMA observations were merged with extensive single-dish maps (Sect. 4.3) and therefore no flux is filtered out in the shown channel maps. This is the source in our sample with the largest angular size. In Fig. 4.9, we show PV diagrams of $^{12}$CO $J = 2-1$ emission, where the presence of an extended component is clearly seen. The total extent of the R Sct nebula is $\sim 40''$, which implies a linear size for the nebula of $\sim 40\,000$ AU. We think that the massive outflow contains most of the total mass; see Sect. 4.5.4.

The PV diagram along the equator (Fig. 4.9 *left*) reveals the presence of a central clump in the inner region that probably represents the emission from the Keplerian disk, which remains unresolved in our maps. This interpretation is particularly uncertain in the case of R Sct, because its classification as binary post-AGB star is not confirmed (see Sect. 4.2.4), and the central peak in its single-dish profiles is less prominent than in other sources (Bujarrabal et al., 2013a). However, the velocity dispersion of the CO emission from its unresolved central condensation is very similar to those found in other disk-containing post-AGB nebulae (see e.g. Bujarrabal et al., 2016, 2017, and data on other sources in previous sections, including a significant lack of blueshifted emission). This kind of phenomenon is frequently observed in this type of source as a result of self-absorption of inner warmer regions by colder gas in expansion in front of them along the line of sight, as mentioned in Sect. 4.4.2.

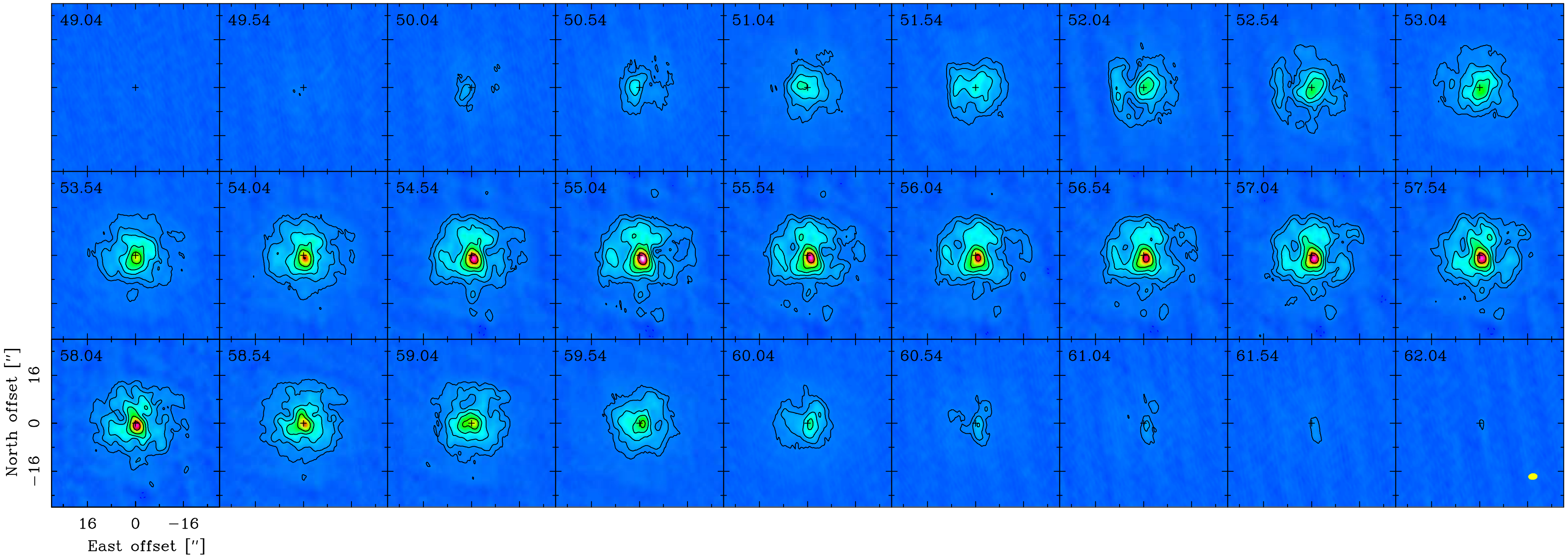

Figure 4.8: Maps per velocity channel of $^{12}$CO $J = 2-1$ emission from R Sct. The beam size (HPBW) is $3\farcs12 \times 2\farcs19$, the major axis oriented at $PA = -185°$. The contour spacing is logarithmic: $\pm 50$, 100, 200, 400, 800 and 1600 mJy beam$^{-1}$ with a maximum emission peak of 2.4 Jy beam$^{-1}$. The LSR velocity is indicated in the upper left corner of each velocity-channel panel and the beam size is shown in the last panel in yellow.

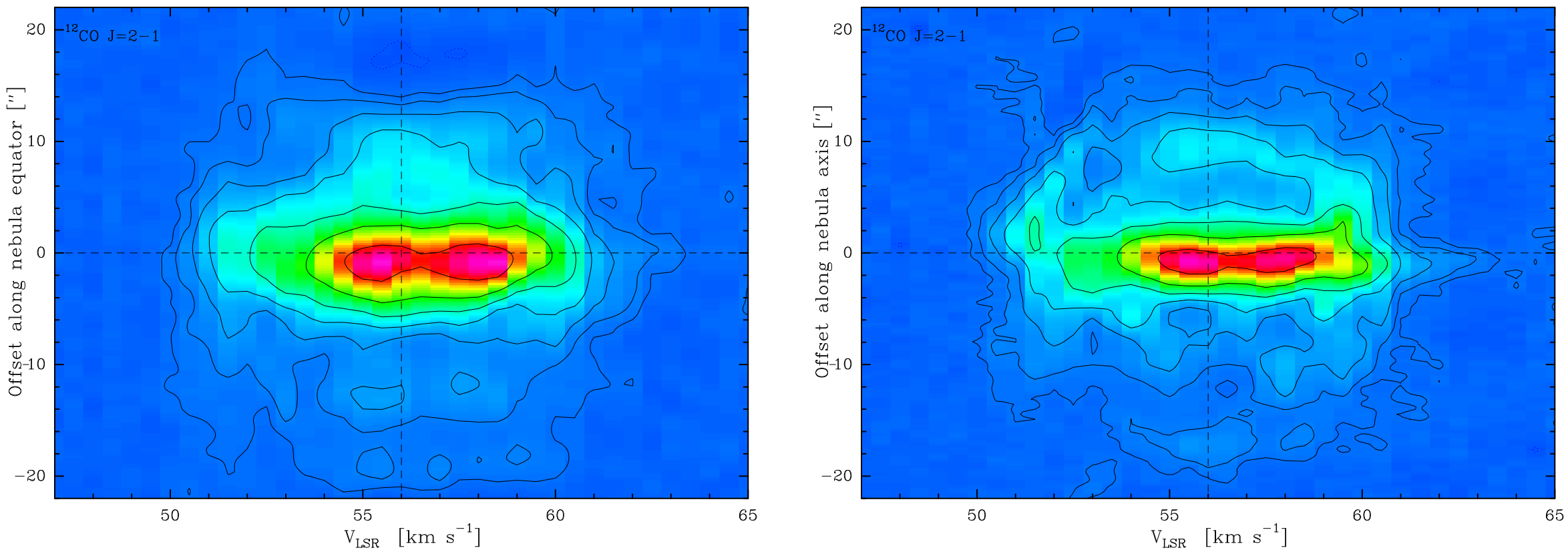

Figure 4.9: *Left:* PV diagram from our maps of $^{12}$CO $J = 2-1$ in R Sct along the direction $PA = 0°$, corresponding to the nebula equator. The contour spacing is logarithmic: ± 25, 50, 100, 200, 400, 800, and 1600 mJy beam$^{-1}$ with a maximum emission peak of 2.4 Jy beam$^{-1}$. The dashed lines show the approximate central position and systemic velocity. *Right*: Same as in *left* but along the perpendicular direction $PA = 90°$.

We also see (Sect. 4.5.4) that models of CO emission from a small disk with very reasonable properties satisfactorily explain the observed emission from the center. Additionally, we find that the line profile from the central compact component (see Fig. 4.23) shows the characteristic double peak of rotating disks (Sect. 4.4 and Sect. 4.7.2), and therefore it probably arises from the unresolved Keplerian disk. Nevertheless, we must keep in mind that we cannot discern the emission of the central and rotating component from the emission and absorption from the most inner and dense gas of the outflow. We conclude that the presence of a disk cannot be demonstrated in a conclusive way, but the central condensation found in the innermost region of R Sct is probably an unresolved rotating disk.

We must pay attention to the PV diagram along the nebula axis (Fig. 4.9 *right*), with $PA = 90°$, where the nebular structure is more evident. We can see two big cavities centered at about ± 10″. This kind of structure is often present in pPNe, such as M 1−92 (Alcolea et al., 2007) or M 2−56 (Castro-Carrizo et al., 2002).

## 4.5 Model fitting of our NOEMA maps

For modeling the sources, we used models and codes similar to those described in our previous mm-wave interferometric works (Bujarrabal et al., 2015, 2016, 2017, 2018b, see also Sect. 4.1). Our model consists of a rotating disk surrounded by an outer outflow that can present different shapes (no sign of an expanding disk is found in our data). We consider axial and equatorial symmetry for our four nebulae. The model only includes CO-rich regions, because if there is a really extended halo poor in CO, it will not be detectable. We assume LTE populations for the rotational levels involved in the studied emission. This assumption is reasonable for low-$J$ rotational levels of CO transitions, at least for gas densities over $n \geq 10^4$ cm$^{-3}$, because Einstein coefficients are much smaller than the typical collisional rates (Bujarrabal and Alcolea, 2013; Bujarrabal et al., 2016). Therefore, the characteristic rotational temperature used here is in most relevant cases equal to the kinetic temperature (see Sect. 4.7.4 for more details). The use of LTE simplifies the calculations significantly and provides an easier interpretation of the fitting parameters of the model.

The inputs for the model are the nebular shape and distributions of the density, temperature, macroscopic velocity, and local turbulence. We assume constant $^{12}$CO and $^{13}$CO abundances. With these elements, the code calculates the emission and absorption coefficients of the observed lines ($^{12}$CO $J=2-1$ and $^{13}$CO in $J=2-1$). These are computed for a large number of projected velocities and for a large number of elemental cells that fill the nebula. Typically, around $10^6$ cells are used in our models. We solve the radiative transfer equation in the direction of the telescope. We then obtain a brightness distribution as a function of the coordinates and of the projected velocity. For this, we take into account the assumed orientation of the nebula axis with respect to the line of sight. Finally, the derived brightness distribution is numerically convolved with the interferometric clean beam.

Here, we present our best-fit model for the four analyzed sources. We stress that the nebula models are complex and have a very large number of parameters. We consider simple laws for the density ($n$) and characteristic rotational temperature ($T$):

$$n = n_0 \left(\frac{r_0}{r}\right)^{\kappa_{\rm n}}, \tag{4.1}$$

$$T = T_0 \left(\frac{r_0}{r}\right)^{\kappa_{\rm T}}, \tag{4.2}$$

where $r_0$ takes the value:

$$r_0 = \begin{cases} \frac{R_{\rm K}}{2}, & \text{if } h \leq h_{\rm K} \text{ and } r \leq R_{\rm K} \\ \frac{h_{\rm outflow}}{2}, & \text{if } h > h_{\rm K} \text{ and } r > R_{\rm K} \end{cases}, \tag{4.3}$$

where $r$ represents the distance to the center, $h$ is the distance to the equator, and $n_0$ and $T_0$ are the values of the density and temperature at $r=r_0$. Here, $\kappa_{\rm n}$ and $\kappa_{\rm T}$ are the values of the slopes in the potential law for density and temperature, respectively, and $R_{\rm K}$ and $h_{\rm K}$ represent the radius and height, respectively, of the disk with Keplerian dynamics. We assume pure Keplerian rotation ($V_{\rm rot_K}$) in the disk (Eq. 4.4) and radial expansion velocity ($V_{\rm exp}$) in the outflow (Eq. 4.5):

$$V_{\rm rot_K} = V_{\rm rot_{K_0}} \sqrt{\frac{10^{16}}{r}}, \tag{4.4}$$

$$V_{\rm exp} = V_{\rm exp_0} \frac{r}{10^{16}}, \tag{4.5}$$

where $V_{\rm rot_{K_0}}$ is the tangential velocity of the Keplerian disk at $10^{16}$ cm. The parameter $V_{\rm exp_0}$ represents the expansion velocity of the ouflow at a distance of $10^{16}$ cm.

We assumed values of the relative abundances compatible with previous estimates. In particular, we adopted X($^{13}$CO) $\sim 2\times10^{-5}$ to ease the comparison of our mass estimates with those by Bujarrabal et al. (2013a) and a low X($^{12}$CO) / X($^{13}$CO) $\sim$ 10, as usually found in those low-mass post-AGB nebulae (Bujarrabal et al., 1990, 2013a).

In Sects. 4.5.1, 4.5.2, 4.5.3, and 4.5.4 we present the results obtained from the model fitting of the observations for the four sources studied in this paper.

### 4.5.1 AC Her

The model for the structure of AC Her (see Fig. 4.11 and Table 4.3) is similar to the one we developed for the Red Rectangle and for IRAS 08544−4431 (see Bujarrabal

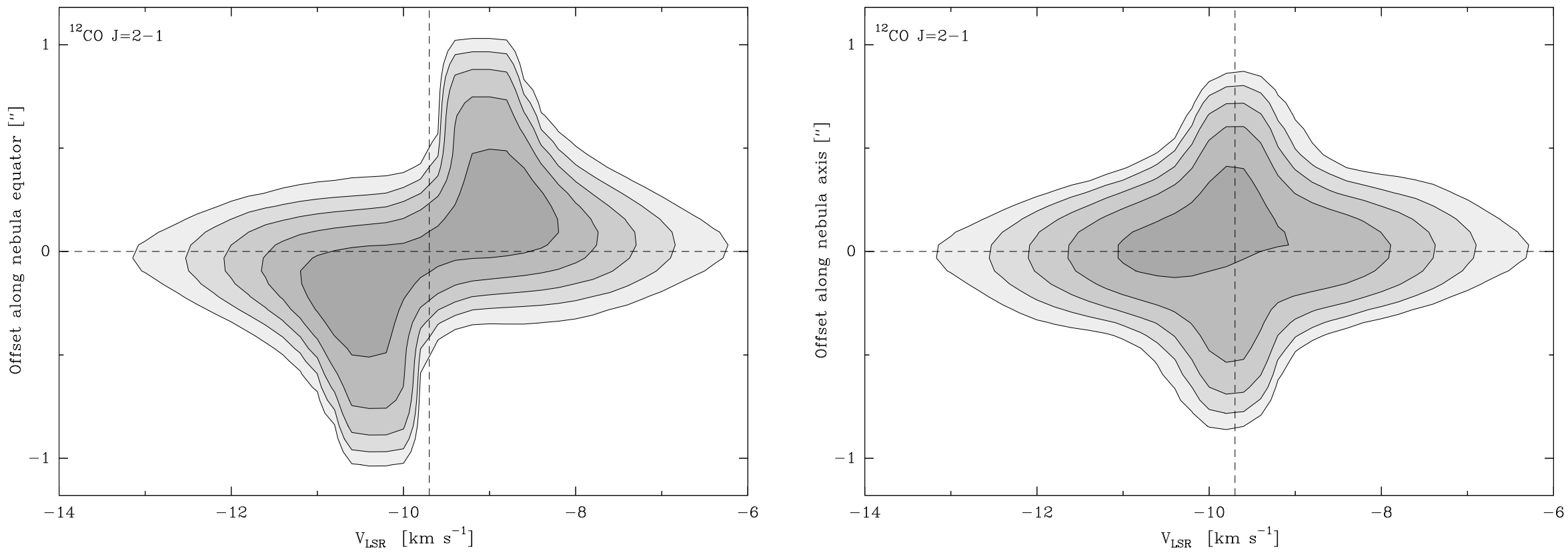

Figure 4.10: *Left:* Synthetic PV diagram from our best-fit model of $^{12}$CO $J = 2-1$ in AC Her with an outflow. For ease of comparison with Fig. 4.2 the scales and contours are the same. *Right*: Same as in *left* but along $PA = 46.1°$.

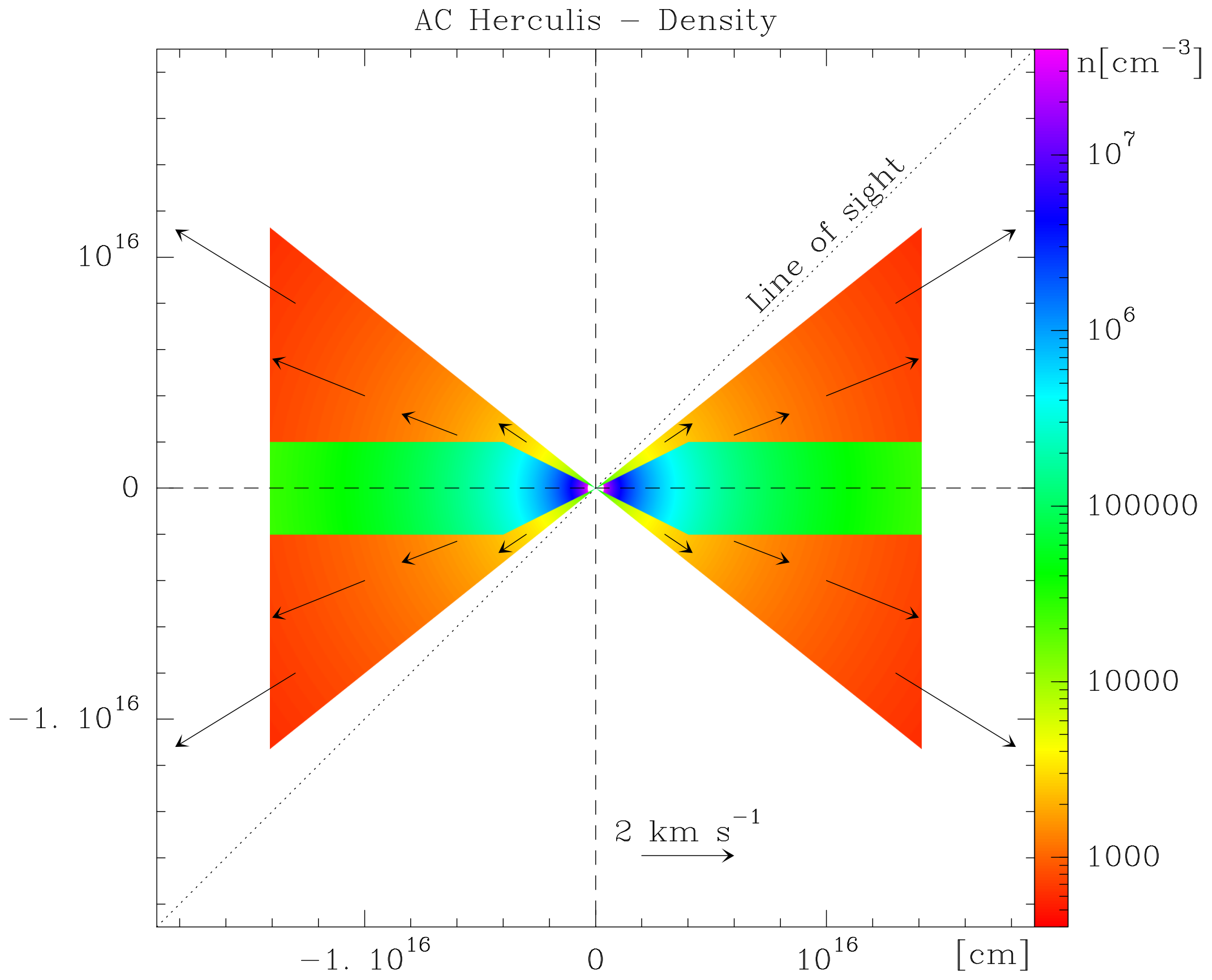

Figure 4.11: Structure and distribution of the density of our best-fit model for the disk and outflow of AC Her. The Keplerian disk presents density values $\geq 10^5$ cm$^{-3}$. The expansion velocity is represented with arrows.

et al., 2013b, 2016, 2018b). Our maps, and mainly the PV diagram at $PA = 136.1°$ (Figs. 4.1 and 4.2), and following a discussion similar to that in Bujarrabal et al. (2015), confirm that the detected CO emission from AC Her comes from an orbiting disk with Keplerian rotation. The presence of an outflow is doubtful, because its emission is very weak (see Sect. 4.4.1 and Sect. 4.7.3). However, the slight asymmetry of the emission in the most extreme angular offsets that we see in the right panel of Fig. 4.2 leads us

Table 4.3 Physical conditions in the molecular disk and outflow of AC Her derived from the model fitting of the CO data.

| Parameter | Disk | Outflow |
|---|---|---|
| Radius [cm] | $1.4 \times 10^{16}$ | $R_{\rm max} = 1.4 \times 10^{16}$ |
| Height [cm] | $2.0 \times 10^{15}$ | $1.1 \times 10^{16}$ |
| Density [cm$^{-3}$] | $n_0 = 2.0 \times 10^5$ | $n_0 = 1.1 \times 10^3$ |
| | $\kappa_{\rm n} = 2.0$ | $\kappa_{\rm n} = 1.0$ |
| Temperature [K] | $T_0 = 39$ | $T_0 = 100$ |
| | $\kappa_{\rm T} = 1.0$ | $\kappa_{\rm T} = 0$ |
| Rot. Vel. [km s$^{-1}$] | 1.0 | – |
| Exp. Vel. [km s$^{-1}$] | – | 2 |
| X($^{12}$CO) | $2 \times 10^{-4}$ | $2 \times 10^{-4}$ |
| $^{12}$CO / $^{13}$CO | 10 | 10 |
| Inclination [°] | | 45 |
| Position angle [°] | | 46.1 |

**Notes.** Parameters and their values used in the best-fit model. $R_{\rm max}$ indicates the maximum radius of the outflow. The density and temperature follow the potential laws of Eq. 4.1 and Eq. 4.2. $V_{{\rm rot}_{\rm K_0}}$ and $V_{{\rm exp}_0}$ are the values of the velocity of the disk and outflow at $10^{16}$ cm in Eqs. 4.4 and 4.5. We show the inclination of the nebula symmetry axis with respect to the line of sight and the position angle of its projection on the plane of the sky.

to think that an outflow could be present.

Our best-fitting model is described in Table 4.3. We have taken the $^{12}$CO abundance to be X($^{12}$CO) $= 2 \times 10^{-4}$, the same value as in Bujarrabal et al. (2013a). We assume an inclination for the nebula axis with respect to the line of sight of 45°, in agreement with Bujarrabal et al. (2015). The density and temperature laws in the disk and outflow are assumed to vary with the distance and they follow potential laws (see Eqs. 4.1 and 4.2), according to the parameters of Table 4.3. Our model shows densities between $10^5$ and $10^7$ cm$^{-3}$ and temperatures between 20 and 200 K (according to Eqs. 4.1 and 4.2 of Sect. 4.5, and parameters of Table 4.3). We find that the observations are compatible with Keplerian rotation in the disk (Eq. 4.4) for a central total stellar mass of $\sim 1$ M$_\odot$.

A representation of our model nebula and predicted results can be seen in Figs. 4.10, 4.11, and 4.37 (see also Figs. 4.27 and 4.28). The main goal of our work is to study the outflow that appears to surround the Keplerian disk. Our modeling of the outflow is based on a linear law for the density and a constant value for the temperature of 100 K, which is a value typical of other outflows around Keplerian disks (Bujarrabal et al., 2016, 2017, 2018b). We find that an increase in the outflow density by 50% yields results that are totally incompatible with observations. With these premises, we have found an upper limit to the mass of the outflow that is consistent with the observations of $\lesssim 2.0 \times 10^{-5}$ M$_\odot$. The derived mass for the nebula of AC Her, including the extended component, is $8.3 \times 10^{-4}$ M$_\odot$, and it is therefore clearly a disk-dominated post-AGB nebula; we find that the mass of outflow represents $\lesssim 3\%$ of the total mass.

Additionally, we present predictions from a model for the nebula of AC Her, in which only emission of the disk is considered (see Figs. 4.38 and 4.39). As we see, these predictions are almost equal to the ones of our standard model with the outflow emission. This is expected because the outflow contribution is very weak as we see in the observational data (see Sect. 4.4.1 and Sect. 4.7.3 for more details).

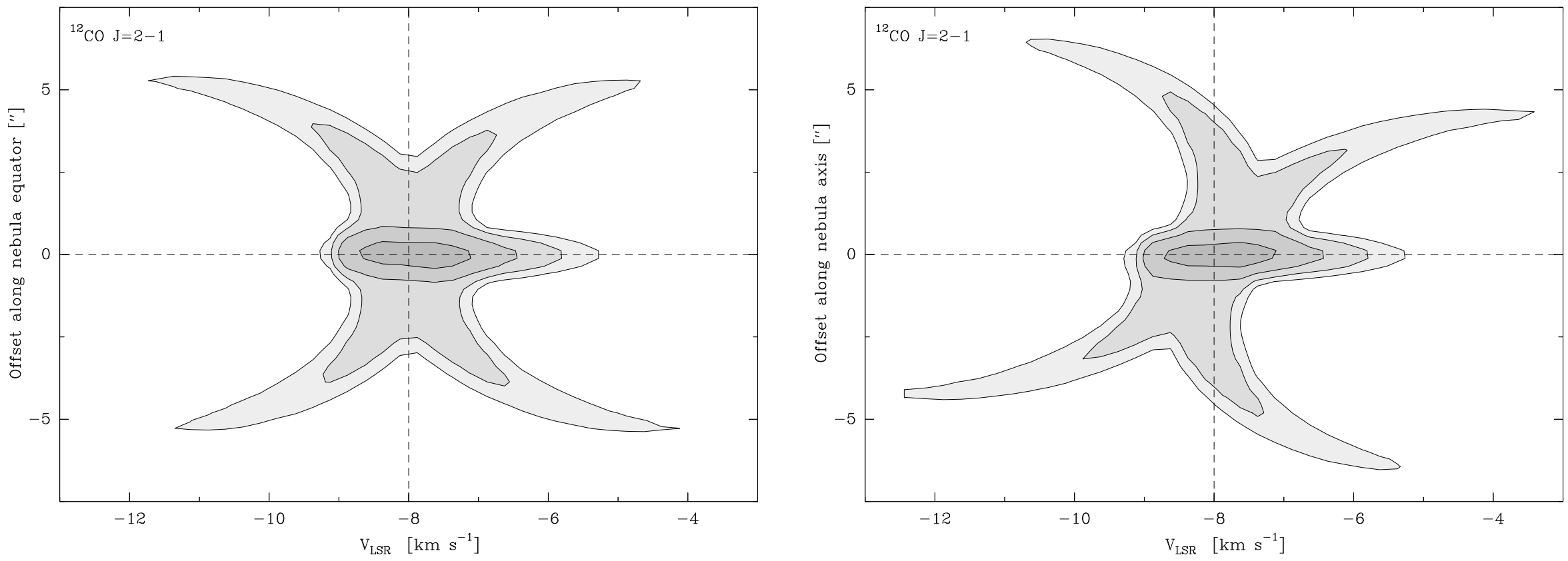

Figure 4.12: *Left:* Synthetic PV diagram from our best-fit model of $^{12}$CO $J = 2-1$ in 89 Her along the direction $PA = 150°$. To be compared with the left panel of Fig. 4.4, the scales and contours are the same. *Right*: Same as in *left* but along $PA = 60°$.

The analysis of the mass of the outflow in AC Her is very uncertain, mainly because of the lack of information on its main properties. In an attempt to quantify these uncertainties, we present an alternative model for AC Her (see Sect. 4.7.6), where the size of the outflow is somewhat larger than that presented immediately above (but still compatible with our general ideas on such outflows). This alternative model (see Fig. 4.47) presents the same characteristics for the disk. We see in Figs. 4.40 and 4.41 that the increase in the outflow size is still consistent with the observational data.

The total mass derived from our alternative model for the nebula is $8.4 \times 10^{-4}$ M$_{\odot}$, of which the mass of this more extended outflow is $3.2 \times 10^{-5}$ M$_{\odot}$. Therefore, the mass of this alternative and larger extended component represents just $\lesssim$4% of the total mass. The increase in the percentage of the outflow mass does not change significantly despite this significant increase in the size of the extended component. This alternative model leads us to conclude that the mass of the outflow must be $\lesssim$5% of the total mass, but we are aware of the uncertain analysis.

We tried to check the presence of the outflow calculating the derived residual emission from a comparison between observations and predictions of the disk-only model (Sect. 4.7.6). Nevertheless, uncertainties in our best-fit models are too large and the residual emission is comparable to the outflow emission we are discussing, because the putative outflow is just $\sim$5% of the maximum emission. Therefore, this method is not useful to prove the existence of the outflow. Finally, we also tried to estimate the excess emission by comparing PV diagrams obtained for different position angles (see Sect. 4.7.3). We find in this way a tentative detection of the outflow emission, which would contain a mass again $\lesssim$5% of the total mass.

### 4.5.2 89 Her

We adopted a nebula model (see Fig. 4.14 and Table 7.1) based on reasonable assumptions and predictions and compatible with the observational data and other works of 89 Her (see Sect. 4.4.2 and Bujarrabal et al., 2007). We assumed the presence of a disk in the central regions of the nebula and a large hourglass-shaped wind. See a representation of our model nebula and predicted results in Figs. 4.12, 4.13, and 4.42.

The inclination of the nebular symmetry axis with respect to the line of sight is 15°, with a $PA$ of 40° (150° for the equatorial direction). These results are directly obtained

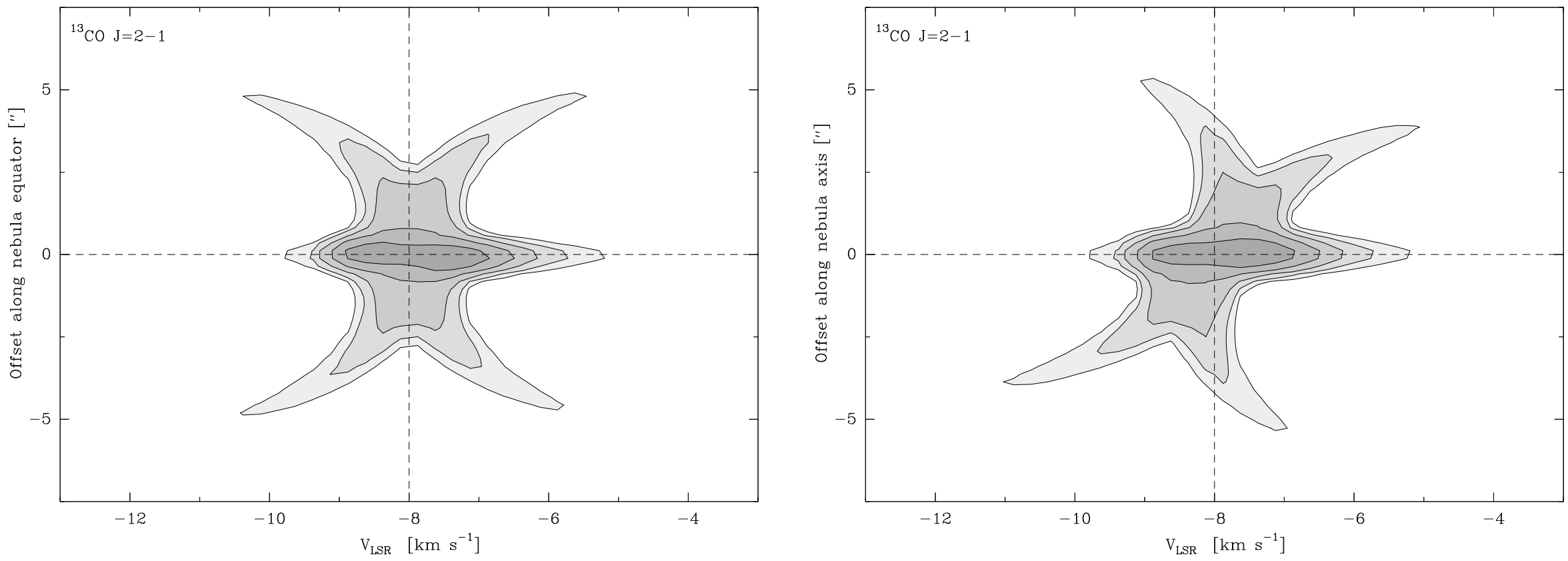

Figure 4.13: Same as in Fig. 4.12 but for $^{13}$CO $J = 2 - 1$. To be compared with Fig. 4.5, the scales and contours are the same.

Table 4.4 Physical conditions in the molecular disk and outflow of 89 Her derived from the model fitting of the CO data.

| Parameter | Disk | Outflow |
|---|---|---|
| Radius [cm] | $5.0 \times 10^{15}$ | $R_{max} = 8.6 \times 10^{16}$<br>$W_o = 0.7 \times 10^{16}$ |
| Height [cm] | $1.0 \times 10^{15}$ | $7.2 \times 10^{16}$ |
| Density [cm$^{-3}$] | $n_0 = 2.0 \times 10^7$<br>$\kappa_\mathrm{n} = 2.0$ | $n_0 = 3.0 \times 10^3$<br>$\kappa_\mathrm{n} = 1.8$ |
| Temperature [K] | $T_0 = 75$<br>$\kappa_\mathrm{T} = 2.5$ | $T_0 = 10$<br>$\kappa_\mathrm{T} = 0$ |
| Rot. Vel. [km s$^{-1}$] | 1.5 | – |
| Exp. Vel. [km s$^{-1}$] | – | 1.2 |
| X($^{12}$CO) | $2.0 \times 10^{-4}$ | $2.0 \times 10^{-4}$ |
| $^{12}$CO / $^{13}$CO | 10 | 10 |
| Inclination [°] | | 15 |
| Position angle [°] | | 60 |

**Notes.** Parameters and their values used in the best-fit model. $R_\mathrm{max}$ indicates the maximum radius of the outflow. $W_\mathrm{o}$ is the width of the outflow walls. The density and temperature follow the potential laws of Eqs. 4.1 and 4.2. $V_{\mathrm{rot}_{\mathrm{K}_0}}$ and $V_{\mathrm{exp}_0}$ are the values of the velocity of the disk and outflow at $10^{16}$ cm in Eqs. 4.4 and 4.5. We show the inclination of the nebula symmetry axis with respect to the line of sight and the position angle of its projection on the plane of the sky.

from our observational data and modelling. Position-velocity diagrams in Figs. 4.4 and 4.5 suggest strong self-absorption effects at negative velocities (Sect. 4.4.2).

The rotation of the disk is assumed to be Keplerian (Eq. 4.4) and is compatible with a central stellar mass of 1.7 $M_\odot$. The density and temperature of the outflow are assumed to vary with the distance to the center of the nebula following potential laws; see Eqs. 4.1 and 4.2. We find reliable results for density and temperature laws with high slope values. We also find expansion velocity in the outflow, according to Eq. 4.5. We can see the predictions from these considerations and the model parameters in the synthetic velocity maps and PV diagrams in Figs. 4.12, 4.13, and 4.42 (see also Figs. 4.29, 4.30, 4.31, and 4.32).

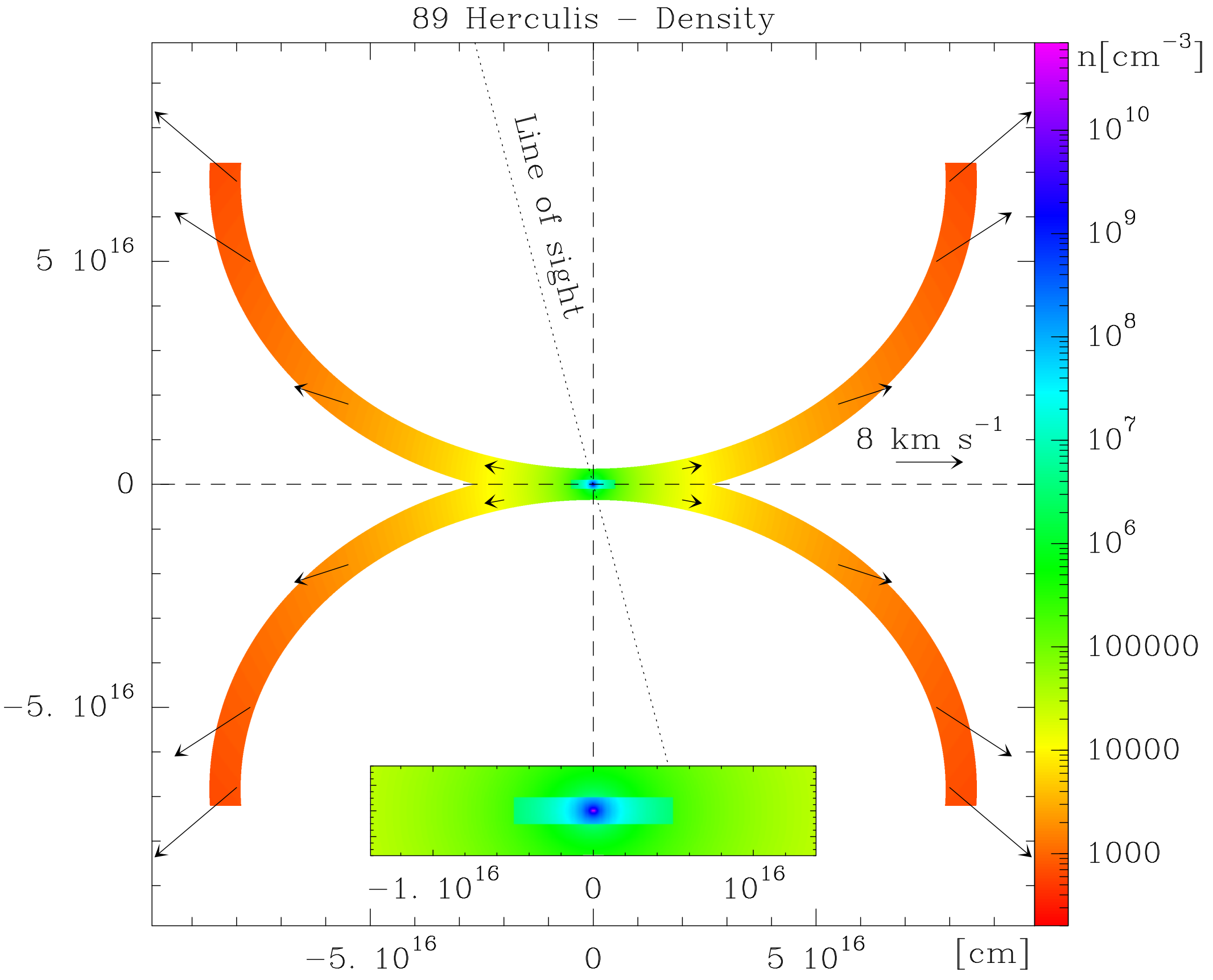

Figure 4.14: Structure and distribution of the density of our best-fit model for the disk and outflow of 89 Her. The lower inset shows a zoom into the inner region of the nebula where the Keplerian disk presents density values $\geq 10^7\,\mathrm{cm}^{-3}$. The expansion velocity is represented with arrows.

Our model reproduces the NOEMA maps and yields a total mass for the nebula of $\sim 1.1 \times 10^{-2}\,\mathrm{M}_\odot$ with a disk mass of $\sim 60\%$. This standard model includes the clear missed flux of the interferometric process with respect to the single-dish flux (see Sect. 4.7.1). This implies that the derived mass is a moderate lower limit, and because of the observed geometry could mostly apply to the extended outflow. To quantify this effect, we considered that the spatially extended emission that is filtered out by the interferometer partially fills the shells, as in the cases of outflows in young stellar sources (e.g., Gueth et al., 1996). In this alternative model, the width of the outflow walls is increased by $\sim 70\%$ and yields a good fit of the $^{13}$CO $J = 2-1$ single-dish profile in Fig. 4.22. With this, we derive a total mass of the nebula of $\sim 1.4 \times 10^{-2}\,\mathrm{M}_\odot$, of which $\sim 6.4 \times 10^{-3}\,\mathrm{M}_\odot$ corresponds to the disk mass. We take these values as the most probable ones. We do not develop this topic in more detail, because we are discussing parts of the outflow structure whose emission is not detected. In the future, we plan to perform on-the-fly observations using the 30 m IRAM telescope and new observations with NOEMA.

We conclude that $41-53\%$ of the total nebular mass is contained in the outflow. We consider the last value to be the most probable, because this result takes into account the lost flux, which very probably has its origin in the hourglass-shaped extended outflow. Therefore, it is an intermediate case between disk- and outflow-dominated post-AGB nebulae.

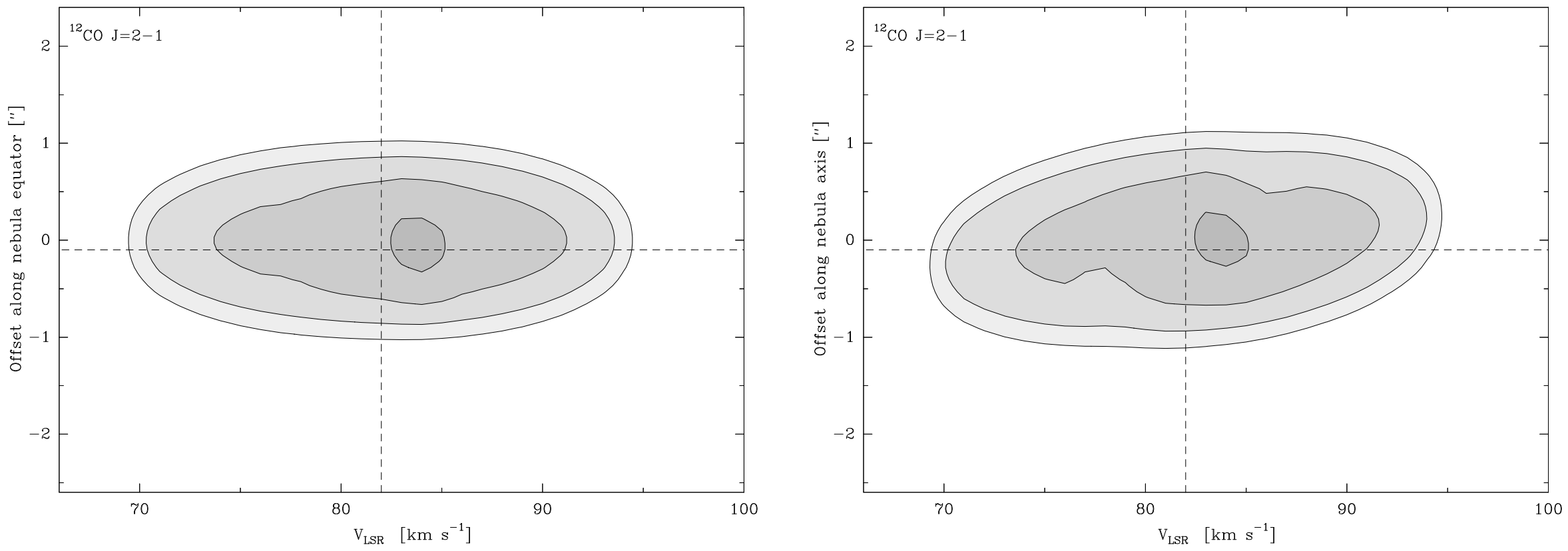

Figure 4.15: *Left:* Synthetic PV diagram from our best-fit model of $^{12}$CO $J = 2-1$ in IRAS 19125+0343 along the direction $PA = -40°$. To be compared with the left panel of Fig. 4.7, the scales and contours are the same. *Right*: Same as in *left* but along $PA = 50°$.

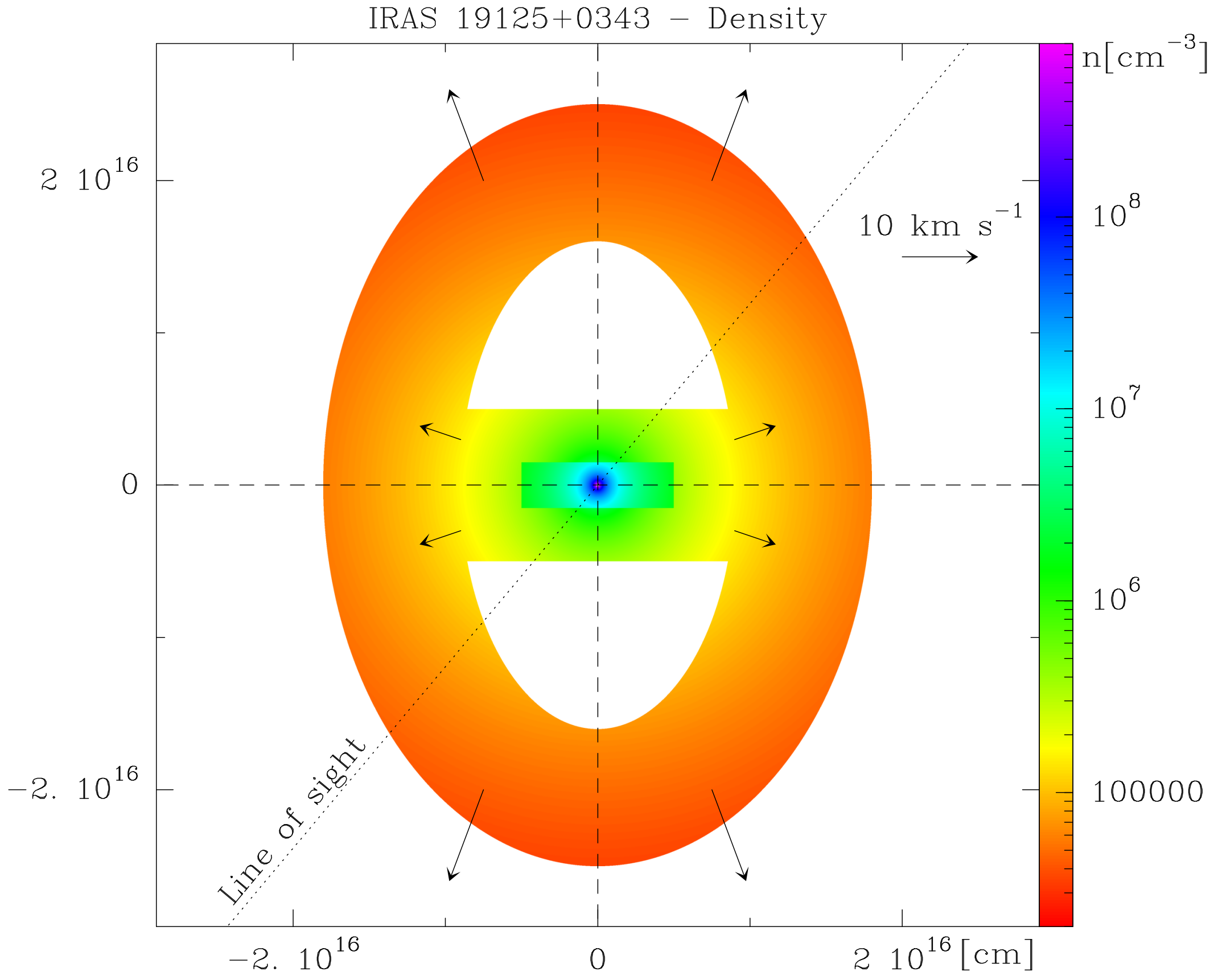

Figure 4.16: Structure and distribution of the density of our best-fit model for the disk and outflow of IRAS 19125+0343. The Keplerian disk presents density values $\geq 10^6$ cm$^{-3}$. The expansion velocity is represented with arrows.

### 4.5.3 IRAS 19125+0343

We present our best nebula model for IRAS 19125+0343 in Fig. 4.16 and Table 4.5. As the angular resolution is relatively poor ($0\farcs7 \times 0\farcs7$), we cannot resolve the disk, but we do know some features of the extended component. We propose a model consisting of a rotating disk and an outflow with cavities along the axis. These cavities are parameterized with variables $h_o$ and $W_o$, where $h_o$ describes the height of the central

Table 4.5 Physical conditions in the molecular disk and outflow of IRAS 19125+0343 derived from the model fitting of the CO data.

| Parameter | Disk | Outflow |
|---|---|---|
| Radius [cm] | $5.0\times10^{15}$ | $R_{max}=1.8\times10^{16}$<br>$W_o=9.0\times10^{15}$ |
| Height [cm] | $1.5\times10^{15}$ | $2.5\times10^{16}$<br>$h_o=5.0\times10^{15}$ |
| Density [$\mathrm{cm^{-3}}$] | $n_0=1.0\times10^{7}$<br>$\kappa_\mathrm{n}=1.5$ | $n_0=2.0\times10^{5}$<br>$\kappa_\mathrm{n}=1.5$ |
| Temperature [K] | $T_0=100$<br>$\kappa_\mathrm{T}=0.5$ | $T_0=14$<br>$\kappa_\mathrm{T}=0.4$ |
| Rot. Vel. [$\mathrm{km\,s^{-1}}$] | 1.2 | – |
| Exp. Vel. [$\mathrm{km\,s^{-1}}$] | – | 5.6 |
| X($^{12}$CO) | $2\times10^{-4}$ | $2\times10^{-4}$ |
| $^{12}$CO / $^{13}$CO | 10 | 10 |
| Inclination [°] | | 40 |
| Position angle [°] | | 50 |

**Notes.** Parameters and their values used in the best-fit model. $R_\mathrm{max}$ indicates the maximum radius of the outflow. The density and temperature follow the potential laws of Eqs. 4.1 and 4.2. $V_{\mathrm{rot_{K_0}}}$ and $V_{\mathrm{exp_0}}$ are the values of the velocity of the disk and outflow at $10^{16}$ cm in Eqs. 4.4 and 4.5. We show the inclination of the nebula symmetry axis with respect to the line of sight and the position angle of its projection on the plane of the sky.

region to the cavity and $W_\mathrm{o}$ describes the width of the outflow walls. This kind of shape is present in many pPNe (see Sect. 4.4.4). As the disk is not resolved, our main goal for this source is to study the outflow. For that purpose, the PV diagram along the nebula axis is the most relevant result.

The final nebula model yields results that are absolutely compatible with the observations, as we can see in Figs. 4.6 and 4.7 (see also Figs. 4.33 and 4.34). An inclination for the nebula axis with respect to the line of sight of 40° is compatible with the data. We analyzed PV diagrams along different $PA$, and we find that the PV diagram along $PA=50°$ is the best to show the velocity position gradient characteristic of the expansion dynamics of the post-AGB and pPNe outflows. This fact implies that the PV diagram along $PA=-40°$ should show the rotation of the equatorial disk (Figs. 4.7 and 4.15); due to the relatively low resolution and the confusion with the intense extended component, the presumed Keplerian rotation of the disk is not detectable. However, in view of the CO line profiles compatible with those of well-identified disks, we think that there must be a rotating disk surrounded by an outflow.

The density and temperature laws of the disk and outflow are assumed to vary with the distance to the center of the nebula following potential laws (see Eqs. 4.1 and 4.2). The velocity law of the disk is Keplerian (Eq. 4.4) and compatible with a central stellar mass of 1.1 $\mathrm{M_\odot}$. In the outflowing component, we assume radial velocity with a modulus linearly increasing with the distance to the center (Eq. 4.5). We find a total mass value of $1.1\times10^{-2}\,\mathrm{M_\odot}$ of which $7.9\times10^{-3}\,\mathrm{M_\odot}$ corresponds to the outflow mass. This means that IRAS 19125+0343 has a Keplerian disk surrounded by an outflow, the mass of which constitutes ∼71% of the total mass.

IRAS 19125+0343 cannot present an hourglass-shaped extended component similar to the one in 89 Her (Fig. 4.14) because predictions are incompatible with the maps. However, we cannot exclude that cavities were smaller. Accordingly, we propose an alternative model to the extended component, in order to check the validity of our best model. We present an outflow without extended cavities in the nebula axis (Fig. 4.48). We show the synthetic velocity maps and relevant PV diagrams (Fig. 4.45) from our alternative model. The predictions from this model are also compatible with the observational data. The shape of the third level of emission at $\sim 78\,\mathrm{km\,s^{-1}}$ and $\sim 85\,\mathrm{km\,s^{-1}}$ at $\sim \pm 0''.4$ in the right panel of Fig. 4.7 can only be reproduced after including cavities in the nebula model. We do not exclude models with small cavities or without them, but predictions of our alternative model are slightly worse than those from our best-fit model. This alternative model also serves to demonstrate our uncertainties.

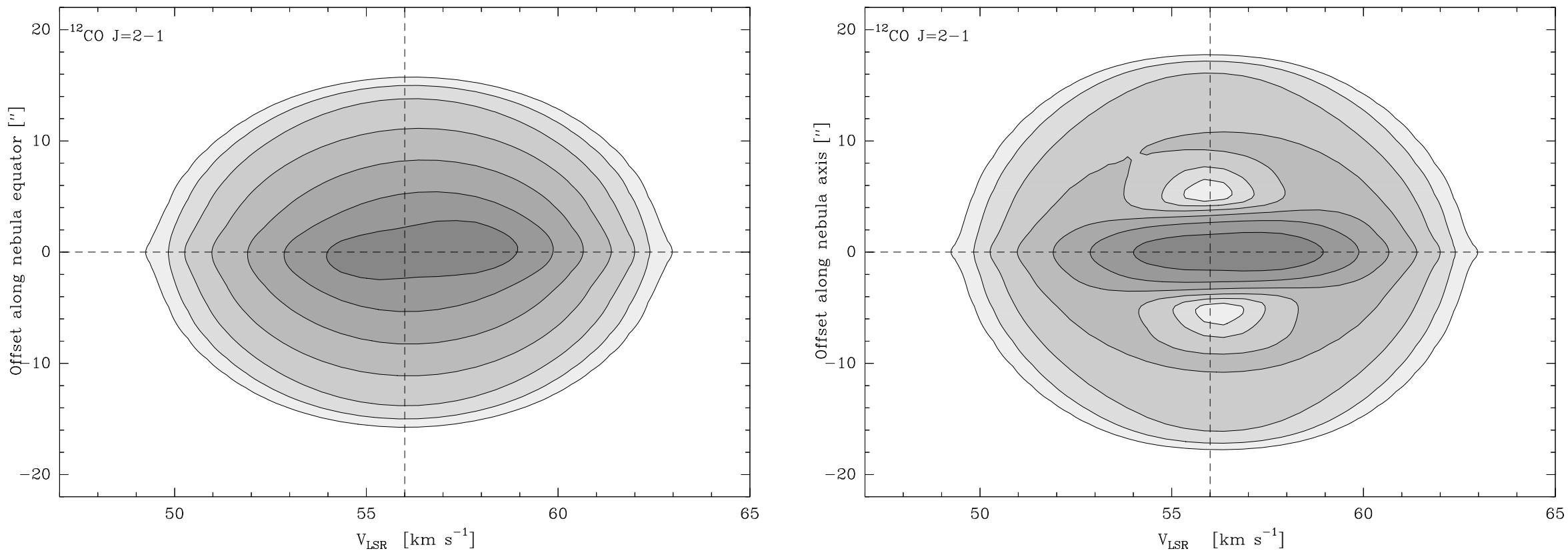

Figure 4.17: *Left:* Synthetic PV diagram from our best-fit model of $^{12}$CO $J = 2-1$ in R Sct along the direction $PA = 0°$. To be compared with the left panel of Fig. 4.9, the scales and contours are the same. *Right*: Same as in *left* but along $PA = 90°$.

### 4.5.4 R Sct

Fig. 4.18 and Table 4.6 present our best-fit model for R Sct. According to the PV diagram shown in the right panel of Fig. 4.9, the structure of R Sct is extended and contains two big cavities along the axis at about $\pm 10''$ (see Fig. 4.18). Similar structures often appear in other similar objects (Sects. 4.5 and 4.4.4).

The parameters that describe our best-fit model are shown in Table 4.6. We adopted a relative abundance with respect to the total number of particles of X($^{12}$CO) = $10^{-4}$ and X($^{13}$CO) = $2 \times 10^{-5}$. We find a low [$^{12}$C] / [$^{13}$C] abundance ratio for this source ($\sim$5), the same as that found by Bujarrabal et al. (1990).

An inclination for the nebula axis with respect to the line of sight of 85° is compatible with the data. We can see the predicted PV diagrams along the equator and along the nebula axis in Fig. 4.17. The agreement between the observations and predictions is reasonable. We note the good fitting of the self-absorption (also present in the line profiles Bujarrabal et al., 2013a, see Sect. 4.4.4), but we stress that the central disk is barely detected due to the angular resolution ($3''.12 \times 2''.19$). In view of the satisfactory model fitting and following the arguments already presented in Sect. 4.4.4 (see Sect. 4.7.2), we think that there is probably a rotating disk with Keplerian dynamics in the center of the nebula.

The outflow cavities are also present in our predictions, as we can see in the velocity maps and in the PV diagram along the nebula axis (Fig. 4.17 *right*). The outflow shape is parameterized with variables $W_o$ and $h_o$, where $W_o$ describes the width of the outflow walls, and $h_o$ describes the height of the central region to the cavity (see Table 4.6 and Fig. 4.18).

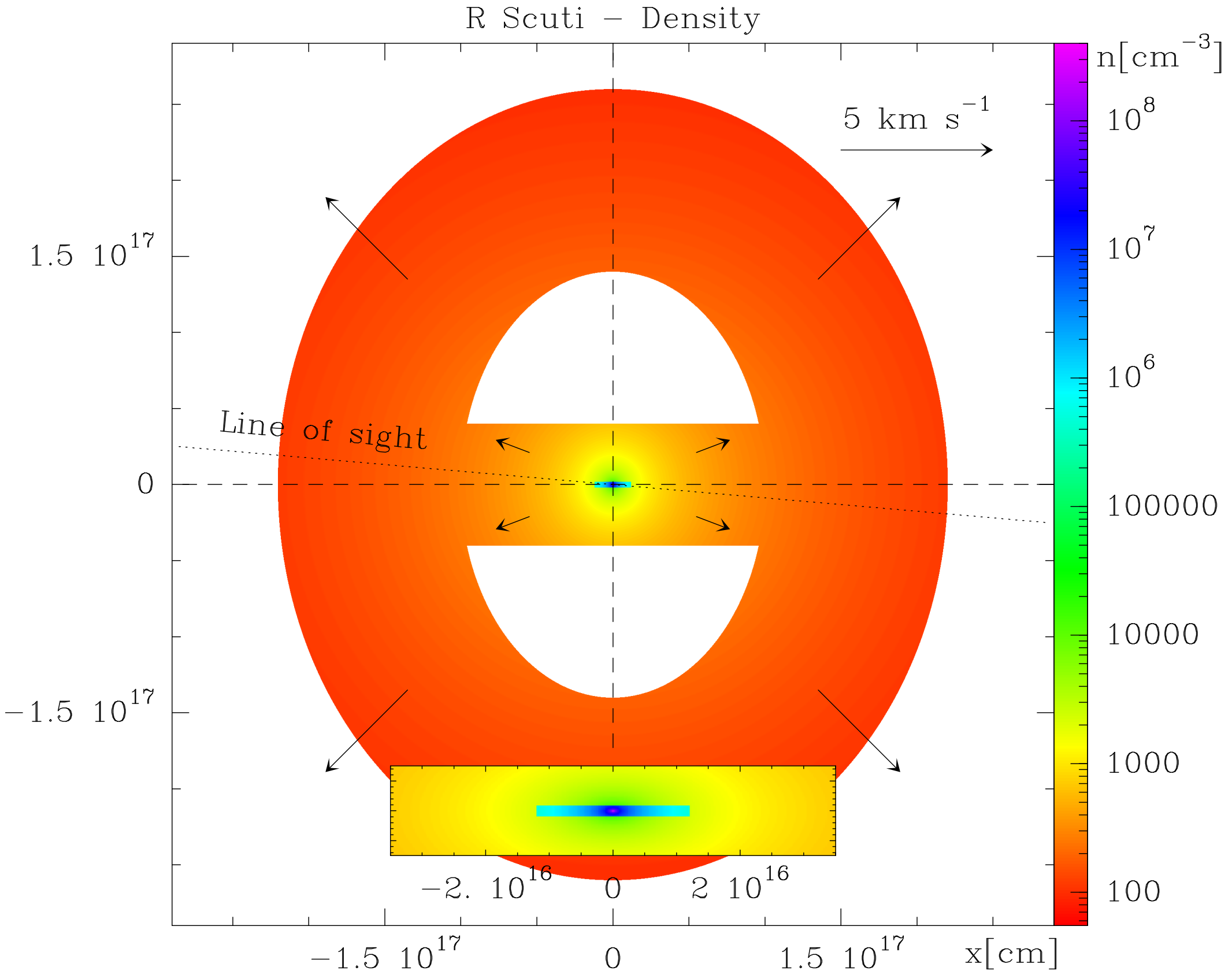

Figure 4.18: Structure and distribution of the density of our best-fit model for the disk and outflow of R Sct. The lower inset shows a zoom into the inner region of the nebula where the Keplerian disk presents density values $\geq 10^6$ cm$^{-3}$. The expansion velocity is represented with arrows.

As we see, we predict a very diffuse and cold outflow. The predicted temperatures may appear very low, but as we see in Sect. 4.7.4, the kinetic temperature ($T_K$) is expected to be higher than excitation temperature ($T_{\rm ex}$) for these low densities because of the expected underpopulations of relevant levels. In reality, we expect $T_{\rm K} \gtrsim 15$ K in the outer regions of R Sct.

The Keplerian velocity of the disk (Eq. 4.4) is compatible with a central stellar mass of 1.7 M$_\odot$. The total mass derived from our best-fit model is $\sim 3.2 \times 10^{-2}$ M$_\odot$ with a disk mass of $\sim 8.5 \times 10^{-3}$ M$_\odot$.

We considered other options for the shape of the outflow, for instance similar to the adopted model for 89 Her, which includes an hourglass-like expanding component surrounding the rotating disk (Sect. 4.5.2). However, we ruled out this option because regardless of the orientation we choose for the hourglass, the result is not consistent with the observational data. Significant changes in the size of the empty regions in the outflow lobes also lead to unacceptable predictions.

Table 4.6 Physical conditions in the molecular disk and outflow of R Sct derived from the model fitting of the CO data.

| Parameter | Disk | Outflow |
|---|---|---|
| Radius [cm] | $1.1 \times 10^{16}$ | $R_{max} = 2.2 \times 10^{17}$<br>$W_o = 1.2 \times 10^{17}$ |
| Height [cm] | $1.9 \times 10^{15}$ | $2.6 \times 10^{17}$<br>$h_o = 4.0 \times 10^{16}$ |
| Density [$cm^{-3}$] | $n_0 = 3 \times 10^6$<br>$\kappa_n = 1.5$ | $n_0 = 4 \times 10^2$<br>$\kappa_n = 1.0$ |
| Temperature [K] | $T_0 = 150$<br>$\kappa_T = 0.2$ | $T_0 = 7$<br>$\kappa_T = 1.0$ |
| Rot. Vel. [km $s^{-1}$] | 1.5 | – |
| Exp. Vel. [km $s^{-1}$] | – | 0.2 |
| X($^{12}$CO) | $1 \times 10^{-4}$ | $1 \times 10^{-4}$ |
| $^{12}$CO / $^{13}$CO | 5 | 5 |
| Inclination [°] | | 85 |
| Position angle [°] | | 90 |

**Notes.** Parameters and their values used in the best-fit model. $R_{\rm max}$ indicates the maximum radius of the ellipsoidal outflow walls. $W_0$ is the width of the outflow walls. The density and temperature follow the potential laws of Eqs. 4.1 and 4.2. $V_{{\rm rot}_{\rm K_0}}$ and $V_{{\rm exp}_0}$ are the values of the velocity of the disk and outflow at $10^{16}$ cm in Eqs. 4.4 and 4.5. We show the inclination of the nebula symmetry axis with respect to the line of sight and the position angle of its projection on the plane of the sky.

### 4.5.5 Uncertainties of the model parameters

Due to the relatively low angular resolution of the observational data, some of the nebula properties are not precisely determined. In particular, owing to the insufficient angular resolution and the inclination of the disk with respect to the plane of the sky, the height of the rotating disk is not well determined in all cases, the angular resolution being the basic estimate of the uncertainty for these parameters. On the contrary, the radius of the disk is better measured. This parameter is derived from the observational data and is confirmed with the best-fit model. The width of the disk (twice the height) is limited by the angular resolution and, often more restrictively, by the disk radius, because the disk width in real cases is significantly smaller than its equatorial size. For instance, we know that the disk width of 89 Her must be $\lesssim 3 \times 10^{15}$ cm (equivalent to $0''.2$, smaller than the angular resolution).

The structure of the outflow is well determined in the case of 89 Her, because the hourglass-shaped structure is absolutely clear in the observational data. In the case of R Sct, its extended outflow structure with cavities is clear according to the velocity maps. The maps of IRAS 19125+0343 do not allow us to determine the shape of the outflow in detail, and only the extent along the symmetry axis is actually measured. We present a simple model to explain the observations of this source, as well as an alternative model similar to those used in R Sct and many pPNe which is also compatible with the data and serves to demonstrate the uncertainties in the structure of IRAS 19125+0343. The case of the outflow of AC Her is complex. We detect weak emission of the extended component in the PV diagrams, but the shape of the outflow is not well determined. For this reason we assume properties and shape similar to those

Table 4.7 Mass and size values for the disk and outflow of our post-AGB stars sample.

| Source | Total mass [M$_\odot$] | Disk mass [M$_\odot$] | Outflow mass [M$_\odot$] | $\frac{Disk}{Total}$ [%] | $\frac{Outflow}{Total}$ [%] | Radius disk [cm] | Outflow size [cm] |
|---|---|---|---|---|---|---|---|
| AC Herculis | $8.3\times10^{-4}$ | $8.1\times10^{-4}$ | $\lesssim2.0\times10^{-5}$ | $\gtrsim95$ | $\lesssim5$ | $1.4\times10^{16}$ | – |
| 89 Herculis | $1.4\times10^{-2}$ | $6.4\times10^{-3}$ | $7.1\times10^{-3}$ | 48 | 53 | $\leq6.0\times10^{15}$ | $1.5\times10^{17}$ |
| IRAS 19125+0343 | $1.1\times10^{-2}$ | $3.3\times10^{-3}$ | $7.9\times10^{-3}$ | 29 | 71 | $\leq6.0\times10^{15}$ | $5.0\times10^{16}$ |
| R Scuti | $3.2\times10^{-2}$ | $8.5\times10^{-3}$ | $2.4\times10^{-2}$ | 27 | 73 | $\leq1.5\times10^{16}$ | $5.2\times10^{17}$ |

**Notes.** Masses calculated for the disk and outflow. These results have been calculated based on the distances of Table 4.1. Together with the masses, we show the value of the Keplerian disk radius and the outflow size in the axial direction.

of the Red Rectangle as they are similar objects. These assumptions clearly increase the uncertainty of our results and we can only give an approximate upper limit to the mass of the outflow.

Other uncertainties in the modeling also affect our main results, and particularly crucial for our discussion is the determination of the mass of the various components. From a comparison of several alternative models, we see that the assumed nebular gas distribution affects the derived mass values only slightly. The derived density distribution and mass are inversely proportional to the assumed CO abundances. The values we adopted, X($^{12}$CO) $\sim 1-2\times10^{-4}$ and X($^{13}$CO) $\sim 2\times10^{-5}$, have been extensively studied in previous works on those post-AGB sources (Bujarrabal et al., 1990, 2013a, 2016) and are accurate to within less than a factor of two. The dependence of the mass on the gas temperature is lower, less than proportional (see detailed discussion in Bujarrabal et al., 2013a). Our temperature distributions are again very similar to those usually found in these objects and are constrained by the observed brightness in optically thick areas (where the brightness is almost equal to the gas temperature after correcting for dilution within the beam). The effects of the temperature on the mass determination uncertainty are therefore a minor contribution.

The distances of these objects are very uncertain and affect other parameters of the models. The size of the nebula determined by the model scales linearly with distance $d$. The density varies with $d^{-1}$ because the column density must be conserved to yield the same optical depth in all lines of sight. A change of distance or density implies that the volume of the nebula also changes, which implies that the mass of the nebula varies. The mass scales with $d^2$. According to the model, the typical values of temperature $T$ and velocity field are not affected by a change in the distance. These dependencies on the distance are those usually encountered when modeling nebulae.

## 4.6 Conclusions

We present interferometric NOEMA maps of $^{12}$CO and $^{13}$CO $J=2-1$ in 89 Her and $^{12}$CO $J=2-1$ in AC Her, IRAS 19125+0343, and R Sct. These objects belong to the binary post-AGB stars sample with NIR excess, and are thought to be surrounded by rotating disks and outflowing disk winds that show axial and equatorial symmetry (see Sect. 4.1). We carefully modeled our observations and find that they are always compatible with this paradigm. The masses and sizes derived from our model fitting are given in Table 4.7.

- AC Her: Our maps are more sensitive than previously published ones (Bujarrabal et al., 2015). The goal of this work is to study the presence of an extended outflowing component whose presence is not obvious in the observation. The PV diagram along $PA = 136.1°$ (nebula equator) is exactly coincident with Keplerian dynamics. To study the presence of the outflow we analyzed the observational data in detail and tentatively detect its emission: $\lesssim 10\,\mathrm{mJy\,beam^{-1}\,km\,s^{-1}}$. The disk radius is $1.4 \times 10^{16}$ cm. We find densities between $10^5$ and $10^7\,\mathrm{cm^{-3}}$ and temperatures between 20 and 200 K. The rotational velocity field is purely Keplerian in the disk and is compatible with a central (total) stellar mass of $\sim 1\,\mathrm{M_\odot}$. The mass of the disk is $8.1 \times 10^{-4}\,\mathrm{M_\odot}$, the same value as in previous works: $\sim 8 \times 10^{-4}\,\mathrm{M_\odot}$ from single-dish observations (Bujarrabal et al., 2013a) with $d = 1100$ pc (same as in this work); $1.5 \times 10^{-3}\,\mathrm{M_\odot}$ from mm-wave interferometric observations (Bujarrabal et al., 2015) with $d = 1600$ pc (which implies $\sim 8 \times 10^{-4}\,\mathrm{M_\odot}$ rescaling to the distance of this work). We modeled an extended structure that surrounds the disk, assuming a structure similar to the one found in the Red Rectangle. We find that the mass of the outflow must be $\lesssim 5\%$ of the total mass. We conclude that AC Her is clearly a binary post-AGB star surrounded by a disk-dominated nebula, because $\gtrsim 95\%$ of the total mass corresponds to the disk. However, even more sensitive maps will be necessary to confirm our model of the low-mass and extended component of AC Her.

- 89 Her: We can see an extended hourglass-like structure in the velocity maps and PV diagrams. The detected outflow emission corresponds to a size of $1.5 \times 10^{17}$ cm. We can also see a central clump of strong emission with relatively low dispersion of velocity, which we think that a Keplerian disk must be responsible for. Due to the limited spatial resolution, we cannot resolve the inner structure of the disk. We find that the line profile from the most central compact component, which arises from the unresolved Keplerian disk and very inner outflow, shows the characteristic double-peak shape of rotating disks (see Sect. 4.4.2 and Sect. 4.7.2). The PV diagram along the nebula equator, $PA \sim 150°$, is compatible with characteristics of Keplerian dynamics, and we derive the main properties of the rotating disk directly from the model fitting. We find that the disk radius must be $\leq 6 \times 10^{15}$ cm and Keplerian rotation that is compatible with a central stellar mass of $1.7\,\mathrm{M_\odot}$. The nebula of 89 Her contains a total mass of $1.4 \times 10^{-2}\,\mathrm{M_\odot}$, of which the outflow is responsible for $41 - 53\%$ (see Sect. 4.5.2).

- IRAS 19125+0343: We cannot resolve the disk structure in our data. We developed a model with a rotating disk with Keplerian dynamics surrounded by an outflow $\sim 5.0 \times 10^{16}$ cm in size in the axial direction, which is completely consistent with the observations. We find that the disk radius must be $\leq 6.0 \times 10^{15}$ cm. The Keplerian rotation of IRAS 19125+0343 is compatible with a central stellar mass of $1.1\,\mathrm{M_\odot}$. However, we note that the disk structure and dynamics are particularly difficult to study from our observations and that the results on this object are less reliable; the existence of an inner rotating disk is mostly based on our experience in the analysis of single-dish profiles and the satisfactory model fitting (see Sect. 4.5.3). Observations with higher resolution are required in order to confirm these results. The total mass of the nebula of IRAS 19125+0343 is $1.1 \times 10^{-2}\,\mathrm{M_\odot}$ ($1.3 \times 10^{-2}\,\mathrm{M_\odot}$ in the alternative model), of which $3.3 \times 10^{-3}\,\mathrm{M_\odot}$ corresponds to the Keplerian disk mass, meaning that $\sim 71\%$ ($\sim 74\%$) of the total mass conforms the outflow escaping from the Keplerian disk. Therefore, IRAS 19125+0343 is an outflow-dominated post-AGB nebula.

· R Sct: Velocity maps and PV diagrams show strong emission from an inner region, which has a relatively low velocity dispersion. This central clump is not resolved in our maps and the interpretation of the central emission in this source is less reliable than for our other sources; see Sect. 4.4.4. Nevertheless, we think that this condensation is probably an unresolved rotating disk, as also argued in Sect. 4.4.4, according to the velocity dispersion and slight redshift, comparable to those observed in similar sources, and that our models easily reproduce its observational properties. In addition, we find a line profile coming from the very central region of the nebula that shows the characteristic double peak shape of rotating disks (see Sect. 4.4.4 and Sect. 4.7.2). The nature of R Sct is not yet clear (Sect. 4.2.4), but our observations strongly suggest that R Sct is also a post-AGB star surrounded by a Keplerian disk (and by a high-mass extended outflow). Therefore, we suspect a binary nature for the central star of R Sct. We present the model composed of a compact rotating disk with Keplerian dynamics and an outflow. The outflow shows two cavities that are clearly identified in the observations and are very extended, $\sim 5.2 \times 10^{17}$ cm. The main properties of the disk derived from the model are necessarily uncertain. The disk radius is $\leq 1.5 \times 10^{16}$ cm and the Keplerian rotation is compatible with a central stellar mass of 1.7 $M_{\odot}$. The total mass of the nebula is $\sim 3.2 \times 10^{-2}$ $M_{\odot}$. Taking into account that the disk represents 26.5% of the total mass and the large size of the outflow (compared with other similar objects), it is clear that R Sct is an outflow-dominated post-AGB nebula. Although the presented results are consistent with the observational data, it would be necessary to observe R Sct with higher resolution to resolve the disk and firmly conclude on the nature of the source.

Based on the presented results for 89 Her, IRAS 19125+0343, and R Sct, we conclude that there is a new subclass of binary post-AGB stars with NIR excess: the outflow-dominated nebulae. These present massive outflows, even more massive than their disks. These outflows are mostly composed of cold gas. There are other single-dish observed sources, such as AI CMi and IRAS 20056+1834, that present narrow CO line profiles with strong wings. These could also belong to this new subclass. We also observed AC Her, where the outflow is barely detected. We estimate an upper limit to its mass of $\lesssim 5\%$ of the total mass, which is even smaller than those of the disk-dominated subclass (the Red Rectangle, IW Carinae, and IRAS 08544−4431). 89 Her is an intermediate source in between disk-dominated and outflow-dominated nebula.

## 4.7 Supporting material

### 4.7.1 Flux comparison

#### NOEMA vs. 30 m

Here, we show the comparison between the interferometric flux and the single-dish integrated flux (data taken from Bujarrabal et al., 2013a). In the case of AC Her, no significant amount of flux was filtered out in the interferometric data (Fig. 4.19). The flux loss in $^{12}$CO $J = 2 - 1$ is $\sim$30% and $\sim$50% in the wings of $^{13}$CO $J = 2 - 1$ for 89 Her (Figs. 4.20 and 4.21). In the case of IRAS 19125+0343, the interferometric visibilities were merged with zero-spacing data obtained with the 30 m IRAM telescope, which guarantees that there is not lost flux in final maps. There is no flux loss for R Sct, because our NOEMA observations were merged with large single-dish maps.

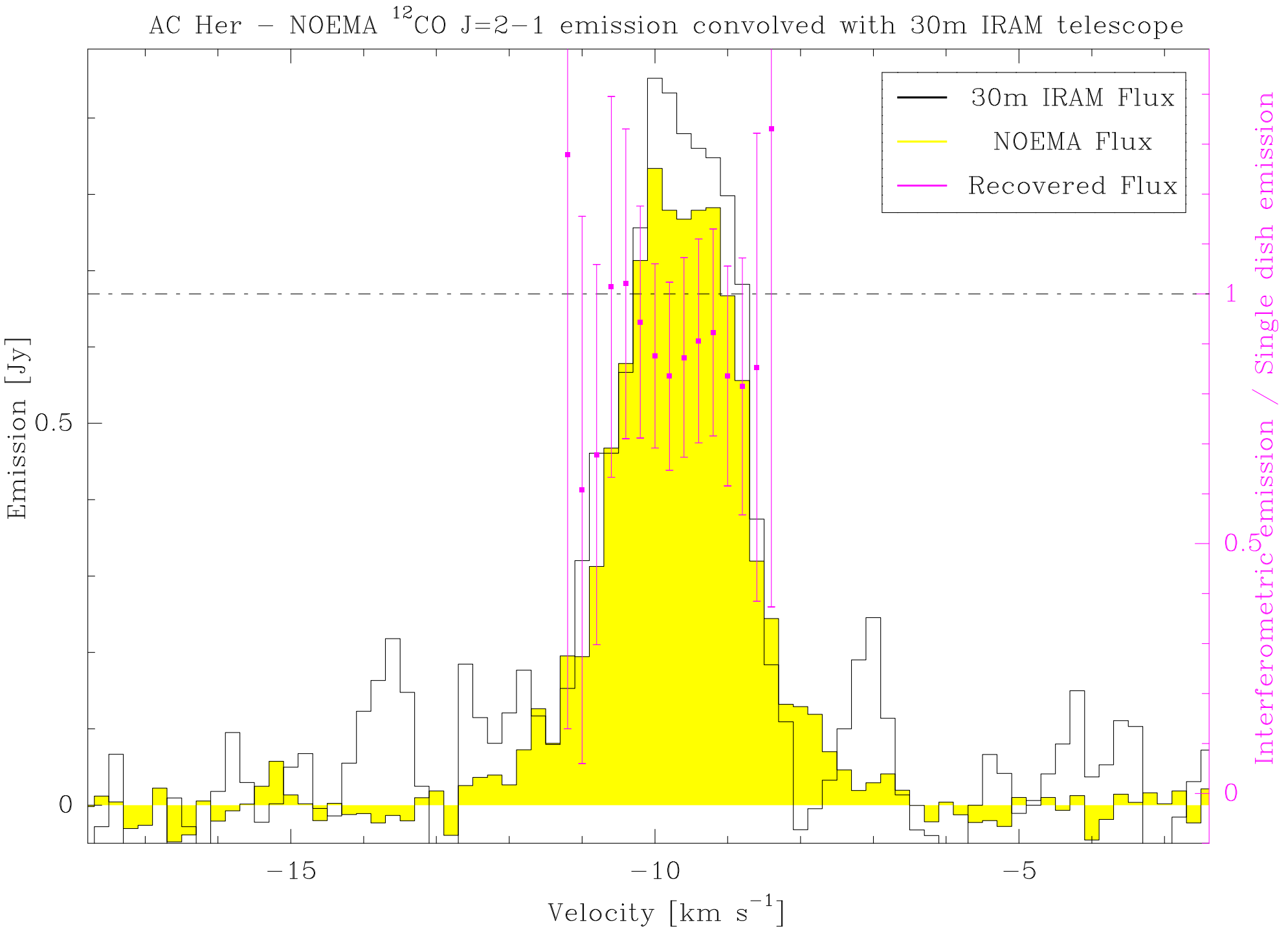

Figure 4.19: Comparison of the NOEMA flux (yellow histogram) of $^{12}$CO $J = 2 - 1$ with 30 m observation (white histogram) for AC Her. The points show the recovered flux across velocity channels (right vertical axis), with error bars including uncertainties in the calibration of both instruments.

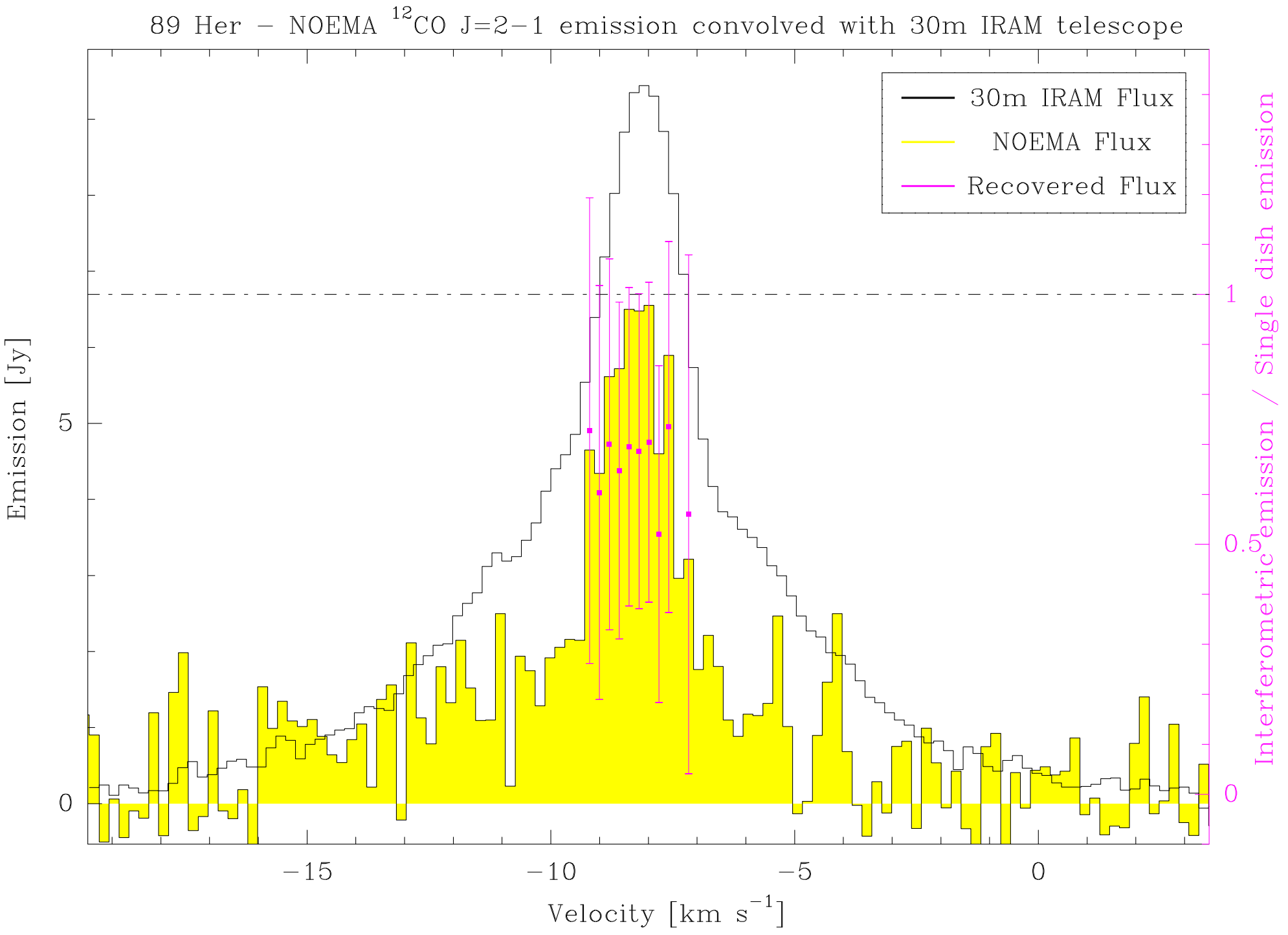

Figure 4.20: Comparison of the NOEMA flux (yellow histogram) of $^{12}$CO $J = 2 - 1$ with 30 m observation (white histogram) for 89 Her. The points show the recovered flux across velocity channels (right vertical axis), with error bars including uncertainties in the calibration of both instruments.

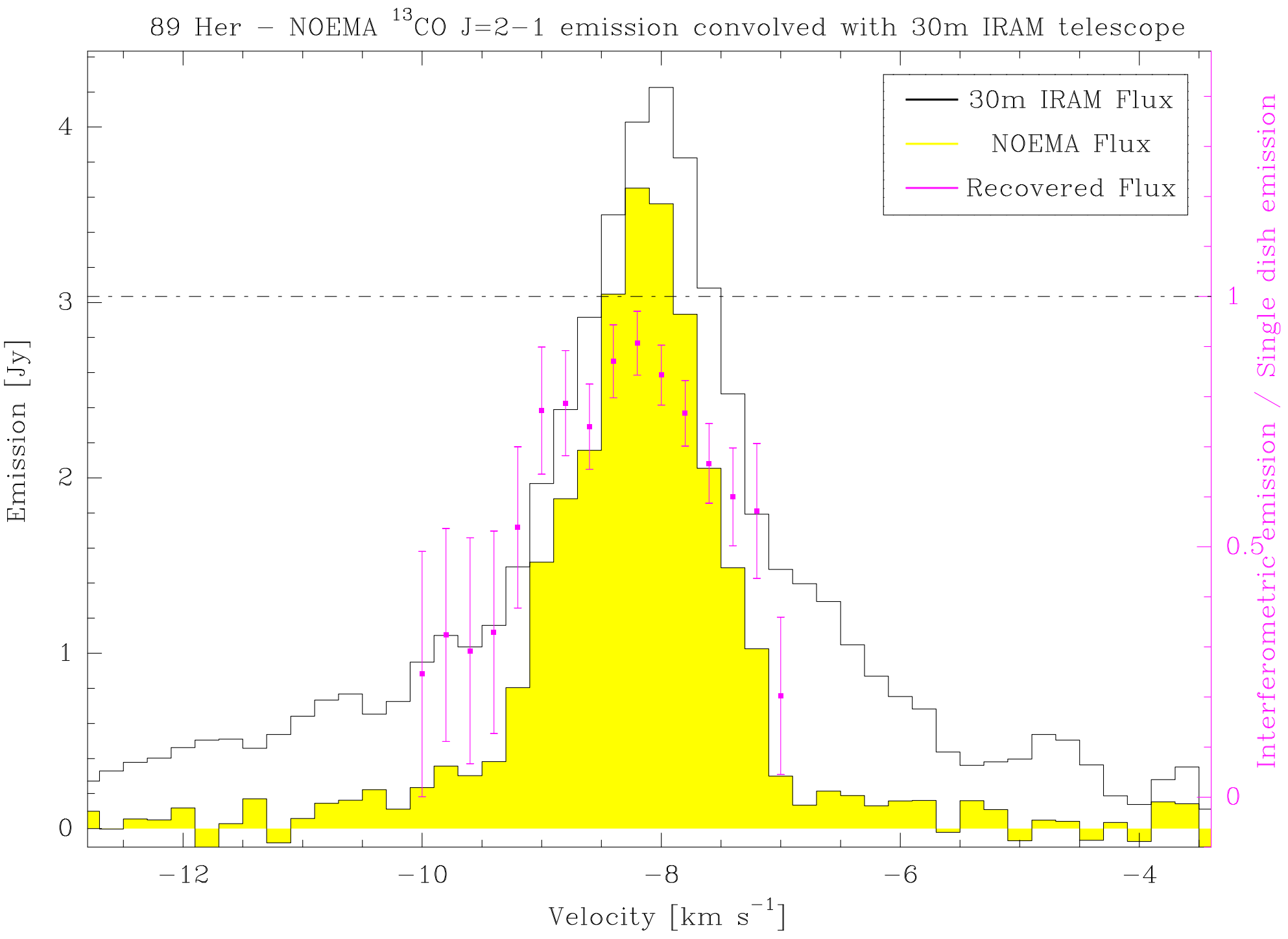

Figure 4.21: Comparison of the NOEMA flux (yellow histogram) of $^{13}CO$ $J = 2-1$ with 30 m observation (white histogram) for 89 Her. The points show the recovered flux across velocity channels (right vertical axis), with error bars including uncertainties in the calibration of both instruments.

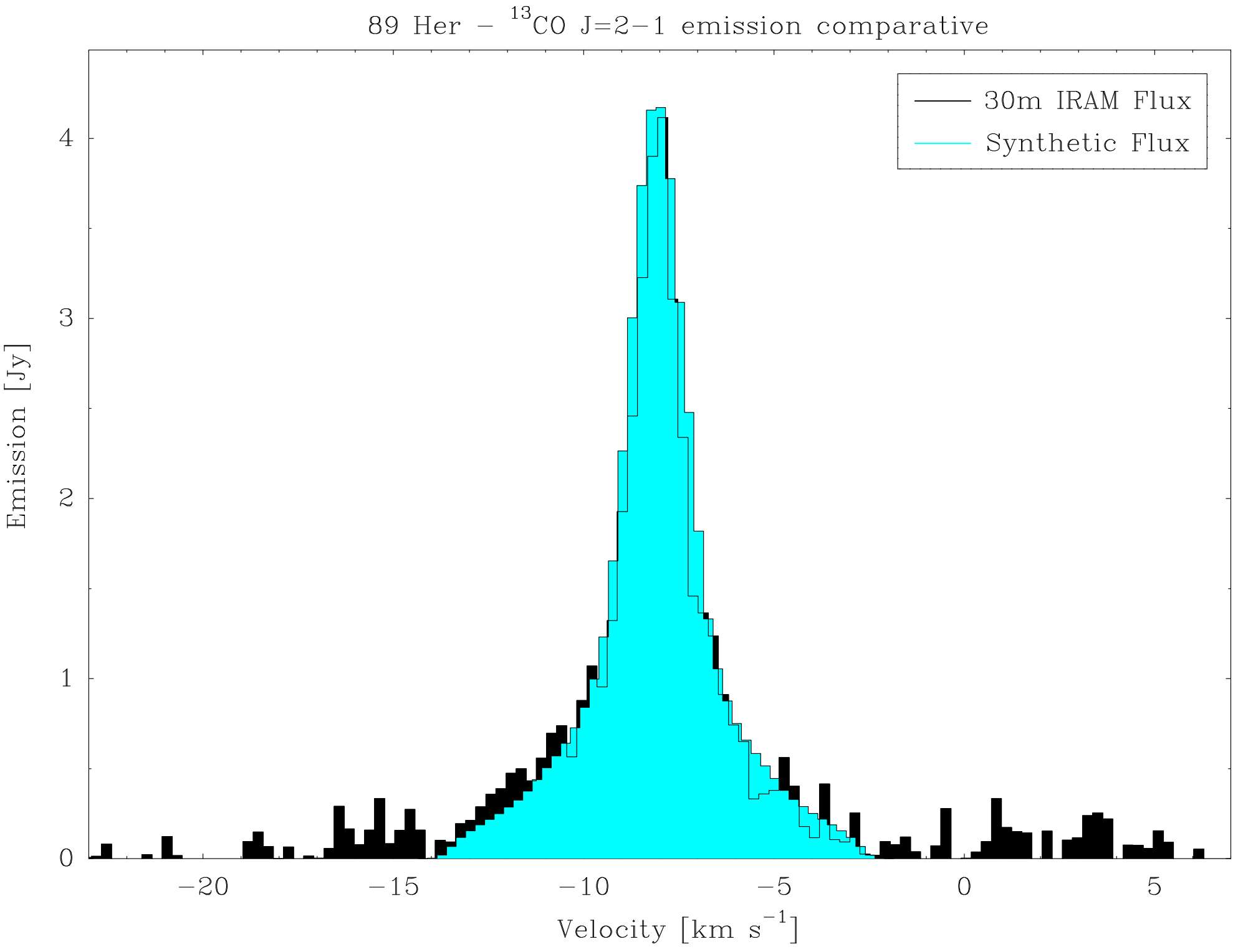

Figure 4.22: Comparison between the 30 m IRAM $^{13}CO$ $J = 2-1$ line profile (black histogram) and the synthetic one (cyan histogram). The synthetic line profile corresponds to the model after increasing the width of the hourglass of the standard model.

#### Flux loss in 89 Her: implications for the mass value

In this Sect., we present the synthetic $^{13}$CO $J = 2 - 1$ emission convolved with the 30 m IRAM telescope for 89 Her (see Fig. 4.22) compared with the observational data (obtained from Bujarrabal et al., 2013a). The flux derived of the standard model (see Table 7.1) is undervalued. We focus our analysis on $^{13}$CO $J = 2 - 1$, which is a transition that is more sensitive to density than $^{12}$CO $J = 2 - 1$. After increasing the width of the outflow walls of our standard model by $\sim$70%, we see that we can recover the lost flux. Therefore, we conclude that the mass derived from this modified model, $1.4 \times 10^{-2}$ M$_{\odot}$, is realistic (see Sect. 4.5.2 and Table 7.1).

### 4.7.2 Analyzing the inner region of the nebulae

In this section, we show the observational line profile from the central component of the nebula of 89 Her, IRAS 19125+0343, and R Sct (see Fig. 4.23). The lines of 89 Her and IRAS 19125+0343 come from maps in which longer baselines are favored. The line of R Sct come from interferometric maps without the 30 m IRAM telescope contribution. In addition, all these selected spectra come from the most inner region because we select a central region with the size of the beam. For these reasons, only a fraction of the total flux can be seen. We tentatively see the double peak line profile in the case of 89 Her and R Sct. We must clarify that it is impossible to distinguish the disk emission from emission and absorption of the innermost and dense outflow gas, which limits the reach of our discussion. In the case of IRAS 19125+0343 we cannot see the double-peak shape in the line profile. We think that the inclination of the disk with the line of sight (130°, see Sect. 4.5.3) must be a key factor to explain the absence of that double peak in this case because this profile shape tends to disappear for face-on disks. We remind the reader that AC Her, for instance, clearly presents a disk with Keplerian rotation and there is no sign of that effect in its single-dish CO line profiles (see Bujarrabal et al., 2013a).

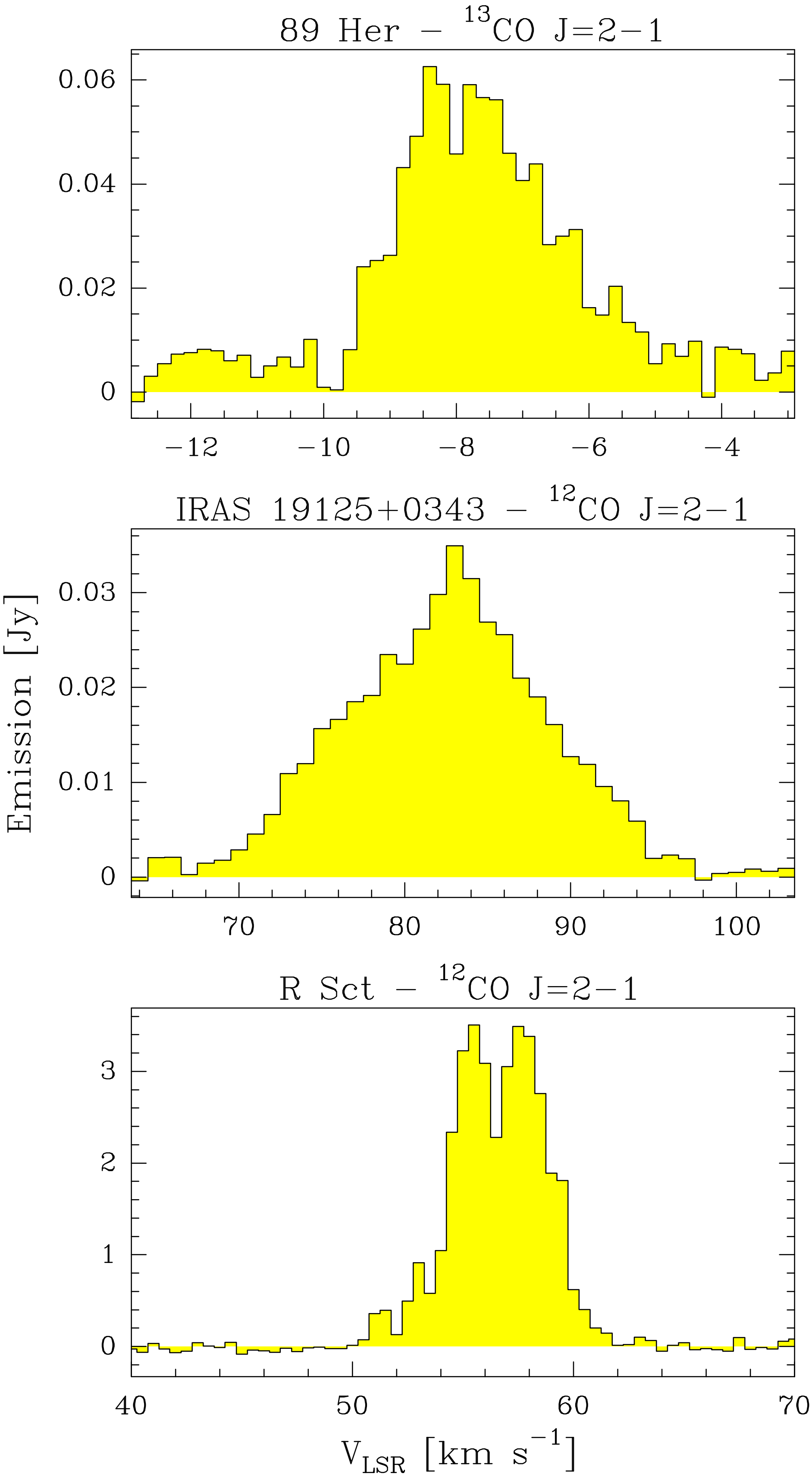

Figure 4.23: Observational line profiles of 89 Her (*Top*), IRAS 19125+0343 (*Middle*), and R Sct (*Bottom*) from the innermost region of each nebulae. See main text for details.

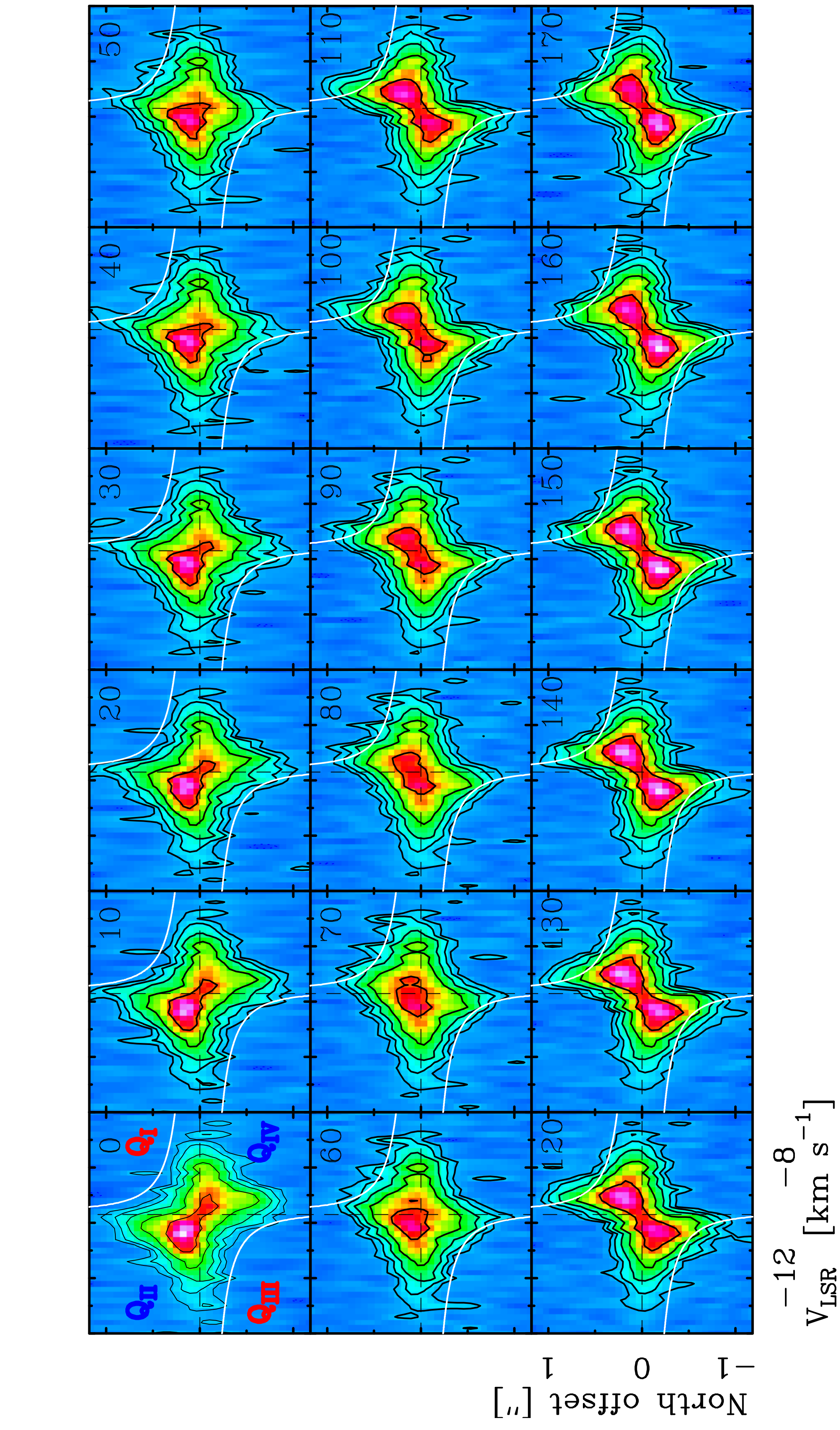

Figure 4.24: AC Her position–velocity diagrams found along different position angles from 0° to 170° with a step of 10°. The *PA* is indicated in each panel in the top right corner. The name of each quadrant is indicated in the panel showing the first velocity channel. Contours are the same as in the corresponding channel maps (Fig. 4.1). To help in the identification of the Keplerian dynamics, we show hyperbolic functions in each panel.

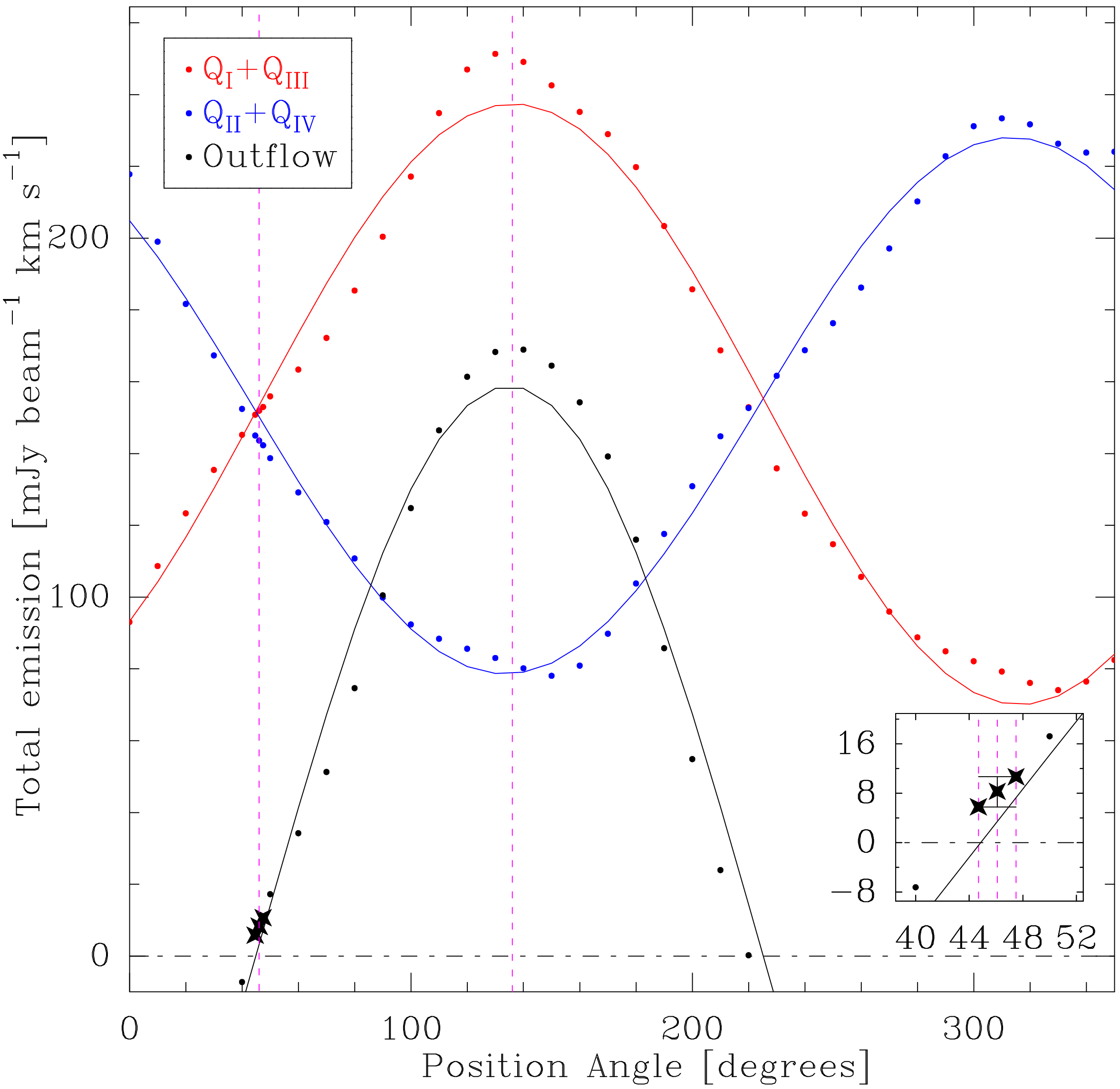

Figure 4.25: Variation of the emission of the first and third quadrants ($Q_{\rm I}$ and $Q_{\rm III}$ in red dots) and the second and fourth quadrants ($Q_{\rm II}$ and $Q_{\rm IV}$ in blue dots) with the $PA$ for our observation of AC Her (see Fig. 4.24). The maximum emission of the mean of $Q_{\rm I}$ and $Q_{\rm III}$ indicates the $PA$, for which the rotational effect of the Keplerian dynamics along the nebula equator are best detected. The angle found is: $PA = 136.1°$. The subtraction of what we name $Q_{\rm I} + Q_{\rm III}$ and $Q_{\rm II} + Q_{\rm IV}$ shows the excess emission of the putative outflow along the nebula axis (black dots). We use the same colors as in Fig. 4.24. The magenta dotted lines represents $PA = 46.1°$ and $PA = 136.1°$. These $PA$ values represent the cut along the nebula axis and equator, respectively. The emission of the putative outflow is represented as a star at $PA = 46.1° \pm 1.4°$. The error bar is the same for each point. The emission at the $PA$ along the nebula axis is shown in the inset of the right corner.

### 4.7.3 Detailed analysis of AC Her maps

This section explains how we estimate the exact position angle that best reveals the Keplerian dynamics of the rotating disk. We also calculate an upper limit to the outflow emission through the observational data. We developed a code that allows us to analyze the emission of a rotating disk and is adapted to the case of disk-dominated pPNe such as AC Her (see Sect. 4.4.1 and previous works).

Fig. 4.24 shows PV diagrams of AC Her along different $PA$. The $PA$ is indicated in the top right corner of each panel. We have drawn hyperbolic functions ($r^{-1/2}$) in each panel to help in the visual inspection. Additionally, we divided each panel in four quadrants ($Q_{\rm i}$), which are enumerated in order (counter clockwise starting from the upper right quadrant).

A disk with Keplerian dynamics must show the typical and well-known "butterfly" shape in the PV diagram along the equatorial direction. Therefore, in an ideal situation (pure Keplerian rotation and infinite spatial resolution), the emission should only appear in two opposite quadrants. The same applies for the case of a bipolar outflow with linear velocity gradient, but when we take the PV diagram along the symmetry axis. On the contrary, in an isotropically expanding envelope, the emission should be the same in all four quadrants regardless of the orientation. In our case here, a source dominated by the central rotating disk, we can use these properties to infer the orientation of the major axes of the nebula, equatorial and axial, by finding the $PA$ that maximizes the emission in two opposite quadrants, $Q_{\rm I}$ and $Q_{\rm III}$ for instance, with respect to the other two. The emission in each quadrant $Q_{\rm i}$ has been spatially averaged and added in velocity. This emission along different $PA$ is shown in Fig. 4.25 with red dots ($Q_{\rm I} + Q_{\rm III}$ emission). We derived this $PA$ by the numerical fitting of a sinusoidal function. The optimal value is found to be $136.1° \pm 1.4°$.

The presence of a putative outflow should be present in the PV diagram along the nebula axis. The $PA$ along the nebula axis will be 46.1° because the $PA$ along the equator disk is 136.1°. The theoretical PV diagram along the nebula axis in the presence of a rotating disk must show emission with a form similar to a rhombus with equal emission in all four quadrants. The subtraction of $Q_{\rm I} + Q_{\rm III}$ emission (red dots) from $Q_{\rm II} + Q_{\rm IV}$ emission (blue dots) removes the disk contribution emission ("Outflow" in black dots in Fig. 4.25) and reveals excess emission. The emission of an outflow will be the one of "Outflow" at $PA = 46.1°$. The excess emission at this $PA$ is 8.3 mJy beam$^{-1}$ km s$^{-1}$. Taking into account the uncertainty of the $PA$, the different values of $PA = 44.7°$ and $PA = 47.5°$ are measurements of the derived outflow intensity. Thus, we find that the excess emission that we attribute to the outflow emission is $8.3^{+2.4}_{-2.5}$ mJy beam$^{-1}$ km s$^{-1}$. We must highlight that the three different $PA$s present positive emission. In addition, the formal error for the trigonometrical fitting yields an uncertainty of 2.4 mJy beam$^{-1}$ km s$^{-1}$, which is highly consistent with the previous calculated uncertainty.

There is another option to estimate the emission of the putative outflow through the PV diagram along the nebula axis. We see how the emission at central velocities is inclined in the right panel of Fig. 4.2. We see an excess emission in quadrants $Q_{\rm I}$ and $Q_{\rm III}$, with central velocities of $\pm 0.5$ km s$^{-1}$and extreme offsets of $\pm 0''.8$. This excess emission could be explained by the presence of an outflow. Taking into account the emission ($\sim$10 mJy beam$^{-1}$) and the width of that emission in terms of velocity ($\sim$0.75 km s$^{-1}$), the putative outflow presents an emission of $\sim$10 mJy beam$^{-1}$ km s$^{-1}$. The uncertainty for this value, $\Delta E$, is:

$$\Delta E = rms \frac{\sqrt{n_{\rm V}}}{\sqrt{n_{\rm Y}}}, \tag{4.6}$$

where $rms$ is the noise of the map (4.8 mJy beam$^{-1}$), $n_{\rm V}$ is the number of velocity channels used in that region ($\sim$0.75 km s$^{-1}$ / 0.2 km s$^{-1}$ = 3.8), and $n_{\rm Y}$ is the number of position channels in that region ($0\farcs5\,/\,0\farcs07 = 7.1$). We find $\Delta E = 3.5\,\mathrm{mJy\,beam^{-1}\,km\,s^{-1}}$. Therefore, we conclude that the emission of the outflow is $10 \pm 3.5\,\mathrm{mJy\,beam^{-1}\,km\,s^{-1}}$.

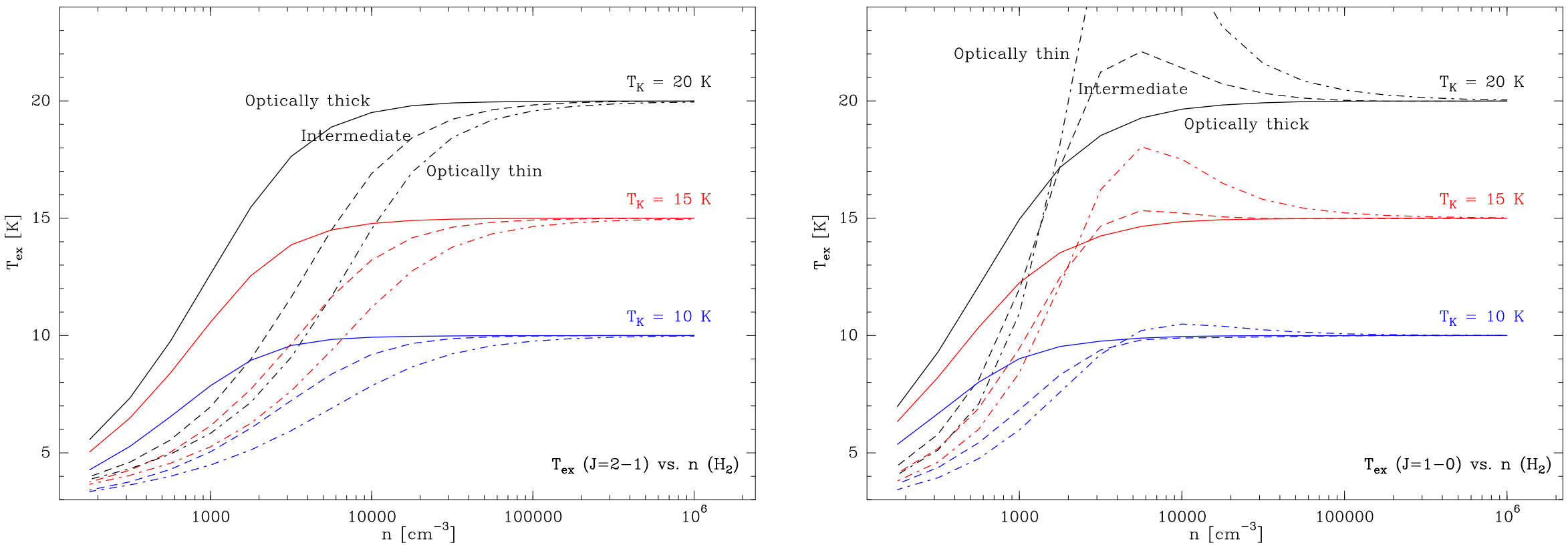

Figure 4.26: Estimates of the equivalent excitation temperature as a function of the kinetic temperature and the density for $J = 2-1$ (*Top*) and $J = 1-0$ (*Bottom*) transitions.

We obtained very close values for the outflow emission through two very different methods, which reinforces our conclusions about the presence of an extended outflow in AC Her. Therefore, we consider the outflow tentatively detected with an emission of $\lesssim 10\,\mathrm{mJy\,beam^{-1}\,km\,s^{-1}}$.

### 4.7.4 Outflow kinetic and rotational temperatures

The sources studied in this work present relatively massive and extended outflows, which show low rotational temperatures (see Sect. 4.5.2, 4.5.3, and 4.5.4). When densities are very low, these rotational temperatures may significantly depart from the kinetic ones. We must remember that "excitation temperature" is always defined as an equivalent temperature for a given line, while "rotational temperature" is defined as the typical value or average of the excitation temperatures of the relevant rotational lines (in our case, low-$J$ transitions including the observed one), and is used to approximately calculate the partition function and the absorption and emission coefficients. In the limit of thermalization (LTE), which is attained for sufficiently high densities, all excitation temperatures are the same, and equal to $T_{\rm rot}$ and $T_{\rm K}$.

In Fig. 4.26 we present theoretical estimates of the equivalence between kinetic temperatures ($T_{\rm K}$) and excitation temperatures ($T_{\rm ex}$) for relevant lines as a function of density. (In the case of very low excitation temperatures, $< 10$ K, only the three lowest levels of CO are significantly populated.) To perform these calculations, we used a standard LVG code very similar to that described for example by Bujarrabal and Alcolea (2013), with the simplest treatment of the velocity field, taking logarithmic velocity gradient to be equal to 1. Three cases are considered, very low optical depths, optically thick case, and calculations with intermediate optical depths. In the optically thick and intermediate opacity cases, we varied the characteristic lengths of the considered regions at the same time as the density, in order to keep opacities of $\sim 10$ and 1, respectively.

As we can see, for larger density values we find that $T_{\rm ex} = T_{\rm K}$. However, for lower density values we find $T_{\rm ex} < T_{\rm K}$, as expected. In some cases, the difference is significant; for the very diffuse extended layers around R Sct, a rotational temperature of $\sim 7$ K corresponds to kinetic temperatures of over $\sim 15$ K.

### 4.7.5 Comparison between observational data and model results from our best-fit models

In this section, we present the observational data together with the synthetic results of our best-fit models for our four sources. We show the observational and synthetic velocity maps and PV diagrams of AC Her for $^{12}$CO $J = 2 - 1$ in Figs. 4.27 and 4.28. We present the observational and synthetic velocity maps and PV diagrams of 89 Her for $^{12}$CO and $^{13}$CO $J = 2 - 1$ emission lines in Figs. 4.29, 4.30, 4.31, and 4.32. The observational and synthetic velocity maps and PV diagrams of IRAS 19125+0343 for $^{12}$CO $J = 2 - 1$ are showed in Figs. 4.33 and 4.34. Finally, our observational and synthetic velocity maps and PV diagrams of R Sct for $^{12}$CO $J = 2 - 1$ are presented in Figs. 4.35 and 4.36. In each case, the scales and contours are the same, so the combined representation of the observational interferometric data in color and the results derived from our best-fit models in white allows an easy comparison (see Sects. 4.4 and 4.5.)

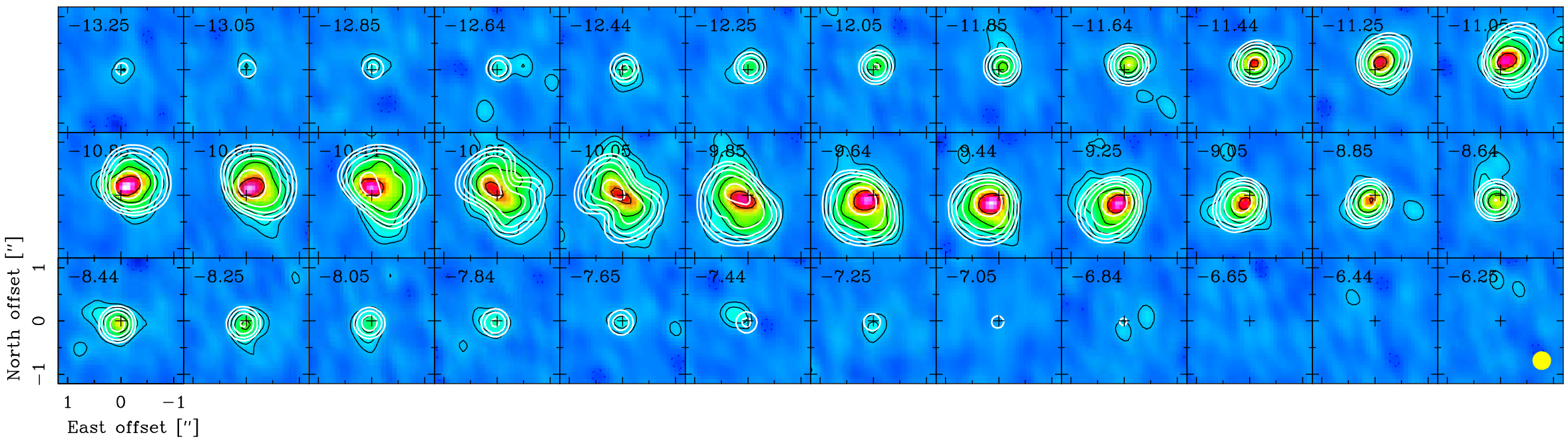

Figure 4.27: NOEMA maps per velocity channel of AC Her in $^{12}$CO $J = 2 - 1$ emission. The contours are $\pm 9$, 18, 36, 76, and 144 mJy beam$^{-1}$ with a maximum emission of 230 mJy beam$^{-1}$. The LSR velocities are indicated in each panel (upper-left corner) and the beam size, $0''\!.35 \times 0''\!.35$, is shown in the last panel at the bottom right corner (yellow ellipse). We also show the synthetic maps from our best-fit model in white contours, to be compared with observational data; the scales and contours are the same.

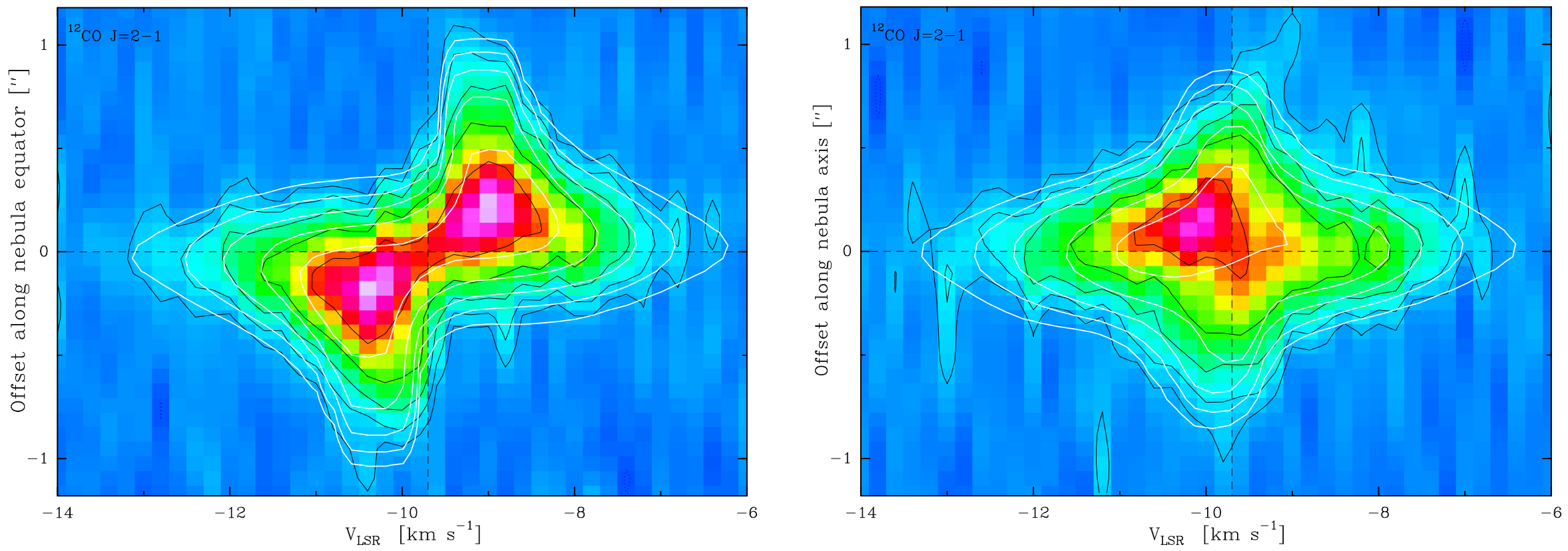

Figure 4.28: *Left*: PV diagram from our NOEMA maps of $^{12}$CO $J = 2 - 1$ in AC Her along the equatorial direction ($PA = 136.1°$). The contours are the same as in Fig. 4.27. The dashed lines show the approximate central position and systemic velocity. Additionally, we show the synthetic PV diagram from our best-fit model in white contours, to be compared with observational data; the scales and contours are the same. *Right*: Same as in *left* but along the axis direction of the nebula ($PA = 46.1°$).

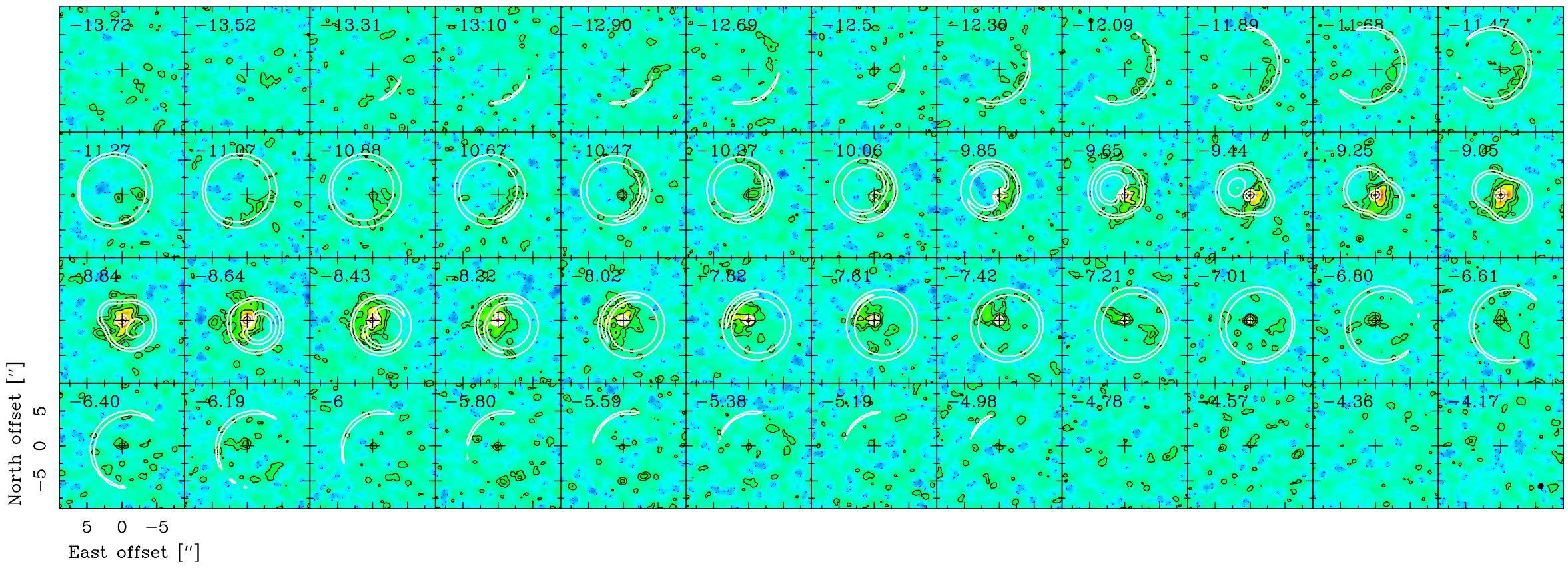

Figure 4.29: NOEMA maps per velocity channel of 89 Her in $^{12}$CO $J = 2 - 1$ emission. The contours are $\pm$70, 140, 280, and 560 mJy beam$^{-1}$ with a maximum emission peak of 870 mJy beam$^{-1}$. The LSR velocities are indicated in each panel (upper-left corner) and the beam size, $1\farcs02 \times 0\farcs83$, is shown in the last panel at the bottom right corner (black ellipse). We also show the synthetic maps from our best-fit model in white contours, to be compared with observational data; the scales and contours are the same.

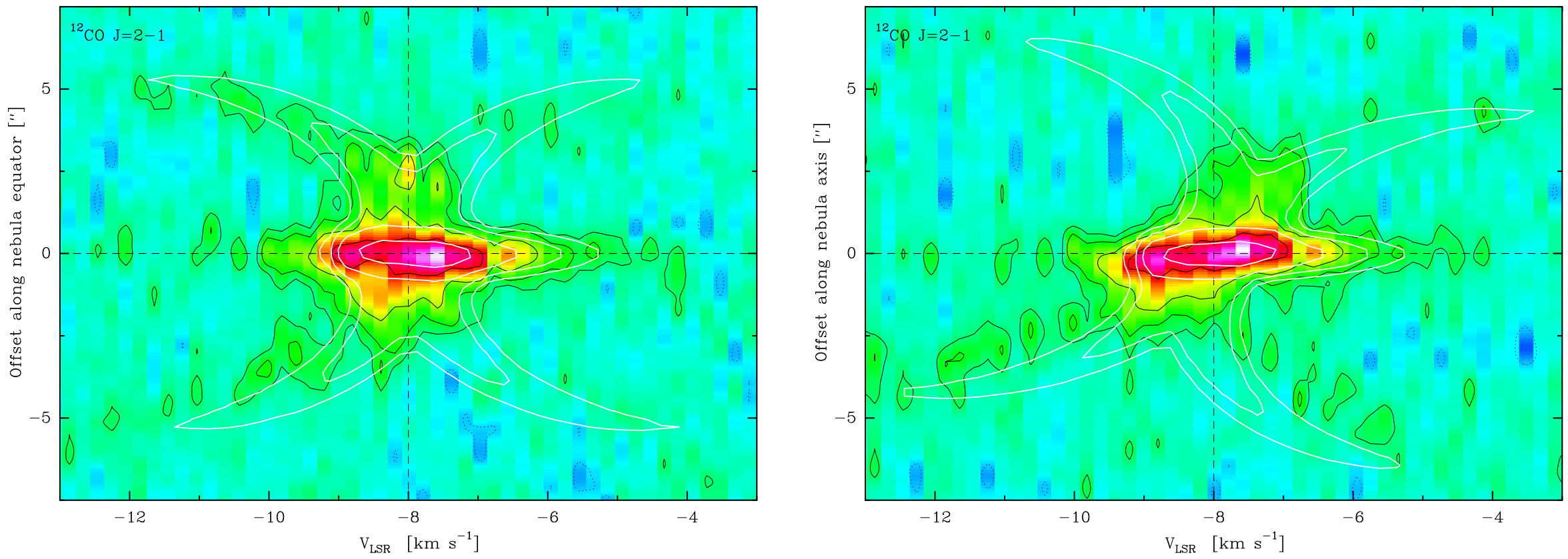

Figure 4.30: *Left*: PV diagram from our NOEMA maps of $^{12}$CO $J = 2 - 1$ in AC Her along the equatorial direction ($PA = 150°$). The contours are $\pm$70, 140, 280, and 560 mJy beam$^{-1}$ with a maximum emission peak of 870 mJy beam$^{-1}$. The dashed lines show the approximate central position and systemic velocity. Additionally, we show the synthetic PV diagram from our best-fit model in white contours, to be compared with observational data; the scales and contours are the same. *Right*: Same as in *left* but along the axis direction of the nebula ($PA = 60°$).

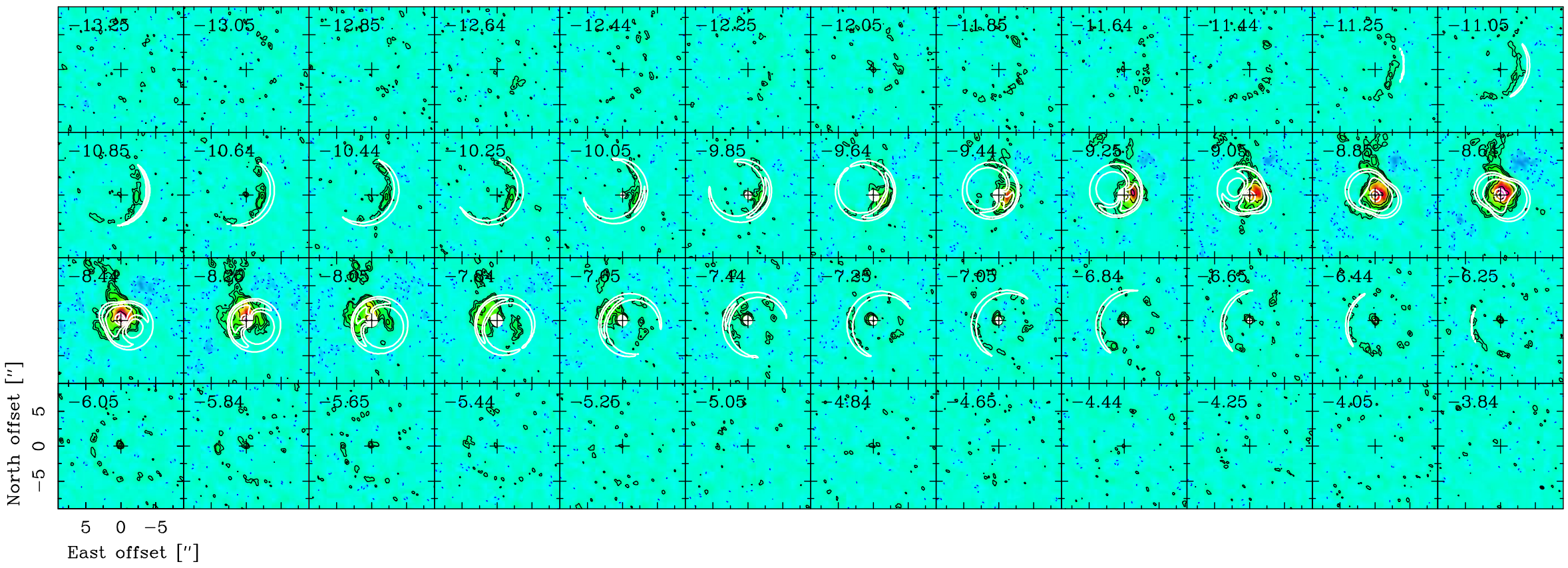

Figure 4.31: *Top*: NOEMA maps per velocity channel of 89 Her in $^{13}$CO $J = 2-1$ emission. The contours are $\pm$11, 22, 44, 88, and 144 mJy beam$^{-1}$ with a maximum emission peak of 225 mJy beam$^{-1}$. The LSR velocities are indicated in each panel (upper-left corner) and the beam size, $0\farcs74 \times 0\farcs56$, is shown in the last panel at the bottom right corner (black ellipse). We also show the synthetic maps from our best-fit model in white contours, to be compared with observational data; the scales and contours are the same.

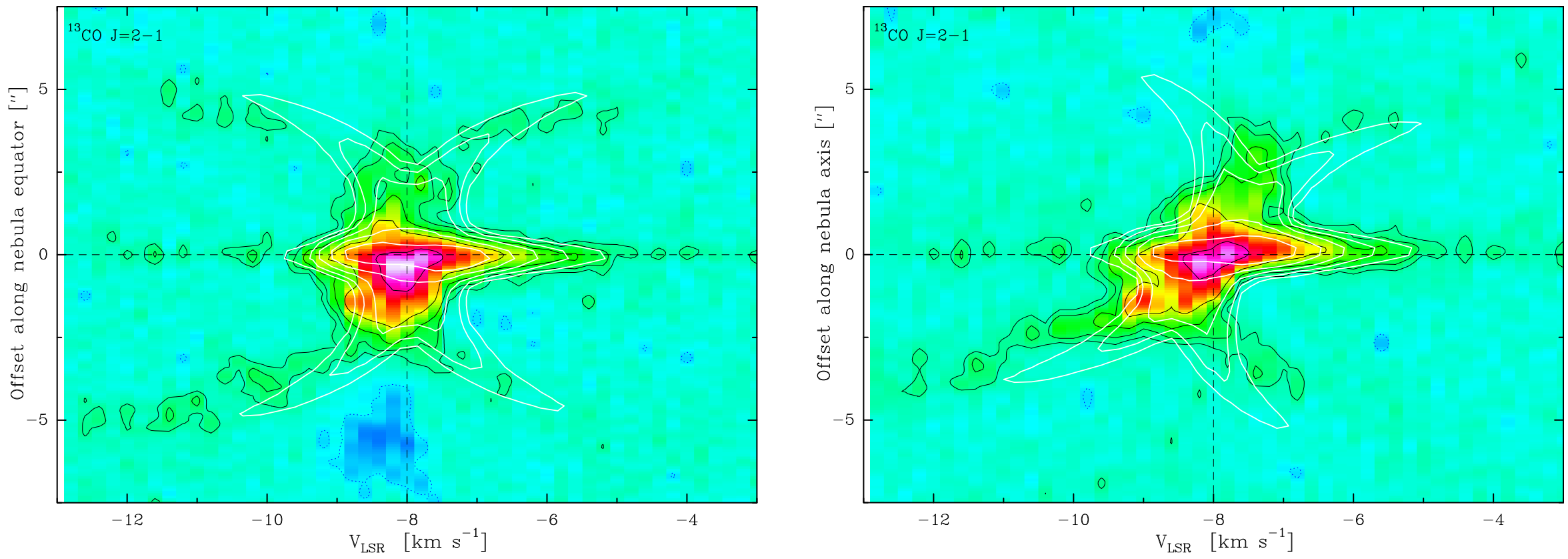

Figure 4.32: *Left:* Same as Fig. 4.30 (*left*) but for $^{13}$CO $J = 2-1$ emission. The contours are $\pm$11, 22, 44, 88, and 144 mJy beam$^{-1}$ with a maximum emission peak of 225 mJy beam$^{-1}$. The dashed lines show the approximate centroid in velocity and position. *Right*: Same as in *left* but along the perpendicular direction $PA = 60°$.

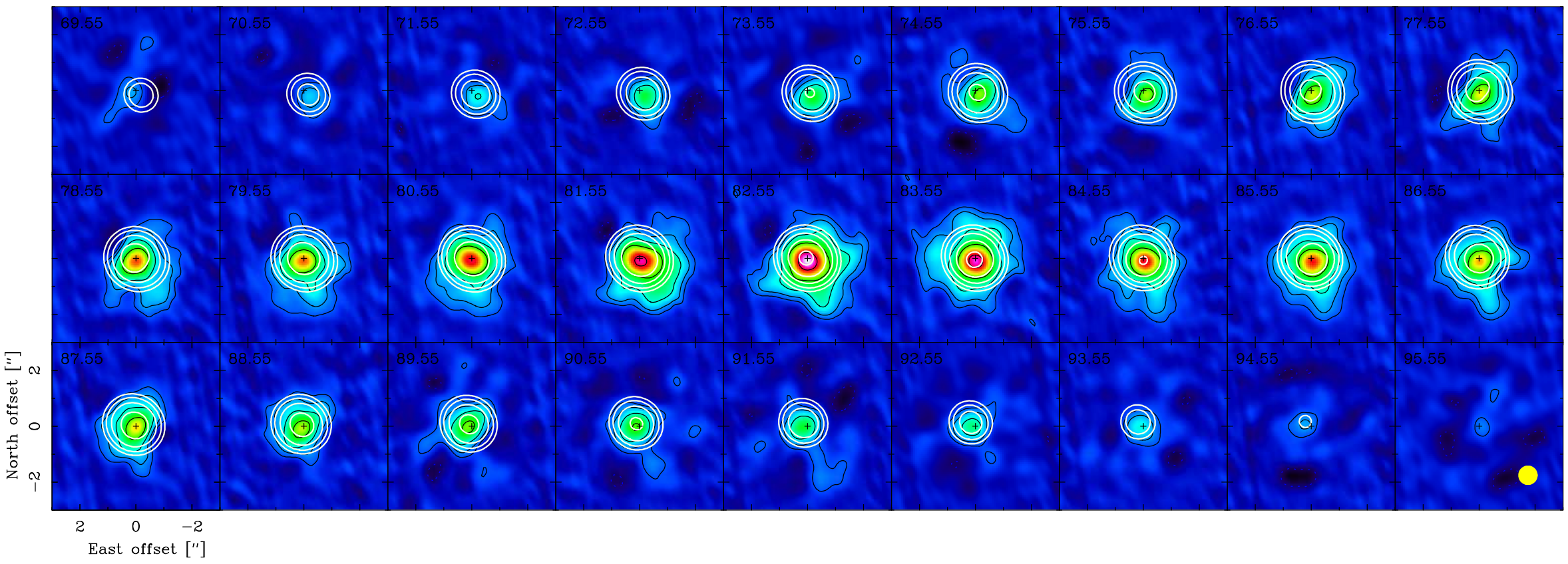

Figure 4.33: Maps per velocity channel of IRAS 19125+0343 in $^{12}$CO $J = 2-1$ emission. The contours are $\pm$20, 40, 80, 160, and 320 mJy beam$^{-1}$ with a maximum emission peak of 413 mJy beam$^{-1}$. The LSR velocities are indicated in each panel (upper-left corner) and the beam size, 3.″12 × 2.″19, is shown in the last panel at the bottom right corner (yellow ellipse). We also show the synthetic maps from our best-fit model in white contours, to be compared with observational data; the scales and contours are the same.

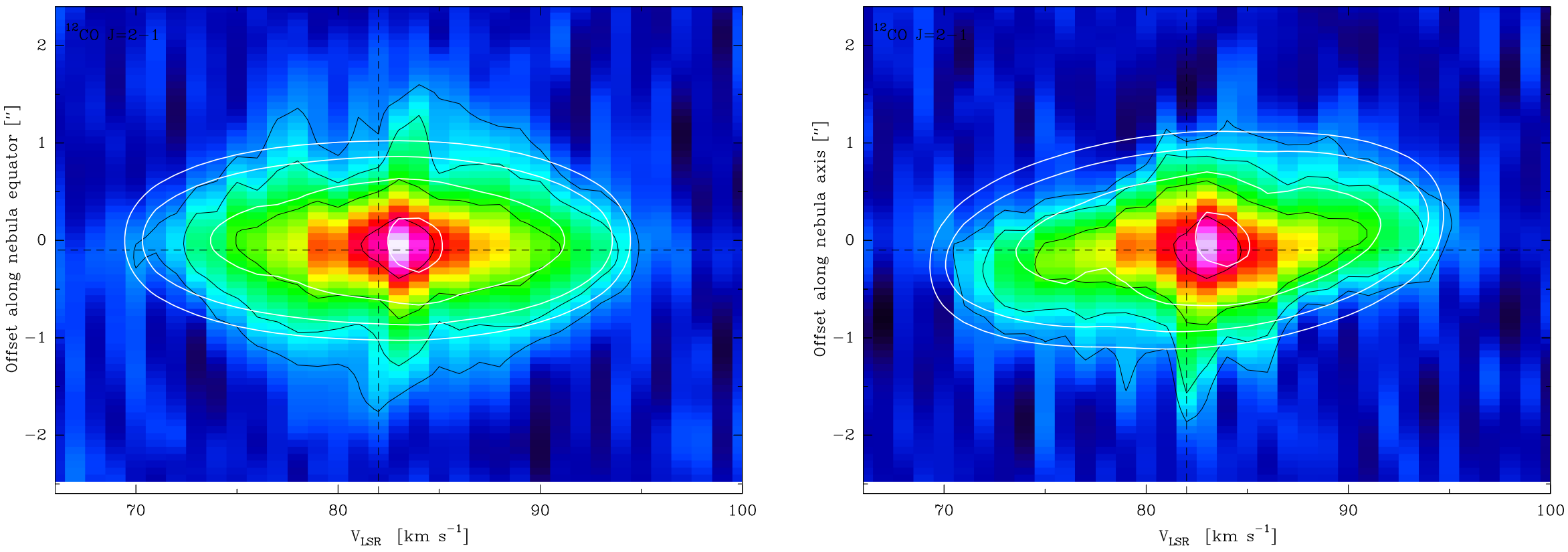

Figure 4.34: *Left*: PV diagram from our NOEMA maps of $^{12}$CO $J = 2-1$ in IRAS 19125+0343 along the equatorial direction ($PA = -40°$). The contours are $\pm$40, 80, 160, and 320 mJy beam$^{-1}$. The dashed lines show the approximate central position and systemic velocity. Additionally, we show the synthetic PV diagram from our best-fit model in white contours, to be compared with observational data; the scales and contours are the same. *Right*: Same as in *left* but along the axis direction of the nebula ($PA = 50°$).

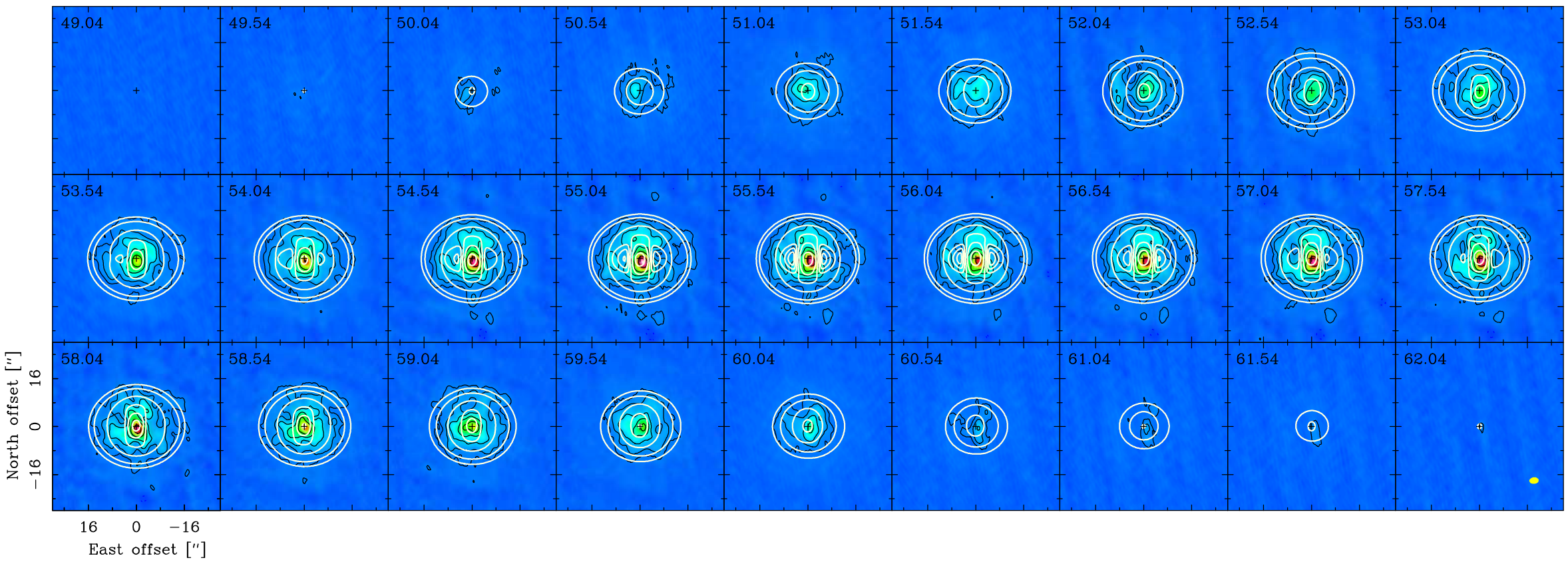

Figure 4.35: Maps per velocity channel of R Sct in $^{12}$CO $J = 2-1$ emission. The contours are ±50, 100, 200, 400, 800 and 1600 mJy beam$^{-1}$ with a maximum emission of 2.4 mJy beam$^{-1}$. The LSR velocities are indicated in each panel (upper-left corner) and the beam size, 3.″12 × 2.″19, is shown in the last panel at the bottom right corner (yellow ellipse). We also show the synthetic maps from our best-fit model in white contours, to be compared with observational data; the scales and contours are the same.

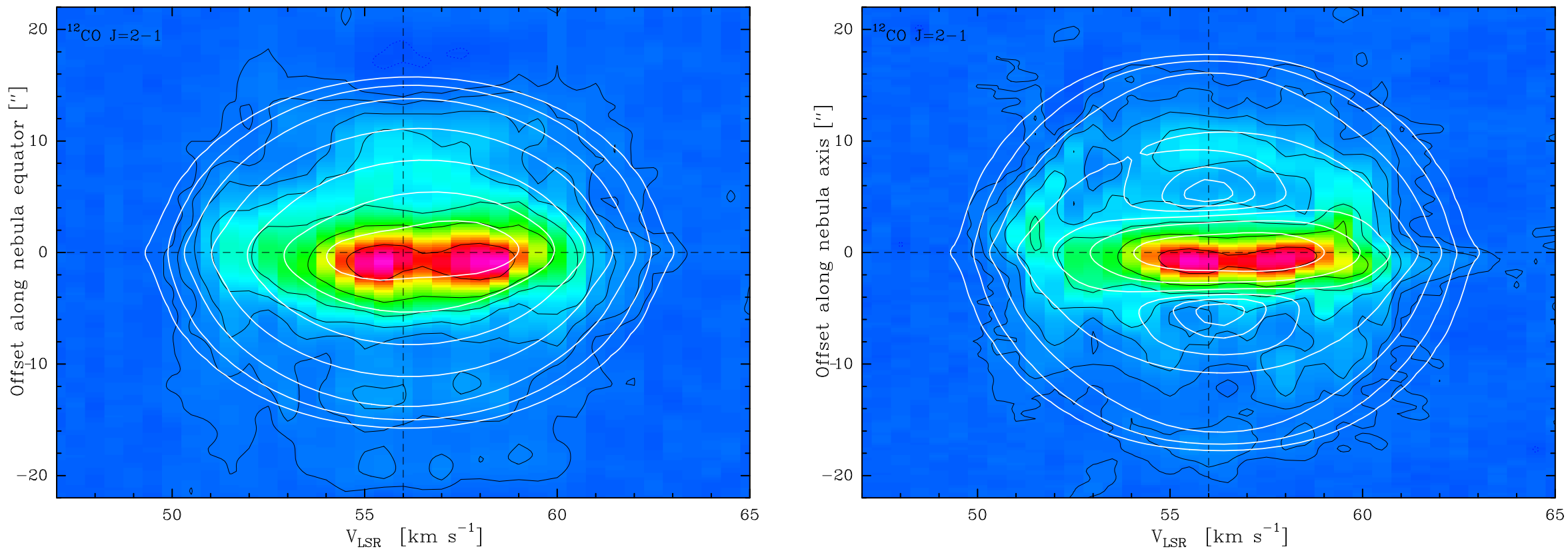

Figure 4.36: *Left*: PV diagram from our maps of $^{12}$CO $J = 2-1$ in R Sct along the equatorial direction ($PA = 0°$). The contours are ±25, 50, 100, 200, 400, 800 and 1600 mJy beam$^{-1}$. The dashed lines reveal the systemic velocity and the central position of the source. Additionally, we show the synthetic PV diagram from our best-fit model in white contours, to be compared with observational data; the scales and contours are the same. *Right*: Same as in *left* but along the axis direction ($PA = 90°$).

### 4.7.6 Additional figures of model calculation

In this section, we show the maps and PV diagrams derived from models for the four analyzed sources. We present the synthetic velocity maps of AC Her for $^{12}$CO $J = 2-1$ (see Fig. 4.38) for our best-fit model (Sect. 4.5.1). Additionally, we show synthetic velocity maps and PV diagrams along the equator and nebula axis for a disk-only model (the standard model of Table 4.3 and Fig. 4.11 without the outflow) in Figs. 4.38 and 4.39. We also show an alternative model of AC Her where the size of the outflow is somewhat larger than the standard one (see Figs. 4.47, 4.40, and 4.41), both cases being compatible with observations.

We present synthetic velocity maps of 89 Her for $^{12}$CO and $^{13}$CO $J = 2-1$ emission for our best-fit model (see Sect. 4.5.2 and Fig. 4.42). The hourglass-shaped structure is present in the synthetic velocity maps for $^{12}$CO and $^{13}$CO $J = 2-1$. We show velocity maps of IRAS 19125+0343 for $^{12}$CO $J = 2-1$ (see Fig. 4.43) of our best-fit model (see Sect. 4.5.3). Additionally, we present synthetic velocity maps for the alternative model of IRAS 19125+0343 and PV diagrams along the equator disk and nebula axis (see Fig. 4.44, 4.45, and 4.48).

We present synthetic velocity maps of R Sct for $^{12}$CO $J = 2-1$ for our best-fit model (see Fig. 4.46 and Sect. 4.5.4). These maps are similar to those of Fig. 4.8. We note that cavities of the extended outflow are also present.

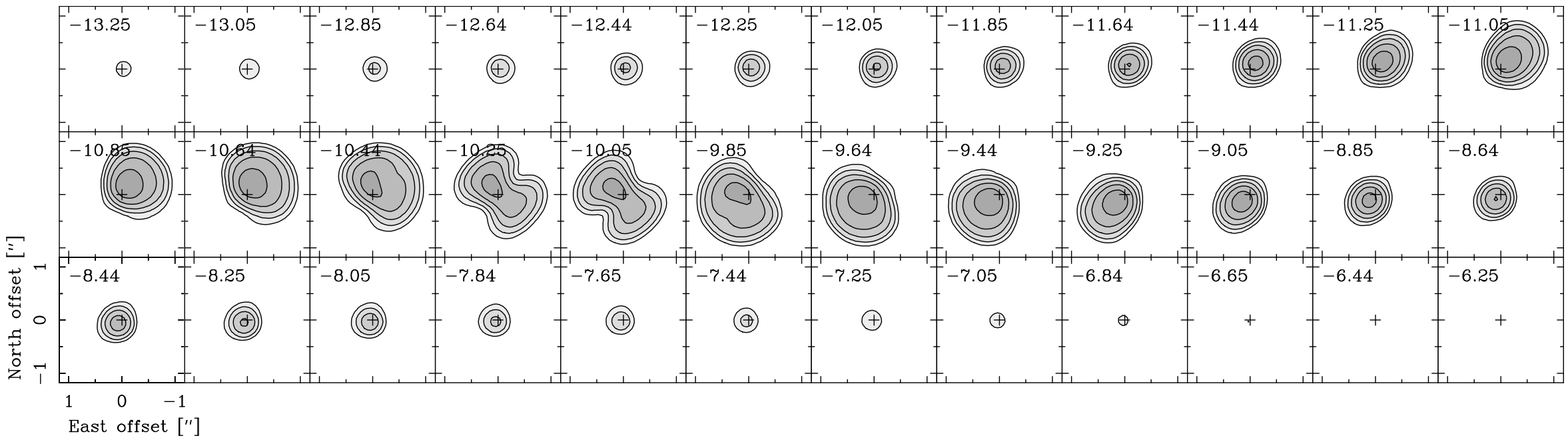

Figure 4.37: Synthetic maps predicted by the alternative model of the $^{12}$CO $J = 2-1$ line emission for the nebula around AC Her. To be compared with Fig. 4.1, the scales and contours are the same.

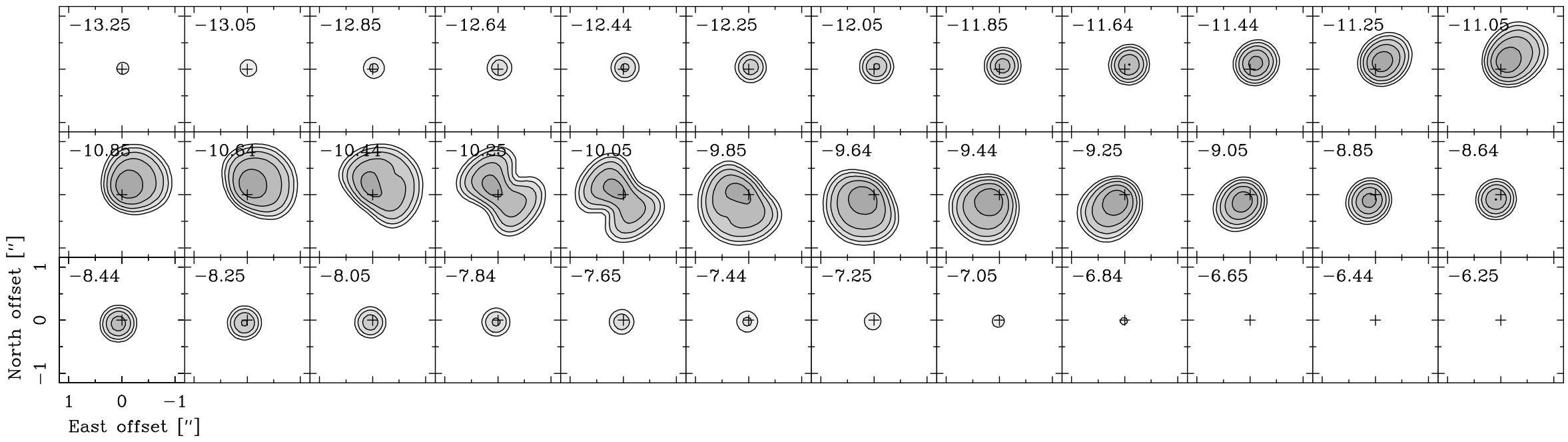

Figure 4.38: Synthetic maps predicted by the disk-only model of the $^{12}$CO $J = 2-1$ line emission for the nebula around AC Her. To be compared with Fig. 4.1, the scales and contours are the same.

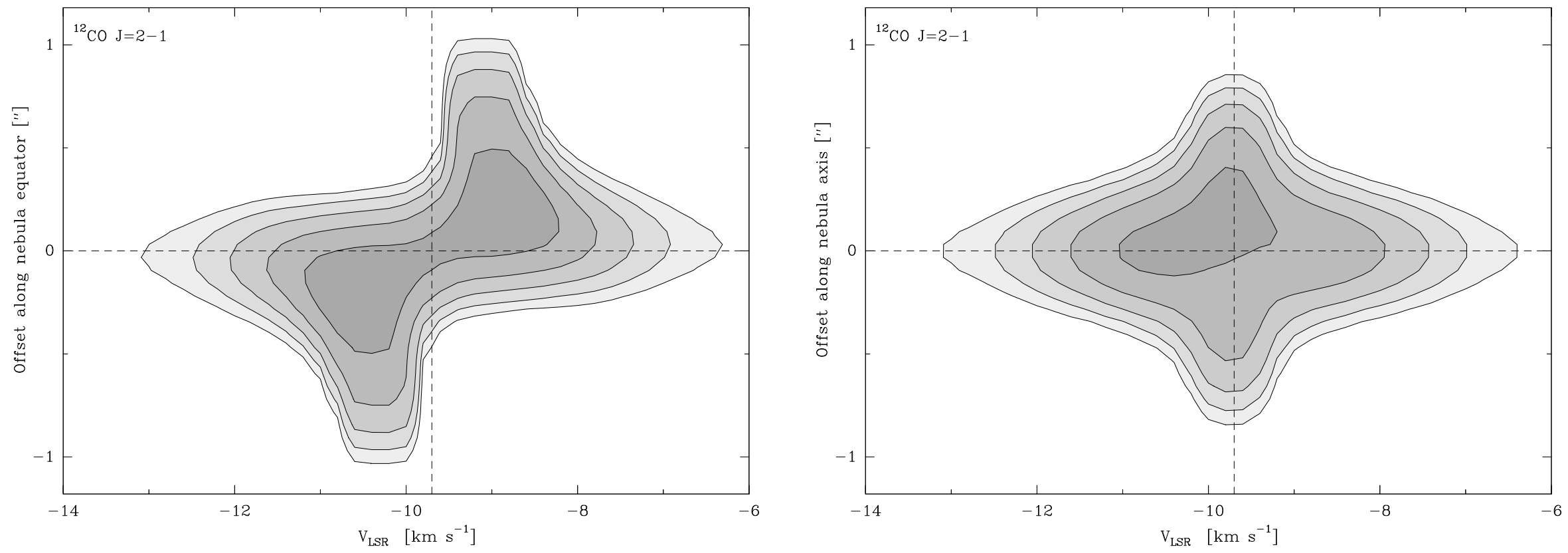

Figure 4.39: *Left:* Synthetic position-velocity diagram from our disk-only best-fit model of $^{12}$CO $J = 2-1$ in AC Her along the direction $PA = 136.1°$. To be compared with the left panel of Fig. 4.2, the scales and contours are the same. *Right*: Same as in *left* but along $PA = 46.1°$.

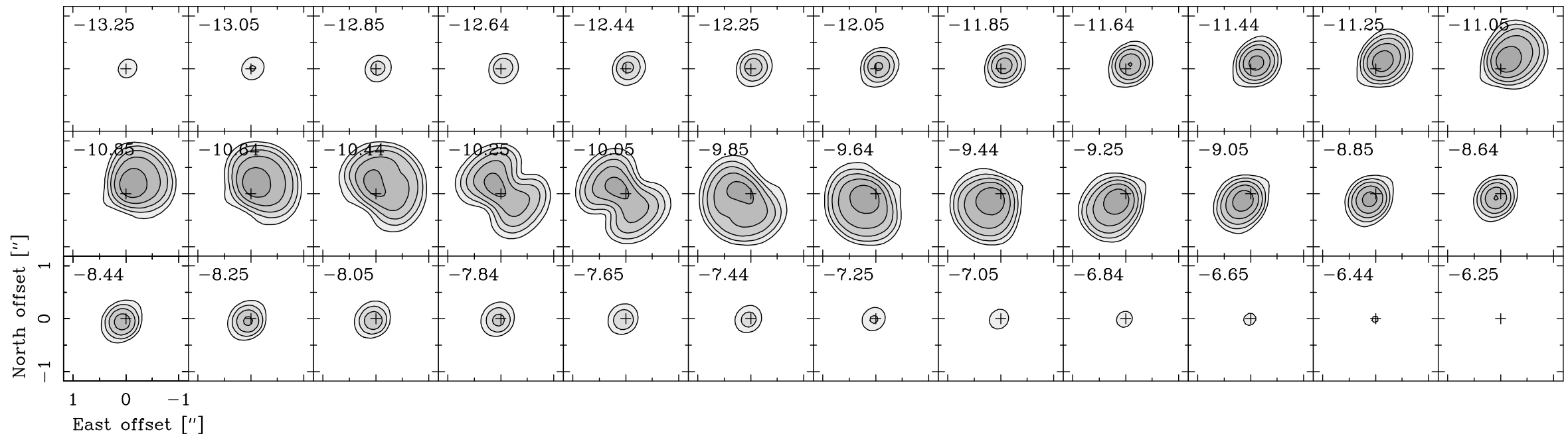

Figure 4.40: Synthetic maps predicted by the alternative model of the $^{12}$CO $J = 2-1$ line emission for the nebula around AC Her. To be compared with Fig. 4.1, the scales and contours are the same.

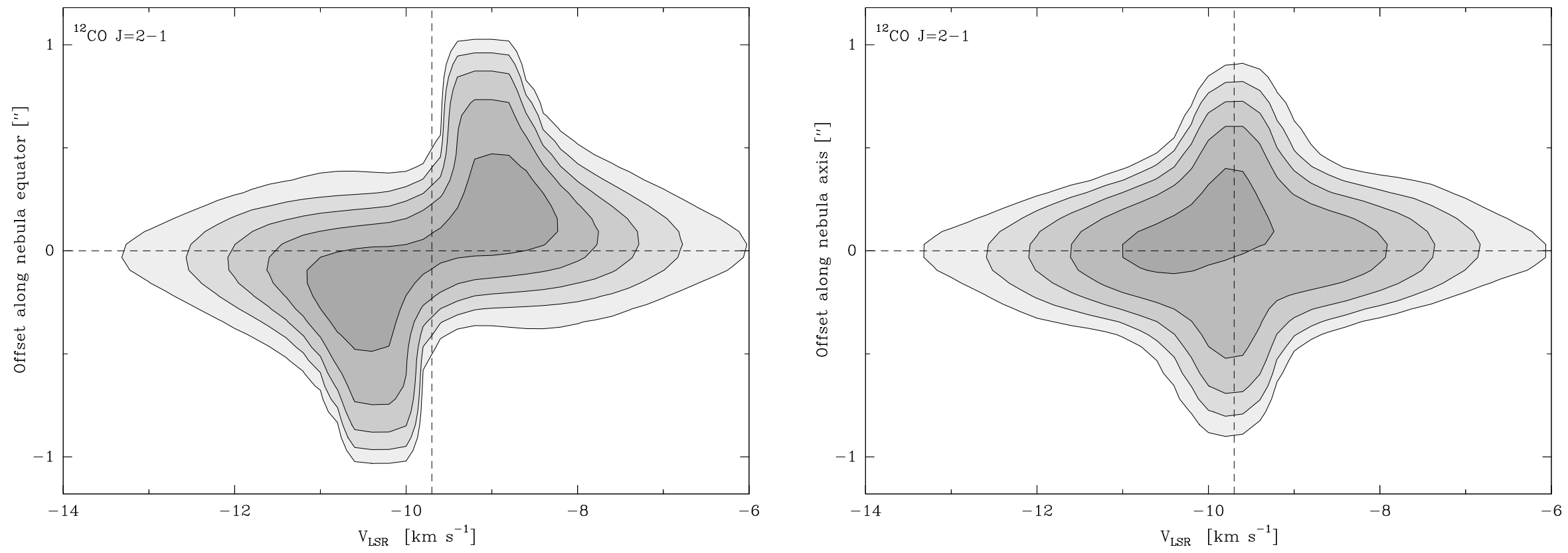

Figure 4.41: *Left:* Synthetic position-velocity diagram from our best-fit alternative model of $^{12}$CO $J = 2-1$ in AC Her with an outflow. To be compared with Fig. 4.2, the scales and contours are the same. *Right*: Same as in *left* but along $PA = 46.1°$.

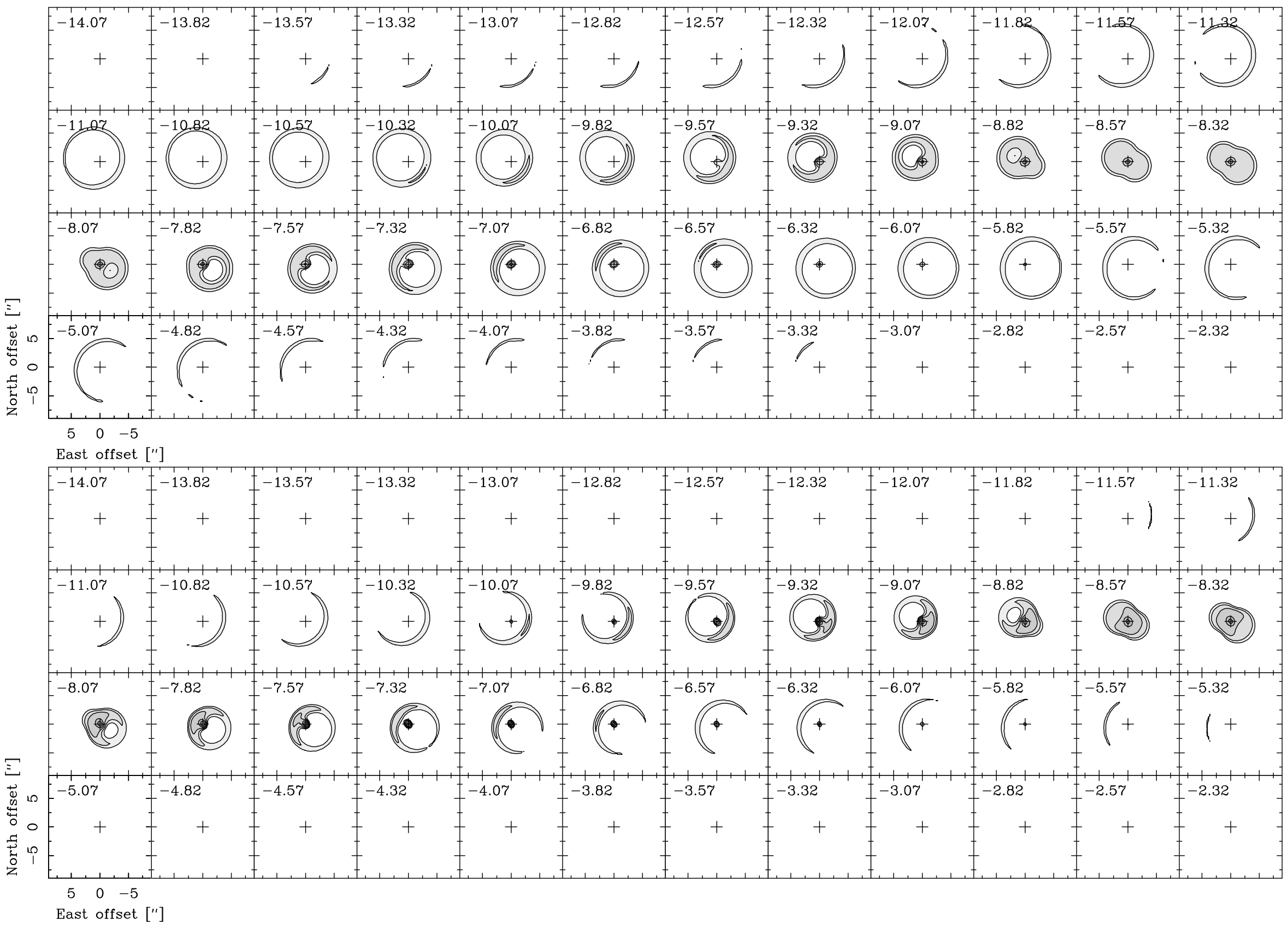

Figure 4.42: Synthetic maps predicted by the model of the $^{12}$CO $J = 2 - 1$ line emission (*Top*) and $^{13}$CO $J = 2 - 1$ (*Bottom*) for the nebula around 89 Her. To be compared with Fig. 4.3, the scales and contours are the same.

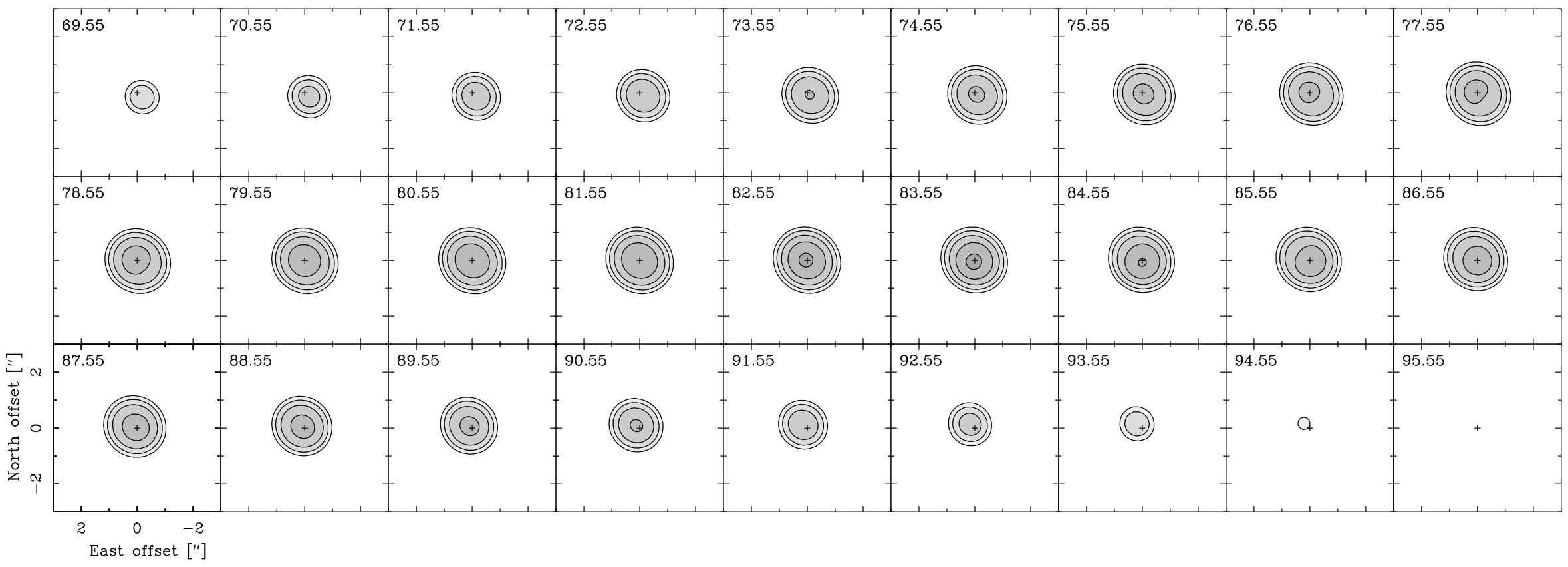

Figure 4.43: Synthetic maps predicted by the model of the $^{12}$CO $J = 2 - 1$ line emission for the nebula around IRAS 19125+0343. To be compared with Fig. 4.6, the scales and contours are the same.

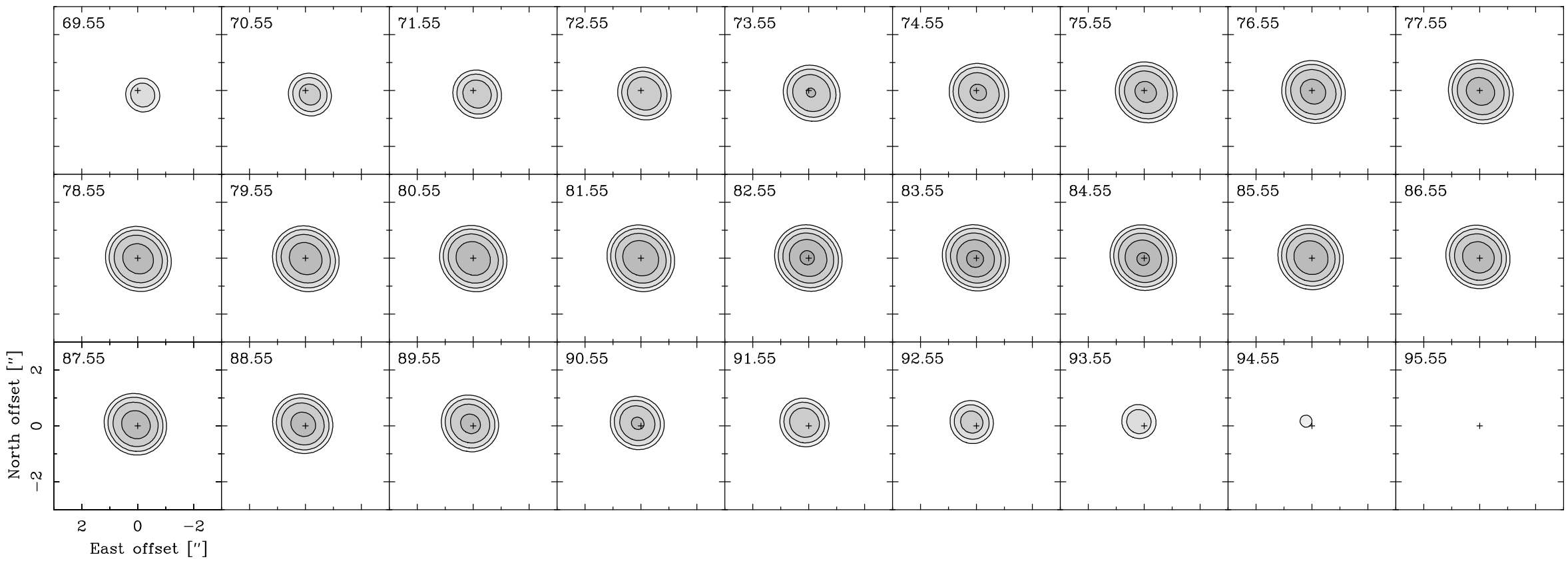

Figure 4.44: Same as in Fig. 4.43 but for the alternative model. To be compared with Fig. 4.6, the scales and contours are the same.

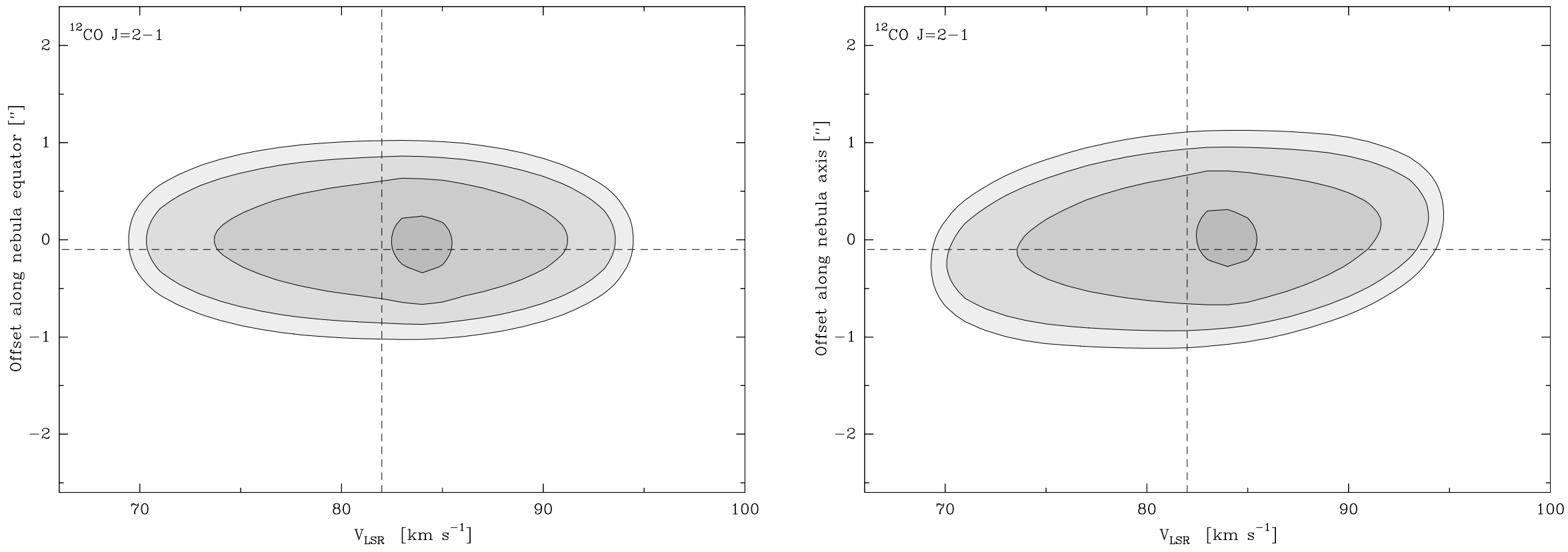

Figure 4.45: *Left:* Synthetic position-velocity diagram from our alternative best-fit model of $^{12}$CO $J = 2-1$ in IRAS 19125+0343 along the direction $PA = -40°$. To be compared with the left panel of Fig. 4.7, the scales and contours are the same. *Right*: Same as in *left* but along $PA = 50°$.

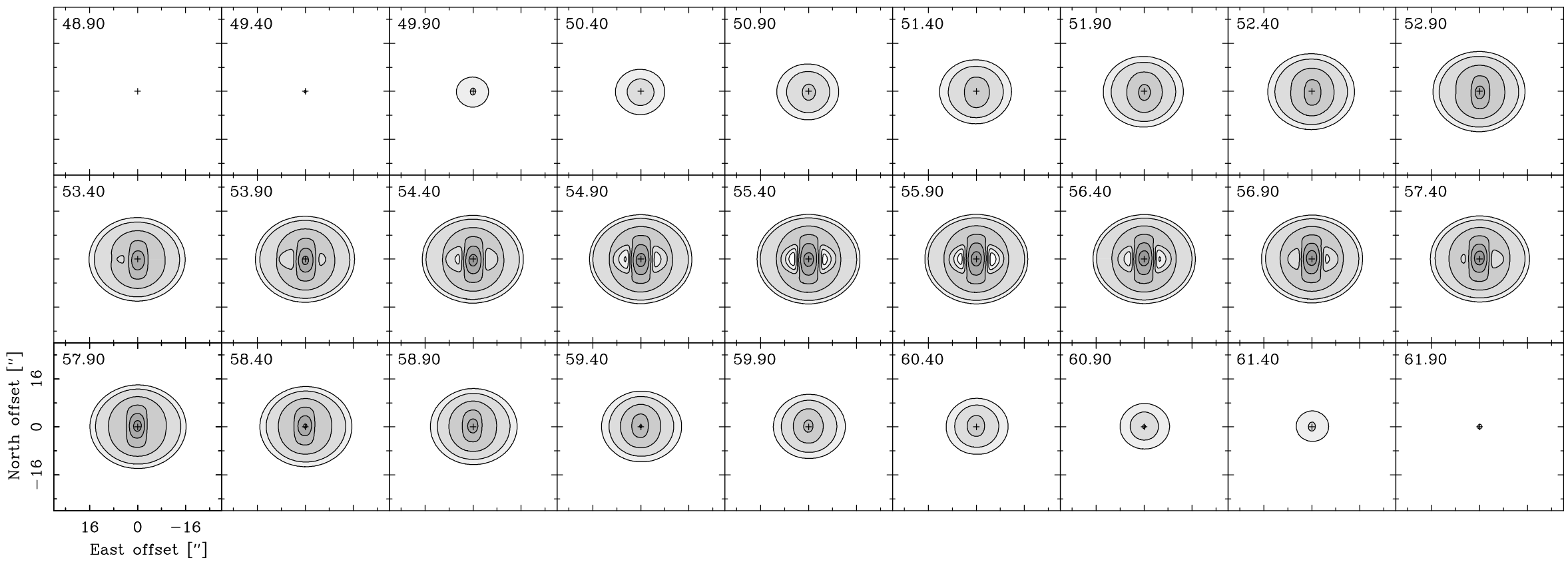

Figure 4.46: Synthetic maps predicted by the model of the $^{12}$CO $J = 2-1$ line emission for the nebula around R Sct. To be compared with Fig. 4.8, the scales and contours and units are the same.

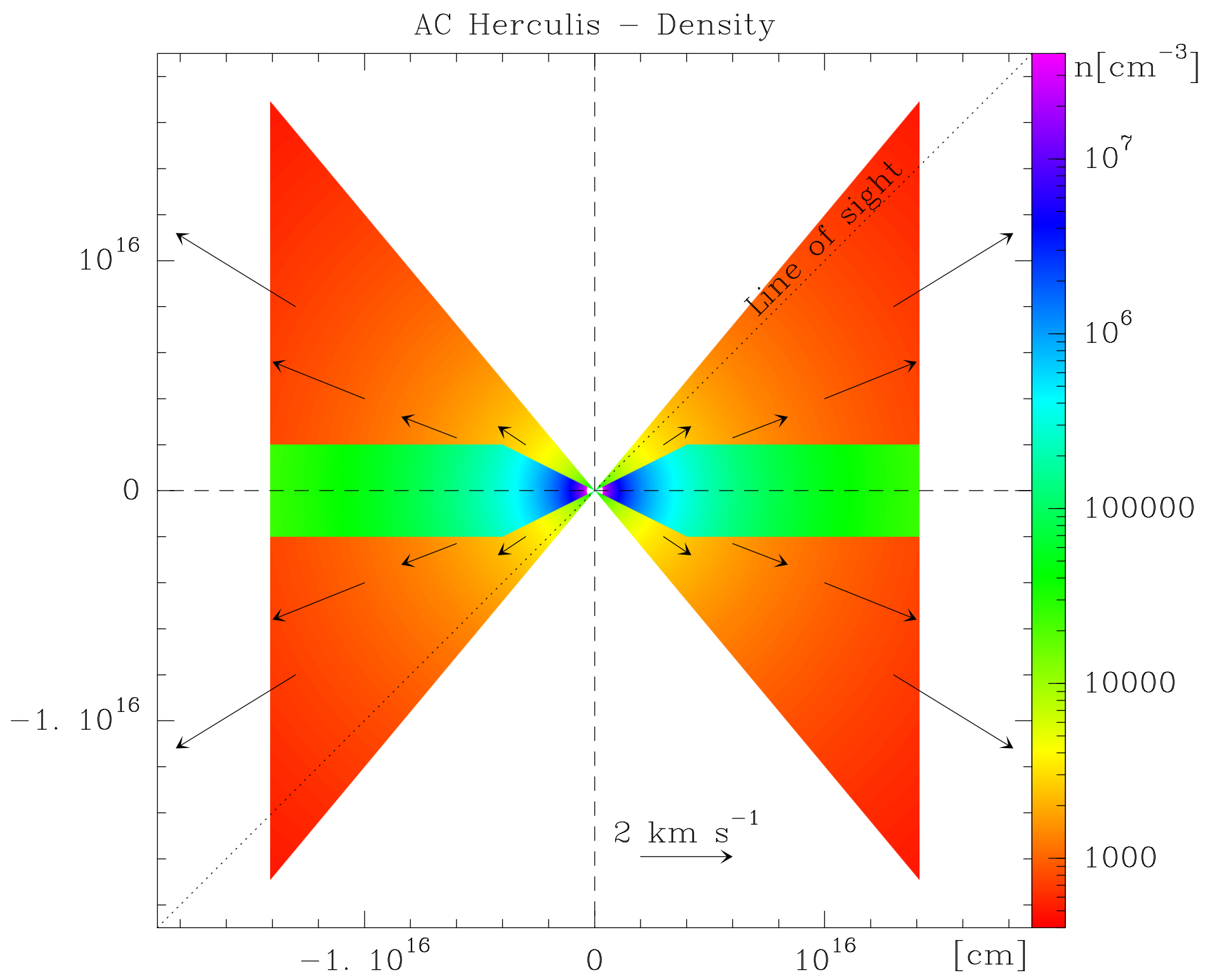

Figure 4.47: Structure and distribution of the density of our alternative best-fit model for the disk and outflow of AC Her. The Keplerian disk presents density values $\geq 10^5\,\mathrm{cm}^{-3}$. The expansion velocity is represented with arrows.

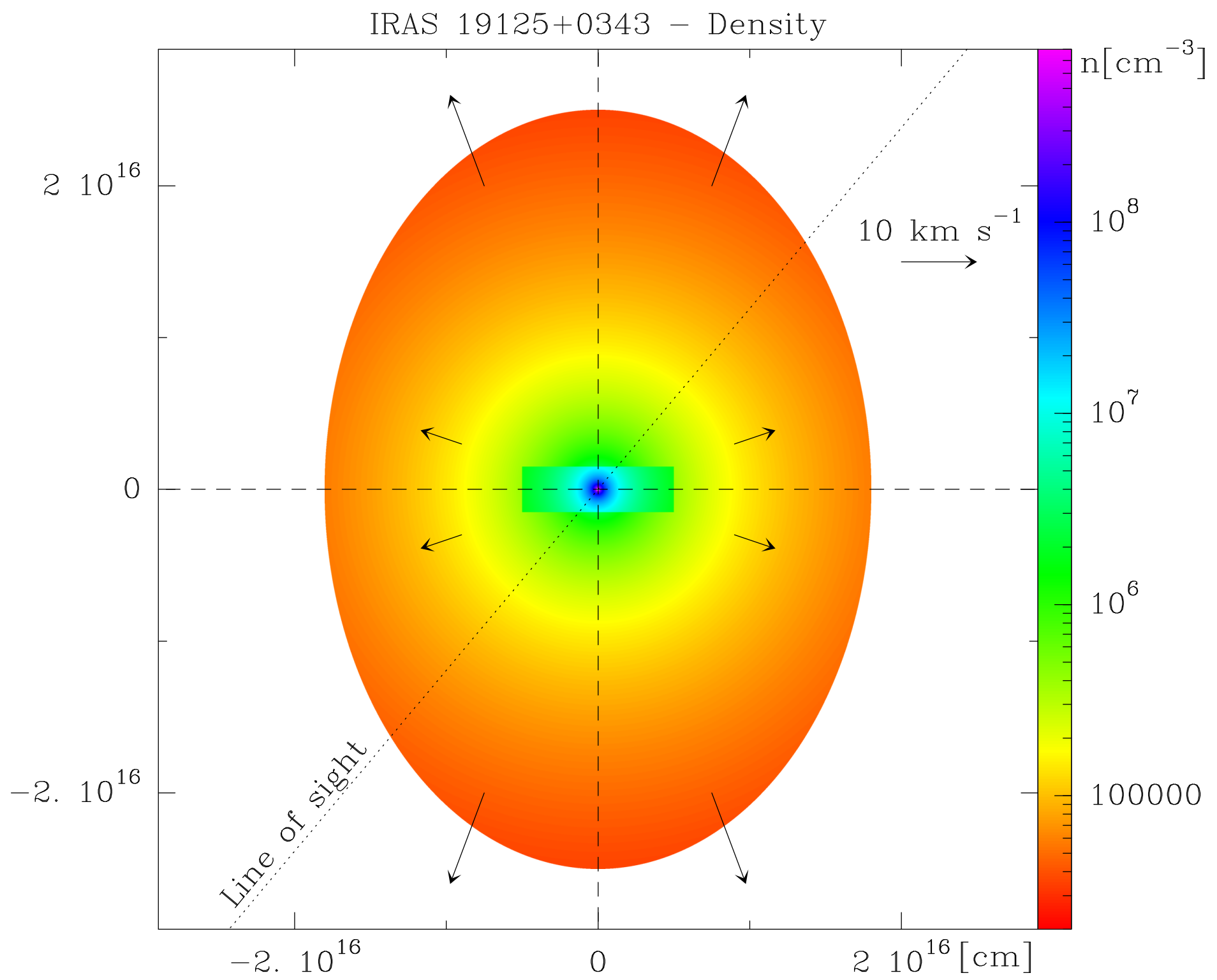

Figure 4.48: Structure and distribution of the density of our best-fit alternative model for the disk and outflow of IRAS 19125+0343. The Keplerian disk presents density values $\geq 10^6\,\mathrm{cm}^{-3}$. The expansion velocity is represented with arrows.

*La química es un amante exigente.*
Belén Liñero Saro

# 5

# Chemistry of nebulae around binary post-AGB stars: A molecular survey of mm-wave lines

*The molecular content of nebulae around binary post-AGB stars was practically unknown, apart from systematic observations of $^{12}CO$ and $^{13}CO$ aimed to reveal the presence of characteristic of rotating disks. Additionally, optical and NIR observations were performed in a few sources. The O-/C-rich dichotomy, which is the key parameter that determines the chemistry in the nebulae around evolved stars, was also unknown.*
*In this chapter, we present the first systematic chemical study of nebulae around binary post-AGB stars. Here, we present a very deep single-dish radio molecular survey of over 500 hours of telescope time. Our complete analysis allows us to study the molecular richness of our sources, the abundance of molecules, and to determine their oxygen/carbon nature. Additionally, we also study the abundance of the $^{12}CO$ and $^{13}CO$ through this isotopic ratio, and the $^{17}O/^{18}O$ ratio, that allows us to investigate the initial stellar mass. From this work, we conclude that the binary post-AGB nebulae are particularly poor in molecular content other than CO. This molecular deficiency is even more severe for the case of the disk-dominated sources. The content of this chapter is adapted from Gallardo Cava et al. (2022c).*

## Abstract

Context: There is a class of binary post-asymptotic giant branch (post-AGB) stars that exhibit remarkable near-infrared (NIR) excess. Such stars are surrounded by Keplerian or quasi-Keplerian disks, as well as extended outflows composed of gas escaping from the disk. This class can be subdivided into disk- and outflow-dominated sources, depending on whether it is the disk or the outflow that represents most of the nebular mass, respectively. The chemistry of this type of source has been practically unknown thus far.

Aims: Our objective is to study the molecular content of nebulae around binary post-AGB stars that show disks with Keplerian dynamics, including molecular line intensities, chemistry, and abundances.

Methods: We focused our observations on the 1.3, 2, 3 mm bands of the 30 m IRAM telescope and on the 7 and 13 mm bands of the 40 m Yebes telescope. Our observations add up ∼600 hours of telescope time. We investigated the integrated intensities of pairs of molecular transitions for CO, other molecular species, and IRAS fluxes at 12, 25, and 60 $\mu$m. Additionally, we studied isotopic ratios, in particular $^{17}O/^{18}O$, to analyze the initial stellar mass, as well as $^{12}CO/^{13}CO$, to study the line and abundance ratios.

Results: We present the first single-dish molecular survey of mm-wave lines in nebulae around binary post-AGB stars. We conclude that the molecular content is relatively low in nebulae around binary post-AGB stars, as their molecular lines and abundances are especially weaker compared with AGB stars. This fact is very significant in those sources where the Keplerian disk is the dominant component of the nebula. The study of their chemistry allows us to classify nebulae around AC Her, the Red Rectangle, AI CMi, R Sct, and IRAS 20056+1834 as O-rich, while that of 89 Her is probably C-rich. The calculated abundances of the detected species other than CO are particularly low compared with AGB stars. The initial stellar mass derived from the $^{17}O/^{18}O$ ratio for the Red Rectangle and 89 Her is compatible with the central total stellar mass derived from previous mm-wave interferometric maps. The very low $^{12}CO/^{13}CO$ ratios found in binary post-AGB stars reveal a high $^{13}CO$ abundance compared to AGB and other post-AGB stars.

## 5.1 Introduction

The spectacular late evolution of low- and intermediate-mass stars (main sequence masses in the approximate range of $0.8 - 8\,M_{\odot}$) is characterized by a very copious mass loss during the asymptotic giant branch (AGB) phase. These kinds of stars experience mass loss rates up to $10^{-4}\,M_{\odot}\,a^{-1}$ that dominate the evolution in this phase[1]. The ejected material creates an expanding circumstellar envelope (CSE). These stars are considered one of the most relevant contributors of dust and enriched material to the Interstellar Medium (ISM); see Gehrz (1989); Matsuura et al. (2009). This environment is favorable for the formation of simple molecules and dust. We find three different kinds of envelopes depending on the C/O ratio: the M-type stars with $C/O < 1$, the S-type with $C/O \approx 1$, and C-type with $C/O > 1$. The kind of molecules and dust grains found in CSEs is determined by the C/O ratio. We find O-bearing molecules, such as SiO or $H_2O$, and silicate dust in M-type stars, making the envelope of these stars O-rich (Engels, 1979; Velilla Prieto et al., 2017). On the contrary, we find C-bearing molecules, such as HCN and CS, often alongside SiS, in C-type stars (Olofsson et al., 1993; Cernicharo et al., 2000). Amorphous carbon dust formed out of primary carbon is dominant in C-rich envelopes, while silicon carbide and magnesium sulphide are minor widespread dust components (Zhukovska and Gail, 2008; Massalkhi et al., 2020).

Molecular lines were analysed in a large sample of evolved stars (Bujarrabal et al., 1994b,a). The authors investigate line ratios between CO and other O- and C-bearing species to analyze the molecular intensities of these sources. They also study O- and C-bearing molecule ratios to find the necessary criteria to discern between O- and C-rich envelopes around evolved stars. They found, in particular, that SiO and HCN lines can be as intense as CO lines in O- and C-rich stars, respectively.

[1]We follow the recommendations for units of the IAU Style Manual (Wilkins, 1990, 1995). Therefore, we use the term annus, abbreviated as "a", for year.

When the star attains the post-asymptotic giant branch (post-AGB) phase, the expanding envelope becomes a pre-planetary nebula (pPN). The final stage of the evolution of the (low- or intermediate-mass) star is heralded when the exposed core, a white dwarf, ionizes the circumstellar nebula and becomes a planetary nebula (PN). The chemical behaviour in pPNe is more complex than for normal circumstellar envelopes around AGB stars; since O- and C-rich pPNe could present differences in the line ratios compared with O- and C-rich stars (Bujarrabal et al., 1994b). For example, SiO is systematically weaker in pPNe than in evolved stars, while HCN appears to show the opposite trend. The chemical evolution during the AGB to pPN to PN phases has been investigated in several surveys (Bujarrabal et al., 1994b,a; Bujarrabal, 2006; Cernicharo et al., 2011). The star experiences striking changes in the molecular composition at the pPN phase (Pardo et al., 2007; Park et al., 2008; Zhang et al., 2013). However, the chemistry in pPN is poorly studied, even molecular line surveys of PNe are relatively scarce (Zhang, 2017).

There is a kind of binary post-AGB stars (binary systems including a post-AGB star) that systematically shows evidence for the presence of disks orbiting the central stars (Van Winckel, 2003; de Ruyter et al., 2006; Bujarrabal et al., 2013a; Hillen et al., 2017). All of them present remarkable near-infrared (NIR) excess and the narrow CO line profiles characteristic of rotating disks. The presence of hot dust close to the stellar system is suspected from their spectral energy distribution (SED) and its disk-like shape has been confirmed by interferometric IR data (Hillen et al., 2017; Kluska et al., 2019). The IR spectra of these sources reveal the presence of highly processed grains, which implies that their disks must be stable structures (Jura, 2003; Sahai et al., 2011; Gielen et al., 2011a). The low-$J$ rotational lines of CO have been well studied in sources with NIR excess (Bujarrabal et al., 2013a). They show narrow CO lines and relatively wide wings. These line profiles are similar to those in young stars surrounded by a rotating disk made of remnants of ISM and those expected from disk-emission (Bujarrabal et al., 2005; Guilloteau et al., 2013). These results indicate that the CO emission lines of our sources come from Keplerian (or quasi-Keplerian) disks.

Due to the relatively small size of these disks, deep studies of Keplerian disks around post-AGB stars need high angular- and spectral-resolution. To date, there have been only four well-resolved Keplerian disks identified: in the Red Rectangle (Bujarrabal et al., 2013b, 2016), AC Her (Bujarrabal et al., 2015; Gallardo Cava et al., 2021), IW Car (Bujarrabal et al., 2017), and IRAS 08544−4431 (Bujarrabal et al., 2018b). These sources are disk-dominated, because the disks contain most of the material of the nebula ($\sim$ 90% of the total mass), while the rest of the mass ($\sim$10%), corresponds to an extended bipolar outflow surrounding the disk. Very detailed studies (Gallardo Cava et al., 2021) also show strong indications of the presence of Keplerian disks in 89 Her, IRAS 19125+0343, and R Sct. The nebulae around IRAS 19125+0343 and R Sct are dominated by the expanding component, instead of the Keplerian disk, because $\sim$ 75% of the total mass is located in the outflow. These sources belong to the outflow-dominated subclass. 89 Her is in an intermediate case in between the disk- and the outflow-dominated sources, because $\sim$ 50% of the material of the nebula is in the Keplerian disk. There are other binary post-AGB stars that are surrounded by a Keplerian

Table 5.1 Binary post-AGB stars observed in this work.

| Source | $M_{\rm neb}$ [$M_\odot$] | $\frac{\rm Disk}{\rm Total}$ [%] | $V_{\rm LSR}$ [km s$^{-1}$] | $d$ [pc] |
|---|---|---|---|---|
| AC Herculis | $8.3 \times 10^{-4}$ | $\gtrsim 95$ | $-9.7$ | 1100 |
| Red Rectangle | $1.4 \times 10^{-2}$ | 90 | 0 | 710 |
| 89 Herculis | $1.4 \times 10^{-2}$ | 50 | $-8.0$ | 1000 |
| HD 52961 | $1.3 \times 10^{-2}$ | $\sim 50$ | $-7$ | 2800 |
| IRAS 19157$-$0257 | $1.4 \times 10^{-2}$ | $\sim 50$ | 46 | 2900 |
| IRAS 18123+0511 | $4.7 \times 10^{-2}$ | $\sim 30$ | 99 | 3500 |
| IRAS 19125+0343 | $1.1 \times 10^{-2}$ | 30 | 82.0 | 1500 |
| AI Canis Minoris | $1.9 \times 10^{-2}$ | 25 | 29 | 1500 |
| IRAS 20056+1834 | $1.0 \times 10^{-1}$ | 25 | $-9$ | 3000 |
| R Scuti | $3.2 \times 10^{-2}$ | 25 | 56.1 | 1000 |

**Notes.** Sources are ordered based on their outflow / disk mass ratio. Nebular masses ($M_{\rm neb}$) and velocities ($V_{\rm LSR}$) are derived from our single-dish and interferometric observations (Bujarrabal et al., 2013a, 2016; Gallardo Cava et al., 2021), except for HD 52961 (Gallardo Cava et al. in prep). Distances ($d$) are adopted from Bujarrabal et al. (2013a) except for HD 52961 (Oomen et al., 2019).

disk that have only been studied with low angular resolution, all of them presenting NIR excess and narrow CO line profiles characteristic of rotating disks (Bujarrabal et al., 2013a).

The chemistry of this class of binary post-AGB sources with Keplerian disks has been practically unknown, with only CO having been very well observed. Furthermore, $H^{13}CN$, C I, C II were also detected in the Red Rectangle and two maser emission lines were found in AI CMi. In this work, we present a deep and wide survey of radio lines in ten of these sources. All of them have been observed in the 7 and 13 mm bands, and most of them have also been observed at 1.3, 2, and 3 mm. The scarce literature about the molecular composition in very evolved objects shows that pPNe display dramatic changes in the chemical composition in the molecular gas with respect to that of their AGB circumstellar progenitors. Nevertheless, this kind of nebulae surrounding binary post-AGB stars seem to present a relatively low molecular content. Here we carry out an analysis to explore whether this lower abundance depends on the specific subclass of the binary post-AGB star (disk- or outflow-dominated sources). Additionally, we want to investigate the possibility of classifying some of our objects as O- / C-rich, based on line ratios of different molecules.

This paper is laid out as follows. We present our binary post-AGB star sample in Sect. 5.2, where relevant results from previous observations are given. Technical information on our observations is presented in Sect. 5.3 for both telescopes (30 m IRAM and 40 m Yebes). We provide our spectra in Sect. 5.4. Detailed discussions about molecular intensities, chemistry, abundances, and isotopic ratio analysis can be found in Sect. 5.5. Finally, we summarize our conclusions in Sect. 5.6.

## 5.2 Description of the sources and previous results

We performed single-dish observations of ten sources (Table 5.1) using the 30 m IRAM and the 40 m Yebes telescopes. These sources are identified as binary post-AGB stars with far-infrared (FIR) excess that is indicative of material ejected by the star. All of them also show significant NIR excess (see Fig. 5.18) characteristic of rotating disks (Oomen et al., 2018). They have been poorly studied in the overall search for molecules other than CO. We adopted the distances used in Bujarrabal et al. (2013a). We refrained from measuring distances via parallax measurements, because in the case of binary stars, it is overly complex (Dominik et al., 2003). The velocities are derived from the single-dish and interferometric observations (Bujarrabal et al., 2013a, 2015, 2016, 2017, 2018b; Gallardo Cava et al., 2021).

### 5.2.1 AC Her

AC Her is a binary post-AGB star (Oomen et al., 2019). Recent mm-wave interferometric observations confirm that AC Her presents a Keplerian disk that clearly dominates the nebula. Observational data and models are compatible with a very diffuse outflowing component surrounding the rotating component. The nebula presents a total mass of $8.3 \times 10^{-3}\,M_{\odot}$ and we find that the mass of the outflow must be $\lesssim 5\%$. The rotation of the disk is compatible with a central total stellar mass of $\sim 1\,M_{\odot}$ (Gallardo Cava et al., 2021). The molecular content of this source was unknown, except for the well-studied CO lines (Bujarrabal et al., 2013a; Gallardo Cava et al., 2021).

### 5.2.2 Red Rectangle

The Red Rectangle is the best-studied object of our sample. It is a binary post-AGB star (Oomen et al., 2019), with an accretion disk that emits in the UV (Witt et al., 2009; Thomas et al., 2013). This UV emission excite a central H II region (Jura et al., 1997; Bujarrabal et al., 2016) and must also yield a Photo Dissociation Region (or Photon-Dominated Regions, or PDR) between the H II region and the extended Keplerian disk.

Its mm-wave interferometric maps reveal that the nebula contains a rotating disk, containing 90% of the total nebular mass ($1.4 \times 10^{-2}\,M_{\odot}$, see Bujarrabal et al., 2016), together with an expanding low-mass component. The CO line profiles of the Red Rectangle are narrow and present weak wings (Bujarrabal et al., 2013a). Bujarrabal et al. (2016) find high excitation lines: $C^{17}O$ $J = 6 - 5$ and $H^{13}CN$ $J = 4 - 3$. The $C^{17}O$ line is useful to study regions closer than 60 AU from the central binary star. The presence of the $H^{13}CN$ line means that this molecule is significantly abundant in the central region of the disk, more precisely, closer than 60 AU. The detection of C I, C II, $H^{13}CN$, and FIR lines is consistent with the presence of a PDR (see e.g. Agúndez et al., 2008), because these lines are the best tracers of these regions.

### 5.2.3 89 Her

89 Her is a binary post-AGB star with NIR excess that implies the presence of hot dust (de Ruyter et al., 2006). The dust is in a stable structure where large dust grains form and settle to the midplane (Hillen et al., 2013, 2014). 89 Her was studied in detail in mm-wave interferometric maps by Gallardo Cava et al. (2021). The nebula contains an extended hourglass-like component and a rotating disk in its innermost region. The total nebular mass is $1.4 \times 10^{-2}\,M_{\odot}$ and the outflow mass represents, at least, 50%.

The molecular content of this source was very poorly studied, except for the single-dish CO observations, which show narrow lines similar to those of the Red Rectangle, but with more prominent wings (Bujarrabal et al., 2013a).

### 5.2.4 HD 52961

This source is a binary post-AGB star (Gielen et al., 2011b; Oomen et al., 2018). It presents relatively narrow CO line profiles. Its molecular content was poorly known. Gielen et al. (2011b) found $CO_2$ and the fullerene $C_{60}$. According to these authors, this source could be O-rich based on its strong similarities to an other post-AGB disk source (EP Lyr).

### 5.2.5 IRAS 19157−0257 and IRAS 18123+0511

IRAS 19157−0257 and IRAS 18123+0511 are also binary post-AGB stars (Oomen et al., 2019; Scicluna et al., 2020).

The variability of IRAS 19157−0257 is cataloged as irregular (Kiss et al., 2007). Its CO line profiles show relatively narrow lines that are similar to those of 89 Her. The nebular mass is $1.3 \times 10^{-2}\,M_{\odot}$ (Bujarrabal et al., 2013a). Its molecular content, apart from CO, was unknown.

IRAS 18123+0511 show wide CO line profiles (similar to those of IRAS 19125+0343). Its nebular mass is $4.7 \times 10^{-2}\,M_{\odot}$ (Bujarrabal et al., 2013a). No molecular species apart from CO have been detected in these sources despite numerous previous observations (Gómez et al., 1990; Lewis, 1997; Deguchi et al., 2012; Liu and Jiang, 2017). Our new data improve the rms values of other works.

### 5.2.6 IRAS 19125+0343

IRAS 19125+0343 is a binary post-AGB star (Gielen et al., 2008), which also belongs to this class of binary post-AGB star with remarkable NIR excess (Oomen et al., 2019). Recent mm-wave interferometric observations and models reveal that the nebula around this binary post-AGB star is composed of a rotating disk with Keplerian dynamics and an extended outflowing component around it (Gallardo Cava et al., 2021). The total mass of the nebula is $1.1 \times 10^{-2}\,M_{\odot}$ and the outflow mass represents 70%. The CO lines of this source are narrow but present prominent wings (Bujarrabal et al., 2013a). Apart from CO, the molecular content of this source was unknown.

### 5.2.7 AI CMi

AI CMi is an irregular pulsating star with variable amplitude and multiperiodicity that is in an early transition phase from the AGB to the post-AGB stage. The spectral type at maxima is G5 − G8 I (Arkhipova et al., 2017).

According to the analysis of the CO lines, AI CMi presents a nebula ($1.9 \times 10^{-2}\,M_{\odot}$) in which the expanding shell, with a velocity of $\sim 4\,\mathrm{km\,s^{-1}}$, would dominate the whole structure (Bujarrabal et al., 2013a; Arkhipova et al., 2017).

AI CMi shows a detached dust shell with $T \sim 200\,\mathrm{K}$ and it also presents TiO absorption bands that are formed in the upper cool layers of the extended atmosphere (Arkhipova et al., 2017). This source has been previously observed in radio lines. te Lintel Hekkert et al. (1991) discovered the OH maser at 1 612 MHz in a survey of IRAS sources. The $H_2O$ maser at 22 235.08 MHz was discovered by Engels and Lewis (1996). Suárez et al. (2007); Yoon et al. (2014) studied this source looking for SiO maser emission, but without success. In view of the above results, AI CMi most likely is an O-rich source (see also Arkhipova et al., 2017).

### 5.2.8 IRAS 20056+1834

This source has been studied in CO by Bujarrabal et al. (2013a) and the analysis yields a nebular mass of $10^{-1}$ $M_{\odot}$, of which $\sim$22% corresponds to the mass of the disk. Apart from CO, there is no other molecular line detection in this source.

### 5.2.9 R Sct

The bright RV Tauri variable R Sct shows very irregular pulsations with variable amplitude (Kalaee and Hasanzadeh, 2019) and a small IR excess (Kluska et al., 2019), indicating that the SED is not clearly linked to the presence of a circumbinary disk. Nevertheless, the CO integrated flux of the innermost region of the nebula around R Sct shows a characteristic double peak, suggesting the presence of rotation. Models for a disk with Keplerian dynamics and a very extended outflow are consistent with mm-wave interferometric observational data (see Gallardo Cava et al., 2021, for more details). The nebular mass is $3.2 \times 10^{-2}$ $M_{\odot}$, where $\sim$75% corresponds to the mass of the outflow. The hypothesis of a rotating disk is reinforced by interferometric data in the $H$-band showing a very compact ring (Kluska et al., 2019). The binarity of R Sct is still questioned, but as Keplerian disks are only detected around binaries in the case of evolved stars, it could well be binary as well.

In particular, R Sct presents composite CO line profiles including a narrow component, which very likely represents emission from the rotating disk (Bujarrabal et al., 2013a; Gallardo Cava et al., 2021). The chemistry of this source has been studied in IR by Matsuura et al. (2002) and Yamamura et al. (2003), who find that the IR spectra of R Sct is dominated by molecular emission features, especially from $H_2O$. These molecules are probably located in a spherical extended atmosphere. In addition, SiO, $CO_2$, and CO bands have been identified. Lebre and Gillet (1991) detected photospheric absorption Fe I lines, with the double absorption and emission Ti I profiles observed, as well as TiO.

## 5.3 Observations and data reduction

Our observations were performed using the 30 m IRAM telescope (Granada, Spain) and the 40 m Yebes telescope (Guadalajara, Spain). We observed at the 1.3, 2, 3, 7, and 13 mm bands (see Table 5.2). Our observations required a total telescope time of $\sim$600 hours distributed over two the telescopes and for several projects (see Table 5.2).

The data reduction was carried out with the software CLASS[2] within the GILDAS[3]

[2]Continuum and Line Analysis Single-dish Software (CLASS) is part of the GILDAS software package

[3]GILDAS is a software package focused in reducing and analysing mainly millimeter observations from single-dish and interferometric telescopes. It is developed and maintained by IRAM,

software package. For each source, we applied the standard procedure of data reduction that consists of rejecting bad scans, averaging the good ones, and subtracting a baseline of a first-order polynomial.

### 5.3.1 30 m IRAM telescope

We observed at the 1.3, 2, and 3 mm bands using the 30 m IRAM telescope. Our observations were performed in four different projects. Observations of 89 Her, R Sct, and the Red Rectangle were obtained between 13 and 19 February 2019 under project 179-18 for 80 hours. The data of project 055-19 were obtained between 29 June and 1 July 2019, when we observed the Red Rectangle for an additional 62 hours. IRAS 19125+0343 and IRAS 20056+1834 were observed between 9 September and 14 September 2020 for 42 hours within project 042-20. As a continuation of the last project, we also observed AI CMi and HD 52961 between 24 and 29 November 2020 for 45 hours. Finally, we observed AC Her, IRAS 19157−0257, and IRAS 18123+0511 between 21 and 27 July 2021, and 27 August 2021 under project E04-20 for 90 hours.

We connected the Fast Fourier Transform Spectrometer (FTS) units to the EMIR receiver with a resolution of 200 kHz per channel. The Half Power Beam Width (HPBW) of the 30 m IRAM telescope is 11″, 17″, and 28″ at 224, 145, and 86 GHz, respectively. The sources were observed using the wobbler-switching mode, which provides flat baselines. The subreflector was shifted every 2 s with a throw of ±120″ in the azimuth direction. We obtained spectra for the vertical and horizontal linear polarization receivers. Both polarizations were obtained simultaneously and we did not find significant differences in their relative calibration.

The calibration was derived using the chopper-wheel method observing the sky, and hot and cold loads. The procedure was repeated every 15 − 20 minutes, depending on the weather conditions. The observed peak emission values have been re-scaled (if applicable) comparing with the intensity of calibration lines observed in NGC 7027 and CW Leo (IRC+10216). The absolute scale accuracy is of the order of 10%, 10%, and 20% in the 3, 2, and 1.3 mm receives, respectively.

### 5.3.2 40 m Yebes telescope

We observed at the 7 and 13 mm bands using the 40 m Yebes telescope, $Q$- and $K$-band, respectively. We performed our observations in the course of two projects. We observed all the sources of our sample at 7 mm between 29 May and 7 June 2020 for 50 hours under project 20A009. In the same project, we also observed at 13 mm between 20 and 27 May 2020 for 15 hours. The data of project 20B006 were obtained in several epochs, between 6 and 13 June, 24 and 26 November 2020, and 14 and 17 January 2021. We observed AI CMi, IRAS 20056+1834, the Red Rectangle, 89 Her, AC Her, HD 52961 and IRAS 19125+0343 at 7 mm for 70 hours. We also observed IRAS 20056 and HD 52961 at 13 mm for 12 hours between 24 and 29 October 2020.

LAOG/Université de Grenoble, LAB/Observatoire de Bordeaux, and LERMA/Observatoire de Paris. For further details, see https://www.iram.fr/IRAMFR/GILDAS

Table 5.2 Observed frequency ranges

| Source | $\lambda = 1.3\,$mm | $\lambda = 2\,$mm | $\lambda = 3\,$mm | $\lambda = 7\,$mm | $\lambda = 13\,$mm |
|---|---|---|---|---|---|
| AC Herculis | ✓ | ✓ | ✓ | ✓ | ✓ |
| Red Rectangle | ✓ | ✓ | ✓ | ✓ | ✓ |
| 89 Herculis | ✓ | ✓ | ✓ | ✓ | ✓ |
| HD 52961 | ✓ | ✓ | ✓ | ✓ | ✓ |
| IRAS 19157−0257 | | ✓ | ✓ | ✓ | ✓ |
| IRAS 18123+0511 | ✓ | ✓ | ✓ | ✓ | ✓ |
| IRAS 19125+0343 | ✓ | | ✓ | ✓ | ✓ |
| AI Canis Minoris | ✓ | ✓ | ✓ | ✓ | ✓ |
| IRAS 20056+1834 | ✓ | | ✓ | ✓ | ✓ |
| R Scuti | ✓ | ✓ | ✓ | ✓ | ✓ |

Table 5.3 Molecular transitions detected in this work.

| Molecule | Transition | | $\nu$ |
|---|---|---|---|
| | Vibrational | Rotational | [MHz] |
| $C^{17}O$ | $v = 0$ | $J = 2-1$ | 224713.53 |
| $C^{18}O$ | $v = 0$ | $J = 2-1$ | 219560.36 |
| $^{28}SiO$ | $v = 0$ | $J = 1-0$ | 43423.85 |
| | | $J = 2-1$ | 86846.99 |
| | | $J = 5-4$ | 217104.98 |
| | $v = 1$ | $J = 1-0$ | 43122.08 |
| | | $J = 2-1$ | 86243.37 |
| | $v = 2$ | $J = 1-0$ | 42820.59 |
| | $v = 6$ | $J = 1-0$ | 41617.40 |
| $^{29}SiO$ | $v = 0$ | $J = 1-0$ | 42879.95 |
| | | $J = 2-1$ | 85759.19 |
| $^{30}SiO$ | $v = 0$ | $J = 1-0$ | 42373.34 |
| | | $J = 2-1$ | 84746.17 |
| | $v = 1$ | $J = 1-0$ | 42082.47 |
| HCN | $v = 0$ | $J = 1-0$ | 88630.42 |
| $HCO^+$ | $v = 0$ | $J = 1-0$ | 89188.52 |
| CS | $v = 0$ | $J = 3-2$ | 146969.00 |
| SiS | $v = 0$ | $J = 5-4$ | 90771.56 |
| | $v = 1$ | $J = 8-7$ | 144520.36 |
| SO | $v = 0$ | $J_N = 2_2 - 1_1$ | 86093.95 |
| | | $J_N = 6_5 - 5_4$ | 219949.44 |
| $SO_2$ | $v = 0$ | $J_{Ka,\,Kc} = 4_{2,\,2} - 4_{1,\,3}$ | 146605.52 |
| $H_2O$ | $v = 0$ | $J_{Ka,\,Kc} = 6_{1,\,6} - 5_{2,\,3}$ | 22235.08 |

**Notes.** Rest frequencies are taken from Cologne Database for Molecular Spectroscopy (CDMS) and Jet Propulsion Laboratory (JPL).

The received signal was detected using the Fast Fourier Transform Spectrometer (FFTS) backend units. The bandwidth of the $Q$-band is 18 GHz, the spectral resolution is 38 kHz, and the HPBW is $37-49''$. In this band, our sources were observed using the position-switching method, always using a separation of around $300''$ in the azimuthal direction, obtaining flat baselines. We obtained spectra for the vertical and horizontal linear polarization simultaneously and we do not find significantly differences in their flux density calibration; see Tercero et al. (2021) for technical details. The bandwidth of the $K$-band is 100 MHz, the spectral resolution is 6.1 kHz, and the HPBW is $79''$. Our sources were observed using the position-switching method, using separation of about $600''$ in the azimuthal direction, obtaining flat baselines. The observations at 13 mm were carried out with dual circular polarization with small calibrations differences of $\leq 20\%$. Pointing and focus were checked in both $Q$- and $K$-bands every hour through pseudo-continuum observations of the SiO $v = 1$ $J = 1-0$ and $H_2O$ maser emission, respectively, towards evolved stars close to our sources that show intense masers; see de Vicente et al. (2016). The pointing errors were always within $5''-7''$.

Table 5.4 Detected and tentatively detected lines in this work.

| Source | Molecule | | Transition | I (peak) [Jy] | $\sigma$ [Jy] | $\int I\,dV$ [Jy km s$^{-1}$] | $\sigma\left(\int I\,dV\right)$ [Jy km s$^{-1}$] | Sp. Res. [km s$^{-1}$] | $V_{\rm LSR}$ [km s$^{-1}$] | Comments |
|---|---|---|---|---|---|---|---|---|---|---|
| AC Her | SiO | $v=1$ | $J=2-1$ | 1.7E-02 | 5.1E-03 | 2.1E-02 | 9.1E-03 | 0.68 | 0.41 | Tentative |
| Red Rectangle | $C^{18}O$ | $v=0$ | $J=2-1$ | 1.7E-01 | 2.4E-02 | 9.5E-01 | 6.5E-02 | 0.53 | $-0.64$ | |
| | $C^{17}O$ | $v=0$ | $J=2-1$ | 1.8E-01 | 2.4E-02 | 7.9E-01 | 4.8E-02 | 0.52 | $-0.59$ | |
| | HCN | $v=0$ | $J=1-0$ | 1.3E-02 | 3.9E-03 | 8.6E-02 | 2.8E-02 | 2.64 | 7.86 | Tentative |
| | SO | $v=0$ | $J_N=6_5-5_4$ | 3.5E-02 | 6.7E-03 | 3.3E-01 | 6.6E-02 | 4.26 | $-1.15$ | |
| | $H_2O$ | $v=0$ | $J_{Ka,\,Kc}=6_{1,\,6}-5_{2,\,3}$ | 5.3E-02 | 1.1E-02 | 6.2E-01 | 9.3E-02 | 3.29 | 8.27 | Tentative |
| 89 Her | $C^{18}O$ | $v=0$ | $J=2-1$ | 1.2E-01 | 1.8E-02 | 2.0E-01 | 2.0E-02 | 0.27 | -8.36 | |
| | $C^{17}O$ | $v=0$ | $J=2-1$ | 1.3E-02 | 5.1E-03 | 6.1E-02 | 5.2E-02 | 4.17 | -11.10 | Tentative |
| | HCN | $v=0$ | $J=1-0$ | 1.2E-02 | 2.8E-03 | 2.1E-01 | 2.5E-02 | 2.64 | -14.76 | Tentative |
| | SiS | $v=0$ | $J=5-4$ | 1.5E-02 | 3.8E-03 | 3.3E-03 | 6.7E-03 | 0.64 | -11.57 | |
| | CS | $v=0$ | $J=3-2$ | 4.9E-02 | 8.1E-03 | 9.7E-02 | 1.4E-02 | 0.40 | -7.42 | |
| AI CMi | SiO | $v=0$ | $J=1-0$ | 2.0E-02 | 5.4E-03 | 7.5E-02 | 2.0E-02 | 1.05 | 29.11 | |
| | | | $J=2-1$ | 1.1E-01 | 1.8E-02 | 8.8E-01 | 7.0E-02 | 0.67 | 29.02 | |
| | | | $J=5-4$ | 2.9E-01 | 3.7E-02 | 2.0E+00 | 1.0E-01 | 0.54 | 29.00 | |
| | | $v=1$ | $J=1-0$ | 1.4E+00 | 9.2E-03 | 2.9E+00 | 1.8E-02 | 0.27 | 28.87 | |
| | | | $J=2-1$ | 1.9E+00 | 1.7E-02 | 4.7E+00 | 4.8E-02 | 0.68 | 28.81 | |
| | | $v=2$ | $J=1-0$ | 4.3E-01 | 8.9E-03 | 6.4E-01 | 1.6E-02 | 0.27 | 29.85 | |
| | $^{29}SiO$ | $v=0$ | $J=1-0$ | 7.7E-03 | 3.9E-03 | 4.9E-02 | 1.5E-02 | 2.13 | 35.04 | Tentative |
| | | | $J=2-1$ | 2.6E-02 | 7.5E-03 | 2.5E-01 | 5.7E-02 | 2.73 | 28.52 | Tentative |
| | $^{30}SiO$ | $v=0$ | $J=1-0$ | 1.2E-02 | 3.2E-03 | 8.4E-02 | 1.7E-02 | 2.16 | 29.03 | Tentative |
| | SiS | $v=1$ | $J=8-7$ | 1.9E-01 | 4.4E-02 | 1.9E-01 | 5.4E-02 | 0.41 | 32.36 | Maser? |
| | SO | $v=0$ | $J_N=2_2-1_1$ | 3.8E-02 | 9.0E-03 | 3.1E-01 | 4.1E-02 | 1.36 | 28.86 | |
| | | | $J_N=6_5-5_4$ | 5.0E-01 | 3.5E-02 | 4.3E+00 | 1.1E-01 | 0.53 | 28.88 | |
| | $SO_2$ | $v=0$ | $J_{Ka,\,Kc}=4_{2,\,2}-4_{1,\,3}$ | 1.8E-01 | 2.9E-02 | 1.1E+00 | 9.0E-02 | 0.80 | 28.68 | |
| | $H_2O$ | $v=0$ | $J_{Ka,\,Kc}=6_{1,\,6}-5_{2,\,3}$ | 1.6E+00 | 5.4E-02 | 9.8E-01 | 5.8E-02 | 0.16 | 27.18 | |
| IRAS 20056+1834 | SiO | $v=1$ | $J=1-0$ | 1.5E-02 | 3.2E-03 | 4.0E-02 | 6.7E-03 | 0.53 | -22.40 | |
| | | $v=2$ | $J=1-0$ | 2.9E-02 | 3.9E-03 | 1.1E-01 | 1.1E-02 | 0.53 | -20.56 | |
| R Sct | SiO | $v=0$ | $J=1-0$ | 7.9E-02 | 8.8E-03 | 0.29 | 2.1E-02 | 0.53 | 56.02 | |
| | | | $J=2-1$ | 3.5E-01 | 6.8E-03 | 1.6E+00 | 2.8E-02 | 0.67 | 57.02 | |
| | | | $J=5-4$ | 1.3E+00 | 5.6E-02 | 5.5E+00 | 1.4E-01 | 0.27 | 57.50 | |
| | | $v=1$ | $J=1-0$ | 9.8E-02 | 7.4E-03 | 0.21 | 1.7E-02 | 0.53 | 54.32 | |
| | | | $J=2-1$ | 5.4E-01 | 7.3E-03 | 1.0E+00 | 2.2E-02 | 0.68 | 53.50 | |
| | | $v=2$ | $J=1-0$ | 2.9E-01 | 4.4E-02 | 0.64 | 7.3E-02 | 0.53 | 55.32 | |
| | | $v=6$ | $J=1-0$ | 2.3E-02 | 6.7E-03 | 8.2E-02 | 1.6E-02 | 1.10 | 52.88 | First $v=6$ detection |
| | $^{29}SiO$ | $v=0$ | $J=1-0$ | 2.9E-02 | 7.8E-03 | 2.6E-02 | 1.1E-02 | 0.53 | 56.31 | Tentative |
| | | | $J=2-1$ | 8.0E-02 | 6.3E-03 | 3.5E-01 | 2.0E-02 | 0.68 | 57.98 | |
| | $^{30}SiO$ | $v=0$ | $J=1-0$ | 2.6E-02 | 7.7E-03 | 2.6E-02 | 1.2E-02 | 0.54 | 54.58 | Tentative |
| | | | $J=2-1$ | 6.1E-02 | 6.5E-03 | 2.7E-01 | 1.7E-02 | 0.69 | 57.35 | |
| | $HCO^+$ | $v=0$ | $J=1-0$ | 4.2E-02 | 5.3E-03 | 1.8E-01 | 1.3E-02 | 0.66 | 56.68 | |
| | SO | $v=0$ | $J_N=6_5-5_4$ | 3.0E-02 | 8.3E-03 | 1.9E-01 | 4.4E-02 | 2.13 | 56.94 | |
| | $H_2O$ | $v=0$ | $J_{Ka,\,Kc}=6_{1,\,6}-5_{2,\,3}$ | 1.7E+00 | 3.0E-02 | 2.0E+00 | 4.7E-02 | 0.41 | 56.75 | |

**Notes.** For complete tables including upper limits, see Sect. 5.7.3.

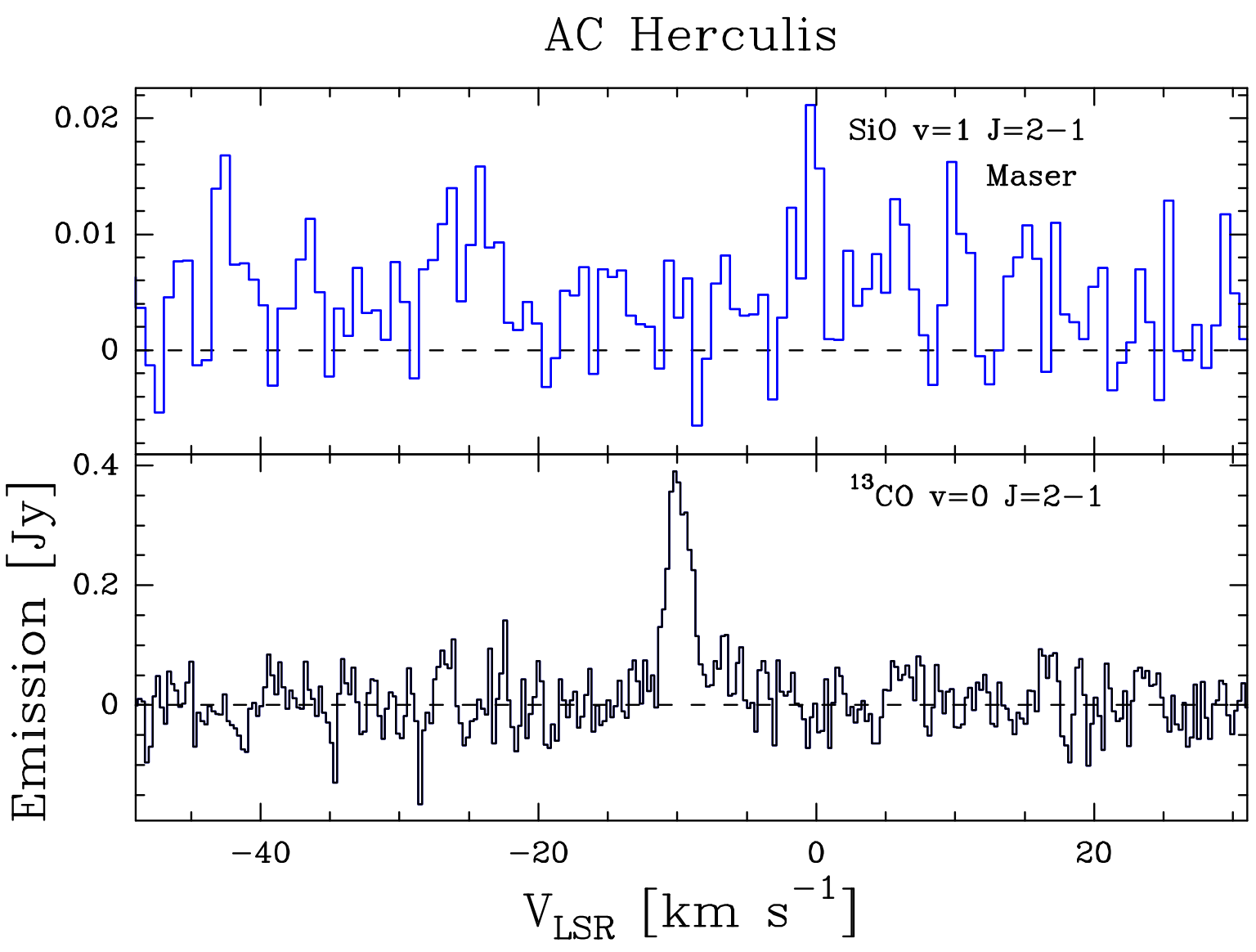

Figure 5.1: Spectra of the newly tentative maser detection in AC Her. For comparison purposes, we also show the $^{13}$CO $J = 2 - 1$ line in black (Bujarrabal et al., 2013a). The x-axis indicates velocity with respect to the local standard of rest ($V_{\rm LSR}$) and the y-axis represents the detected flux measured in Jansky.

## 5.4 Observational results

We compiled a list of the detected molecular transitions, together with upper limits for undetected lines, summarized in Sect. 5.7.3. Additionally, we summarized the main results from detected and tentatively detected lines in Table 5.4. The tables list, for the main transitions, the peak intensity flux (I [Jy]), its associated noise ($\sigma$ [Jy]), the integrated line intensity ($\int I\,{\rm d}V$ [Jy km s$^{-1}$]), its associated uncertainty ($\sigma\left(\int I\,{\rm d}V\right)$ [Jy km s$^{-1}$]), the spectral resolution (Sp. Res. [km s$^{-1}$]), and the velocity centroid of the line measured with respect to the local standard of rest ($V_{\rm LSR}$ [km s$^{-1}$]).

All our sources were previously detected in CO lines (Bujarrabal et al., 2013a). We present the first detection of other mm-wave lines (Table 5.3) in these sources and this work is one of the few that systematically surveyed pPNe in the search for molecules. Our detected lines at 1.3, 2, 3, 7, and 13 mm (Table 5.4) provide new data for the ten binary post-AGB stars of our survey (Table 4.1). We detected new lines in the Red Rectangle, 89 Her, AI CMi, IRAS 20056+1834, and R Sct. We also present significant upper limits for a large number of lines (Sect. 5.7.3), which often include valuable information.

### 5.4.1 AC Her

We observed this source at 1.3, 2, 7, and 13 mm, resulting in no positive detections, attaining an rms of 15, 9, 7, and 36 mJy, respectively (Table 5.9). We also observed at 3 mm, where we attained a rms value of 4 mJy, and we tentatively detected the $v = 1$ $J = 2 - 1$ SiO maser (Fig. 5.1). This tentative detection is centered at $\sim$0 km s$^{-1}$, which is shifted respect to the CO line profiles (Gallardo Cava et al., 2021). We assume that the SiO maser emission would have come from the innermost region of the disk, which shows the highest velocity shifts, explaining the observed velocity.

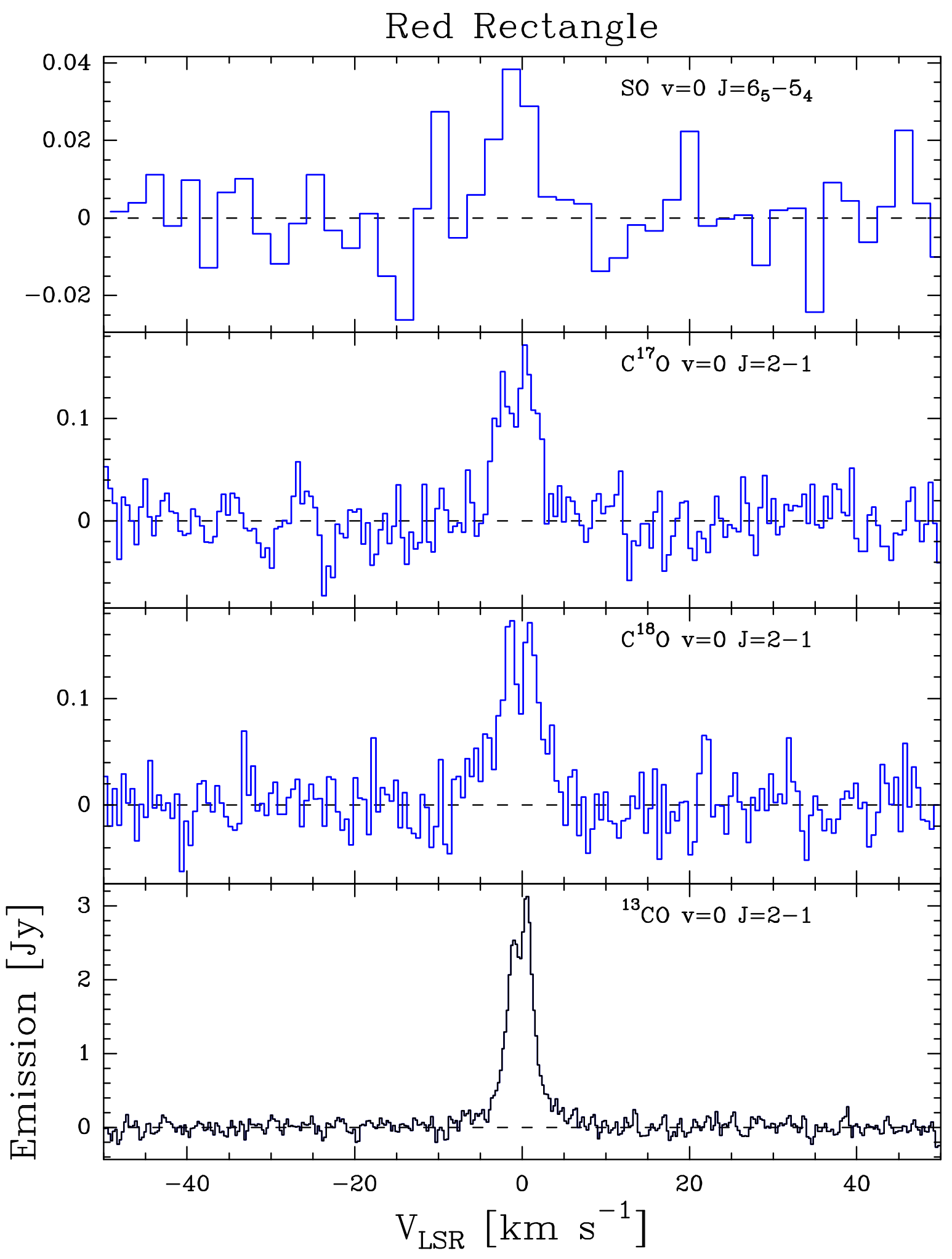

Figure 5.2: Spectra of the newly detected transitions in the Red Rectangle. For comparison purposes, we also show the $^{13}$CO $J = 2 - 1$ line in black (Bujarrabal et al., 2013a). The x-axis indicates velocity with respect to the local standard of rest ($V_{\rm LSR}$) and the y-axis represents the detected flux measured in Jansky.

### 5.4.2 Red Rectangle

The Red Rectangle is the best-studied object of our sample (Bujarrabal et al., 2013a,b, 2016). We present our results in Tables 5.10 and 5.4, and in Figs. 5.2 and 5.3 (and 5.19). We have detected the rarest CO isotopic species, namely C$^{17}$O and C$^{18}$O $J = 2 - 1$ . The values for the line peak and integrated intensity of the two lines are similar. We have tentatively detected HCN $J = 1 - 0$ with signal-to-noise ratio (S/N > 3). This detection is consistent with the presence of H$^{13}$CN $J = 4 - 3$ (Bujarrabal et al., 2016). The presence of these molecules, together with the C I, C II lines, and PAHs, could be due to the development of a PDR in dense disk regions close to the central stellar binary system. The origin of HCN could be photoinduced chemistry, because it has been detected in the innermost regions of several disks in young stars. This fact could be also present in relatively evolved PNe with UV excess (Bublitz et al., 2019), and is predicted by theoretical modelling of the PDR chemistry (e.g., Agúndez et al., 2008).

The line profile of these transitions shows the double peak characteristic of rotating disks (Guilloteau and Dutrey, 1998; Bujarrabal et al., 2013a; Guilloteau et al., 2013).

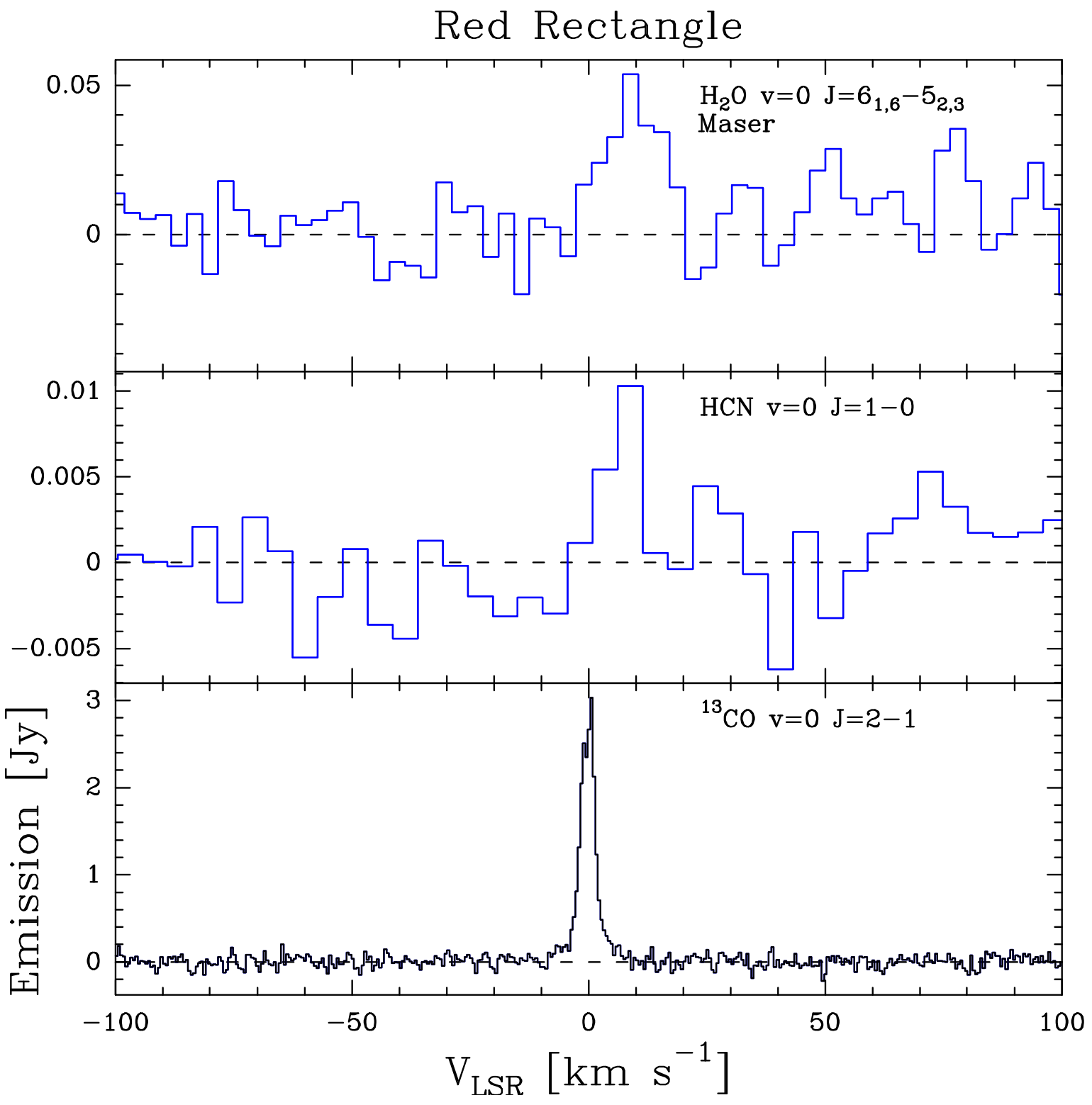

Figure 5.3: Spectra of the newly tentative detections in the Red Rectangle. For comparison purposes, we also show the $^{13}$CO $J = 2 - 1$ line in black (Bujarrabal et al., 2013a). The x-axis indicates velocity with respect to the local standard of rest ($V_{\rm LSR}$) and the y-axis represents the detected flux measured in Jansky.

It implies that the emission of these molecules comes from the innermost region of the rotating component of the nebula.

In addition, we detected SO $J_N = 6_6 - 5_4$. The line profile of the SO line is narrow, as the CO lines, which suggests that oxygen rich material could be located in the Keplerian disk. We also tentatively detected emission of $H_2O$ $J_{Ka,Kc} = 6_{1,6} - 5_{2,3}$. The centroid of this line is 8.3 km s$^{-1}$, and this shift in velocity could have its origin in maser emission. We have confirmed the detection of $H_2O$ maser emission at 325 GHz in our new ALMA maps (private communication). We also improved the rms at 90 GHz given by Liu and Jiang (2017) by a factor of 13 (see Table 5.10).

### 5.4.3 89 Her

We present our results in Tables 5.11 and 5.4, and in Figs. 5.4 and 5.5. We detected C$^{18}$O $J = 2 - 1$ and tentatively C$^{17}$O $J = 2 - 1$ . We also detected CS $J = 3 - 2$ and SiS $J = 5 - 4$. All these lines show narrow profiles. Additionally, we tentatively detected HCN $J = 1 - 0$ . Based on the shape of the observed profiles, we are confident that the emission of these molecules comes from the Keplerian disk and not from the extended outflow.

### 5.4.4 HD 52961

We did not detect any emission at the observed frequencies, but we highlight that we present significant upper limits (see Table 5.12). We attained a rms of 40, 45, and 18 mJy at 1.3, 2, and 3 mm, respectively. The observation at 7 and 13 mm was focused in the detection of SiO and $H_2O$ maser emission and we attained a rms of 6 and 31.5 mJy, respectively.

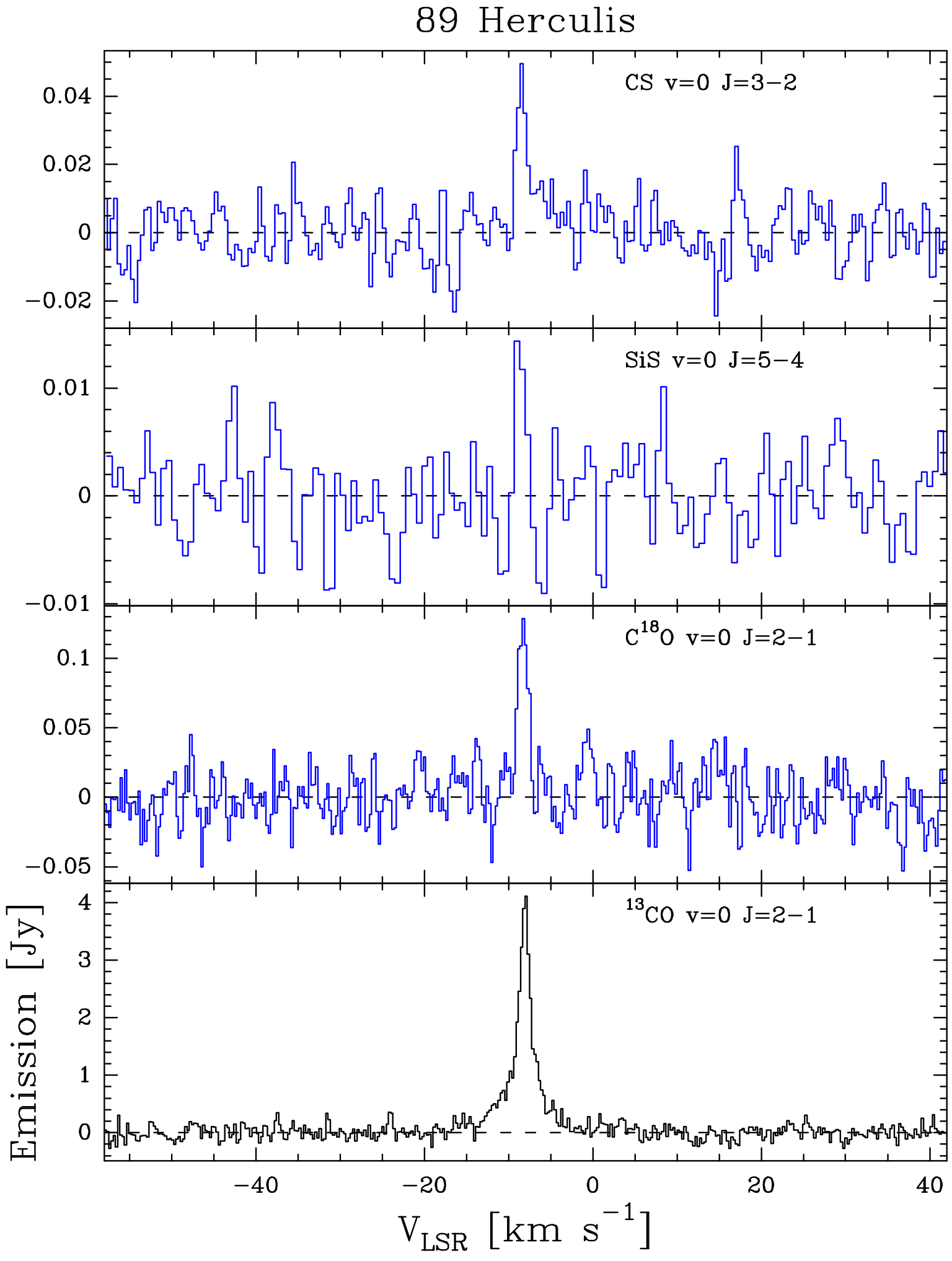

Figure 5.4: Spectra of the newly detected transitions in 89 Her. For comparison purposes, we also show the $^{13}$CO $J = 2-1$ line in black (Bujarrabal et al., 2013a). The x-axis indicates velocity with respect to the local standard of rest ($V_{\rm LSR}$) and the y-axis represents the detected flux measured in Jansky.

### 5.4.5 IRAS 19157−0257 and IRAS 18123+0511

We focus our mm-wave single-dish observations in the detection of SiO and $H_2O$ maser emission at 7 and 13 mm. We have not confirmed the lines in IRAS 19157−0257 and IRAS 18123+0511, but we provide upper limits (we present our results in Tables 5.13 and 5.14). Some of them clearly improve the rms achieved by previous works by other authors.

We attained an rms in IRAS 19157−0257 of 50, 40, 14 and 34 mJy at 2, 3, 7, and 13 mm, respectively. We attain an rms in IRAS 18123+0511 of 90, 55, 30, 17 and 24 mJy at 1.3, 2, 3, 7, and 13 mm, respectively.

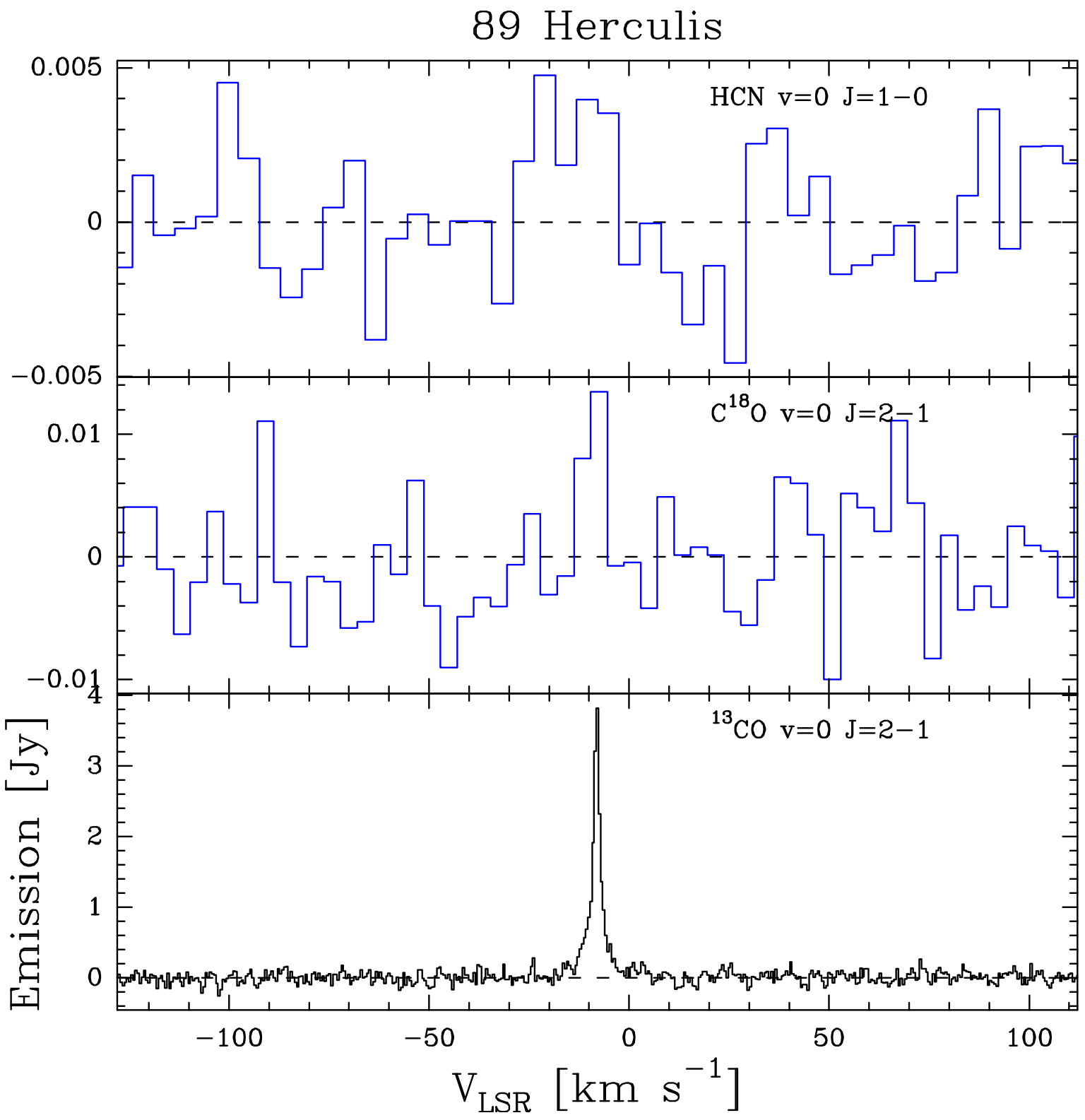

Figure 5.5: Spectra of the newly tentative detections in 89 Her. For comparison purposes, we also show the $^{13}$CO $J = 2-1$ line in black (Bujarrabal et al., 2013a). The x-axis indicates velocity with respect to the local standard of rest ($V_{\rm LSR}$) and the y-axis represents the detected flux measured in Jansky.

### 5.4.6 IRAS 19125+0343

We did not detect any emission in IRAS 19125+0343 at the observed frequencies, but we highlight that we present significant upper limits (see Table 5.15). We have attained a rms of 50, 20, 10, and 23 mJy at 1.3, 3, 7, and 13 mm, respectively.

### 5.4.7 AI CMi

We present our results in Tables 5.16 and 5.4, and in Figs. 5.8, 5.6, and 5.7. We detected several lines of thermal SiO: $J = 1-0$ , $J = 2-1$ , and $J = 5-4$. We also detected SiO isotopic species, $^{29}$SiO $J = 1-0$ and $J = 2-1$ and $^{30}$SiO $J = 1-0$ . We find very intense SiO maser emission for transitions $v = 1$ $J = 1-0$ and $J = 2-1$ and $v = 2$ $J = 1-0$ . We find wide composite profiles of SO $J_N = 2_2 - 1_1$ and $J_N = 6_5 - 5_4$ and $SO_2$ $J_{Ka,\,Kc} = 4_{2,\,2} - 4_{1,\,3}$. In addition, we detected $H_2O$ $J_{Ka,\,Kc} = 6_{1,\,6} - 5_{2,\,3}$ (this line was previously detected, see Sect. 5.2.7). We detected SiS $v = 1$ $J = 8-7$, which, if confirmed, will also be the first detection of this species in AI CMi. However, we note that we did not detect any SiS $v = 0$ in the observed bands, which is quite surprising but not impossible (SiS $v = 1$ emission could be due to some weak maser emission).

All the detected thermal lines show composite profiles with a narrow component, in a similar way to what we find in the CO line profiles of this source and in R Sct. We think that both sources could be very similar.

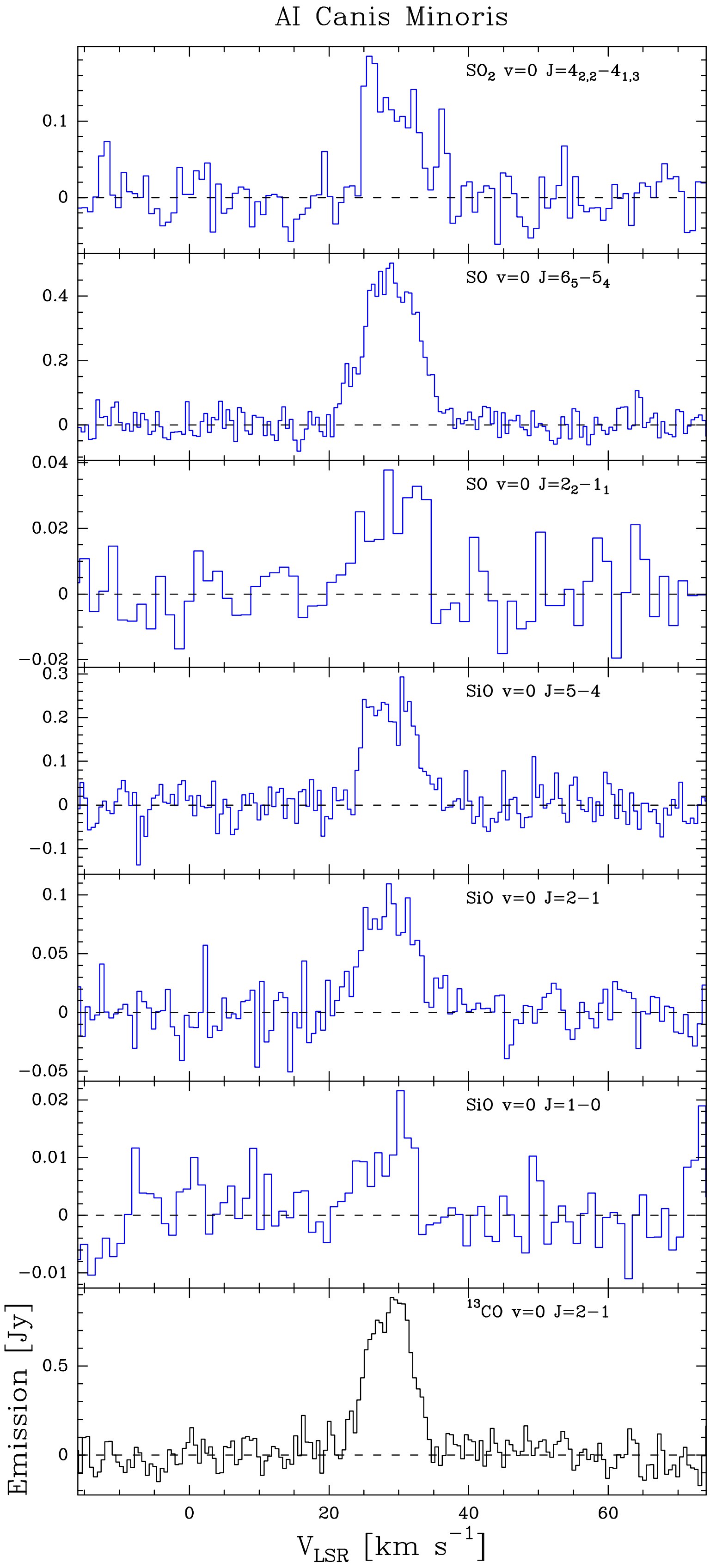

Figure 5.6: Spectra of the newly detected transitions in AI CMi. For comparison purposes, we also show the $^{13}$CO $J = 2 - 1$ line in black (Bujarrabal et al., 2013a). The x-axis indicates velocity with respect to the local standard of rest ($V_{\rm LSR}$) and the y-axis represents the detected flux measured in Jansky.

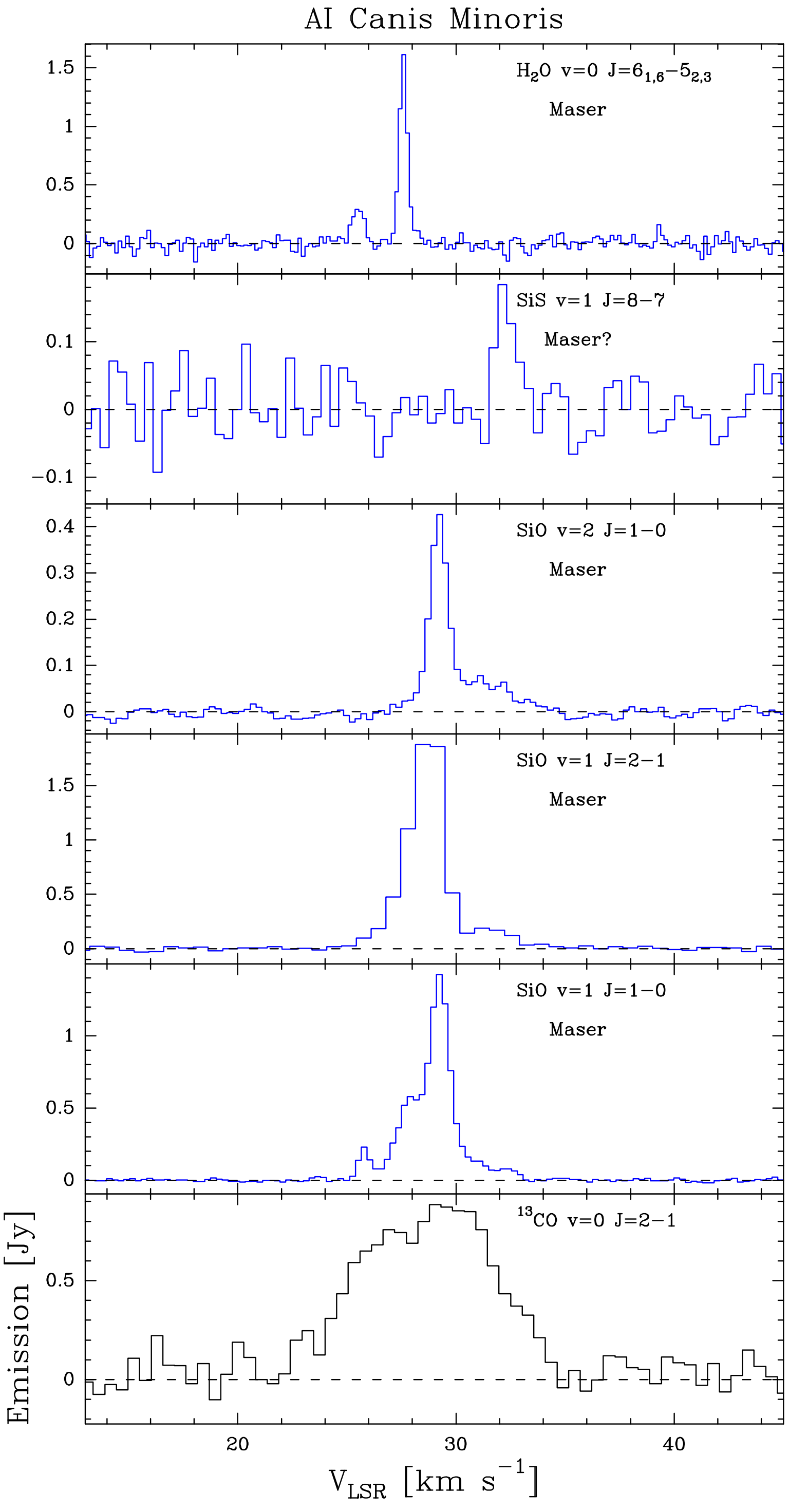

Figure 5.7: Spectra of the newly maser detections in AI CMi. For comparison purposes, we also show the $^{13}$CO $J = 2 - 1$ line in black (Bujarrabal et al., 2013a). The x-axis indicates velocity with respect to the local standard of rest ($V_{\rm LSR}$) and the y-axis represents the detected flux measured in Jansky.

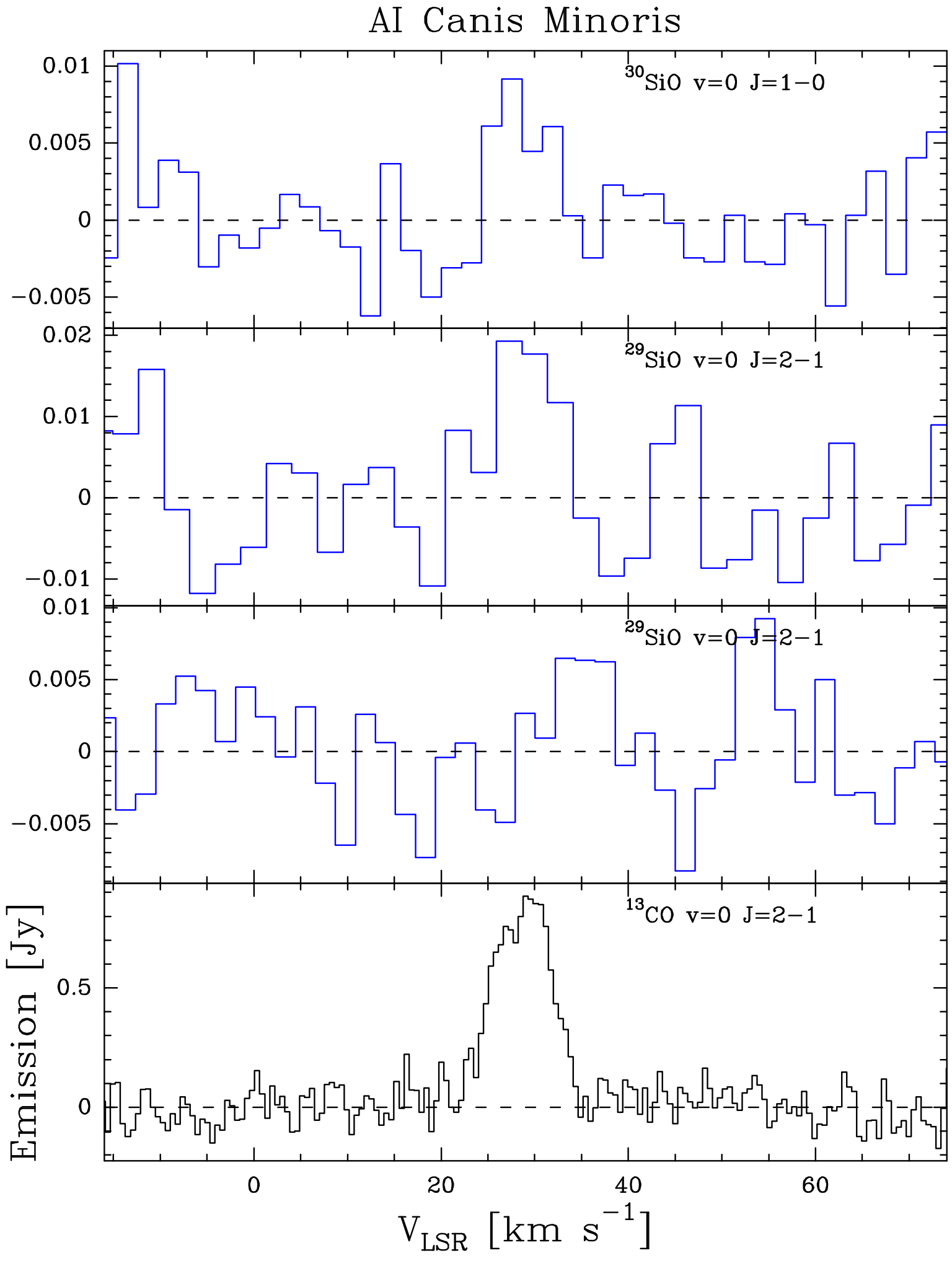

Figure 5.8: Spectra of the newly tentative detections in AI CMi. For comparison purposes, we also show the $^{13}$CO $J = 2 - 1$ line in black (Bujarrabal et al., 2013a). The x-axis indicates velocity with respect to the local standard of rest ($V_{\rm LSR}$) and the y-axis represents the detected flux measured in Jansky.

### 5.4.8 IRAS 20056+1834

We observed this source in 1.3 and 3 mm, resulting in no positive detections and we attained rms values of 45 and 18 mJy, respectively. We also observed this source at 7 mm, where we detected maser emission, SiO $v = 1\, J = 1 - 0$ and $v = 2\, J = 1 - 0$ (see Fig. 5.9). There are no signs of $H_2O$ maser emission at 13 mm, but we provide a rms of 27 mJy (see Table 5.17 for further details).

The SiO maser emission is centered at $\sim -20$ km s$^{-1}$, which is shifted with respect to the CO line profiles by $\sim$10 km s$^{-1}$. This fact is supported by Klochkova et al. (2007), who highlight that velocities measured from lines formed in the photosphere are variable. They reveal differential line shifts up 10 km s$^{-1}$. Shifts between thermal and maser lines are in any case not usual.

We think that IRAS 20056+1834 could be similar to R Sct and AI CMi (a binary system surrounded by a Keplerian disk and an extended outflow that dominates the whole nebula), which would explain the chemical similarities between these sources.

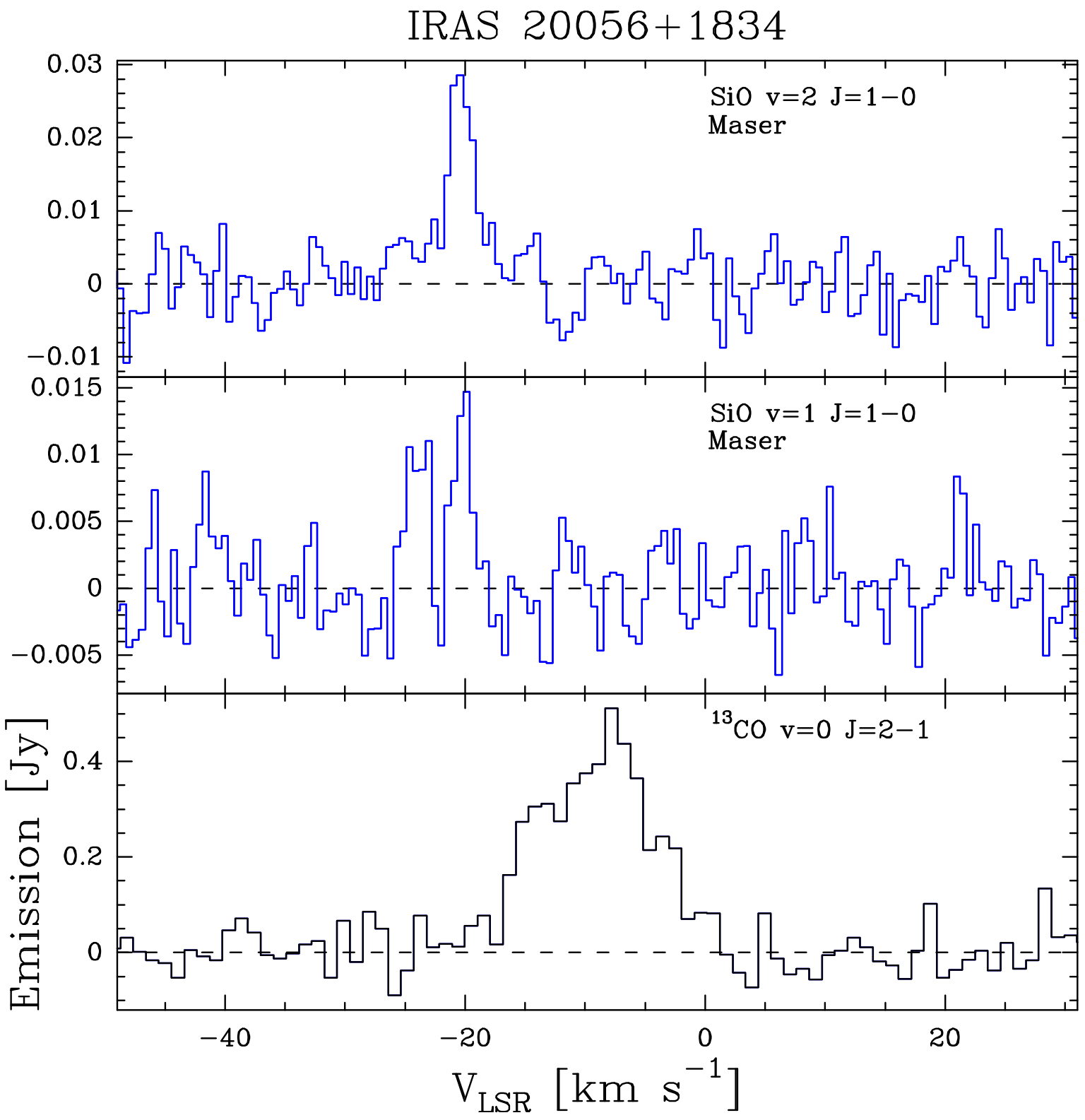

Figure 5.9: Spectra of the newly maser detections in IRAS 20056+1834. For comparison purposes, we also show the $^{13}$CO $J = 2-1$ line in black (Bujarrabal et al., 2013a). The x-axis indicates velocity with respect to the local standard of rest ($V_{\rm LSR}$) and the y-axis represents the detected flux measured in Jansky.

### 5.4.9 R Sct

We present our results in Tables 5.18 and 5.4, and in Figs. 5.12, 5.11, and 5.10. We detected several lines of thermal SiO: $J = 1-0$ , $J = 2-1$ , and $J = 5-4$. We also detected SiO isotopic species, $^{29}$SiO and $^{30}$SiO $J = 1-0$ and $J = 2-1$ . All of them show a narrow profile, which is coincident with the narrow component of the composited CO line profiles. We also detected SiO maser emission for transitions $v = 1$ $J = 1-0$ and $J = 2-1$ , $v = 2$ $J = 1-0$ , and the first-ever detection of the $v = 6$ $J = 1-0$. We detected SO $J_N = 6_5 - 5_4$, and the $H_2O$ $J_{Ka,Kc} = 6_{1,6} - 5_{2,3}$ maser emission (see Sect. 5.2.9).

In addition, we also detected $HCO^+$ $J = 1-0$ . This molecule presents weak emission in AGB stars when detected (Bujarrabal et al., 1994b; Pulliam et al., 2011). Furthermore, $HCO^+$ is detected in other standard pPNe, such as CRL 618 (Sánchez Contreras and Sahai, 2004), OH 231.8+4.2 (Sánchez Contreras et al., 2015), or M1−92 (Alcolea et al., 2019). The detection of $HCO^+$ in the envelope of R Sct implies that the disk is (partially) ionized. The most probable reason for the presence of $HCO^+$ is that the central star, which is now in the post-AGB phase, can partially ionize the closest and densest regions. These regions correspond to the disk, which would justify the narrow profile of this line (see Fig. 5.11). Another way to explain the presence of ionized gas could be the presence of shocks, in which the abundance of SO is also favored.

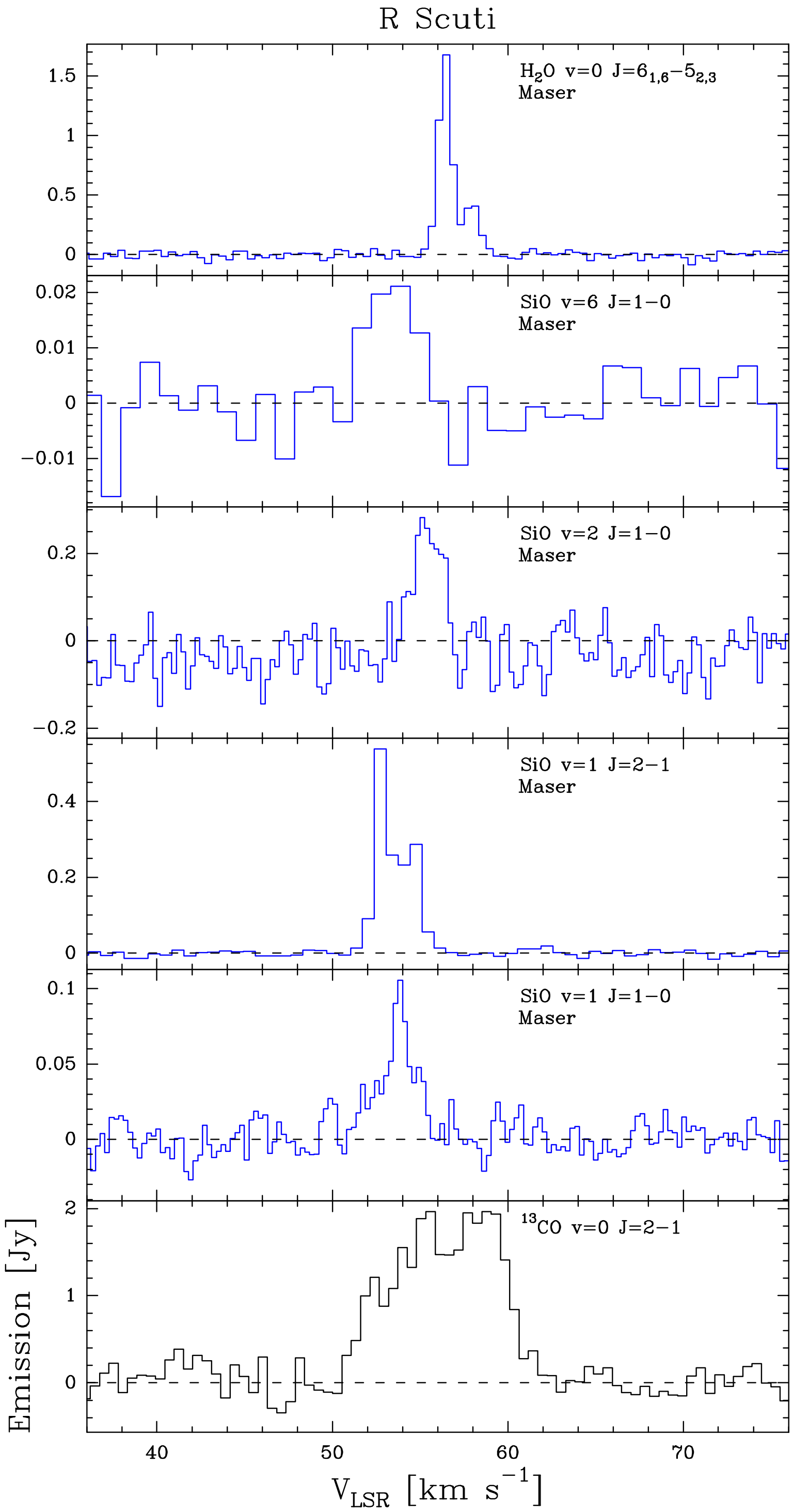

Figure 5.10: Spectra of the newly maser detections in R Sct. For comparison purposes, we also show the $^{13}$CO $J = 2-1$ line in black (Bujarrabal et al., 2013a). The x-axis indicates velocity with respect to the local standard of rest ($V_{\rm LSR}$) and the y-axis represents the detected flux measured in Jansky.

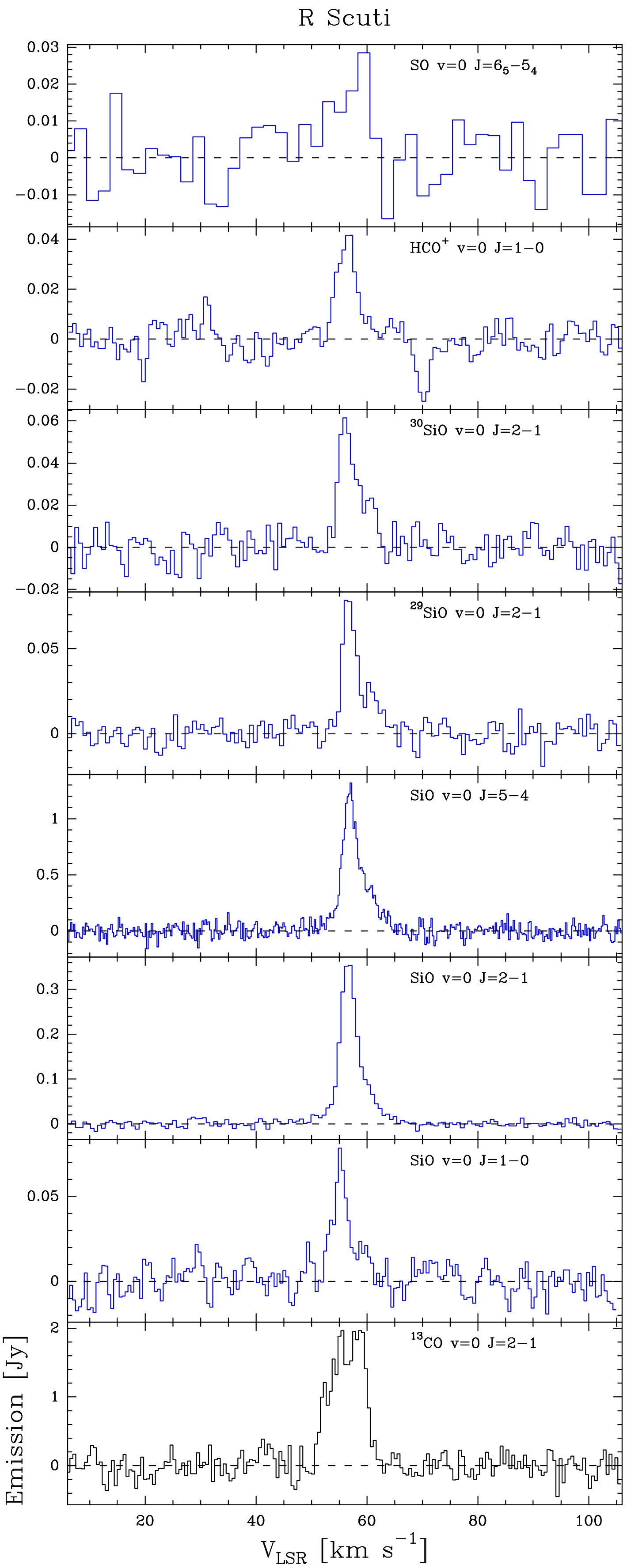

Figure 5.11: Spectra of the newly detected transitions in R Sct. For comparison purposes, we also show the $^{13}$CO $J = 2 - 1$ line in black (Bujarrabal et al., 2013a). The x-axis indicates velocity with respect to the local standard of rest ($V_{\rm LSR}$) and the y-axis represents the detected flux measured in Jansky.

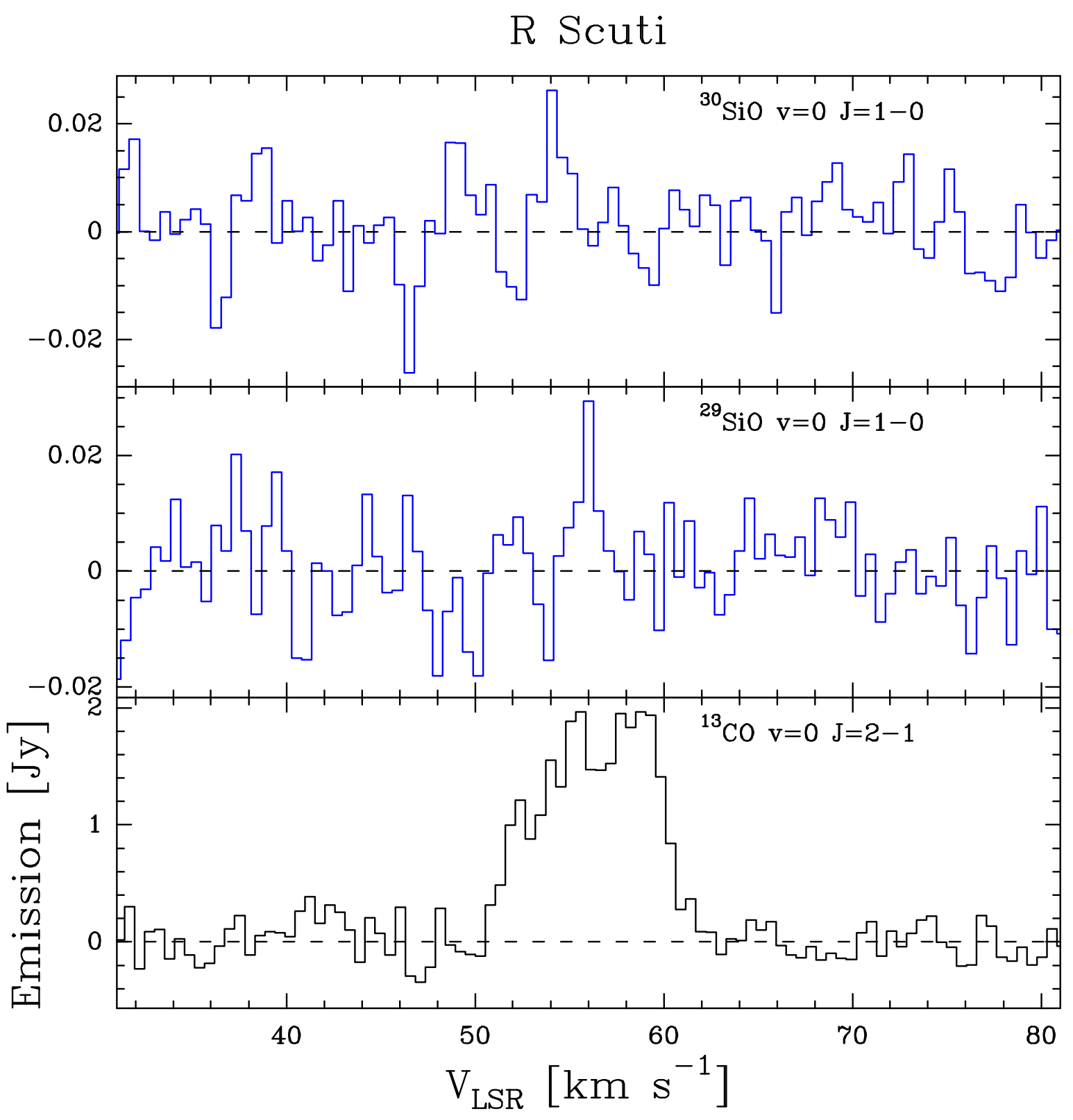

Figure 5.12: Spectra of the newly tentative detections in R Sct. For comparison purposes, we also show the $^{13}$CO $J = 2 - 1$ line in black (Bujarrabal et al., 2013a). The x-axis indicates velocity with respect to the local standard of rest ($V_{\rm LSR}$) and the y-axis represents the detected flux measured in Jansky.

## 5.5 Discussion

### 5.5.1 Molecular richness

Prior to this work, the presence of molecules other than CO in these objects was practically unknown. We compared our results with standard AGB stars (see e.g., Bujarrabal et al., 1994b), which are the precursors of our objects, representing a homogeneous group that has been studied in great detail and whose envelopes are rich in molecular emission. Additionally, photodissociation effects in AGB envelopes only appear in outer layers. The comparison of our results with PNe is more difficult, because these objects present large differences in the molecular gas content (Huggins et al., 1996; Santander-García et al., 2021), which is also present in young PNe (Bujarrabal et al., 1988, 1992). This fact is most likely due to molecular photodissociation, which strongly depends on the different evolutionary state of each source (Bachiller et al., 1997a,b; Bublitz et al., 2019; Ziurys et al., 2020). An added difficulty is that there is no complete survey of PNe or young PNe to compare with.

In Fig. 5.13, we show integrated intensity ratios between the main molecules (SO, SiO, SiS, CS, and HCN) and CO ($^{13}$CO $J = 2 - 1$ and with $^{12}$CO $J = 1 - 0$). Additionally, we compare these molecular integrated intensities with infrared emission at 12, 25, and 60 $\mu$m (from the *IRAS* mission, see Neugebauer et al., 1984). We used these three IRAS wavelengths because the shape of the SEDs seems to be strongly related with the evolutionary state of the source (see Fig. 5.18 and Kwok et al., 1989;

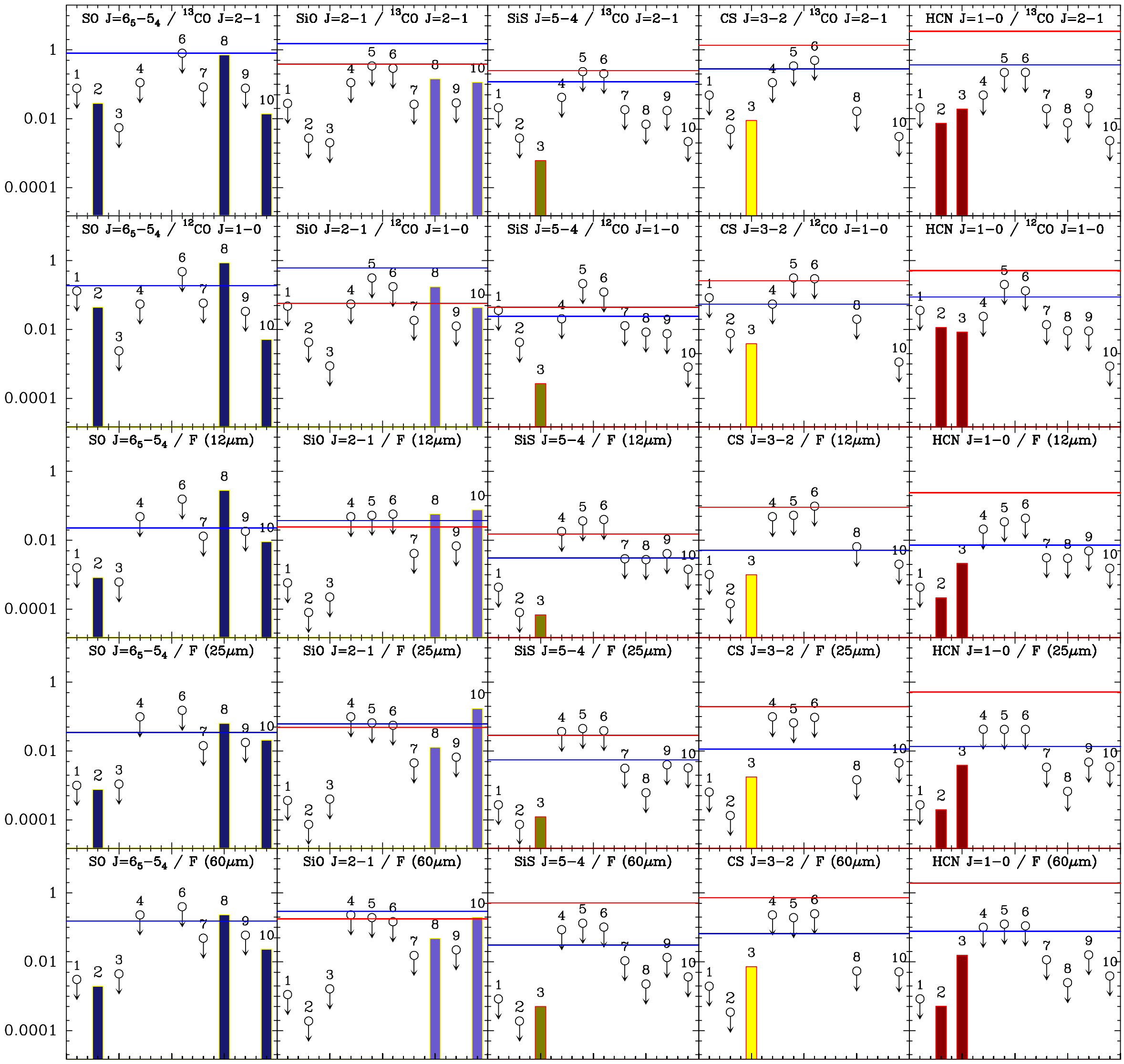

Figure 5.13: Ratios of integrated intensities of molecules (SO, SiO, SiS, CS, and HCN) and infrared emission (12, 25, and 60 $\mu$m) in binary post-AGB stars. Upper limits are represented with empty circles. Blue and red lines represent the averaged values for O- and C-rich AGB CSEs. Sources are ordered by increasing outflow dominance and enumerated as follows: 1 – AC Her, 2 – Red Rectangle, 3 – 89 Her, 4 – HD 52961, 5 – IRAS 19157−0257, 6 – IRAS 18123+0511, 7 – IRAS 19125+0343, 8 – AI CMi, 9 – IRAS 20056+1834, and 10 – R Sct. Sources 1 and 2 are disk-dominated binary post-AGB stars, sources 6 to 10 are outflow-dominated, while sources 3 and 4, and 5 are intermediate cases. We note the broad range (on a logarithmic scale) of intensity ratios. We always find uncertainties lower than $\sim$20% in these ratios, which are basically dominated by that of the integrated intensity of the molecules other than CO (see main text for details).

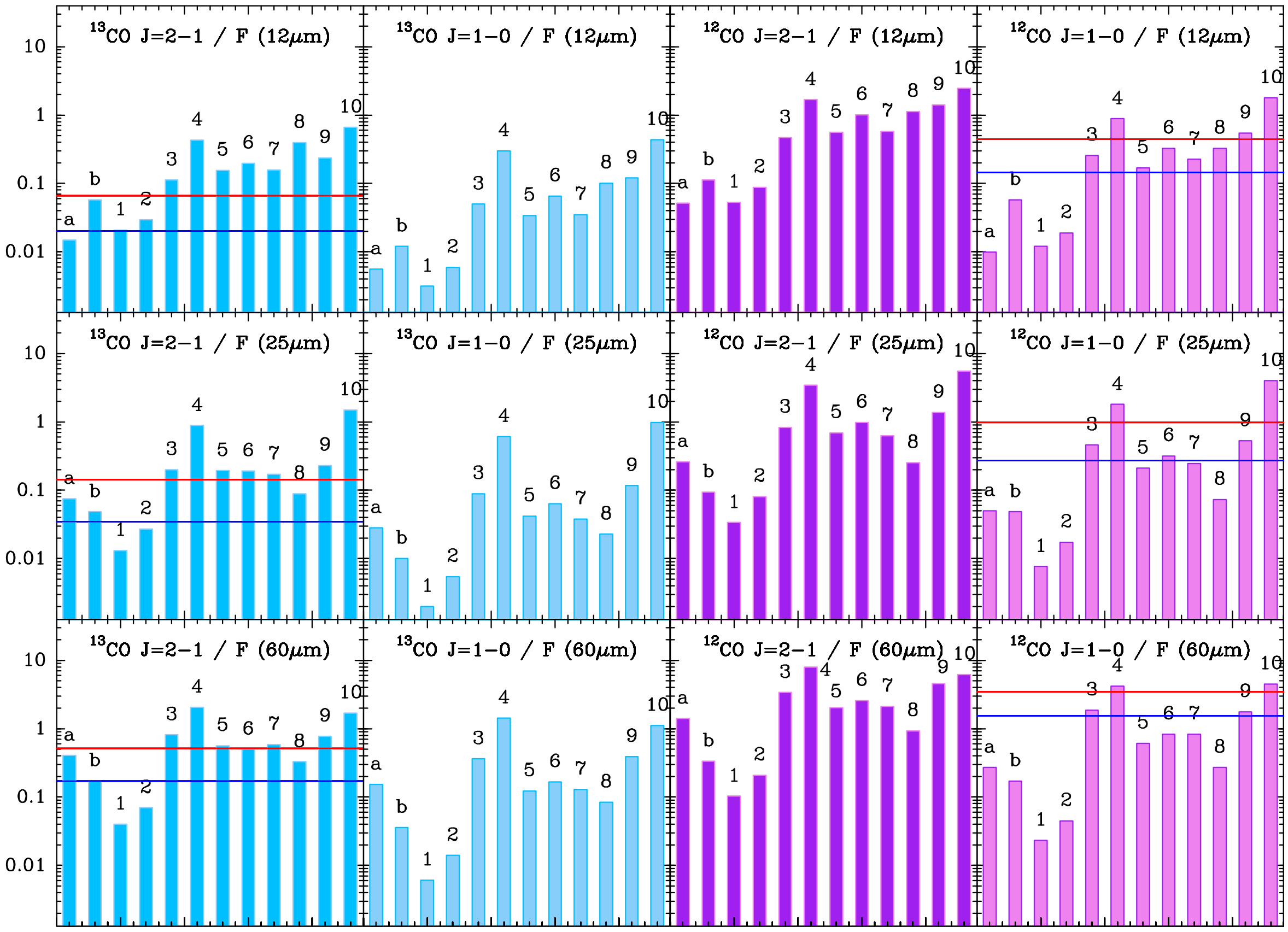

Figure 5.14: Ratios of integrated intensities of $^{12}$CO (blue) and $^{13}$CO (purple) $J = 2-1$ and $J = 1-0$ and infrared emission (12, 25, and 60 $\mu$m) in binary post-AGB stars. Blue and red lines represent the averaged values for O- and C-rich AGB CSEs. Sources are are ordered by increasing outflow dominance and enumerated as follows: a – HR 4049, b – DY Ori, 1 – AC Her, 2 – Red Rectangle, 3 – 89 Her, 4 – HD 52961, 5 – IRAS 19157−0257, 6 – IRAS 18123+0511, 7 – IRAS 19125+0343, 8 – AI CMi, 9 – IRAS 20056+1834, and 10 – R Sct. Souces a and b are not included in this mm-wave survey. Sources 1, 2, a, and b are disk-dominated binary post-AGB stars, sources 6 to 10 are outflow-dominated, and sources 3, 4, and 5 are intermediate cases. We note the broad range (on a logarithmic scale) of intensity ratios. We find uncertainties less than ∼10% in these ratios, which are basically dominated by that of the integrated intensity of CO (see Bujarrabal et al., 2013a).

Hrivnak et al., 1989; van der Veen et al., 1995; Oomen et al., 2018). Molecular emission in AGB stars is usually compared to 60 $\mu$m, since this band adequately represents the bulk of the circumstellar emission. Many pPNe present a bimodal SED with an excess at $25 - 60\,\mu$m and less emission at $5 - 10\,\mu$m, because of a lack of dust characterized by an intermediate temperature. On the contrary, our objects tend to present a significant NIR excess at 12 and 25 $\mu$m, and these two bands are probably better tracing most of the nebular material. However, since we want to study the gast-to-dust ratios in our sources and compare them with those in other objects, we needed to use all these three CO-to-IRAS ratios. We always find uncertainties less than ∼20% in our ratios, which basically are dominated by that of the integrated intensity of the molecules other than CO (see tables of Sect. 5.7.3 for further details). These uncertainties are too small to be represented in the figures; thus, we note the very broad range (on a logarithmic scale) of the intensity ratios. Our results were compared with molecular emission of AGB stars. Blue and red horizontal lines represent averaged values of the molecular emission for O- and C-rich AGB stars, respectively. We note that the emission of molecules other than CO in our sources is low. This fact is especially remarkable in those sources which are dominated by their rotating disk (AC Her and the Red Rectangle) or intermediate cases (such as 89 Her).

In Fig. 5.14, we show, $^{12}$CO - to - IR and $^{13}$CO - to - IR ratios for the integrated intensity of $J = 2 - 1$ and $J = 1 - 0$ . For transitions compared with the IR fluxes, the $^{12}$CO emission is relatively low with respect to the levels exhibited by AGB stars. Again, we find two subclasses in our sample. Those sources in which the Keplerian disk dominates the nebula tend to present a lower relative intensity of $^{12}$CO compared to the outflow-dominated ones. On the contrary, we do not see this effect in $^{13}$CO emission, which seems to be relatively intense compared with $^{12}$COand comparable to that from AGB stars. The difference found in this ratio in the disk- and the outflow-dominated sources could be due to higher optical depths in the $^{12}$CO $J = 2 - 1$ line. The emission of the outflow-dominated sources comes from relatively extensive and very diffuse areas, thus they are expected to show low optical depths in the $^{12}$CO $J = 2 - 1$ line. For example, in the case of R Sct, the emitting area exceeds $2 \times 10^{17}$ cm. However, in the case of the very well known disk-dominated sources, such as AC Her, the emitting area does not exceed $2 \times 10^{16}$ cm, which means that it is ten times smaller than in the case of R Sct. Thus, based on this discussion and the results (see Fig. 5.13), we find a real difference in the relative abundances that seems to confirm that the outflow-dominated sources present relative higher $^{13}$CO abundance. This fact was previously noted in R Sct, which is an outflow-dominated source (Gallardo Cava et al., 2021; Bujarrabal et al., 1990), where a very low $^{12}$C/$^{13}$C ratio was reported.

### 5.5.2 Discrimination between O- and C-rich envelopes

In general, evolved stars can be classified as O- and C-rich according to their relative O/C elemental abundance. It is known that this difference, even if the O/C abundance ratio is not always shown to vary much from 1, has important effects on the molecular abundances. SiO, SO are O-bearing molecules and their lines are much more intense in O-rich envelopes than in the C-rich ones, while HCN and CS ($HC_3N$ and HNC) are C-bearing molecules (but also SiS), whose lines are more intense in C-rich envelopes than in O-rich (see e.g. Bujarrabal et al., 1994b,a). Additionally, SiO maser emission is detected in M- and S-type AGB stars, and $H_2O$ maser emission is only seen in M-type stars. In any case, SiO and $H_2O$ maser emission is seen exclusively towards O-rich envelopes (Kim et al., 2019). The analysis of integrated intensities of pairs of molecular transitions is very useful to distinguish between O- and C-rich environments (see Fig. 5.15). Integrated intensities are larger for O-rich than for C-rich objects when an O-bearing molecule is compared with a C-bearing one. Our results are compared with CSEs around AGB stars, which are deeply studied objects and they are prototypical of environments rich in molecules (for this comparison, AGB data is taken from Bujarrabal et al., 1994b,a, which represent a wide sample of CSEs around evolved stars in the search molecules other than CO). In Fig. 5.15, blue and red lines represent averaged values for these ratios in O- and C-rich AGB CSEs, respectively.

In the case of AI CMi and R Sct, the integrated intensities of SO and SiO compared with C-bearing molecules are remarkably larger than for C-rich evolved stars. Both sources present SiO and $H_2O$ maser emission (AI CMi also presents OH maser emission according to te Lintel Hekkert et al., 1991). Additionally, the IR spectra of R Sct is

Table 5.5 Chemistry of the envelopes around binary post-AGB stars observed in this work.

| Source | Chemistry | Comments |
|---|---|---|
| AC Her | O-rich | Tentative SiO maser emission |
| Red Rectangle | O-rich | See Fig. 5.15 and Sect. 5.5.4 |
| 89 Herculis | C-rich | See Fig. 5.15 |
| HD 52961 | O-rich | See Gielen et al. (2011b) |
| AI CMi | O-rich | See Fig. 5.15<br>SiO, $H_2O$, and OH maser emission<br>See Arkhipova et al. (2017) |
| IRAS 20056+1834 | O-rich | SiO maser emission |
| R Scuti | O-rich | See Fig. 5.15<br>SiO and $H_2O$ maser emission<br>$H_2O$ emission in IR spectra |

**Notes.** We did not detect any thermal emission of O-bearing molecules in AC Her and IRAS 20056+1834, but we detected SiO maser emission (see Sects. 5.4.1 and 5.4.8). HD 52961 is cataloged as O-rich, based on discussions by Gielen et al. (2011b).

mainly dominated by $H_2O$ emission (Matsuura et al., 2002; Yamamura et al., 2003). These facts lead us to classify AI CMi and R Sct as O-rich. The Red Rectangle presents O- and C-bearing molecular emission. This case will be examined in more detail in Sect. 5.5.4, but according to Fig. 5.15 and the tentative $H_2O$ maser detection, we catalog the Red Rectangle as O-rich too. There are no signs of O-bearing molecules in 89 Her and we detected HCN, CS, and SiS, so the chemistry of this source is absolutely compatible with a C-rich nebula. Finally, we did not detect any thermal SiO emission in IRAS 20056+1834, but we have clearly confirmed the existence of SiO maser emission and tentative SiO maser emission in AC Her(see Sects. 5.4.1 and 5.4.8), which is exclusive of O-rich environments.

Based on the integrated intensities ratios in Fig. 5.15 and the maser detection of O-bearing molecules, we can classify some of our sources as O-/C-rich (see a summary of our results in Table 5.5). 89 Her presents a $O/C<1$ chemistry. On the contrary, AC Her, AI CMi, IRAS 20056+1834, and R Sct present a $O/C>1$ chemistry. The chemistry of the Red Rectangle could be uncertain, but we also classified this source as O-rich (see Sect. 5.5.4 for further details). In any case, we recall that our sources are still poorly studied and that additional work, both observational and theoretical, is needed to confirm our findings on the general chemistry of these sources.

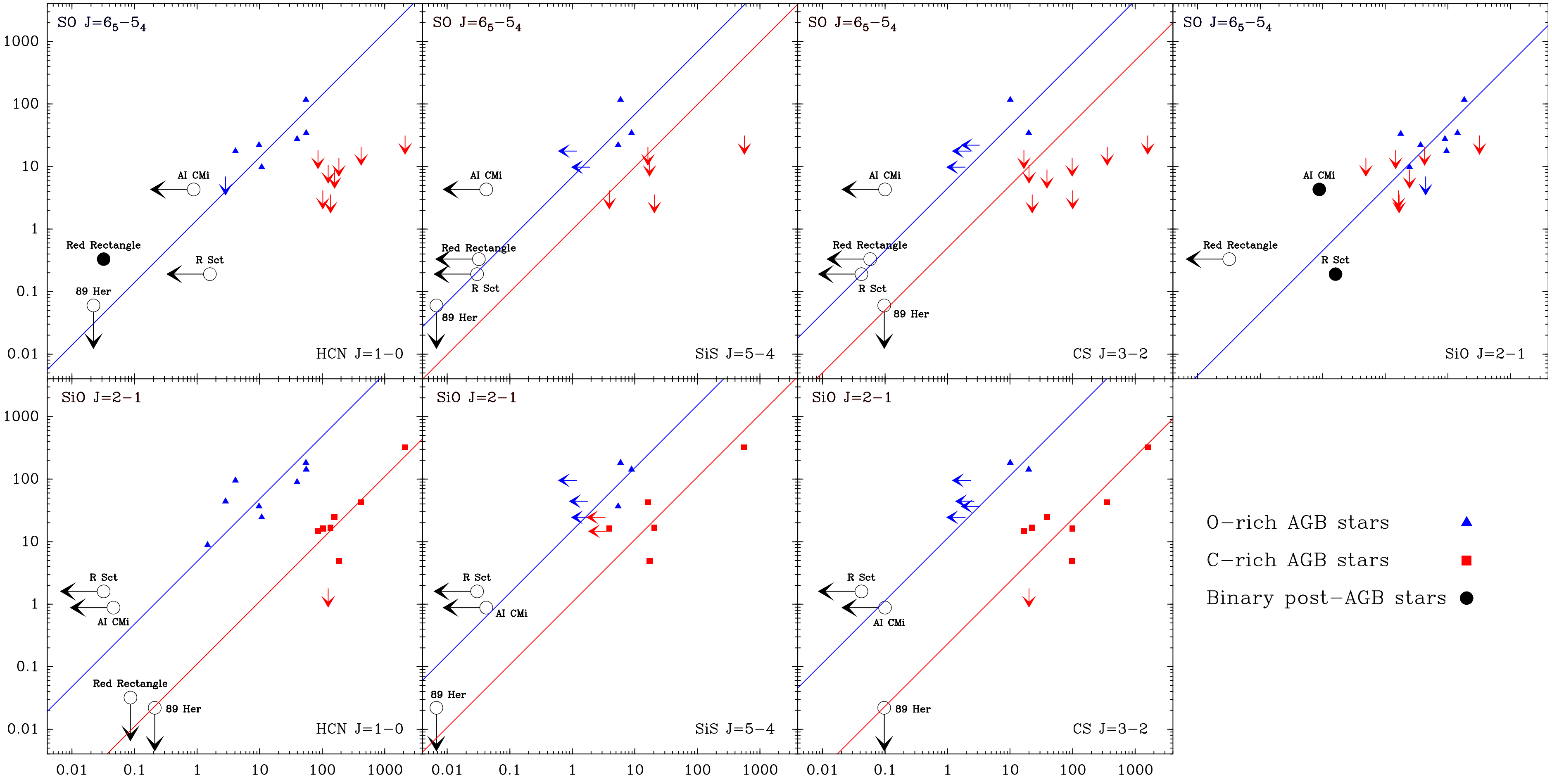

Figure 5.15: Integrated intensities of pairs of molecular transitions in binary post-AGB stars (black), as well in O- and C-rich stars (blue and red). Blue and red lines represent the averaged values for O- and C-rich AGB CSEs (only in the case of detections). The X and Y axes represent the integrated intensities of the observed transitions in logarithmic scale and units of Jy km s$^{-1}$. The upper limits are represented bv arrows.

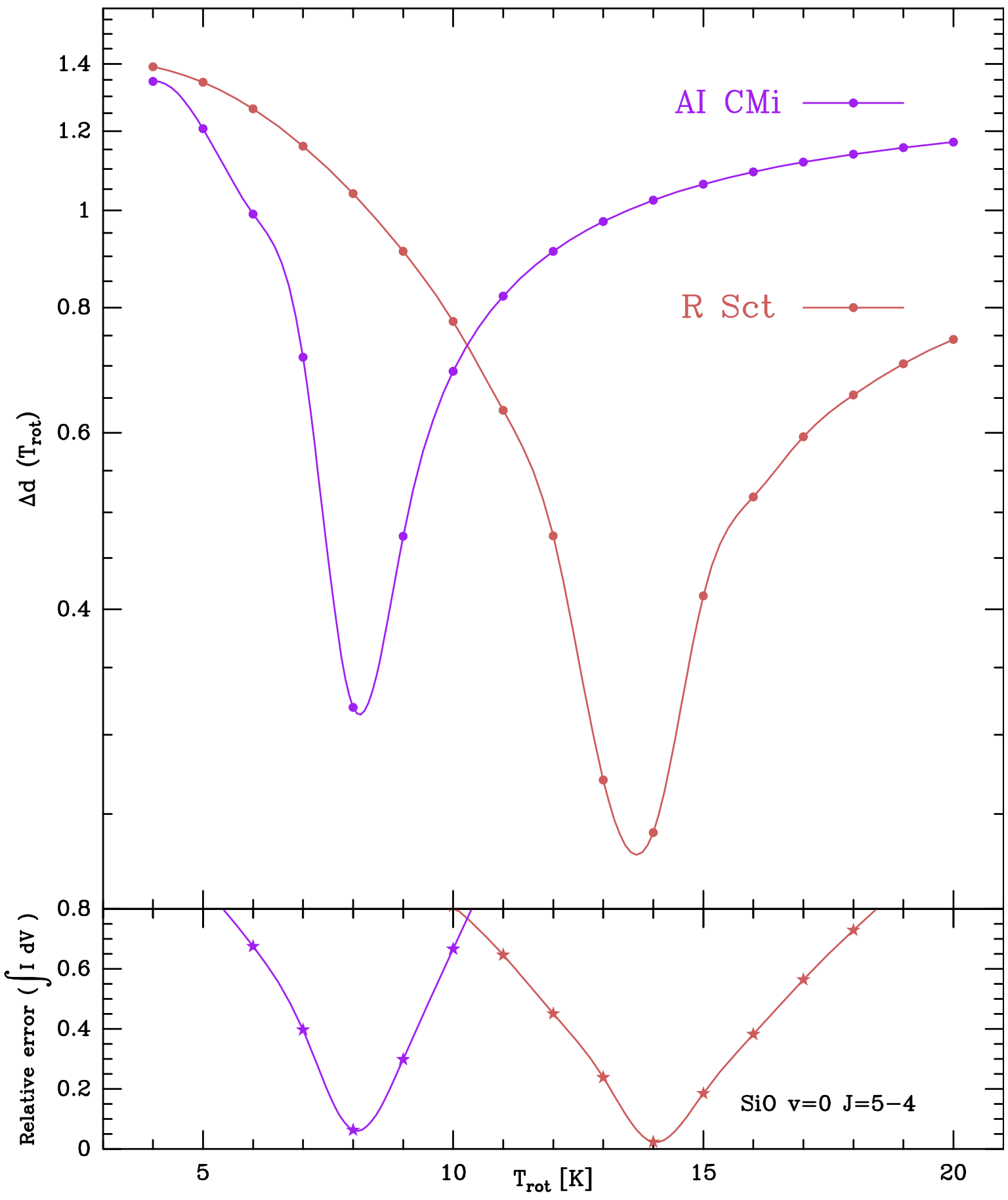

Figure 5.16: *Top*: Variations of the relative difference of the abundances produced by the line-ratio method as a function of the rotational temperature $\Delta d\,(T_{rot})$ in the case of AI CMi and R Sct. The minimum relative difference corresponds to the best-fit value of $T_{rot}$. This procedure can only be performed for those sources in which several transitions of the same molecule have been observed. See text and Sect. 5.7.2 for more details. *Bottom*: Variations of the relative error of the observational integrated intensity line and the predicted one as a function of the rotational temperature for SiO $J = 5-4$. The predicted integrated intensity lines are derived from Eq. 5.11 using the estimated abundance and rotational temperature.

### 5.5.3 Abundance estimates

The methodology to estimate abundances is similar to that used in Bujarrabal et al. (2001); Quintana-Lacaci et al. (2007); Bujarrabal et al. (2013a) and is described in Sect. 5.7.2 in detail. This procedure assumes optically thin emission, which is justified in our case, because the molecular lines present very weak emission (see Sect. 5.5.1). To estimate the characteristic rotational temperature ($T_{\rm rot}$), it is necessary to observe (at least) two transitions of the same molecule. We apply this procedure, using Eq. 5.11, to AI CMi and R Sct for SiO and its rare isotopic substitutions. We assume nebular masses derived from previous works (see Table 4.1, Bujarrabal et al., 2013a; Gallardo Cava et al., 2021).

After computing the abundances, we checked that our optically thin assumption is valid. We estimated the line opacities expected for the derived values of abundances and $T_{\rm rot}$. These values are shown in Table 5.8. We can see that in all cases, $\tau \ll 1$.

We estimate the rotational temperature ($T_{\rm rot}$) from the observed line ratios. Different values of the abundance are calculated for each transition as a function of $T_{\rm rot}$ (see Sect. 5.7.2 for further details). The minimum relative difference of the abundances yields the best-fit rotational temperature (see Fig. 5.16 and Eqs. 5.12 and 5.13). For SiO, the best-fit $T_{\rm rot}$ is 8 and 14 K in AI CMi and R Sct, respectively. We find relative errors lower than 5% between the integrated line intensity and the predicted one derived from Eq. 5.11 (see Fig. 5.16 *bottom*). This relative error between the integrated intensity line and the predicted one considerably increases if we consider small variations with respect to our best-fit $T_{\rm rot}$. We can assume the rotational temperature of SiO for $^{29}$SiO and $^{30}$SiO, as well as for $HCO^+$, because the lines show similar excitation properties and line profiles (see Fig. 5.11).

We see that most of the molecular line shapes are very similar to that of the CO lines (except the masers, whose lines have not been considered for the calculation of abundances; see Sect. 5.4). We estimated the molecular mass of the envelopes of these sources based on the CO lines. Therefore, we can calculate relative abundances of molecules other than CO assuming that their emission comes from the same molecule-rich zone for which mass values were derived. This is not the case of R Sct, whose thermal emission of molecules is narrower than that of CO. We think that the emission of these molecules very probably comes from the rotating disk and the high density region of the outflow (see Figs. 5.12 and 5.11). We present abundance calculations for the total nebular mass, $\langle X \rangle$, and for that of the central region of the nebula, $\langle X_c \rangle$, which represents the rotating gas and the innermost regions of the outflow. In other words, $\langle X \rangle$ represents the relative abundance considering that the species are located in the whole nebula, while $\langle X_{\rm c} \rangle$ is the value obtained assuming that the species are located only in the densest part of the nebula. According to our previous reasoning, for R Sct our best option is to compute relative abundances with respect to the CO mass in these central regions, which in this case account for 40% of the total nebular mass. The shapes of the SiO, SO, and $SO_2$ line profiles of AI CMi are very similar to that in CO (see Fig. 5.6); therefore we think that the molecular emission of AI CMi must arise from the entire nebula. We recall that in any case, these estimates are uncertain by a factor of 2 (in addition to uncertainties for other reasons) in view of the assumed fraction of involved material. To estimate relative abundances assuming other fractions of the mass of the molecular rich component, $f$, with respect to the total nebular mass, we only need to multiply $\langle X \rangle$ by $f^{-1}$.

In the case of the Red Rectangle and 89 Her there is only one detected transition per molecule, so we can only make very crude estimations of the abundance by assuming the obtained $T_{\rm rot}$ of SiO for AI CMi and R Sct and the nebular masses derived from previous works (Bujarrabal et al., 2016; Gallardo Cava et al., 2021). We also estimated abundances for the whole nebula and for the inner regions: in the case of the Red Rectangle $\sim$90% is located in the disk; in the case of 89 Her, 60% of the total molecular mass corresponds to the rotating disk and the inner regions of the outflow. Based on the same reasoning as in the case of AI CMi, our best option for Red Rectangle and 89 Her is to consider the abundance corresponding to the total molecular mass.

The calculated abundances for AI CMi, R Sct, Red Rectangle, and 89 Her can be seen in Table 5.6 (we also show estimated optical depths in Table 5.8). We note the very low abundances, compared with standard AGB stars, which reinforces our conclusion, set out in Sect. 5.5.1. We see again how the presence of molecules in disk-dominated or intermediate sources is much weaker than that in outflow-dominated ones.

Table 5.6 Abundances estimated.

| Source | Molecule | $T_{\rm rot}$ [K] | $\langle X \rangle$ | $\langle X_{\rm c} \rangle$ |
|---|---|---|---|---|
| AI CMi | SiO | 8 | $1.3\times10^{-8}$ * | $3.1\times10^{-8}$ |
| | $^{29}$SiO | 8 | $3.7\times10^{-9}$ * | $9.3\times10^{-8}$ |
| | $^{30}$SiO | 8 | $<1.1\times10^{-8}$ * | $<2.7\times10^{-8}$ |
| R Scuti | SiO | 14 | $5.6\times10^{-9}$ | $1.5\times10^{-8}$ * |
| | $^{29}$SiO | 14 | $5.0\times10^{-10}$ | $4.2\times10^{-9}$ * |
| | $^{30}$SiO | 14 | $1.1\times10^{-9}$ | $2.7\times10^{-9}$ * |
| | $HCO^+$ | 14 | $3.8\times10^{-10}$ | $1.0\times10^{-9}$ * |
| Red Rectangle | HCN | $8-14$ | $(1.4\pm0.5)\times10^{-9}$ * | – |
| 89 Herculis | HCN | $8-14$ | $(1.1\pm0.2)\times10^{-10}$ * | $(1.9\pm0.2)\times10^{-9}$ |
| | SiS | $8-14$ | $(3.7\pm0.5)\times10^{-11}$ * | $(6.3\pm0.8)\times10^{-11}$ |
| | CS | $8-14$ | $(5.9\pm0.8)\times10^{-10}$ * | $(5.1\pm3.9)\times10^{-10}$ |

**Notes.** Averaged abundances of the emitting gas for SiO, $^{29}$SiO, and $^{30}$SiO in the case of AI CMi and R Sct. $\langle X \rangle$ is the abundance considering the total mass of the envelope. $\langle X_{\rm c} \rangle$ is the abundance considering only the central region of the envelope and it represents $\sim$40%, $\sim$40%, $\sim$100%, and $\sim$60% of the total mass in the case of AI CMi, R Sct, the Red Rectangle, and 89 Her, respectively. The asterisk represents our preference for each case.

### 5.5.4 Dual nature of the Red Rectangle

The chemistry of these kinds of objects was barely known even for the case of the Red Rectangle, which is by far the best-studied object. Apart from CO, Bujarrabal et al. (2016) discovered H$^{13}$CN $J = 4-3$, C I $J = 2-1$ and $J = 1-0$, and C II $J = 2-1$. In this work, we detected C$^{17}$O and C$^{18}$O $J = 2-1$, and SO $J_N = 6_6-5_4$, and we also tentatively detected HCN $J = 1-0$ and $H_2O$ $J_{Ka,Kc} = 6_{1,6}-5_{2,3}$. Both SO and $H_2O$ (at 22 GHz) are good tracers of O-rich environments (see Sect. 5.5.2). Additionally, we attained very significant SiO and SiS emission upper limits. We estimated the upper limits on the abundance of these molecules following the method described in Sect. 5.7.2, assuming a $3\sigma$ limit for intensity and integrated intensity lines and $T_{\rm rot} = 11\pm3$ K (see Table 5.10 and Sect. 5.5.3). We find $\langle X\rangle_{\rm SiO} < (5.9\pm0.8)\times10^{-9}$ and $\langle X\rangle_{\rm SiS} < (2.3\pm0.1)\times10^{-10}$ (see Table 5.6 for comparison).

The solar abundance of Si is $\sim6.5\times10^{-5}$ (see e.g., Asplund et al., 2021). In the case of O-rich AGB stars, SiO abundances vary in between $4\times10^{-5}$ and $5\times10^{-6}$, depending on the state of depletion onto dust grains (Verbena et al., 2019). These values for the abundances are considerably larger than the estimated upper limits for SiO and SiS. Taking into account the fact that the abundance of SiO decreases as grains of dust are formed (Gobrecht et al., 2016), it seems that the depletion process is very efficient, at least in the Red Rectangle and practically all Si could be depleted onto the grains. This conclusion is reinforced if we consider that the molecular emission of the Red Rectangle practically comes from the Keplerian disk, where depletion should be more effective than in the case of an expanding and very large envelope, as for AGB stars (since, in a rotating disk structure, there is more time for the grains to grow bigger).

C II emission must have its origin in the PDR, because it is the best tracer of these regions. C I is also very often associated with PDRs. The presence of H$^{13}$CN is also

Table 5.7 O-isotopic abundance ratio, initial stellar mass estimation compared with the central total stellar mass and the total nebula mass.

| Source | $^{17}O/^{18}O$ | $M_i$ [$M_\odot$] | $M_{Tc}$ [$M_\odot$] | $M_{neb}$ [$M_\odot$] |
|---|---|---|---|---|
| Red Rectangle | 0.8 | 1.5 | 1.7 | $1.4 \times 10^{-2}$ |
| 89 Herculis | 0.3 | 1.2 | 1.7 | $1.4 \times 10^{-2}$ |

**Notes.** $M_i$ represents the initial stellar mass of the post-AGB component, $M_{Tc}$ is of the total central stellar mass, and $M_{neb}$ is the total mass of the nebula (mass values for the Red Rectangle and 89 Her taken from Bujarrabal et al., 2016; Gallardo Cava et al., 2021).

compatible with the development of a PDR in the dense disk region that is close to the stellar system, which is a binary system with a secondary star and an accretion disk that emits in the UV (Witt et al., 2009; Thomas et al., 2013). The UV emission can excite a zone in between the H II region and the Keplerian molecule-rich disk (Bujarrabal et al., 2016). According to models of the PDR chemistry, the origin of the HCN can be the photoinduced formation (Agúndez et al., 2008) and it has been detected in the innermost regions of several disks around young stars. This fact is also consistent with the detection of HCN (see Fig. 5.2).

In conclusion, we think that the Red Rectangle most likely presents a O-rich gas chemistry, but it also presents a PDR in the innermost region of the Keplerian disk. The UV excess can explain the detection of HCN (and $H^{13}CN$, C I, and C II) together with PAHs, even in an O-rich environment. We note that this is the typical situation in star-forming regions and that these are thought to be O-rich despite the presence of C-bearing species, including PAHs.

### 5.5.5 Isotopic ratios

#### $^{17}O/^{18}O$

Complex stellar evolution models reveal that the $^{17}O/^{18}O$ ratio remains approximately constant over the entire thermally pulsating AGB phase, and the initial stellar mass ($M_i$) can be directly obtained from this ratio (De Nutte et al., 2017). We can estimate the $^{17}O/^{18}O$ abundance ratio using the $C^{17}O$ and $C^{18}O$ $J = 2-1$ line intensities corrected for different beam widths and Einstein coefficients, because both molecules present almost identical radial depth dependence (Visser et al., 2009), low optical depths, and very similar and easily predictable excitations conditions.

According to our results in Tables 5.10 and 5.11, we estimate the abundance ratio for the Red Rectangle and 89 Her, respectively. Initial stellar masses for the post-AGBs stars are shown in Table 5.7. We additionally show the total stellar mass ($M_{Tc}$) derived from the Keplerian dynamics and the total nebula mass ($M_{neb}$), which includes the mass of the disk and outflow (Bujarrabal et al., 2016; Gallardo Cava et al., 2021). We note that the derived values of $M_i$ are consistent with $M_{Tc}$, which must be higher because of the presence of a companion. The $^{17}O/^{18}O$ ratio for 89 Her is typical of O-rich, while its chemistry is typical of C-rich environments. We note that abundance models are set up for single stars and not for binary stars where it is very likely that mass transfer happen after the results of the evolution of the individual stars.

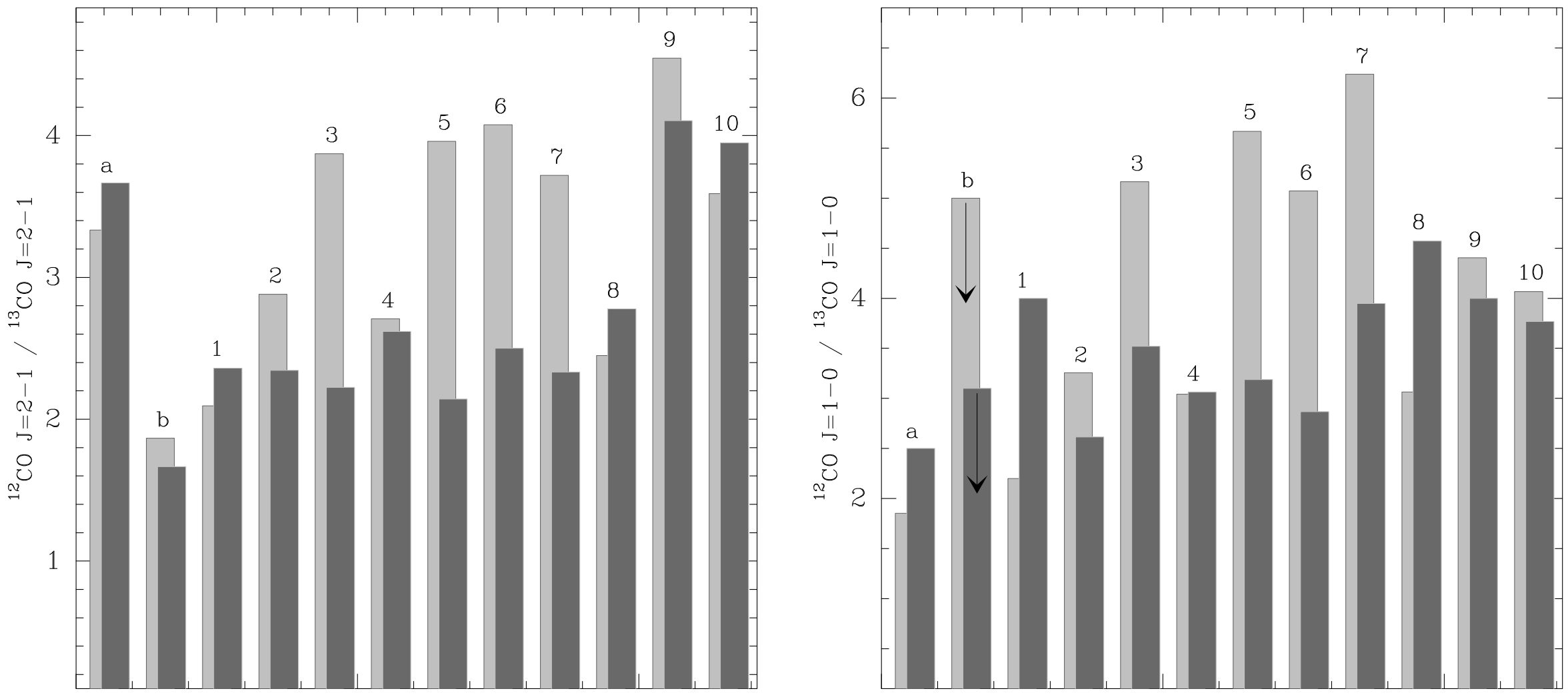

Figure 5.17: Ratios of integrated intensities of $^{12}$CO/$^{13}$CO in binary post-AGB stars in light grey. We also show line-peak ratios in dark grey ($J = 2-1$ at *left*; $J = 1-0$ at *right*). Sources are ordered by outflow dominance and enumerated as follows: a – HR 4049, b – DY Ori, 1 – AC Her, 2 – Red Rectangle, 3 – 89 Her, 4 – HD 52961, 5 – IRAS 19157−0257, 6 – IRAS 18123+0511, 7 – IRAS 19125+0343, 8 – AI CMi, 9 – IRAS 20056+1834, and 10 – R Sct. Sources a and b are not included in this mm-wave survey. Sources 1, 2, a, and b are disk-dominated binary post-AGB stars, sources 6 to 10 are outflow-dominated, and sources 3, 4, and 5 are intermediate cases.

## $^{12}$CO/$^{13}$CO

For our binary post-AGB stars, we find a $^{12}$CO/$^{13}$CO integrated-line ratio (see Fig. 5.17) with a mean value of 4.1 and a standard deviation of 1.4 for the transition $J = 1-0$ ($3.3 \pm 0.9$ for $J = 2-1$). This ratio can be compared with intensity line ratios of other post-AGB stars, in which we find $10.4 \pm 7.6$ (see Palla et al., 2000; Balser et al., 2002; Sánchez Contreras and Sahai, 2012). Although these ratios must be corrected for different beam widths, Einstein coefficients, and opacity (especially for the $^{12}$CO $J = 2-1$ lines), our objects seem to present higher $^{13}$CO abundance compared with other post-AGB stars. As stated before, the $^{12}$CO/$^{13}$CO abundance ratio was previously derived for seven of our sources (see Bujarrabal et al., 2016, 2017, 2018b; Gallardo Cava et al., 2021). From these works, we are able to find a $^{12}$CO/$^{13}$CO mean value of $8.6 \pm 2.4$. We can compare this ratio to that of AGB stars: $12.7 \pm 5.8$, $28.0 \pm 16.2$, and $31.3 \pm 28.2$ for M-, S-, and C-type stars, respectively (see Ramstedt and Olofsson, 2014). Again, our objects present a higher $^{13}$CO abundance compared with any sub-class of AGB stars, and are only similar to those of M-type stars presenting very low abundance ratios.

Our results support the suggestion by Ramstedt and Olofsson (2014) that the binary nature of the sources is the origin of these very low $^{12}$CO/$^{13}$CO ratios found in other post-AGB stars, because all our sources are bona-fide binary systems in which a strong interaction between the stars and the circumbinary disk is expected (Van Winckel, 2003).

## 5.6 Conclusions

We present a very deep survey of mm-wave lines in binary post-AGB stars, which is the first systematic study of this kind. We show our single-dish observational results in Sect. 5.4, where detected lines together with very relevant upper limits are presented.

Our molecular study allows us to verify that our sources present lower molecular intensities of all species than those typical of AGB sources (see Sect. 5.5.1). This fact is very significant in those binary post-AGB stars for which the Keplerian disk is the dominant component (like the Red Rectangle and AC Her) or at least it has a significant weight in the whole nebula, such as 89 Her. This lower molecular richness is also present in CO, and it is again very remarkable in disk-dominated sources. On the other hand, we find some overabundance of $^{13}$CO, especially in our outflow-dominated objects.

Our analysis of integrated intensities ratios of pairs of molecules other than CO leads us to classify the chemistry of (some of) our sources as C- or O-rich (see Sect. 5.5.2). We cataloged the nebulae around AC Her, the Red Rectangle, AI CMi, R Sct, and IRAS 20056+1834 as O-rich environments. On the contrary, the nebula around 89 Her is a C-rich environment.

We estimated abundances of the different species using the method described in detail in Sect. 5.7.2. As expected, our results reveal that our sources present very low abundances in molecules other than CO (see Sect. 5.5.3) compared with standard evolved stars. The chemistry of these objects is yet to be studied from the theoretical point of view.

We also studied the isotopic ratios of our sources (see Sect. 5.5.5). We deduced the initial stellar mass for the Red Rectangle and 89 Her via the $^{17}$O/$^{18}$O ratio, which is consistent with the current central total stellar mass derived of our mm-wave interferometric maps and models (see Table 5.7). We studied the $^{12}$CO/$^{13}$CO ratios and we find that our sources, which are all post-AGB stars, tend to present an overabundance of $^{13}$CO compared to AGB and other post-AGB stars.

## 5.7 Supporting material

### 5.7.1 Additional figures

In this section, we show other relevant figures. In Fig. 5.19 we present our tentative detection of HCN $J=1-0$ in the Red Rectangle. We also show in black H$^{13}$CN $J=1-0$, C I $J=1-0$, C II $J=1-0$, and C II $J=2-1$ (ALMA data taken from Bujarrabal et al., 2016).

In Fig. 5.18, we show the color-color diagram where we represent [25]−[60] versus [12]−[25] IRAS colors. We also show the regions populated stars with similar properties (see van der Veen and Habing, 1988, for more details). These kinds of plots show the evolutionary sequence of AGB / post-AGB stars. We note that the sources of this work, the binary post-AGB stars, are in between of AGB stars and pPNe and young PNe, which confirms the nature of our sources.

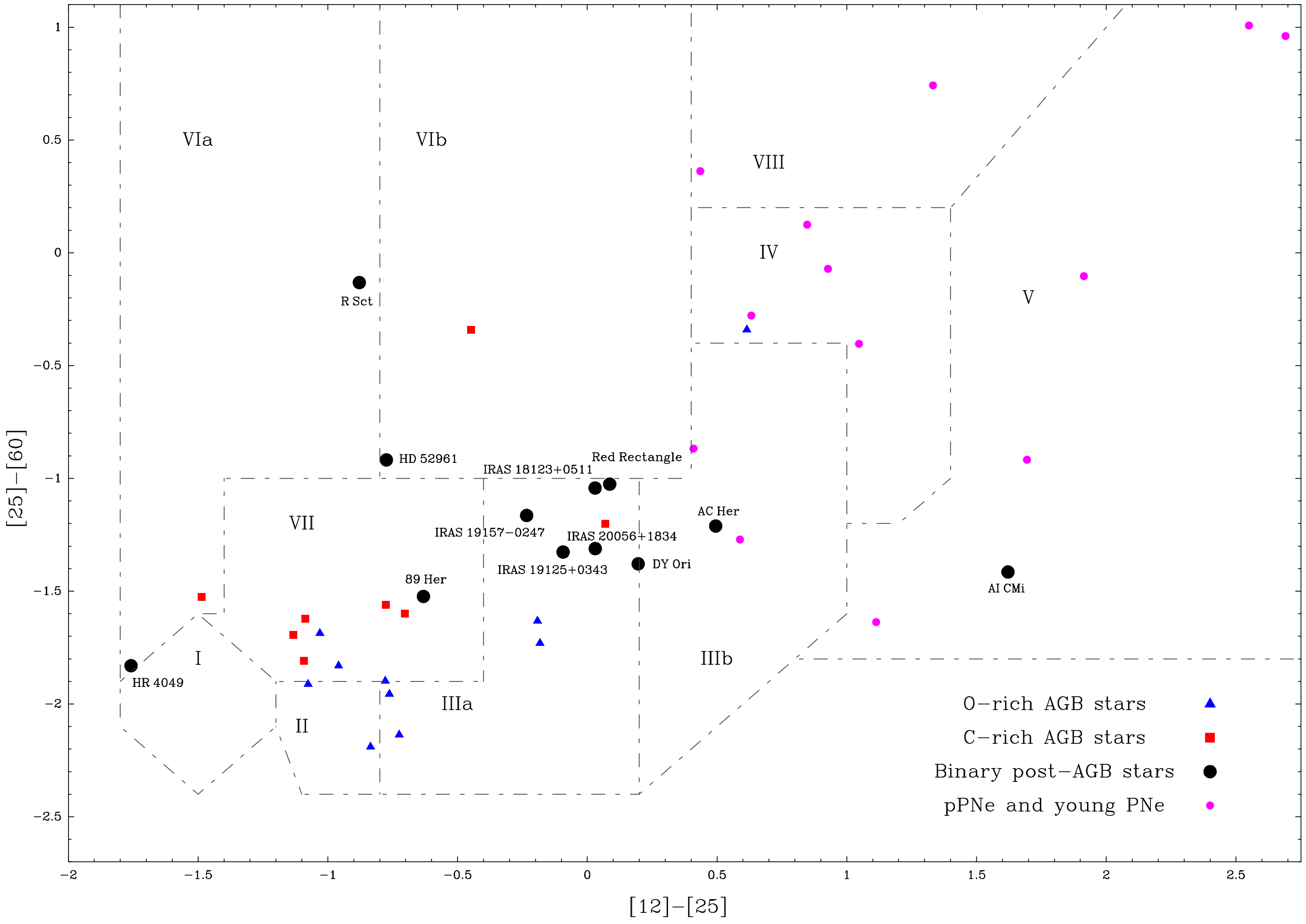

Figure 5.18: Diagram of [25]−[60] vs. [12]−[25] IRAS colors. The IRAS color-color diagram is divided in different regions where sources with similar characteristics are located (see van der Veen and Habing, 1988, for details). Standard AGB stars are represented with blue triangles (O-rich) and red squares (C-rich), pre–planetary and young planetary nebulae are represented with purple circles and the binary post-AGB stars of this work are shown with black circles.

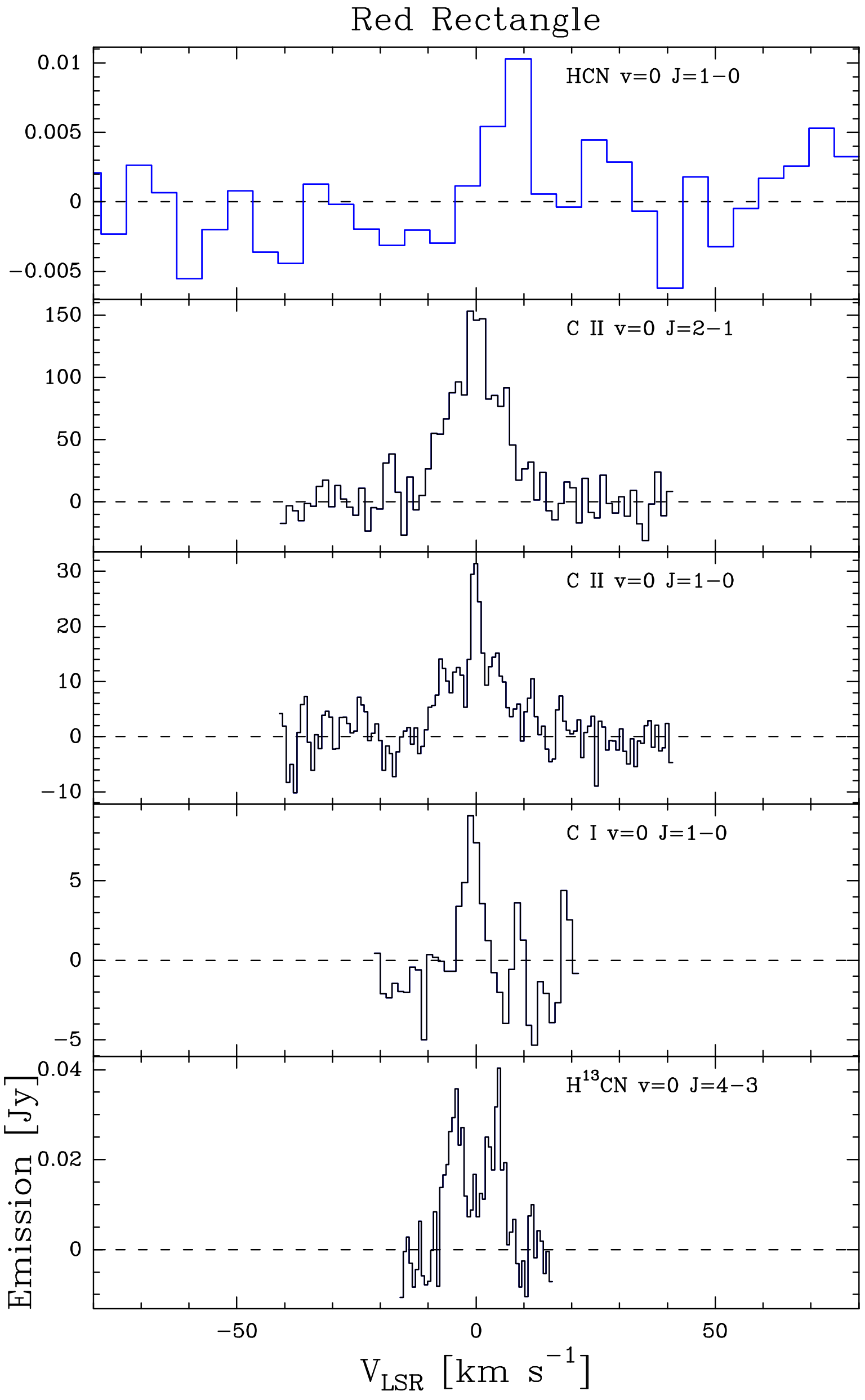

Figure 5.19: Spectra for HCN $J = 1 - 0$ detected in the Red Rectangle. We also show other PDR-tracing lines in black taken from Bujarrabal et al. (2016). The x-axis indicates velocity with respect to the local standard of rest ($V_{\rm LSR}$) and the y-axis represents the emitted emission measured in Jansky.

### 5.7.2 Estimation of abundances: Methodology

In this section, we show in detail the procedure used for the estimation of the abundance of molecules with more than one detected transition, assuming optically thin emission. In our case, we use this procedure to estimate the abundance of SiO for R Sct and AI CMi (see Sects. 5.4.7 and 5.4.9).

The observed main-beam temperature for a specific molecular transition is expressed as:

$$T_{\rm mb}\left(V_{\rm LSR}\right) = S_{\nu}\,\tau\,\phi\left(V_{\rm LSR}\right)\frac{\Omega_{\rm s}}{\Omega_{\rm mb}}, \tag{5.1}$$

where $\tau$ is the characteristic optical depth, $\phi\,(V_{\rm LSR})$ is the normalized line profile in terms of Doppler-shifted velocity, and $\Omega_{\rm s}$ and $\Omega_{\rm mb}$ are the solid angles subtended by the source and the telescope main beam, respectively. These are given by:

$$\Omega_{\rm mb} = \frac{\pi}{4\ln 2}\,\theta^2_{\rm HPBW} \cong 1.133\,\theta^2_{\rm HPBW}, \tag{5.2}$$

$$\Omega_{\rm s} = 2\pi\,(1-\cos\alpha) \cong \pi\alpha^2 \cong \pi\frac{r^2}{D^2} = \frac{S}{D^2}, \tag{5.3}$$

where $\theta_{\rm HPBW}$ is the Half Power Beam Width (HPBW) and $\alpha$ is the angle formed by the radius $r$ of the source from the observer at a distance $D$. In the limit of small angle $\alpha$, we assume that $\tan\alpha \cong \frac{r}{D}$ and $\cos\alpha \cong 1-\frac{\alpha^2}{2}$. The ratio $\frac{\Omega_{\rm s}}{\Omega_{\rm mb}}$ is known as the dilution factor, which can only take values lower than 1 (for $\Omega_{\rm s}$ larger than $\Omega_{\rm mb}$, namely, spatially resolved sources, the dilution factor is set to one).

The source function is given by:

$$S_\nu = \frac{h\nu}{k_{\rm B}}\left(\frac{1}{e^{\frac{h\nu}{k_{\rm B}T_{\rm ex}}}-1} - \frac{1}{e^{\frac{h\nu}{k_{\rm B}T_{\rm BG}}}-1}\right), \tag{5.4}$$

where $T_{\rm ex}$ is the excitation temperature of the transition and $T_{\rm BG} = 2.7\,{\rm K}$ is the background temperature. The optical depth, $\tau$, can be expressed as:

$$\tau = \frac{c^3}{8\pi\nu^3}\,A_{\rm ul}\,g_{\rm u}\,(x_{\rm l}-x_{\rm u})\,n_{\rm T}\,X\,l, \tag{5.5}$$

where $X$ is the abundance of the studied molecule, $n_T$ is the total number density of particles, and $l$ is the typical length of the source along the line of sight. The subindexes $u$ and $l$ represent the upper and lower levels of the transition, respectively. $A_{ul}$ is the Einstein coefficient of the transition, $g_{\rm u}$ is the statistical weight of the upper limit expressed as

$$g_J = 2\,J+1, \tag{5.6}$$

and $x_{\rm u}$ and $x_{\rm l}$ are:

$$x_J \sim \frac{e^{-\frac{E_J}{T_{\rm rot}}}}{F\,(T_{\rm rot})}, \tag{5.7}$$

where $E$ is the energy of the considered levels, in terms of temperature, and for simple linear rotors it can be calculated as:

$$E_J = B_{\rm rot}\,J\,(J+1), \tag{5.8}$$

where $B_{\rm rot}$ is the rotational constant of the molecule. $T_{\rm rot}$ is the rotational temperature and it is defined as the typical value or average of the excitation temperatures of the relevant rotational lines (in our case, low-$J$ transitions) and is used to approximately calculate the partition function $F\,(T_{\rm rot})$:

$$F\,(T_{\rm rot}) = \sum_J g_J\,e^{-\frac{E_J}{T_{\rm rot}}} = \frac{T_{\rm rot}}{B_{\rm rot}}. \tag{5.9}$$

We can express $\tau$ (Eq. 5.5) in terms of the mass, taking into account:

$$M = n_{\rm T}\,l\,S = n\,M_{\rm mol}\,l\,S. \tag{5.10}$$

Integrating (Eq. 5.1) in terms of velocity, we get:

$$\int T_{\rm mb}\,(V_{\rm LSR})\,{\rm d}V = S_\nu \frac{c^3}{8\pi\nu^3} A_{\rm ul}\, g_{\rm u}\,(x_{\rm l}-x_{\rm u})\, X \frac{4\ln 2}{\pi}\,\theta_{\rm HPBW}^{-2}\, M_{\rm mol}\, D^{-2} \tag{5.11}$$

We estimate $T_{\rm rot}$ from the line intensity ratio for those molecules, where we observed more than one transition (SiO, see Sect. 5.4). We calculated several abundances for the different transitions ($X_{J={\rm u}-{\rm l}}$, $X'_{J={\rm u'}-{\rm l'}}$) of the same molecule and then we estimated the parameter $d_i$ for each pair of transitions:

$$d_{\rm i}\,(T_{\rm rot}) = \frac{\left|X_{J={\rm u}-{\rm l}} - X'_{J={\rm u'}-{\rm l'}}\right|}{(X_{J={\rm u}-{\rm l}} + X'_{J={\rm u'}-{\rm l'}})/2}. \tag{5.12}$$

We can quantify the quality of the fit through the parameter $\Delta d\,(T_{\rm rot})$:

$$\Delta d\,(T_{\rm rot}) = \left(\prod_i^N d_{\rm i}\,(T_{\rm rot})\right)^{\frac{1}{N}}, \tag{5.13}$$

where $N$ indicates the number of transitions used in each case. The value of $T_{\rm rot}$ that minimizes the value of $\Delta d\,(T_{\rm rot})$ in Eq. 5.13 will be the best-fit rotational temperature (see Fig. 5.16). Finally, knowing $T_{\rm rot}$, we can estimate the abundance through Eq. 5.11.

Additionally, using Eq. 5.1, we can express the optical depth as:

$$\tau \sim \frac{T_{\rm mb}}{T_{\rm ex}}\frac{\Omega_{\rm mb}}{\Omega_{\rm s}}. \tag{5.14}$$

and we can express the averaged optical depth over the beam of the telescope as:

$$\overline{\tau} \sim \frac{T_{\rm mb}}{T_{\rm ex}}. \tag{5.15}$$

The averaged optical depths values can be seen in Table 5.8. We note that these $\overline{\tau}$ values are very small, and that for the dilution factors expected for our sources, always lower than 10 (see Gallardo Cava et al., 2021; Bujarrabal et al., 2016, 2013a), implying that the lines are optically thin ($\tau < 1$). Additionally, we note that always compare our detections with $^{13}$CO $J = 2-1$ , which is an optically thin line and shows a similar beam size.

Table 5.8 Optical depths estimated.

| Source | Molecule | $T_{\rm rot}$ [K] | $\overline{\tau}$ |
|---|---|---|---|
| AI CMi | SiO | 8 | $7.3 \times 10^{-3}$ |
| | $^{29}$SiO | 8 | $6.5 \times 10^{-4}$ |
| | $^{30}$SiO | 8 | $< 3.0 \times 10^{-4}$ |
| R Scuti | SiO | 14 | $1.9 \times 10^{-2}$ |
| | $^{29}$SiO | 14 | $1.1 \times 10^{-3}$ |
| | $^{30}$SiO | 14 | $8.7 \times 10^{-4}$ |
| | $HCO^+$ | 14 | $6.0 \times 10^{-4}$ |
| Red Rectangle | HCN | $8 - 14$ | $(2.6 \pm 0.7) \times 10^{-4}$ |
| 89 Herculis | HCN | $8 - 14$ | $(2.4 \pm 0.6) \times 10^{-4}$ |
| | SiS | $8 - 14$ | $(2.9 \pm 0.8) \times 10^{-4}$ |
| | CS | $8 - 14$ | $(9.6 \pm 2.6) \times 10^{-4}$ |

**Notes.** The averaged optical depth, $\overline{\tau}$, is the value of $\tau$ averaged over the telescope beam, see Eqs. 5.14 and 5.15.

### 5.7.3 Tables

In this section, we show the most relevant line parameters for AC Her, 89 Her, IRAS 19125+0343, and R Sct(see Sect. 5.4). We note that most of the sources have been observed in the 1.3, 2, 3, 7, and 13 mm bands (see Table 5.2).

Table 5.9 Radio molecular line survey of AC Her.

| Molecule | Transition | I (peak) [Jy] | $\sigma$ [Jy] | $\int I\,\mathrm{d}V$ [Jy km s$^{-1}$] | $\sigma\left(\int I\,\mathrm{d}V\right)$ [Jy km s$^{-1}$] | Sp. Res. [km s$^{-1}$] | $V_{\rm LSR}$ [km s$^{-1}$] | Comments |
|---|---|---|---|---|---|---|---|---|
| $C^{18}O$ | $v=0\ J=2-1$ | ≤ 4.6E-02 | 1.5E-02 | ≤ 1.0E-01 | 3.4E-02 | 0.53 | – | Not detected |
| $C^{17}O$ | $v=0\ J=2-1$ | ≤ 4.6E-02 | 1.5E-02 | ≤ 1.0E-01 | 3.4E-02 | 0.52 | – | Not detected |
| SiO | $v=0\ J=1-0$ | ≤ 2.2E-02 | 7.3E-03 | ≤ 4.9E-02 | 1.6E-02 | 0.53 | – | Not detected |
| | $J=2-1$ | ≤ 1.6E-02 | 5.2E-03 | ≤ 3.5E-02 | 1.2E-02 | 0.67 | – | Not detected |
| | $J=5-4$ | ≤ 5.4E-02 | 1.8E-02 | ≤ 1.2E-01 | 4.0E-02 | 0.54 | – | Not detected |
| | $v=1\ J=1-0$ | ≤ 2.4E-02 | 7.9E-03 | ≤ 5.3E-02 | 1.8E-02 | 0.53 | – | Not detected |
| | $J=2-1$ | 1.7E-02 | 5.1E-03 | 2.1E-02 | 9.1E-03 | 0.68 | 0.41 | Tentative |
| | $v=2\ J=1-0$ | ≤ 2.6E-02 | 8.6E-03 | ≤ 5.8E-02 | 1.9E-02 | 0.53 | – | Not detected |
| | $J=2-1$ | ≤ 1.5E-02 | 5.1E-03 | ≤ 3.4E-02 | 1.1E-02 | 0.68 | – | Not detected |
| | $v=6\ J=1-0$ | ≤ 2.4E-02 | 7.9E-03 | ≤ 5.3E-02 | 1.8E-02 | 0.55 | – | Not detected |
| $^{29}SiO$ | $v=0\ J=1-0$ | ≤ 3.0E-02 | 9.9E-03 | ≤ 6.7E-02 | 2.2E-02 | 0.53 | – | Not detected |
| | $J=1-0$ | ≤ 1.6E-02 | 5.4E-03 | ≤ 3.6E-02 | 1.2E-02 | 0.68 | – | Not detected |
| $^{30}SiO$ | $v=0\ J=1-0$ | ≤ 2.4E-02 | 8.0E-03 | ≤ 5.4E-02 | 1.8E-02 | 0.54 | – | Not detected |
| | $J=2-1$ | ≤ 1.8E-02 | 6.0E-03 | ≤ 4.0E-02 | 1.3E-02 | 0.69 | – | Not detected |
| HCN | $v=0\ J=1-0$ | ≤ 1.2E-02 | 4.1E-03 | ≤ 2.7E-02 | 9.2E-03 | 0.66 | – | Not detected |
| HNC | $v=0\ J=1-0$ | ≤ 1.0E-02 | 3.5E-03 | ≤ 2.3E-02 | 7.7E-03 | 0.65 | – | Not detected |
| $HC_3N$ | $v=0\ J=10-9$ | ≤ 1.3E-02 | 4.2E-03 | ≤ 2.8E-02 | 9.4E-03 | 0.64 | – | Not detected |
| $H^{13}CN$ | $v=0\ J=4-3$ | ≤ 1.6E-02 | 5.4E-03 | ≤ 3.6E-02 | 1.2E-02 | 0.68 | – | Not detected |
| $HCO^+$ | $v=0\ J=1-0$ | ≤ 1.2E-02 | 4.1E-03 | ≤ 2.8E-02 | 9.3E-03 | 0.66 | – | Not detected |
| SiS | $v=0\ J=2-1$ | ≤ 1.8E-02 | 5.9E-03 | ≤ 4.0E-02 | 1.3E-02 | 0.63 | – | Not detected |
| | $J=5-4$ | ≤ 1.2E-02 | 4.1E-03 | ≤ 2.8E-02 | 9.2E-03 | 0.65 | – | Not detected |
| | $J=8-7$ | ≤ 3.0E-02 | 9.8E-03 | ≤ 6.6E-02 | 2.2E-02 | 0.81 | – | Not detected |
| | $v=1\ J=8-7$ | ≤ 2.6E-02 | 8.8E-03 | ≤ 5.9E-02 | 2.0E-02 | 0.81 | – | Not detected |
| SO | $v=0\ J_N=2_2-1_1$ | ≤ 1.6E-02 | 5.4E-03 | ≤ 3.6E-02 | 1.2E-02 | 0.68 | – | Not detected |
| | $J_N=6_5-5_4$ | ≤ 4.5E-02 | 1.5E-02 | ≤ 1.0E-01 | 3.3E-02 | 0.53 | – | Not detected |
| $SO_2$ | $v=0\ J_{Ka,\,Kc}=4_{2,\,2}-4_{1,\,3}$ | ≤ 2.9E-02 | 9.6E-03 | ≤ 6.4E-02 | 2.1E-02 | 0.80 | – | Not detected |
| CS | $v=0\ J=1-0$ | ≤ 3.9E-02 | 1.3E-02 | ≤ 8.7E-02 | 2.9E-02 | 0.93 | – | Not detected |
| | $J=3-2$ | ≤ 2.8E-02 | 9.2E-03 | ≤ 6.2E-02 | 2.1E-02 | 0.80 | – | Not detected |
| $H_2O$ | $v=0\ J_{Ka,\,Kc}=6_{1,\,6}-5_{2,\,3}$ | ≤ 1.1E-01 | 3.6E-02 | ≤ 2.4E-01 | 8.1E-02 | 0.82 | – | Not detected |

Table 5.10 Radio molecular line survey of the Red Rectangle.

| Molecule | Transition | I (peak) [Jy] | $\sigma$ [Jy] | $\int I\,\mathrm{d}V$ [Jy km s$^{-1}$] | $\sigma\left(\int I\,\mathrm{d}V\right)$ [Jy km s$^{-1}$] | Sp. Res. [km s$^{-1}$] | $V_{\rm LSR}$ [km s$^{-1}$] | Comments |
|---|---|---|---|---|---|---|---|---|
| $C^{18}O$ | $v=0\;J=2-1$ | 1.7E-01 | 2.4E-02 | 9.5E-01 | 6.5E-02 | 0.53 | −0.64 | |
| $C^{17}O$ | $v=0\;J=2-1$ | 1.8E-01 | 2.4E-02 | 7.9E-01 | 4.8E-02 | 0.52 | −0.59 | |
| SiO | $v=0\;J=1-0$ | ≤ 2.4E-02 | 8.0E-03 | ≤ 5.3E-02 | 1.8E-02 | 0.53 | – | Not detected |
| | $J=2-1$ | ≤ 2.2E-02 | 7.5E-03 | ≤ 4.9E-02 | 1.6E-02 | 0.67 | – | Not detected |
| | $J=5-4$ | – | – | – | – | – | – | Not observed |
| | $v=1\;J=1-0$ | ≤ 2.7E-02 | 8.9E-03 | ≤ 5.8E-02 | 1.9E-02 | 0.53 | – | Not detected |
| | $J=2-1$ | ≤ 2.7E-02 | 9.1E-03 | ≤ 6.0E-02 | 2.0E-02 | 0.68 | – | Not detected |
| | $v=2\;J=1-0$ | ≤ 2.5E-02 | 8.5E-03 | ≤ 5.5E-02 | 1.8E-02 | 0.53 | – | Not detected |
| | $J=2-1$ | ≤ 2.8E-02 | 9.4E-03 | ≤ 6.1E-02 | 2.0E-02 | 0.68 | – | Not detected |
| | $v=6\;J=1-0$ | ≤ 2.6E-02 | 8.7E-03 | ≤ 5.7E-02 | 1.9E-02 | 0.55 | – | Not detected |
| $^{29}SiO$ | $v=0\;J=1-0$ | ≤ 2.7E-02 | 9.2E-03 | ≤ 6.0E-02 | 2.0E-02 | 0.53 | – | Not detected |
| | $J=2-1$ | ≤ 2.4E-02 | 7.8E-03 | ≤ 5.1E-02 | 1.7E-02 | 0.68 | – | Not detected |
| $^{30}SiO$ | $v=0\;J=1-0$ | ≤ 2.6E-02 | 8.6E-03 | ≤ 5.6E-02 | 1.9E-02 | 0.54 | – | Not detected |
| | $J=2-1$ | ≤ 2.9E-02 | 9.6E-03 | ≤ 6.3E-02 | 2.1E-02 | 0.69 | – | Not detected |
| HCN | $v=0\;J=1-0$ | 1.3E-02 | 3.9E-03 | 8.6E-02 | 2.8E-02 | 2.64 | 7.86 | Tentative |
| HNC | $v=0\;J=1-0$ | ≤ 1.9E-02 | 6.3E-03 | ≤ 4.1E-02 | 1.4E-02 | 0.65 | – | Not detected |
| $HC_3N$ | $v=0\;J=10-9$ | ≤ 2.2E-02 | 7.4E-03 | ≤ 4.9E-02 | 1.6E-02 | 0.64 | – | Not detected |
| $H^{13}CN$ | $v=0\;J=4-3$ | 2≤ .3E-02 | 7.5E-03 | ≤ 4.9E-02 | 1.6E-02 | 0.68 | – | Not detected |
| $HCO^+$ | $v=0\;J=1-0$ | ≤ 2.3E-02 | 7.6E-03 | ≤ 5.0E-02 | 1.7E-02 | 0.66 | – | Not detected |
| SiS | $v=0\;J=2-1$ | ≤ 2.1E-02 | 6.9E-03 | ≤ 4.5E-02 | 1.5E-02 | 0.63 | – | Not detected |
| | $J=5-4$ | ≤ 2.2E-02 | 7.4E-03 | ≤ 4.8E-02 | 1.6E-02 | 0.65 | – | Not detected |
| | $J=8-7$ | ≤ 4.1E-02 | 1.4E-02 | ≤ 9.0E-02 | 3.0E-02 | 0.81 | – | Not detected |
| | $v=1\;J=8-7$ | ≤ 4.1E-02 | 1.4E-02 | ≤ 8.9E-02 | 3.0E-02 | 0.81 | – | Not detected |
| SO | $v=0\;J_N=2_2-1_1$ | ≤ 2.8E-02 | 9.2E-03 | ≤ 5.9E-02 | 1.9E-02 | 0.68 | – | Not detected |
| | $J_N=6_5-5_4$ | 3.5E-02 | 6.7E-03 | 3.3E-01 | 6.6E-02 | 4.26 | −1.15 | |
| $SO_2$ | $v=0\;J_{Ka,Kc}=4_{2,2}-4_{1,3}$ | ≤ 4.3E-02 | 1.4E-02 | ≤ 9.5E-02 | 3.2E-02 | 0.80 | – | Not detected |
| CS | $v=0\;J=1-0$ | ≤ 4.3E-02 | 1.4E-02 | ≤ 9.4E-02 | 3.1E-02 | 0.93 | – | Not detected |
| | $J=3-2$ | ≤ 4.0E-02 | 1.3E-02 | ≤ 8.7E-02 | 2.9E-02 | 0.80 | – | Not detected |
| $H_2O$ | $v=0\;J_{Ka,Kc}=6_{1,6}-5_{2,3}$ | 5.3E-02 | 1.1E-02 | 6.2E-01 | 9.3E-02 | 3.29 | 8.27 | Tentative |

Table 5.11 Radio molecular line survey of 89 Her.

| Molecule | Transition | I (peak) [Jy] | $\sigma$ [Jy] | $\int I\,\mathrm{d}V$ [Jy km s$^{-1}$] | $\sigma\left(\int I\,\mathrm{d}V\right)$ [Jy km s$^{-1}$] | Sp. Res. [km s$^{-1}$] | $V_{\mathrm{LSR}}$ [km s$^{-1}$] | Comments |
|---|---|---|---|---|---|---|---|---|
| $C^{18}O$ | $v=0\ J=2-1$ | 1.2E-01 | 1.8E-02 | 2.0E-01 | 2.0E-02 | 0.27 | -8.36 | |
| $C^{17}O$ | $v=0\ J=2-1$ | 1.3E-02 | 5.1E-03 | 6.1E-02 | 5.2E-02 | 4.17 | -11.10 | Tentative |
| SiO | $v=0\ J=1-0$ | ≤ 2.3E-02 | 7.8E-03 | ≤ 4.3E-02 | 1.4E-02 | 0.53 | – | Not detected |
| | $J=2-1$ | ≤ 1.7E-02 | 5.8E-03 | ≤ 3.2E-02 | 1.1E-02 | 0.67 | – | Not detected |
| | $J=5-4$ | ≤ 9.4E-02 | 3.1E-02 | ≤ 1.7E-01 | 5.8E-02 | 0.54 | – | Not detected |
| | $v=1\ J=1-0$ | ≤ 2.2E-02 | 7.5E-03 | ≤ 4.1E-02 | 1.4E-02 | 0.53 | – | Not detected |
| | $J=2-1$ | ≤ 1.8E-02 | 5.9E-03 | ≤ 3.3E-02 | 1.1E-02 | 0.68 | – | Not detected |
| | $v=2\ J=1-0$ | ≤ 2.2E-02 | 7.2E-03 | ≤ 4.0E-02 | 1.3E-02 | 0.53 | – | Not detected |
| | $J=2-1$ | ≤ 1.8E-02 | 6.1E-03 | ≤ 3.4E-02 | 1.1E-02 | 0.68 | – | Not detected |
| | $v=6\ J=1-0$ | ≤ 2.1E-02 | 7.1E-03 | ≤ 3.9E-02 | 1.3E-02 | 0.55 | – | Not detected |
| $^{29}SiO$ | $v=0\ J=1-0$ | ≤ 2.4E-02 | 7.9E-03 | ≤ 4.4E-02 | 1.5E-02 | 0.53 | – | Not detected |
| | $J=2-1$ | ≤ 2.0E-02 | 6.6E-03 | ≤ 3.6E-02 | 1.2E-02 | 0.68 | – | Not detected |
| $^{30}SiO$ | $v=0\ J=1-0$ | ≤ 2.4E-02 | 7.9E-03 | ≤ 4.4E-02 | 1.5E-02 | 0.54 | – | Not detected |
| | $J=2-1$ | ≤ 2.0E-02 | 6.5E-03 | ≤ 3.6E-02 | 1.2E-02 | 0.69 | – | Not detected |
| HCN | $v=0\ J=1-0$ | 1.2E-02 | 2.8E-03 | 2.1E-01 | 2.5E-02 | 2.64 | -14.76 | Tentative |
| HNC | $v=0\ J=1-0$ | ≤ 1.4E-02 | 4.7E-03 | ≤ 2.6E-02 | 8.6E-03 | 0.65 | – | Not detected |
| $HC_3N$ | $v=0\ J=10-9$ | ≤ 1.3E-02 | 4.5E-03 | ≤ 2.5E-02 | 8.2E-03 | 0.64 | – | Not detected |
| $H^{13}CN$ | $v=0\ J=4-3$ | ≤ 2.0E-02 | 6.6E-03 | ≤ 3.6E-02 | 1.2E-02 | 0.68 | – | Not detected |
| $HCO^+$ | $v=0\ J=1-0$ | ≤ 1.3E-02 | 4.3E-03 | ≤ 2.4E-02 | 8.0E-03 | 0.66 | – | Not detected |
| SiS | $v=0\ J=2-1$ | ≤ 1.9E-02 | 6.2E-03 | ≤ 3.4E-02 | 1.1E-02 | 0.63 | – | Not detected |
| | $J=5-4$ | 1.5E-02 | 3.8E-03 | 3.3E-03 | 6.7E-03 | 0.64 | -11.57 | |
| | $J=8-7$ | ≤ 2.0E-02 | 6.8E-03 | ≤ 3.7E-02 | 1.2E-02 | 0.81 | – | Not detected |
| | $v=1\ J=8-7$ | ≤ 2.4E-02 | 7.9E-03 | ≤ 4.3E-02 | 1.4E-02 | 0.81 | – | Not detected |
| SO | $v=0\ J_N=2_2-1_1$ | ≤ 2.1E-02 | 6.9E-03 | ≤ 3.8E-02 | 1.3E-02 | 0.68 | – | Not detected |
| | $J_N=6_5-5_4$ | ≤ 4.9E-02 | 1.6E-02 | ≤ 9.1E-02 | 3.0E-02 | 0.53 | – | Not detected |
| $SO_2$ | $v=0\ J_{Ka,Kc}=4_{2,2}-4_{1,3}$ | ≤ 2.2E-02 | 7.5E-03 | ≤ 4.1E-02 | 1.4E-02 | 0.40 | – | Not detected |
| CS | $v=0\ J=1-0$ | ≤ 5.2E-02 | 1.7E-02 | ≤ 9.6E-02 | 3.2E-02 | 0.47 | – | Not detected |
| | $J=3-2$ | 4.9E-02 | 8.1E-03 | 9.7E-02 | 1.4E-02 | 0.40 | -7.42 | |
| $H_2O$ | $v=0\ J_{Ka,Kc}=6_{1,6}-5_{2,3}$ | ≤ 7.3E-02 | 2.4E-02 | ≤ 1.3E-01 | 4.5E-02 | 0.82 | – | Not detected |

Table 5.12 Radio molecular line survey of HD 52961.

| Molecule | Transition | I (peak) [Jy] | $\sigma$ [Jy] | $\int I\,dV$ [Jy km s$^{-1}$] | $\sigma\left(\int I\,dV\right)$ [Jy km s$^{-1}$] | Sp. Res. [km s$^{-1}$] | $V_{LSR}$ [km s$^{-1}$] | Comments |
|---|---|---|---|---|---|---|---|---|
| $C^{18}O$ | $v=0\ J=2-1$ | $\leq$ 1.3E-01 | 4.3E-02 | $\leq$ 3.3E-01 | 1.1E-01 | 0.53 | – | Not detected |
| $C^{17}O$ | $v=0\ J=2-1$ | $\leq$ 1.3E-01 | 4.4E-02 | $\leq$ 3.3E-01 | 1.1E-01 | 0.52 | – | Not detected |
| SiO | $v=0\ J=1-0$ | $\leq$ 1.8E-02 | 6.0E-03 | $\leq$ 4.5E-02 | 1.5E-02 | 0.53 | – | Not detected |
| | $J=2-1$ | $\leq$ 7.7E-02 | 2.6E-02 | $\leq$ 1.9E-01 | 6.4E-02 | 0.67 | – | Not detected |
| | $J=5-4$ | $\leq$ 1.5E-01 | 4.9E-02 | $\leq$ 3.7E-01 | 1.2E-01 | 0.54 | – | Not detected |
| | $v=1\ J=1-0$ | $\leq$ 1.9E-02 | 6.4E-03 | $\leq$ 4.8E-02 | 1.6E-02 | 0.53 | – | Not detected |
| | $J=2-1$ | $\leq$ 8.1E-02 | 2.7E-02 | $\leq$ 2.0E-01 | 6.7E-02 | 0.68 | – | Not detected |
| | $v=2\ J=1-0$ | $\leq$ 1.9E-02 | 6.4E-03 | $\leq$ 4.8E-02 | 1.6E-02 | 0.53 | – | Not detected |
| | $J=2-1$ | $\leq$ 7.4E-02 | 2.5E-02 | $\leq$ 1.9E-01 | 6.2E-02 | 0.68 | – | Not detected |
| | $v=6\ J=1-0$ | $\leq$ 2.0E-02 | 6.6E-03 | $\leq$ 5.0E-02 | 1.7E-02 | 0.55 | – | Not detected |
| $^{29}SiO$ | $v=0\ J=1-0$ | $\leq$ 1.9E-02 | 6.5E-03 | $\leq$ 4.8E-02 | 1.6E-02 | 0.53 | – | Not detected |
| | $J=2-1$ | $\leq$ 7.8E-02 | 2.6E-02 | $\leq$ 2.0E-01 | 6.5E-02 | 0.68 | – | Not detected |
| $^{30}SiO$ | $v=0\ J=1-0$ | $\leq$ 2.1E-02 | 7.0E-03 | $\leq$ 5.3E-02 | 1.8E-02 | 0.54 | – | Not detected |
| | $J=2-1$ | $\leq$ 8.7E-02 | 2.9E-02 | $\leq$ 2.2E-01 | 7.2E-02 | 0.69 | – | Not detected |
| HCN | $v=0\ J=1-0$ | $\leq$ 5.8E-02 | 1.9E-02 | $\leq$ 1.4E-01 | 4.8E-02 | 0.66 | – | Not detected |
| HNC | $v=0\ J=1-0$ | $\leq$ 5.3E-02 | 1.8E-02 | $\leq$ 1.3E-01 | 4.4E-02 | 0.65 | – | Not detected |
| $HC_3N$ | $v=0\ J=10-9$ | $\leq$ 5.4E-02 | 1.8E-02 | $\leq$ 1.3E-01 | 4.5E-02 | 0.64 | – | Not detected |
| $H^{13}CN$ | $v=0\ J=4-3$ | $\leq$ 7.1E-02 | 2.4E-02 | $\leq$ 1.8E-01 | 5.9E-02 | 0.68 | – | Not detected |
| $HCO^+$ | $v=0\ J=1-0$ | $\leq$ 4.9E-02 | 1.6E-02 | $\leq$ 1.2E-01 | 4.1E-02 | 0.66 | – | Not detected |
| SiS | $v=0\ J=2-1$ | $\leq$ 1.5E-02 | 5.1E-03 | $\leq$ 3.8E-02 | 1.3E-02 | 0.63 | – | Not detected |
| | $J=5-4$ | $\leq$ 5.0E-02 | 1.7E-02 | $\leq$ 1.3E-01 | 4.2E-02 | 0.65 | – | Not detected |
| | $J=8-7$ | $\leq$ 1.5E-01 | 4.9E-02 | $\leq$ 3.7E-01 | 1.2E-01 | 0.81 | – | Not detected |
| | $v=1\ J=8-7$ | $\leq$ 1.3E-01 | 4.2E-02 | $\leq$ 3.2E-01 | 1.1E-01 | 0.81 | – | Not detected |
| SO | $v=0\ J_N=2_2-1_1$ | $\leq$ 7.3E-02 | 2.4E-02 | $\leq$ 1.8E-01 | 6.1E-02 | 0.68 | – | Not detected |
| | $J_N=6_5-5_4$ | $\leq$ 1.3E-01 | 4.4E-02 | $\leq$ 3.3E-01 | 1.1E-01 | 0.53 | – | Not detected |
| $SO_2$ | $v=0\ J_{Ka,Kc}=4_{2,2}-4_{1,3}$ | $\leq$ 1.3E-01 | 4.4E-02 | $\leq$ 3.3E-01 | 1.1E-01 | 0.80 | – | Not detected |
| CS | $v=0\ J=1-0$ | $\leq$ 3.5E-02 | 1.2E-02 | $\leq$ 8.7E-02 | 2.9E-02 | 0.93 | – | Not detected |
| | $J=3-2$ | $\leq$ 1.4E-01 | 4.5E-02 | $\leq$ 3.4E-01 | 1.1E-01 | 0.80 | – | Not detected |
| | $v=1\ J=3-2$ | $\leq$ 1.4E-01 | 4.7E-02 | $\leq$ 3.6E-01 | 1.2E-01 | 0.80 | – | Not detected |

Table 5.13 Radio molecular line survey of IRAS 19157−0257.

| Molecule | Transition | I (peak) [Jy] | $\sigma$ [Jy] | $\int I\,\mathrm{d}V$ [Jy km s$^{-1}$] | $\sigma\left(\int I\,\mathrm{d}V\right)$ [Jy km s$^{-1}$] | Sp. Res. [km s$^{-1}$] | $V_{\mathrm{LSR}}$ [km s$^{-1}$] | Comments |
|---|---|---|---|---|---|---|---|---|
| SiO | $v=0$ $J=1-0$ | ≤ 4.2E-02 | 1.4E-02 | ≤ 1.3E-01 | 4.4E-02 | 0.53 | – | Not detected |
| | $J=2-1$ | ≤ 1.5E-01 | 5.1E-02 | ≤ 4.7E-01 | 1.6E-01 | 0.67 | – | Not detected |
| | $v=1$ $J=1-0$ | ≤ 4.1E-02 | 1.4E-02 | ≤ 1.3E-01 | 4.2E-02 | 0.53 | – | Not detected |
| | $J=2-1$ | ≤ 1.5E-01 | 5.0E-02 | ≤ 4.6E-01 | 1.5E-01 | 0.68 | – | Not detected |
| | $v=2$ $J=1-0$ | ≤ 3.9E-02 | 1.3E-02 | ≤ 7.2E-02 | 2.4E-02 | 0.53 | – | Not detected |
| | $J=2-1$ | ≤ 1.6E-01 | 5.2E-02 | ≤ 4.8E-01 | 1.6E-01 | 0.68 | – | Not detected |
| | $v=6$ $J=1-0$ | ≤ 4.3E-02 | 1.4E-02 | ≤ 1.3E-01 | 4.4E-02 | 0.55 | – | Not detected |
| $^{29}$SiO | $v=0$ $J=1-0$ | ≤ 4.3E-02 | 1.4E-02 | ≤ 1.3E-01 | 4.4E-02 | 0.53 | – | Not detected |
| | $J=1-0$ | ≤ 1.3E-01 | 4.3E-02 | ≤ 4.0E-01 | 1.3E-01 | 0.68 | – | Not detected |
| $^{30}$SiO | $v=0$ $J=1-0$ | ≤ 4.6E-02 | 1.5E-02 | ≤ 1.4E-01 | 4.7E-02 | 0.54 | – | Not detected |
| | $J=2-1$ | ≤ 1.9E-01 | 6.3E-02 | ≤ 5.8E-01 | 1.9E-01 | 0.69 | – | Not detected |
| HCN | $v=0$ $J=1-0$ | ≤ 9.8E-02 | 3.3E-02 | ≤ 3.0E-01 | 1.0E-01 | 0.66 | – | Not detected |
| HNC | $v=0$ $J=1-0$ | ≤ 7.2E-02 | 2.4E-02 | ≤ 2.4E-02 | 8.0E-02 | 1.29 | – | Not detected |
| $HC_3N$ | $v=0$ $J=10-9$ | ≤ 1.0E-01 | 3.5E-02 | ≤ 3.2E-01 | 1.1E-01 | 0.64 | – | Not detected |
| $H^{13}CN$ | $v=0$ $J=4-3$ | ≤ 1.3E-01 | 4.4E-02 | ≤ 4.1E-01 | 1.4E-01 | 0.68 | – | Not detected |
| $HCO^+$ | $v=0$ $J=1-0$ | ≤ 9.4E-02 | 3.1E-02 | ≤ 2.9E-01 | 9.7E-02 | 0.66 | – | Not detected |
| SiS | $v=0$ $J=2-1$ | ≤ 2.8E-02 | 9.3E-03 | ≤ 8.7E-02 | 2.9E-02 | 0.63 | – | Not detected |
| | $J=5-4$ | ≤ 1.0E-01 | 3.5E-02 | ≤ 3.2E-01 | 1.1E-01 | 0.65 | – | Not detected |
| | $J=8-7$ | ≤ 1.7E-01 | 5.6E-02 | ≤ 5.2E-01 | 1.7E-01 | 0.81 | – | Not detected |
| | $v=1$ $J=8-7$ | ≤ 1.8E-01 | 6.1E-02 | ≤ 5.6E-01 | 1.9E-01 | 0.81 | – | Not detected |
| SO | $v=0$ $J_N=2_2-1_1$ | ≤ 1.5E-01 | 4.8E-02 | ≤ 4.5E-01 | 1.5E-01 | 0.68 | – | Not detected |
| $SO_2$ | $v=0$ $J_{Ka,\,Kc}=4_{2,\,2}-4_{1,\,3}$ | ≤ 1.5E-01 | 5.1E-02 | ≤ 4.7E-01 | 1.6E-01 | 0.80 | – | Not detected |
| CS | $v=0$ $J=1-0$ | ≤ 8.7E-02 | 2.9E-02 | ≤ 2.7E-01 | 8.9E-02 | 0.47 | – | Not detected |
| | $J=3-2$ | ≤ 1.5E-01 | 5.1E-02 | ≤ 4.7E-01 | 1.6E-01 | 0.80 | – | Not detected |
| $H_2O$ | $v=0$ $J_{Ka,\,Kc}=6_{1,\,6}-5_{2,\,3}$ | ≤ 1.0E-01 | 3.4E-02 | ≤ 3.2E-01 | 1.1E-01 | 0.82 | – | Not detected |

Table 5.14 Radio molecular line survey of IRAS 18123+0511.

| Molecule | Transition | I (peak) [Jy] | $\sigma$ [Jy] | $\int I\,dV$ [Jy km s$^{-1}$] | $\sigma\left(\int I\,dV\right)$ [Jy km s$^{-1}$] | Sp. Res. [km s$^{-1}$] | $V_{LSR}$ [km s$^{-1}$] | Comments |
|---|---|---|---|---|---|---|---|---|
| $C^{18}O$ | $v=0$ $J=2-1$ | $\leq$2.6E-01 | 8.7E-02 | $\leq$1.6E+00 | 5.4E-01 | 0.53 | – | Not detected |
| $C^{17}O$ | $v=0$ $J=2-1$ | $\leq$2.8E-01 | 9.3E-02 | $\leq$1.7E+00 | 5.8E-01 | 0.52 | – | Not detected |
| SiO | $v=0$ $J=1-0$ | $\leq$5.0E-02 | 1.7E-02 | $\leq$3.1E-01 | 1.0E-01 | 0.53 | – | Not detected |
| | $J=2-1$ | $\leq$1.0E-01 | 3.3E-02 | $\leq$6.2E-01 | 2.1E-01 | 0.67 | – | Not detected |
| | $J=5-4$ | $\leq$3.2E-01 | 1.1E-01 | $\leq$2.0E+00 | 6.5E-01 | 0.54 | – | Not detected |
| | $v=1$ $J=1-0$ | $\leq$2.6E-02 | 8.8E-03 | $\leq$1.2E-02 | 4.0E-02 | 2.12 | – | Not detected |
| | $J=2-1$ | $\leq$1.0E-01 | 3.3E-02 | $\leq$6.2E-01 | 2.1E-01 | 0.68 | – | Not detected |
| | $v=2$ $J=1-0$ | $\leq$4.9E-02 | 1.6E-02 | $\leq$3.1E-01 | 1.0E-01 | 0.53 | – | Not detected |
| | $J=2-1$ | $\leq$1.1E-01 | 3.5E-02 | $\leq$6.6E-01 | 2.2E-01 | 0.68 | – | Not detected |
| | $v=6$ $J=1-0$ | $\leq$5.0E-02 | 1.7E-02 | $\leq$3.1E-01 | 1.0E-01 | 0.55 | – | Not detected |
| $^{29}SiO$ | $v=0$ $J=1-0$ | $\leq$5.3E-02 | 1.8E-02 | $\leq$3.3E-01 | 1.1E-01 | 0.53 | – | Not detected |
| | $J=1-0$ | $\leq$1.0E-01 | 3.4E-02 | $\leq$6.4E-01 | 2.1E-01 | 0.68 | – | Not detected |
| $^{30}SiO$ | $v=0$ $J=1-0$ | $\leq$4.9E-02 | 1.6E-02 | $\leq$3.1E-01 | 1.0E-01 | 0.54 | – | Not detected |
| | $J=2-1$ | $\leq$1.2E-01 | 4.1E-02 | $\leq$7.6E-01 | 2.5E-01 | 0.69 | – | Not detected |
| HCN | $v=0$ $J=1-0$ | $\leq$7.4E-02 | 2.5E-02 | $\leq$4.6E-01 | 1.5E-01 | 0.66 | – | Not detected |
| HNC | $v=0$ $J=1-0$ | $\leq$5.9E-02 | 2.0E-02 | $\leq$3.7E-01 | 1.2E-01 | 0.65 | – | Not detected |
| $HC_3N$ | $v=0$ $J=10-9$ | $\leq$7.0E-02 | 2.3E-02 | $\leq$4.4E-01 | 1.5E-01 | 0.64 | – | Not detected |
| $H^{13}CN$ | $v=0$ $J=4-3$ | $\leq$9.8E-02 | 3.3E-02 | $\leq$6.1E-01 | 2.0E-01 | 0.68 | – | Not detected |
| $H^{13}CN$ | $v=0$ $J=4-3$ | $\leq$5.9E-02 | 2.0E-02 | $\leq$3.7E-01 | 1.2E-01 | 0.66 | – | Not detected |
| SiS | $v=0$ $J=2-1$ | $\leq$3.5E-02 | 1.2E-02 | $\leq$2.2E-01 | 7.3E-02 | 0.63 | – | Not detected |
| | $J=5-4$ | $\leq$6.8E-02 | 2.3E-02 | $\leq$4.2E-01 | 1.4E-01 | 0.65 | – | Not detected |
| | $J=8-7$ | $\leq$1.6E-01 | 5.4E-02 | $\leq$1.0E+00 | 3.3E-01 | 0.81 | – | Not detected |
| | $v=1$ $J=8-7$ | $\leq$1.6E-01 | 5.4E-02 | $\leq$1.0E+00 | 3.4E-01 | 0.81 | – | Not detected |
| SO | $v=0$ $J_N=2_2-1_1$ | $\leq$1.1E-01 | 3.6E-02 | $\leq$6.7E-01 | 2.2E-01 | 0.68 | – | Not detected |
| | $J_N=6_5-5_4$ | $\leq$2.7E-01 | 9.0E-02 | $\leq$1.7E+00 | 5.6E-01 | 0.53 | – | Not detected |
| $SO_2$ | $v=0$ $J_{Ka,Kc}=4_{2,2}-4_{1,3}$ | $\leq$1.6E-01 | 5.4E-02 | $\leq$1.0E+00 | 3.4E-01 | 0.80 | – | Not detected |
| CS | $v=0$ $J=1-0$ | $\leq$8.1E-02 | 2.7E-02 | $\leq$5.1E-01 | 1.7E-01 | 0.93 | – | Not detected |
| | $J=3-2$ | $\leq$1.7E-01 | 5.6E-02 | $\leq$1.1E+00 | 3.5E-01 | 0.80 | – | Not detected |
| $H_2O$ | $v=0$ $J_{Ka,Kc}=6_{1,6}-5_{2,3}$ | $\leq$7.3E-02 | 2.4E-02 | $\leq$4.5E-01 | 1.5E-01 | 0.82 | – | Not detected |

Table 5.15 Radio molecular line survey of IRAS 19125+0343.

| Molecule | Transition | I (peak) [Jy] | $\sigma$ [Jy] | $\int I\,dV$ [Jy km s$^{-1}$] | $\sigma\left(\int I\,dV\right)$ [Jy km s$^{-1}$] | Sp. Res. [km s$^{-1}$] | $V_{LSR}$ [km s$^{-1}$] | Comments |
|---|---|---|---|---|---|---|---|---|
| $C^{18}O$ | $v=0\ J=2-1$ | $\leq$ 1.7E-01 | 5.8E-02 | $\leq$ 6.0E-01 | 2.0E-01 | 0.53 | – | Not detected |
| $C^{17}O$ | $v=0\ J=2-1$ | $\leq$ 1.5E-01 | 4.9E-02 | $\leq$ 5.1E-01 | 1.7E-01 | 0.52 | – | Not detected |
| SiO | $v=0\ J=1-0$ | $\leq$ 2.1E-02 | 7.2E-03 | $\leq$ 7.4E-02 | 2.5E-02 | 0.53 | – | Not detected |
| | $J=2-1$ | $\leq$ 5.4E-02 | 1.8E-02 | $\leq$ 1.8E-01 | 6.1E-02 | 0.67 | – | Not detected |
| | $J=5-4$ | $\leq$ 2.2E-01 | 7.2E-02 | $\leq$ 7.5E-01 | 2.5E-01 | 0.54 | – | Not detected |
| | $v=1\ J=1-0$ | $\leq$ 2.4E-02 | 8.1E-03 | $\leq$ 8.5E-02 | 2.8E-02 | 0.53 | – | Not detected |
| | $J=2-1$ | $\leq$ 5.4E-02 | 1.8E-02 | $\leq$ 1.9E-01 | 6.2E-02 | 0.68 | – | Not detected |
| | $v=2\ J=1-0$ | $\leq$ 1.1E-02 | 3.7E-03 | $\leq$ 1.8E-02 | 1.8E-02 | 2.14 | – | Not detected |
| | $J=2-1$ | $\leq$ 5.4E-02 | 1.8E-02 | $\leq$ 1.9E-01 | 6.3E-02 | 0.68 | – | Not detected |
| | $v=6\ J=1-0$ | $\leq$ 2.3E-02 | 7.7E-03 | $\leq$ 8.0E-02 | 2.7E-02 | 0.55 | – | Not detected |
| $^{29}SiO$ | $v=0\ J=1-0$ | $\leq$ 2.2E-02 | 7.5E-03 | $\leq$ 7.8E-02 | 2.6E-02 | 0.53 | – | Not detected |
| | $J=2-1$ | $\leq$ 5.2E-02 | 1.7E-02 | $\leq$ 1.8E-01 | 6.0E-02 | 0.68 | – | Not detected |
| $^{30}SiO$ | $v=0\ J=1-0$ | $\leq$ 2.4E-02 | 8.1E-03 | $\leq$ 8.4E-02 | 2.8E-02 | 0.54 | – | Not detected |
| | $J=2-1$ | $\leq$ 5.3E-02 | 1.8E-02 | $\leq$ 1.9E-01 | 6.2E-02 | 0.69 | – | Not detected |
| HCN | $v=0\ J=1-0$ | $\leq$ 3.9E-02 | 1.3E-02 | $\leq$ 1.4E-01 | 4.5E-02 | 0.66 | – | Not detected |
| HNC | $v=0\ J=1-0$ | $\leq$ 3.6E-02 | 1.2E-02 | $\leq$ 1.2E-01 | 4.1E-02 | 0.65 | – | Not detected |
| $HC_3N$ | $v=0\ J=10-9$ | $\leq$ 3.7E-02 | 1.2E-02 | $\leq$ 1.3E-01 | 4.2E-02 | 0.64 | – | Not detected |
| $H^{13}CN$ | $v=0\ J=4-3$ | $\leq$ 5.3E-02 | 1.8E-02 | $\leq$ 1.8E-01 | 6.1E-02 | 0.68 | – | Not detected |
| $HCO^+$ | $v=0\ J=1-0$ | $\leq$ 3.8E-02 | 1.3E-02 | $\leq$ 1.3E-01 | 4.4E-02 | 0.66 | – | Not detected |
| SiS | $v=0\ J=2-1$ | $\leq$ 1.8E-02 | 6.1E-03 | $\leq$ 6.3E-02 | 2.1E-02 | 0.63 | – | Not detected |
| | $J=5-4$ | $\leq$ 3.6E-02 | 1.2E-02 | $\leq$ 1.3E-01 | 4.2E-02 | 0.65 | – | Not detected |
| SO | $v=0\ J_N=2_2-1_1$ | $\leq$ 5.3E-02 | 1.8E-02 | $\leq$ 1.8E-01 | 6.1E-02 | 0.68 | – | Not detected |
| | $J_N=6_5-5_4$ | $\leq$ 1.7E-01 | 5.5E-02 | $\leq$ 5.7E-01 | 1.9E-01 | 0.53 | – | Not detected |
| CS | $v=0\ J=1-0$ | $\leq$ 5.5E-02 | 1.8E-02 | $\leq$ 1.9E-01 | 6.3E-02 | 0.47 | – | Not detected |
| $H_2O$ | $v=0\ J_{Ka,Kc}=6_{1,6}-5_{2,3}$ | $\leq$ 7.0E-02 | 2.3E-02 | $\leq$ 2.4E-01 | 8.1E-02 | 0.82 | – | Not detected |

Table 5.16 Radio molecular line survey of AI CMi.

| Molecule | Transition | I (peak) [Jy] | $\sigma$ [Jy] | $\int I\,dV$ [Jy km s$^{-1}$] | $\sigma\left(\int I\,dV\right)$ [Jy km s$^{-1}$] | Sp. Res. [km s$^{-1}$] | $V_{LSR}$ [km s$^{-1}$] | Comments |
|---|---|---|---|---|---|---|---|---|
| $C^{18}O$ | $v=0\ J=2-1$ | ≤ 1.0E-01 | 3.5E-02 | ≤ 1.9E-01 | 6.4E-02 | 0.53 | – | Not detected |
| $C^{17}O$ | $v=0\ J=2-1$ | ≤ 9.5E-02 | 3.2E-02 | ≤ 1.7E-01 | 5.8E-02 | 0.52 | – | Not detected |
| SiO | $v=0\ J=1-0$ | 2.0E-02 | 5.4E-03 | 7.5E-02 | 2.0E-02 | 1.05 | 29.11 | |
| | $J=2-1$ | 1.1E-01 | 1.8E-02 | 8.8E-01 | 7.0E-02 | 0.67 | 29.02 | |
| | $J=5-4$ | 2.9E-01 | 3.7E-02 | 2.0E+00 | 1.0E-01 | 0.54 | 29.00 | |
| | $v=1\ J=1-0$ | 1.4E+00 | 9.2E-03 | 2.9E+00 | 1.8E-02 | 0.27 | 28.87 | |
| | $J=2-1$ | 1.9E+00 | 1.7E-02 | 4.7E+00 | 4.8E-02 | 0.68 | 28.81 | |
| | $v=2\ J=1-0$ | 4.3E-01 | 8.9E-03 | 6.4E-01 | 1.6E-02 | 0.27 | 29.85 | |
| | $J=2-1$ | ≤ 4.6E-02 | 1.5E-02 | ≤ 8.4E-02 | 2.8E-02 | 0.68 | – | Not detected |
| | $v=6\ J=1-0$ | ≤ 2.2E-02 | 7.4E-03 | ≤ 4.1E-02 | 1.4E-02 | 0.55 | – | Not detected |
| $^{29}SiO$ | $v=0\ J=1-0$ | 7.7E-03 | 3.9E-03 | 4.9E-02 | 1.5E-02 | 2.13 | 35.04 | Tentative |
| | $J=2-1$ | 2.6E-02 | 7.5E-03 | 2.5E-01 | 5.7E-02 | 2.73 | 28.52 | Tentative |
| $^{30}SiO$ | $v=0\ J=1-0$ | 1.2E-02 | 3.2E-03 | 8.4E-02 | 1.7E-02 | 2.16 | 29.03 | Tentative |
| | $J=2-1$ | 6.0E-02 | 2.0E-02 | 1.1E-01 | 3.7E-02 | 0.69 | – | Not detected |
| HCN | $v=0\ J=1-0$ | ≤ 3.8E-02 | 1.3E-02 | ≤ 6.9E-02 | 2.3E-02 | 0.66 | – | Not detected |
| HNC | $v=0\ J=1-0$ | ≤ 3.6E-02 | 1.2E-02 | ≤ 6.6E-02 | 2.2E-02 | 0.65 | – | Not detected |
| $HC_3N$ | $v=0\ J=10-9$ | ≤ 3.5E-02 | 1.2E-02 | ≤ 6.4E-02 | 2.1E-02 | 0.64 | – | Not detected |
| $H^{13}CN$ | $v=0\ J=4-3$ | ≤ 4.8E-02 | 1.6E-02 | ≤ 8.9E-02 | 3.0E-02 | 0.68 | – | Not detected |
| $HCO^+$ | $v=0\ J=1-0$ | ≤ 3.1E-02 | 1.0E-02 | ≤ 5.8E-02 | 1.9E-02 | 0.66 | – | Not detected |
| SiS | $v=0\ J=2-1$ | 1.2E-02 | 4.1E-03 | 4.1E-02 | 1.4E-02 | 1.26 | 27.59 | Tentative |
| | $J=5-4$ | ≤ 3.4E-02 | 1.1E-02 | ≤ 6.2E-02 | 2.1E-02 | 0.65 | – | Not detected |
| | $J=8-7$ | ≤ 9.7E-02 | 3.2E-02 | ≤ 1.8E-01 | 6.0E-02 | 0.81 | – | Not detected |
| | $v=1\ J=8-7$ | 1.9E-01 | 4.4E-02 | 1.9E-01 | 5.4E-02 | 0.41 | 32.36 | Tentative |
| SO | $v=0\ J_N=2_2-1_1$ | 3.8E-02 | 9.0E-03 | 3.1E-01 | 4.1E-02 | 1.36 | 28.86 | |
| | $J_N=6_5-5_4$ | 5.0E-01 | 3.5E-02 | 4.3E+00 | 1.1E-01 | 0.53 | 28.88 | |
| $SO_2$ | $v=0\ J_{Ka,Kc}=4_{2,2}-4_{1,3}$ | 1.8E-01 | 2.9E-02 | 1.1E+00 | 9.0E-02 | 0.80 | 28.68 | |
| CS | $v=0\ J=1-0$ | ≤ 4.8E-02 | 1.6E-02 | ≤ 8.9E-02 | 3.0E-02 | 0.47 | – | Not detected |
| | $J=3-2$ | ≤ 8.3E-02 | 2.8E-02 | ≤ 1.5E-01 | 5.1E-02 | 0.80 | – | Not detected |
| $H_2O$ | $v=0\ J_{Ka,Kc}=6_{1,6}-5_{2,3}$ | 1.6E+00 | 5.4E-02 | 9.8E-01 | 5.8E-02 | 0.16 | 27.18 | |

Table 5.17 Radio molecular line survey of IRAS 20056+1834.

| Molecule | Transition | I (peak) [Jy] | $\sigma$ [Jy] | $\int I\,\mathrm{d}V$ [Jy km s$^{-1}$] | $\sigma\left(\int I\,\mathrm{d}V\right)$ [Jy km s$^{-1}$] | Sp. Res. [km s$^{-1}$] | $V_{\mathrm{LSR}}$ [km s$^{-1}$] | Comments |
|---|---|---|---|---|---|---|---|---|
| $C^{18}O$ | $v=0$ $J=2-1$ | $\leq$ 1.4E-01 | 4.7E-02 | $\leq$ 4.6E-01 | 1.5E-01 | 0.53 | – | Not detected |
| $C^{17}O$ | $v=0$ $J=2-1$ | $\leq$ 1.3E-01 | 4.2E-02 | $\leq$ 4.1E-01 | 1.4E-01 | 0.52 | – | Not detected |
| SiO | $v=0$ $J=1-0$ | $\leq$ 1.2E-02 | 3.9E-03 | $\leq$ 3.8E-02 | 1.3E-02 | 0.53 | – | Not detected |
| | $J=2-1$ | $\leq$ 5.4E-02 | 1.8E-02 | $\leq$ 1.7E-01 | 5.8E-02 | 0.67 | – | Not detected |
| | $J=5-4$ | $\leq$ 1.4E-01 | 4.7E-02 | $\leq$ $\leq$ 4.6E-01 | 1.5E-01 | 0.54 | – | Not detected |
| | $v=1$ $J=1-0$ | 1.5E-02 | 3.2E-03 | 4.0E-02 | 6.7E-03 | 0.53 | -22.40 | |
| | $J=2-1$ | $\leq$ 5.5E-02 | 1.8E-02 | $\leq$ 1.8E-01 | 6.0E-02 | 0.68 | – | Not detected |
| | $v=2$ $J=1-0$ | 2.9E-02 | 3.9E-03 | 1.1E-01 | 1.1E-02 | 0.53 | -20.56 | |
| | $J=2-1$ | $\leq$ 5.4E-02 | 1.8E-02 | $\leq$ 1.8E-01 | 5.8E-02 | 0.68 | – | Not detected |
| | $v=6$ $J=1-0$ | $\leq$ 1.1E-02 | 3.7E-03 | $\leq$ 3.6E-02 | 1.2E-02 | 0.55 | – | Not detected |
| $^{29}SiO$ | $v=0$ $J=1-0$ | $\leq$ 1.0E-02 | 3.4E-03 | $\leq$ 3.3E-02 | 1.1E-02 | 0.53 | – | Not detected |
| | $J=2-1$ | $\leq$ 5.4E-02 | 1.8E-02 | $\leq$ 1.8E-01 | 5.8E-02 | 0.68 | – | Not detected |
| $^{30}SiO$ | $v=0$ $J=1-0$ | $\leq$ 1.0E-02 | 3.3E-03 | $\leq$ 3.2E-02 | 1.1E-02 | 0.54 | – | Not detected |
| | $J=2-1$ | $\leq$ 5.5E-02 | 1.8E-02 | $\leq$ 1.8E-01 | 6.0E-02 | 0.69 | – | Not detected |
| HCN | $v=0$ $J=1-0$ | $\leq$ 4.0E-02 | 1.3E-02 | $\leq$ 1.3E-01 | 4.3E-02 | 0.66 | – | Not detected |
| HNC | $v=0$ $J=1-0$ | $\leq$ 3.2E-02 | 1.1E-02 | $\leq$ 1.0E-01 | 3.5E-02 | 0.65 | – | Not detected |
| $HC_3N$ | $v=0$ $J=10-9$ | $\leq$ 3.1E-02 | 1.0E-02 | $\leq$ 1.0E-01 | 3.4E-02 | 0.64 | – | Not detected |
| $H^{13}CN$ | $v=0$ $J=4-3$ | $\leq$ 5.4E-02 | 1.8E-02 | $\leq$ 1.8E-01 | 5.9E-02 | 0.68 | – | Not detected |
| $HCO^+$ | $v=0$ $J=1-0$ | $\leq$ 3.6E-02 | 1.2E-02 | $\leq$ 1.2E-01 | 3.8E-02 | 0.66 | – | Not detected |
| SiS | $v=0$ $J=2-1$ | $\leq$ 9.2E-03 | 3.1E-03 | $\leq$ 3.0E-02 | 9.9E-03 | 0.63 | – | Not detected |
| | $J=5-4$ | $\leq$ 3.4E-02 | 1.1E-02 | $\leq$ 1.1E-01 | 3.6E-02 | 0.65 | – | Not detected |
| SO | $v=0$ $J_N=2_2-1_1$ | $\leq$ 5.2E-02 | 1.7E-02 | $\leq$ 1.7E-01 | 5.6E-02 | 0.68 | – | Not detected |
| | $J_N=6_5-5_4$ | $\leq$ 1.4E-01 | 4.8E-02 | $\leq$ 4.7E-01 | 1.6E-01 | 0.53 | – | Not detected |
| $SO_2$ | $v=0$ $J=68-68$ | $\leq$ 3.6E-02 | 1.2E-02 | $\leq$ 1.2E-01 | 3.9E-02 | 0.64 | – | Not detected |
| CS | $v=0$ $J=1-0$ | $\leq$ 2.4E-02 | 8.1E-03 | $\leq$ 7.9E-02 | 2.6E-02 | 0.47 | – | Not detected |
| $H_2O$ | $v=0$ $J_{Ka,Kc}=6_{1,6}-5_{2,3}$ | $\leq$ 8.1E-02 | 2.7E-02 | $\leq$ 2.6E-01 | 8.8E-02 | 0.82 | – | Not detected |

Table 5.18 Radio molecular line survey of R Sct.

| Molecule | Transition | I (peak) [Jy] | $\sigma$ [Jy] | $\int I\,dV$ [Jy km s$^{-1}$] | $\sigma\left(\int I\,dV\right)$ [Jy km s$^{-1}$] | Sp. Res. [km s$^{-1}$] | $V_{LSR}$ [km s$^{-1}$] | Comments |
|---|---|---|---|---|---|---|---|---|
| $C^{18}O$ | $v=0$ $J=2-1$ | ≤ 4.4E-02 | 1.5E-02 | ≤ 1.3E-01 | 4.2E-02 | 1.07 | – | Not detected |
| $C^{17}O$ | $v=0$ $J=2-1$ | ≤ 5.7E-02 | 1.9E-02 | ≤ 1.7E-01 | 5.5E-02 | 0.52 | – | Not detected |
| SiO | $v=0$ $J=1-0$ | 7.9E-02 | 8.8E-03 | 0.29 | 2.1E-02 | 0.53 | 56.02 | |
| | $J=2-1$ | 3.5E-01 | 6.8E-03 | 1.6E+00 | 2.8E-02 | 0.67 | 57.02 | |
| | $J=5-4$ | 1.3E+00 | 5.6E-02 | 5.5E+00 | 1.4E-01 | 0.27 | 57.50 | |
| | $v=1$ $J=1-0$ | 9.8E-02 | 7.4E-03 | 0.21 | 1.7E-02 | 0.53 | 54.32 | |
| | $J=2-1$ | 5.4E-01 | 7.3E-03 | 1.0E+00 | 2.2E-02 | 0.68 | 53.50 | |
| | $v=2$ $J=1-0$ | 2.9E-01 | 4.4E-02 | 0.64 | 7.3E-02 | 0.53 | 55.32 | |
| | $J=2-1$ | ≤ 1.9E-02 | 6.4E-03 | ≤ 5.6E-02 | 1.9E-02 | 0.68 | – | Not detected |
| | $v=6$ $J=1-0$ | 2.3E-02 | 6.7E-03 | 8.2E-02 | 1.6E-02 | 1.10 | 52.88 | Tentative |
| $^{29}SiO$ | $v=0$ $J=1-0$ | 2.9E-02 | 7.8E-03 | 2.6E-02 | 1.1E-02 | 0.53 | 56.31 | Tentative |
| | $J=2-1$ | 8.0E-02 | 6.3E-03 | 3.5E-01 | 2.0E-02 | 0.68 | 57.98 | |
| $^{30}SiO$ | $v=0$ $J=1-0$ | 2.6E-02 | 7.7E-03 | 2.6E-02 | 1.2E-02 | 0.54 | 54.58 | Tentative |
| | $J=2-1$ | 6.1E-02 | 6.5E-03 | 2.7E-01 | 1.7E-02 | 0.69 | 57.35 | |
| HCN | $v=0$ $J=1-0$ | ≤ 1.6E-02 | 5.4E-03 | ≤ 4.7E-02 | 1.6E-02 | 0.66 | – | Not detected |
| HNC | $v=0$ $J=1-0$ | ≤ 1.4E-02 | 4.5E-03 | ≤ 3.9E-02 | 1.3E-02 | 0.65 | – | Not detected |
| $HC_3N$ | $v=0$ $J=10-9$ | ≤ 1.8E-02 | 5.9E-03 | ≤ 5.1E-02 | 1.7E-02 | 0.64 | – | Not detected |
| $H^{13}CN$ | $v=0$ $J=4-3$ | ≤ 2.2E-02 | 7.3E-03 | ≤ 6.4E-02 | 2.1E-02 | 0.68 | – | Not detected |
| $HCO^+$ | $v=0$ $J=1-0$ | 4.2E-02 | 5.3E-03 | 1.8E-01 | 1.3E-02 | 0.66 | 56.68 | |
| SiS | $v=0$ $J=2-1$ | ≤ 2.3E-02 | 7.8E-03 | ≤ 6.8E-02 | 2.3E-02 | 0.63 | – | Not detected |
| | $J=5-4$ | ≤ 1.6E-02 | 5.3E-03 | ≤ 4.6E-02 | 1.5E-02 | 0.65 | – | Not detected |
| | $J=8-7$ | ≤ 2.2E-02 | 7.5E-03 | ≤ 6.5E-02 | 2.2E-02 | 0.81 | – | Not detected |
| | $v=1$ $J=8-7$ | ≤ 2.4E-02 | 7.8E-03 | ≤ 6.8E-02 | 2.3E-02 | 0.81 | – | Not detected |
| SO | $v=0$ $J_N=2_2-1_1$ | ≤ 1.3E-02 | 4.2E-03 | ≤ 3.9E-02 | 1.1E-02 | 1.36 | – | Not detected |
| | $J_N=6_5-5_4$ | 3.0E-02 | 8.3E-03 | 1.9E-01 | 4.4E-02 | 2.13 | 56.94 | |
| $SO_2$ | $v=0$ $J_{Ka,Kc}=4_{2,2}-4_{1,3}$ | ≤ 1.9E-02 | 6.4E-03 | ≤ 5.6E-02 | 1.9E-02 | 0.80 | – | Not detected |
| CS | $v=0$ $J=1-0$ | ≤ 5.8E-02 | 1.9E-02 | ≤ 1.7E-01 | 5.6E-02 | 0.93 | – | Not detected |
| | $J=3-2$ | ≤ 2.2E-02 | 7.4E-03 | ≤ 6.5E-02 | 2.2E-02 | 0.80 | – | Not detected |
| $H_2O$ | $v=0$ $J_{Ka,Kc}=6_{1,6}-5_{2,3}$ | 1.7E+00 | 3.0E-02 | 2.0E+00 | 4.7E-02 | 0.41 | 56.75 | |

*With great power comes great responsibility.*
Uncle Ben – Spiderman

# 6
# Rotating and expanding gas in binary post-AGB stars

*In this chapter, we present a very detailed study of rotating disks and outflows around binary post-AGB stars, together with very detailed analysis of the chemistry of this class. The content of this chapter is based on articles already published (Gallardo Cava et al., 2021, 2022c) and this scientific material was presented at the Asymmetrical Post-Main-Sequence Nebulae 8 (APN 8) congress. The meeting was dedicated to the shaping effects of stellar outflows from evolved stars in the formation of asymmetrical post-main-sequence nebulae, because most planetary nebulae exhibit complex structures. This is the case of the outflows of our sample of nebulae around binary post-AGB stars, which are far from the spherical shape. The contribution defended in the congress and the new and valuable contents were published after a referee process. Therefore, the content of this chapter is adapted from Gallardo Cava et al. (2022b).*

## Abstract

There is a particular class of binary post-AGB stars (binary system including a post-AGB star) that are surrounded by Keplerian disks and outflows resulting from gas escaping from the disk. To date, there are seven sources that have been studied in detail through interferometric millimeter-wave maps of CO lines (ALMA/NOEMA). For the cases of the Red Rectangle, IW Carinae, IRAS 08544-4431, and AC Herculis, it is found that around ≥ 85% of the total nebular mass is located in the disk with Keplerian dynamics. The remainder of the nebular mass is located in an expanding component. This outflow is probably a disk wind consisting of material escaping from the rotating disk. These sources are the disk-dominated nebulae. On the contrary, our maps and modeling of 89 Herculis, IRAS 19125+0343, and R Scuti, which allowed us to study their morphology, kinematics, and mass distribution, suggest that in these sources the outflow clearly is the dominant component of the nebula (∼ 75% of the total nebular mass), resulting in a new subclass of nebula around binary post-AGB stars: the outflow-dominated sources. Besides CO, the chemistry of this type of source has been practically unknown thus far. We also present a very deep single-dish radio molecular survey in the 1.3, 2, 3, 7, and 13 mm bands (∼600 hours of telescope time). Our results and detections allow us to classify our sources as O- or / C-rich. We also conclude that the calculated abundances of the detected molecular species other than CO are particularly low, compared with AGB stars. This fact is very significant in those sources where the rotating disk is the dominant component of the nebula.

## 6.1 Introduction

There is a kind of post-AGB stars characterized by their spectral energy distributions (SEDs), which shows a near-infrared (NIR) excess indicating the presence of hot dust close to the the stellar system (Van Winckel, 2003). Their IR spectra reveal the presence of highly processed dust grains, so the the dust might be located in stables structures (Gielen et al., 2011a; Jura, 2003; Sahai et al., 2011). All this suggests the presence of circumbinary disks. Their disk-like shape has been confirmed by interferometric IR data (see e.g. Hillen et al., 2017; Kluska et al., 2019). Their radial velocity curves reveal that the post-AGB stars are part of a binary system (see e.g. Oomen et al., 2018). The systematic detection of binary systems in these objects strongly suggest that the angular momentum of the disks come from the stellar system.

Observations of $^{12}$CO and $^{13}$CO in the $J = 2-1$ and $J = 1-0$ lines (230.538 and 220.398 GHz, respectively) have been well analyzed in sources with such a NIR excess (Bujarrabal et al., 2013a). There are two kinds of CO line profiles. (a) narrow CO line profiles characteristic of rotating disks and weak wings, which implies that most of the nebular mass is contained in the Keplerian disk. And (b), composite CO line profiles including a narrow component, which very probably represents emission from the rotating disk, and strong wings, which represents emission from the outflow, which could dominate the nebula (Bujarrabal et al., 2013a). These kind of line profiles are also found in young stars surrounded by a rotating disk made of remnants of interstellar medium (ISM) and those expected from disk-emission modelling (see e.g. Bujarrabal et al., 2013a; Guilloteau et al., 2013). These results indicate that the CO emission lines of our sources come from from disk with Keplerian or quasi-Keplerian rotation.

The study of the chemistry of this class of binary post-AGB stars, together with the very detailed kinematic analysis of Keplerian disks and outflows around these sources, is based on published articles (see Gallardo Cava et al., 2021, 2022c).

This paper is organized as follows. Technical information of our observations is given in Sect. 6.2. In Sect. 6.2, we present mm-wave interferometric maps and models of the most representative cases of a disk-dominated nebula (AC Herculis), an outflow-dominated nebula (R Scuti), and an intermediate case in between the disk- and the outflow-dominated nebula (89 Herculis). We present the first molecular survey in this kind of objects in Sect. 6.4, together with discussions about molecular intensities and chemistry. Finally, we summarize our conclusions in Sect. 6.5.

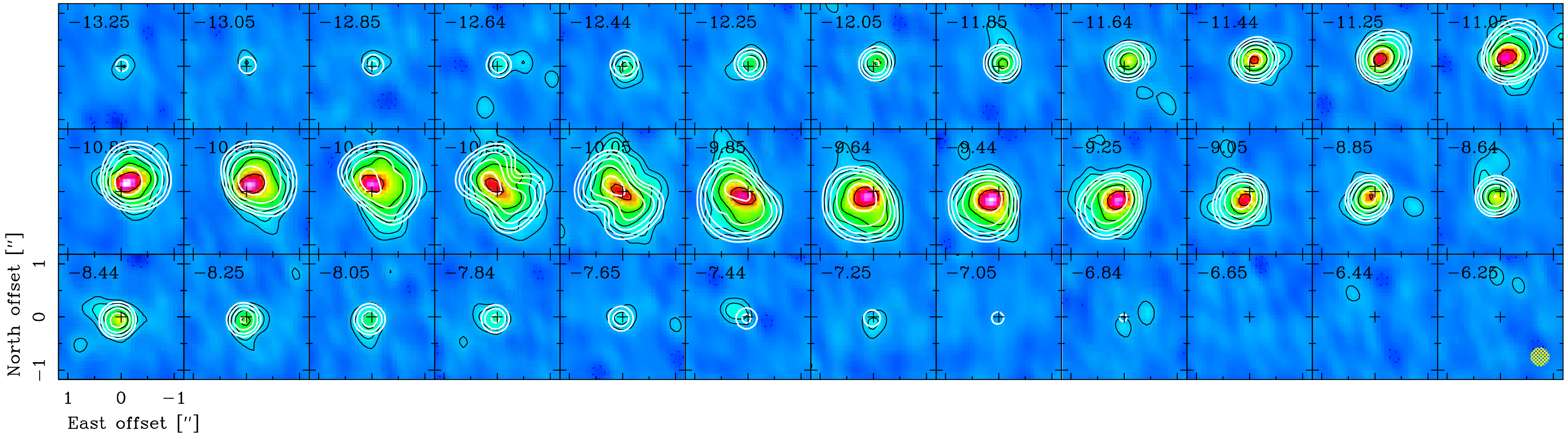

Figure 6.1: NOEMA maps per velocity channel of AC Her in $^{12}$CO $J = 2 - 1$ emission. The contours are $\pm 9$, 18, 36, 76, and 144 mJy beam$^{-1}$ with a maximum emission of 230 mJy beam$^{-1}$. The LSR velocities are indicated in each panel (upper-left corner) and the beam size, 0.″35 $\times$ 0.″35, is shown in the last panel at the bottom right corner (yellow ellipse). We also show the synthetic maps from our best-fit model in white contours, to be compared with observational data; the scales and contours are the same.

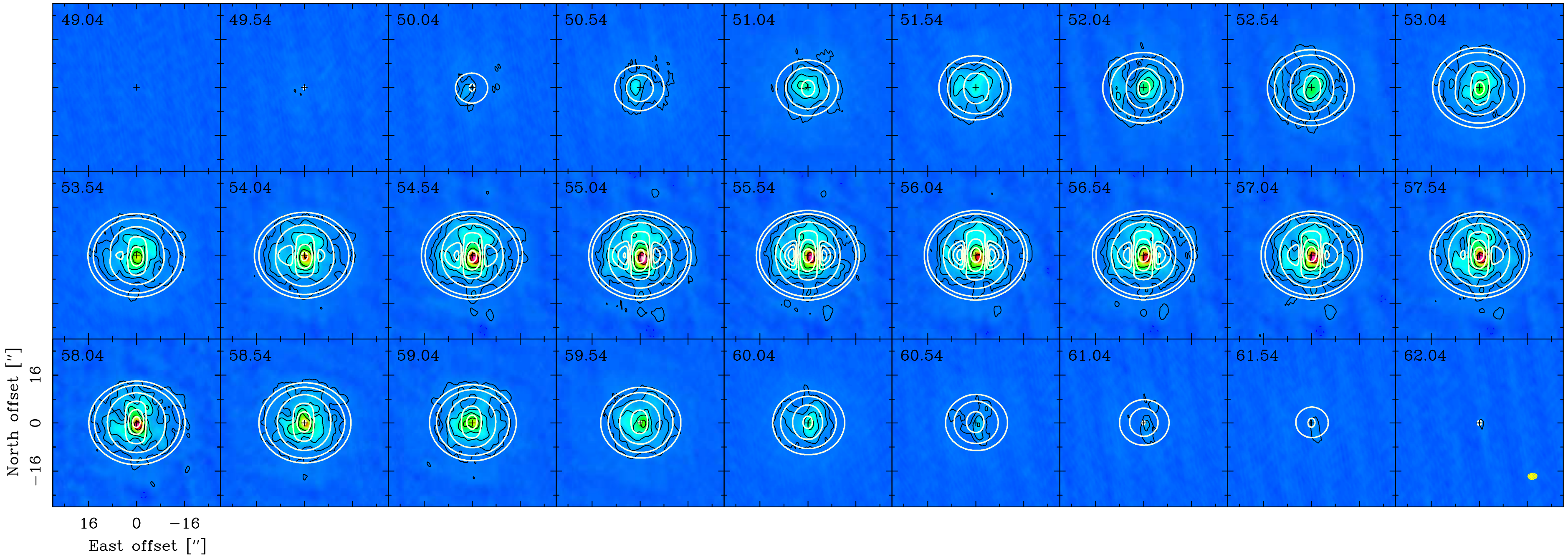

Figure 6.2: Maps per velocity channel of R Sct in $^{12}$CO $J = 2-1$ emission. The contours are $\pm 50$, 100, 200, 400, 800 and 1600 mJy beam$^{-1}$ with a maximum emission of 2.4 Jy beam$^{-1}$. The LSR velocities are indicated in each panel (upper-left corner) and the beam size, 3.″12 $\times$ 2.″19, is shown in the last panel at the bottom right corner (yellow ellipse). We also show the synthetic maps from our best-fit model in white contours, to be compared with observational data; the scales and contours are the same.

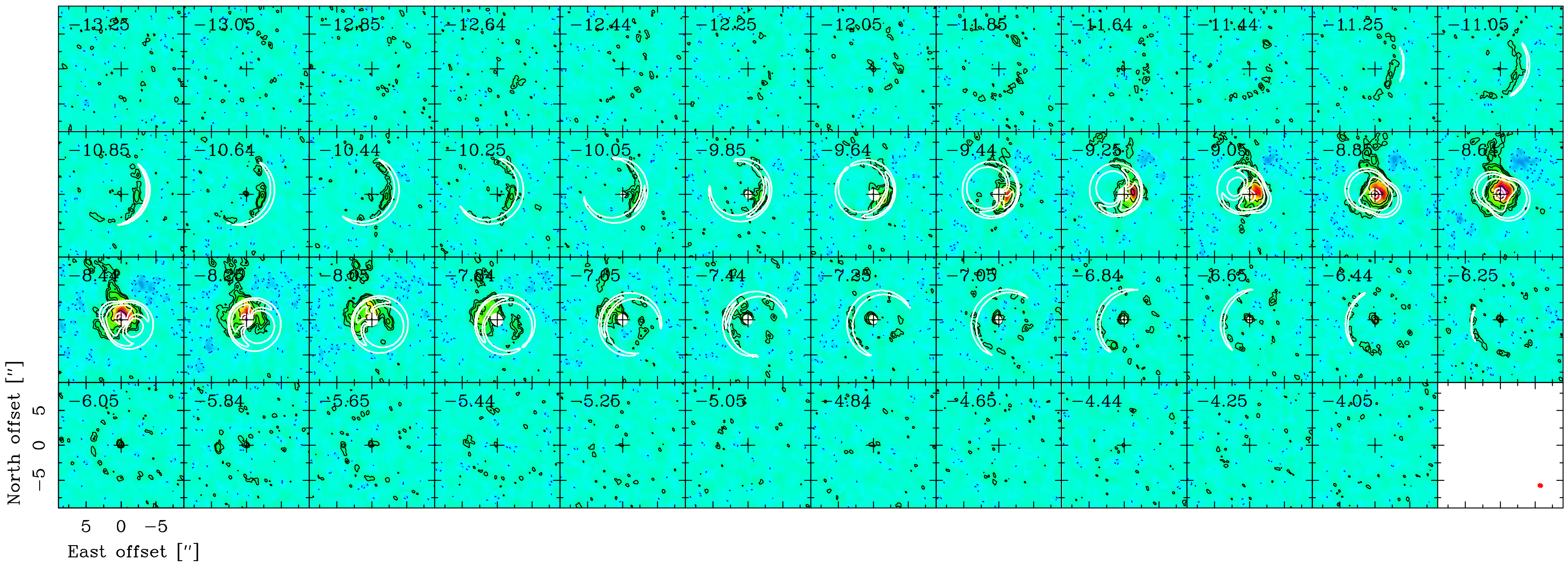

Figure 6.3: NOEMA maps per velocity channel of 89 Her in $^{13}$CO $J = 2-1$ emission. The contours are $\pm 11$, 22, 44, 88, and 144 mJy beam$^{-1}$ with a maximum emission of 225 mJy beam$^{-1}$. The LSR velocities are indicated in each panel (upper-left corner) and the beam size, 0.''74 × 0.''56, is shown in the last panel at the bottom right corner (red ellipse). We also show the synthetic maps from our best-fit model in white contours, to be compared with observational data; the scales and contours are the same.

## 6.2 Observations

We show interferometric maps of our sources using the NOEMA interferometer. Observations of the $^{12}$CO $J = 2-1$ rotational transition were carried out towards AC Herculis, R Scuti, and 89 Herculis. Observations of the $^{13}$CO $J = 2-1$ rotational transition were also obtained for 89 Herculis.

Our single-dish observations were performed using the 30 m IRAM radiotelescope (Granada, Spain) and the 40 m Yebes telescope (Guadalajara, Spain). We observed at the 1.3, 2, 3, 7, and 13 mm bands. Our observations required a total telescope time of $\sim$600 hours distributed over the two telescopes and for several projects to observe the nebula around the next binary post-AGB stars: AC Her, the Red Rectangle, HD 52961, IRAS 19157−0257, IRAS 18123+0511, IRAS 19125+0343, AI CMi, IRAS 20056+1834, and R Sct.

## 6.3 NOEMA observations

In this section, we present the results directly obtained from the observations for AC Her, R Sct, and 89 Her (Gallardo Cava et al., 2021). We show our NOEMA maps per velocity channel (see Figs. 6.1, 6.2, and 6.3) and position-velocity (PV) diagrams along the the equatorial rotating disk and along the axis of the nebula (see Figs. 6.4, 6.5, and 6.6).

### 6.3.1 AC Herculis

The $^{12}$CO $J = 2-1$ mm-wave interferometric maps are presented in Fig. 6.1. We can see in the left panel of Fig. 6.4 the PV diagram along the equatorial direction and it very clearly shows the characteristic signature of rotation with Keplerian dynamics. On the contrary, the analysis of the PV diagram along the nebula axis direction help us to detect the presence of an axially outflowing component (see Fig. 6.4 *right*). A theoretical PV diagram along the axis direction in the presence of a disk with presents Keplerian dynamics might show emission with a form close similar to a rhombus, with equal or very similar emission in all the four quadrants of the PV diagram. Nevertheless, we do not see equal emission in the four quadrants: we see slightly inclined emission at central velocities at around $\pm 1''$. This fact can be explained by the existence of an expanding component that surrounds the rotating disk.

### 6.3.2 R Scuti

We present combined NOEMA maps and 30 m maps of R Sct in $^{12}$CO $J = 2-1$ emission in Fig. 6.2 and PV diagrams in Fig. 6.5. In both figures, we clearly see two components: an intense inner region; and an extended component of around $\sim 40''$ surrounding the inner region. This extended and expanding component contains most of the total nebular mass (see Sect. 4.5.4). The PV diagram along the equatorial direction shows an intense central clump in the innermost region of the nebula that may represent the unresolved rotating disk (see Fig. 4.9 *left*). The velocity dispersion from the inner (and unresolved) central condensation is similar to other post-AGB nebulae with disks, including a significant lack of blueshifted emission (see e.g. Bujarrabal et al., 2016). The PV diagram along the nebula axis reveals the structure of the nebula (see Fig. 4.9 *right*): we clearly see tow big cavities at about $\pm 10''$. We see this kind of structure in other pPNe, such as M 2−56 (Castro-Carrizo et al., 2002) or M 1−92 (Alcolea et al., 2007).

### 6.3.3 89 Her

We present NOEMA maps and PV diagrams of 89 Her of $^{13}$CO $J = 2-1$ emission (and $^{12}$CO $J = 2-1$, see Gallardo Cava et al., 2021, for further details) in Figs. 6.3 and 6.6. We see an intense central clump and an extended hourglass-shaped structure surrounding this central clump. For a distance of 1 kpc, the size of the hourglass-like structure is, at least, 10 000 AU.

### 6.3.4 Models

Our models consist of a disk with present Keplerian dynamics and an extended and expanding component escaping from the rotating disk and surrounding it. The outflowing component can present different shapes, such as an hourglass, an ellipsoid, etc. We assume LTE populations, which is a reasonable assumption for low-$J$ rotational levels of CO transitions. We consider potential laws for the density ($n$) and rotational temperature ($T$). Additionally, we also consider Keplerian dynamics in the rotating disk ($V_{\mathrm{rot_K}}$) and radial expansion in the extended component ($V_{\mathrm{exp}}$). We must highlight that our code produces results that can be quantitatively compared to observations.

$$n = n_0 \left(\frac{r_0}{r}\right)^{\kappa_{\mathrm{n}}}, \tag{6.1}$$

$$T = T_0 \left(\frac{r_0}{r}\right)^{\kappa_{\mathrm{T}}}, \tag{6.2}$$

$$V_{\mathrm{rot_K}} = V_{\mathrm{rot_{K_0}}} \sqrt{\frac{10^{16}}{r}}, \tag{6.3}$$

$$V_{\mathrm{exp}} = V_{\mathrm{exp_0}} \frac{r}{10^{16}}. \tag{6.4}$$

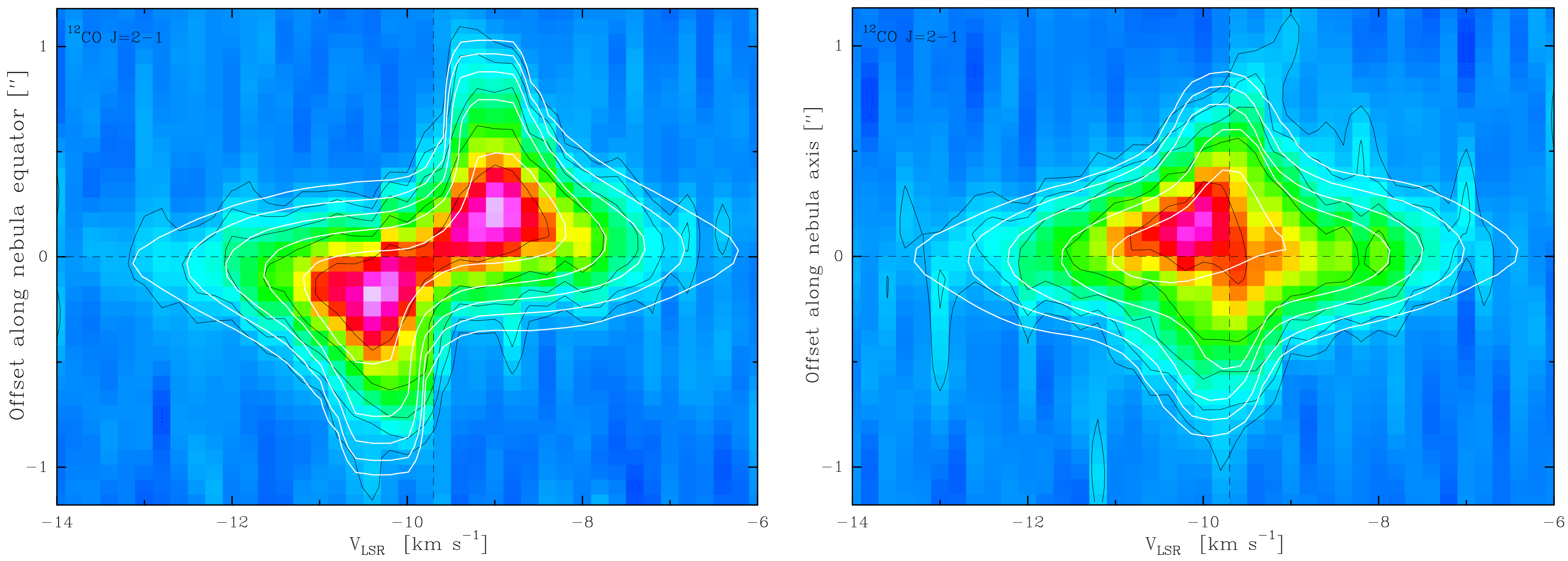

Figure 6.4: *Left:* PV diagram from our NOEMA maps of $^{12}$CO $J = 2-1$ in AC Her along the equatorial direction of the disk ($PA = 136.1°$). The contours are the same as in Fig. 6.1. The dashed lines reveal the systemic velocity and the central position of the source. Additionally, we show the synthetic PV diagram from our best-fit model in white contours, to be compared with observational data; the scales and contours are the same. *Right*: Same as in *left* but along the axis direction of the nebula ($PA = 46.1°$).

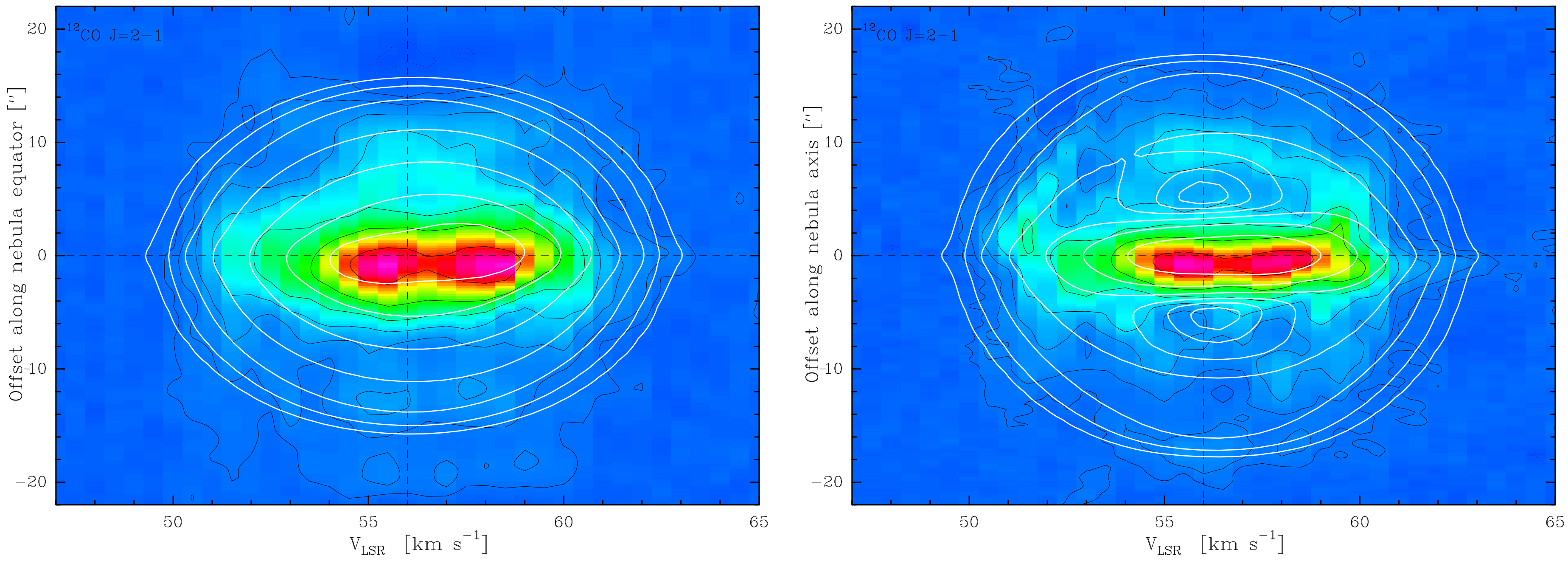

Figure 6.5: *Left:* PV diagram from our maps of $^{12}$CO $J = 2-1$ in R Sct along the equatorial direction ($PA = 0°$). The contours are the same as in Fig. 6.2. The dashed lines reveal the systemic velocity and the central position of the source. Additionally, we show the synthetic PV diagram from our best-fit model in white contours, to be compared with observational data; the scales and contours are the same. *Right*: Same as in *left* but along the axis direction ($PA = 90°$).

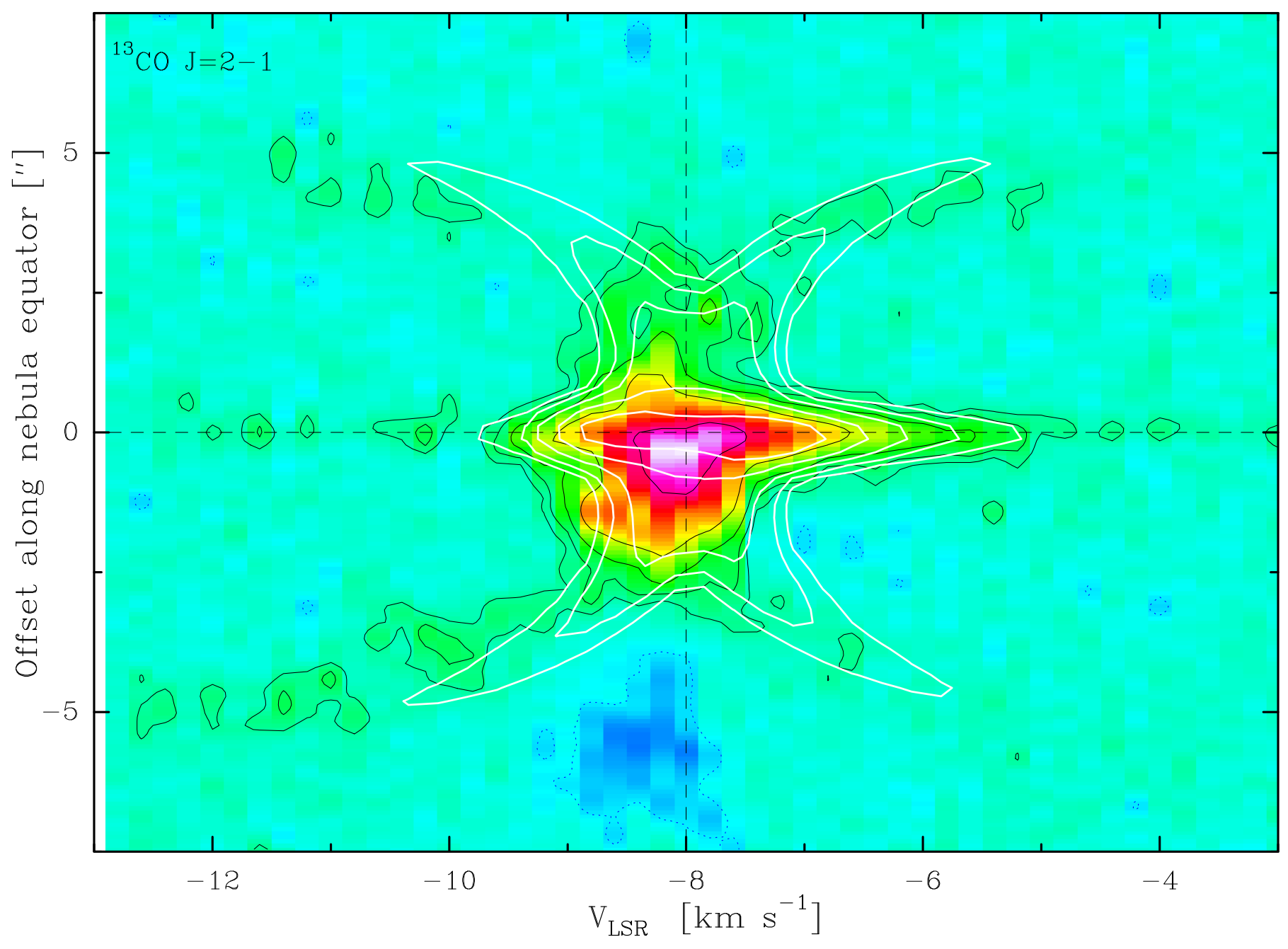

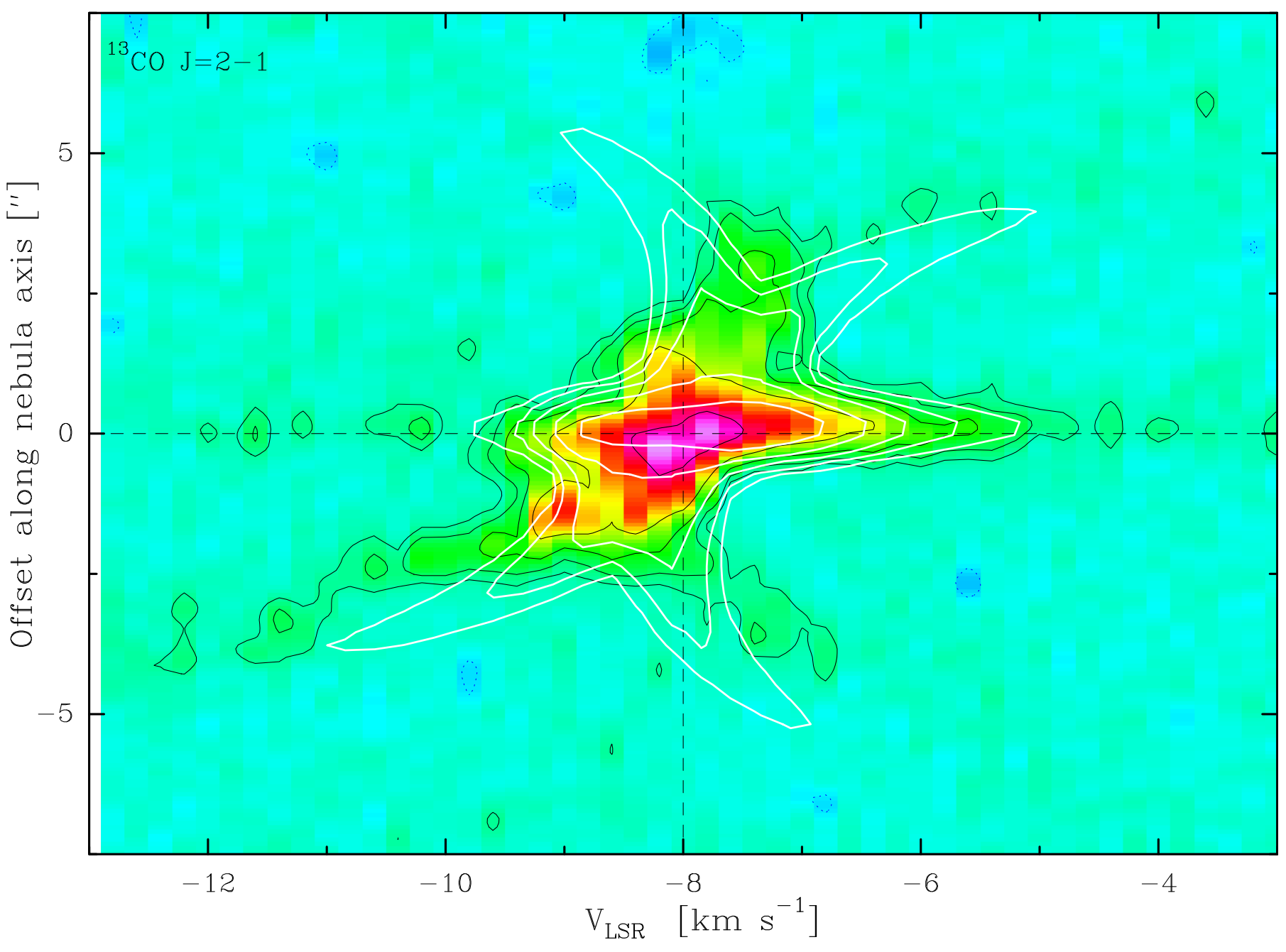

Figure 6.6: *Left:* PV diagram from our NOEMA maps of $^{13}$CO $J = 2 - 1$ in 89 Her along the equatorial direction of the disk ($PA = 150°$). The contours are the same as in Fig. 6.3. The dashed lines reveal the systemic velocity and the central position of the source. Additionally, we show the synthetic PV diagram from our best-fit model in white contours, to be compared with observational data; the scales and contours are the same. *Right*: Same as in *left* but along the axis direction of the nebula ($PA = 60°$).

### AC Her

Our proposed model for the structure of AC Her (Fig. 6.7, see also Figs. 6.1 and 6.4) is very similar to the one found for the Red Rectangle, IRAS 08544−4431, and IW Car (see Bujarrabal et al., 2016, 2017, 2018b). The total mass of the nebula is $8.3 \times 10^{-4}\,M_\odot$. Our model predicts that the mass of the outflow must be $\leq 5\%$ of the total mass. Thus, AC Her is clearly a binary post-AGB star surrounded by a disk-dominated nebula, due to the mass of the Keplerian disk is, at least, 19 times larger than that of the outflow. The Keplerian rotation velocity field of the disk is compatible with a central total stellar mass of $\sim 1\,M_\odot$.

### R Sct

Our proposed model for the structure of R Sct (Fig. 6.7, see also Figs. 6.2 and 6.5). There are slightly differences between both PV diagrams, but all of them are accounted for in our uncertainties. We note that our best-fit model cannot be very different from other models). The nature of R Sct is not yet clear, but our interferometric maps firmly suggest that this source is also a binary post-AGB star surrounded by a disk with Keplerian dynamics and by a high-mass extended and expanding component. The mass of the nebula is found to be $\sim 3.2 \times 10^{-2}\,M_\odot$ and about $\sim 25\%$ of the nebular material would be placed in the rotating disk. This fact, together with the large size of the outflow, allows us to classify this source as an outflow-dominated post-AGB nebula. The disk with Keplerian dynamics is compatible with a central stellar mass of $1.7\,M_\odot$.

### 89 Her

The total mass of the nebula around 89 Her is $1.4 \times 10^{-2}\,M_\odot$ and our proposed model predicts that the mass of the hourglass must be $\sim 50\%$ of the total mass (Fig. 6.7, see also Figs. 6.3 and 6.6). Thus, this source is in between of the disk- and outflow-dominated sources. We find that the disk with Keplerian dynamics is compatible with a central stellar mass of $1.7\,M_\odot$.

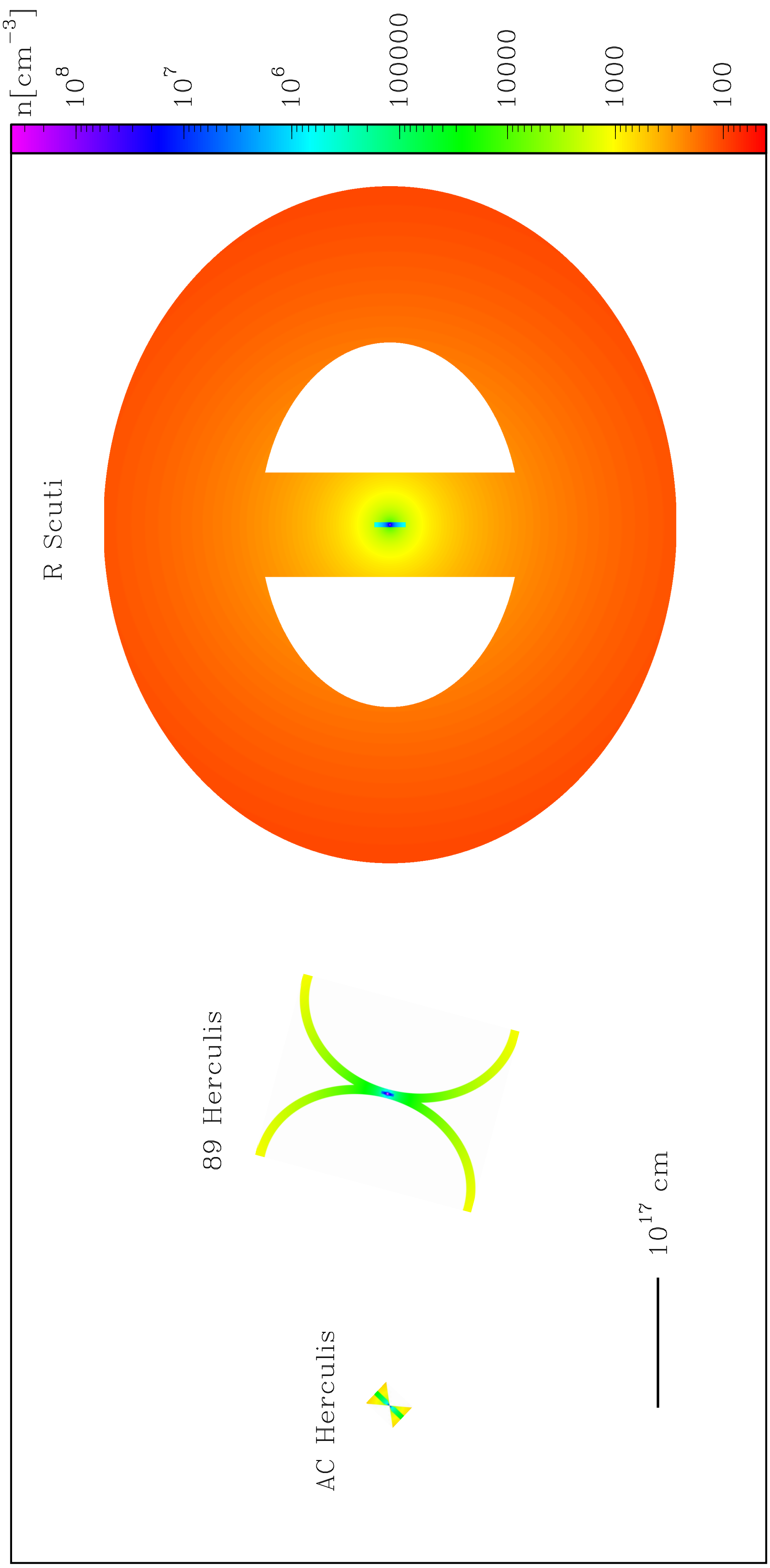

Figure 6.7: Density distribution and structure of our best-fit model for the disk and outflow of AC Her, 89 her, and R Sct as seen from the observer. All models are shown with the same scale so we can see their relative sizes.

Table 6.1 Molecular transitions detected in this work.

| O-bearing molecules | | | | C-bearing molecules | | | |
|---|---|---|---|---|---|---|---|
| Specie | Transition | | $\nu$ [MHz] | Specie | Transition | | $\nu$ [MHz] |
| $^{28}$SiO | $v=0$ | $J=1-0$ | 43423.85 | HCN | $v=0$ | $J=1-0$ | 88630.42 |
| | $v=0$ | $J=2-1$ | 86846.99 | CS | $v=0$ | $J=3-2$ | 146969.00 |
| | $v=0$ | $J=5-4$ | 217104.98 | SiS | $v=0$ | $J=5-4$ | 90771.56 |
| | $v=1$ | $J=1-0$ | 43122.08 | | | | |
| | $v=1$ | $J=2-1$ | 86243.37 | | | | |
| | $v=2$ | $J=1-0$ | 42820.59 | | | | |
| SO | $v=0$ | $J_N=6_5-5_4$ | 219949.44 | | | | |
| $H_2O$ | $v=0$ | $J_{Ka,Kc}=6_{1,6}-5_{2,3}$ | 22235.08 | | | | |

# 6.4 First molecular survey in binary post-AGB stars

The chemistry of this kind of binary post-AGB sources with rotating disks was practically unknown. We present a very deep and wide survey of radio lines in ten of our sources: AC Her, the Red Rectangle, 89 Her, HD 52961, IRAS 19157−0257 IRAS 18123+0511, IRAS 19125+0343, AI CMi, IRAS 20056+1834, and R Sct. All of them have been observed at 7 and 13 mm, and most of them have been also observed in the 1.3, 2, and 3 mm bands; see Table 6.1 (see Gallardo Cava et al., 2022c).

## 6.4.1 Molecular richness

We show in Fig. 6.8 integrated intensity ratios between the main rare molecules (SO, SiO, SiS, CS, and HCN) and CO ($^{13}$CO $J=2-1$ and $^{12}$CO $J=1-0$). Additionally, we also compare these molecular integrated intensities with the 12, 25, and 60 $\mu$m IR emission. The averaged of our results are represented with black horizontal lines. We compare our results with the molecular emission of AGB stars (blue and red horizontal lines represent averaged values of the molecular emission for O- and C-rich AGB stars, respectively). We note that the average of our results always presents low molecular emission in molecules other than CO. Note the large range (logarithmic scale) of intensity ratios. This low intensities are more remarkable in the disk-dominated sources, such as AC Her and the Red Rectangle (see Gallardo Cava et al., 2021, 2022c).

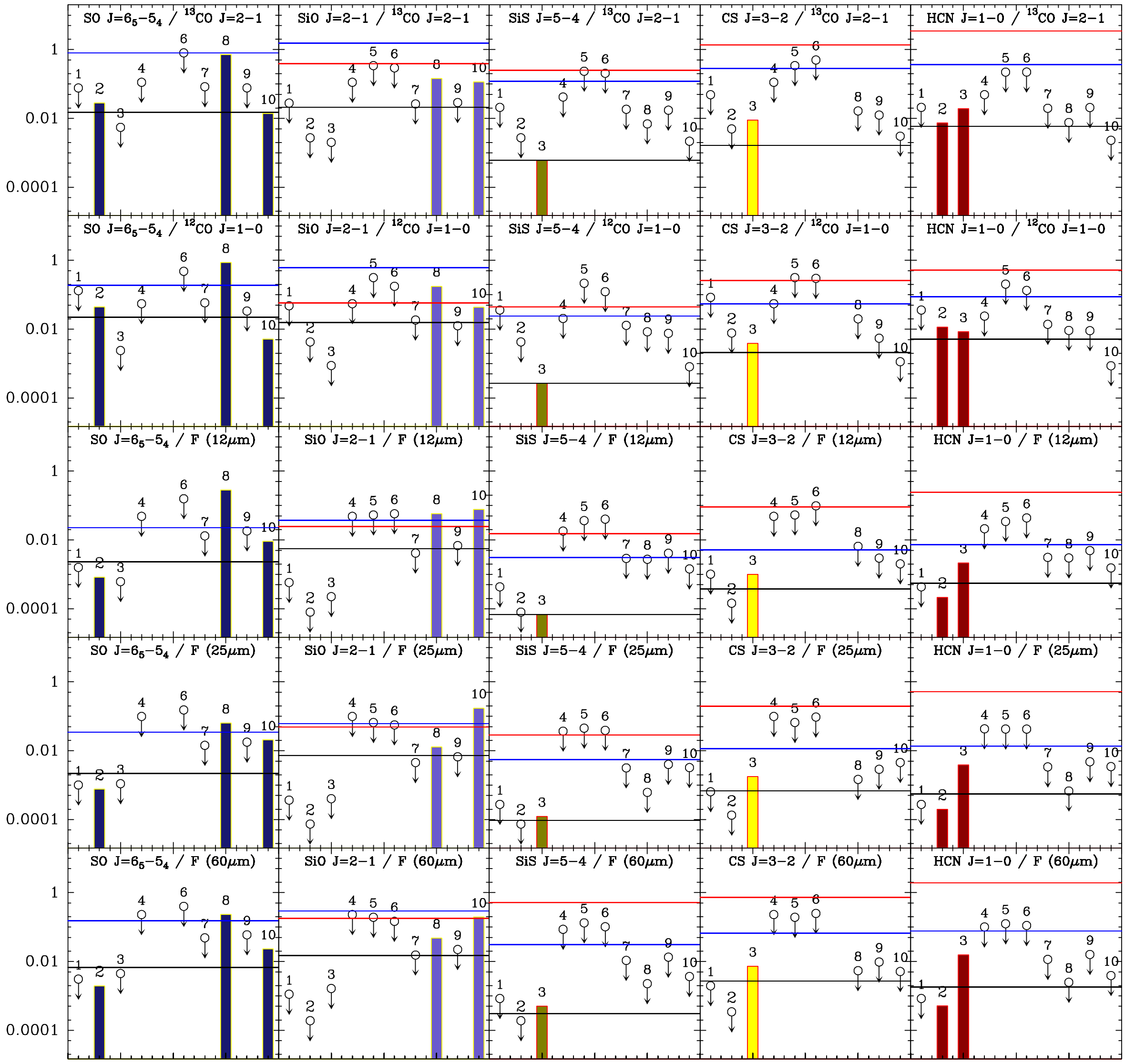

Figure 6.8: Ratios of integrated intensities of molecules (SO, SiO, SiS, CS, and HCN) and IR emission (12, 25, and 60 $\mu$m) in our sources. The binary post-AGB stars are ordered by increasing outflow dominance: 1 – AC Her, 2 – Red Rectangle, 3 – 89 Herculis, 4 – HD 52961, 5 – IRAS 19157−0257, 6 – IRAS 18123+0511, 7 – IRAS 19125+0343, 8 – AI CMi, 9 – IRAS 20056+1834, and 10 – R Sct. Empty circles with arrows represent upper limits. Our results are averaged (black lines) and are compared with averaged values for O- and C-rich AGB CSEs (blue and red lines, values taken from Bujarrabal et al., 1994b,a). Our sample of nebulae around binary post-AGB stars clearly present low molecular emission in molecules other than CO.

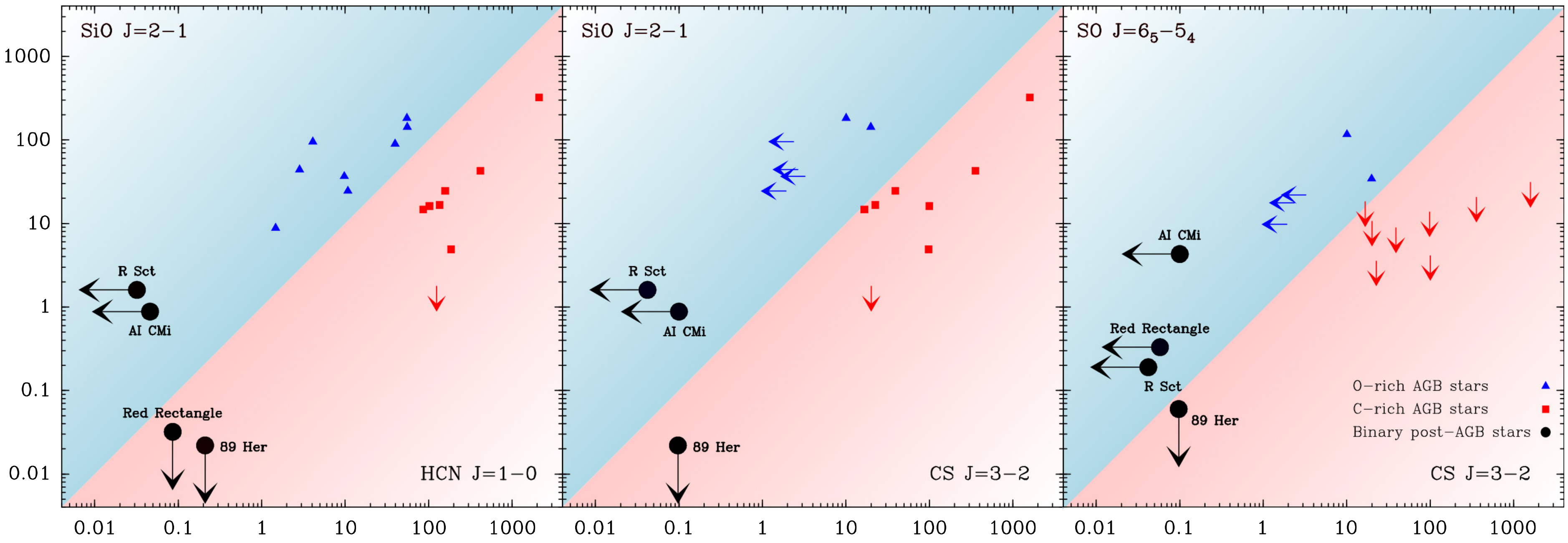

Figure 6.9: Integrated intensities of pairs of molecular transitions in binary post-AGB stars (black circles, upper limits are represented with arrows), as well in O- and C-rich AGB CSEs (blue and red squares). O- and C-rich environment areas are represented in blue and red, respectively. The integrated intensities of the transitions are expressed in $\mathrm{Jy\,km\,s^{-1}}$ and in logarithmic scale.

### 6.4.2 The discrimination between O- and C-rich envelopes

Evolved stars present a O/C > 1 and O/C < 1 chemistry, so they can be classified as O- and C-rich environments, respectively. The O/C abundance ratio has important effects on the molecular abundances. The lines of O-bearing molecules are much more intense in O-rich environments than in the C-rich ones (such as SiO and SO). On the contrary, the lines of C-bearing molecules are much more intense in C-rich environments (see e.g. Bujarrabal et al., 1994b,a). Additionally, SiO and $H_2O$ maser emission is exclusive of O-rich environments (see e.g. Kim et al., 2019)

We analyze integrated intensities of pairs of molecular transitions, and this analysis is crucial to distinguish between O- and C-rich environments (Fig. 6.9). When an O-bearing molecule is compared with a C-bearing one, we find that integrated intensities are larger in O-rich than in C-rich sources. Our results are compared with CSEs around AGB stars, because they are prototypical environments rich in molecules.

Based on the maser detection of O-bearing molecules (SiO maser emission in R Sct, AI CMi, and IRAS 20056+1834; $H_2O$ maser emission in R Sct and AI CMi), we classify some of our sources as O-rich. Contrarily, based on the integrated intensities ratios, the nebula around 89 Her seems to be C-rich (Fig. 6.9). Therefore, the nebula around AC Her, the Red Rectangle, AI CMi, IRAS 20056+1834, and R Sct presents a O/C > 1 chemistry, while 89 Her presents a O/C < 1 environment (Gallardo Cava et al., 2022c).

## 6.5 Conclusions

There is a class of post-AGB star that is part of a binary system with a significant NIR excess that is surrounded by a disk with Keplerian dynamics and an extended and expanding component composed of gas escaping from the disk and surrounding it.

Based on our observational data and model results, we find disk-dominated sources that present ≥ 85% of the total nebular mass located in the Keplerian disk. This is the case of AC Herculis. We also find a subclass of these binary post-AGB stars, in which the disk contains ∼25% of the total mass of the nebula, such as R Scuti. The extended component of these outflow-dominated sources are mainly composed of cold gas. Moreover, our NOEMA maps and modeling suggest that the nebula around 89 Her is in an intermediate case in between both the disk- and the outflow-dominated sources, since around 50% of the nebular mass is located in the rotating disk. See Sect. 6.3 for further details. HD 52961 and IRAS 1957−0247 would also belong to this intermediate case. However, the existence of this intermediate type is not clear, because these objects were classified as intermediate sources under high uncertainties and they could belong to either subclass: the disk- or the outflow-dominated sources. In the case of 89 Her, our new 30 m IRAM on-the-fly observations recover all the filtered out flux. These maps show a larger hourglass-like structure compared to that in NOEMA maps. According to these new maps and preliminary results, the hourglass-like structure around 89 Her could contain most of the material (see Sect. 7).

We present the first survey in the search of molecules other than CO in binary post-AGB stars surrounded by Keplerian disks (see Sect. 6.4). The emission of molecules other than CO in our sources is low and this fact is especially remarkable in the disk-dominated nebulae. Additionally and according to our analysis, we catalogue the chemistry of 89 Her as C-rich. On the contrary, we find O-rich environments in AC Her, the Red Rectangle, AI CMi, IRAS 20056+1834, and R Sct.

*Caed mil, bloede dh'oine... Hocus-Pocus... Abracadabra... Arse blathanna.*

Geralt of Rivia – The Witcher III

# 7

# The nebulae around the binary post-AGB star 89 Herculis

*89 Herculis is a binary post-AGB star surrounded by a very extended hourglass shaped expanding component and a Keplerian disk in the innermost region. According to the previous NOEMA maps and models, the mass of the nebula was $1.4 \times 10^{-2}\,M_{\odot}$, of which ∼ 50% corresponds to the outflow. However, the emission of the outflow was underestimated, because ∼ 50% of the flux is filtered out in the line wings and, additionally, our previous NOEMA and 30 m observations are limited by the HPBW of the 30 m dish. In this chapter, we present high combined, NOEMA + OTF 30 m IRAM, resolution maps of $^{12}CO$ and $^{13}CO$ $J = 2-1$ emission lines in 89 Her. These maps contain all detectable flux from the source, so we can completely study the large hourglass-like structure that surrounds the disk around 89 Her. As a result of this study, 89 Her is now classified as an outflow-dominated source.*

*We have also revised the disk and outflow contributions in the whole sample of binary post-AGB sources. According to these new mass calculations, the disk-mass ratio in these sources is found to follow a clear bimodal distribution: we find no intermediate sources in between the disk- and the outflow-dominated ones. Consequently, we suggest that there is not an evolutionary link between both the disk- and the outflow-dominated nebulae. The content of this chapter is adapted from Gallardo Cava et al. (2022a).*

## Abstract

Context: There is a class of binary post-asymptotic giant branch (post-AGB) stars that exhibit remarkable near-infrared (NIR) excess. These stars are surrounded by disks with Keplerian, or quasi-Keplerian, dynamics and outflows composed of gas escaping from the disk. Depending on the dominance of these components there are two subclasses of binary post-AGB stars: the disk- and the outflow-dominated ones.

Aims: We aim to properly study the hourglass-like structure that surrounds the Keplerian disk around 89 Her.

Methods: We present total-power on-the-fly maps of $^{12}$CO and $^{13}$CO $J = 2-1$ emission lines in 89 Her. Previous studies are known to suffer from flux losses in the most extended components. We merge these total-power maps with previous NOEMA maps. The resulting combined maps are expected to detect the whole nebula extent of the source.

Results: Our new combined maps contain all detectable flux of the source but at the same time have the high spatial resolution provided by the interferometric observations. We find that the hourglass-like extended outflow around the rotating disk is larger and more massive compared to previous works. The total nebular mass of this very extended nebula is $1.8 \times 10^{-2}$ M$_\odot$, of which $\sim$65% comes from the outflow. The observational data and model results lead us to classify the envelope around 89 Her as an outflow-dominated nebula, together with R Sct and IRAS 19125+0343 (and very probably AI CMi, IRAS 20056+1834, and IRAS 18123+0511). The updated statistics on the masses of the two post-AGB main components reveals that there are two distinct subclasses of nebulae around binary post-AGB stars depending on which component is the dominant one. We speculate that the absence of an intermediate subclass of sources is due to different initial conditions of the stellar system and not because both subclasses are in different stages of the post-AGB evolution.

## 7.1 Introduction

Circumstellar envelopes around AGB stars (CSE-AGBs) are often found to be spherical and in slow expansion, while their descendants, pre-planetary and planetary nebulae (pPNe and PNe) present different morphologies: pPNe present strongly aspherical (often axisymmetrical) shapes resulting from the interaction of axial fast winds with the CSE formed in the AGB stage (Ueta et al., 2000; Sahai et al., 2007; Castro-Carrizo et al., 2010); PNe tend to present (quasi-)spherical, axisymmetrical, or irregular shapes (Sahai et al., 2011; Stanghellini et al., 2016). This evolution is very rapid and takes place in a very short time, $\sim$1 000 years timescale (see Balick and Frank, 2002; Van Winckel, 2003). This spectacular transformation is thought to be due to magnetocentrifugal launching of outflows from rotating disks, which implies the presence of a stellar companion to provide the necessary amounts of angular momentum (see e.g. Frank and Blackman, 2004; Bujarrabal et al., 2016).

There is a class of binary post-AGB stars (binary systems including a post-AGB star) that systematically presents circumbinary disks with Keplerian dynamics. These sources tend to present remarkable near-infrared excess and narrow CO line profiles characteristic of rotating disks (Van Winckel, 2003; Bujarrabal et al., 2013a, and references therein). The IR data of these sources reveal the presence of highly processed dust grains, which implies that these disks must be stable structures (Jura, 2003; Sahai et al., 2011; Gielen et al., 2011a).

In the four sources in which the disk rotation has been well observed (the Red Rectangle, AC Herculis, IW Carinae, IRAS 08544−4431), the disk contains most of the mass, but there is also gas in expansion, a disk-wind that is extracted from the disk and represents $\lesssim$15% of the total nebular mass (Bujarrabal et al., 2016, 2017, 2018b; Gallardo Cava et al., 2021). In contrast, there is a subclass of these binary post-AGB stars whose nebulae are dominated by the expanding component, instead of the Keplerian disk: the outflow-dominated subclass (see Gallardo Cava et al., 2021). In this latter class of sources, the nebula emission is quite different from that of the other well studied cases (the disk-dominated nebulae), because their mm-wave interferometric data confirm the presence and dominance of an extended and expanding component that contains most of the detected nebular material. This is the case of R Scuti and IRAS 19125+0343, where $\sim$75% of the total nebular mass corresponds to the extended and expanding component (Gallardo Cava et al., 2021). AI CMi, IRAS 20056+1834, and IRAS 18123+0511 very probably belong to this subclass too. As we show in this paper, the last member of this (outflow-dominated) subclass is 89 Herculis.

89 Herculis is a binary post-AGB star, with warm dust located in a stable structure and large dust grains formed and settle onto the midplane (see Shenton et al., 1995; de Ruyter et al., 2006; Hillen et al., 2014). This source has been well studied through single-dish observations. It shows narrow CO line profiles similar to those of the Red Rectangle, but with prominent wings, suggesting a significant contribution of the extended component (Bujarrabal et al., 2013a). Observations at 1.3, 2, and 3 mm reveal the presence of $C^{18}O$, $C^{17}O$, CS, SiS, and HCN and a very detailed analysis reveals that the nebula around 89 Her is C-rich (see Gallardo Cava et al., 2022c). According to Gallardo Cava et al. (2021), the nebula around 89 Her is composed of a Keplerian disk and an hourglass-shaped structure; see also Alcolea and Bujarrabal (1995); Fong et al. (2006); Bujarrabal et al. (2007). The mass of the nebula was deduced to be $1.4 \times 10^{-2}$ M$_\odot$, of which $6.4 \times 10^{-3}$ M$_\odot$ corresponds to the Keplerian disk ($\sim$50%). Previous studies suggest that 89 Her was the only source of our sample in which there was the same mass in the disk and in the disk-wind. However, the emission of the outflow was found to be underestimated because the NOEMA maps presented a significant amount of filtered out flux in the line wings, in comparison with the single-dish 30 m IRAM profile taken towards the center of the source. Furthermore, the NOEMA maps show that the total extent of the outflow has a size comparable to the Half Power Beam Width (HPBW) of the 30 m IRAM at the frequency of the CO $J = 2 - 1$ transitions. Therefore, it is also possible that the lost flux problem of the interferometric observation is even more severe, as the 30 m IRAM single-dish data may not contain all the flux emitted by the source, as NOEMA maps demonstrate that the source is larger than the beam. This implies that the mass derived for the outflow could be largely underestimated.

To overcome this problem and properly derive the mass of the extended outflow component in a definitive way, here we present new single-dish total-power maps of 89 Her. These maps, that probe 89 Her in its full extent, are merged with the previous interferometric data, resulting in high resolution maps of 89 Her containing all the flux for all the nebular components detected in molecular gas.

## 7.2 Observations and observational results

As we have just stated, to solve the problem of the underestimation of the mass of the extended outflow component, it is necessary to perform maps of the source including the large spatial scales and not only those provided by the interferometric observations, which only probed structures of 5″ or smaller. To do this we have performed new total-power single-dish maps 92″ in diameter, that we have later combined with the previously obtained interferometric maps already presented by Gallardo Cava et al. (2021).

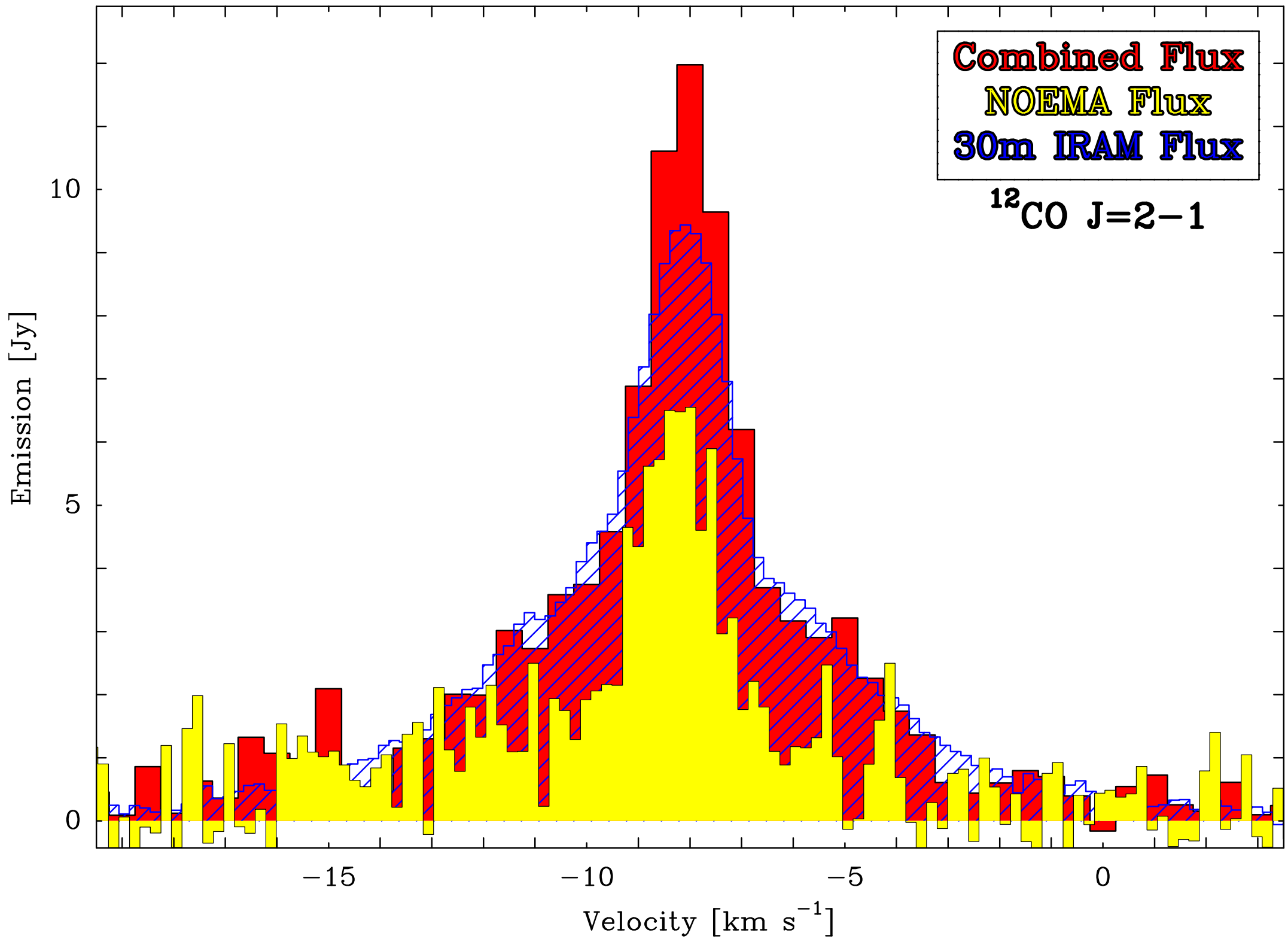

Figure 7.1: Histogram showing the flux comparison between old NOEMA (yellow), old 30 m IRAM single-pointed (blue dashed), and after the merging of NOEMA and the new total-power maps (red), for $^{12}$CO $J = 2 - 1$ .

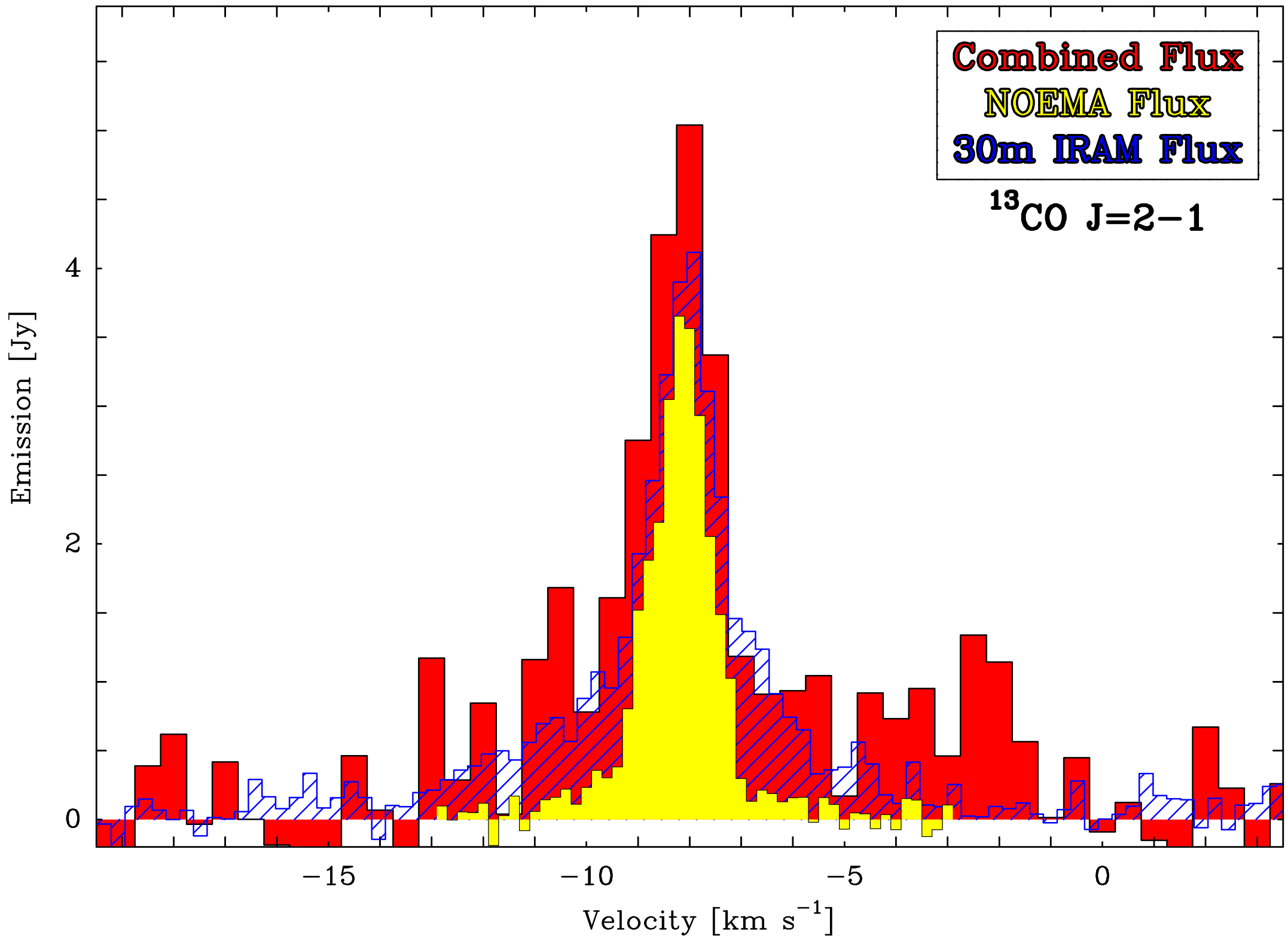

Figure 7.2: Histogram showing the flux comparison between old NOEMA (yellow), old 30 m IRAM single-pointed (blue dashed), and after the merging of NOEMA and the new total-power maps (red), for $^{13}$CO $J = 2 - 1$ .

### 7.2.1 Total-power data

These new single-dish 89 Her observations were performed using the 30 m IRAM telescope at Pico Veleta (Granada, Spain). The observations were carried out between June 2 and June 8 2021, for a total of 65 hours of telescope time (project 061-21). We focus our observations on the $^{12}$CO and $^{13}$CO $J = 2 - 1$ emission lines (230.5 and 220.4 GHz), but we also simultaneously observed the $^{12}$CO and $^{13}$CO $J = 1 - 0$ lines. However, the S/N attained for the $J = 1 - 0$ lines did not result on their detection, and therefore these lines are not discussed in this paper. The HPBW of the telescope is around 11.″3 at 230.5 and 220.4 GHz. During the observing time the weather conditions were good, with typical 225 GHz zenital opacities between 0.1 and 0.3.

We have preformed $92'' \times 92''$ maps every 20 min. The scanning direction of these maps was done in six different directions, at paralactic angles (PA) 0°, 30°, 60°, 90°, 120° and 150°, to minimize interleaving and weaving patterns in the resulting final averaged map. The observations were performed in the on-the-fly mode (OTF), with 23 parallel scans per map, with a separation between scans of 4″, a scanning velocity of $4''\mathrm{s}^{-1}$, and a dump rate of 1 s, resulting in a $4'' \times 4''$ grid (to be compared with the HPBW of 11.″3, i.e. a factor 3 oversampling). The observations were performed in the on-off mode, observing the reference position (600″ away in the R.A. direction) every three scans ($\sim$70 s). The calibration was performed at the beginning of each individual map, using the chopper-wheel method, observing the sky and both hot and cold loads. Observations of NGC 7027 were also performed to check the consistency of intensity scale: since we did not found calibration changes larger that 20%, no re-scale has been applied to the data.

We connected the Fast Fourier Transform Spectrometer (FTS) units to the EMIR receiver with a spectral velocity resolution of 0.25 km s$^{-1}$ ($\sim$200 kHz) per channel. We obtained spectra for vertical and horizontal linear polarization receivers; since no significant changes were found between the two polarizations both maps were averaged together.

We applied the standard data reduction procedure consisting of the removal of baselines, polarization average, and data resampling to the desired final spectral resolution. Individual OTF maps were inspected for the consistency of the pointing: we found no significant differences in the position of the central peak, so all maps were averaged without applying corrections.

The OTF maps for the $J = 2 - 1$ lines are shown in Figs. 7.9 and 7.10. As we can see in Fig. 7.9, the extent of the nebula is restricted to a region of 20″ in diameter. This is fully compatible with the extent of the nebula in the interferometric maps considering the resolution of the OTF data. This relatively large extent also justifies the flux loss in the NOEMA data. Moreover, this extent is larger than the HPBW of the 30 m IRAM, which explain the larger total flux in the new OTF observations in comparison with the old single-pointed 30 m data. On the contrary, this size is smaller than the HPBW of the primary beam of the individual NOEMA antennas at 1.3 mm ($\sim$22″). Nevertheless, the primary beam shape attenuation is taken into account, because it is corrected before the process of merging with the total-power data, following the standard procedure of the GILDAS software[1].

[1]GILDAS is a software package focused in reducing and analyzing mainly mm observations from single-dish and interferometric telescopes. It is developed and maintained by IRAM, LAOG/Université de Grenoble, LAB/Observatoire de Bordeaux, and LERMA/Observatoire de Paris. See https://www.iram.fr/IRAMFR/GILDAS

### 7.2.2 Interferometric data

Interferometric observations of the $^{12}$CO and $^{13}$CO $J=2-1$ rotational transitions were carried out toward 89 Her with the IRAM NOEMA interferometer at Plateau de Bure (Grenoble, France). $^{12}$CO $J=2-1$ observations were performed under project name P05E, while $^{13}$CO $J=2-1$ observations were performed under project names X073 and W14BT. These maps were published by Gallardo Cava et al. (2021), where the complete technical description can be found.

### 7.2.3 Combined maps

In this work, we present NOEMA maps of $^{12}$CO and $^{13}$CO $J=2-1$ emission lines, which include short-spacing pseudovisibilities obtained from the total-power maps with the 30 m IRAM through OTF mode. Therefore, our combined maps contain all detectable flux because: (a) they recover the lost flux of the extended component filtered out by the interferometer; (b) they include large areas that single-dish single-pointed observations cannot detect, due to the limitations of the beam of the telescope; see Figs. 7.1 and 7.2.

To include the large extent scales probed by the OTF maps into the interferometric data, these total-power maps should be first converted into pseudo-visibility data cubes with exactly the same velocity configuration (same number of channels, velocity spacing and velocity value for the reference channel) than the NOEMA data. This has been done by resampling the OTF maps into a spectral resolution of 0.5 km s$^{-1}$ and adopting a LSR velocity of $-7.94$ km s$^{-1}$ for the central channel. Then, data cubes with a pixel size of $2'' \times 2''$ were built from the OTF scans, which are Fourier transformed for obtaining the corresponding pseudo-visibilities to be merged with the interferometric visibilities from the corresponding NOEMA observations. Before the merging, these interferometric data are corrected for the NOEMA primary beam attenuation, following the standard procedure of GILDAS.

After the merging of the $uv$ data sets, we have verified that the calibration of the two instruments, 30 m IRAM and NOEMA, is compatible. Then the new maps were produced by first mapping and then cleaning the merged uv data sets. For the $^{12}$CO $J=2-1$ maps we have used robust weighing and a $uv$ tapper of 200 m, resulting in a HPBW synthetic clean beam of $1\farcs95 \times 1\farcs45$. For the case of $^{13}$CO $J=2-1$ , since the interferometric data were of better quality, we used robust weighting and no tappering. This resulted in a HPBW synthetic clean beam of $0\farcs76 \times 0\farcs59$. The resulting new merged maps for $^{12}$CO $J=2-1$ and $^{13}$CO $J=2-1$ are presented in Figs. 7.3 and 7.4, respectively. We have verified that the new maps are compatible with the old interferometric data and that the total flux is the same as in the new OTF maps, i.e. no flux is missing in the new data. We therefore conclude that our new combined maps contain all the detectable flux since they are not affected by any interferometric losses and include the full extent of the source. This new total flux is larger than that obtained from the interferometric observation alone and that provided by single-dish but single-pointed observations; see Figs. 7.1 and 7.2. In addition, position-velocity (PV) diagrams for $PA = 150°$ are shown in Figs. 7.5, 7.6.

In these velocity maps and equatorial PV diagrams, we see an intense central clump and an expanding component. On the one hand, the central clump corresponds to an unresolved rotating disk very probably with Keplerian dynamics. On the other hand, we also see an expanding component that very nicely shows hourglass-like features.

Following the same reasoning as in the previous work (Gallardo Cava et al., 2021), we find that PV diagrams along 150° is the best $PA$ that reveals the presence of the rotating gas (see Figs. 7.5 and 7.6). Therefore, the PV diagrams with $PA = 150°$ suggest that a rotating disk with moderate velocity dispersion is responsible for this compact and central clump. Thanks to our new combined maps, we confirm the presence of asymmetrical velocities in both maps, because more emission is present at positive velocities than at negative ones. This phenomenon is a result of self-absorption by cold gas in expansion located in front of the rotating component and is often observed in our disk-containing sources, such as the Red Rectangle or R Sct, a disk- and an outflow-dominated source, respectively; see Bujarrabal et al. (2016); Gallardo Cava et al. (2021).

Our combined maps, with all possible detected flux, show an hourglass outflow, whose symmetry axis is practically in our line of sight. These maps also reveal that the size of this outflowing component is larger than we thought in the previous work, both in the height of the hourglass and in the width of its expanding walls. This recovered flux has a relevant impact on the mass of the outflow that surrounds the disk (see Sect. 7.3).

Apart from the rotating disk and the very large hourglass-like expanding component, we do not detect any other structures (such as a halo, for instance). We have verified this by performing radial averages of the emission in the OTF maps. The resulting intensity vs. offset profiles obtained for both $^{12}$CO and $^{13}$CO $J=2-1$ maps are compatible with the 89 Her nebula consisting only in the two structures mentioned before. Therefore, we conclude that the molecular envelope of 89 Her is only conformed by an hourglass-shaped outflow and a rotating disk with Keplerian dynamics.

We note that S/N is higher in the new combined maps, which allows de detection of the outflow at slightly larger expansion velocities. The only new feature worth mentioning is the tentative detection of a spiral like pattern seen at some receding velocities ($-5.95\,\mathrm{km\,s^{-1}}$) in the $^{12}$CO $J=2-1$ maps. If real, this structure cannot be explained as a result of the orbital motions in the binary system since it would imply a period much larger (1 800 years for an expansion velocity of $\sim 10\,\mathrm{km\,s^{-1}}$ and a distance of 1 000 pc) than that derived from the radial velocity curve of the primary (289.1 days, see Oomen et al., 2018). This structure could be instead due to the effect of a precessing jet launched from the compact companion (see Raga et al., 2009; Velázquez et al., 2011). However this should be better investigated using more sensitive maps. This is similar to the case of the pattern seen in the Red Rectangle, also suggesting a periodicity much larger than the orbital period in the system (see Cohen et al., 2004; Waelkens et al., 1996).

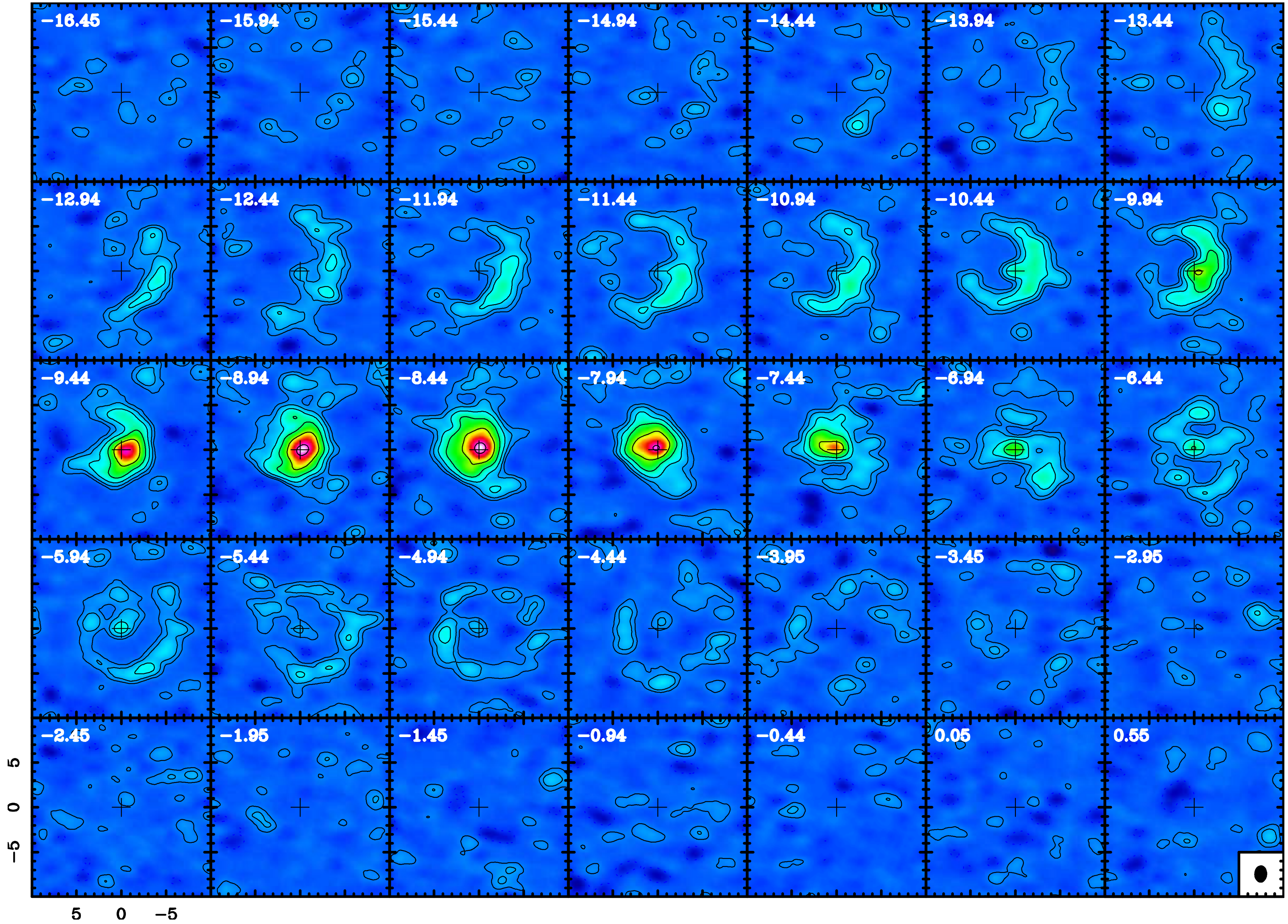

Figure 7.3: Merged (NOEMA + 30 m) maps per velocity channel of $^{12}$CO $J = 2 - 1$ emission from 89 Her. The contours are −40, 40, 80, 160, 320, 640, and 1280 mJy beam$^{-1}$, with a maximum emission peak of 1.5 Jy beam$^{-1}$. The resulting clean-beam has a HPBW of 1.″95 × 1.″45, the major axis oriented at PA = 92°. The LSR velocity is indicated in the upper right corner of each velocity-channel panel and the beam size is shown in the last panel. The FOV of each panel is 20″ × 20″. X and Y units are offsets in arc-sec in the East and North directions respectively.

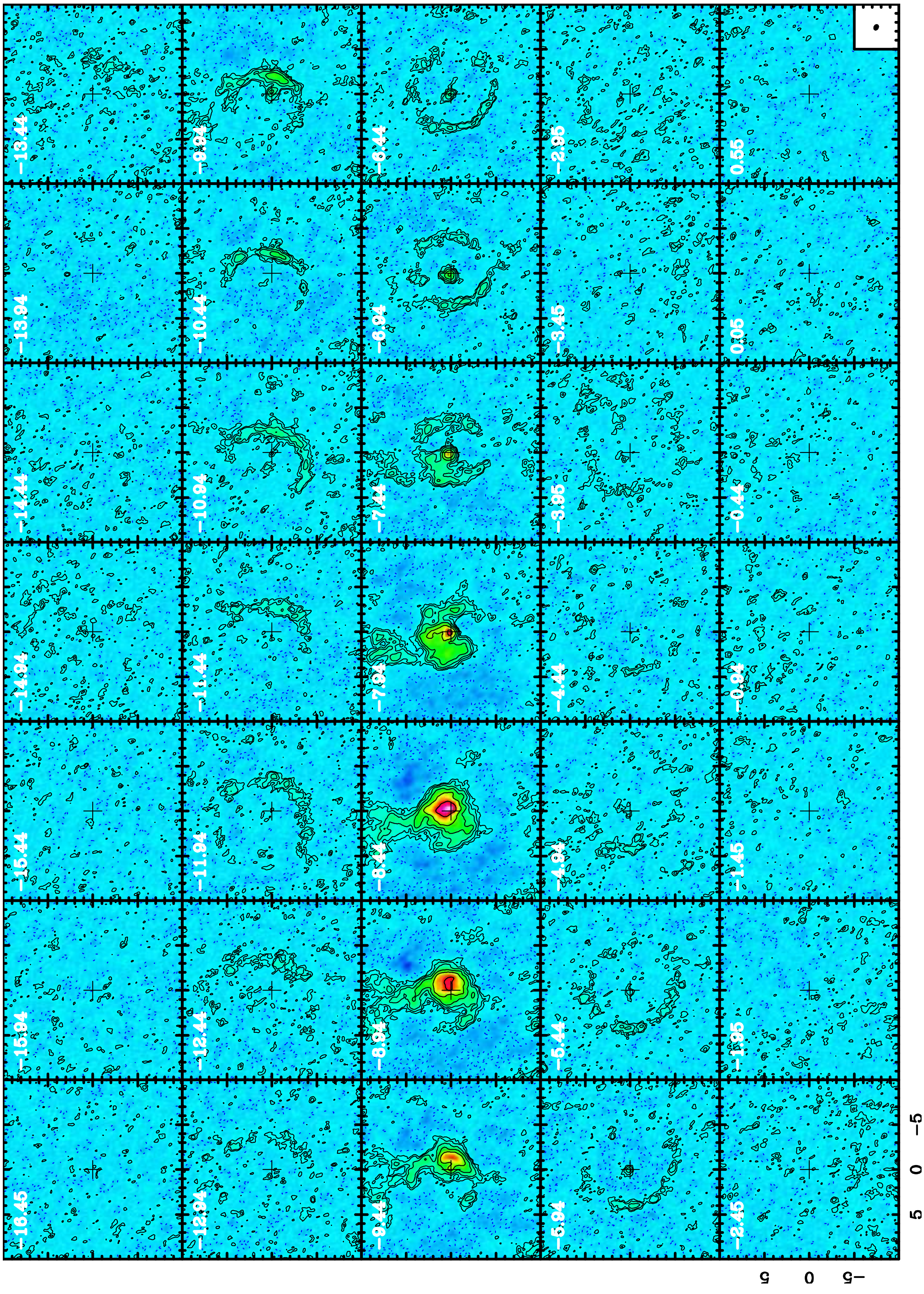

Figure 7.4: Merged (NOEMA + 30 m) maps per velocity channel of $^{13}$CO $J = 2 - 1$ emission from 89 Her. The contours are: −5, 5, 10, 20, 40, 80, and 160 mJy beam$^{-1}$, with a maximum emission peak of 240 mJy beam$^{-1}$. The resulting clean-beam has a HPBW of $0\farcs76 \times 0\farcs59$, the major axis oriented at PA = 27°. The LSR velocity is indicated in the upper right corner of each velocity-channel panel and the beam size is shown in the last panel. The FOV of each panel is $20'' \times 20''$. X and Y units are offsets in arc-sec in the East and North directions respectively.

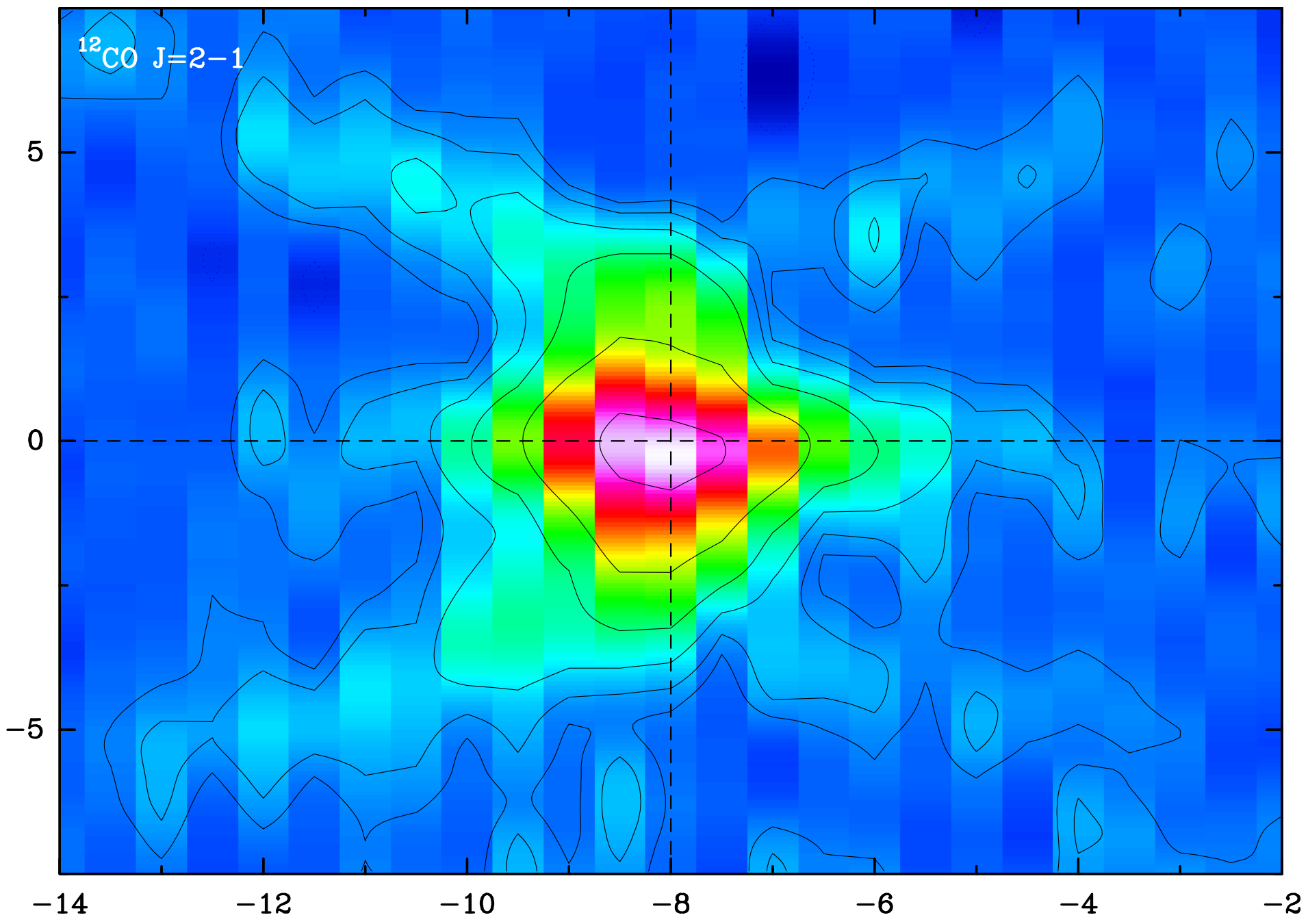

Figure 7.5: Position-velocity diagram from our merged maps of $^{12}$CO $J = 2 - 1$ in 89 Her along the direction $PA$ = 150°, corresponding to the nebula equator. The contours are: −40, 40, 80, 160, 320, 640, and 1280 mJy beam$^{-1}$, with a maximum emission peak of 1.45 Jy beam$^{-1}$. The dashed lines show the central position and systemic velocity.

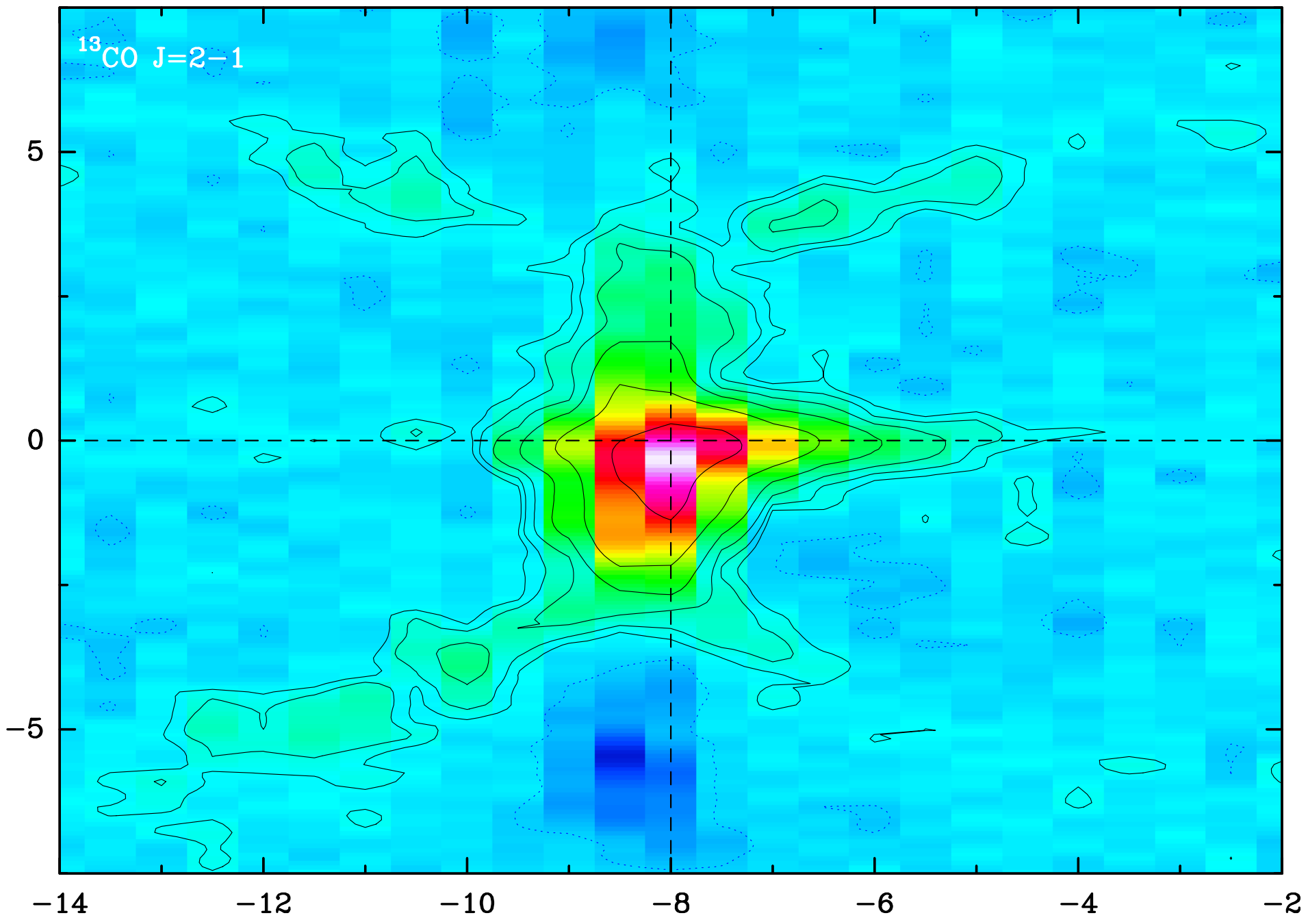

Figure 7.6: Position-velocity diagram from our merged maps of $^{13}$CO $J = 2 - 1$ in 89 Her along the direction $PA$ = 150°, corresponding to the nebula equator. The contours are: −5, 5, 10, 20, 40, 80, and 160 mJy beam$^{-1}$, with a maximum emission peak of 240 mJy beam$^{-1}$. The dashed lines show the central position and systemic velocity.

Table 7.1 Physical conditions in the molecular Keplerian disk and expanding hourglass-like component of 89 Her derived from our best-fit model of the CO data.

| Parameter | Disk | Outflow | Outflow (old) |
|---|---|---|---|
| Radius [cm] | $5.0 \times 10^{15}$ | $R_{\rm max} = 9.5 \times 10^{16}$<br>$W_{\rm o} = 0.8 \times 10^{16}$ | $8.6 \times 10^{16}$<br>$0.7 \times 10^{16}$ |
| Height [cm] | $1.0 \times 10^{15}$ | $12 \times 10^{16}$ | $7.2 \times 10^{16}$ |
| Density [cm$^{-3}$] | $n_0 = 2.0 \times 10^7$<br>$\kappa_{\rm n} = 2.0$ | $n_0 = 1.5 \times 10^3$<br>$\kappa_{\rm n} = 2.0$ | $n_0 = 3.0 \times 10^3$<br>$\kappa_{\rm n} = 1.8$ |
| Temperature [K] | $T_0 = 75$<br>$\kappa_{\rm T} = 2.5$ | $T_0 = 5.5$<br>$\kappa_{\rm T} = 0.2$ | $T_0 = 10$<br>$\kappa_{\rm T} = 0$ |
| Rot. Vel. [km s$^{-1}$] | 1.5 | – | |
| Exp. Vel. [km s$^{-1}$] | – | 1.2 | |
| X($^{12}$CO) | $2.0 \times 10^{-4}$ | $2.0 \times 10^{-4}$ | |
| $^{12}$CO / $^{13}$CO | 10 | 10 | |
| Inclination [°] | | 15 | |
| Position angle [°] | | 60 | |

**Notes.** Parameters and their values used in our best-fit model for the disk and outflow. Note that we also include those parameters from the previous work (old) that have changed. See Gallardo Cava et al. (2021) for a complete description of the model and its parameters.

## 7.3 Model fitting: the mass of the nebula

To derive the physical parameters of the nebula from the new data here presented, we have adopted a nebular model based on a rotating disk with Keplerian dynamics surrounded by an extended and expanding hourglass-shaped structure. We use the code described in Gallardo Cava et al. (2021), where a complete description can be found. As in the previous work, we adopt a distance of 1 000 pc. We assume LTE populations for the $J = 2-1$ CO lines, which is a reasonable assumption for low-$J$ rotational levels, and simplifies the calculations and provides a more comprehensible interpretation of the fitting parameters. We adopt the same relative abundance values as in the previous work: X($^{13}$CO) = $2 \times 10^{-5}$ and an abundance ratio of X($^{12}$CO)/X($^{13}$CO) = 10. These abundance values are usually found in nebulae around binary post-AGB stars.

We assume the presence of a rotating disk in the innermost region of the nebula and a large and wide hourglass-like extended and expanding component. We confirm the inclination of the nebular symmetry axis with respect to the line of sight, together with $PA = 150°$ for the equatorial direction. We also corroborate the self-absorption effects at low negative expansion velocities.

The main parameters of our new best model can be seen in Table 7.1 and can be described as follows. As in the first work, we find the same reliable results for density and temperature laws with high slope values. The density of the rotating disk is assumed to vary with the distance to the binary system following a potential law, $n \propto r^{-2}$, with a value of $2.0 \times 10^7$ cm$^{-3}$ at $2.5 \times 10^{15}$ cm. The temperature varies as $T \propto r^{-2.5}$, with a temperature of 425 K at $5 \times 10^{15}$ cm (the disk radius). Moreover, we assume Keplerian rotation in the disk, with 1.5 km s$^{-1}$ at $10^{16}$ cm, that is compatible with a central total stellar mass of 1.7 M$_\odot$.

In the case of the outflow/disk-wind, we assume a radial velocity with a modulus that increases linearly with the distance to the center, and we find moderate velocities $\sim 10\,\mathrm{km\,s}^{-1}$. We assume an hourglass 67% larger with walls 10% wider compared to the previous work (see Fig. 7.7). Therefore, the total height of the hourglass is $2.4 \times 10^{17}$ cm and its walls are $\sim 8 \times 10^{15}$ cm wide. The density law varies as $n \propto r^{-2}$, with values in between $10^5\,\mathrm{cm}^{-3}$, in the zones closest to the rotating disk, and $\leq 10^3\,\mathrm{cm}^{-3}$, in the most external regions of the outflow. The temperature of the outflow varies as $T \propto r^{-0.2}$, with values $\leq 10$ K in the external regions of the outflow. So, we find a very extended outflow composed of cold gas, same as in the case of R Sct or IRAS 19125+0343, which are similar objects (see Gallardo Cava et al., 2021).

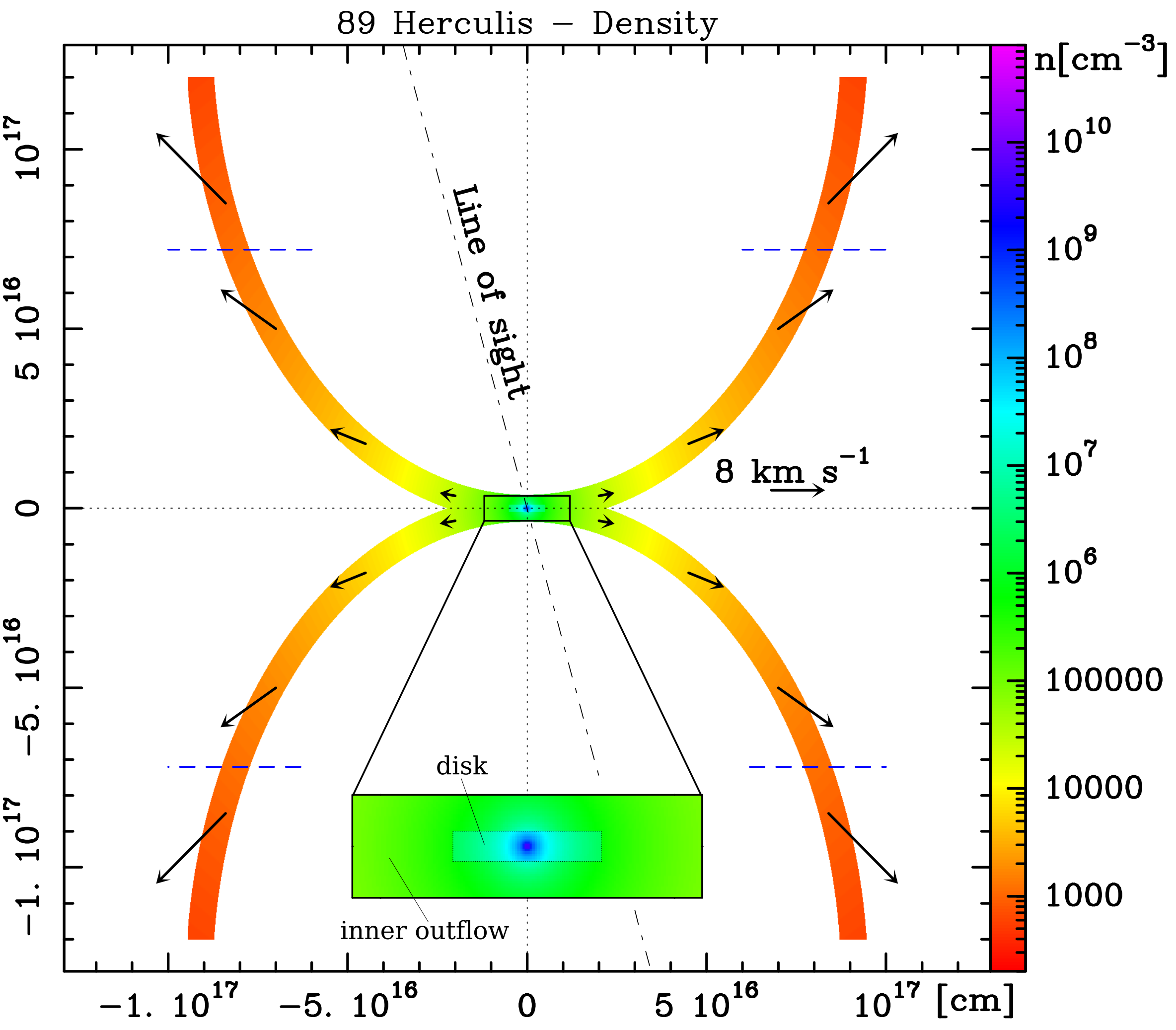

Figure 7.7: Structure and distribution of the density of our best-fit model for the disk and outflow of 89 Her. The lower inset shows a zoom into the innermost region of the outflow together with the rotating disk, with density values of $\sim 10^6$ and $\geq 10^7\,\mathrm{cm}^{-3}$, respectively. The rest of the outflow presents density values $\leq 10^5\,\mathrm{cm}^{-3}$. The expansion velocity is represented with arrows. The four horizontal blue dashed lines mark the former extent of the nebula in our previous model (see Gallardo Cava et al., 2021).

We show the predictions from our best model in our synthetic velocity maps (Figs. 7.11 and 7.12) and synthetic PV diagrams (Figs. 7.13 and 7.14) and our new nebula model for 89 Her in Fig. 7.7. We have also checked that our model satisfactorily reproduces previous interferometric maps of the $^{12}$CO $J = 1 - 0$ emission line (data taken from Fong et al., 2006, whose maps also show no flux loss). The model reproduces the observational data and yields a total mass for the nebula of $1.8 \times 10^{-2}\,M_{\odot}$, of which $1.2 \times 10^{-2}\,M_\odot$ corresponds to the outflow and $6.4 \times 10^{-3}\,M_\odot$ to the rotating disk. This means that 89 Her presents a Keplerian disk surrounded by a very extended and expanding outflow that represents $\sim$65% of the total mass of the nebula.

Some of the nebula properties are not accurately determined because of the relatively low resolution of the interferometric data, such as the height of the Keplerian disk. On the contrary, the outflow structure is well determined, because the hourglass-like shape is clear in our maps. See Gallardo Cava et al. (2021), for further details on the uncertainties of our model.

## 7.4 Disk-mass ratio in binary post-AGB stars

In our previous work, we classified 89 Her as an intermediate subclass nebula, because both the disk and its outflow presented very similar masses, but those maps suffered a significant amount of filtered out flux (see Sect. 7.1). According to our new merged maps and the new model, the pPN around 89 Her is dominated by its disk-wind, because $\sim$65% of the total mass is located in the very large hourglass-like expanding component.

After this update on the masses of both outflow and disk in 89 Her, this source is more in line with what is found in those classified as outflow-dominated. In fact, if we only consider sources for which these masses are relative well determined (see Table 7.2 and Fig. 7.8), it seems that there is not an intermediate case in between the disk- and the outflow-dominated subclass. However, before drawing any conclusion, it is better to revise any problem in the estimation of the masses of the two components, such as the impact of and hypothetical loss of flux (as in the case of 89 Her we have just reviewed), or the impact of other nebular components, in particular the neutral and ionized atomic gas, or the contribution of the layers where molecules have been photo-dissociated by the interstellar UV field.

We remind that the main goal of this work is to study the circumbinary disk and the outflow/disk-wind that is escaping from the rotating component during the post-AGB phase. Other nebular components, such as collimated high-velocity stellar winds (launched by the companion) or larger haloes (formed during the previous AGB phase, which may exist and whose detection could be quite difficult), are not part of this work. Anyhow, we discuss the relevance of these structures and their (possible) contribution to the total nebular mass.

### 7.4.1 The effects of the interferometric flux losses

Seven sources have been well studied through interferometric observations: AC Her, the Red Rectangle, IRAS 08544−4431, IW Car, 89 Her, IRAS 19125+0343, and R Sct. Here, we discuss the flux loss present in the maps used to estimate the molecular mass in each source of our sample and update the values of the mass of the outflow and consequently the disk-to-outflow mass ratio. In the following, we will assume that the filtered out flux is due to the outflow, since no other large scale structures have been detected in these kind of objects.

Table 7.2 Sample of binary post-AGB stars, in which the disk-mass ratio has been analysed.

| Source | $M_{\rm neb}$ [$M_\odot$] | Outflow [%] | $d$ [pc] | Subclass |
|---|---|---|---|---|
| AC Herculis | $8.3 \times 10^{-4}$ | 5 | 1100 | Disk-dominated |
| Red Rectangle | $1.4 \times 10^{-2}$ | 11 | 710 | Disk-dominated |
| IRAS 08544−4431 | $2.0 \times 10^{-2}$ | ≤12 | 1100 | Disk-dominated |
| IW Carinae | $4.0 \times 10^{-3}$ | ≤14 | 1000 | Disk-dominated |
| 89 Herculis | $1.8 \times 10^{-2}$ | 65 | 1000 | Outflow-dominated |
| IRAS 18123+0511* | $4.7 \times 10^{-2}$ | 70 | 3500 | Outflow-dominated |
| IRAS 19125+0343 | $1.1 \times 10^{-2}$ | 71 | 1500 | Outflow-dominated |
| R Scuti | $3.2 \times 10^{-2}$ | 73 | 1000 | Outflow-dominated |
| AI Canis Minoris* | $1.9 \times 10^{-2}$ | 75 | 1500 | Outflow-dominated |
| IRAS 20056+1834* | $1.0 \times 10^{-1}$ | 75 | 3000 | Outflow-dominated |

**Notes.** Sources marked with asterisk have been studied only via single-dish observations and the values of their masses and outflow-mass ratios can present higher uncertainties.

### AC Herculis

The $^{12}$CO $J = 2 - 1$ mm-wave interferometric observations of this source does not present a significant amount of filtered out flux. In addition, its angular size is around $2''$, which is too small compared with the beam size. Therefore, we can definitely discard the presence of a very extended and not detected component. According to this, the total molecular nebular mass is $8.3 \times 10^{-4}$ M$_\odot$ and the derived disk-mass ratio of AC Her, $\sim$95%, is well determined. See Gallardo Cava et al. (2021) for further details.

### The Red Rectangle

This source is the best studied of our sample. The nebular mass was derived from model fitting of maps of lines $^{12}$CO $J = 3 - 2$ and $J = 6 - 5$, $^{13}$CO $J = 3 - 2$, C$^{17}$O $J = 6 - 5$, and H$^{13}$CN $J = 4 - 3$. The total nebular mass is $1.4 \times 10^{-2}$ M$_\odot$, of which $\sim$9% corresponds to the outflow that surrounds the Keplerian disk. Bujarrabal et al. (2016) mentioned that they found 25% of flux loss in the line wings. If we assume that this filtered out flux corresponds to the outflow, the contribution of the outflow to the total molecular mass of the nebula will be up to 11%. Therefore, the Red Rectangle remains as a disk-dominated source.

### IRAS 08544−4431

The nebular mass of this source is $2.0 \times 10^{-2}$ M$_\odot$ (Bujarrabal et al., 2018b). According to the authors, the mass of the outflow is 10%. The $^{12}$CO $J = 3 - 2$ maps present a small fraction of flux loss ($<$ 20%), and the $^{13}$CO $J = 3 - 2$ maps, with a less extended brightness distribution, do not show a significant amount of filtered out flux. Assuming that this flux loss is due to the outflow, its contribution to the total molecular mass will be $\leq$ 12%. Additionally, the angular size of this source is around $4''$, which is considerably smaller than the beam size.

#### IW Carinae

This source, of $4''$ in angular size, presents a total nebular mass of $4 \times 10^{-3}\,M_\odot$, of which 11% is located in the expanding component (see Bujarrabal et al., 2017, for details). The authors found a small fraction of filtered out flux (25%) in the $^{12}$CO $J = 3 - 2$ maps and can be higher in the line wings. In the case that this flux loss corresponded to the expanding component, it mass contribution will be $\leq 14\%$ of the total nebular mass. Additionally, Bujarrabal et al. (2013a) found, via single-dish observations, very narrow CO line profiles characteristic of rotating disk and weak wings that correspond to the outflow. This source would still be a disk-dominated source even in the worst flux loss scenario.

#### 89 Herculis

In this work, we present interferometric maps that were merged with large-scale single-dish maps. So, these combined maps do not present filtered out flux (see Sect. 4.4). Our new combined maps and model allow us to estimate the mass of the nebula that surrounds 89 Her. We find a total molecular mass of $1.8 \times 10^{-2}\,M_\odot$, in which the hourglass-shaped outflow represents 65% of the mass (see Sect. 7.3).

#### IRAS 19125+0343

This nebula presents a total mass of $1.1 \times 10^{-2}\,M_\odot$ and 71% corresponds to the outflow (see Gallardo Cava et al., 2021, for details). The interferometric visibilities were merged with zero-spacing data obtained with the 30 m IRAM telescope. Therefore, there is not filtered out flux in the $^{12}$CO $J = 2 - 1$ final maps, so this source is clearly an outflow-dominated source.

#### R Scuti

The nebular mass is $3.2 \times 10^{-2}\,M_\odot$, of which 73% is located in the extended outflow that surrounds the Keplerian disk (see Gallardo Cava et al., 2021, for complete description). The $^{12}$CO $J = 2 - 1$ maps do not present filtered out flux, because they were merged with short-spacing pseudo-visibilities obtained from on-the-fly maps with the 30 m IRAM telescope to compensate the extended component filtered out by the interferometer, since this source present a large angular size. This case is similar to that of 89 Her.

### 7.4.2 Atomic mass

Some sources of our sample have been studied in the C II line (157.14 $\mu$m). The flux of this line can be used to measure the mass of the low-excitation atomic component in pPNe, because this transition is useful to estimate the total mass of Photo Dissociation Regions (PDRs). PDRs practically coincide with the cold atomic gas regions, because they are very clearly delimited zones in between the molecular and the H II regions. In addition, most of our sources show relatively low stellar temperatures, $<10\,000$ K, so H II regions are not expected in them. We note that in Gallardo

Cava et al. (2022c) no radio recombination lines were detected in these spectral surveys. Castro-Carrizo et al. (2001) and Fong et al. (2001) found that the PDR masses for AC Her, the Red Rectangle, 89 Her, and R Sct (rescaled to our distances, see Table 7.2) are $< 10^{-2}$, $< 10^{-2}$, $< 2 \times 10^{-3}$, and $< 3 \times 10^{-2}\,M_\odot$. Bujarrabal et al. (2016) detected C II in the Red Rectangle and, using the method described in Castro-Carrizo et al. (2001), the mass of the PDR derived from the measured flux would represent $< 10^{-3}\,M_\odot$ that would mostly located in the rotating disk.

We conclude that there are not indications of a significant contribution of PDRs or H II regions in the masses of the disk-containing nebulae around binary post-AGB stars. Therefore, there is no need to consider these small contributions in the disk-mass ratio analysis.

### 7.4.3 post-AGB stellar winds

Systematic high-resolution radial velocity studies reveal that stellar jets or winds, are a common phenomenon in binary post-AGB stars (see Bollen et al., 2021, and references therein). A collimated low-density and high-velocity jet, launched by the companion, often operates during the late AGB and early post-AGB stages (see Bollen et al., 2022, and their Fig. 1). We note that this is a nebular component not studied in this work (focused on rotating disks and disk-winds). However, we argue that the mass of this component is small.

This kind of stellar wind is present in the best studied source of our sample, the Red Rectangle, and can be clearly seen in the optical image (Cohen et al., 2004). At first sight, the H$\alpha$ emission of the famous X-shaped structure seems to dominate the whole nebula. Nevertheless, according to the high velocities and mass-loss found in this stellar wind ($\sim 200\,\mathrm{km\,s^{-1}}$ and $10^{-6}\,M_\odot\,\mathrm{a}^{-1}$, respectively[2], see Thomas et al., 2013; Bollen et al., 2019), we deduce that this jet is $\lesssim 1\,000$ years old and its mass is $\lesssim 10^{-3}\,M_\odot$. Although this stellar wind are not part of our study, its mass is small compared with the mass derived in this work for the rotating disk and, and slight lower than that of the molecular outflow. In addition, we highlight that the X-like structure seen in Cohen et al. (2004) only shows the gas that is closest to the axis, while CO interferometric observations reveal that the optical X-shaped structure represents the inner layer of the outflowing biconical component (Bujarrabal et al., 2016). Moreover, Men'shchikov et al. (2002), from a detailed model of this nebula in polar direction, found that the gas and dust density values sharply decay from $2''$. This is roughly consistent with the strong decrease in brightness seen in Cohen et al. (2004) at $\pm 2''$, as the optical brightness of the outermost regions of the Red Rectangle ($\pm 10''$) is more than 50 times weaker than that at $\pm 2''$.

In view of what is stated in Sect. 7.4.1, the rotating disk dominates the Red Rectangle nebula. As for the expanding gas, we clearly differentiate between an outflow, composed of low-velocity expanding gas escaping from the disk, and these collimated high-velocity stellar winds that do not contribute notably to the total nebular mass (and their study is not part of this work). Therefore, we prefer not to take this contribution into account, as it does not significantly affect our results. We expect a similar or even more favorable situation for all the sources in our study. For instance, the size of 89 Her via CO data seems to be compatible with that in the optical/IR image.

[2]We follow the recommendations for units of the IAU Style Manual (Wilkins, 1990, 1995). Therefore, we use the term annus, abbreviated as "a", for year.

### 7.4.4 Possible presence of haloes

It is though that $\lesssim 1\,M_\odot$ of mass-loss from the AGB phase is necessary to reach the post-AGB stage. However, the masses detected in this study are small compared to what is predicted from theoretical works. One may ask if there is a large component surrounding the material that we detect.

The origin of these large quasi-spherical fossil shells requires successive ejecta of stellar material when the star reached the tip of the AGB phase (Guerrero and Manchado, 1999). This structure can be difficult to detect and its mass can vary between $10^{-4}$ and $\sim 0.1\,M_\odot$, being its contribution to the total mass perhaps not negligible (Chu et al., 1991; Bujarrabal et al., 2001; Guerrero et al., 2003; Cox et al., 2012; Van de Steene et al., 2015). Models of dust emission from our sources and standard pre-planetary nebulae show significant differences, with no hint in our disk-containing objects of extended shells that are characteristic of standard pPNe, see e.g. Gezer et al. (2015); Hrivnak et al. (1989) (as in general detected in CO lines). In particular, from far infrared (FIR) data in Gezer et al. (2015), in particular the emission excess at $20 - 100\,\mu$m in their Fig. 2, one finds that the mass of the undetected outer-shell in 89 Her must be more than 100 times lower than that of the well-known pPN HD 161796. We remind that the *Gaia* distance of HD 161796 is $\sim 2$ kpc and that its nebula is supposed to be placed at $1 - 3 \times 10^{17}$cm, just beyond the nebula we detect in 89 Her. Since the mass of the HD 161796 nebula is around $0.1\,M_\odot$ (from CO data in Bujarrabal et al., 2001, and again accounting for the new *Gaia* distance), we derive a lower limit for the mass for any very outer shell around 89 Her, $\lesssim 10^{-3}\,M_\odot$, which is a small value compared with the mass values derived in this paper for the rotating disk and disk wind.

The objects of our sample could present the same well-known missing mass problem as the rest of evolved sources (see e.g. Santander-García et al., 2021, and references therein). Is this missing mass in a hard-to-detect halo? Future ultra-sensitive observations and theoretical works will be needed to answer this question. In any case, these haloes, which by definition arise from mass loss during the AGB stage, have nothing to do with the post-AGB structures studied in this work (rotating disks and outflows).

### 7.4.5 Contribution of outer layers of the disk-wind

We remind that, according to our previous discussion, the contribution of very wide components is not significant, in particular in the case of 89 Her (see also Hrivnak et al., 1989; Gezer et al., 2015).

In our analysis, the contribution of very outer layers, where CO can be photodissociated by interstellar UV, has not been included. This is a very general problem in the study of circumstellar envelopes, because these cool, dissociated layers are very difficult to detect and analyse. We think that the contribution of such outer haloes is likely negligible. General studies of FIR dust emission (see e.g. Gezer et al., 2015; Vickers et al., 2015), which is known to be useful to identify very outer layers beyond the CO photodissociation radius, do not reveal the presence of sizable very extended components in binary post-AGB stars.

Additionally, Men'shchikov et al. (2002) modeled with great detail the dust emission from the Red Rectangle, in regions far from the poles, including an extended component in expansion, and found that the density decreases very steeply from a radius of 1 000 – 2 000 AU, very similar to that derived from CO maps (Bujarrabal et al., 2016). Seemingly, the contribution of outer photodissociated shells to the total mass budget should be negligible in our sources.

Moreover, from the CO maps, we know that the kinetic time of the CO outflows in binary post-AGB stars is of a few thousand years. According to the evolutionary models, this time scale is comparable to the total expected time spent by a star in its transition from the AGB to the PN phase (see Bloecker and Schoenberner, 1990; Miller Bertolami, 2016). High-mass Keplerian disks are a typical post-AGB phenomenon, being very rare or absent in AGB stars, and the ejection of the observed outflow from the disk must be also a post-AGB phenomenon.

To conclude, as far as we know we cannot expect much more extended outflows in the studied objects. Although, we recognize that the analysis of these difficult-to-detect layers is uncertain. Even in the hypothetical case that the mass of outer and undetected layers of the outflow would be comparable to that of the detected ones (i.e. a factor of 2), the contribution of the disk-wind to the total nebular mass would be always $< 25\%$ in the case of the disk-dominated sources. On the contrary, this outflow mass ratio would be $\gtrsim 80\%$ in the case of the outflow-dominated ones. Thus, even under this hypothetical scenario, our main conclusions of the disk-to-mass ratio fraction will not change significantly.

### 7.4.6 Bimodal distribution of the wind contribution

After this very detailed analysis of each object, we have updated the total mass and the contribution of the outflow (i.e. the disk-wind) for the sample. These updated values can be seen in Table 7.2 and Fig. 7.8. In particular, the figure shows that there is a clear bimodal distribution of the contribution of the disk-wind to the total nebular mass around binary post-AGB stars. There are two types of sources; the disk- and the outflow-dominated sources. Note that the segregation between the two subclasses is very clear. The disk-dominated sources always present disk-mass percentages above 15%, whereas in the outflow-dominated ones this value is always below 40%. More precisely, the disk-dominated sources present an outflow contribution of $10 \pm 5\%$, while in the outflow-dominated ones this figure is $70 \pm 10\%$.

We note that there are not intermediate sources in between both subclasses; see Table 7.2 and Fig. 7.8. The absence of intermediate sources indicates that the two subclasses of sources are probably not related via an evolutionary link, otherwise one should expect to find intermediate sources because of the similar outflow and post-AGB lifetimes (Sect. 7.4.5). Therefore, we suggest that the existence of these two subclasses will be related to the initial stellar properties and configuration of the binary system, resulting in the ability of launching a more or a less massive wind from the rotating disk. Of course, such a reasoning would have to be revised if the post-AGB life scales of our sources is demonstrated to be much longer than for standard pPNe, at least by an order of magnitude, which cannot be excluded; to explain the bimodal distribution we would then have to assume that the mass loss takes place in a critical small fraction of the post-AGB evolution.

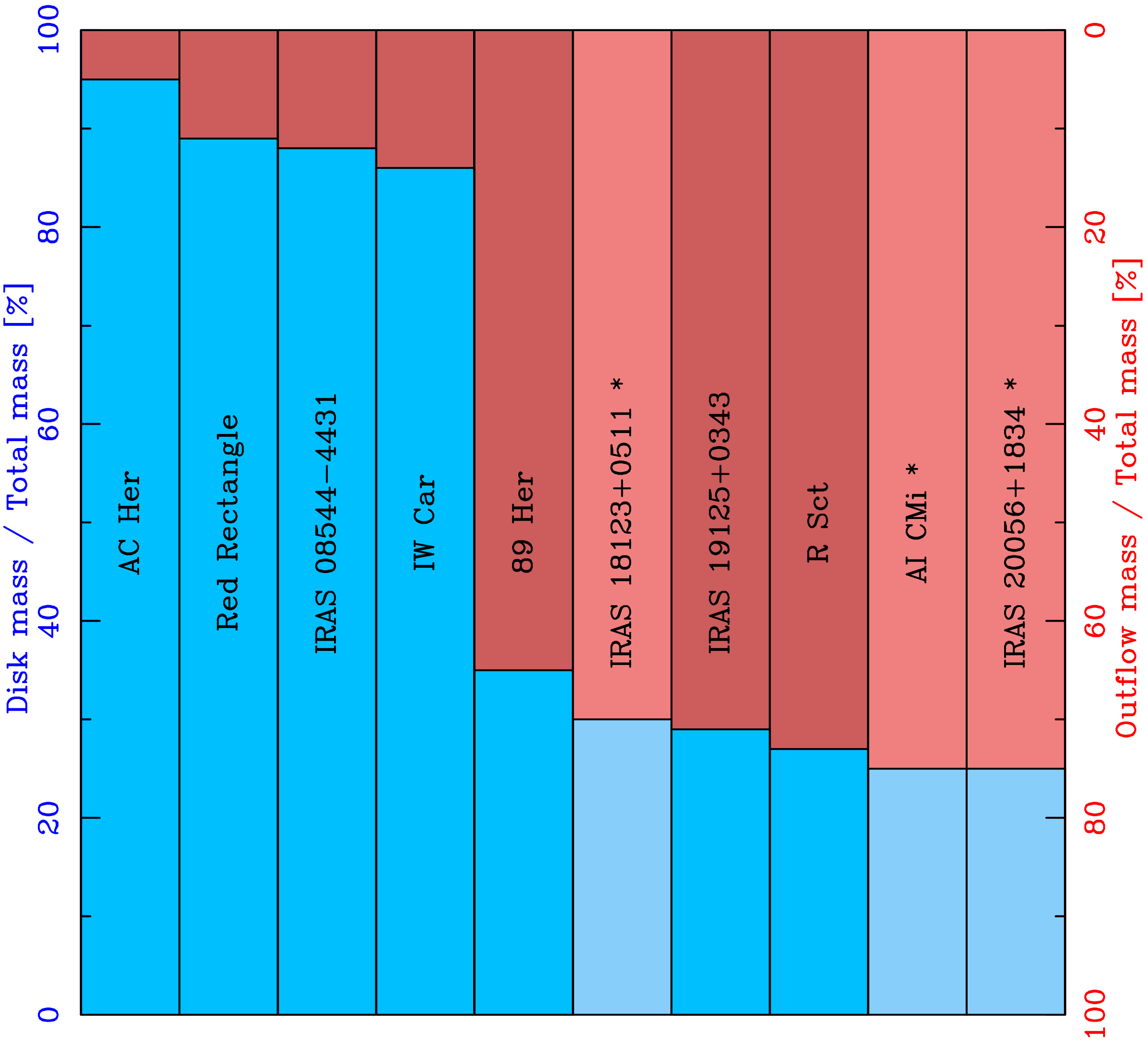

Figure 7.8: Histogram showing the mass percentage of the disk (blue) and outflow (red) components for our sample of nebulae around binary post-AGB stars. Two subclasses are clearly differentiated: the disk- and the outflow-dominated sources. Sources marked with asterisk (and lighter colours) have been studied only through single-dish observations, and their values are less certain.

## 7.5 Conclusions

The new combined maps of 89 Her with no flux loss, lead us to study the very large hourglass-like extended component that surrounds the Keplerian disk. We find a total nebular mass of $1.8 \times 10^{-2}$ $M_\odot$, of which 65% is located in the outflow. According to these new mass estimations, 89 Her has been re-classified as an outflow-dominated source.

After a very detailed analysis of each object of our sample of nebulae around binary post-AGB stars, we clearly find two-subclasses: the disk- and outflow-dominated sources. We present in Fig. 7.8, an histogram that shows the mass percentage of the rotating disk and outflow components for our sample. The disk-dominated sources contain $85-95\%$ of their mass located in the rotating component, while the outflow-dominated sources only contain $25-35\%$ (HD 52961 and IRAS 19157−0247 have not been considered in this work, because they were classified as intermediate sources under high uncertainties, and they could belong to either subclass).

The nebulae around binary post-AGB stars conform a bimodal distribution (see Fig. 7.8). The existence of these two very different subclasses does not support an evolutionary relationship between them, since the post-AGB evolution is believed to be very fast, comparable to the time required to form these disk-winds. On the contrary, this could be due to a different initial configuration of the binary system or different initial masses. We propose that the outflow-dominated sources could result from the presence, for a yet unknown reason, of a very efficient mass loss from the disk in certain objects. It is possible that the outflow would be particularly conspicuous in objects that have spent a relatively long time in the post-AGB phase and then had time to form high-mass very extended outflows.

## 7.6 Supporting material

### 7.6.1 $^{12}$CO and $^{13}$CO $J=2-1$ total-power maps

Here we present total-power maps for 89 Her of the $^{12}$CO and $^{13}$CO $J=2-1$ emission lines, see Figs. 7.9 and 7.10. These observations were performed through the OTF mode using the 30 m IRAM telescope. We detect emission at angular distances of around 20″ in diameter. Our new total-power maps present expansion velocities, which reveal a larger expanding component. See Sect. 7.2.1 for complete description.

### 7.6.2 Model results

In this appendix, we show the synthetic velocity maps of 89 Her for $^{12}$CO and $^{13}$CO $J=2-1$ emission for our best-fit model (see Figs. 7.11 and 7.12). The large hourglass-like structure is present in our synthetic velocity maps. We also show the synthetic PV diagrams along the equatorial direction for $^{12}$CO and $^{13}$CO $J=2-1$ line emission (see Figs. 7.13 and 7.14). See Sect. 7.3 for further details.

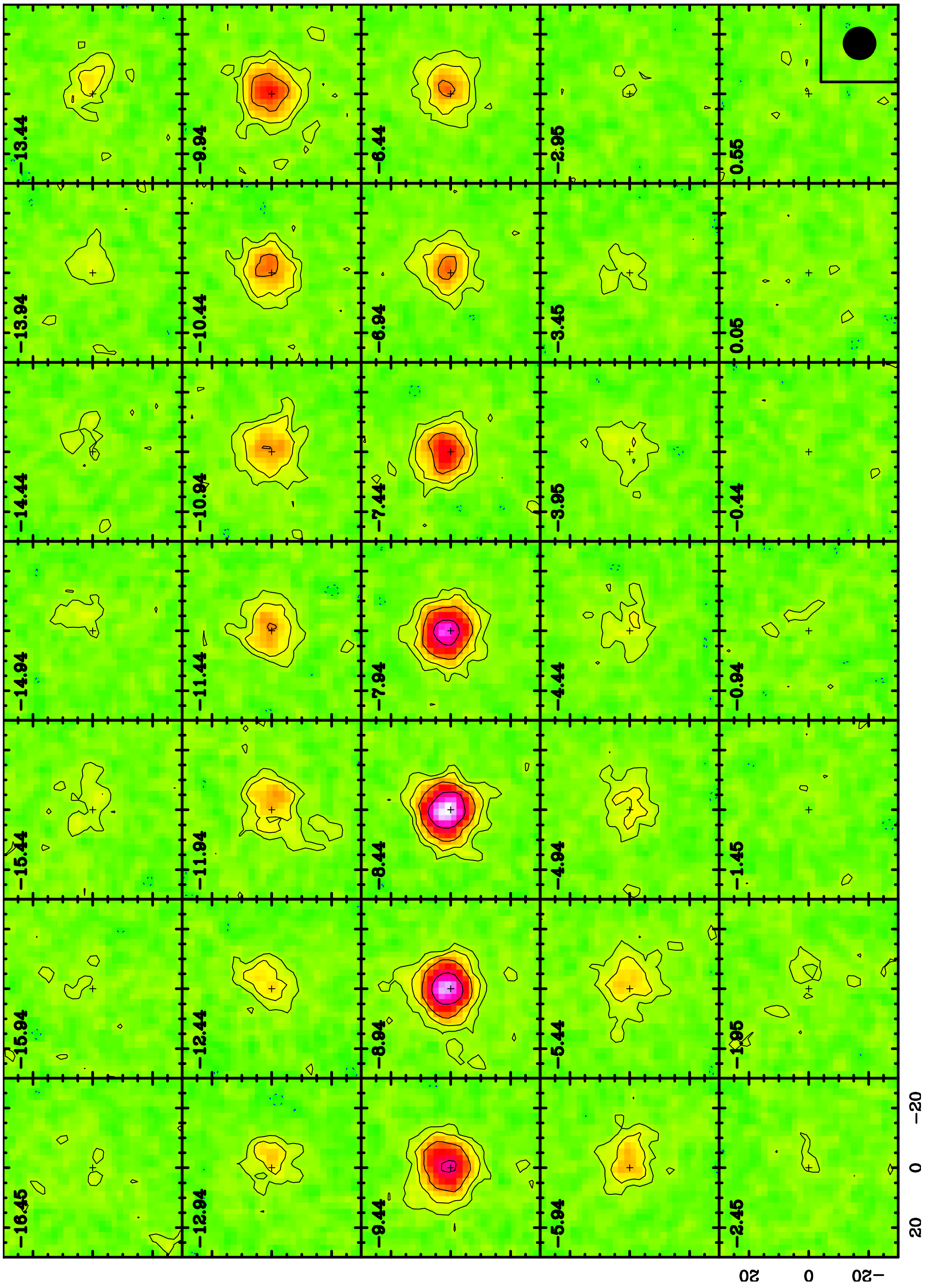

Figure 7.9: Total-power maps per velocity channel of $^{12}$CO $J = 2 - 1$ emission from 89 Her. The contours are −0.1, 0.1, 0.2, 0.4, and 0.8 K in main-beam scale. The beam size (HPBW) is $11.''25 \times 11.''25$. The LSR velocity is indicated in the upper right corner of each velocity-channel panel and the beam size is shown in the last panel. The FOV of each pannel is $60'' \times 60''$.

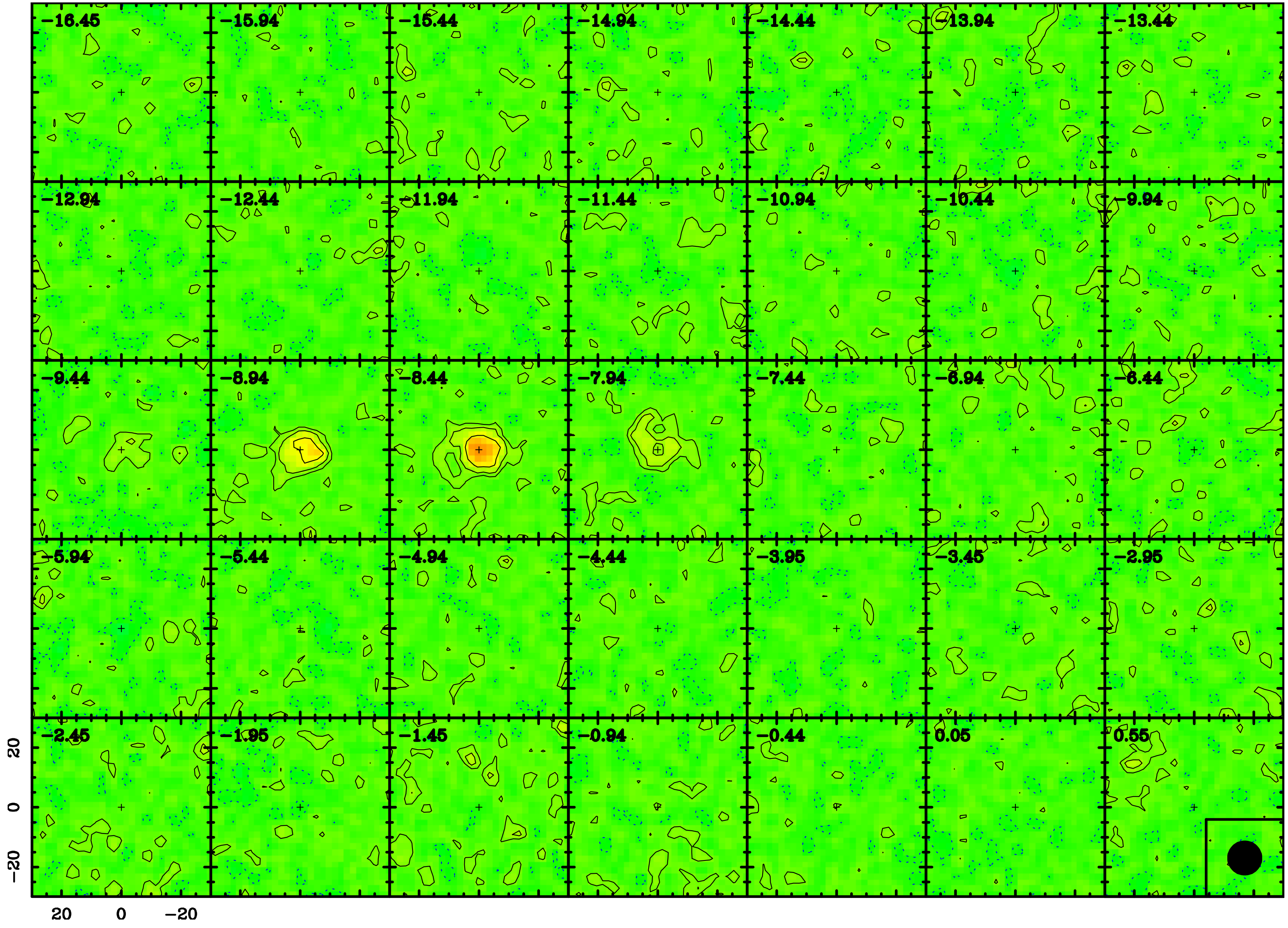

Figure 7.10: Total-power maps per velocity channel of $^{13}$CO $J = 2 - 1$ emission from 89 Her. The contours are −0.08, 0.08, 0.16, and 0.32 K in main-beam scale. The beam size (HPBW) is 11.″77 × 11.″77. The LSR velocity is indicated in the upper right corner of each velocity-channel panel and the beam size is shown in the last panel. The FOV of each pannel is 60″ × 60″.

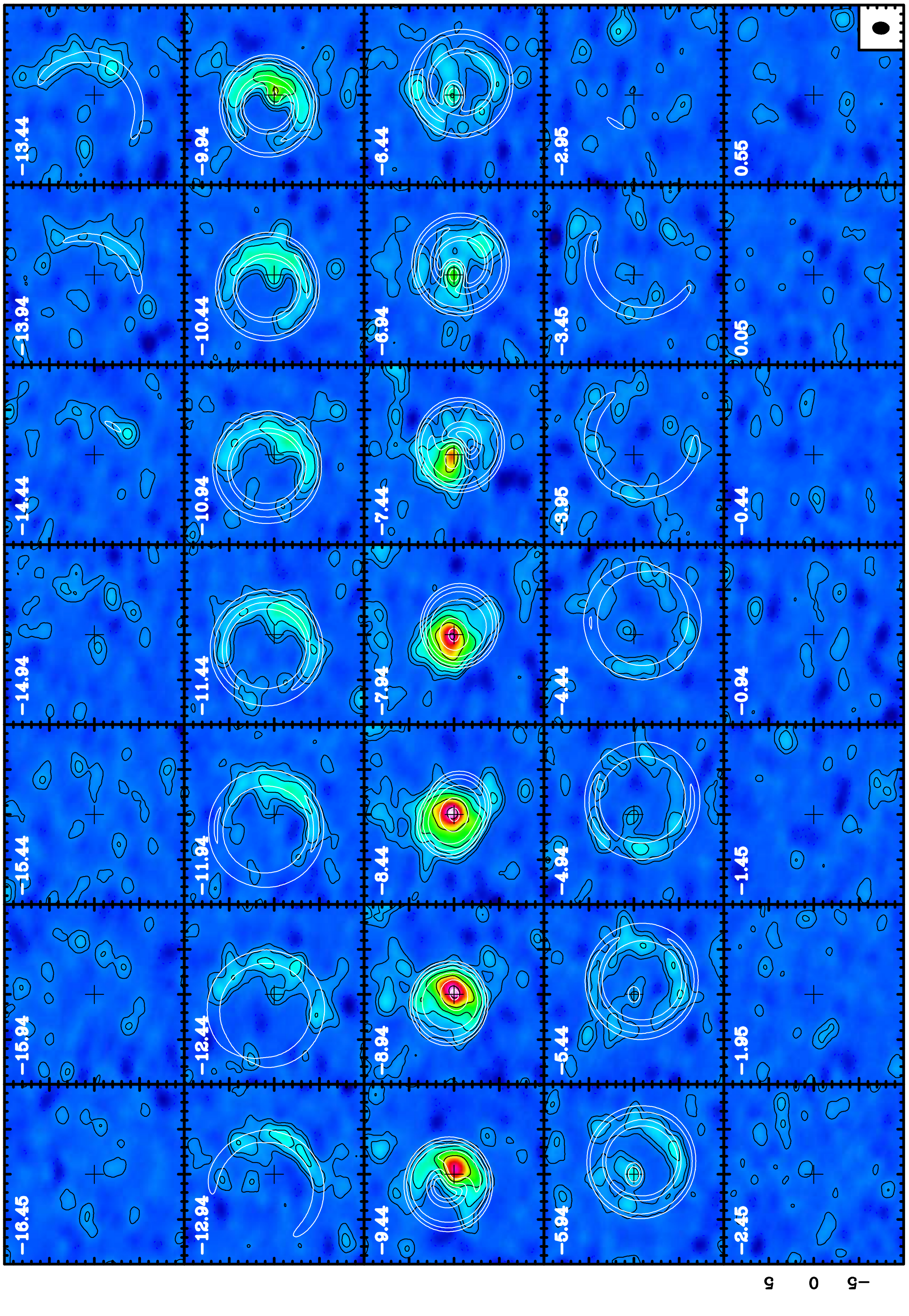

Figure 7.11: Synthetic maps predicted by the model of the $^{12}$CO $J = 2 - 1$ line emission for the nebula around 89 Her in white colours. They are superimposed to the observational ones. The contours for both observational data and model are: −40 (observational data only), 40, 80, 160, 320, 640, and 1280 mJy beam$^{-1}$.

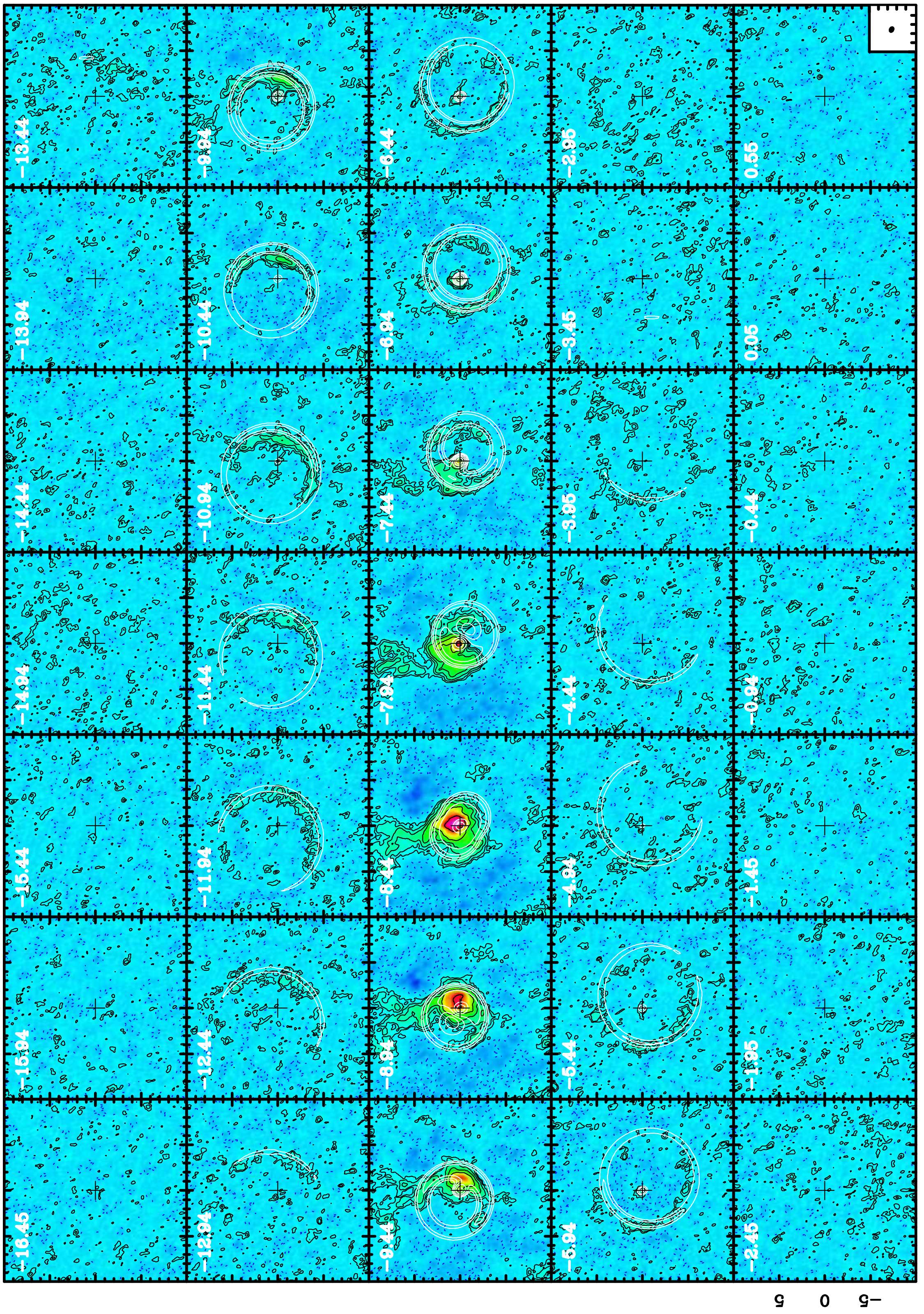

Figure 7.12: Synthetic maps predicted by the model of the $^{13}$CO $J = 2-1$ line emission for the nebula around 89 Her in white colours. They are superimposed to the observational ones. The contours for both observational data and model are: −5 (observational data only), 5, 10, 20, 40, 80, and 160 mJy beam$^{-1}$.

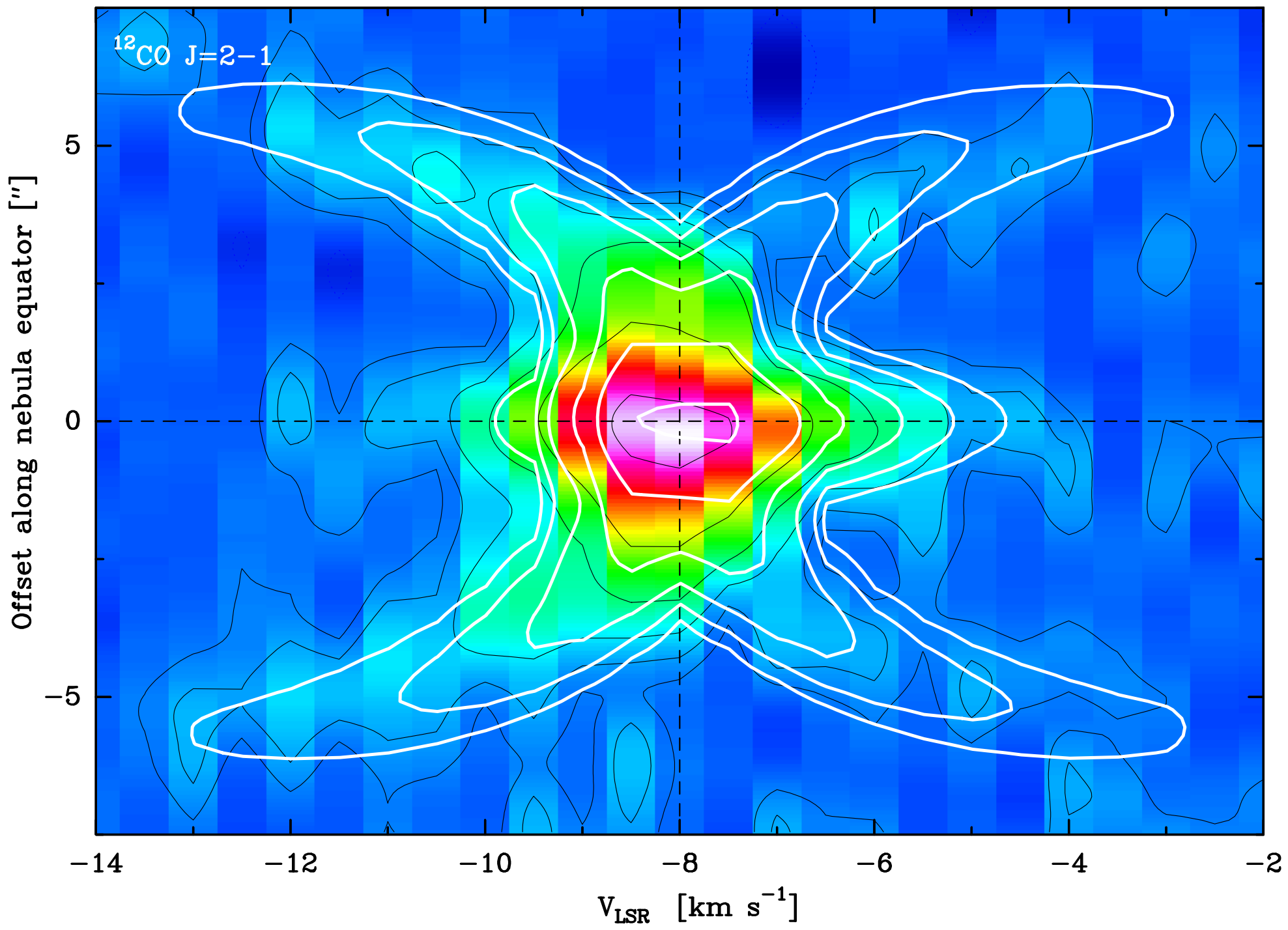

Figure 7.13: Synthetic PV diagram from our best-fit model of $^{12}$CO $J = 2-1$ in 89 Her along the equatorial direction ($PA = 150°$) in white contours. It is superimposed to the observational one. The contours for both observational data and model are: −40 (observational data only), 40, 80, 160, 320, 640, and 1280 mJy beam$^{-1}$.

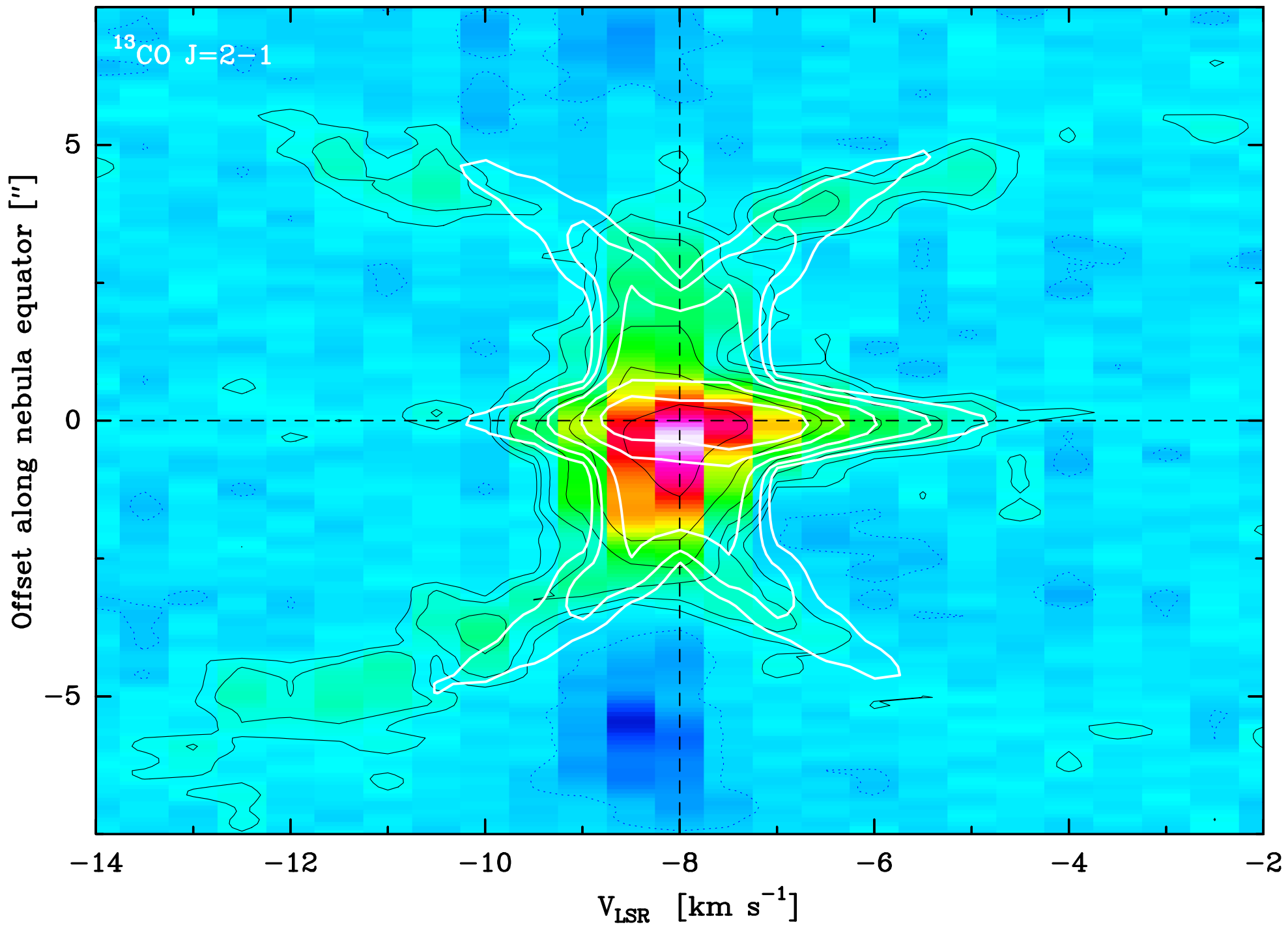

Figure 7.14: Synthetic PV diagram from our best-fit model of $^{13}$CO $J = 2-1$ in 89 Her along the equatorial direction ($PA = 150°$) in white contours. It is superimposed to the observational one. The contours for both observational data and model are: −5 (observational data only), 5, 10, 20, 40, 80, and 160 mJy beam$^{-1}$.

# Part III

# Conclusions and future work

*But what I do I do because I like to do.*
Anthony Burgess – A Clockwork Orange

# 8
# Summary and conclusions

*In this chapter, we present a summary and the main conclusions of this work. The results and conclusiones present in this Ph.D. thesis contribute to expand the knowledge of the binary post-AGB stars surrounded by Keplerian disks and outflows.*

## 8.1 Summary

In this thesis work, we have presented and analysed interferometric and single-dish observational data at millimetre wavelengths of a sample of binary post-AGB stars surrounded by disks with Keplerian dynamics and outflows. Our objective was twofold: to study the nebular emission of molecular envelopes around binary post-AGB stars through intermerometric observations, and to study the chemical content present in these sources through single-dish observations. We summarize our results (see Chapters 4, 5, 6 and 7, for further information):

- In Gallardo Cava et al. (2021), we analyse NOEMA maps of $^{12}$CO and $^{13}$CO $J = 2 - 1$ in 89 Her and $^{12}$CO $J = 2 - 1$ in AC Her, IRAS 19125+0343, and R Sct. In the case of AC Her, we detect Keplerian dynamics in the disk, but the presence of an outflow is doubtful. Nevertheles, our new maps and models allow us to estimate an upper limit to the mass of the outflow, which represents $\lesssim 5\%$ of the total nebular mass ($8.3 \times 10^{-4}$ M$_\odot$). For IRAS 19125+0343 and R Sct, we find a total nebular mass of $1.1 \times 10^{-2}$ M$_\odot$ and $3.2 \times 10^{-2}$ M$_\odot$, respectively, whose $\sim 25\%$ is located in the rotating component. The remaining 75% of the nebular mass corresponds to an extended outflow that surrounds the disk. We find that the hourglass-like extended component that surrounds the Keplerian disk of 89 Her represents around 50% of the total nebular mass ($1.4 \times 10^{-2}$ M$_\odot$). However, this percentage is just a lower limit, because our NOEMA maps present a significant amount of filtered out flux, mainly for the outflow emission. See also Gallardo Cava et al. (2022b).

- Gallardo Cava et al. (2022a): To solve the problem of the flux loss and the mass underestimation of the extended component of the nebula around 89 Her, we perform single-dish on-the-fly observations that include large spatial scales. We perform these observations using the 30 m IRAM telescope and focus our observations on the $^{12}$CO and $^{13}$CO $J = 2 - 1$ emission lines. We combine our total-power maps with our NOEMA maps. These new merged maps contain all detectable flux of the source and, at the same time, have a high spatial resolution provided by the interferometric observations. According to our new maps and model, we find that the hourglass-like structure of 89 Her is larger and more massive than in the previous work. The total nebular mass is $1.8 \times 10^{-2}$ M$_{\odot}$, and 65% of the mass corresponds to the extended and expanding hourglass-shaped structure.

- In Gallardo Cava et al. (2022c), we focus on the chemistry of this type of sources and present the first single-dish molecular survey of mm-wave lines in nebulae around binary post-AGB stars. We perform a deep molecular survey of mm-wave lines at 1.3, 2, 3, 7, and 13 mm using the IRAM 30 m MRT and IGN 40 m RT. Our observations took around 600 hours of telescope time. This chemical analysis reveals that the molecular content is low, except for CO, specially in the sources where the disk dominates the nebula. Additionally, based on our analysis of integrated intensity ratios (or the presence of maser detection of O-bearing molecules), we classify some sources as O-/C-rich. See also Gallardo Cava et al. (2022b).

## 8.2 Conclusions

The deep analysis and interpretation of the results of this Ph.D. thesis lead us to conclude the following key findings:

- There is a class of binary post-AGB stars with remarkable near-infrared excess that are surrounded by disks with Keplerian dynamics and extended outflows composed of gas escaping from the disk.

- Based on the contribution of the rotating component and the extended outflow, we differentiate two kinds of disk-containing nebulae around binary post-AGB stars: the disk- and the outflow-dominated sources.

- On the one hand, AC Her, the Red Rectangle, IW Car, and IRAS 08544−4431 belongs to the disk-dominated sources. On the other hand, R Sct, IRAS 19125+0343, and 89 Her (and very probably AI CMi, IRAS 20056+1834, and IRAS 18123+0511) are part of the outflow-dominated sources.

- Those outflow-dominated post-AGB nebulae present massive outflows that are mainly composed of cold gas, compared to the disk-dominated sources.

- Regarding the importance of the two components (disk and outflow), nebulae around binary post-AGB stars conform a bimodal distribution without intermediate objects. The absence of an intermediate class suggest that the two kinds of sources are not related via an evolutionary path. On the contrary, it seems that the prevalence of the disk or of the outflow component will be related to the configuration of the binary system (masses of the stars, orbital parameters, mass loss and accretion rates, . . . ).

· The Keplerian velocity field found in the disk of AC Her is compatible with a central total stellar mass of around 1 $M_{\odot}$. The Keplerian rotation of the unresolved disks of 89 Her, IRAS 19125+0343, and R Sct are compatible with central total stellar masses of 1.7, 1.1, and 1.7 $M_{\odot}$, respectively. Previous studies of the Red Rectangle, IRAS 08544−4431, and IW Car reveal that their central total stellar masses are 1.7, 1.8, and 1 $M_{\odot}$, respectively.

· The emission of rare species other than CO in our disk-containing nebulae around binary post-AGB stars is low. This fact is very significant in those sources in which the disk is the dominant component of the nebula. The lower molecular richness can be also seen in $^{12}$CO, where it is very remarkable in the disk-dominated sources. Nevertheless, we find overabundance of $^{13}$CO in our sources and it is remarkable in those sources dominated by the outflow.

· The estimated abundances of molecules other than CO are very low in our sources. The molecular abundances are weaker in those sources dominated by the disk, compared to the outflow-dominated ones.

· The nebulae around AC Her, the Red Rectangle, HD 52961, AI CMi, IRAS 20056+1834, and R Sct are O-rich environments. However, 89 Her present a C-rich nebula.

· We estimated the initial stellar mass for the Red Rectangle and 89 Her as 1.5 and 1.2 $M_{\odot}$, respectively. We deduced these values via the $^{17}$O/$^{18}$O ratio. These values are consistent with the central total stellar mass derived of our maps and models.

· The $^{17}$O/$^{18}$O ratio for 89 Her is typical of O-rich sources, while its chemistry is probably C-rich. This apparent contradiction suggest that the transfer of material between the two stars in the binary system may have altered the chemical composition of the present nebula.

*You see in this world there's two kinds of people, my friend. Those with loaded guns, and those who dig. You dig.*

Blondie – The Good, the Bad and the Ugly

# 9

# Current status and future work

*The work presented in this thesis will help and will be taken as a reference in the study of nebulae around binary post-AGB stars. This thesis work opens up several lines of future research. So, in this final chapter, we introduce the current status and the possible actions to be taken as a continuation of this Ph.D. thesis research work.*

## 9.1 Current work

The disk-containing nebulae conform a bimodal distribution where we find the disk- and the outflow-dominated sources. We show evidence to support that there is not a evolutionary link between both subclasses. Given our massive amount of CO emission data, we perform a statistical analysis to search for relationships between CO emission and other observational parameters such as IR emission. On the one hand, we analyze the traits of these objects. On the other hand, we study the characteristics of each subclass.

## 9.2 Future work

We have performed a broad analysis covering some very relevant topics such as the dynamics and chemistry of these objects (both the disk- and the outflow-dominated sources). Nevertheless, we present several plausible lines of research:

- As stated before, to extend this study to other sources, so to increase the statistical significance of the results, and know about the common characteristics they have and also to know how the two subgroups differ. Additionally, to perform more single-dish observations using other telescopes at frequencies different from those already observed. In this way, we could detect additional rotational lines of the same molecules, more molecules, or other types of compounds, such as neutral atoms (like C I via ALMA observations), ionized gas, or dust.

· Performing high resolution observations in those sources dominated by their outflow. The are many evidences of the presence of these Keplerian disks in the innermost regions of the outflow-dominated sources, but resolved images of the central parts are still missing. These observations will help us to clearly map these rotating structures. This is the case of 89 Her, R Sct, or IRAS 19125+0343 (and very probably AI CMi, IRAS 20056+1834, and IRAS 18123+0511), whose rotating disks are $\lesssim 1''$. Therefore, NOEMA or ALMA observations are needed to map their rotating disks.

· To observe those molecules other than CO detected with the 30 m IRAM telescope through ALMA observations. These observations will allow us to know the origin of this molecular emission, and to discern whether it comes from the rotating disk or the outflow.

· To further study the very weak detected emission of the outflow that surrounds the Keplerian disk in AC Her. We have even modeled this outflow assuming some properties of this component (based on similarities with other disk-dominated sources). Thus, high sensitivity observations are needed to clearly map this slow expanding component.

· To perform very high resolution observations in those dusty disk-dominated nebulae, looking for the detection of the formation of a second generation of planets or planetoids in these stable structures that surround binary post-AGB stars.

· To perform observations looking for rotating disks orbiting white dwarfs. We consider that there could be an evolutionary link between disks around binary AGB/post-AGB stars, and around white dwarfs that form part of a binary system. The main difference between both classes are that disks around AGB/post-AGB stars come from the ejecta of the evolved star, while the material around white dwarfs would be originated from, under the usually accepted scenario, remnants of planetary material that survived the stellar evolution. Nevertheless, we propose that disks around white dwarfs would be the natural evolution of those around post-AGB stars, both being finally formed from gas and dust ejected during the AGB and early post-AGB stages.

· To perform high resolution interferometric observations in other kinds of nebulae that present two distinct structures: high-velocity bipolar outflows and slowly-expanding dense disks in the equatorial regions from which the lobes emerge. The shape of these large-scale structures of these objects suggests the presence of a binary star. In addition, the velocity modulus decreases significantly for short distances to the axis of the disk, so this low expansion velocity structure may hide a rotating component. Models suggest the presence of circumbinary disks or, at least, disks around the companion star. This is the case of OH 231.8+4.2, which probably presents a rotating circumbinary disk that was (selectively) traced by NaCl, KCl, and $H_2O$. This criterion can be applied to other objects. Therefore, nebulae of this kind, such as M 1−92, could also present rotating disks. This pPN contains fast bipolar outflows and a confirmed flat disk with low expansion velocity. Observations of SiO and other good tracers of the innermost regions of these objects with ALMA, with a resolution under $0\farcs1$, would be necessary to detect the presence of rotating disks in these nebulae.

*I tried to be perfect*
*It just wasn't worth it*
*Nothing could ever be so wrong*
*It's hard to believe me*
*It never gets easy*
*I guess I knew that all along*

Pieces – Sum 41

# A

# List of publications and contributions

*Here, I list all the refereed published works and contributions to congresses related to this Ph.D. thesis. Additionally, I list all the observational projects, whose data have be consulted in the articles included in this thesis.*

## A.1 Refereed publications

- "Keplerian disks and outflows in post-AGB stars: AC Herculis, 89 Herculis, IRAS 19125+0343, and R Scuti". **Gallardo Cava, I.**, Gómez-Garrido, M., Bujarrabal, V., Castro-Carrizo, A., Alcolea, J., and Van Winckel, H. Astronomy & Astrophysics, 648, A93 (2021).

- "Chemistry of nebulae around binary post-AGB stars: A molecular survey of mm-wave lines". **Gallardo Cava, I.**, Bujarrabal, V., Alcolea, J., Gómez-Garrido, M., and Santander-García, M. Astronomy & Astrophysics, 659, A134 (2022).

- "Rotating and expanding gas in binary post-AGB stars". **Gallardo Cava, I.**, Bujarrabal, V., Alcolea, J., Gómez-Garrido, M., Castro-Carrizo, A., Van Winckel, H., and Santander-García, M. Astronomy, 1(2), 84 (2022).

- "The nebulae around the binary post-AGB star 89 Herculis". **Gallardo Cava, I.**, Alcolea, J., Bujarrabal, V., Gómez-Garrido, M., and Castro-Carrizo, A. Astronomy & Astrophysics, submitted.

## A.2 Conference contributions

- "Keplerian disks and outflows around binary post-AGB stars". **Gallardo Cava, I.**, Bujarrabal, V., Alcolea, J., Gómez-Garrido, M., Castro-Carrizo, A., Van Winckel, H., and Santander-García, M. Oral contribution presented at the 50$^{th}$ Young European Radio Astronomers Conference (YERAC), Grenoble, France, August 2021 (virtual edition).

- “Keplerian disks and outflows around binary post-AGB stars”. **Gallardo Cava, I.**, Bujarrabal, V., Alcolea, J., Gómez-Garrido, M., Castro-Carrizo, A., Van Winckel, H., and Santander-García, M. Oral contribution presented at the Asymmetrical Planetary Nebulae 8 (APN 8) – The Shaping of the Circumstellar Matter conference, Granada, Spain, October 2021 (virtual edition).
- “Keplerian disks and outflows around binary post-AGB stars”. **Gallardo Cava, I.**, Bujarrabal, V., Alcolea, J., Gómez-Garrido, M., Castro-Carrizo, A., Van Winckel, H., and Santander-García, M. Oral contribution presented at the Ph-Day 2021, Madrid, Spain, October 2021.
- “Keplerian disks and outflows around binary post-AGB stars”. **Gallardo Cava, I.**, Bujarrabal, V., Alcolea, J., Gómez-Garrido, M., Castro-Carrizo, A., Van Winckel, H., and Santander-García, M. Poster contribution presented at The Origin of Outflows in Evolved Stars – IAU Symposium 366 conference, Leuven, Belgium, November 2021 (virtual edition).
- “The nebulae around the binary post-AGB star 89 Herculis: a Keplerian disk and an hourglass-like expanding component”. **Gallardo Cava, I.**, Alcolea, J., Bujarrabal, V., Gómez-Garrido, M., and Castro-Carrizo, A. Oral contribution presented at the 51$^{th}$ Young European Radio Astronomers Conference (YERAC), Helsinki, Finland, August 2022.
- “Keplerian disks and outflows around binary post-AGB stars”. **Gallardo Cava, I.**, Bujarrabal, V., Alcolea, J., Gómez-Garrido, M., Castro-Carrizo, A., Van Winckel, H., and Santander-García, M. Poster contribution presented at XV Reunión de la Sociedad Española de Astronomía (SEA), Tenerife, Spain, September 2022.

*A wise man can learn more from a foolish question than a fool can learn from a wise answer.*

Bruce Lee

# Bibliography

Adams, F. C., Bodenheimer, P., and Laughlin, G. (2005). M dwarfs: planet formation and long term evolution. *Astronomische Nachrichten*, 326(10):913–919.

Agúndez, M., Cernicharo, J., and Goicoechea, J. R. (2008). Formation of simple organic molecules in inner T Tauri disks. *A&A*, 483(3):831–837.

Alcolea, J. (1993). PhD thesis.

Alcolea, J., Agúndez, M., Bujarrabal, V., Castro Carrizo, A., Desmurs, J.-F., Sánchez-Contreras, C., and Santander-García, M. (2019). M 1-92 revisited: the chemistry of a common envelope nebula? *IAU Symposium*, 343:343–344.

Alcolea, J. and Bujarrabal, V. (1991). The post-AGB evolution of low-mass stars. *A&A*, 245:499.

Alcolea, J. and Bujarrabal, V. (1995). The molecular envelope around 89 Herculis. *A&A*, 303:L21.

Alcolea, J., Bujarrabal, V., Sánchez Contreras, C., Neri, R., and Zweigle, J. (2001). The highly collimated bipolar outflow of OH 231.8+4.2. *A&A*, 373:932–949.

Alcolea, J., Neri, R., and Bujarrabal, V. (2007). Minkowski's footprint revisited. Planetary nebula formation from a single sudden event? *A&A*, 468(3):L41–L44.

Andrews, S. M. (2020). Observations of protoplanetary disk structures. *arXiv preprint arXiv:2001.05007*.

Andriesse, C. D. (1974). Far-infrared properties of interstellar grains. *A&A*, 37(2):257–262.

Arkhipova, V. P., Ikonnikova, N. P., Esipov, V. F., and Komissarova, G. V. (2017). VizieR Online Data Catalog: AI CMi UBV light curves (Arkhipova+, 2017). *VizieR Online Data Catalog*, page J/PAZh/43/460.

Asplund, M., Amarsi, A. M., and Grevesse, N. (2021). The chemical make-up of the Sun: A 2020 vision. *A&A*, 653:A141.

Baars, J. W. (2007). *The paraboloidal reflector antenna in radio astronomy and communication*, volume 348. Springer.

Bachiller, R., Forveille, T., Huggins, P. J., and Cox, P. (1997a). The chemical evolution of planetary nebulae. *A&A*, 324:1123–1134.

Bachiller, R., Fuente, A., Bujarrabal, V., Colomer, F., Loup, C., Omont, A., and de Jong, T. (1997b). A survey of CN in circumstellar envelopes. *A&A*, 319:235–243.

Bailer-Jones, C. A. L., Rybizki, J., Fouesneau, M., Mantelet, G., and Andrae, R. (2018). Estimating Distance from Parallaxes. IV. Distances to 1.33 Billion Stars in Gaia Data Release 2. *AJ*, 156(2):58.

Balick, B. and Frank, A. (2002). Shapes and Shaping of Planetary Nebulae. *ARA&A*, 40:439–486.

Balser, D. S., McMullin, J. P., and Wilson, T. L. (2002). CO Isotopes in Planetary Nebulae. *ApJ*, 572(1):326–334.

Beasant, P., Gallart Gual, E., Díaz de Mendívil Pérez, J. M., and Serret, M. (1995). *1001 Secretos del espacio*. Larousse Planeta.

Blackman, E. G. and Lucchini, S. (2014). Using kinematic properties of pre-planetary nebulae to constrain engine paradigms. *MNRAS*, 440:L16–L20.

Bladh, S., Höfner, S., Aringer, B., and Eriksson, K. (2015). Exploring wind-driving dust species in cool luminous giants. III. Wind models for M-type AGB stars: dynamic and photometric properties. *A&A*, 575:A105.

Bloecker, T. and Schoenberner, D. (1990). On the fading of massive AGB remnants. *A&A*, 240:L11–L14.

Bollen, D., Kamath, D., Van Winckel, H., and De Marco, O. (2019). A spatio-kinematic model for jets in post-AGB stars. *A&A*, 631:A53.

Bollen, D., Kamath, D., Van Winckel, H., De Marco, O., Verhamme, O., Kluska, J., and Wardle, M. (2022). The structure of jets launched from post-AGB binary systems. *A&A*, 666:A40.

Bollen, D., Kamath, D., Van Winckel, H., De Marco, O., and Wardle, M. (2021). Jet parameters for a diverse sample of jet-launching post-AGB binaries. *MNRAS*, 502(1):445–462.

Bonačić Marinović, A. A., Glebbeek, E., and Pols, O. R. (2008). Orbital eccentricities of binary systems with a former AGB star. *A&A*, 480(3):797–805.

Bowers, P. F., Johnston, K. J., and de Vegt, C. (1989). Envelope Structures and Optical/Radio Positions of Cool Stars. *ApJ*, 340:479.

Bublitz, J., Kastner, J. H., Santander-García, M., Bujarrabal, V., Alcolea, J., and Montez, R. (2019). A new radio molecular line survey of planetary nebulae. HNC/HCN as a diagnostic of ultraviolet irradiation. *A&A*, 625:A101.

Bujarrabal, V. (2006). Molecular line emission from planetary and protoplanetary nebulae. In Barlow, M. J. and Méndez, R. H., editors, *Planetary Nebulae in our Galaxy and Beyond*, volume 234, pages 193–202.

Bujarrabal, V., Agúndez, M., Gómez-Garrido, M., Kim, H., Santander-García, M., Alcolea, J., Castro-Carrizo, A., and Mikołajewska, J. (2021). Structure and dynamics of the inner nebula around the symbiotic stellar system R Aquarii. *A&A*, 651:A4.

Bujarrabal, V. and Alcolea, J. (2013). Warm gas in the rotating disk of the Red Rectangle: accurate models of molecular line emission. *A&A*, 552:A116.

Bujarrabal, V., Alcolea, J., and Bachiller, R. (1990). The 12C/13C abundance ratio in the envelope of R Scuti. *A&A*, 234:355.

Bujarrabal, V., Alcolea, J., Mikołajewska, J., Castro-Carrizo, A., and Ramstedt, S. (2018a). High-resolution observations of the symbiotic system R Aqr. Direct imaging of the gravitational effects of the secondary on the stellar wind. *A&A*, 616:L3.

Bujarrabal, V., Alcolea, J., and Neri, R. (1998a). The Structure and Dynamics of the Proto-Planetary Nebula M1-92. *ApJ*, 504(2):915–920.

Bujarrabal, V., Alcolea, J., and Planesas, P. (1992). The molecular emission of young preplanetary nebulae. *A&A*, 257:701–714.

Bujarrabal, V., Alcolea, J., Sahai, R., Zamorano, J., and Zijlstra, A. A. (1998b). The shock structure in the protoplanetary nebula M1-92: imaging of atomic and H_2 line emission. *A&A*, 331:361–371.

Bujarrabal, V., Alcolea, J., Sánchez Contreras, C., and Sahai, R. (2002). HST observations of the protoplanetary nebula OH 231.8+4.2: The structure of the jets and shocks. *A&A*, 389:271–285.

Bujarrabal, V., Alcolea, J., Van Winckel, H., Santander-García, M., and Castro-Carrizo, A. (2013a). Extended rotating disks around post-AGB stars. *A&A*, 557:A104.

Bujarrabal, V., Castro-Carrizo, A., Alcolea, J., and Neri, R. (2005). The orbiting gas disk in the Red Rectangle. *A&A*, 441(3):1031–1038.

Bujarrabal, V., Castro-Carrizo, A., Alcolea, J., and Sánchez Contreras, C. (2001). Mass, linear momentum and kinetic energy of bipolar flows in protoplanetary nebulae. *A&A*, 377:868–897.

Bujarrabal, V., Castro-Carrizo, A., Alcolea, J., Santand er-García, M., van Winckel, H., and Sánchez Contreras, C. (2016). Further ALMA observations and detailed modeling of the Red Rectangle. *A&A*, 593:A92.

Bujarrabal, V., Castro-Carrizo, A., Alcolea, J., and Van Winckel, H. (2015). Detection of Keplerian dynamics in a disk around the post-AGB star AC Herculis. *A&A*, 575:L7.

Bujarrabal, V., Castro-Carrizo, A., Alcolea, J., Van Winckel, H., Sánchez Contreras, C., and Santander-García, M. (2017). A second post-AGB nebula that contains gas in rotation and in expansion: ALMA maps of IW Carinae. *A&A*, 597:L5.

Bujarrabal, V., Castro-Carrizo, A., Alcolea, J., Van Winckel, H., Sánchez Contreras, C., Santander-García, M., Neri, R., and Lucas, R. (2013b). ALMA observations of the Red Rectangle, a preliminary analysis. *A&A*, 557:L11.

Bujarrabal, V., Castro-Carrizo, A., Van Winckel, H., Alcolea, J., Sánchez Contreras, C., Santander-García, M., and Hillen, M. (2018b). High-resolution observations of IRAS 08544-4431. Detection of a disk orbiting a post-AGB star and of a slow disk wind. *A&A*, 614:A58.

Bujarrabal, V., Fuente, A., and Omont, A. (1994a). Molecular observations of O- and C-rich circumstellar envelopes. *A&A*, 285:247–271.

Bujarrabal, V., Fuente, A., and Omont, A. (1994b). The Discrimination between O- and C-rich Circumstellar Envelopes from Molecular Observations. *ApJ*, 421:L47.

Bujarrabal, V., Gomez-Gonzalez, J., Bachiller, R., and Martin-Pintado, J. (1988). Proto-planetary nebulae : the case of CRL 618. *A&A*, 204:242–252.

Bujarrabal, V., van Winckel, H., Neri, R., Alcolea, J., Castro-Carrizo, A., and Deroo, P. (2007). The nebula around the post-AGB star 89 Herculis. *A&A*, 468(3):L45–L48.

Castor, J. I. (1970). Spectral line formation in Wolf-Rayet envelopes. *MNRAS*, 149:111.

Castro-Carrizo, A., Bujarrabal, V., Fong, D., Meixner, M., Tielens, A., Latter, W., and Barlow, M. (2001). Low-excitation atomic gas around evolved stars-ii. iso observations of o-rich nebulae. *Astronomy & Astrophysics*, 367(2):674–693.

Castro-Carrizo, A., Bujarrabal, V., Sánchez Contreras, C., Alcolea, J., and Neri, R. (2002). The structure and dynamics of the molecular envelope of M 2-56. *A&A*, 386:633–645.

Castro-Carrizo, A., Quintana-Lacaci, G., Neri, R., Bujarrabal, V., Schöier, F. L., Winters, J. M., Olofsson, H., Lindqvist, M., Alcolea, J., Lucas, R., and Grewing, M. (2010). Mapping the $^{12}$CO J = 1-0 and J = 2-1 emission in AGB and early post-AGB circumstellar envelopes. I. The COSAS program, first sample. *A&A*, 523:A59.

Cernicharo, J., Agúndez, M., and Guélin, M. (2011). Spectral Line Surveys of Evolved Stars. In Cernicharo, J. and Bachiller, R., editors, *The Molecular Universe*, volume 280, pages 237–248.

Cernicharo, J., Guélin, M., and Kahane, C. (2000). A lambda 2 mm molecular line survey of the C-star envelope IRC+10216. *A&AS*, 142:181–215.

Cernicharo, J., Marcelino, N., Agúndez, M., and Guélin, M. (2015). Molecular shells in IRC+10216: tracing the mass loss history. *A&A*, 575:A91.

Chang, R. (2007). *Chemistry*. McGraw-Hill Higher Education.

Chu, Y.-H., Manchado, A., Jacoby, G. H., and Kwitter, K. B. (1991). The Multiple-Shell Structure of the Planetary Nebula NGC 6751. *ApJ*, 376:150.

Clark, B. G. (1980). An efficient implementation of the algorithm 'CLEAN'. *A&A*, 89(3):377.

Cohen, M., Van Winckel, H., Bond, H. E., and Gull, T. R. (2004). Hubble Space Telescope Imaging of HD 44179, The Red Rectangle. *AJ*, 127(4):2362–2377.

Condon, J. J. and Ransom, S. M. (2016). Essential radio astronomy. In *Essential Radio Astronomy*. Princeton University Press.

Cornwell, T. J. (2008). Multiscale clean deconvolution of radio synthesis images. *IEEE Journal of selected topics in signal processing*, 2(5):793–801.

Cox, N. L. J., Kerschbaum, F., van Marle, A. J., Decin, L., Ladjal, D., Mayer, A., Groenewegen, M. A. T., van Eck, S., Royer, P., Ottensamer, R., Ueta, T., Jorissen, A., Mecina, M., Meliani, Z., Luntzer, A., Blommaert, J. A. D. L., Posch, T., Vandenbussche, B., and Waelkens, C. (2012). A far-infrared survey of bow shocks and detached shells around AGB stars and red supergiants. *A&A*, 537:A35.

Dayal, A. and Bieging, J. H. (1995). The Distribution of HCN, H 13CN, and CN in IRC +10216. *ApJ*, 439:996.

De Marco, O. (2009). The Origin and Shaping of Planetary Nebulae: Putting the Binary Hypothesis to the Test. *PASP*, 121(878):316.

De Marco, O. and Izzard, R. G. (2017). Dawes Review 6: The Impact of Companions on Stellar Evolution. *PASA*, 34:e001.

De Nutte, R., Decin, L., Olofsson, H., Lombaert, R., de Koter, A., Karakas, A., Milam, S., Ramstedt, S., Stancliffe, R. J., Homan, W., and Van de Sande, M. (2017). Nucleosynthesis in AGB stars traced by oxygen isotopic ratios. I. Determining the stellar initial mass by means of the $^{17}$O/$^{18}$O ratio. *A&A*, 600:A71.

de Ruyter, S., van Winckel, H., Dominik, C., Waters, L. B. F. M., and Dejonghe, H. (2005). Strong dust processing in circumstellar discs around 6 RV Tauri stars. Are dusty RV Tauri stars all binaries? *A&A*, 435(1):161–166.

de Ruyter, S., van Winckel, H., Maas, T., Lloyd Evans, T., Waters, L. B. F. M., and Dejonghe, H. (2006). Keplerian discs around post-AGB stars: a common phenomenon? *A&A*, 448(2):641–653.

de Vicente, P., Bujarrabal, V., Díaz-Pulido, A., Albo, C., Alcolea, J., Barcia, A., Barbas, L., Bolaño, R., Colomer, F., Diez, M. C., Gallego, J. D., Gómez-González, J., López-Fernández, I., López-Fernández, J. A., López-Pérez, J. A., Malo, I., Moreno, A., Patino, M., Serna, J. M., Tercero, F., and Vaquero, B. (2016). $^{28}$SiO v = 0 J = 1-0 emission from evolved stars. *A&A*, 589:A74.

Deguchi, S., Sakamoto, T., and Hasegawa, T. (2012). Kinematics of Red Variables in the Solar Neighborhood I. Basic Data Obtained by an SiO Maser Survey. *PASJ*, 64:4.

Deroo, P., Acke, B., Verhoelst, T., Dominik, C., Tatulli, E., and van Winckel, H. (2007). AMBER and MIDI interferometric observations of the post-AGB binary IRAS 08544-4431: the circumbinary disc resolved. *A&A*, 474(3):L45–L48.

Deroo, P., van Winckel, H., Min, M., Waters, L. B. F. M., Verhoelst, T., Jaffe, W., Morel, S., Paresce, F., Richichi, A., Stee, P., and Wittkowski, M. (2006). Resolving the compact dusty discs around binary post-AGB stars using N-band interferometry. *A&A*, 450(1):181–192.

Diamond, P. J., Kemball, A. J., Junor, W., Zensus, A., Benson, J., and Dhawan, V. (1994). Observation of a Ring Structure in SiO Maser Emission from Late-Type Stars. *ApJ*, 430:L61.

Dominik, C., Dullemond, C. P., Cami, J., and van Winckel, H. (2003). The dust disk of HR 4049. Another brick in the wall. *A&A*, 397:595–609.

Draine, B. T. and Lee, H. M. (1984). Optical Properties of Interstellar Graphite and Silicate Grains. *ApJ*, 285:89.

Dutrey, A., Semenov, D., Chapillon, E., Gorti, U., Guilloteau, S., Hersant, F., Hogerheijde, M., Hughes, M., Meeus, G., Nomura, H., et al. (2014). Physical and chemical structure of planet-forming disks probed by millimeter observations and modeling. *Protostars and Planets VI*, pages 317–338.

Engels, D. (1979). Catalogue of late-type stars with OH, $H_2O$ or SiO maser emission. *A&AS*, 36:337–345.

Engels, D. and Lewis, B. M. (1996). A survey for 22GHz water maser emission from the Arecibo set of OH/IR stars. *A&AS*, 116:117–155.

Feder, J., Russell, K. C., Lothe, J., and Pound, G. M. (1966). Homogeneous nucleation and growth of droplets in vapours. *Advances in Physics*, 15(57):111–178.

Ferreira, J., Dougados, C., and Cabrit, S. (2006). Which jet launching mechanism(s) in T Tauri stars? *A&A*, 453(3):785–796.

Fong, D., Meixner, M., Castro-Carrizo, A., Bujarrabal, V., Latter, W. B., Tielens, A. G. G. M., Kelly, D. M., and Sutton, E. C. (2001). Low-excitation atomic gas around evolved stars. I. ISO observations of C-rich nebulae. *A&A*, 367:652–673.

Fong, D., Meixner, M., Sutton, E. C., Zalucha, A., and Welch, W. J. (2006). Evolution of the Circumstellar Molecular Envelope. I. A BIMA CO Survey of Evolved Stars. *ApJ*, 652(2):1626–1653.

Frank, A. and Blackman, E. G. (2004). Application of Magnetohydrodynamic Disk Wind Solutions to Planetary and Protoplanetary Nebulae. *ApJ*, 614(2):737–744.

Gallardo Cava, I., Alcolea, J., Bujarrabal, V., Gómez-Garrido, M., and Castro-Carrizo, A. (2022a). The nebulae around the binary post-AGB star 89 Herculis. *A&A, submitted*.

Gallardo Cava, I., Bujarrabal, V., Alcolea, J., Gómez-Garrido, M., Castro-Carrizo, A., Van Winckel, H., and M., S. (2022b). Rotating and expanding gas in binary post-agb stars. *Astronomy*, 1(2):84–92.

Gallardo Cava, I., Bujarrabal, V., Alcolea, J., Gómez-Garrido, M., and Santander-García, M. (2022c). Chemistry of nebulae around binary post-AGB stars: A molecular survey of mm-wave lines. *A&A*, 659:A134.

Gallardo Cava, I., Gómez-Garrido, M., Bujarrabal, V., Castro-Carrizo, A., Alcolea, J., and Van Winckel, H. (2021). Keplerian disks and outflows in post-AGB stars: AC Herculis, 89 Herculis, IRAS 19125+0343, and R Scuti. *A&A*, 648:A93.

Gehrz, R. (1989). Sources of Stardust in the Galaxy. In Allamandola, L. J. and Tielens, A. G. G. M., editors, *Interstellar Dust*, volume 135, page 445.

Gezer, I., Van Winckel, H., Bozkurt, Z., De Smedt, K., Kamath, D., Hillen, M., and Manick, R. (2015). The WISE view of RV Tauri stars. *MNRAS*, 453(1):133–146.

Gielen, C., Bouwman, J., van Winckel, H., Lloyd Evans, T., Woods, P. M., Kemper, F., Marengo, M., Meixner, M., Sloan, G. C., and Tielens, A. G. G. M. (2011a). Silicate features in Galactic and extragalactic post-AGB discs. *A&A*, 533:A99.

Gielen, C., Cami, J., Bouwman, J., Peeters, E., and Min, M. (2011b). Carbonaceous molecules in the oxygen-rich circumstellar environment of binary post-AGB stars. $C_{60}$ fullerenes and polycyclic aromatic hydrocarbons. *A&A*, 536:A54.

Gielen, C., van Winckel, H., Min, M., Waters, L. B. F. M., and Lloyd Evans, T. (2008). SPITZER survey of dust grain processing in stable discs around binary post-AGB stars. *A&A*, 490(2):725–735.

Gobrecht, D., Cherchneff, I., Sarangi, A., Plane, J. M. C., and Bromley, S. T. (2016). Dust formation in the oxygen-rich AGB star IK Tauri. *A&A*, 585:A6.

Goldreich, P. and Scoville, N. (1976). OH-IR stars. I. Physical properties of circumstellar envelopes. *ApJ*, 205:144–154.

Gómez, Y., Moran, J. M., and Rodríguez, L. F. (1990). $H_2O$ and SiO maser emission in OH/IR stars. *Rev. Mexicana Astron. Astrofis.*, 20:55.

González Delgado, D., Olofsson, H., Kerschbaum, F., Schöier, F. L., Lindqvist, M., and Groenewegen, M. A. T. (2003). "Thermal" SiO radio line emission towards M-type AGB stars: A probe of circumstellar dust formation and dynamics. *A&A*, 411:123–147.

Gorlova, N., Van Winckel, H., Gielen, C., Raskin, G., Prins, S., Pessemier, W., Waelkens, C., Frémat, Y., Hensberge, H., Dumortier, L., Jorissen, A., and Van Eck, S. (2012). Time-resolved spectroscopy of BD+46°442: Gas streams and jet creation in a newly discovered evolved binary with a disk. *A&A*, 542:A27.

Groenewegen, M. A. T. (1994). The mass loss rates of OH/IR 32.8-0.3 and OH/IR 44.8-2.3. *A&A*, 290:544–552.

Guelin, M., Lucas, R., and Cernicharo, J. (1993). MgNC and the carbon-chain radicals in IRC +10216. *A&A*, 280:L19–L22.

Guerrero, M. A., Chu, Y.-H., Manchado, A., and Kwitter, K. B. (2003). Physical Structure of Planetary Nebulae. I. The Owl Nebula. *AJ*, 125(6):3213–3221.

Guerrero, M. A. and Manchado, A. (1999). On the Chemical Abundances of Multiple-Shell Planetary Nebulae with Halos. *ApJ*, 522(1):378–386.

Gueth, F., Guilloteau, S., and Bachiller, R. (1996). A precessing jet in the L1157 molecular outflow. *A&A*, 307:891–897.

Guilloteau, S., Di Folco, E., Dutrey, A., Simon, M., Grosso, N., and Piétu, V. (2013). A sensitive survey for $^{13}$CO, CN, $H_2$CO, and SO in the disks of T Tauri and Herbig Ae stars. *A&A*, 549:A92.

Guilloteau, S. and Dutrey, A. (1998). Physical parameters of the Keplerian protoplanetary disk of DM Tauri. *A&A*, 339:467–476.

Guilloteau, S., Reboussin, L., Dutrey, A., Chapillon, E., Wakelam, V., Piétu, V., Di Folco, E., Semenov, D., and Henning, T. (2016). Chemistry in disks. X. The molecular content of protoplanetary disks in Taurus. *A&A*, 592:A124.

Habing, H. J. and Olofsson, H. (2004). *Asymptotic Giant Branch Stars*.

Han, Z., Podsiadlowski, P., Maxted, P. F. L., and Marsh, T. R. (2003). The origin of subdwarf B stars - II. *MNRAS*, 341(2):669–691.

Han, Z., Podsiadlowski, P., Maxted, P. F. L., Marsh, T. R., and Ivanova, N. (2002). The origin of subdwarf B stars - I. The formation channels. *MNRAS*, 336(2):449–466.

Henning, T. and Semenov, D. (2013). Chemistry in protoplanetary disks. *Chemical Reviews*, 113(12):9016–9042.

Hillen, M., de Vries, B. L., Menu, J., Van Winckel, H., Min, M., and Mulders, G. D. (2015). The evolved circumbinary disk of AC Herculis: a radiative transfer, interferometric, and mineralogical study. *A&A*, 578:A40.

Hillen, M., Kluska, J., Le Bouquin, J. B., Van Winckel, H., Berger, J. P., Kamath, D., and Bujarrabal, V. (2016). Imaging the dust sublimation front of a circumbinary disk. *A&A*, 588:L1.

Hillen, M., Menu, J., Van Winckel, H., Min, M., Gielen, C., Wevers, T., Mulders, G. D., Regibo, S., and Verhoelst, T. (2014). An interferometric study of the post-AGB binary 89 Herculis. II. Radiative transfer models of the circumbinary disk. *A&A*, 568:A12.

Hillen, M., Van Winckel, H., Menu, J., Manick, R., Debosscher, J., Min, M., de Wit, W. J., Verhoelst, T., Kamath, D., and Waters, L. B. F. M. (2017). A mid-IR interferometric survey with MIDI/VLTI: resolving the second-generation protoplanetary disks around post-AGB binaries. *A&A*, 599:A41.

Hillen, M., Verhoelst, T., Van Winckel, H., Chesneau, O., Hummel, C. A., Monnier, J. D., Farrington, C., Tycner, C., Mourard, D., ten Brummelaar, T., Banerjee, D. P. K., and Zavala, R. T. (2013). An interferometric study of the post-AGB binary 89 Herculis. I. Spatially resolving the continuum circumstellar environment at optical and near-IR wavelengths with the VLTI, NPOI, IOTA, PTI, and the CHARA Array. *A&A*, 559:A111.

Hinkle, K. H., Hall, D. N. B., and Ridgway, S. T. (1982). Time series infrared spectroscopy of the mira variable KHI Cyg. *ApJ*, 252:697–714.

Höfner, S., Bladh, S., Aringer, B., and Ahuja, R. (2016). Dynamic atmospheres and winds of cool luminous giants. I. $Al_2O_3$ and silicate dust in the close vicinity of M-type AGB stars. *A&A*, 594:A108.

Höfner, S. and Olofsson, H. (2018). Mass loss of stars on the asymptotic giant branch. Mechanisms, models and measurements. *A&A Rev.*, 26(1):1.

Högbom, J. A. (1974). Aperture Synthesis with a Non-Regular Distribution of Interferometer Baselines. *A&AS*, 15:417.

Homan, W., Richards, A., Decin, L., Kervella, P., de Koter, A., McDonald, I., and Ohnaka, K. (2017). ALMA observations of the nearby AGB star $L_2$ Puppis. II. Gas disk properties derived from $^{12}$CO and $^{13}$CO J = 3-2 emission. *A&A*, 601:A5.

Hrivnak, B. J., Kwok, S., and Volk, K. M. (1989). A Study of Several F and G Supergiant-like Stars with Infrared Excesses as Candidates for Proto–Planetary Nebulae. *ApJ*, 346:265.

Hrivnak, B. J., Van de Steene, G., Van Winckel, H., Sperauskas, J., Bohlender, D., and Lu, W. (2017). Where are the Binaries? Results of a Long-term Search for Radial Velocity Binaries in Proto-planetary Nebulae. *ApJ*, 846(2):96.

Huggins, P. J., Bachiller, R., Cox, P., and Forveille, T. (1996). The molecular envelopes of planetary nebulae. *A&A*, 315:284–302.

Humphreys, E. M. L., Gray, M. D., Yates, J. A., Field, D., Bowen, G. H., and Diamond, P. J. (2002). Numerical simulations of stellar SiO maser variability. Investigation of the effect of shocks. *A&A*, 386:256–270.

Humphreys, E. M. L., Immer, K., Gray, M. D., De Beck, E., Vlemmings, W. H. T., Baudry, A., Richards, A. M. S., Wittkowski, M., Torstensson, K., De Breuck, C., Møller, P., Etoka, S., and Olberg, M. (2017). Simultaneous 183 GHz $H_2O$ maser and SiO observations towards evolved stars using APEX SEPIA Band 5. *A&A*, 603:A77.

Iben, I., J. (1985). The life and times of an intermediate mass star - In isolation/in a close binary. *QJRAS*, 26:1–39.

Jura, M. (2003). A Flared, Orbiting, Dusty Disk around HD 233517. *ApJ*, 582(2):1032–1035.

Jura, M., Turner, J., Balm, and S. P. (1997). Big Grains in the Red Rectangle? *ApJ*, 474(2):741–748.

Justtanont, K., Skinner, C. J., and Tielens, A. G. G. M. (1994). Molecular Rotational Line Profiles from Oxygen-rich Red Giant Winds. *ApJ*, 435:852.

Kalaee, M. J. and Hasanzadeh, A. (2019). A time series analysis of light curve of R Scuti star from 1970 to 2017. *New A*, 70:57–63.

Katz, J. L. (1970). Condensation of a Supersaturated Vapor. I. The Homogeneous Nucleation of the n-Alkanes. *J. Chem. Phys.*, 52(9):4733–4748.

Keady, J. J., Hall, D. N. B., and Ridgway, S. T. (1988). The IRC +10216 Circumstellar Envelope. I. Models for the Dust and Gas. *ApJ*, 326:832.

Kervella, P., Homan, W., Richards, A. M. S., Decin, L., McDonald, I., Montargès, M., and Ohnaka, K. (2016). ALMA observations of the nearby AGB star $L_2$ Puppis. I. Mass of the central star and detection of a candidate planet. *A&A*, 596:A92.

Kim, H., Trejo, A., Liu, S.-Y., Sahai, R., Taam, R. E., Morris, M. R., Hirano, N., and Hsieh, I. T. (2017). The large-scale nebular pattern of a superwind binary in an eccentric orbit. *Nature Astronomy*, 1:0060.

Kim, J., Cho, S. H., Bujarrabal, V., Imai, H., Dodson, R., Yoon, D. H., and Zhang, B. (2019). Time variations of $H_2O$ and SiO masers in the proto-Planetary Nebula OH 231.8+4.2. *MNRAS*, 488(1):1427–1445.

Kiss, L. L., Derekas, A., Szabó, G. M., Bedding, T. R., and Szabados, L. (2007). Defining the instability strip of pulsating post-AGB binary stars from ASAS and NSVS photometry. *MNRAS*, 375(4):1338–1348.

Klochkova, V. G., Panchuk, V. E., Chentsov, E. L., and Yushkin, M. V. (2007). The evolutionary status of the semiregular variable QY Sge. *Astrophysical Bulletin*, 62(3):217–234.

Kluska, J., Hillen, M., Van Winckel, H., Manick, R., Min, M., Regibo, S., and Royer, P. (2018). The perturbed sublimation rim of the dust disk around the post-AGB binary IRAS08544-4431. *A&A*, 616:A153.

Kluska, J., Van Winckel, H., Coppée, Q., Oomen, G. M., Dsilva, K., Kamath, D., Bujarrabal, V., and Min, M. (2022). A population of transition disks around evolved stars: Fingerprints of planets. Catalog of disks surrounding Galactic post-AGB binaries. *A&A*, 658:A36.

Kluska, J., Van Winckel, H., Hillen, M., Berger, J. P., Kamath, D., Le Bouquin, J. B., and Min, M. (2019). VLTI/PIONIER survey of disks around post-AGB binaries. Dust sublimation physics rules. *A&A*, 631:A108.

Kramer, C. (1997). *Calibration of spectral line data at the IRAM 30m radio telescope.*

Kwok, S. (1975). Radiation pressure on grains as a mechanism for mass loss in red giants. *ApJ*, 198:583–591.

Kwok, S. (1993). Proto-planetary nebulae. *ARA&A*, 31:63–92.

Kwok, S., Volk, K. M., and Hrivnak, B. J. (1989). A 21 Micron Emission Feature in Four Proto–Planetary Nebulae. *ApJ*, 345:L51.

Lattanzio, J. (2003). Nucleosynthesis in AGB Stars: the Role of Dredge-Up and Hot Bottom Burning (invited review). In Kwok, S., Dopita, M., and Sutherland, R., editors, *Planetary Nebulae: Their Evolution and Role in the Universe*, volume 209, page 73.

Lebre, A. and Gillet, D. (1991). The bright RV Tauri star R Scuti during an exceptional irregular light phase. *A&A*, 246:490.

Lewis, B. M. (1989). The Chronological Sequence of Circumstellar Masers: Identifying Proto–Planetary Nebulae. *ApJ*, 338:234.

Lewis, B. M. (1997). Main-line OH Observations of the Arecibo Set of OH/IR Stars. *ApJS*, 109(2):489–515.

Lin, M.-K. (2019). Dust settling against hydrodynamic turbulence in protoplanetary discs. *MNRAS*, 485(4):5221–5234.

Liu, J. and Jiang, B. (2017). On the Relation of Silicates and SiO Maser in Evolved Stars. *AJ*, 153(4):176.

Mangum, J. G. and Shirley, Y. L. (2015). How to Calculate Molecular Column Density. *PASP*, 127(949):266.

Marigo, P. and Girardi, L. (2007). Evolution of asymptotic giant branch stars. I. Updated synthetic TP-AGB models and their basic calibration. *A&A*, 469(1):239–263.

Massalkhi, S., Agúndez, M., Cernicharo, J., and Velilla-Prieto, L. (2020). The abundance of S- and Si-bearing molecules in O-rich circumstellar envelopes of AGB stars. *A&A*, 641:A57.

Matsuura, M., Barlow, M. J., Zijlstra, A. A., Whitelock, P. A., Cioni, M. R. L., Groenewegen, M. A. T., Volk, K., Kemper, F., Kodama, T., Lagadec, E., Meixner, M., Sloan, G. C., and Srinivasan, S. (2009). The global gas and dust budget of the Large Magellanic Cloud: AGB stars and supernovae, and the impact on the ISM evolution. *MNRAS*, 396(2):918–934.

Matsuura, M., Yamamura, I., Zijlstra, A. A., and Bedding, T. R. (2002). The extended atmosphere and evolution of the RV Tau star, R Scuti. *A&A*, 387:1022–1031.

Mauron, N. and Huggins, P. J. (2006). Imaging the circumstellar envelopes of AGB stars. *A&A*, 452(1):257–268.

Mauron, N., Huggins, P. J., and Cheung, C. L. (2013). Deep optical imaging of asymptotic giant branch circumstellar envelopes. *A&A*, 551:A110.

Men'shchikov, A. B., Balega, Y., Blöcker, T., Osterbart, R., and Weigelt, G. (2001). Structure and physical properties of the rapidly evolving dusty envelope of <ASTROBJ>IRC +10 216</ASTROBJ> reconstructed by detailed two-dimensional radiative transfer modeling. *A&A*, 368:497–526.

Men'shchikov, A. B., Schertl, D., Tuthill, P. G., Weigelt, G., and Yungelson, L. R. (2002). Properties of the close binary and circumbinary torus of the <ASTROBJ>Red Rectangle</ASTROBJ>. *A&A*, 393:867–885.

Miller Bertolami, M. M. (2016). New models for the evolution of post-asymptotic giant branch stars and central stars of planetary nebulae. *A&A*, 588:A25.

Neri, R., Kahane, C., Lucas, R., Bujarrabal, V., and Loup, C. (1998). A $^{12}$CO (J=1-0) and (J=2-1) atlas of circumstellar envelopes of AGB and post-AGB stars. *A&AS*, 130:1–64.

Neugebauer, G., Habing, H. J., van Duinen, R., Aumann, H. H., Baud, B., Beichman, C. A., Beintema, D. A., Boggess, N., Clegg, P. E., de Jong, T., Emerson, J. P., Gautier, T. N., Gillett, F. C., Harris, S., Hauser, M. G., Houck, J. R., Jennings, R. E., Low, F. J., Marsden, P. L., Miley, G., Olnon, F. M., Pottasch, S. R., Raimond, E., Rowan-Robinson, M., Soifer, B. T., Walker, R. G., Wesselius, P. R., and Young, E. (1984). The Infrared Astronomical Satellite (IRAS) mission. *ApJ*, 278:L1–L6.

Nie, J. D., Wood, P. R., and Nicholls, C. P. (2012). Predicting the fate of binary red giants using the observed sequence E star population: binary planetary nebula nuclei and post-RGB stars. *MNRAS*, 423(3):2764–2780.

Nixon, C. J., King, A. R., and Pringle, J. E. (2018). The Maximum Mass Solar Nebula and the early formation of planets. *MNRAS*, 477(3):3273–3278.

Nyman, L. A., Olofsson, H., Johansson, L. E. B., Booth, R. S., Carlstrom, U., and Wolstencroft, R. (1993). A molecular radio line survey of the carbon star IRAS 15194-5115. *A&A*, 269:377–389.

Olofsson, H., Eriksson, K., Gustafsson, B., and Carlstroem, U. (1993). A Study of Circumstellar Envelopes around Bright Carbon Stars. II. Molecular Abundances. *ApJS*, 87:305.

Olofsson, H., Johansson, L., Nguyen-Q-Rieu, Sopka, R. J., and Zuckerman, B. (1982). Molecular Line Observations of Envelopes Around Evolved Stars. In *Bulletin of the American Astronomical Society*, volume 14, page 895.

Oomen, G.-M., Van Winckel, H., Pols, O., and Nelemans, G. (2019). Modelling depletion by re-accretion of gas from a dusty disc in post-AGB stars. *A&A*, 629:A49.

Oomen, G.-M., Van Winckel, H., Pols, O., Nelemans, G., Escorza, A., Manick, R., Kamath, D., and Waelkens, C. (2018). Orbital properties of binary post-AGB stars. *A&A*, 620:A85.

Paczynski, B. (1976). Common Envelope Binaries. In Eggleton, P., Mitton, S., and Whelan, J., editors, *Structure and Evolution of Close Binary Systems*, volume 73, page 75.

Palla, F., Bachiller, R., Stanghellini, L., Tosi, M., and Galli, D. (2000). Measurements of $^{12}C/^{13}C$ in planetary nebulae: Implications on stellar and Galactic chemical evolution. *A&A*, 355:69–78.

Pardo, J. R., Cernicharo, J., Goicoechea, J. R., Guélin, M., and Asensio Ramos, A. (2007). Molecular Line Survey of CRL 618 from 80 to 276 GHz and Complete Model. *ApJ*, 661(1):250–261.

Park, J. A., Cho, S.-H., Lee, C. W., and Yang, J. (2008). A Spectral Line Survey of CRL 2688 in the Range of 85-116 GHz. *AJ*, 136(6):2350–2358.

Pijpers, F. P. and Habing, H. J. (1989). Driving the stellar wind of AGB stars by acoustic waves - Exploration of a simple model. *A&A*, 215(2):334–346.

Pijpers, F. P. and Hearn, A. G. (1989). A model for a stellar wind driven by linear acoustic waves. *A&A*, 209(1-2):198–210.

Pollard, K. R., Cottrell, P. L., Lawson, W. A., Albrow, M. D., and Tobin, W. (1997). RV Tauri stars - II. A spectroscopic study. *MNRAS*, 286(1):1–22.

Pols, O. R. (2004). Effects of binarity on asymptotic giant branch evolution. *Mem. Soc. Astron. Italiana*, 75:749.

Pulliam, R. L., Edwards, J. L., and Ziurys, L. M. (2011). Circumstellar Ion-Molecule Chemistry: Observations of $HCO^+$ in the Envelopes of O-rich Stars and IRC+10216. *ApJ*, 743(1):36.

Quintana-Lacaci, G., Bujarrabal, V., Castro-Carrizo, A., and Alcolea, J. (2007). The chemical composition of the circumstellar envelopes around yellow hypergiant stars. *A&A*, 471(2):551–560.

Raga, A. C., Esquivel, A., Velázquez, P. F., Cantó, J., Haro-Corzo, S., Riera, A., and Rodríguez-González, A. (2009). Mirror and Point Symmetries in a Ballistic Jet from a Binary System. *ApJ*, 707(1):L6–L11.

Ramstedt, S. and Olofsson, H. (2014). The $^{12}CO/^{13}CO$ ratio in AGB stars of different chemical type. Connection to the $^{12}C/^{13}C$ ratio and the evolution along the AGB. *A&A*, 566:A145.

Rowan-Robinson, M. and Harris, S. (1982). Radiative transfer in dust clouds. II. Circumstellar dust shells around early M giants and supergiants. *MNRAS*, 200:197–215.

Sahai, R., Claussen, M. J., Schnee, S., Morris, M. R., and Sánchez Contreras, C. (2011). An Expanded Very Large Array and CARMA Study of Dusty Disks and Torii with Large Grains in Dying Stars. *ApJ*, 739(1):L3.

Sahai, R., Dayal, A., Watson, A. M., Trauger, J. T., Stapelfeldt, K. R., Burrows, C. J., Gallagher, John S., I., Scowen, P. A., Hester, J. J., Evans, R. W., Ballester, G. E., Clarke, J. T., Crisp, D., Griffiths, R. E., Hoessel, J. G., Holtzman, J. A., Krist, J., and Mould, J. R. (1999). The Etched Hourglass Nebula MyCn 18. I. Hubble Space Telescope Observations. *AJ*, 118(1):468–476.

Sahai, R., Morris, M., Sánchez Contreras, C., and Claussen, M. (2007). Preplanetary Nebulae: A Hubble Space Telescope Imaging Survey and a New Morphological Classification System. *AJ*, 134(6):2200–2225.

Sahai, R., Sugerman, B. E. K., and Hinkle, K. (2009). Sculpting an Asymptotic Giant Branch Mass-Loss Envelope into a Bipolar Planetary Nebula: High-Velocity Outflows in V Hydrae. *ApJ*, 699(2):1015–1023.

Sahai, R. and Trauger, J. T. (1998). Multipolar Bubbles and Jets in Low-Excitation Planetary Nebulae: Toward a New Understanding of the Formation and Shaping of Planetary Nebulae. *AJ*, 116(3):1357–1366.

Salpeter, E. E. (1974). Nucleation and growth of dust grains. *ApJ*, 193:579–584.

Sánchez Contreras, C., Alcolea, J., Bujarrabal, V., Castro-Carrizo, A., Velilla Prieto, L., Santander-García, M., Quintana-Lacaci, G., and Cernicharo, J. (2018). Through the magnifying glass: ALMA acute viewing of the intricate nebular architecture of OH 231.8+4.2. *A&A*, 618:A164.

Sánchez Contreras, C., Alcolea, J., Cardoso, R. R., Bujarrabal, V., Castro-Carrizo, A., Quintana-Lacaci, G., Velilla-Prieto, L., and Santander-Garcia, M. (2022). Dissecting the central regions of OH 231.8+4.2 with ALMA: A salty rotating disk at the base of a young bipolar outflow. *A&A*, 665:A88.

Sánchez Contreras, C., Bujarrabal, V., Neri, R., and Alcolea, J. (2000). High-resolution observations at lambda = 3 mm of the OH 231.8+4.2 molecular outflow. *A&A*, 357:651–660.

Sánchez Contreras, C. and Sahai, R. (2004). Physical Structure of the Protoplanetary Nebula CRL 618. II. Interferometric Mapping of Millimeter-Wavelength HCN J = 1-0, $HCO^+$ J = 1-0, and Continuum Emission. *ApJ*, 602(2):960–977.

Sánchez Contreras, C. and Sahai, R. (2012). OPACOS: OVRO Post-AGB CO (1-0) Emission Survey. I. Data and Derived Nebular Parameters. *ApJS*, 203(1):16.

Sánchez Contreras, C., Velilla Prieto, L., Agúndez, M., Cernicharo, J., Quintana-Lacaci, G., Bujarrabal, V., Alcolea, J., Goicoechea, J. R., Herpin, F., Menten, K. M., and Wyrowski, F. (2015). Molecular ions in the O-rich evolved star OH231.8+4.2: $HCO^+$, $H^{13}CO^+$ and first detection of $SO^+$, $N_2H^+$, and $H_3O^+$. *A&A*, 577:A52.

Santander-García, M., Jones, D., Alcolea, J., Bujarrabal, V., and Wesson, R. (2021). The ionised and molecular mass of post-common-envelope planetary nebulae. The missing mass problem. *arXiv e-prints*, page arXiv:2110.15261.

Sasao, T. and Fletcher, A. B. (2009). *Radio Telescope Antennas.*

Schirrmacher, V., Woitke, P., and Sedlmayr, E. (2003). On the gas temperature in the shocked circumstellar envelopes of pulsating stars. III. Dynamical models for AGB star winds including time-dependent dust formation and non-LTE cooling. *A&A*, 404:267–282.

Scicluna, P., Kemper, F., Trejo, A., Marshall, J. P., Ertel, S., and Hillen, M. (2020). Rapid grain growth in post-AGB disc systems from far-infrared and sub-millimetre photometry. *MNRAS*, 494(2):2925–2936.

Shenton, M., Evans, A., and Williams, P. M. (1995). Millimetre continuum observations of low-mass carbon-rich stars. *MNRAS*, 273(4):906–912.

Sobolev, V. V. (1960). The Theory of Stellar Evolution. *Soviet Ast.*, 4:372.

Soker, N. (2002). Formation of Bipolar Lobes by Jets. *ApJ*, 568(2):726–732.

Stahler, S. W. and Palla, F. (2004). *The Formation of Stars.*

Stanghellini, L., Shaw, R. A., and Villaver, E. (2016). Compact Galactic Planetary Nebulae: An HST/WFC3 Morphological Catalog, and a Study of Their Role in the Galaxy. *ApJ*, 830(1):33.

Suárez, O., Gómez, J. F., and Morata, O. (2007). New detections of H2O masers in planetary nebulae and post-AGB stars using the Robledo-70 m antenna. *A&A*, 467(3):1085–1091.

Sukhorukov, A. V. and Leenaarts, J. (2017). Partial redistribution in 3D non-LTE radiative transfer in solar-atmosphere models. *A&A*, 597:A46.

Taylor, G., Carilli, C., and Perley, R. (2000). Book review: Synthesis imaging in radio astronomy ii/astronomical society of the pacific, 1999. *Irish Astronomical Journal*, 27:107.

te Lintel Hekkert, P., Caswell, J. L., Habing, H. J., Haynes, R. F., Haynes, R. F., and Norris, R. P. (1991). 1612 MHz OH survey of IRAS point sources. I. Observations made at Dwingloo, Effelsberg and Parkes. *A&AS*, 90:327.

Tercero, F., López-Pérez, J. A., Gallego, J. D., Beltrán, F., García, O., Patino-Esteban, M., López-Fernández, I., Gómez-Molina, G., Diez, M., García-Carreño, P., Malo, I., Amils, R., Serna, J. M., Albo, C., Hernández, J. M., Vaquero, B., González-García, J., Barbas, L., López-Fernández, J. A., Bujarrabal, V., Gómez-Garrido, M., Pardo, J. R., Santander-García, M., Tercero, B., Cernicharo, J., and de Vicente, P. (2021). Yebes 40 m radio telescope and the broad band Nanocosmos receivers at 7 mm and 3 mm for line surveys. *A&A*, 645:A37.

Tessore, B., Lèbre, A., and Morin, J. (2015). Spectropolarimetric study of the cool RV Tauri star R Scuti. In *SF2A-2015: Proceedings of the Annual meeting of the French Society of Astronomy and Astrophysics*, pages 429–433.

Thomas, J. D., Witt, A. N., Aufdenberg, J. P., Bjorkman, J. E., Dahlstrom, J. A., Hobbs, L. M., and York, D. G. (2013). Geometry and velocity structure of HD 44179's bipolar jet. *MNRAS*, 430(2):1230–1237.

Thompson, A., Moran, J., and Swenson, G. W. (2001). Jr. interferometry and synthesis in radio astronomy.

Thompson, A. R., Moran, J. M., and Swenson, G. W. (2017). *Interferometry and synthesis in radio astronomy*. Springer Nature.

Trammell, S. R. and Goodrich, R. W. (2002). Multiple Collimated Outflows in the Proto-Planetary Nebula GL 618. *ApJ*, 579(2):688–693.

Truong-Bach, Morris, D., and Nguyen-Q-Rieu (1991). CO (J=1-0 and 2-1) mapping of IRC +10216 : a hot core model for the gas kinetic temperature distribution and the mass-loss rate. *A&A*, 249:435.

Ueta, T., Meixner, M., and Bobrowsky, M. (2000). A Hubble Space Telescope Snapshot Survey of Proto-Planetary Nebula Candidates: Two Types of Axisymmetric Reflection Nebulosities. *ApJ*, 528(2):861–884.

Uttenthaler, S. (2007). *Nucleosynthesis and mixing processes in Galactic Bulge AGB stars studied with high-resolution spectroscopy*. PhD thesis, University of Vienna, Department of Astronomy; European Southern Observatory, Germany; Katholieke University of Leuven, Astronomical Institute.

Van de Steene, G. C., van Hoof, P. A. M., Exter, K. M., Barlow, M. J., Cernicharo, J., Etxaluze, M., Gear, W. K., Goicoechea, J. R., Gomez, H. L., Groenewegen, M. A. T., Hargrave, P. C., Ivison, R. J., Leeks, S. J., Lim, T. L., Matsuura, M., Olofsson, G., Polehampton, E. T., Swinyard, B. M., Ueta, T., Van Winckel, H., Waelkens, C., and Wesson, R. (2015). Herschel imaging of the dust in the Helix nebula (NGC 7293). *A&A*, 574:A134.

van der Veen, W. E. C. J. and Habing, H. J. (1988). The IRAS two-colour diagram as a tool for studying late stages of stellar evolution. *A&A*, 194:125–134.

van der Veen, W. E. C. J., Omont, A., Habing, H. J., and Matthews, H. E. (1995). The distribution of dust around Asymptotic Giant Branch stars. *A&A*, 295:445–458.

Van Winckel, H. (2003). Post-AGB Stars. *ARA&A*, 41:391–427.

Van Winckel, H. (2018). Binary post-AGB stars as tracers of stellar evolution. *arXiv e-prints*, page arXiv:1809.00871.

Van Winckel, H., Cohen, M., and Gull, T. R. (2002). The ERE of the "Red Rectangle" revisited. *A&A*, 390:147–154.

Van Winckel, H., Jorissen, A., Exter, K., Raskin, G., Prins, S., Perez Padilla, J., Merges, F., and Pessemier, W. (2014). Binary central stars of planetary nebulae with long orbits: the radial velocity orbit of BD+33°2642 (PN G052.7+50.7) and the orbital motion of HD 112313 (PN LoTr5). *A&A*, 563:L10.

Velázquez, P. F., Steffen, W., Raga, A. C., Haro-Corzo, S., Esquivel, A., Cantó, J., and Riera, A. (2011). Shaping the Red Rectangle Proto-planetary Nebula by a Precessing Jet. *ApJ*, 734(1):57.

Velilla Prieto, L., Sánchez Contreras, C., Cernicharo, J., Agúndez, M., Quintana-Lacaci, G., Alcolea, J., Bujarrabal, V., Herpin, F., Menten, K. M., and Wyrowski, F. (2015). New N-bearing species towards OH 231.8+4.2. HNCO, HNCS, $HC_3N$, and NO. *A&A*, 575:A84.

Velilla Prieto, L., Sánchez Contreras, C., Cernicharo, J., Agúndez, M., Quintana-Lacaci, G., Bujarrabal, V., Alcolea, J., Balança, C., Herpin, F., Menten, K. M., and Wyrowski, F. (2017). The millimeter IRAM-30 m line survey toward IK Tauri. *A&A*, 597:A25.

Verbena, J. L., Bujarrabal, V., Alcolea, J., Gómez-Garrido, M., and Castro-Carrizo, A. (2019). Interferometric observations of SiO thermal emission in the inner wind of M-type AGB stars IK Tauri and IRC+10011. *A&A*, 624:A107.

Vickers, S. B., Frew, D. J., Parker, Q. A., and Bojičić, I. S. (2015). New light on Galactic post-asymptotic giant branch stars - I. First distance catalogue. *MNRAS*, 447(2):1673–1691.

Visser, R., van Dishoeck, E. F., and Black, J. H. (2009). The photodissociation and chemistry of CO isotopologues: applications to interstellar clouds and circumstellar disks. *A&A*, 503(2):323–343.

Völschow, M., Banerjee, R., and Hessman, F. V. (2014). Second generation planet formation in NN Serpentis? *A&A*, 562:A19.

Waelkens, C., Van Winckel, H., Waters, L. B. F. M., and Bakker, E. J. (1996). Variability and nature of the binary in the Red Rectangle nebula. *A&A*, 314:L17–L20.

Wilkins, G. A. (1990). The IAU Style Manual (1989): The Preparation of Astronomical Papers and Reports. *Transactions of the International Astronomical Union, Series B*, 20:Siii–S50.

Wilkins, G. A. (1995). Revision of IAU Style Manual. *Vistas in Astronomy*, 39(2):277.

Williams, J. P. and Cieza, L. A. (2011). Protoplanetary Disks and Their Evolution. *ARA&A*, 49(1):67–117.

Wilson, T. L., Rohlfs, K., and Hüttemeister, S. (2013). *Tools of Radio Astronomy.*

Witt, A. N., Vijh, U. P., Hobbs, L. M., Aufdenberg, J. P., Thorburn, J. A., and York, D. G. (2009). The Red Rectangle: Its Shaping Mechanism and Its Source of Ultraviolet Photons. *ApJ*, 693(2):1946–1958.

Yamamura, I., Matsuura, M., Zijlstra, A. A., and Bedding, T. R. (2003). The extended atmosphere and evolution of the RV Tauri star, R Scuti. In Gry, C., Peschke, S., Matagne, J., Garcia-Lario, P., Lorente, R., and Salama, A., editors, *Exploiting the ISO Data Archive. Infrared Astronomy in the Internet Age*, volume 511 of *ESA Special Publication*, page 153.

Yoon, D.-H., Cho, S.-H., Kim, J., Yun, Y. j., and Park, Y.-S. (2014). SiO and $H_2O$ Maser Survey toward Post-asymptotic Giant Branch and Asymptotic Giant Branch Stars. *ApJS*, 211(1):15.

Zhang, Y. (2017). Molecular studies of Planetary Nebulae. In Liu, X., Stanghellini, L., and Karakas, A., editors, *Planetary Nebulae: Multi-Wavelength Probes of Stellar and Galactic Evolution*, volume 323, pages 141–149.

Zhang, Y., Kwok, S., Nakashima, J.-i., Chau, W., and Dinh-V-Trung (2013). A Molecular Line Survey of the Carbon-rich Protoplanetary Nebula AFGL 2688 in the 3 mm and 1.3 mm Windows. *ApJ*, 773(1):71.

Zhukovska, S. and Gail, H. P. (2008). Condensation of MgS in outflows from carbon stars. *A&A*, 486(1):229–237.

Ziurys, L. M. (2006). Interstellar Chemistry Special Feature: The chemistry in circumstellar envelopes of evolved stars: Following the origin of the elements to the origin of life. *Proceedings of the National Academy of Science*, 103(33):12274–12279.

Ziurys, L. M., Schmidt, D. R., and Woolf, N. J. (2020). Carbon Isotope Ratios in Planetary Nebulae: The Unexpected Enhancement of $^{13}C$. *ApJ*, 900(2):L31.